\documentclass{raa07}
\usepackage{natbib}
\usepackage{amssymb,amsmath}
\usepackage{url}
\usepackage{upgreek}
\bibpunct{(}{)}{;}{a}{}{,}
\usepackage{graphicx,times}

\usepackage{hyperref}
\usepackage{breakurl}

\begin{document}

\title{The FAST Galactic Plane Pulsar Snapshot survey:
  \\
  I. Project design and pulsar discoveries\thanks{\href{http://www.raa-journal.org/docs/Supp/2021Newsonline.pdf}{News and views} on this paper}
}

 \volnopage{{\bf 2021} Vol.~{\bf 21} No.~{\bf 5},~107(38pp)~
   {\small  doi: 10.1088/1674-4527/21/5/107}}
   \setcounter{page}{1}

   \author{
     J. L. Han\inst{1,2,3}, Chen Wang\inst{1,2}, P. F. Wang\inst{1,2}, Tao Wang\inst{1,3}, D. J. Zhou\inst{1,3},
     Jing-Hai Sun\inst{1,2}, Yi Yan\inst{1,3},  \\  Wei-Qi Su\inst{1,3}, Wei-Cong Jing\inst{1,3},  Xue Chen\inst{1,3},
     X. Y. Gao\inst{1,2}, Li-Gang Hou\inst{1,2}, Jun Xu\inst{1,2},
     K. J. Lee\inst{1,4},  \\ Na Wang\inst{6,7},  Peng Jiang\inst{1,2}, Ren-Xin Xu\inst{5}, Jun Yan\inst{1,2},
     Heng-Qian Gan\inst{1,2},
     Xin Guan\inst{1,2},
     Wen-Jun Huang\inst{8},
     Jin-Chen Jiang\inst{4,5},
     Hui Li\inst{1,2},
     Yun-Peng Men\inst{4,5},
     Chun Sun\inst{1,2},
     Bo-Jun Wang\inst{4,5},
     H. G. Wang\inst{8}, \\
     Shuang-Qiang Wang\inst{6,7},
     Jin-Tao Xie\inst{6,7},
     Heng Xu\inst{4,5},
     Rui Yao\inst{1,2},
     Xiao-Peng You\inst{9},
     D. J. Yu\inst{1,2},  \\
     Jian-Ping Yuan\inst{6,7},
     Rai Yuen\inst{6,7},
     Chun-Feng Zhang\inst{4,5} \and
     Yan Zhu\inst{1,2}
   }

   \institute{
     National Astronomical Observatories, Chinese Academy of Sciences,  Beijing 100101, China;\\
     \hspace{2mm} {\it hjl@nao.cas.cn; wangchen@nao.cas.cn}\\
     \and
     CAS Key Laboratory of FAST, NAOC, Chinese Academy of Sciences,
     Beijing 100101, China \\
     \and
     School of Astronomy, University of Chinese Academy of Sciences,
     Beijing 100049, China \\
     \and
     Kavli Institute for Astronomy and Astrophysics, Peking University,
     Beijing 100871, China\\
     \and
     Department of Astronomy, Peking University, Beijing 100871, China \\
     \and
     Xinjiang Astronomical Observatory, Chinese Academy of Sciences,
      Urumqi 830011, China \\
     \and
     Key Laboratory of Radio Astronomy, Chinese Academy of Sciences,
     Nanjing 210008, China \\
     \and
     Department of Astronomy, Guangzhou University,
     Guangzhou 510006, China \\
     \and
     School of Physical Science and Technology, Southwest University,
     Chongqing 400715, China\\
    \vs \no
   {\small Received 2021 January 14; accepted 2021 April 20}
}

   \abstract{Discovery of pulsars is one of the main goals for large
     radio telescopes. The Five-hundred-meter Aperture Spherical radio
     Telescope (FAST), that incorporates an L-band 19-beam receiver
     with a system temperature of about 20~K, is the most sensitive
     radio telescope utilized for discovering pulsars. We designed the
     {\it snapshot} observation mode for a FAST key science project,
     the Galactic Plane Pulsar Snapshot (GPPS) survey, in which every
     four nearby pointings can observe {\it a cover} of a sky patch of
     0.1575 square degrees through beam-switching of the L-band
     19-beam receiver. The integration time for each pointing is 300
     seconds so that the GPPS observations for a cover can be made in
     21 minutes. The goal of the GPPS survey is to discover pulsars
     within the Galactic latitude of $\pm10^{\circ}$ from the Galactic
     plane, and the highest priority is given to the inner Galaxy
     within $\pm5^{\circ}$.  Up to now, the GPPS survey has discovered
     201 pulsars, including currently the faintest pulsars which
     cannot be detected by other telescopes, pulsars with extremely
     high dispersion measures (DMs) which challenge the currently
     widely used models for the Galactic electron density
     distribution, pulsars coincident with supernova remnants, 40
     millisecond pulsars, 16 binary pulsars, some nulling and
     mode-changing pulsars and rotating radio transients (RRATs). The
     follow-up observations for confirmation of new pulsars have
     polarization-signals recorded for polarization profiles of the
     pulsars. Re-detection of previously known pulsars in the survey
     data also leads to significant improvements in parameters for 64
     pulsars.
     The GPPS survey discoveries are published and will be
     updated at {\it\url{http://zmtt.bao.ac.cn/GPPS/}}. %
        }

   \authorrunning{{\it J. L. Han et al.}: The FAST GPPS Survey: I. Project Design and Pulsar Discoveries }            
   \titlerunning{{\it J. L. Han et al.}: The FAST GPPS Survey: I. Project Design and Pulsar Discoveries}  
   \maketitle

%
\section{Introduction}           
\label{sect:intro}

Pulsars are fantastic objects with many physically extreme properties
that are not understood yet. It is widely believed that they are
degenerate stars consisting of neutrons but their internal content and
structure are hitherto not known \citep{of16}. Inside neutron stars,
neutrons {could be }mixed with quarks or strangeons
\citep{afo86,xu03,wdx20}. Such physics can be observationally
constrained by the discovery of pulsars with high masses
\citep[e.g.][]{dpr+10,afn+13,cfr+20} or short spin periods
\citep{bkh+82,hrs+06}. Pulsars have strong magnetic fields, in general
around $10^{12}$\,G, but $10^{13}-10^{14}$\,G in magnetars \citep{kb17} or
$10^{8}-10^{9}$\,G in millisecond pulsars (MSPs). Their emission can be
continuous but may null for some periods \citep[e.g.][]{wmj07,whh+20}
or show a giant pulse occasionally \citep[e.g.][]{jpk+10,msb+19}. The
pulse profiles can have one peak, two peaks or many peaks
\citep[e.g.][]{ran93,lm88,mh04}, which may be generated in different
regions of{ the} magnetosphere and come from different parts of an
emission beam \citep{cor78,tho91a,phi92,md99,hm01}. The profiles can
be highly polarized, even 100\% linearly polarized, and the
polarization angle swings follow S-curves
\citep[e.g.][]{lm88,hdvl09}. The circular polarization has diverse
behaviors, often with reversed senses for the central profile peak but
keeping one sense for the shoulder components \citep{hmxq98}. The
periodic emission of pulses from some MSPs is very stable, with
long-term stability even better than atomic clocks \citep{hgc+20},
while the periods of young pulsars occasionally have a glitch
\citep{lss00a, wmp+00, ywml10, elsk11}.

The discovery of pulsars has been a main task for large
telescopes. Immediately after the first discovery \citep{hbp+68}, many
large radio telescopes, such as the Jodrell bank 76~m telescope
\citep{dhl+68,dls73,cl86,clj+92}, the Molonglo Observatory Synthesis
Telescope \citep{lv71,mlt+78}, the Parkes radio telescope
\citep{rcg+68,kac+73}, the 91~m old Green Bank Telescope
\citep{dth78,dbtb82}, and the Arecibo telescope
\citep[e.g.][]{ht74,cfl+06}, have been used for pulsar hunting sooner
or later. The new Green Bank Telescope \citep[GBT,
][]{HRK2008,blr+13,slr+14} and the Giant Metrewave Radio Telescope
\citep[GMRT, ][]{bcm+16,sjm+18,brs+19} also joined in the efforts
later. It has been very clear that instrument improvements are the key
to finding more pulsars. The first pulsar discoveries were made by the
narrow-band signal recording at that time
\citep[e.g.][]{hbp+68}. Multi-channel signal recording and
de-dispersion were big steps forward \citep[e.g.][]{mlt+78}, leading
to many more new discoveries. The fast Fourier transform (FFT)
technology for long-time data-recording and folding is an another step
\citep[e.g.][]{kac+73,dth78}. Later, cooling receivers at the L-band
(i.e. the radio frequency band around 1.4~GHz) with a very wide
bandwidth certainly made sense for improving the sensitivity, which
is the key to detecting distant weak pulsars
\citep{jlm+92,mld+96,lml+98}.  Multi-beam receivers, first mounted on
the Parkes telescope \citep{swb+96}, can not only speed the survey but
also give a much longer integration time so that an unprecedented
sensitivity can be achieved. The pulsar survey of the Galactic plane
by the Parkes telescope
\citep{mlc+01,mhl+02,kbm+03,hfs+04,fsk+04,lfl+06}, and later extended
to mid to high Galactic latitudes \citep{bjd+06,kjv+10}, led to a
great increase in pulsar discoveries.
Recently, low frequency arrays, which incorporate beam-forming
technology, could survey very large sky areas for pulsars
concurrently, also leading to more new discoveries, such as the pulsar
survey by the Low-Frequency Array \citep[LOFAR,][]{scb+19}.
In addition, advances in candidate selection technology, such as
sorting and scoring \citep[e.g.][]{kel+09,ekl+09}, image pattern
recognition \citep{zbm+14} and classification approaches
\citep{lsc+16}, also help pulsar discoveries. Single pulse search
technology \citep{cm03} led to the discovery of RRATs \citep{mll+06}
and fast radio bursts \citep[FRBs, ][]{lbm+07}. In addition, the
acceleration search technology \citep{ran01,eklk13,ar18} led to
discoveries of many binaries such as those in Terzan 5
\citep[e.g.][]{rhj+05} and even a millisecond pulsar in a triple
system \citep{rsa+14}.

Up to now, there are about 3000 pulsars in the updated version of the
Australia Telescope National Facility (ATNF) Pulsar Catalogue
\citep{mhth05}. Most of them are located in the Galactic disk
\citep{ymw17}, and a few tens are in the Magellanic clouds
\citep{mmh+91,ckm+01,mfl+06,rcl+13}. Historically, big increases in
pulsar numbers always come from dedicated pulsar surveys, mostly on
the Galactic disk, for example, the Molonglo2 survey \citep{mlt+78}
which led to the discovery of 155 pulsars, the Parkes Southern Pulsar
Survey \citep{lml+98} which revealed 101 pulsars and the Parkes
multi-beam pulsar survey
\citep{mlc+01,mhl+02,kbm+03,hfs+04,fsk+04,lfl+06} which discovered
more than 700 pulsars.
In addition, the Fermi satellite has recorded
many Gamma-ray pulsars \citep{aaa+13}.
Currently, many pulsar surveys are still going on, for
example,
(1) the Deep Multibeam Survey Processing (DMSP)\footnote{\it\url{http://astro.phys.wvu.edu/dmb/}} using
Parkes telescope, which has discovered 15 new pulsars;
(2) the SUrvey for Pulsars and Extragalactic Radio Bursts
(SUPERB) using Parkes telescope \citep{kbj+18,sfj+20};
(3) the GBT 350-MHz Drift Scan
Survey\footnote{\it\url{http://astro.phys.wvu.edu/GBTdrift350/}},
which has found 35 new pulsars, including 7 MSPs;
(4) the Green Bank North Celestial Cap (GBNCC) pulsar
survey\footnote{\it\url{http://astro.phys.wvu.edu/GBNCC/}}
\citep{slr+14}, which has
discovered 190 pulsars, including 33 MSPs and 24 RRATs;
(5) the LOFAR Pilot Pulsar Survey (LPPS) and the LOFAR
Tied-Array All-Sky Survey (LOTAAS)\footnote{\it\url{https://www.astron.nl/lotaas/}}, which
have revealed 81 new pulsars \citep{scb+19,MBC+2020};
(6) the GMRT High-resolution Southern Sky Survey for Pulsars and Transients
\citep{bcm+16,brs+19};
(7) the Pulsar ALFA (PALFA)
survey\footnote{\it\url{http://www.naic.edu/~palfa/newpulsars/}} by
the Arecibo telescope, which has discovered 208 pulsars
\citep[e.g.][]{cfl+06,lbh+15};
(8) the Arecibo 327 MHz Drift
Survey\footnote{\it\url{http://www.naic.edu/~deneva/drift-search/}}
which has found 95 pulsars including 10 MSPs and 20
RRATs.
Though some discoveries of these surveys have been published in
papers \citep[e.g.][]{ncb+15}, many of the newly discovered pulsars
are just listed in webpages and not yet formally published.

With many pulsars found already, why do we have to discover more?
What new physics or science can be further uncovered? There have
been many exciting progresses on pulsar science, such as the
discovery of the first pulsars \citep{hbp+68} which revealed an
extreme matter state in the Universe and one kind of end product of
stellar evolution, the discovery of the first binary pulsar
\citep{ht75a} and double pulsars \citep{bdp+03,lbk+04} which have
been used for experimental tests of predictions from the general
relativity of gravitational radiation \citep{tw89,ksm+06} and the
discovery of MSPs \citep{bkh+82,hrs+06} which constrains the
equation of state of matter. For new surveys, the most important
goal is to find exotic pulsars, especially pulsars with a very short
spin period or a short orbital period \citep[see e.g. ][]{sfc+18},
or with a large mass \citep[e.g.][]{dpr+10,cfr+20} or in a
neutron-star black-hole binary \citep[see e.g.][]{fl11}. Discovery
of a pulsar with a spin period of less than 1~ms \citep[e.g. efforts
by][]{hmlq04} or a very massive pulsar \citep[see a complete list of
mass measurements in][]{stjf20} would constrain the compositions of
matters in the interior of neutron stars \citep[e.g.][]{hkw+20} and
have implications for quark or strangeon stars
\citep[e.g.][]{afo86,xu03}; discovery of binary pulsars in various
orbits can make breakthroughs in the current knowledge of stellar
evolution \citep[e.g.][]{jldd15,tkf+17,wl20}. Since at least 10
mergers of stellar binary black holes and one merger of binary
neutron stars were detected by the advanced Laser Interferometer
Gravitational wave Observatory \citep[LIGO,
e.g.][]{aaa+16,aaa+19,aaa+20}, it is expected to discover a
neutron-star black-hole binary in the near future
\citep[e.g.][]{sl18,csh+20}, which can better constrain the theory
of gravity and general relativity better than any known double
neutron-star binaries \citep[e.g.][]{ksm+06,sfc+18} or white-dwarf
neutron-star binaries \citep[e.g.][]{fwe+12}. Even if the parameters
of newly discovered pulsars are within the ranges of known pulsars,
more new distant pulsars can be exploited to explore the
interstellar medium in a large unexplored region of the Galactic
disk such as the farther spiral arms for both the electron density
distribution and interstellar magnetic fields \citep[e.g.][]{hmlq02,
ymw17,hmvd18}; more weaker pulsars in the solar vicinity can
  even probe more detailed properties of the interstellar medium
\citep[e.g.][]{hfm04,xh19}. Discoveries of pulsars in other galaxies
\citep{mmh+91,mfl+06,rcl+13,vmk+20} can probe the intergalactic medium
if the contributions from the Milky Way and the host galaxies can be
accounted for properly.

\begin{figure*}
  \center{\includegraphics[width=0.92\textwidth]{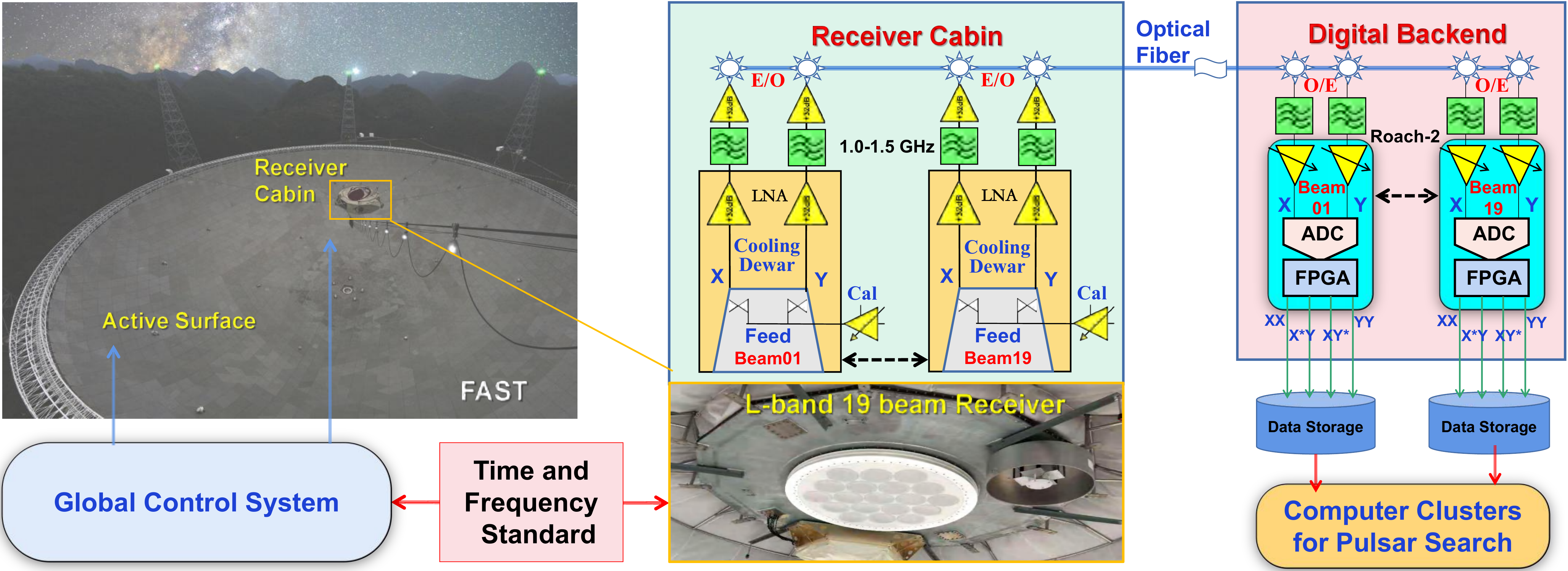}}

  \begin{minipage}{12cm}
    \caption{The observational system of FAST for the GPPS survey. See
      details in text.}
     \label{fig01_system}\end{minipage}
\end{figure*}

The Five-hundred-meter Aperture Spherical radio Telescope
\citep[FAST,][]{nan08,nlj+11} has the largest collecting area for
radio waves, with an aperture of 300~m in diameter. Mounted together
with the 19-beam L-band receiver that has a system temperature of
about 20~K, FAST is currently the most sensitive radio telescope for
discovering pulsars. Construction of FAST was completed in September
2016, and the commissioning was finished in January 2020
\citep{jyg+19}. During the initial commissioning phase when FAST was
not ready to track a source, a drifting survey, the so-called
Commensal Radio Astronomy FAST Survey \citep[CRAFTS,][]{lwq+18},
discovered 124 pulsars as reported on the
webpage\footnote{\it\url{https://crafts.bao.ac.cn/pulsar/}}.  By
viewing the plots produced from the search pipeline, \cite{qpl+19}
discovered the first FAST pulsar. Already published discoveries from
this drift survey include a pulsar with nulling and subpulse drifting
\citep{zlh+19}, an FRB \citep{zll+20} and 11 new pulsars
\citep{clh+20}. Targeted pulsar searches using FAST have also led to
discoveries of a binary millisecond pulsar in the globular cluster M13
\citep{wps+20} and an eclipsing binary millisecond pulsar in M92
\citep{prl+20}. Obviously FAST has great prospects to discover pulsars
\citep{qys+20}. Based on a simulation, \citet{slk+09} estimated that
5200 pulsars can be discovered by FAST, which is probably
over-optimistic according to \citet{lpr+19} and \citet{hh20}, and
there may be much fewer depending on the emission models. The number
of actual discoveries in the Arecibo multibeam pulsar survey
\citep{cfl+06,lbh+15} is much less than what is expected from similar
simulations, mainly because there are much fewer pulsars in the outer
region of the Galactic disk.

With the advantage of its large collection area and sensitive
receivers, FAST is an excellent radio telescope to discover more weak
pulsars, distant pulsars or pulsars in binary systems. Because most
pulsars were born in the Galactic disk and hence the distribution of
pulsars is concentrated in the Galactic plane, we designed the
Galactic Plane Pulsar Snapshot (GPPS) survey, and carried out the
pulsar survey in the Galactic latitude range of $\pm10^{\circ}$ from
the Galactic plane, with the highest priority given to the inner
Galactic disk within the Galactic latitude of $\pm5^{\circ}$. At the
end of 2019, the GPPS project was selected by the FAST science
committee as one of the five FAST high-priority key science projects
for the next few years.  In this paper, we briefly introduce the
observational system and survey strategy in Section~\ref{sect:str},
and introduce the observation status and data-processing in
Section~\ref{sect:Obs}.  Discoveries of new pulsars are presented in
Section~\ref{sect:res}.  Re-detection of known pulsars and improved
parameters are described in Section~\ref{Sect5}. Some perspectives are
discussed in Section~\ref{sect:fin}.

\section{Instruments and survey design}           
\label{sect:str}

The observational system of FAST for the GPPS survey is
illustrated in Figure~\ref{fig01_system}, which consists
of the FAST facility with an active
surface and the mobile receiver cabin controlled by the Global
Control System, the L-band 19-beam receiver, the digital backends,
the data storage system and the computer cluster for data
processing.

\begin{figure}[htp!!!]
  \centering
  \includegraphics[width=75mm]{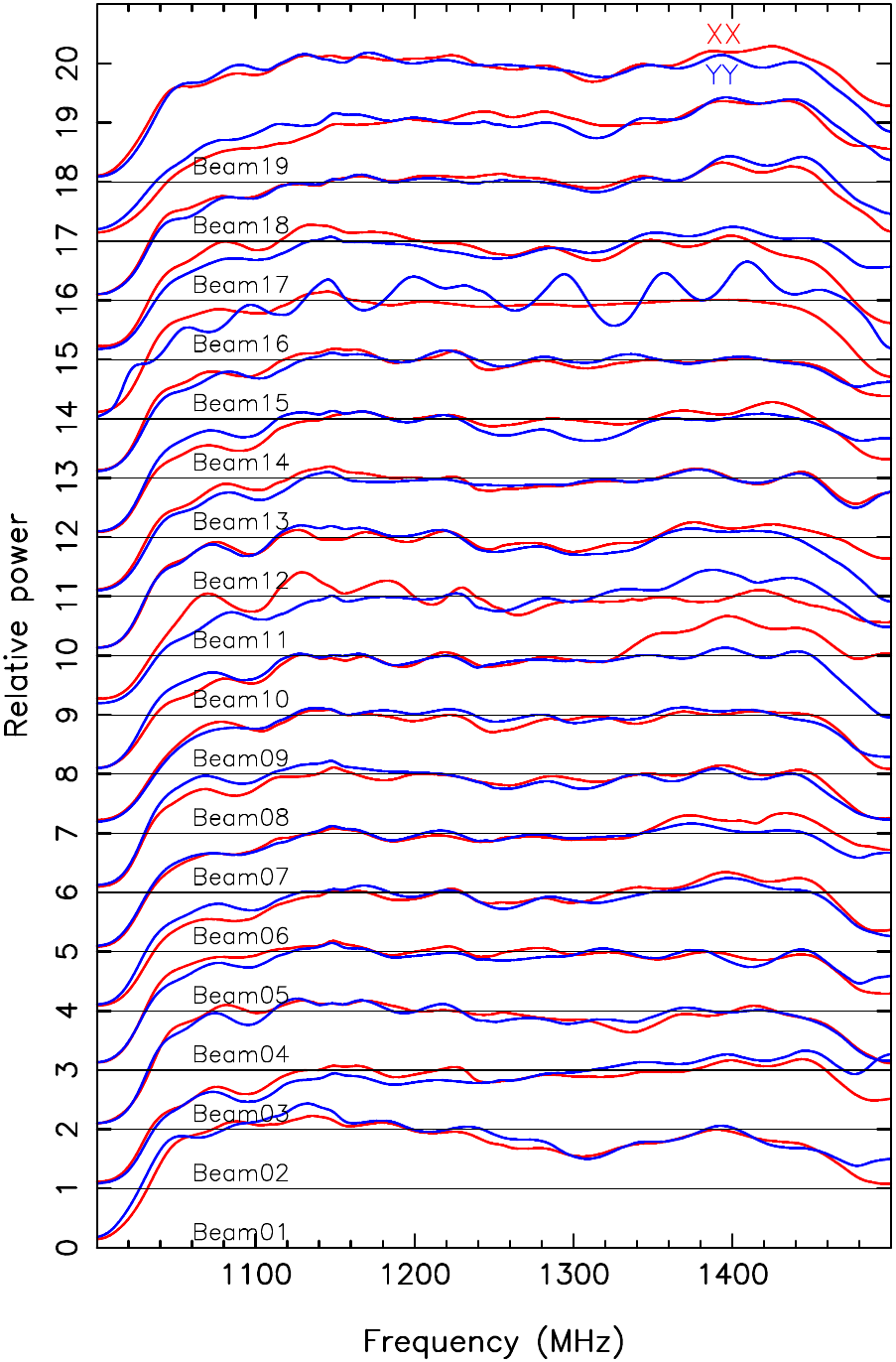}
  \caption{\baselineskip 3.8mm The complete bandpass for the L-band 19-beam receivers,
    represented by the standard deviation of actual data from an
    example of snapshot observations, after narrow-band and wide-band
    RFI are efficiently mitigated by using the ArPLS and SumThreshold
    algorithms \citep[see][for details]{zcl+21}. The bandpass spectrum
    has been measured and is related to the system temperature in some
    parts of the band by \citet{jth+20}, and the curves have a mean
    amplitude around 22{\,}K in general, depending on beam and frequency.}
  \label{fig02_bandpassx19}
\end{figure}

\begin{figure}
   \centering
   \includegraphics[width=80mm,angle=0]{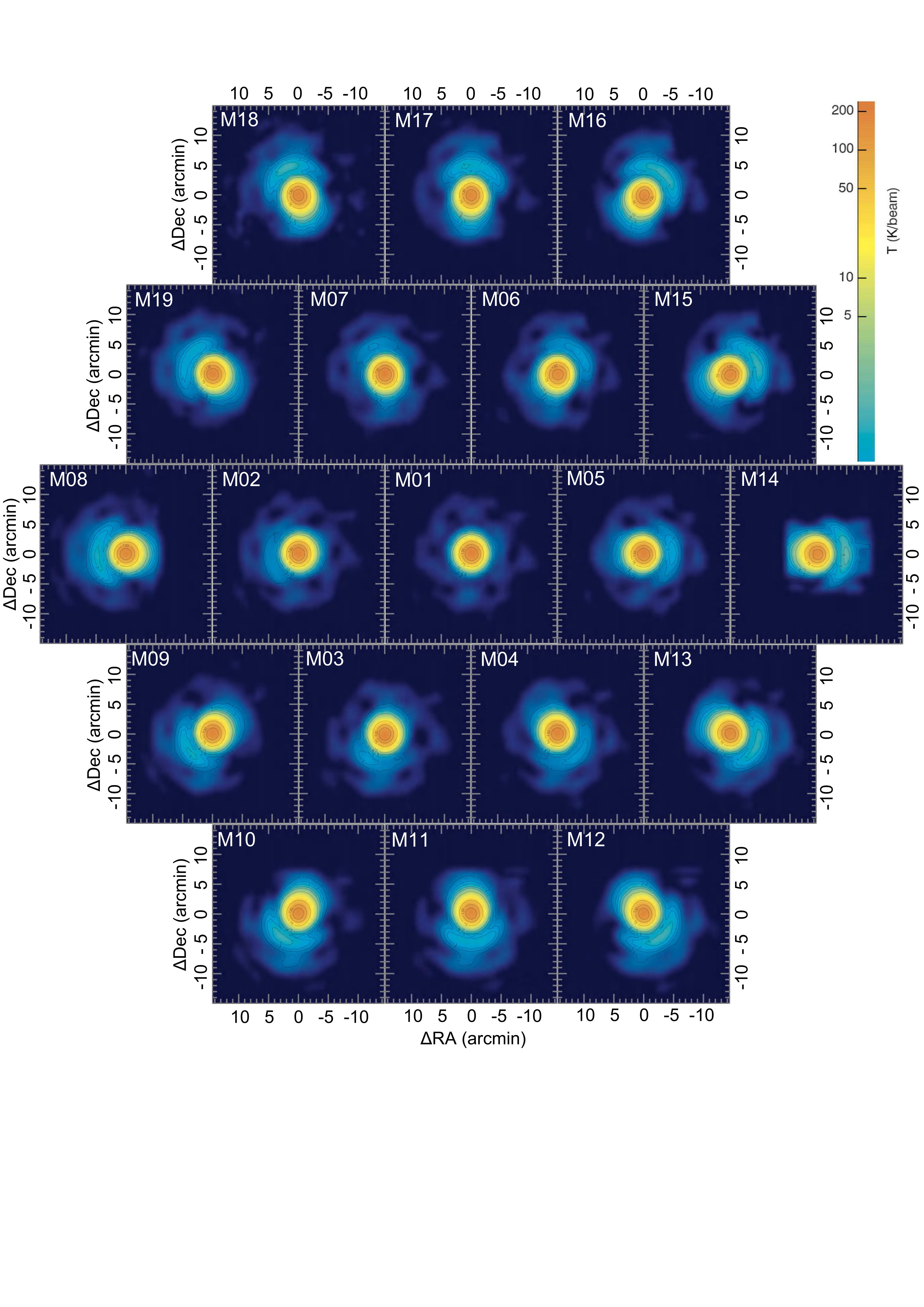}
   \caption{\baselineskip 3.8mm Beam pattern of the L-band 19-beam receiver of
     FAST obtained from the scanning observations of 3C~454.3 and
     integrated in the frequency range of 1050 -- 1450\,MHz. A low
     level of sidelobes is displayed on enhanced log-scales. }
   \label{fig03_19beam}
\end{figure}


\subsection{The FAST and Receivers}

The FAST facility consists of a huge active spherical surface as the
main reflector and the mobile receiver cabin \citep{nlj+11}.  The
reflector has a diameter of 500~m. When FAST observes a source,
according to its location in the sky, part of the spherical surface
with a diameter of 300~m is deformed from the spherical surface to a
quasi-paraboloidal surface and acts as the main reflector, so that
radio waves can focus. This is controlled and monitored by the Global
Control System, and is adjusted by the dynamic-support system,
currently based on the pre-set standard paraboloid model according to
pre-measurements. The deviation from a paraboloid of the so-adjusted
surface, as measured by control nodes, is about 3~mm, so that the
measured efficiency of the 300~m aperture radio telescope is about
60\% at the L-band \citep{jyg+19}. The receiver cabin, which is
suspended and driven by six cables that run over six towers around the
huge reflector, can be moved to any designed focus position with an
accuracy of about 10~mm. When a radio source moves in the sky, the
active part of the main surface and the mobile receiver cabin are
simultaneously adjusted, so that the receiver keeps moving and stays
at the designed focus of the series of paraboloids. This makes the
telescope able to track a radio source properly.

Though the best frequency range to discover distant pulsars several
kpc away in the Galactic disk is about 2 to 3~GHz \citep{xwhh11},
currently the best receiver available in the receiver cabin is the
L-band 19-beam receiver \citep{jth+20}. It formally works in the
frequency range from 1050 to 1450~MHz, but in practice radio signals
from 1000 to 1500~MHz are all received with a degraded band of only 20
-- 40~MHz at each edge, see Figure~\ref{fig02_bandpassx19} for the
bandpasses of the 19-beam receiver. The bandpasses are all very stable
as we have checked the real observational data acquired from the GPPS
survey after narrow- and wide-band radio frequency interference (RFI)
is efficiently mitigated by using the ArPLS and SumThreshold
algorithms \citep[see][for details]{zcl+21}. The RFI is more serious
in daytime, occupying about 10\% of the band; while it is much better
in the nighttime and RFI could sometimes disappear completely. See
figure~24 and figure~25 in \citet{jth+20} to ascertain the RFI
situation in 2019 at the FAST site.

The receiver feeds illuminate a quasi-parabolic area forming an
aperture 300~m in diameter, without spillover radiation due to large
reflectors surrounding that area. The beam size is about $3'$ in the
L-band, varying with frequency in the range of $2.8'$ for 1440~MHz to
$3.5'$ at 1060~MHz \citep[see table 2 in][]{jth+20}. The beams of the
L-band 19-beam receiver are well organized, and the outer beams have
more obvious side lobes and are more aberrated, as illustrated in
Figure~\ref{fig03_19beam}.  For any radio source within 26.4$^{\circ}$
of the zenith, the system temperature is about 20~K and FAST can
track the source by the full illumination area with a full gain {in}
$G${ of} about 16~K~Jy$^{-1}$. The outer beams have a smaller aperture
efficiency, with an additional degradation of about 85\%. Outside the
zenith angle (i.e.  $26.4^\circ<$ZA$<40^\circ$) of the full
illumination, the gain and system temperature further degrade due to
partial illumination.

Therefore, we conclude that three aspects of FAST, the huge surface,
the accurate positioning of the surface and receiver, and the
excellent performance of the L-band 19-beam receiver, make FAST the
most sensitive radio telescope currently in the world to survey
pulsars.

\begin{figure}
   \centering
   \includegraphics[width=70mm,angle=0]{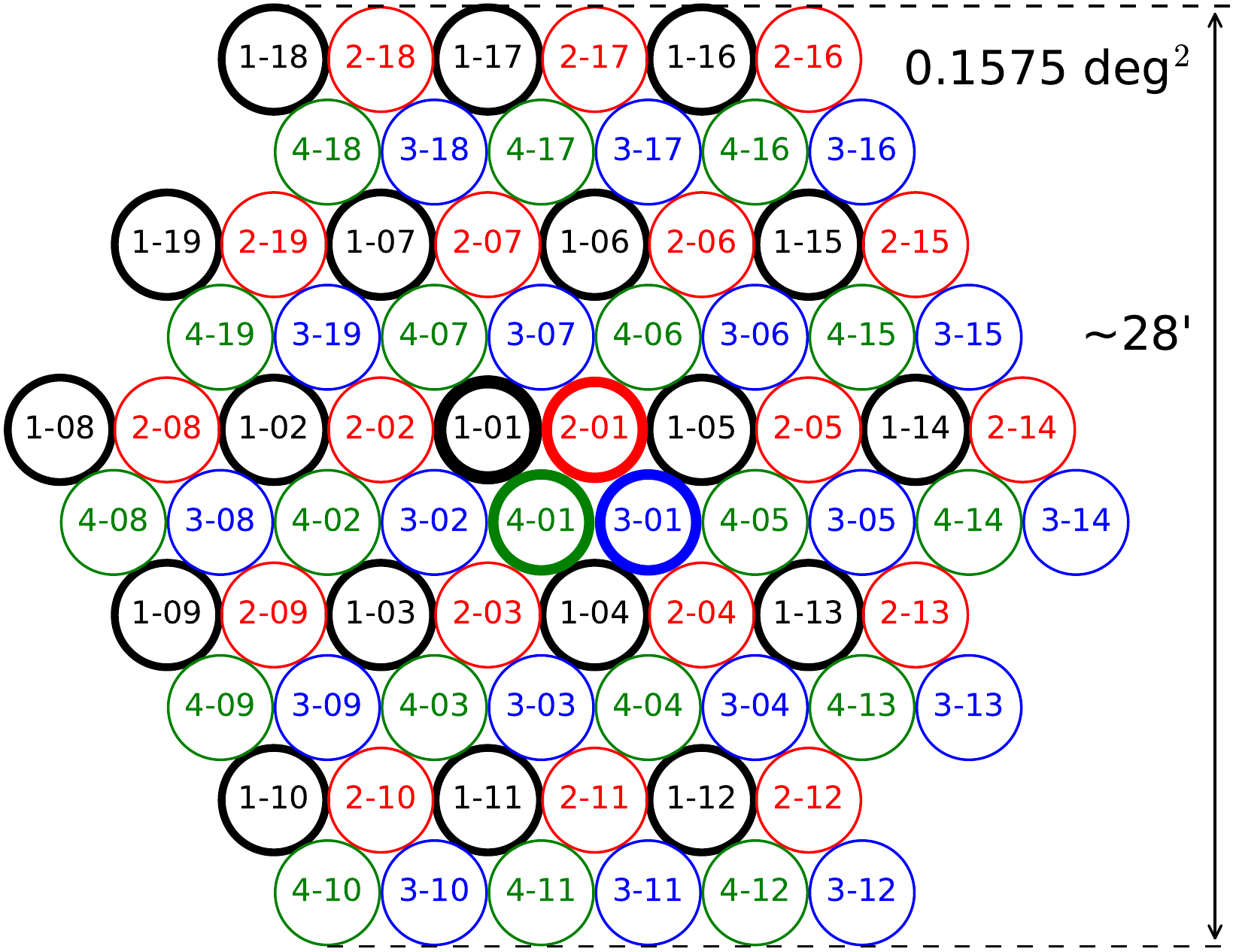}
   \caption{\baselineskip 3.8mm A snapshot made by four pointings via
     three-beam switchings of the 19 beams of the L-band 19-beam
     receiver of FAST which can survey a {\it cover} of a sky area of
     0.1575 square degrees.}
   \label{fig04_4x19beam}
\end{figure}

\subsection{The Snapshot Observation Mode}

As mentioned above, the pointing changes of FAST to a new source are
realized by adjusting the main surface according to the pre-set
standard paraboloid model and by positioning the receiver cabin at
the designed focus. This takes a few minutes for the mechanical
movements and cable stabilization. The pointing accuracy of FAST
is better than $8''$ \citep{jth+20}, according to real measurements
for strong sources.

\begin{figure}
  \centering
  \includegraphics[width=70mm]{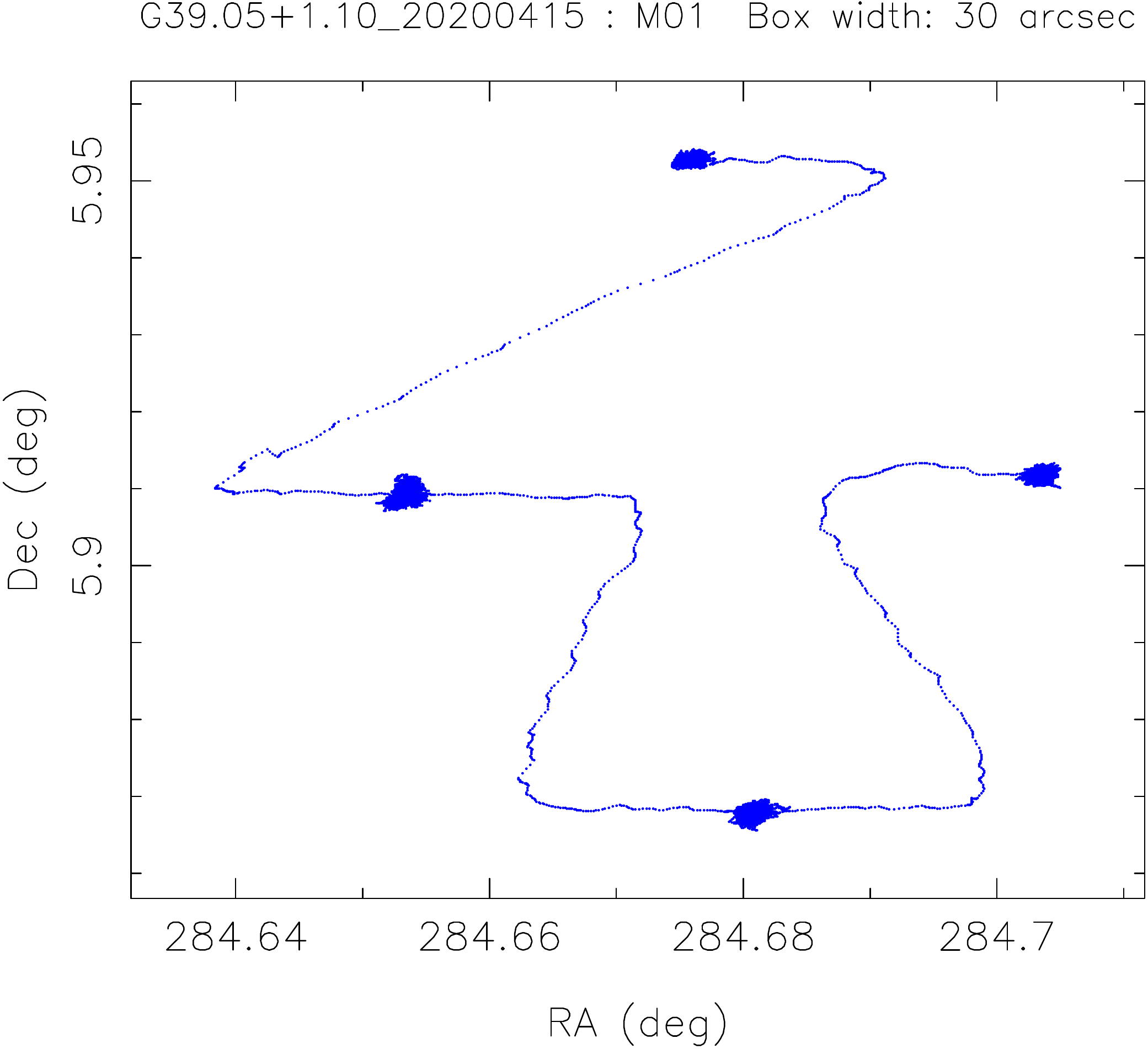}
  \caption{\baselineskip 3.8mm An example of the pointing trajectory in the sky of central
    beam of the L-band 19-beam receiver during GPPS observations
    of a cover.}
  \label{fig05_track}
\end{figure}

\begin{figure}
   \centering
   \includegraphics[width=70mm]{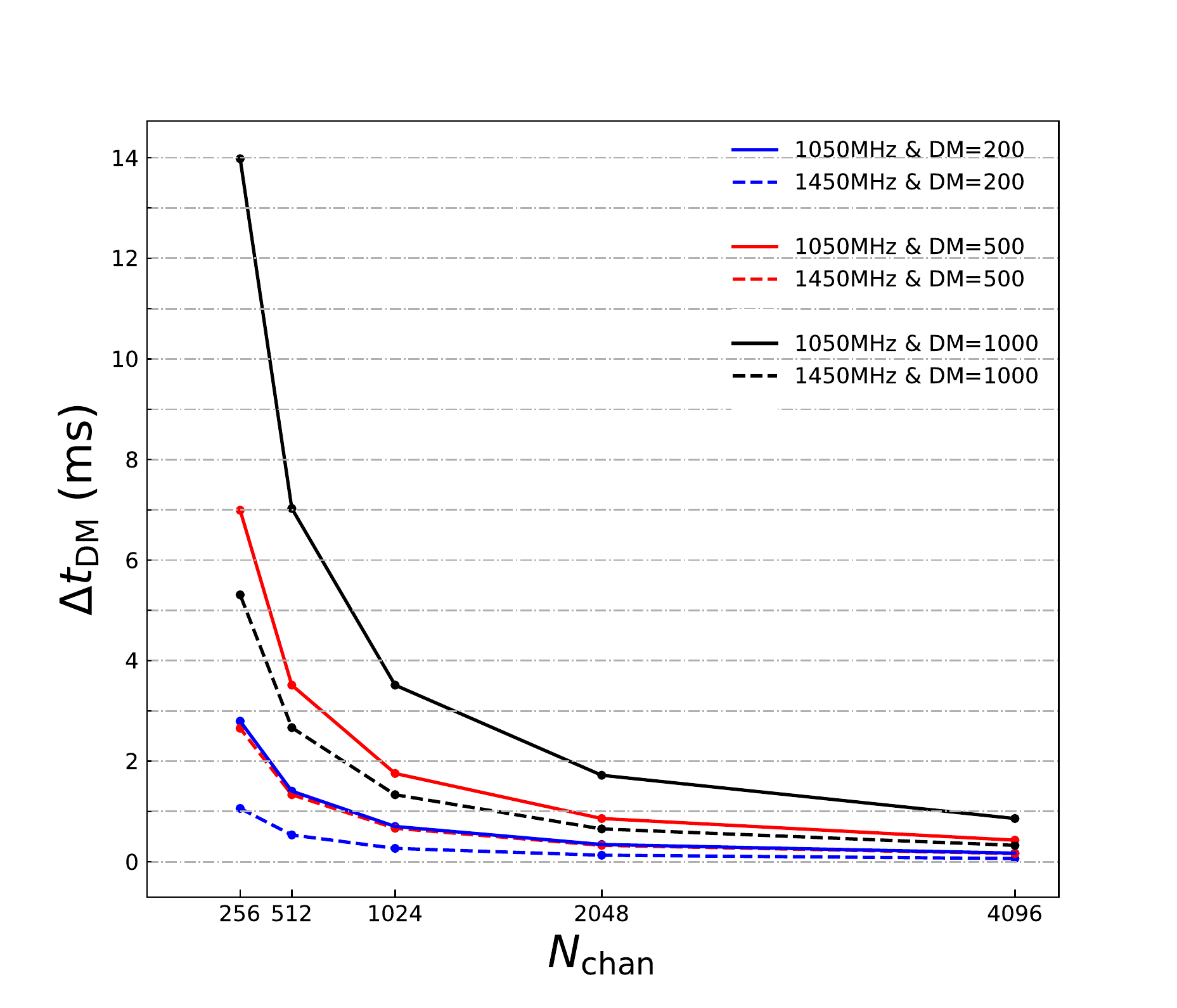}
   \includegraphics[width=70mm]{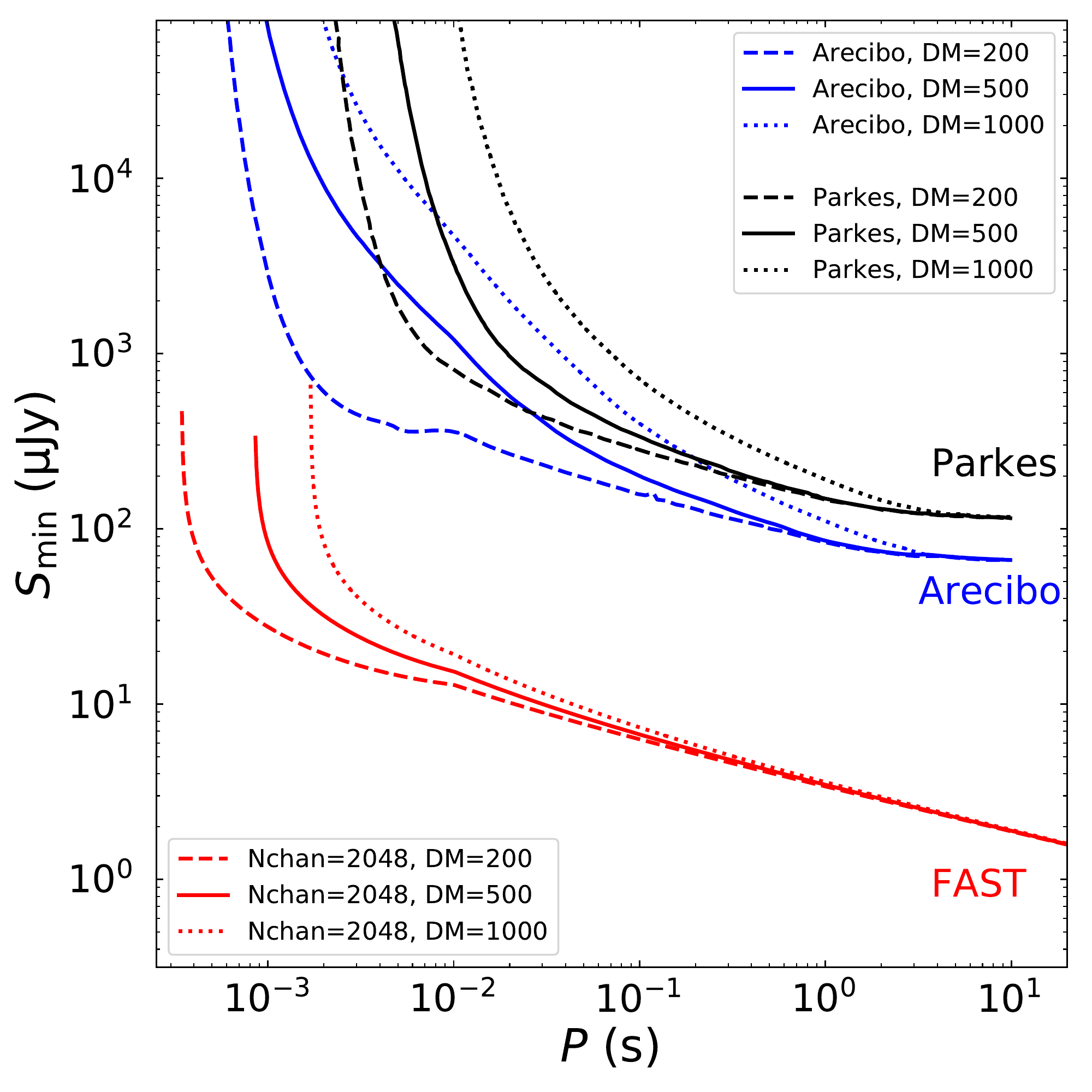}
   \caption{\baselineskip 3.8mm {\it Top}: the channel DM smearing time of two frequency
     channels at $\nu=$~1050~MHz and 1450~MHz, calculated by $83
     (DM/100)(\nu/100~{\rm MHz})^{-3} \Delta\nu$ (in ms) for three
     example DM values. Here $\nu$ and $\Delta\nu$ have the unit of MHz,
     and $\Delta\nu$ is the channel bandwidth. {\it Bottom}: the
     10$\sigma$ detection sensitivity curves of the GPPS survey with
     parameters listed in Table~\ref{gppspara} and an assumed 10\%
     pulse width for a pulsar with a period $P<10$~ms but declining
     with $P^{-1/2}$ when the period $P>10$~ms.  The scattering time of
     $3.2 (DM/100)^{3.5}(\nu/100~{\rm MHz})^{-5} \Delta\nu$ (in ms)
     and the channel DM smearing time as described above has been
     calculated at $\nu=$~1100~MHz and combined with the sampling time
     for drawing the sensitivity curves. Compared to the survey
     sensitivity curves of the Parkes multibeam survey \citep{mlc+01}
     and the Arecibo PALFA survey \citep{cfl+06}, the FAST GPPS survey
     is super sensitive, especially for MSPs (see Fig.~\ref{fig13_fluxP}
     below).}
   \label{fig06_sensiDM}
   \end{figure}

In the receiver cabin, the L-band 19-beam receiver is mounted on a
Stewart manipulator which acts as a stabilizer depressing the
vibration caused by flexible cables \citep{jyg+19}. In the limited
range of a few mm, the receiver can be finely adjusted to the desired
position and tilted to a designed angle within only a few
seconds. Considering this very fine feature which is enough for
adjusting telescope pointing for several arcminutes, we designed the
snapshot observation mode (see Fig.~\ref{fig04_4x19beam}) in four steps: (1)
first a normal pointing is made to a desired position in the sky,
working as pointing No.1, and then tracking observation is carried
out for some time, e.g. 5 minutes for the GPPS survey. All data from
the 19-beam receiver can be recorded; (2) second, the central beam
(certainly also other beams) is offset by $3'$ to the right, working
as pointing No.2, and then tracking observations and
data-recording can be made in the same mode; (3) then it is offset by
$3'$ to the lower-right and tracking, working as pointing No.3; and
(4) finally offset by $3'$ to the left and tracking, working as
pointing No.4. With these four points, a {\it cover} of the sky patch
of about 0.1575 square degrees can be surveyed. An example for the
trajectory of the central beam in the sky plane during observations of
a cover is displayed in Figure~\ref{fig05_track}. The beam-switching between these
pointings can be realized within a few seconds. To ensure the accuracy
of upcoming tracking, 20 seconds are given for each
beam-switching. The GPPS observations for a cover therefore cost only
21 minutes, including the four tracking observations each with an
integration time of 5 minutes plus the three beam-switches each lasting 20
seconds. The snapshot observation mode enables high efficiency for the
usage of telescope time. From the real observation data, we found
that the stability of FAST pointings is better than $4''$.

For the snapshot survey of the Galactic plane, the L-band 19-beam
receiver is rotated to be parallel with the Galactic plane, i.e.,
beams No.08, No.02, No.01, No.05 and No.14 (see
Fig.~\ref{fig04_4x19beam}) are aligned with the Galactic plane, so
that beams observed in different covers can be easily connected
continuously.

\begin{figure*}
   \centering
   \includegraphics[width=150mm]{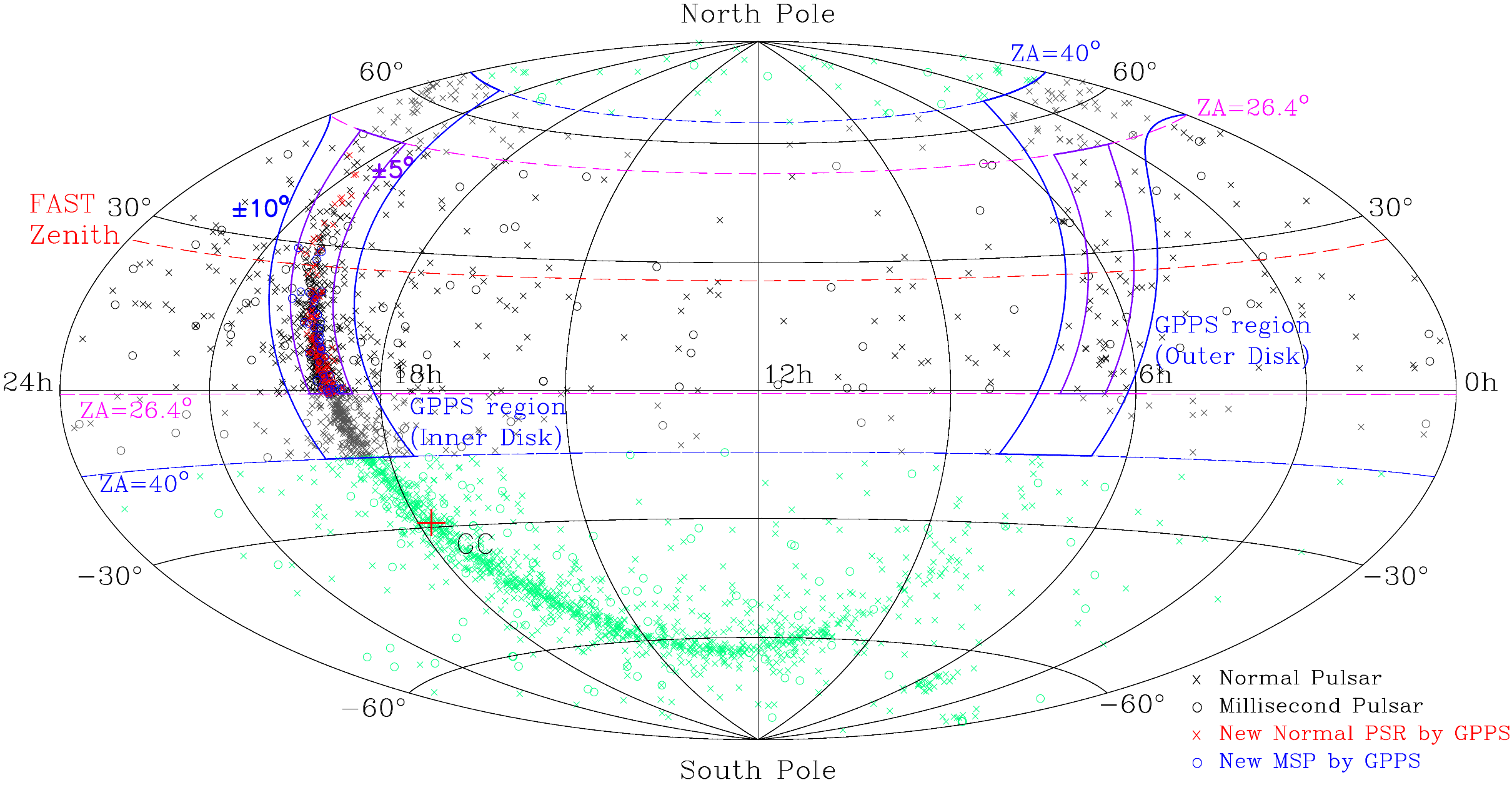}
   \caption{\baselineskip 3.8mm The sky area in the Galactic latitude of $\pm10^{\circ}$
     of the Galactic disk accessible by FAST is planned for the GPPS survey
     as outlined by the blue lines. High priority is given to the area
     of the Galactic latitude of $\pm5^{\circ}$ in the inner disk.
     Observations within the zenith angle ZA$<26.4^{\circ}$ are now
     carried out with the full gain of FAST.}
   \label{fig07_sky}
\end{figure*}

\begin{figure*}
   \centering
   \includegraphics[width=160mm]{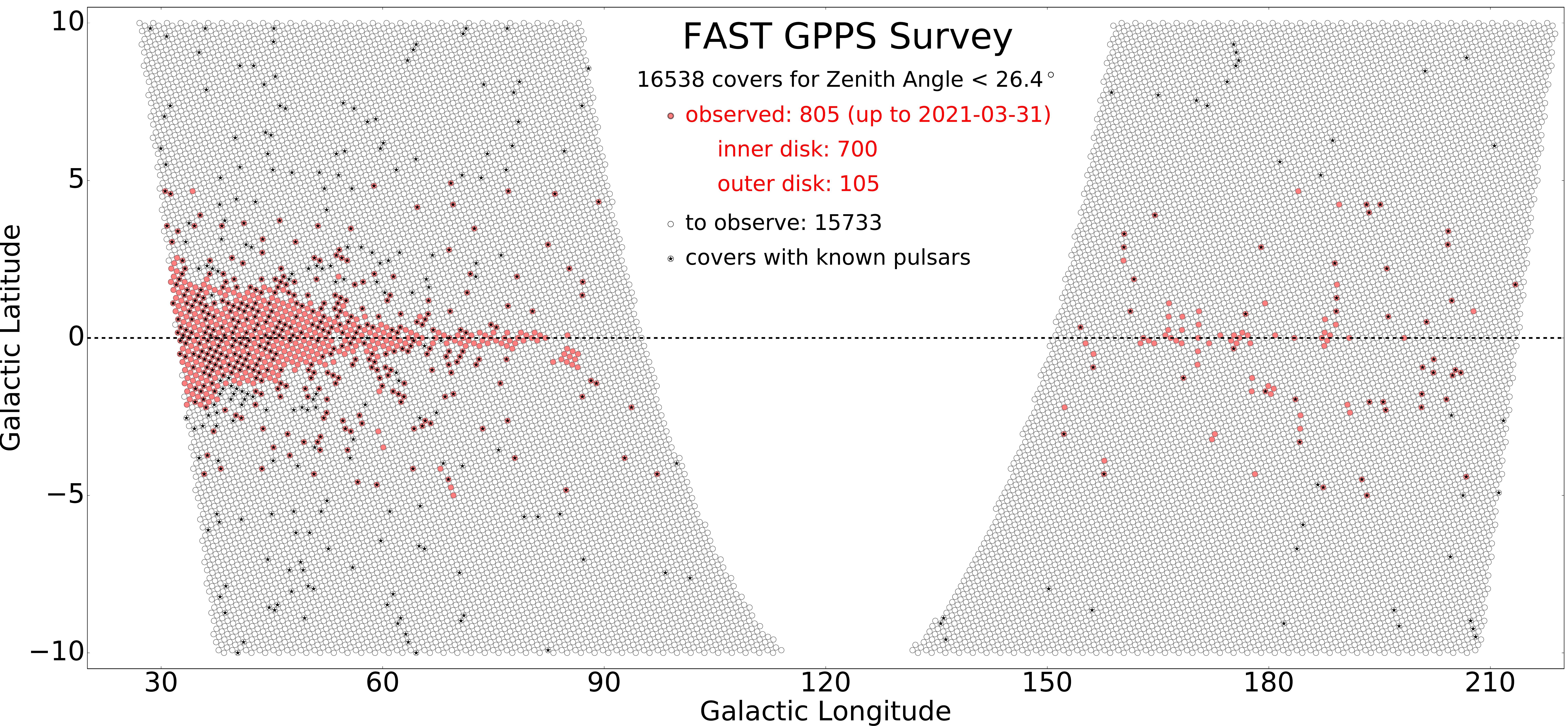}
   \caption{\baselineskip 3.8mm The distribution of covers for the GPPS survey with the
     full gain of FAST, i.e.,  observable within the zenith angle ZA
     $<26.4^{\circ}$. Each small circle is a cover for 76
     beams. Observed covers are marked in red, and covers with any
     known pulsars are marked with a black star inside.}
   \label{fig08_covers}
   \end{figure*}

   \begin{figure*}
   \centering
   \includegraphics[width=152mm]{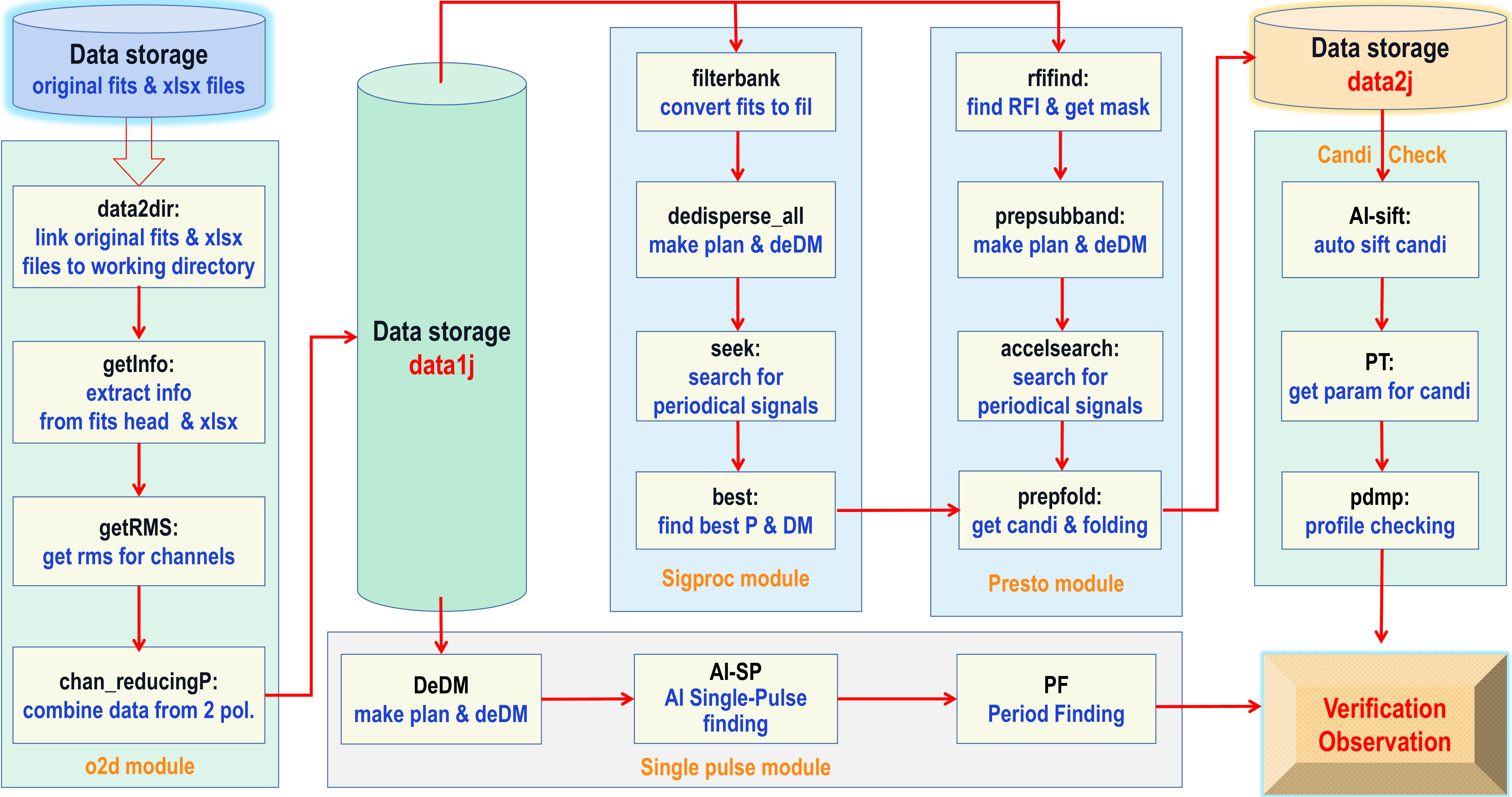}

   \begin{minipage}{6cm}
   \caption{The flowchart for data processing.}
   \label{fig09_process}\end{minipage}
   \end{figure*}

The telescope control parameters and the position of the phase-center
of the central beam receiver are recorded in an xlsx file, with
proper time stamps, forming part of the metadata of FAST
operations. Combining these metadata data and the original fits data
files recorded by the digital backends (see below), one can make a
proper fits file for each beam of every pointing.

In the open risk-share observation session of FAST in 2019, we
successfully designed and tested the snapshot observation mode in
March 2019.

\subsection{The Digital Backends and Data Storage}

The radio signals from two orthogonal linear polarizations $X$ and
$Y$, from each beam of the L-band 19-beam receiver, are amplified and
filtered, and then transferred to the data-recording room via optical
fibers connected with an optical transmitter and an optical receiver
(see Fig.~\ref{fig01_system}). The voltage signals are filtered
against interference and aliasing, and then sampled and channelized by
using a pulsar digital-backend developed on Re-configurable Open
Architecture Computing Hardware-2 (ROACH2). After self-correlation and
cross-correlation and also accumulation in the field-programmable gate
array (FPGA) board, the power data of the four polarization channels
of $XX$, $X^*Y$, $XY^*$ and $YY$ are produced. These data can be
recorded with selected channel numbers, such as 4096, 2048, 1024 or
even 512 channels, with accumulation times of 98.304~$\upmu$s,
49.152~$\upmu$s, 24.563~$\upmu$s and 12.281~$\upmu$s for four
polarization channels or just two polarization channels (i.e., $XX$
and $YY$).

Specifically, for the GPPS survey observations, the digital backends
work for recording data from 19 beams. The accumulation time, i.e.,
the sampling time of each channel, is taken as being $\tau_{\rm
  sampling} = 49.152~\upmu$s. The $XX$ and $YY$ data are recorded for
2048 channels by default, though the data from 4096 channels with four
polarizations were always recorded in the pilot phase of the GPPS
survey before February 2020. These data are stored in fits files for
each beam with proper timestamps, and each original fits file stores
data for 12.885~s for four polarization channels or for 25.770~s for
two polarization channels.

In addition, the amplified voltage signals $X$ and $Y$ are split and
fed to a number of digital spectrometers, which can simultaneously
work for spectral line observations for all 19 beams. We record the
spectral data in the whole band between 1000 -- 1500~MHz by using
1024~K channels and an accumulation time of 1~s as default, so
that anyone who is interested in the spectral lines of the
interstellar medium in the Galactic disk can have spectral data for
free.

\begin{table}
  \centering
\caption{Relevant Parameters for the GPPS Survey}
  \label{gppspara}
\fns
 \begin{tabular}{ll}
\hline
Parameter  & {V}alue  \\
\hline
FAST location: latitude   &  N~$25.647^{\circ}$  \\
FAST location: longitude  &  E~$106.856^{\circ}$ \\
Effective aperture diameter &  300~m         \\
Beam size                 &  $\sim3'$      \\
Aperture efficiency       &  60\%         \\
Beam {n}umber for a cover   & $4\times19$    \\
T$_{\rm sys}$+T$_{\rm sky}$   &  $\sim$25~K         \\
Telescope gain            &  $\sim$16~K\,Jy$^{-1}$ \\
Sky region to obs.        & $GB<10^{\circ}$ \\
Zenith angle limit for a full gain   &   $<26.4^{\circ}$     \\
Covers to obs.            & 16{\,}538  \\
Obs. freq. range          &  1.0--1.5 GHz        \\
Freq. channel number      &  2048               \\
Freq. resolution          &  0.24414 MHz        \\
Effective useful bandwidth &  450 MHz            \\
Polarization channels          & $XX$+$YY$          \\
Sampling time $\tau_{\rm sampling}$  &  49.152~$\upmu$s      \\
Survey integration time t$_{\rm survey~obs}$          &  300~s              \\
Verification integration time t$_{\rm verify~obs}$          &  900~s              \\
\hline
\end{tabular}
\end{table}

\subsection{The Strategy for FAST GPPS Survey}

By referencing the parameters for the GPPS survey listed in
Table~\ref{gppspara} we can calculate the sensitivity for detection
of pulsars with an assumed pulse-width of 10\% of pulsar periods, as
plotted in Figure~\ref{fig06_sensiDM}.  This is the most sensitive
pulsar survey up to now, the first down to a level of $\upmu$Jy.

FAST is located at the latitude of N~$25.647^{\circ}$, the longitude
of E~$106.856^{\circ}$, and can observe the sky area in the
declination range of $-0.9^{\circ}<$Dec$<52.2^{\circ}$ with full gain
in the zenith angle of $<26.4^{\circ}$. The Galactic plane between the
Galactic longitude of about $GL=30^{\circ}$ to 90$^{\circ}$ in the
inner Galaxy and between about $GL=145^{\circ}$ to 215$^{\circ}$ in
the outer disk are visible by FAST (see Fig.~\ref{fig07_sky}). We
currently plan to survey the Galactic disk within the Galactic
latitude of $\pm10^{\circ}$ from the Galactic plane, 76 beams per
cover and 16\,538 covers in the first stage. The highest priority is
given to 4024 covers in the area for Galactic longitude of
$\pm5^{\circ}$ in the inner Galactic disk (see
Fig.~\ref{fig08_covers}).

After the survey for the area accessible within the zenith angle of
$26.4^{\circ}$ is finished, the survey will extend to the area observable
within the zenith angle of $\pm40^{\circ}$, because the control system
will become more sophisticated and the spillover radiation is better
screened in near future.

\section{Observations and Data processing}
\label{sect:Obs}

\subsection{The GPPS Survey Observations}

After initial tests for the snapshot observation mode and data file
storage in March 2019, we successfully carried out a pilot project in the
FAST shared-risk open session in 2019 targeting the outer Galactic
disk. Observations in 2019 have data recorded in ``the standard
format'' for 4096 channels, 19 beams and four polarization channels
($XX$, $X^*Y$, $XY^*$, $YY$) for signals in the radio band from 1000
-- 1500~MHz. To save disk space, since February 2020, only two
polarization channels ($XX$ and $YY$) have been recorded for 2048 channels
for the frequency range. A cover is named by the pointing position in
Galactic coordinates of the central beam and observation date,
such as G184.19--3.30\_20190422. See Figure~\ref{fig08_covers} for the
distribution of covers and observation status, which are updated often
on the GPPS webpage\footnote{\it\url{http://zmtt.bao.ac.cn/GPPS/}}.

To scale the flux of discovered pulsars, at the beginning and end
of each observation session lasting about 2 to 3 hours, data are recorded
for 2 minutes with calibration signals on-off (1~second each) on the
pointing position. The cal-signals and the receiver gain are found to
be very stable in general.

   \begin{figure}
   \centering
   \includegraphics[width=80mm]{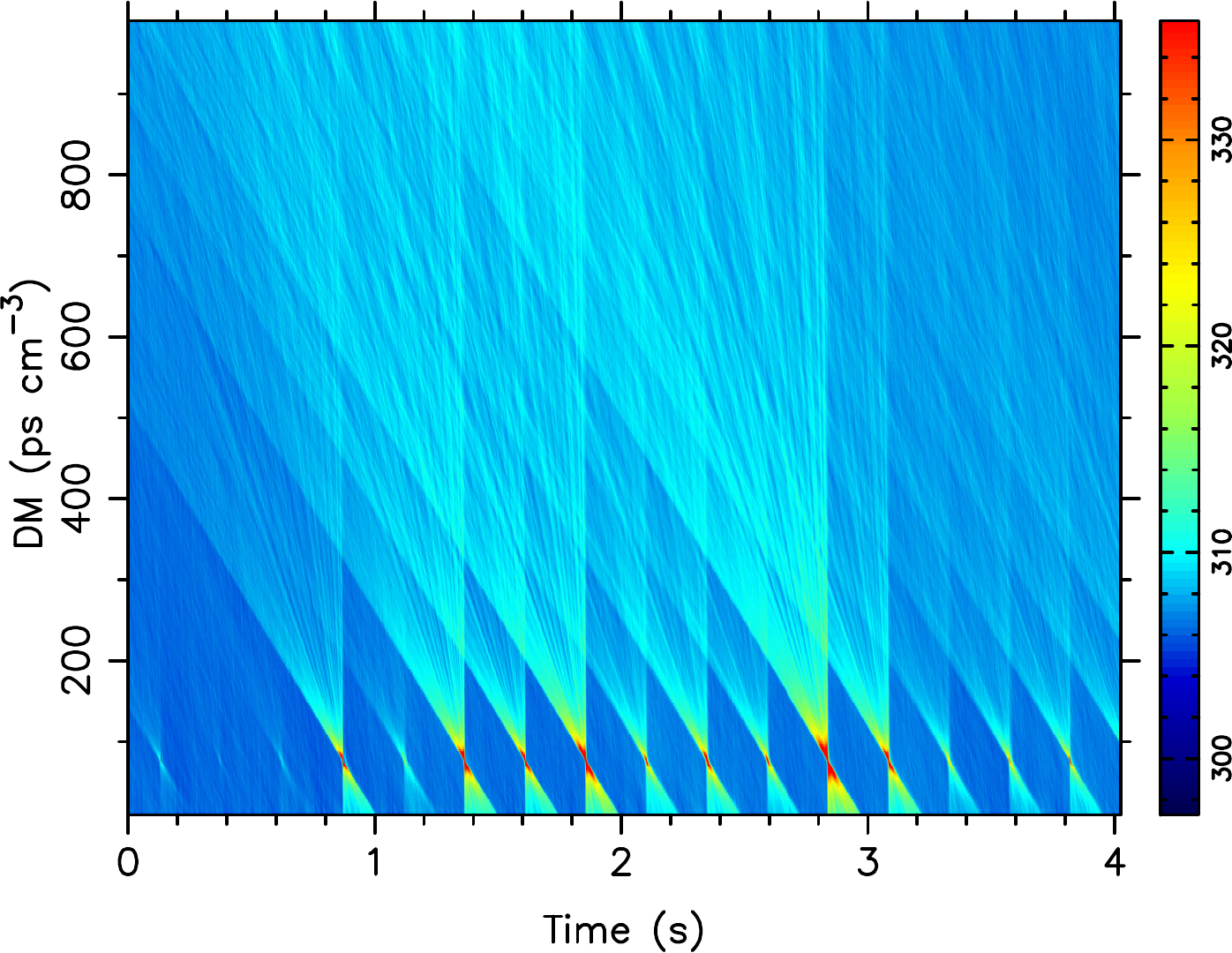}
   \caption{\baselineskip 3.8mm An example of DM-time plots of a strong pulsar, which show
     a large number of peaks as fake pulsar candidates in a very wide
     range of DM values. This example is taken from data of beam No.08
     at pointing No.2 of the GPPS survey cover of
     G184.19--3.30\_20190422, where a strong pulsar, PSR J0543+2329
     with DM=77.7~cm$^{-3}$pc, is in the beam.}
   \label{fig10_DMt}
   \end{figure}

\subsection{Data Processing}

For pulsar search, there are three major steps involved in processing
the data: (1) data preparing; (2) pulsar search; (3) results
evaluation. See Figure~\ref{fig09_process} for a flow-chart of data
processing.

\subsubsection{Data preparing}

A series of original fits files with a given size are recorded for
each beam and stored in the repository with proper timestamps. As
mentioned above, for each observation of a source, there is also an
xlsx file that archives the control parameters and the focus cabin's
position.  It is necessary to combine these metadata with fits files
for pulsar search.

We first make a directory for the cover and link all original
recovered data files in the repository for the snapshot observations
including the fits file and the xlsx file, or tracking observations
for verification, into a sub-directory called {\it ori}. The
calibration files will be linked into a sub-directory called {\it
  cal}. The relevant information from fits files is extracted in the
order of being time-stamped with Modified Julian Date (MJD), as
is the position information on the feed of the central beam from
the xlsx file. These metadata are matched in MJD so that the
coordinates of the telescope pointings of snapshot observations can be
calculated (see Fig.~\ref{fig05_track}). Based on such a list of
coordinates, we can split sub-integration data of different original
fits files for different pointings of every beam in a cover.

\begin{table*}
\centering
\begin{minipage}[]{160mm}
  \caption{A List of Pulsars Discovered by the GPPS Survey (See {\it\url{http://zmtt.bao.ac.cn/GPPS/}} for Updates)}
 \label{gppsPSRtab1}
\end{minipage}

\fns
\setlength{\tabcolsep}{3.5pt}
 \begin{tabular}{cccrccccrrr}
\hline\noalign{\smallskip}
 Name$^*$    &  gpps No.  &  Period    & DM      &  RA(2000)   &   D{ec}(2000)  &   GL  &    GB  &  $S_{\rm 1.25GHz}$  & $D_{\rm NE2001}$ & $D_{\rm YMW16}$ \\
             &            &   (s)   &(cm$^{-3}$pc)&  (hh:mm:ss) & ($\pm$dd:mm) &  ($^{\circ}$) & ($^{\circ}$) & (${\upmu}$Jy) & (kpc)  & (kpc)      \\
 \hline
J1901$+$0659g & gpps0001 & 0.07573 & 126.2 & 19:01:24.8 & $ +$06:59 &  40.2854 & $ +$0.9763 &    51.8 &  3.7 &   3.8  \\
J1924$+$1923g & gpps0002 & 0.68924 & 386.7 & 19:24:20.5 & $ +$19:23 &  53.8632 & $ +$1.7939 &    54.7 & 12.0 &   9.8  \\
J1904$+$0823g & gpps0003 & 1.50773 &  60.4 & 19:04:43.8 & $ +$08:23 &  41.9112 & $ +$0.8895 &    22.0 &  2.8 &   1.8  \\
J1928$+$1915g & gpps0004 & 0.97435 & 201.0 & 19:28:59.6 & $ +$19:15 &  54.2628 & $ +$0.7624 &     5.1 &  6.6 &   4.6  \\
J1838$+$0044g & gpps0005 & 2.20317 & 229.6 & 18:38:10.8 & $ +$00:44 &  32.0861 & $ +$3.2910 &    72.9 &  5.6 &   6.9  \\
J1924$+$1932g & gpps0006 & 0.38886 & 280.3 & 19:24:49.0 & $ +$19:32 &  54.0482 & $ +$1.7661 &    20.8 &  8.9 &   6.6  \\
J1925$+$1629g & gpps0007 & 0.00411 & 214.1 & 19:25:10.1 & $ +$16:29 &  51.4006 & $ +$0.2470 &    87.8 &  6.7 &   5.1  \\
J1905$+$0656g & gpps0008 & 2.51165 &  23.0 & 19:05:45.3 & $ +$06:56 &  40.7393 & $ -$0.0019 &    27.6 &  1.6 &   1.0  \\
J1904$+$0852g & gpps0009 & 0.00619 & 195.1 & 19:04:55.0 & $ +$08:52 &  42.3646 & $ +$1.0716 &    36.0 &  5.3 &   5.9  \\
J1857$+$0214g & gpps0010 & 0.33389 & 986.3 & 18:57:08.2 & $ +$02:14 &  35.5819 & $ -$0.2422 &    56.2 & 14.8 &   7.8  \\
J1947$+$2011g & gpps0011 & 0.00817 & 127.4 & 19:47:47.0 & $ +$20:11 &  57.2512 & $ -$2.6299 &     9.8 &  5.1 &   4.3  \\
J1917$+$1259g & gpps0012 & 0.00563 & 117.0 & 19:17:29.0 & $ +$12:59 &  47.4303 & $ +$0.2353 &    36.1 &  4.3 &   3.6  \\
J1930$+$1403g & gpps0013 & 0.00321 & 150.5 & 19:30:17.9 & $ +$14:03 &  49.8431 & $ -$2.0053 &    94.1 &  5.4 &   4.7  \\
J1852$+$0056g & gpps0014 & 1.17779 & 905.7 & 18:52:14.0 & $ +$00:56 &  33.8644 & $ +$0.2543 &    37.9 & 11.7 &   7.2  \\
J1859$+$0430g & gpps0015 & 0.33629 & 783.8 & 18:59:10.4 & $ +$04:30 &  37.8266 & $ +$0.3389 &    28.0 & 11.9 &   8.9  \\
J1900$+$0405g & gpps0016 & 0.07238 & 634.4 & 19:00:39.6 & $ +$04:05 &  37.6202 & $ -$0.1839 &    32.5 &  9.4 &   6.7  \\
J1906$+$0822g & gpps0017 & 0.43344 & 367.9 & 19:06:37.7 & $ +$08:22 &  42.1147 & $ +$0.4661 &     9.5 &  7.6 &   7.1  \\
J1850$-$0020g & gpps0018 & 1.57459 & 605.4 & 18:50:05.9 & $ -$00:20 &  32.4771 & $ +$0.1433 &    44.2 &  8.1 &   5.7  \\
J2052$+$4421g & gpps0019 & 0.37531 & 547.0 & 20:52:53.2 & $ +$44:21 &  84.8417 & $ -$0.1616 &   280.3 & 50.0 &  25.0  \\
J1854$+$0012g & gpps0020 & 0.00271 & 204.1 & 18:54:18.4 & $ +$00:12 &  33.4516 & $ -$0.5396 &    11.8 &  5.3 &   4.1  \\
J1912$+$1105g & gpps0021 & 0.67071 & 150.4 & 19:12:25.6 & $ +$11:05 &  45.1765 & $ +$0.4480 &    17.7 &  4.4 &   4.4  \\
J1854$-$0033g & gpps0022 & 0.36147 & 617.5 & 18:54:29.6 & $ -$00:33 &  32.7882 & $ -$0.9319 &    17.7 &  9.6 &   8.3  \\
J1917$+$1411g & gpps0023 & 0.44646 & 123.4 & 19:17:25.5 & $ +$14:11 &  48.4969 & $ +$0.8145 &    12.5 &  4.6 &   4.0  \\
J1905$+$0450g & gpps0024 & 0.78330 & 442.3 & 19:05:59.3 & $ +$04:50 &  38.9029 & $ -$1.0168 &     8.1 &  8.3 &   9.8  \\
J1926$+$1857g & gpps0025 & 0.27873 & 424.2 & 19:26:58.9 & $ +$18:57 &  53.7730 & $ +$1.0372 &    30.4 & 12.3 &   9.6  \\
J1855$+$0139g & gpps0026 & 0.44414 & 405.6 & 18:55:13.6 & $ +$01:39 &  34.8429 & $ -$0.0849 &    36.8 &  7.0 &   5.2  \\
J1849$-$0014g & gpps0027 & 0.49171 & 346.6 & 18:49:30.6 & $ -$00:14 &  32.4943 & $ +$0.3174 &   107.3 &  6.3 &   4.9  \\
J1901$+$0712g & gpps0028 & 1.03771 & 332.4 & 19:01:29.8 & $ +$07:12 &  40.4908 & $ +$1.0586 &    43.1 &  6.6 &   8.9  \\
J1859$+$0434g & gpps0029 & 0.45834 & 320.1 & 18:59:41.1 & $ +$04:34 &  37.9367 & $ +$0.2522 &    15.2 &  6.4 &   5.3  \\
J1853$+$0013g & gpps0030 & 0.92853 & 312.4 & 18:53:52.9 & $ +$00:13 &  33.4145 & $ -$0.4392 &    17.0 &  6.2 &   4.7  \\
J1908$+$0811g & gpps0031 & 0.18164 & 300.1 & 19:08:43.6 & $ +$08:11 &  42.1800 & $ -$0.0847 &    19.1 &  6.6 &   5.6  \\
J1924$+$1343g & gpps0032 & 0.00572 &  98.5 & 19:24:20.0 & $ +$13:43 &  48.8687 & $ -$0.8864 &    32.2 &  4.1 &   3.0  \\
J1904$+$0358g & gpps0033 & 0.75154 & 532.0 & 19:04:22.5 & $ +$03:58 &  37.9509 & $ -$1.0557 &    14.2 &  9.3 &  12.1  \\
J1914$+$1029g & gpps0034 & 2.48499 &  59.7 & 19:14:22.9 & $ +$10:29 &  44.8706 & $ -$0.2541 &    11.8 &  2.8 &   1.9  \\
J1848$+$0127g & gpps0035 & 0.53402 &  77.0 & 18:48:19.1 & $ +$01:27 &  33.8770 & $ +$1.3601 &    36.9 &  2.7 &   2.5  \\
J1904$+$0535g & gpps0036 & 0.60376 &  78.4 & 19:04:51.8 & $ +$05:35 &  39.4407 & $ -$0.4236 &    45.4 &  3.1 &   2.4  \\
J1904$+$0519g & gpps0037 & 1.68053 &  80.8 & 19:04:07.6 & $ +$05:19 &  39.1228 & $ -$0.3814 &    77.4 &  3.1 &   2.5  \\
J1858$-$0024g & gpps0038 & 0.40060 & 190.0 & 18:58:46.5 & $ -$00:24 &  33.3996 & $ -$1.8214 &    25.8 &  5.2 &   4.7  \\
J1904$+$0553g & gpps0039 & 0.00491 & 164.2 & 19:04:16.9 & $ +$05:53 &  39.6443 & $ -$0.1556 &   825.6 &  4.5 &   4.0  \\
J1912$+$0934g & gpps0040 & 0.89747 & 143.8 & 19:12:41.2 & $ +$09:34 &  43.8581 & $ -$0.3134 &    30.3 &  2.6 &   4.0  \\
J1933$+$2038g & gpps0041 & 0.04075 & 302.9 & 19:33:42.5 & $ +$20:38 &  56.0116 & $ +$0.4585 &    22.9 &  9.1 &   8.5  \\
J1852$-$0024g & gpps0042 & 0.35545 & 290.0 & 18:52:13.7 & $ -$00:24 &  32.6549 & $ -$0.3640 &    35.2 &  5.9 &   4.5  \\
J1905$+$0758g & gpps0043 & 1.19276 & 200.7 & 19:05:47.8 & $ +$07:58 &  41.6666 & $ +$0.4660 &    14.7 &  5.2 &   4.8  \\
J2017$+$2819g & gpps0044 & 1.83246 &  66.0 & 20:17:19.9 & $ +$28:19 &  67.7232 & $ -$4.0410 &    46.2 &  3.7 &   4.9  \\
J2009$+$3122g & gpps0045 & 0.07654 & 144.1 & 20:09:40.1 & $ +$31:22 &  69.3534 & $ -$0.9742 &    17.1 &  5.8 &   6.5  \\
J1956$+$2826g & gpps0046 & 0.07179 & 112.0 & 19:56:47.4 & $ +$28:26 &  65.3854 & $ -$0.1721 &    13.2 &  4.8 &   6.5  \\
J1924$+$1510g & gpps0047 & 0.49863 & 115.6 & 19:24:01.8 & $ +$15:10 &  50.1028 & $ -$0.1402 &     9.6 &  4.5 &   3.5  \\
J1903$+$0845g & gpps0048 & 0.15314 & 129.1 & 19:03:36.0 & $ +$08:45 &  42.1036 & $ +$1.3030 &    18.1 &  4.2 &   4.3  \\
J1910$+$1054g & gpps0049 & 0.00387 & 139.4 & 19:10:01.4 & $ +$10:54 &  44.7481 & $ +$0.8898 &    15.5 &  4.0 &   4.4  \\
J1930$+$1357g & gpps0050 & 0.32354 & 186.6 & 19:30:12.0 & $ +$13:57 &  49.7499 & $ -$2.0289 &    27.9 &  6.3 &   5.6  \\
J1837$+$0033g & gpps0051 & 0.41814 & 186.9 & 18:37:41.1 & $ +$00:33 &  31.8669 & $ +$3.3179 &    10.1 &  4.7 &   5.5  \\
J1847$+$0133g & gpps0052 & 2.84691 & 193.8 & 18:47:00.9 & $ +$01:33 &  33.8153 & $ +$1.6945 &    10.1 &  5.3 &   4.9  \\
J1926$+$1631g & gpps0053 & 0.67839 & 195.1 & 19:26:24.2 & $ +$16:31 &  51.5693 & $ +$0.0016 &    46.4 &  6.2 &   4.8  \\
J1952$+$2702g & gpps0054 & 0.00414 & 213.1 & 19:52:18.4 & $ +$27:02 &  63.6790 & $ -$0.0424 &    14.0 &  7.4 &   7.8  \\
J1911$+$0751g & gpps0055 & 0.79691 & 220.7 & 19:11:43.2 & $ +$07:51 &  42.2258 & $ -$0.8963 &    51.9 &  5.7 &   5.6  \\
J1915$+$0832g & gpps0056 & 2.71009 &  36.2 & 19:15:05.8 & $ +$08:32 &  43.2318 & $ -$1.3131 &    15.9 &  2.2 &   1.4  \\
J1910$+$1117g & gpps0057 & 1.32152 & 296.5 & 19:10:27.2 & $ +$11:17 &  45.1394 & $ +$0.9745 &    17.5 &  7.0 &   7.9  \\
J1926$+$1452g & gpps0058 & 0.30451 & 298.1 & 19:26:54.6 & $ +$14:52 &  50.1780 & $ -$0.8897 &     5.5 &  8.6 &   6.5  \\
J1924$+$1509g & gpps0059 & 0.23995 & 296.6 & 19:24:33.5 & $ +$15:09 &  50.1512 & $ -$0.2590 &    14.1 &  8.3 &   6.4  \\
J1917$+$1121g & gpps0060 & 0.51031 & 303.5 & 19:17:26.8 & $ +$11:21 &  45.9926 & $ -$0.5138 &     9.2 &  7.6 &   7.1  \\
J1855$+$0511g & gpps0061 & 1.42147 & 307.5 & 18:55:11.6 & $ +$05:11 &  37.9847 & $ +$1.5342 &    29.9 &  6.7 &   9.1  \\
J1929$+$1615g & gpps0062 & 0.04460 & 308.7 & 19:29:36.4 & $ +$16:15 &  51.6945 & $ -$0.8048 &    11.7 &  9.1 &   6.1  \\
J1900$+$0213g & gpps0063 & 0.03209 & 310.5 & 19:00:12.5 & $ +$02:13 &  35.9189 & $ -$0.9318 &    24.7 &  6.2 &   5.7  \\
J1952$+$2836g & gpps0064 & 0.01802 & 313.0 & 19:52:47.7 & $ +$28:36 &  65.0826 & $ +$0.6701 &    95.4 &  9.7 &  10.5  \\
J1936$+$1952g & gpps0065 & 0.00972 & 325.0 & 19:36:00.1 & $ +$19:52 &  55.6078 & $ -$0.3812 &    29.3 &  9.4 &   8.6  \\
J1859$+$0026g & gpps0066 & 0.00857 & 334.1 & 18:59:58.0 & $ +$00:26 &  34.3029 & $ -$1.6933 &    29.1 &  7.2 &   6.5  \\
J1856$+$0615g & gpps0067 & 0.32697 & 333.1 & 18:56:39.1 & $ +$06:15 &  39.0950 & $ +$1.6947 &    38.8 &  7.6 &  11.6  \\
\hline\noalign{\smallskip}
\end{tabular}

\parbox{15.5cm}{``g'' indicates the temporary nature, due to position uncertainty of about 1.5{$'$}. }
\end{table*}
\addtocounter{table}{-1}
\begin{table*}
\centering
\begin{minipage}{3cm}
  \caption[]{{\it --Continued.}}
 \label{gppsPSRtab1}
\end{minipage}
\fns
\setlength{\tabcolsep}{3.5pt}
 \begin{tabular}{cccrccccrrr}
\hline\noalign{\smallskip}
 Name$^*$    &  gpps No.  &  Period    & DM      &  RA(2000)   &   D{ec}(2000)  &   GL  &    GB  &  $S_{\rm 1.25GHz}$  & $D_{\rm NE2001}$ & $D_{\rm YMW16}$ \\
             &            &   (s)   &(cm$^{-3}$pc)& (hh:mm:ss) & ($\pm$dd:mm) &  ($^{\circ}$) & ($^{\circ}$) & (${\upmu}$Jy) & (kpc)  & (kpc)      \\
 \hline
J1853$+$0312g & gpps0068 & 0.43809 & 345.0 & 18:53:14.9 & $ +$03:12 &  35.9921 & $ +$1.0592 &    12.0 &  7.0 &   6.7  \\
J1855$+$0455g & gpps0069 & 0.10101 & 372.9 & 18:55:15.4 & $ +$04:55 &  37.7532 & $ +$1.3980 &    40.3 &  7.5 &  10.3  \\
J1858$+$0609g & gpps0070 & 0.48435 & 381.5 & 18:58:09.5 & $ +$06:09 &  39.1719 & $ +$1.3133 &    14.0 &  7.9 &  11.6  \\
J1914$+$0805g & gpps0071 & 0.45555 & 339.2 & 19:14:05.4 & $ +$08:05 &  42.7106 & $ -$1.3051 &    17.4 &  8.0 &  10.5 \\
J1857$+$0224g & gpps0072 & 0.87595 & 401.0 & 18:57:08.1 & $ +$02:24 &  35.7231 & $ -$0.1693 &    13.1 &  7.0 &   5.3 \\
J1903$+$0534g & gpps0073 & 0.35765 & 407.7 & 19:03:29.3 & $ +$05:34 &  39.2697 & $ -$0.1269 &    41.1 &  7.2 &   5.9 \\
J1928$+$1809g & gpps0074 & 0.29446 & 431.1 & 19:28:06.0 & $ +$18:09 &  53.1982 & $ +$0.4240 &    29.9 & 11.8 &   9.3 \\
J1856$+$0011g & gpps0075 & 0.92847 & 455.8 & 18:56:47.5 & $ +$00:11 &  33.7176 & $ -$1.1013 &     8.6 &  8.1 &   6.9 \\
J2022$+$3845g & gpps0076 & 1.00890 & 487.5 & 20:22:11.4 & $ +$38:45 &  76.9110 & $ +$1.0169 &    54.7 & 50.0 &  17.2 \\
J2005$+$3411g & gpps0077 & 0.65105 & 489.0 & 20:05:45.0 & $ +$34:11 &  71.2811 & $ +$1.2438 &    81.9 & 50.0 &  17.3 \\
J1921$+$1505g & gpps0078 & 0.61190 & 519.9 & 19:21:24.7 & $ +$15:05 &  49.7378 & $ +$0.3812 &    14.5 & 13.6 &  10.6 \\
J1904$+$0415g & gpps0079 & 0.23145 & 521.0 & 19:04:05.7 & $ +$04:15 &  38.1707 & $ -$0.8637 &    53.0 &  8.8 &   9.5 \\
J1859$+$0126g & gpps0080 & 0.95770 & 531.7 & 18:59:15.4 & $ +$01:26 &  35.1076 & $ -$1.0810 &    35.9 &  9.0 &   9.6 \\
J1918$+$1340g & gpps0081 & 0.23299 & 575.9 & 19:18:53.9 & $ +$13:40 &  48.2019 & $ +$0.2541 &    43.7 & 14.3 &  11.9 \\
J1911$+$0939g & gpps0082 & 0.36547 & 597.3 & 19:11:38.3 & $ +$09:39 &  43.8189 & $ -$0.0423 &    25.8 &  8.1 &  10.6 \\
J1852$+$0158g & gpps0083 & 0.18573 & 607.5 & 18:52:43.8 & $ +$01:58 &  34.8424 & $ +$0.6159 &    35.8 &  9.2 &   7.6 \\
J1849$+$0037g & gpps0084 & 0.39649 & 611.5 & 18:49:28.0 & $ +$00:37 &  33.2692 & $ +$0.7265 &    65.8 &  9.0 &   8.0 \\
J2051$+$4434g & gpps0085 & 1.30316 & 616.0 & 20:51:28.6 & $ +$44:34 &  84.8479 & $ +$0.1706 &   341.8 & 50.0 &  25.0 \\
J1920$+$1515g & gpps0086 & 1.60276 & 655.5 & 19:20:22.4 & $ +$15:15 &  49.7623 & $ +$0.6780 &    19.0 & 50.0 &  15.7 \\
J2021$+$4024g & gpps0087 & 0.37054 & 680.5 & 20:21:12.9 & $ +$40:24 &  78.1680 & $ +$2.1153 &    88.1 & 50.0 &  25.0 \\
J1921$+$1340g & gpps0088 & 4.60294 & 754.9 & 19:21:29.7 & $ +$13:40 &  48.4946 & $ -$0.3047 &    30.8 & 50.0 &  25.0 \\
J1853$+$0023g & gpps0089 & 0.57686 & 203.3 & 18:53:09.5 & $ +$00:23 &  33.4769 & $ -$0.2041 &    45.3 &  5.3 &   4.0 \\
J1903$+$0433g & gpps0090 & 14.0499 & 202.6 & 19:03:19.0 & $ +$04:33 &  38.3396 & $ -$0.5584 &    14.3 &  4.9 &   4.4 \\
J1859$+$0239g & gpps0091 & 0.05611 & 250.9 & 18:59:18.4 & $ +$02:39 &  36.1937 & $ -$0.5373 &    10.1 &  5.8 &   4.7 \\
J1906$+$0646g & gpps0092 & 0.35552 & 290.5 & 19:06:15.1 & $ +$06:46 &  40.6461 & $ -$0.1889 &    48.4 &  6.1 &   5.3 \\
J1849$+$0001g & gpps0093 & 0.52560 & 189.4 & 18:49:23.2 & $ +$00:01 &  32.7166 & $ +$0.4659 &    12.7 &  5.1 &   4.0 \\
J1858$+$0026g & gpps0094 & 4.71467 & 415.3 & 18:58:04.9 & $ +$00:26 &  34.0801 & $ -$1.2779 &    41.9 &  7.9 &   6.8 \\
J1850$+$0011g & gpps0095 & 0.16754 & 506.5 & 18:50:03.9 & $ +$00:11 &  32.9428 & $ +$0.3912 &    30.5 &  7.6 &   5.7 \\
J1849$+$0009g & gpps0096 & 1.31855 & 501.5 & 18:49:42.0 & $ +$00:09 &  32.8816 & $ +$0.4624 &    14.1 &  7.6 &   5.8 \\
J1850$-$0002g & gpps0097 & 0.89336 & 543.9 & 18:50:00.0 & $ -$00:02 &  32.7376 & $ +$0.3044 &    30.3 &  7.8 &   5.7 \\
J1852$-$0002g & gpps0098 & 0.24510 & 558.1 & 18:52:10.3 & $ -$00:02 &  32.9763 & $ -$0.1834 &    41.5 &  7.9 &   5.6 \\
J1921$+$1259g & gpps0099 & 0.57316 & 366.3 & 19:21:18.1 & $ +$12:59 &  47.8792 & $ -$0.5795 &    14.6 &  9.3 &   8.2 \\
J1903$+$0839g & gpps0100 & 0.00462 & 166.5 & 19:03:52.3 & $ +$08:39 &  42.0470 & $ +$1.1983 &   181.1 &  4.9 &   5.3 \\
J1903$+$0851g & gpps0101 & 1.23197 &  78.9 & 19:03:04.3 & $ +$08:51 &  42.1353 & $ +$1.4661 &    14.8 &  3.2 &   2.5 \\
J1906$+$0757g & gpps0102 & 0.05719 &  79.0 & 19:06:39.8 & $ +$07:57 &  41.7447 & $ +$0.2648 &    14.5 &  3.2 &   2.5 \\
J1852$+$0018g & gpps0103 & 0.31876 & 452.0 & 18:52:29.5 & $ +$00:18 &  33.3272 & $ -$0.0935 &    29.0 &  7.1 &   5.2 \\
J1904$+$0836g & gpps0104 & 0.00444 &  90.5 & 19:04:35.4 & $ +$08:36 &  42.0824 & $ +$1.0168 &    17.3 &  3.5 &   2.9 \\
J1900$+$0715g & gpps0105 & 0.97044 & 266.1 & 19:00:22.4 & $ +$07:15 &  40.4103 & $ +$1.3306 &    60.5 &  6.1 &   8.7 \\
J2139$+$4738g & gpps0106 & 0.55704 & 138.9 & 21:39:57.7 & $ +$47:38 &  92.8848 & $ -$3.7423 &    12.2 &  5.0 &   4.1 \\
J1838$+$0022g & gpps0107 & 0.00509 & 122.6 & 18:38:24.4 & $ +$00:22 &  31.7733 & $ +$3.0672 &    32.3 &  3.4 &   4.1 \\
J1848$+$0150g & gpps0108 & 3.29015 & 500.8 & 18:48:43.4 & $ +$01:50 &  34.2558 & $ +$1.4403 &    12.4 &  9.7 &  10.0 \\
J1844$+$0028g & gpps0109 & 0.00357 & 181.2 & 18:44:35.5 & $ +$00:28 &  32.5683 & $ +$1.7368 &    53.8 &  5.0 &   4.6 \\
J1853$-$0003g & gpps0110 & 0.17152 & 667.2 & 18:53:24.1 & $ -$00:03 &  33.1063 & $ -$0.4623 &     7.1 &  9.1 &   6.3 \\
J1949$+$2516g & gpps0111 & 0.41034 & 425.7 & 19:49:36.6 & $ +$25:16 &  61.8501 & $ -$0.4234 &     5.3 & 13.0 &  15.1 \\
J1845$-$0028g & gpps0112 & 0.08450 & 301.0 & 18:45:39.7 & $ -$00:28 &  31.8488 & $ +$1.0676 &    10.2 &  6.2 &   5.3 \\
J1901$+$0020g & gpps0113 & 0.21481 & 237.6 & 19:01:17.2 & $ +$00:20 &  34.3569 & $ -$2.0365 &    41.6 &  5.8 &   5.6 \\
J1908$+$1035g & gpps0114 & 0.01069 &  10.9 & 19:08:23.2 & $ +$10:35 &  44.2835 & $ +$1.1015 &    11.6 &  0.6 &   0.7 \\
J1925$+$1532g & gpps0115 & 1.65510 & 145.9 & 19:25:00.3 & $ +$15:32 &  50.5449 & $ -$0.1694 &    20.6 &  5.1 &   4.2 \\
J1849$+$0225g & gpps0116 & 1.47452 & 259.9 & 18:49:13.1 & $ +$02:25 &  34.8363 & $ +$1.5983 &    13.8 &  6.2 &   5.8 \\
J1854$+$0131g & gpps0117 & 2.04385 & 474.9 & 18:54:34.0 & $ +$01:31 &  34.6470 & $ +$0.0001 &   312.9 &  7.5 &   5.5 \\
J2030$+$3944g & gpps0118 & 0.30618 & 937.4 & 20:30:21.5 & $ +$39:44 &  78.6299 & $ +$0.2988 &   218.4 & 50.0 &  25.0 \\
J1849$-$0019g & gpps0119 & 0.91137 & 513.2 & 18:49:15.2 & $ -$00:19 &  32.3966 & $ +$0.3394 &     8.9 &  7.6 &   5.6 \\
J1907$+$0709g & gpps0120 & 0.34410 & 278.1 & 19:07:56.7 & $ +$07:09 &  41.1741 & $ -$0.3886 &    83.8 &  6.1 &   5.4 \\
J1928$+$1816g & gpps0121 & 0.01054 & 346.3 & 19:28:10.0 & $ +$18:16 &  53.3088 & $ +$0.4661 &    26.2 &  9.7 &   7.2 \\
J1914$+$1228g & gpps0122 & 2.27755 & 312.0 & 19:14:55.1 & $ +$12:28 &  46.6871 & $ +$0.5509 &    22.9 &  7.9 &   7.3 \\
J1955$+$2912g & gpps0123 & 0.27951 & 193.0 & 19:55:23.1 & $ +$29:12 &  65.8853 & $ +$0.4914 &     5.5 &  7.0 &   7.4 \\
J1838$+$0027g & gpps0124 & 0.05413 & 205.9 & 18:38:12.7 & $ +$00:27 &  31.8265 & $ +$3.1492 &     7.5 &  5.0 &   5.8 \\
J1858$+$0244g & gpps0125 & 0.00261 & 282.6 & 18:58:01.2 & $ +$02:44 &  36.1246 & $ -$0.2117 &    73.0 &  6.1 &   4.7 \\
J1847$+$0110g & gpps0126 & 0.00653 & 183.4 & 18:47:23.1 & $ +$01:10 &  33.5218 & $ +$1.4405 &    37.4 &  5.1 &   4.6 \\
J1907$+$0658g & gpps0127 & 0.21834 & 523.0 & 19:07:34.0 & $ +$06:58 &  40.9807 & $ -$0.3831 &    54.1 &  9.0 &   7.7 \\
J1908$+$0949g & gpps0128 & 0.00905 & 220.3 & 19:08:07.8 & $ +$09:49 &  43.5742 & $ +$0.8048 &    19.4 &  5.7 &   6.0 \\
J1923$+$1521g & gpps0129 & 1.04876 & 346.0 & 19:23:12.0 & $ +$15:21 &  50.1781 & $ +$0.1273 &    15.5 &  9.5 &   7.0 \\
J1919$+$1527g & gpps0130 & 1.37146 & 697.5 & 19:19:50.2 & $ +$15:27 &  49.8846 & $ +$0.8895 &     7.1 & 50.0 &  16.9 \\
J1859$+$0313g & gpps0131 & 0.00161 & 107.7 & 18:59:35.5 & $ +$03:13 &  36.7345 & $ -$0.3395 &    60.3 &  3.4 &   3.1 \\
J1858$+$0310g & gpps0132 & 0.37275 & 699.5 & 18:58:05.1 & $ +$03:10 &  36.5152 & $ -$0.0293 &   119.0 &  9.9 &   6.7 \\
J1856$+$0243g & gpps0133 & 0.54660 & 178.3 & 18:56:01.4 & $ +$02:43 &  35.8826 & $ +$0.2246 &   114.5 &  5.1 &   4.0 \\
J1909$+$0657g & gpps0134 & 1.24589 &  59.9 & 19:09:13.8 & $ +$06:57 &  41.1462 & $ -$0.7623 &    87.2 &  2.8 &   1.8 \\
J1855$+$0339g & gpps0135 & 1.76134 & 416.2 & 18:55:42.5 & $ +$03:39 &  36.6771 & $ +$0.7203 &    12.3 &  7.5 &   6.9 \\
\hline
\end{tabular}
\end{table*}
\addtocounter{table}{-1}
\begin{table*}[htp!!!]
\centering
\begin{minipage}{3cm}
  \caption[]{{\it --Continued.}}
 \label{gppsPSRtab1}
\end{minipage}

\fns
\setlength{\tabcolsep}{3.5pt}
 \begin{tabular}{cccrccccrrr}
\hline\noalign{\smallskip}
 Name$^*$    &  gpps No.  &  Period    & DM      &  RA(2000)   &   D{ec}(2000)  &   GL  &    GB  &  $S_{\rm 1.25GHz}$  & $D_{\rm NE2001}$ & $D_{\rm YMW16}$ \\
             &            &   (s)   &(cm$^{-3}$pc)& (hh:mm:ss) & ($\pm$dd:mm) &  ($^{\circ}$) & ($^{\circ}$) & (${\upmu}$Jy) & (kpc)  & (kpc)      \\
 \hline
J2011$+$3006g & gpps0136 & 2.50566 &  14.0 & 20:11:12.1 & $ +$30:06 &  68.4718 & $ -$1.9414 &   312.3 &  1.2 &   0.9 \\
J1904$+$0207g & gpps0137 & 0.00504 & 229.9 & 19:04:44.4 & $ +$02:07 &  36.3346 & $ -$1.9911 &    17.9 &  4.2 &   6.0 \\
J1924$+$2027g & gpps0138 & 0.00195 & 211.7 & 19:24:33.5 & $ +$20:27 &  54.8223 & $ +$2.2501 &    82.5 &  7.3 &   5.3 \\
J1855$+$0228g & gpps0139 & 0.25317 & 530.0 & 18:55:45.0 & $ +$02:28 &  35.6253 & $ +$0.1694 &    37.4 &  8.1 &   5.9 \\
J1909$+$1132g & gpps0140 & 0.00680 & 276.1 & 19:09:36.7 & $ +$11:32 &  45.2621 & $ +$1.2707 &    14.8 &  6.9 &   8.1 \\
J1914$+$1054g & gpps0141 & 0.13887 & 418.7 & 19:14:06.8 & $ +$10:54 &  45.2132 & $ -$0.0000 &    33.0 &  8.8 &   7.8 \\
J1913$+$1054g & gpps0142 & 0.45062 & 338.0 & 19:13:07.0 & $ +$10:54 &  45.0905 & $ +$0.2120 &    24.1 &  7.6 &   7.1 \\
J1928$+$1852g & gpps0143 & 0.79281 & 291.6 & 19:28:04.7 & $ +$18:52 &  53.8248 & $ +$0.7703 &    20.0 &  8.8 &   5.8 \\
J1918$+$1540g & gpps0144 & 0.00428 & 271.2 & 19:18:19.1 & $ +$15:40 &  49.9066 & $ +$1.3149 &    15.9 &  8.2 &   6.5 \\
J1918$+$1536g & gpps0145 & 0.10995 & 123.7 & 19:18:23.4 & $ +$15:36 &  49.8602 & $ +$1.2709 &    10.0 &  4.7 &   3.8 \\
J1840$+$0012g & gpps0146 & 0.00534 & 100.9 & 18:40:49.4 & $ +$00:12 &  31.9078 & $ +$2.4571 &   164.9 &  2.9 &   3.5 \\
J1916$+$1030Bg& gpps0147 & 0.34938 & 519.7 & 19:16:46.1 & $ +$10:30 &  45.1641 & $ -$0.7626 &    11.4 & 11.4 &  11.3 \\
J1855$+$0235g & gpps0148 & 0.98303 & 103.3 & 18:55:48.9 & $ +$02:35 &  35.7442 & $ +$0.2121 &    21.8 &  3.3 &   3.0 \\
J1913$+$1037g & gpps0149 & 0.43421 & 437.0 & 19:13:52.6 & $ +$10:37 &  44.9266 & $ -$0.0846 &     8.1 &  9.0 &   7.9 \\
J1939$+$2352g & gpps0150 & 2.14534 & 415.5 & 19:39:48.6 & $ +$23:52 &  59.5212 & $ +$0.8047 &     7.5 & 12.7 &  12.3 \\
J1855$+$0424g & gpps0151 & 2.22025 & 678.5 & 18:55:00.4 & $ +$04:24 &  37.2690 & $ +$1.2201 &    11.7 & 12.8 &  17.4 \\
J2030$+$3929g & gpps0152 & 1.71842 & 491.9 & 20:30:47.2 & $ +$39:29 &  78.4765 & $ +$0.0848 &    35.2 & 50.0 &  25.0 \\
J1905$+$0936g & gpps0153 & 1.63451 & 414.0 & 19:05:24.7 & $ +$09:36 &  43.0608 & $ +$1.2934 &    46.8 &  9.2 &  12.3 \\
J1950$+$2352g & gpps0154 & 0.31976 & 342.5 & 19:50:56.1 & $ +$23:52 &  60.7928 & $ -$1.3979 &    32.4 & 10.9 &  13.3 \\
J1923$+$2022g & gpps0155 & 0.03799 & 175.3 & 19:23:55.9 & $ +$20:22 &  54.6782 & $ +$2.3395 &     8.4 &  6.2 &   4.6 \\
J1923$+$1143g & gpps0156 & 0.37121 & 260.6 & 19:23:38.0 & $ +$11:43 &  47.0176 & $ -$1.6852 &     9.2 &  7.6 &   7.8 \\
J1917$+$1046g & gpps0157 & 0.08773 & 163.1 & 19:17:55.3 & $ +$10:46 &  45.5310 & $ -$0.8894 &    14.1 &  5.1 &   4.9 \\
J1856$+$0211g & gpps0158 & 9.89012 & 113.9 & 18:56:53.1 & $ +$02:11 &  35.5031 & $ -$0.2120 &    10.7 &  3.4 &   3.1 \\
J1933$+$1923g & gpps0159 & 0.37173 &  97.7 & 19:33:58.1 & $ +$19:23 &  54.9486 & $ -$0.1999 &    16.3 &  4.1 &   3.1 \\
J1918$+$1547g & gpps0160 & 0.00376 &  64.6 & 19:18:34.0 & $ +$15:47 &  50.0313 & $ +$1.3133 &    16.5 &  3.3 &   2.2 \\
J1854$+$0704g & gpps0161 & 0.45090 &  10.8 & 18:54:31.0 & $ +$07:04 &  39.5878 & $ +$2.5417 &    34.6 &  1.0 &   0.6 \\
J1859$+$0658g & gpps0162 & 0.00511 & 290.4 & 18:59:04.7 & $ +$06:58 &  40.0033 & $ +$1.4828 &    35.6 &  6.8 &   9.4 \\
J1928$+$1902g & gpps0163 & 0.00580 &  28.9 & 19:28:05.0 & $ +$19:02 &  53.9692 & $ +$0.8474 &     7.1 &  2.3 &   1.5 \\
J1851$+$0056g & gpps0164 & 0.29110 & 332.3 & 18:51:27.0 & $ +$00:56 &  33.7751 & $ +$0.4286 &    13.4 &  6.4 &   5.0 \\
J1847$+$0614g & gpps0165 & 1.66302 & 270.5 & 18:47:33.7 & $ +$06:14 &  38.0557 & $ +$3.7011 &    50.5 &  8.0 &  16.7 \\
J1916$+$0741g & gpps0166 & 0.01122 & 220.1 & 19:16:17.4 & $ +$07:41 &  42.6135 & $ -$1.9711 &    29.8 &  6.2 &   8.4 \\
J1917$+$0743g & gpps0167 & 0.81347 & 199.5 & 19:17:26.3 & $ +$07:43 &  42.7679 & $ -$2.2114 &    22.7 &  5.9 &   8.2 \\
J1904$+$0603g & gpps0168 & 1.97493 & 413.2 & 19:04:52.4 & $ +$06:03 &  39.8565 & $ -$0.2117 &    23.5 &  7.4 &   6.1 \\
J1912$+$1417g & gpps0169 & 0.00317 &  66.6 & 19:12:31.9 & $ +$14:17 &  48.0257 & $ +$1.9063 &    39.8 &  3.4 &   2.2 \\
J1950$+$2556g & gpps0170 & 2.03864 & 420.4 & 19:50:03.7 & $ +$25:56 &  62.4808 & $ -$0.1695 &    20.1 & 12.7 &  14.5 \\
J1852$+$0309g & gpps0171 & 0.00558 & 358.0 & 18:52:10.9 & $ +$03:09 &  35.8211 & $ +$1.2707 &    89.7 &  7.3 &   7.1 \\
J1852$+$0857g & gpps0172 & 3.77214 &  85.9 & 18:52:37.4 & $ +$08:57 &  41.0548 & $ +$3.8125 &    21.5 &  3.4 &   3.4 \\
J1852$-$0033g & gpps0173 & 1.36903 & 320.7 & 18:52:25.9 & $ -$00:33 &  32.5466 & $ -$0.4766 &    18.9 &  6.2 &   4.7 \\
J1852$-$0044g & gpps0174 & 0.00241 & 272.9 & 18:52:27.4 & $ -$00:44 &  32.3907 & $ -$0.5634 &   556.1 &  5.8 &   4.5 \\
J1916$+$0748g & gpps0175 & 0.86791 & 153.1 & 19:16:59.9 & $ +$07:48 &  42.7917 & $ -$2.0758 &    14.3 &  4.9 &   6.4 \\
J1929$+$1937g & gpps0176 & 0.56373 & 458.1 & 19:29:10.0 & $ +$19:37 &  54.6040 & $ +$0.9018 &    17.0 & 13.4 &  10.0 \\
J1852$-$0039g & gpps0177 & 0.80291 & 361.7 & 18:52:10.6 & $ -$00:39 &  32.4255 & $ -$0.4670 &    85.5 &  6.5 &   4.9 \\
J1909$+$0905g & gpps0178 & 1.49488 & 250.6 & 19:09:39.3 & $ +$09:05 &  43.0852 & $ +$0.1271 &    12.0 &  5.9 &   5.4 \\
J1928$+$1839g & gpps0179 & 2.26091 &  70.0 & 19:28:38.5 & $ +$18:39 &  53.7001 & $ +$0.5507 &     8.2 &  3.4 &   2.5 \\
J1855$+$0327g & gpps0180 & 0.78282 & 298.6 & 18:55:45.1 & $ +$03:27 &  36.5008 & $ +$0.6178 &    13.7 &  6.3 &   5.3 \\
J1911$+$1252g & gpps0181 & 0.02724 &  68.8 & 19:11:28.4 & $ +$12:52 &  46.6561 & $ +$1.4829 &    26.5 &  3.3 &   2.2 \\
J1905$+$0920g & gpps0182 & 0.17047 & 396.6 & 19:05:08.7 & $ +$09:20 &  42.7949 & $ +$1.2302 &     9.6 &  8.8 &  11.6 \\
J1927$+$1430g & gpps0183 & 0.20288 & 207.2 & 19:27:39.6 & $ +$14:30 &  49.9334 & $ -$1.2284 &    12.8 &  6.6 &   5.6 \\
J1849$+$0339g & gpps0184 & 1.66672 & 349.5 & 18:49:54.7 & $ +$03:39 &  36.0225 & $ +$2.0100 &    55.2 &  7.6 &  10.0 \\
J1902$+$0809g & gpps0185 & 0.19023 & 436.9 & 19:02:19.7 & $ +$08:09 &  41.4253 & $ +$1.3075 &    10.0 &  9.2 &  13.9 \\
J2013$+$3100g & gpps0186 & 0.36855 & 158.4 & 20:13:56.6 & $ +$31:00 &  69.5489 & $ -$1.9394 &    26.9 &  6.2 &   6.8 \\
J2030$+$3818g & gpps0187 & 0.13372 & 596.7 & 20:30:08.5 & $ +$38:18 &  77.4491 & $ -$0.5085 &    17.3 & 50.0 &  25.0 \\
J1853$+$0237g & gpps0188 & 0.42739 & 725.5 & 18:53:57.3 & $ +$02:37 &  35.5518 & $ +$0.6355 &    38.1 & 10.8 &  10.0 \\
J2018$+$3418g & gpps0189 & 2.19160 & 317.1 & 20:18:15.0 & $ +$34:18 &  72.8022 & $ -$0.8473 &     4.9 & 10.4 &  10.4 \\
J1953$+$1844g & gpps0190 & 0.00444 & 113.1 & 19:53:44.0 & $ +$18:44 &  56.7087 & $ -$4.5750 &   121.8 &  4.8 &   4.4 \\
J2008$+$2755g & gpps0191 & 1.51926 & 155.1 & 20:08:16.6 & $ +$27:55 &  66.2982 & $ -$2.5868 &   110.5 &  6.3 &   7.4 \\
J1924$+$2037g & gpps0192 & 0.68480 &  82.3 & 19:24:33.5 & $ +$20:37 &  54.9718 & $ +$2.3301 &     3.3 &  3.8 &   2.9 \\
J1853$-$0008g & gpps0193 & 0.00282 & 285.3 & 18:53:12.3 & $ -$00:08 &  33.0086 & $ -$0.4571 &    13.8 &  5.9 &   4.6 \\
J1929$+$1731g & gpps0194 & 3.99540 & 431.1 & 19:29:04.2 & $ +$17:31 &  52.7463 & $ -$0.0848 &    13.4 & 11.6 &   9.2 \\
J1909$+$0930g & gpps0195 & 2.02078 & 479.6 & 19:09:46.6 & $ +$09:30 &  43.4764 & $ +$0.2967 &    16.0 &  8.5 &   8.4 \\
J1925$+$1636g & gpps0196 & 0.04971 &  33.1 & 19:25:47.6 & $ +$16:36 &  51.5723 & $ +$0.1693 &    11.2 &  2.3 &   1.5 \\
J1936$+$2036g & gpps0197 & 0.03292 & 198.8 & 19:36:41.0 & $ +$20:36 &  56.3160 & $ -$0.1689 &    28.6 &  6.7 &   5.0 \\
J1849$+$0016g & gpps0198 & 0.00181 & 271.4 & 18:49:58.2 & $ +$00:16 &  33.008  & $ +$0.4511 &    53.8 &  5.8 &   4.6 \\
J1852$-$0055g & gpps0199 & 0.16403 & 211.7 & 18:52:22.7 & $ -$00:55 &  32.2216 & $ -$0.6281 &    51.9 &  5.3 &   4.1 \\
J1901$+$0315g & gpps0200 & 0.81982 & 410.5 & 19:01:33.7 & $ +$03:15 &  36.9871 & $ -$0.7625 &    26.9 &  7.5 &   6.5 \\
J2023$+$2853g & gpps0201 & 0.01133 &  22.8 & 20:23:15.6 & $ +$28:52 &  68.9127 & $ -$4.8032 &   161.1 &  2.0 &   1.6 \\
\hline
\end{tabular}
\end{table*}

%
\begin{figure*}
\begin{tabular}{rrrrrr}
\includegraphics[width=39mm]{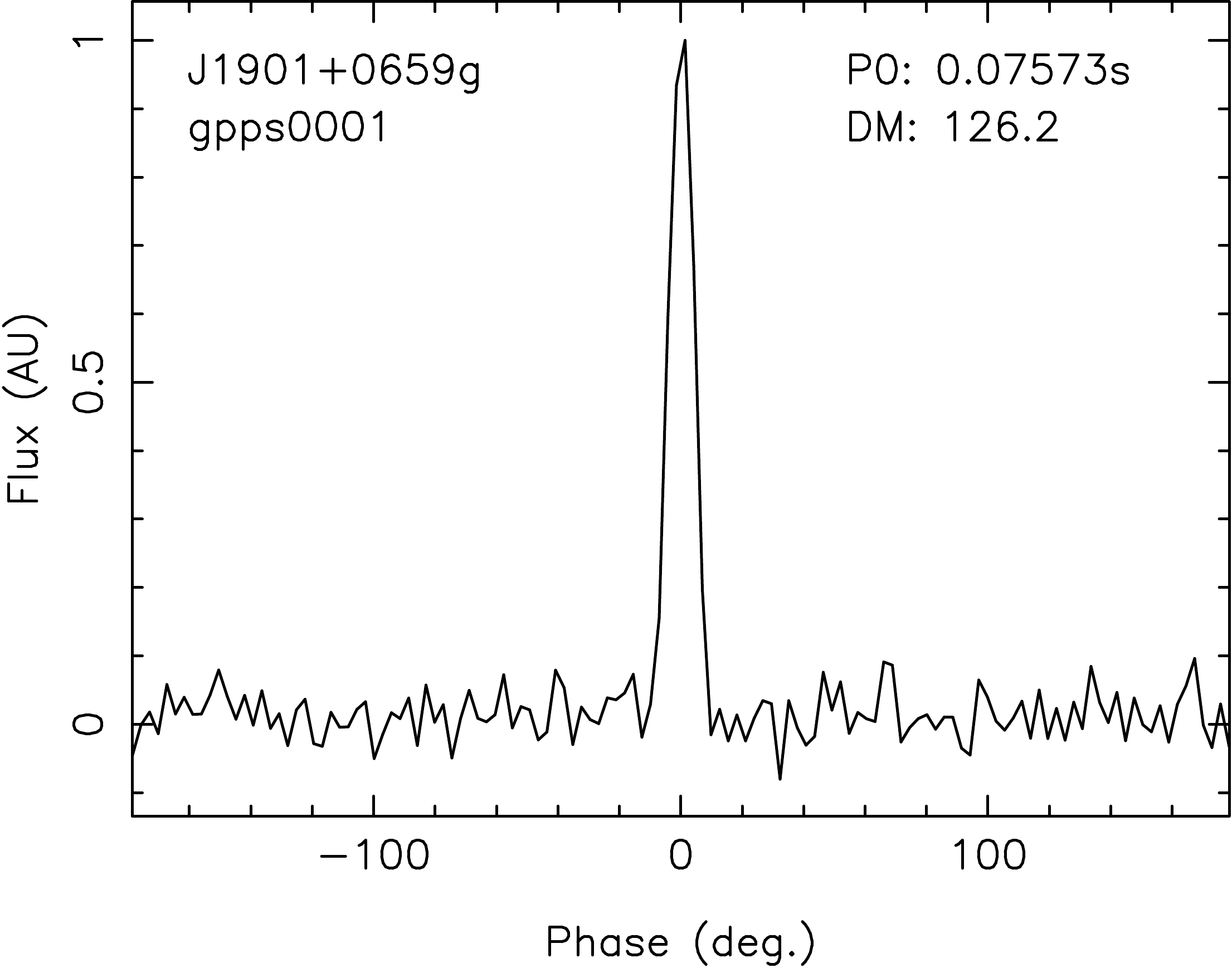}&
\includegraphics[width=39mm]{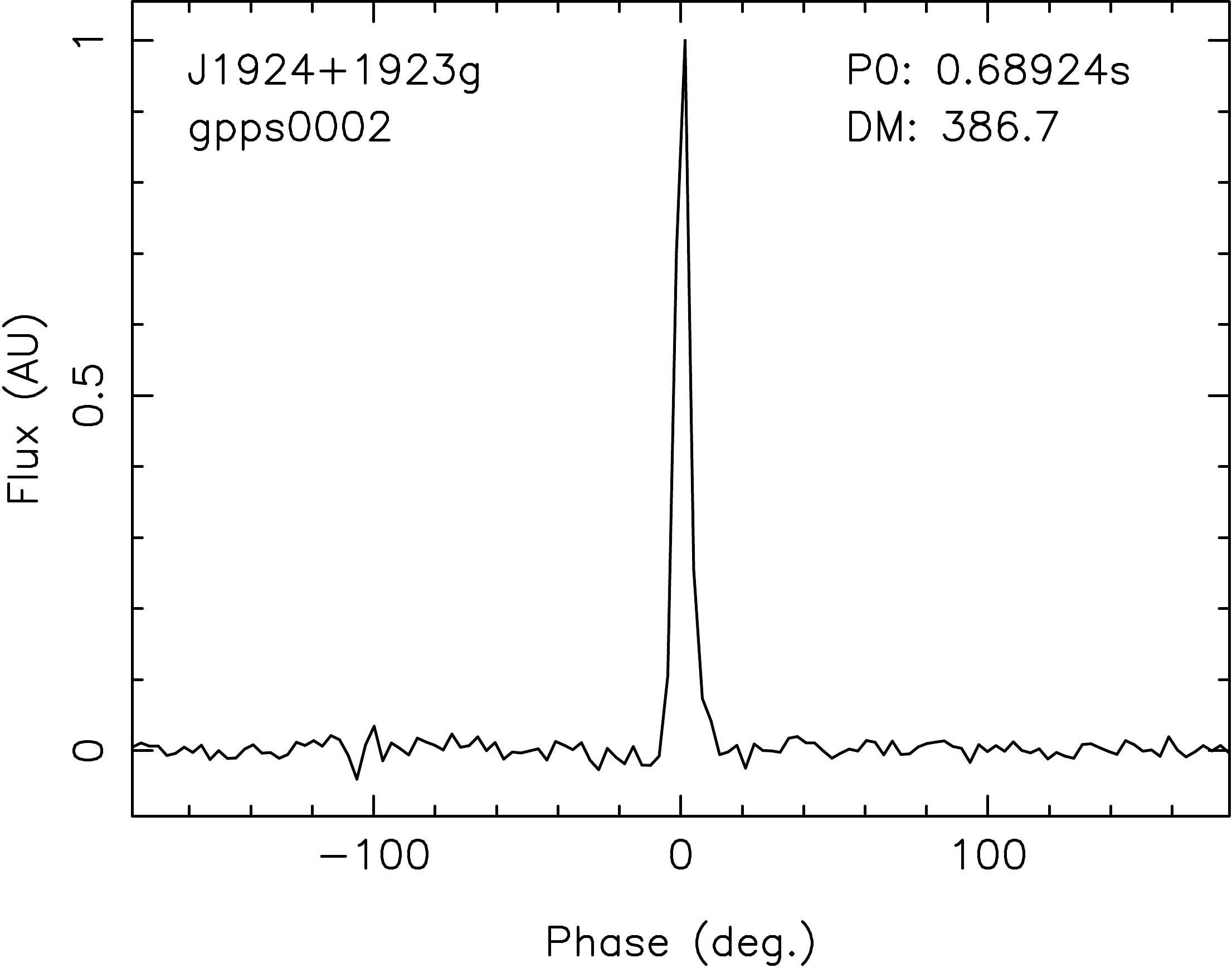}&
\includegraphics[width=39mm]{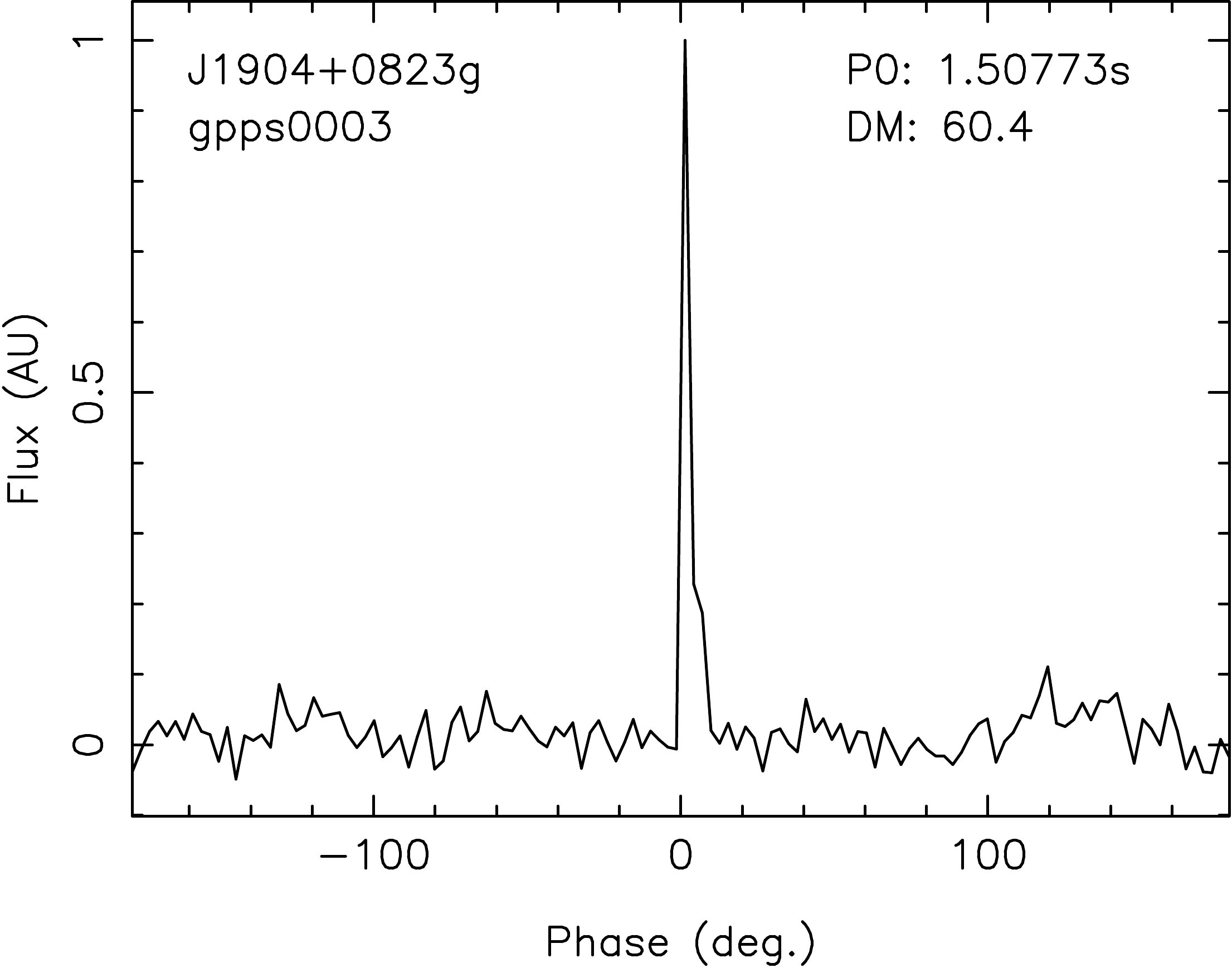}&
\includegraphics[width=39mm]{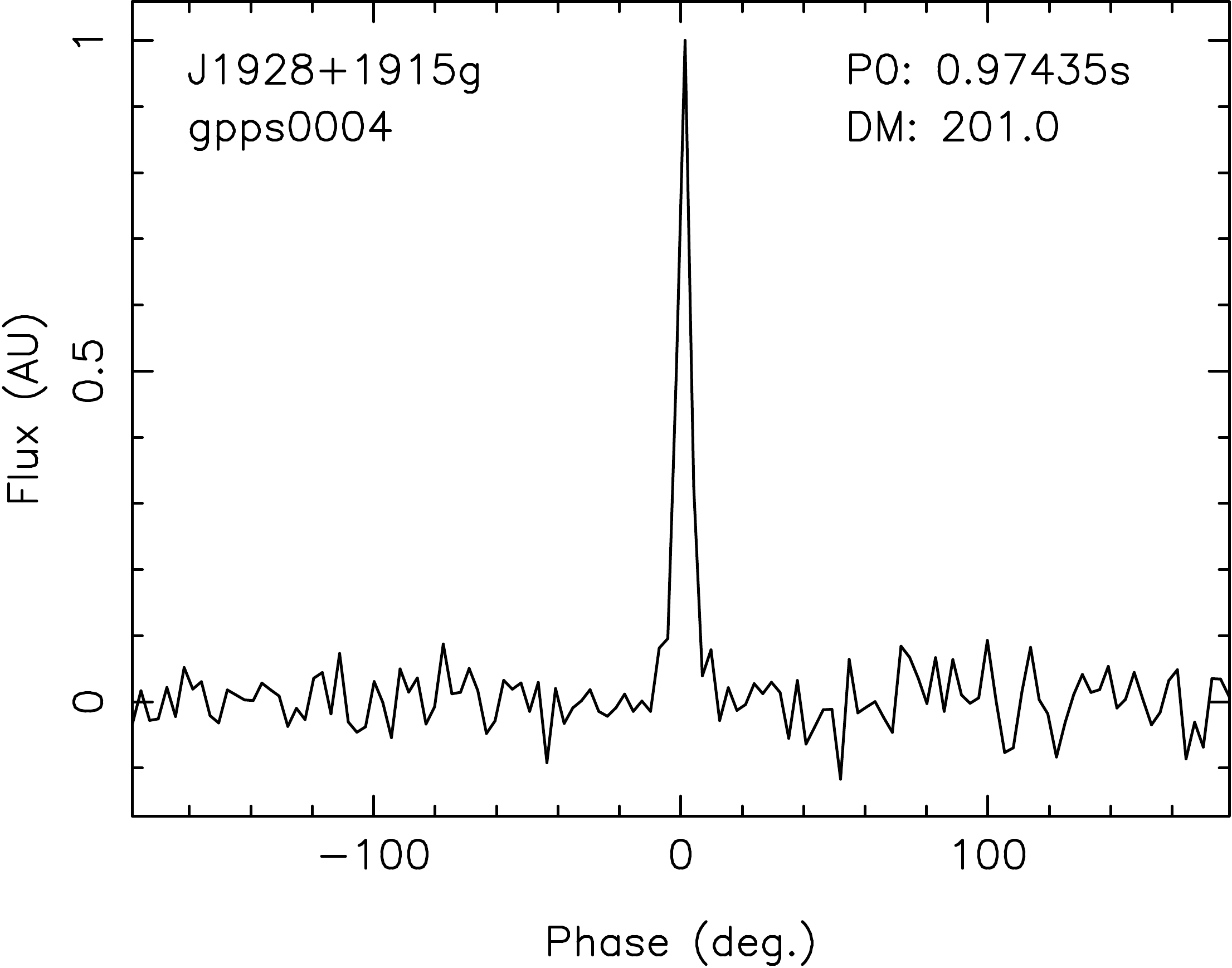}\\[2mm]
\includegraphics[width=39mm]{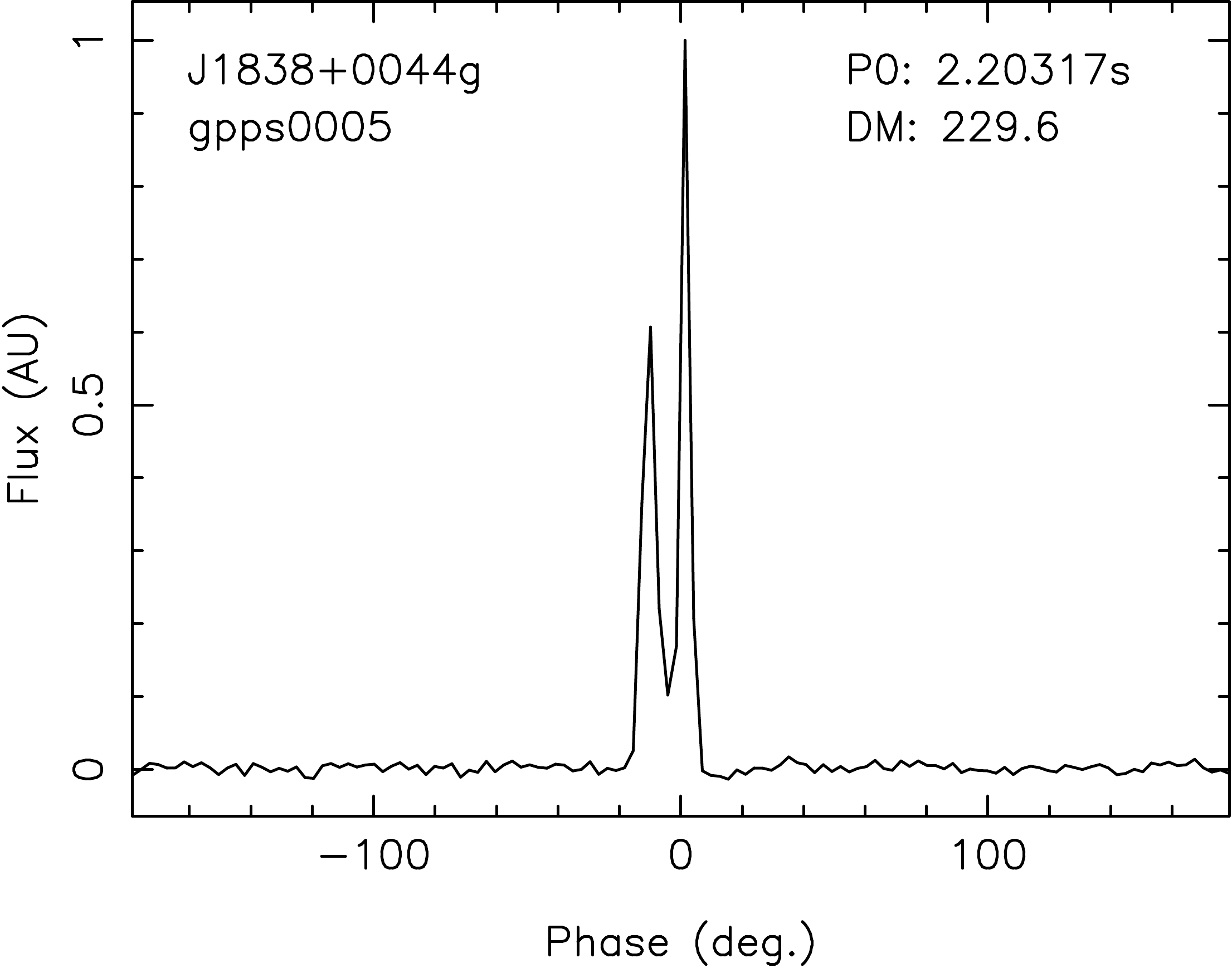}&
\includegraphics[width=39mm]{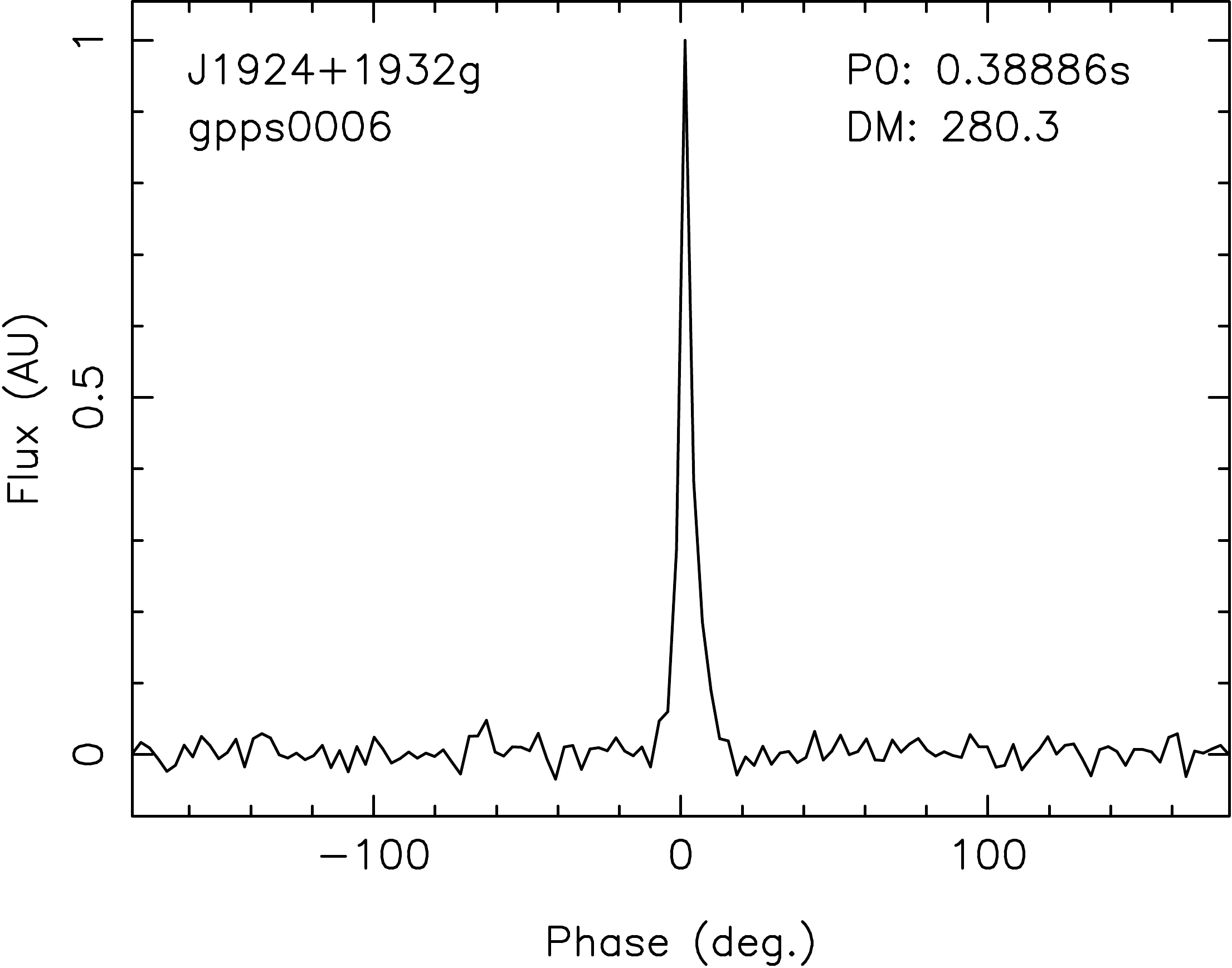}&
\includegraphics[width=39mm]{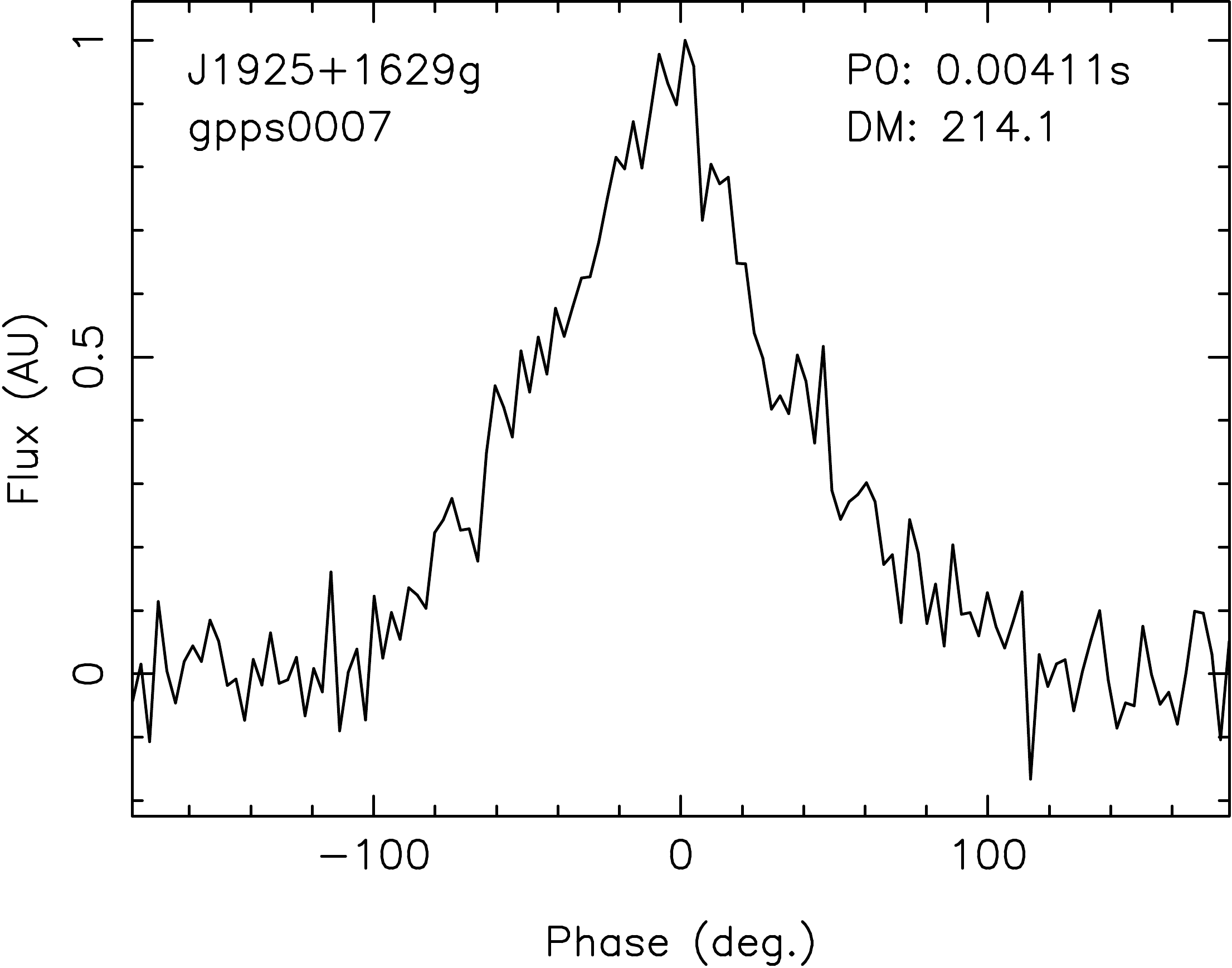}&
\includegraphics[width=39mm]{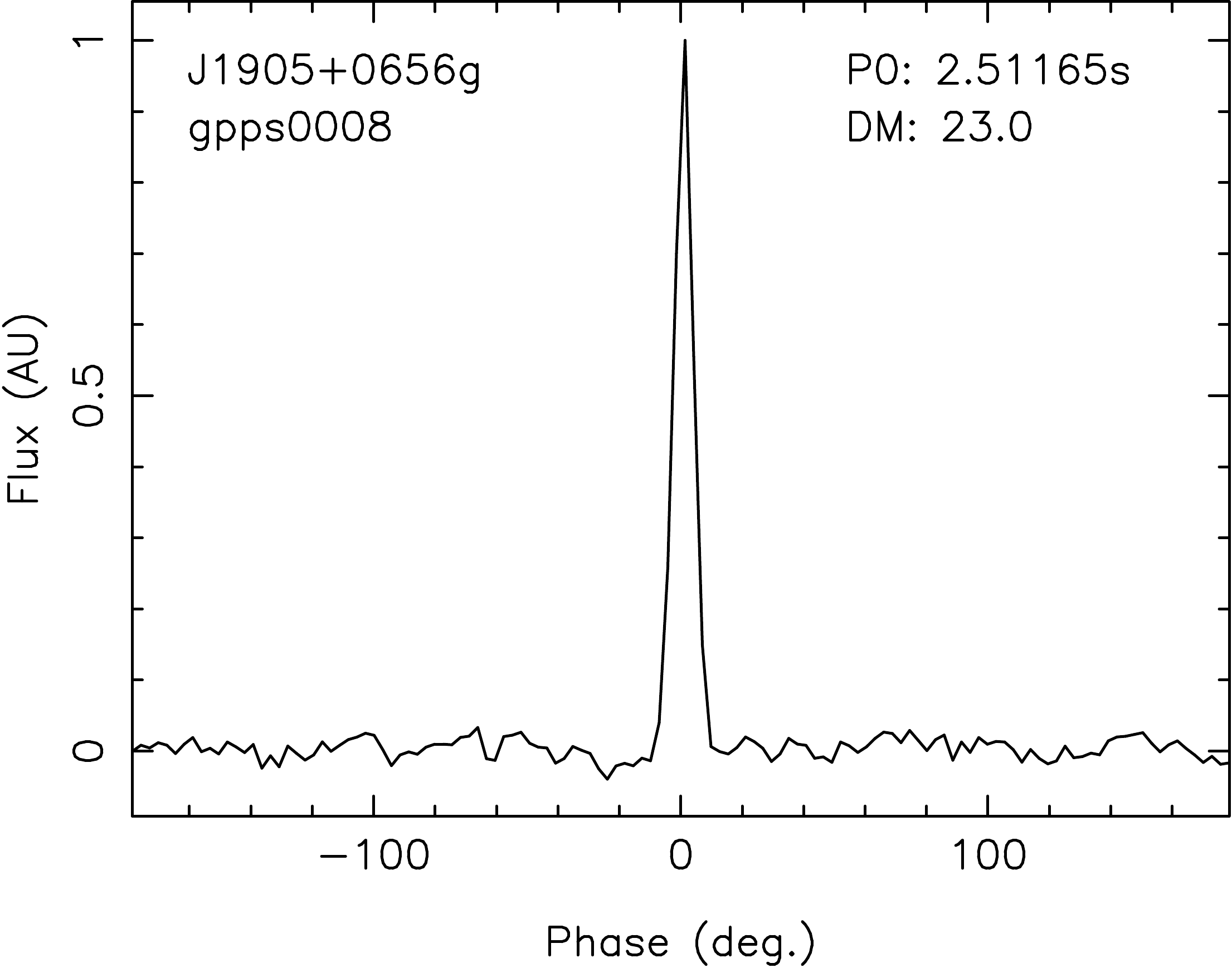}\\[2mm]
\includegraphics[width=39mm]{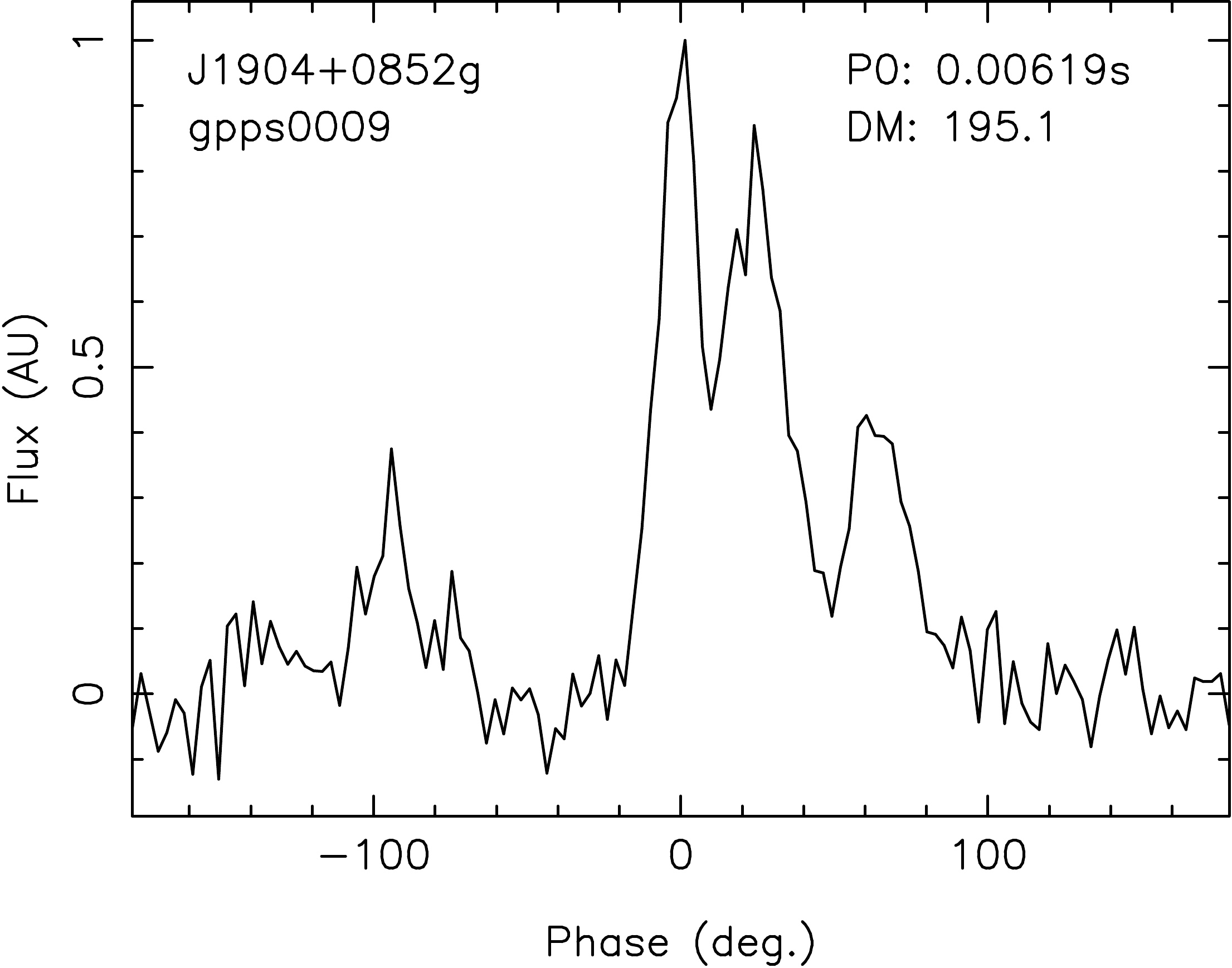}&
\includegraphics[width=39mm]{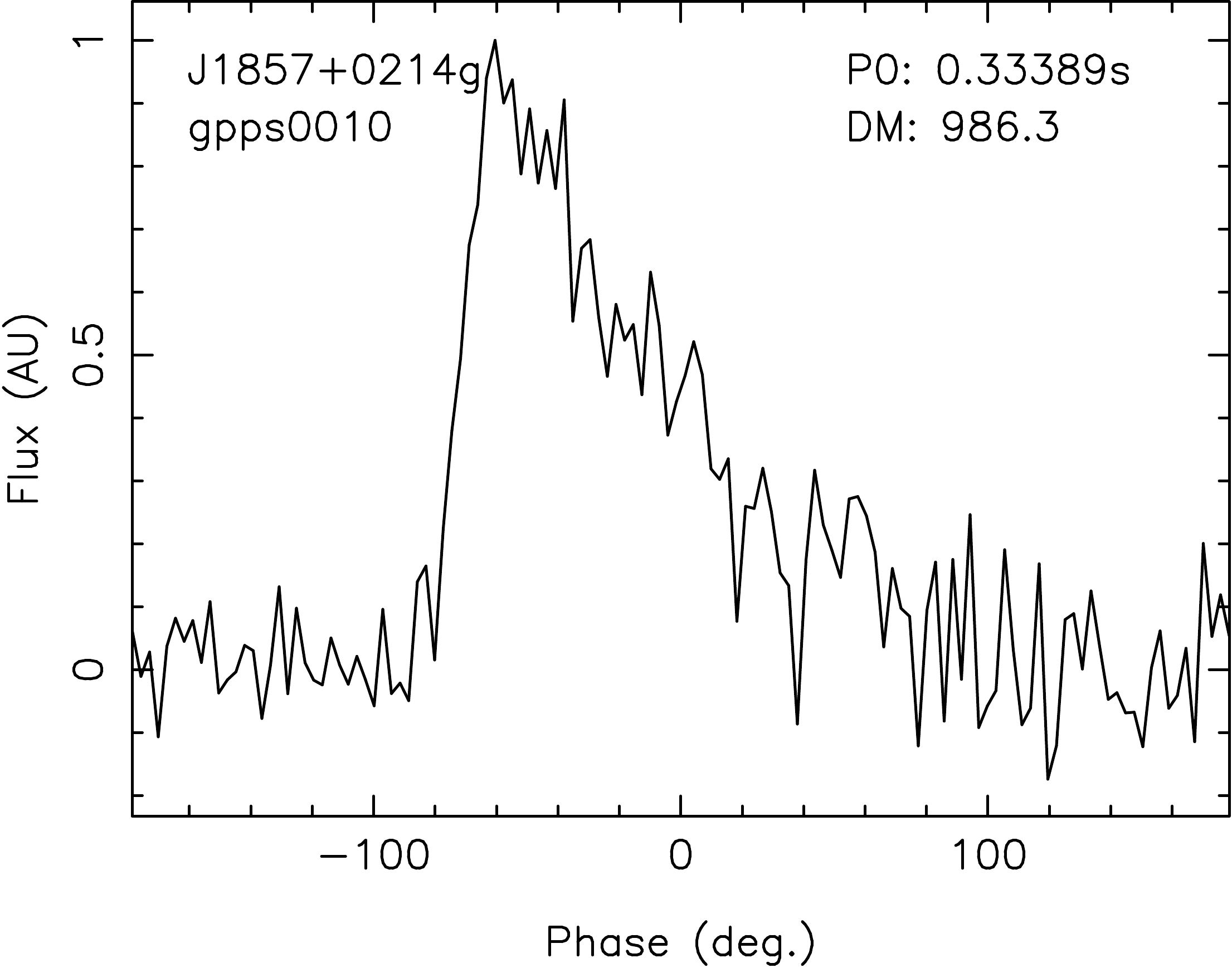}&
\includegraphics[width=39mm]{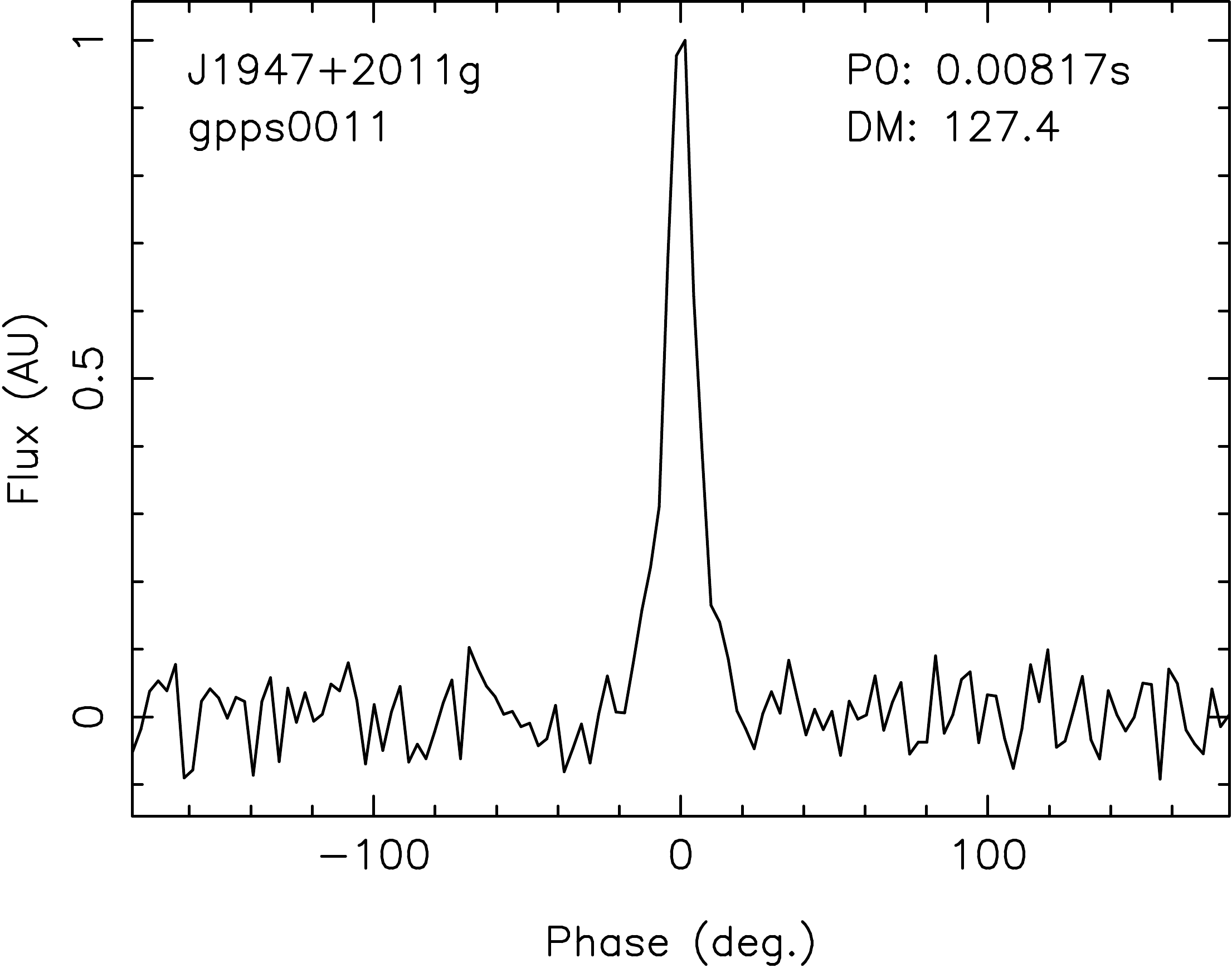}&
\includegraphics[width=39mm]{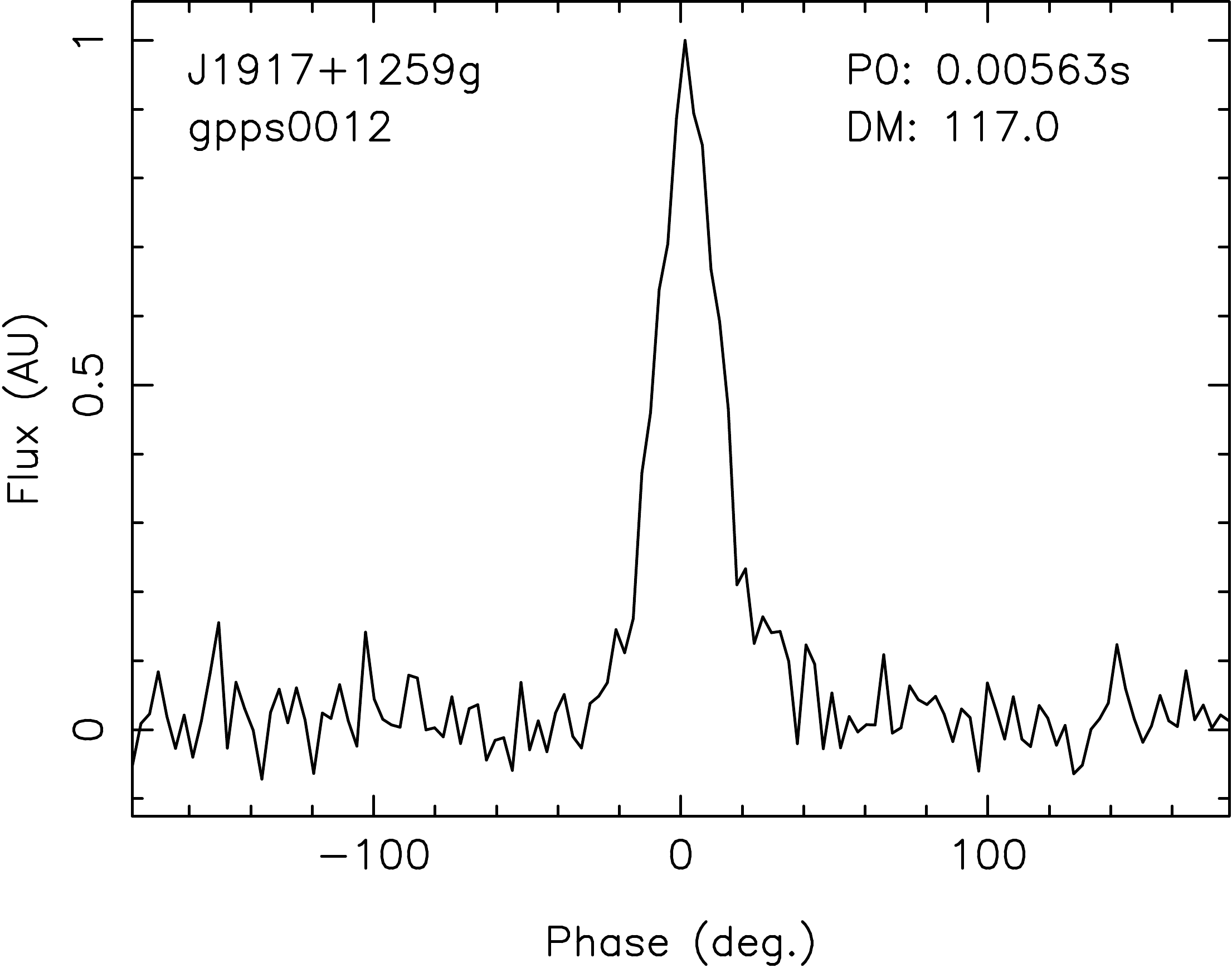}\\[2mm]
\includegraphics[width=39mm]{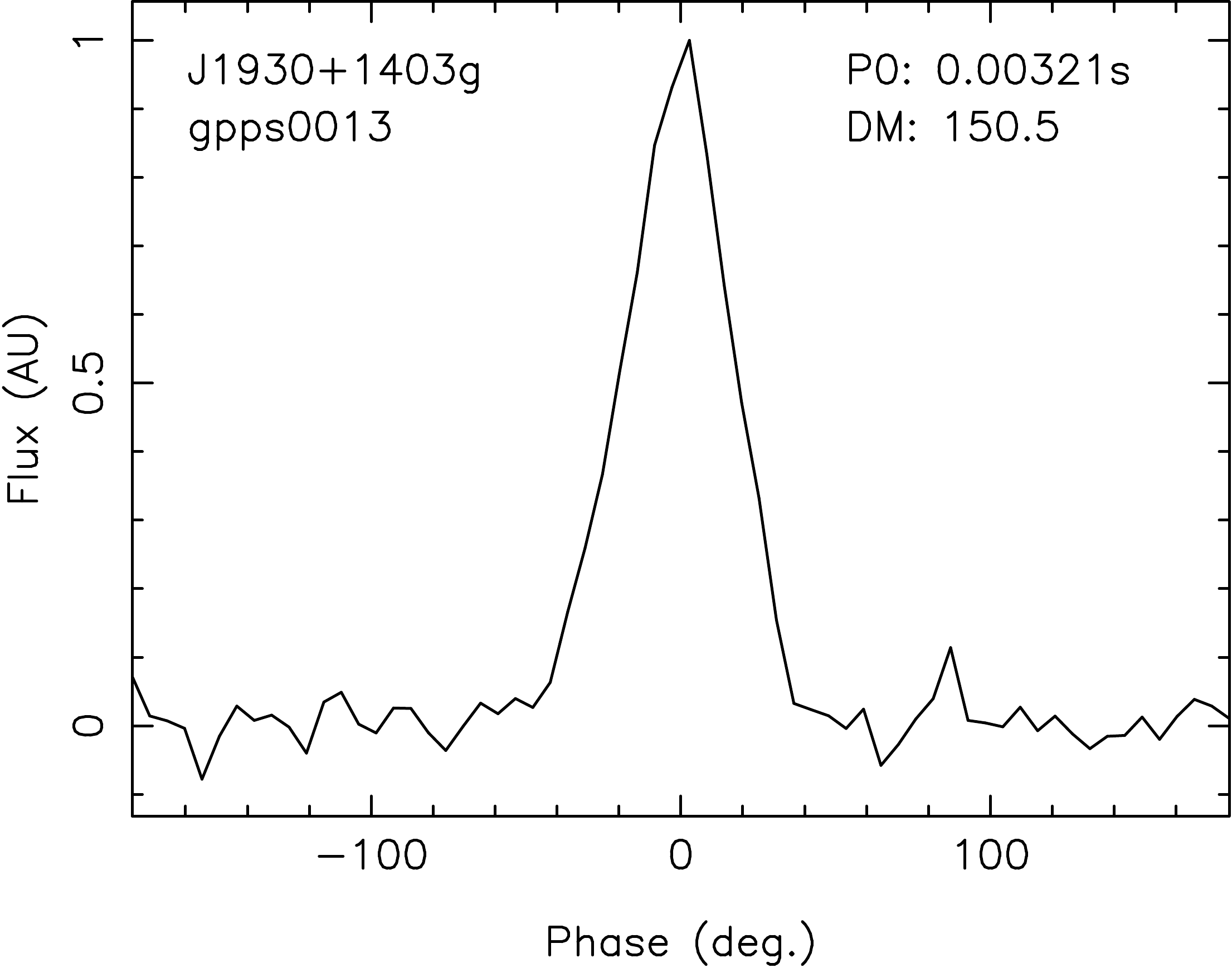}&
\includegraphics[width=39mm]{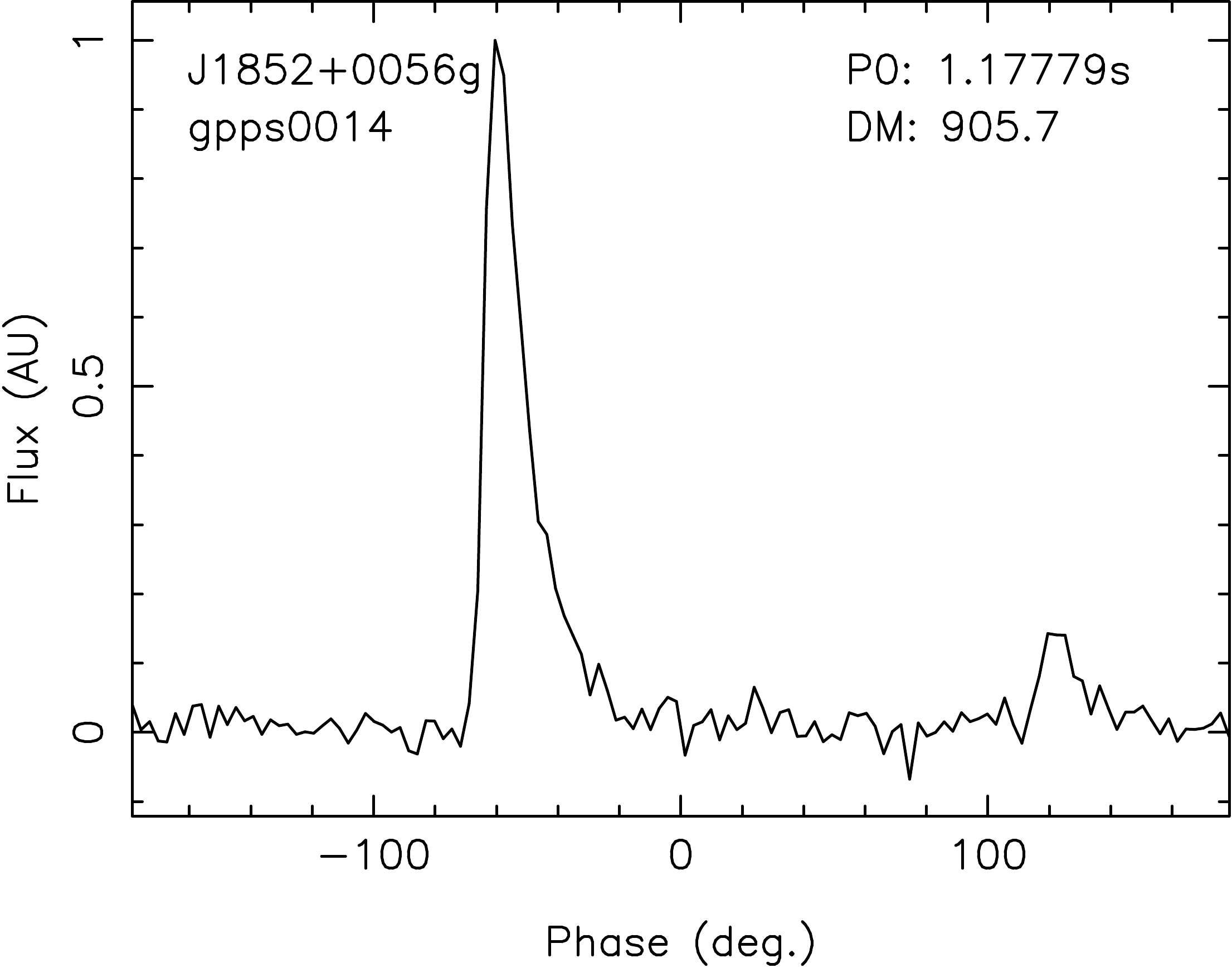}&
\includegraphics[width=39mm]{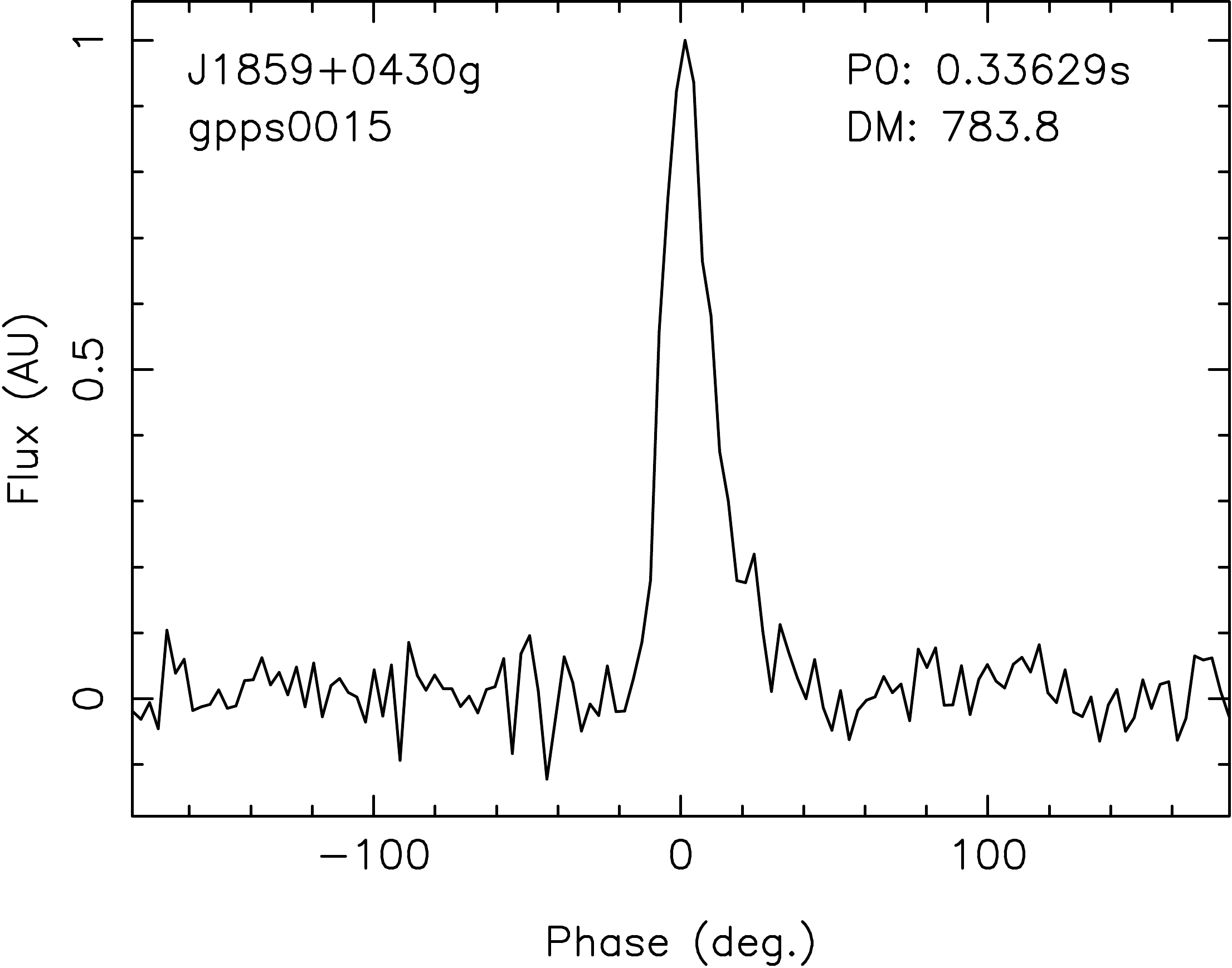}&
\includegraphics[width=39mm]{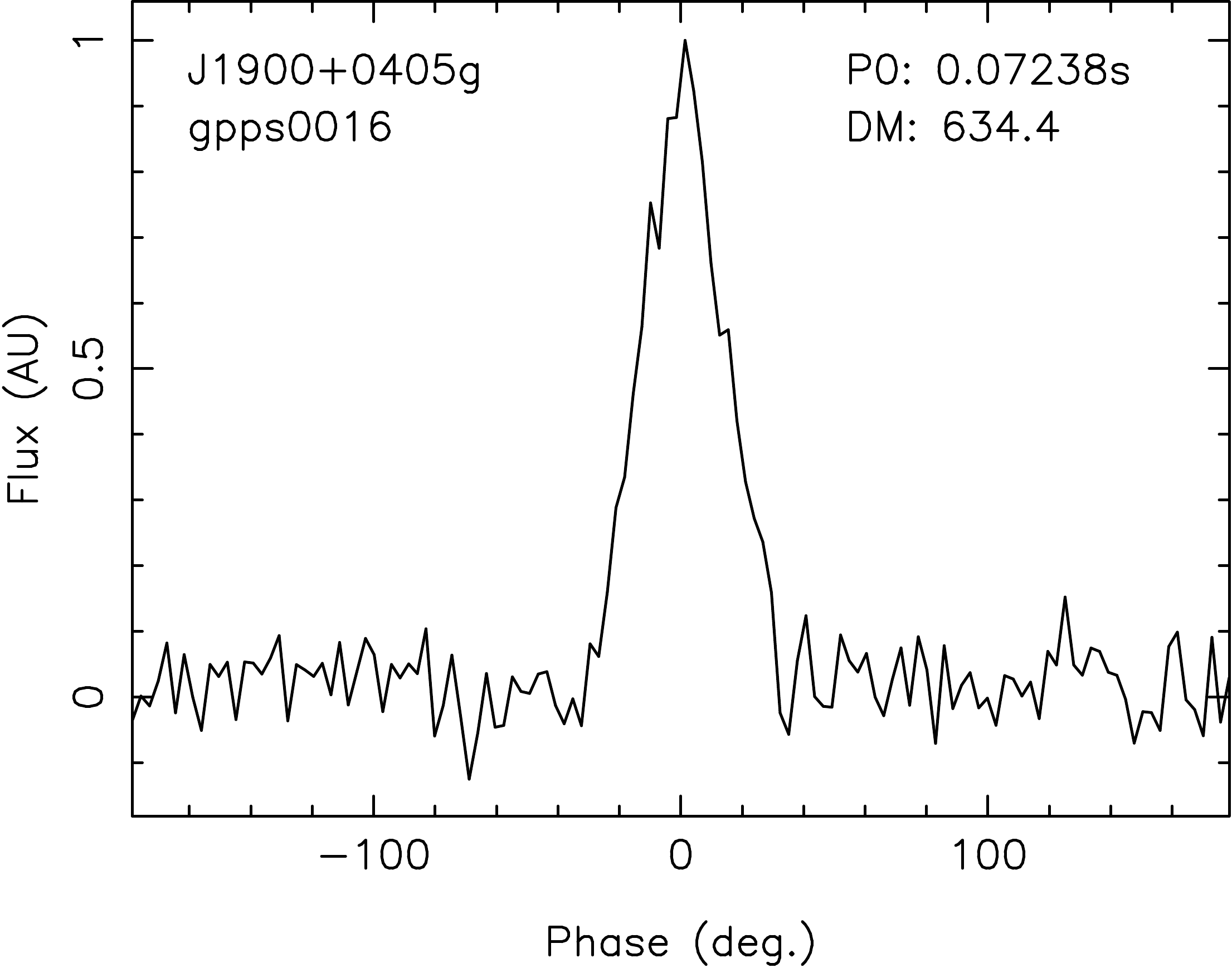}\\[2mm]
\includegraphics[width=39mm]{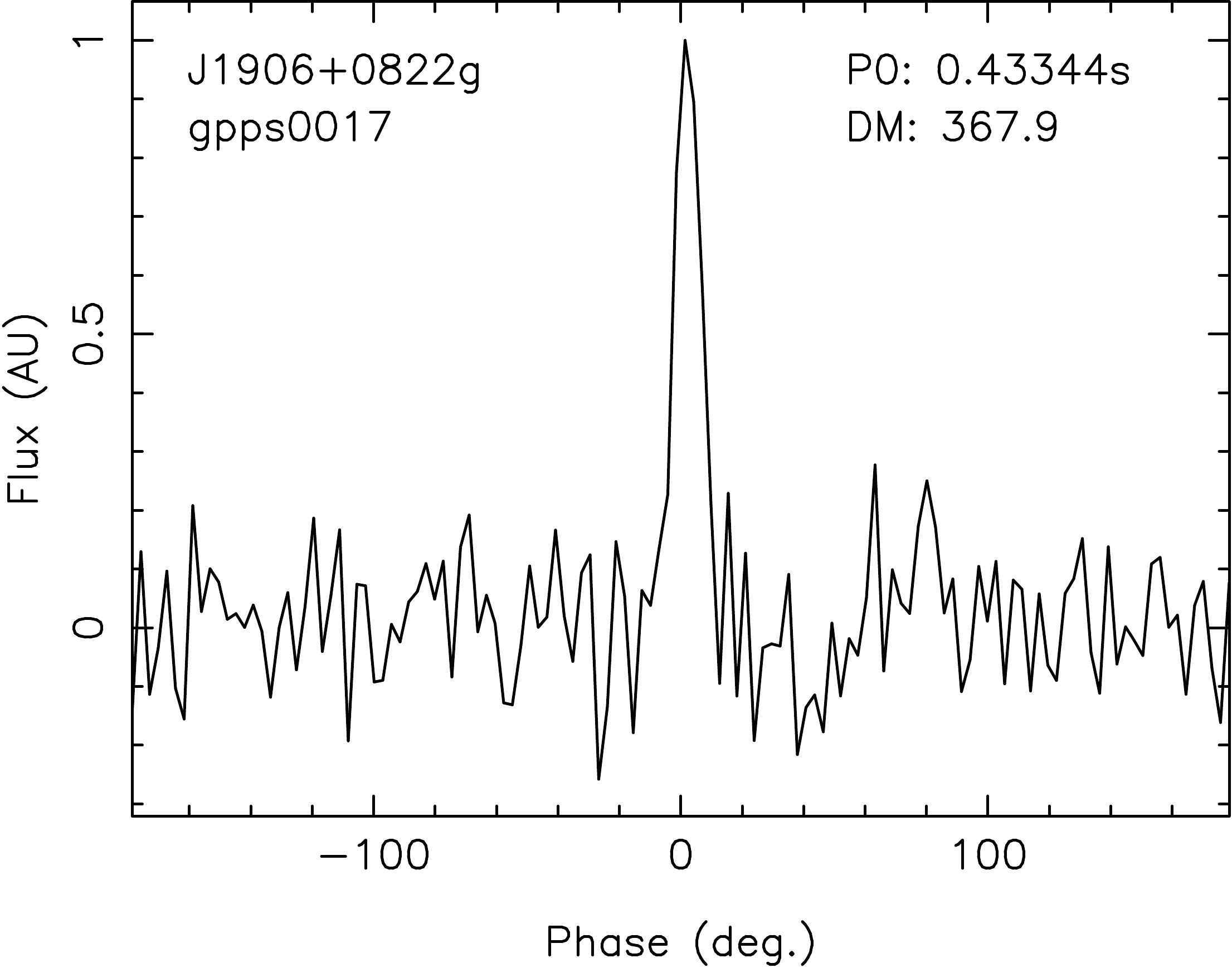}&
\includegraphics[width=39mm]{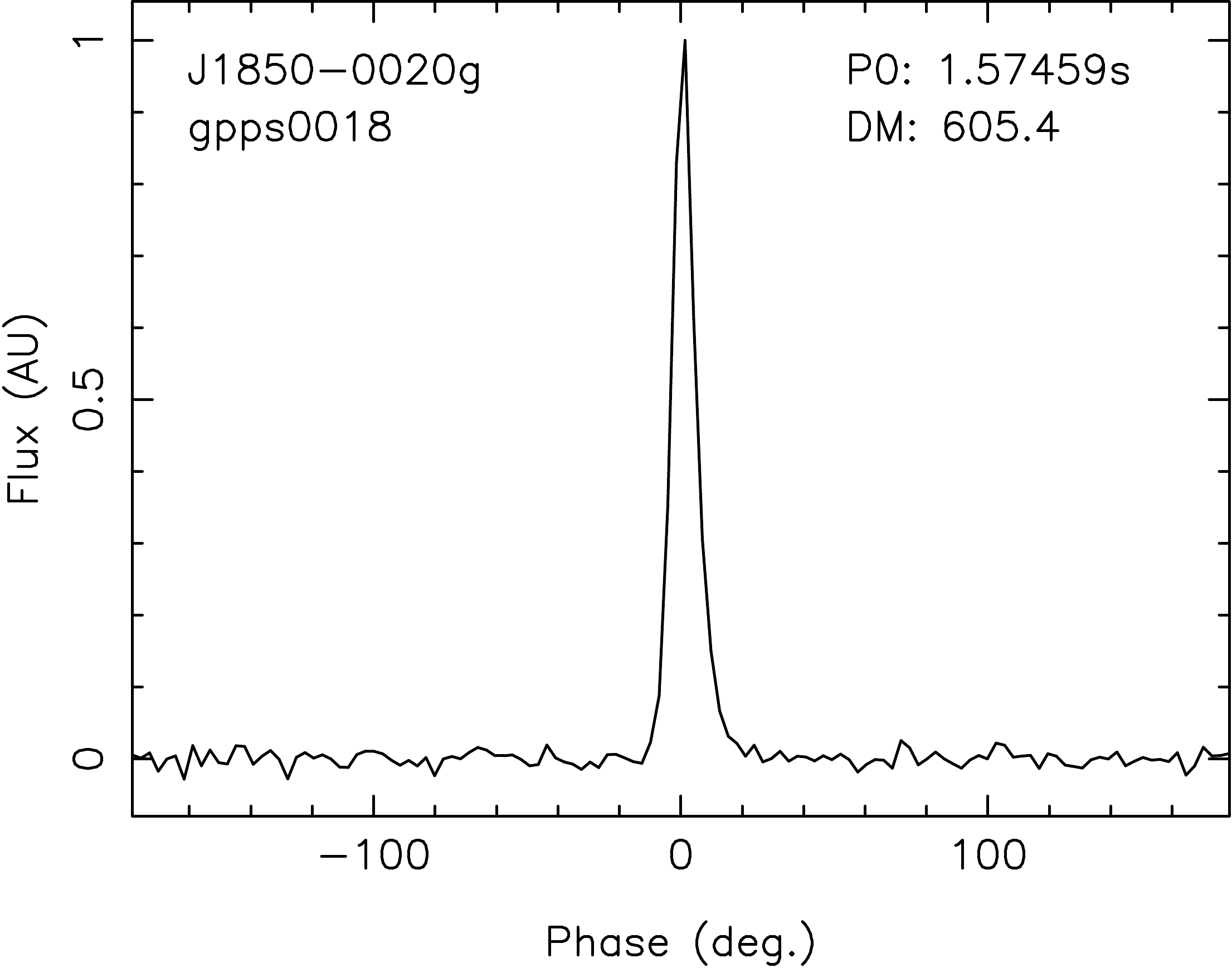}&
\includegraphics[width=39mm]{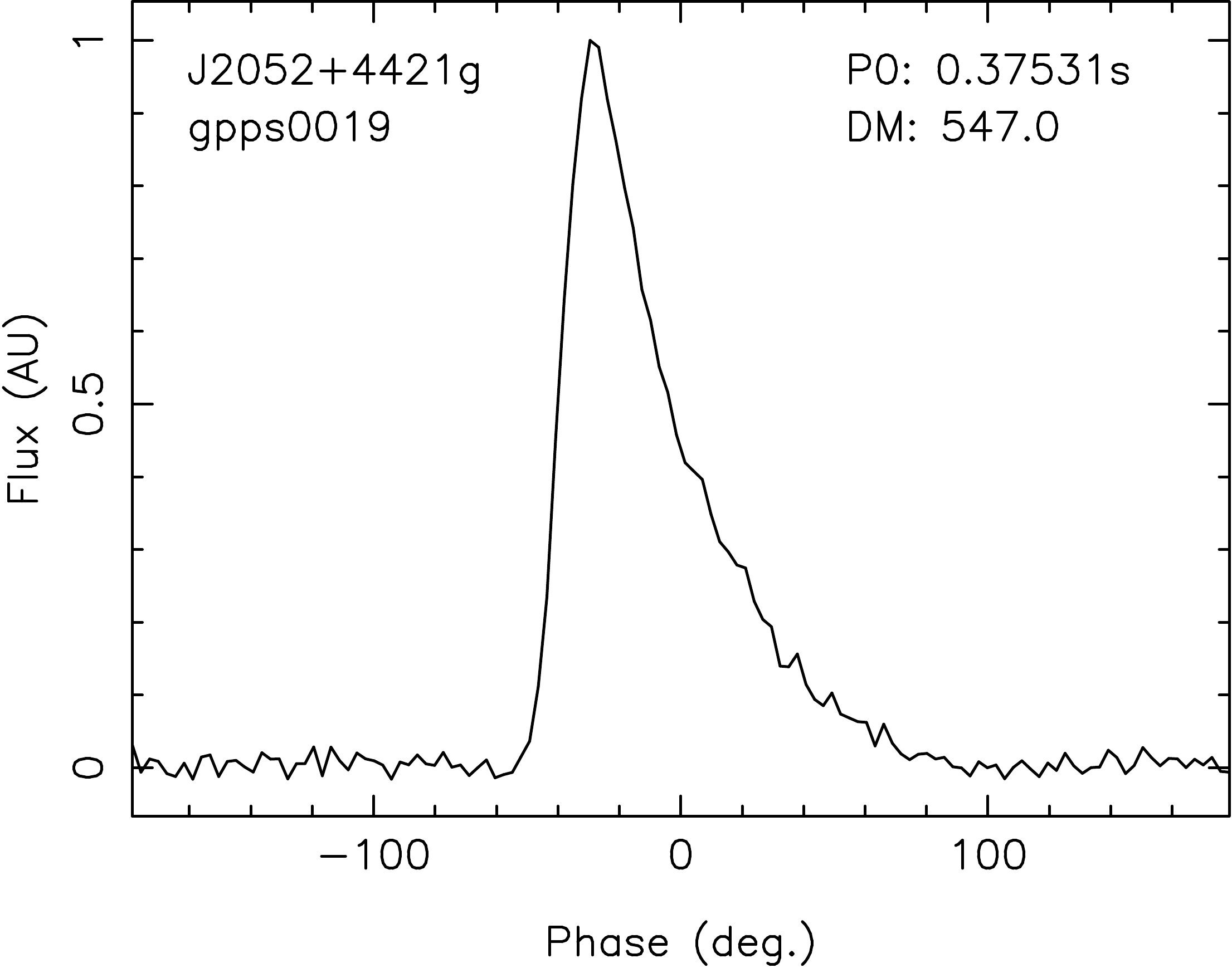}&
\includegraphics[width=39mm]{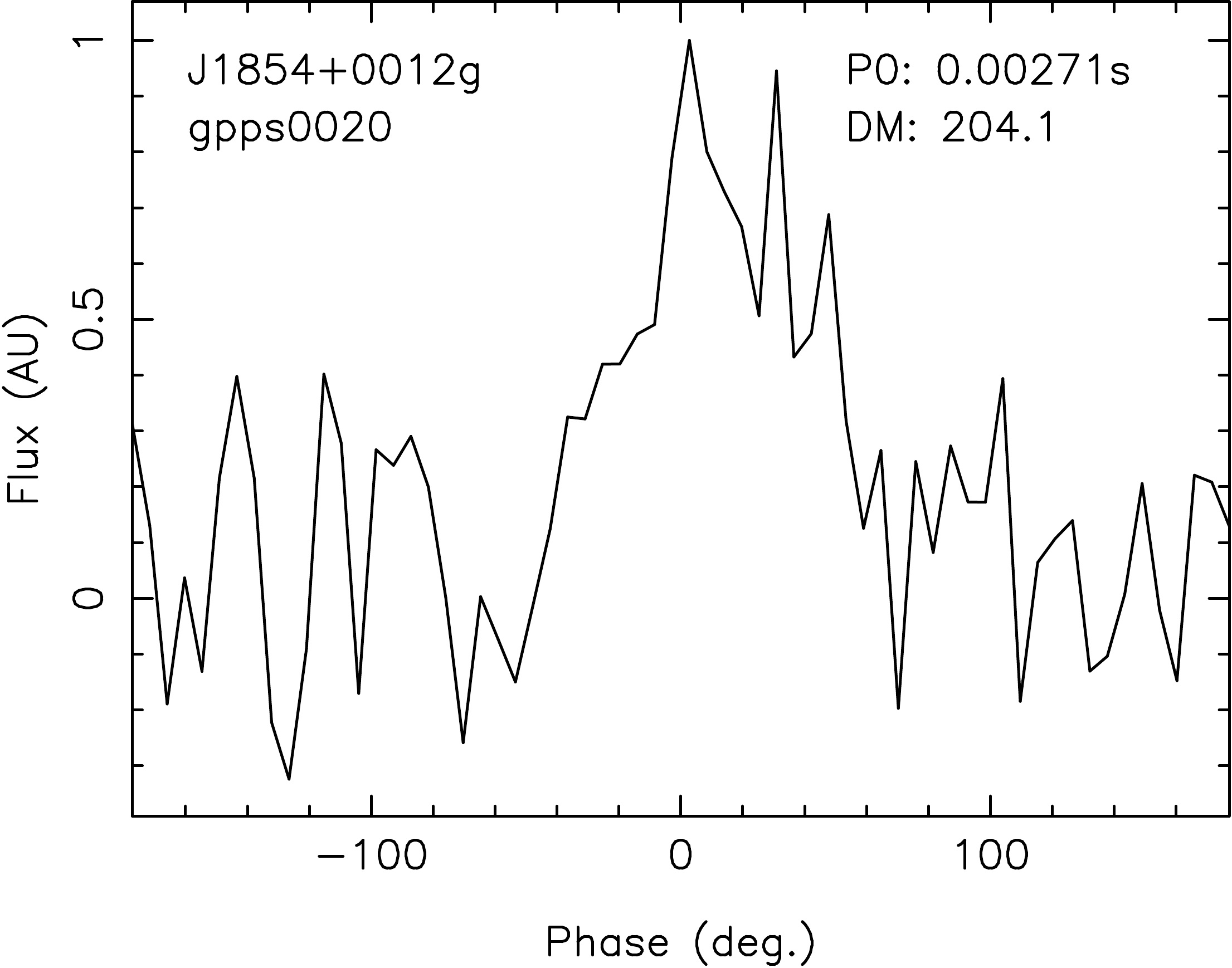}\\[2mm]
\includegraphics[width=39mm]{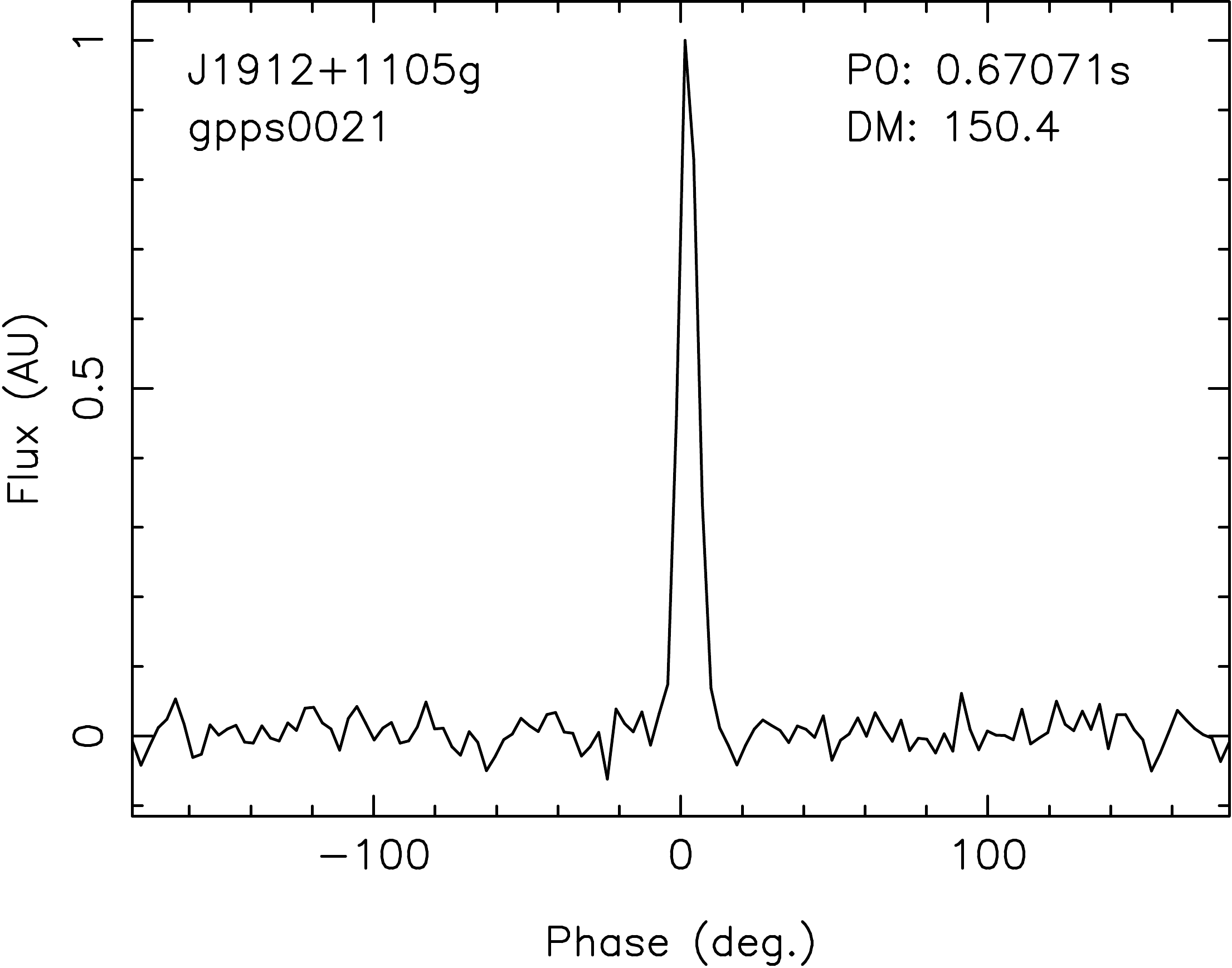}&
\includegraphics[width=39mm]{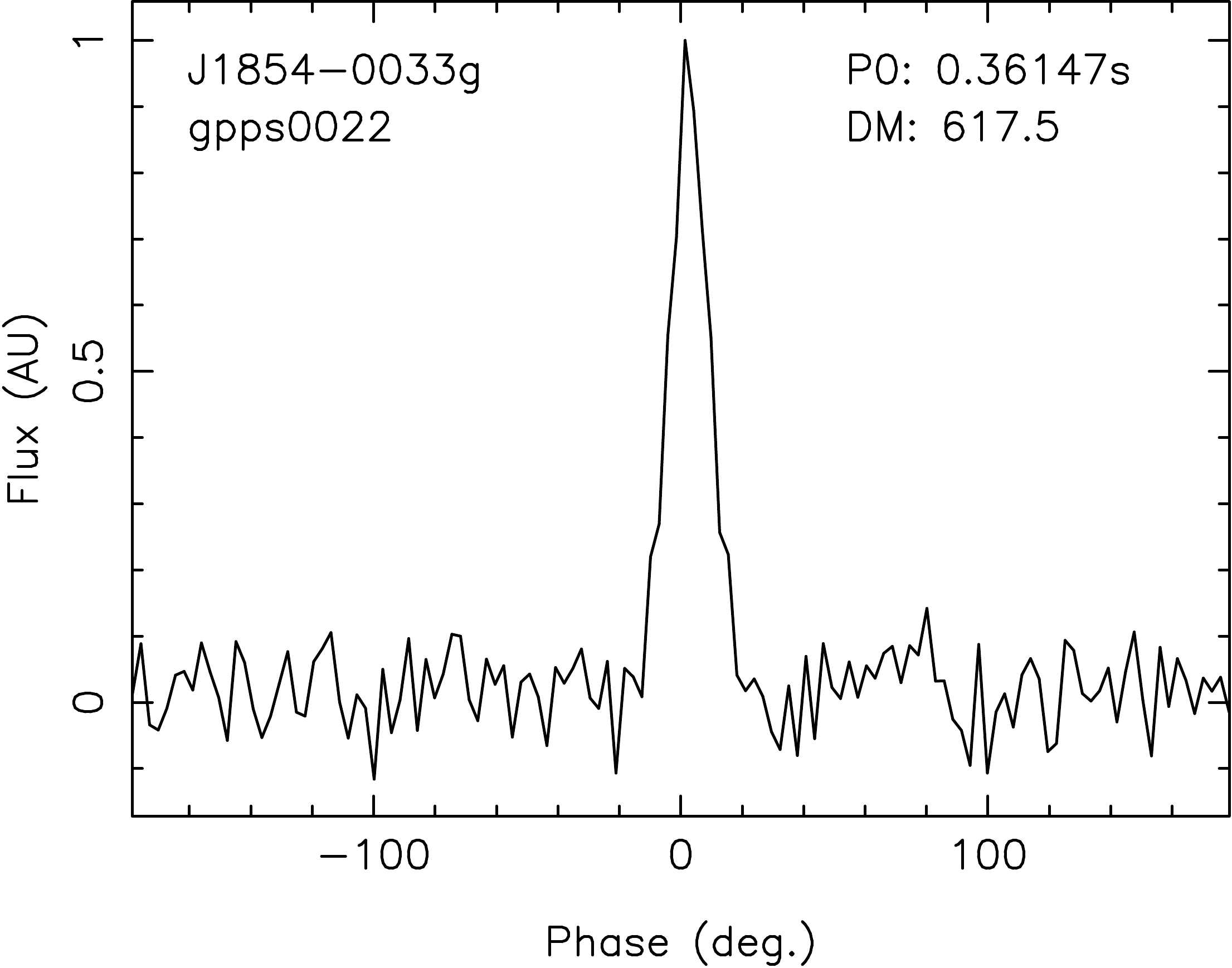}&
\includegraphics[width=39mm]{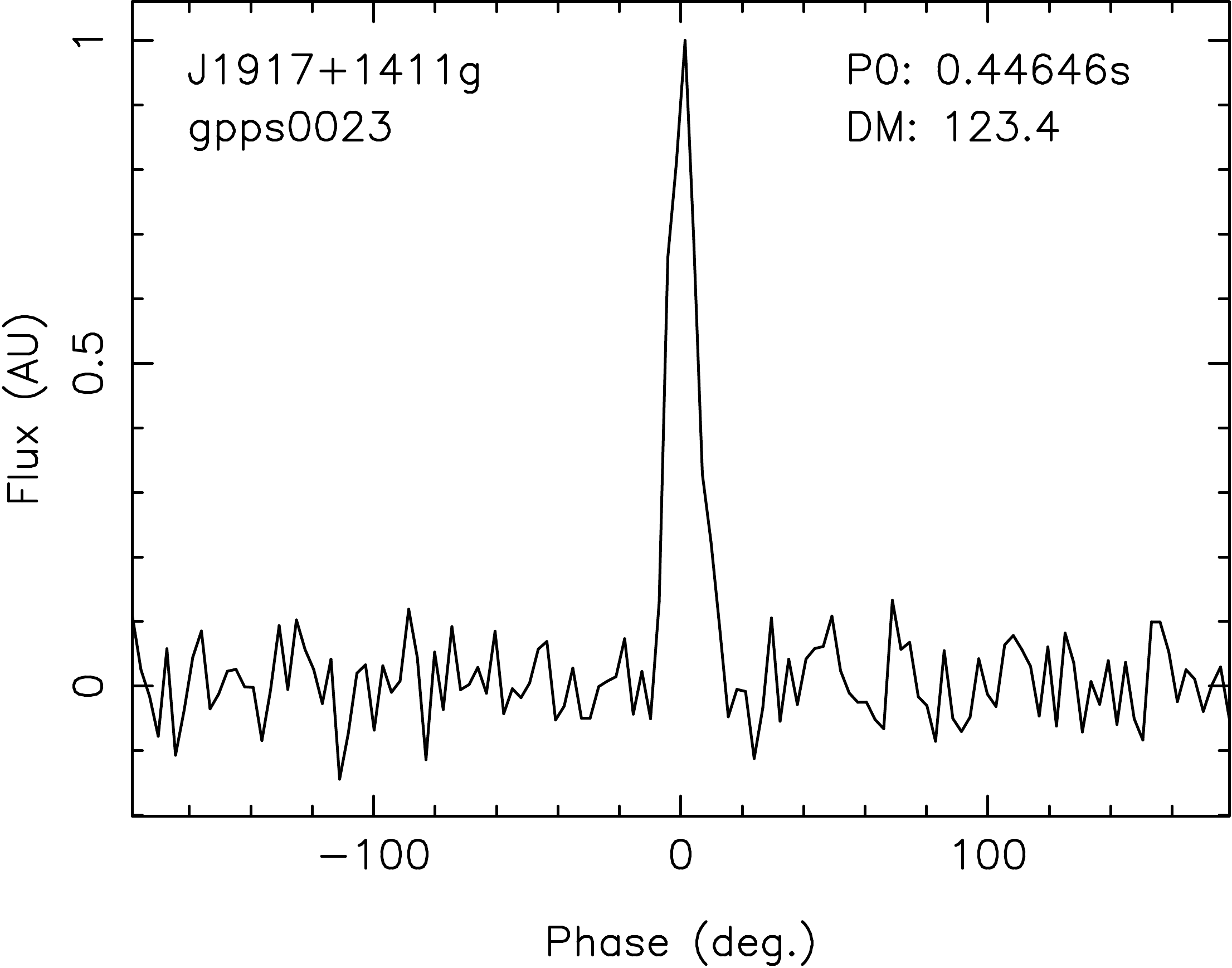}&
\includegraphics[width=39mm]{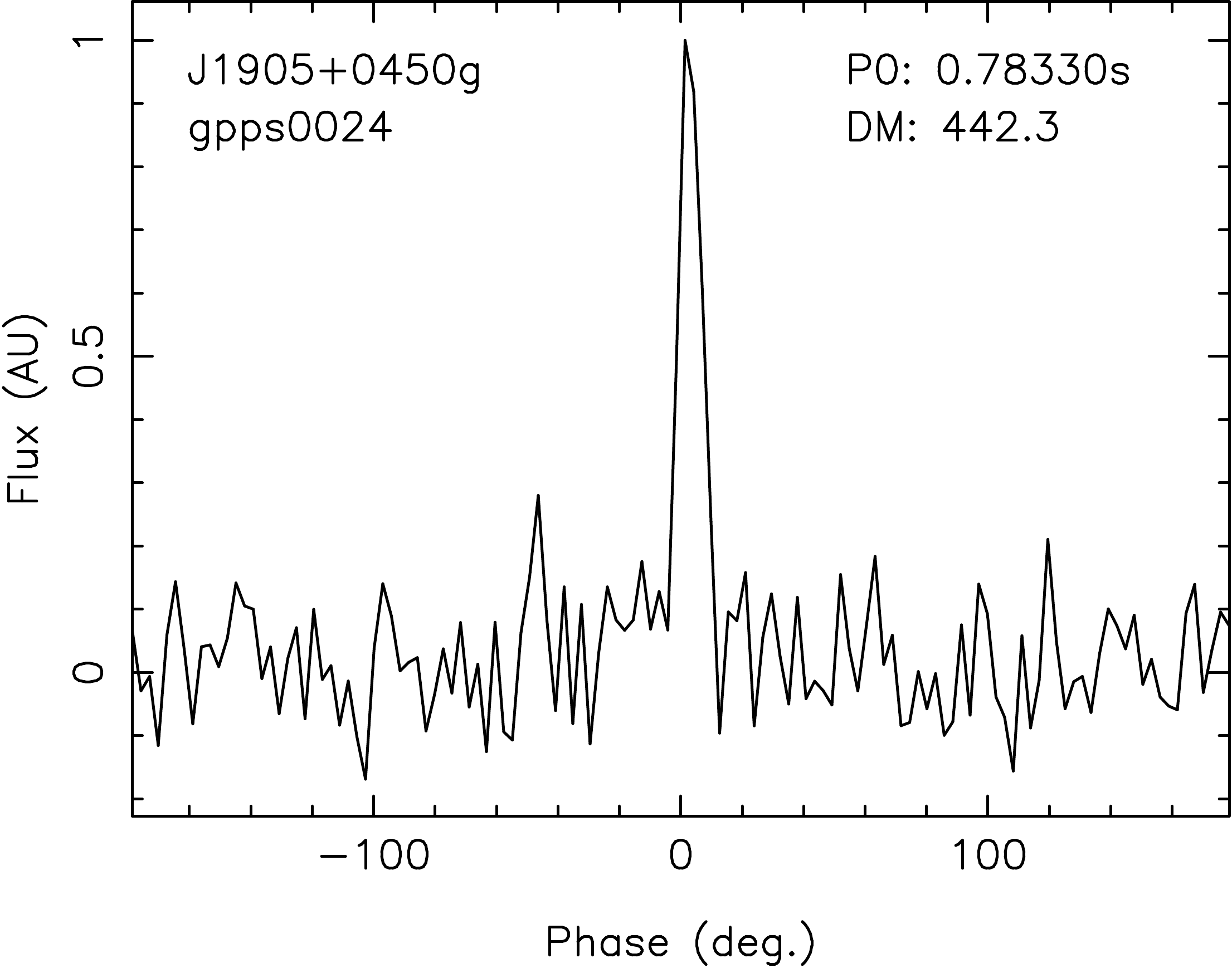}\\[2mm]
\includegraphics[width=39mm]{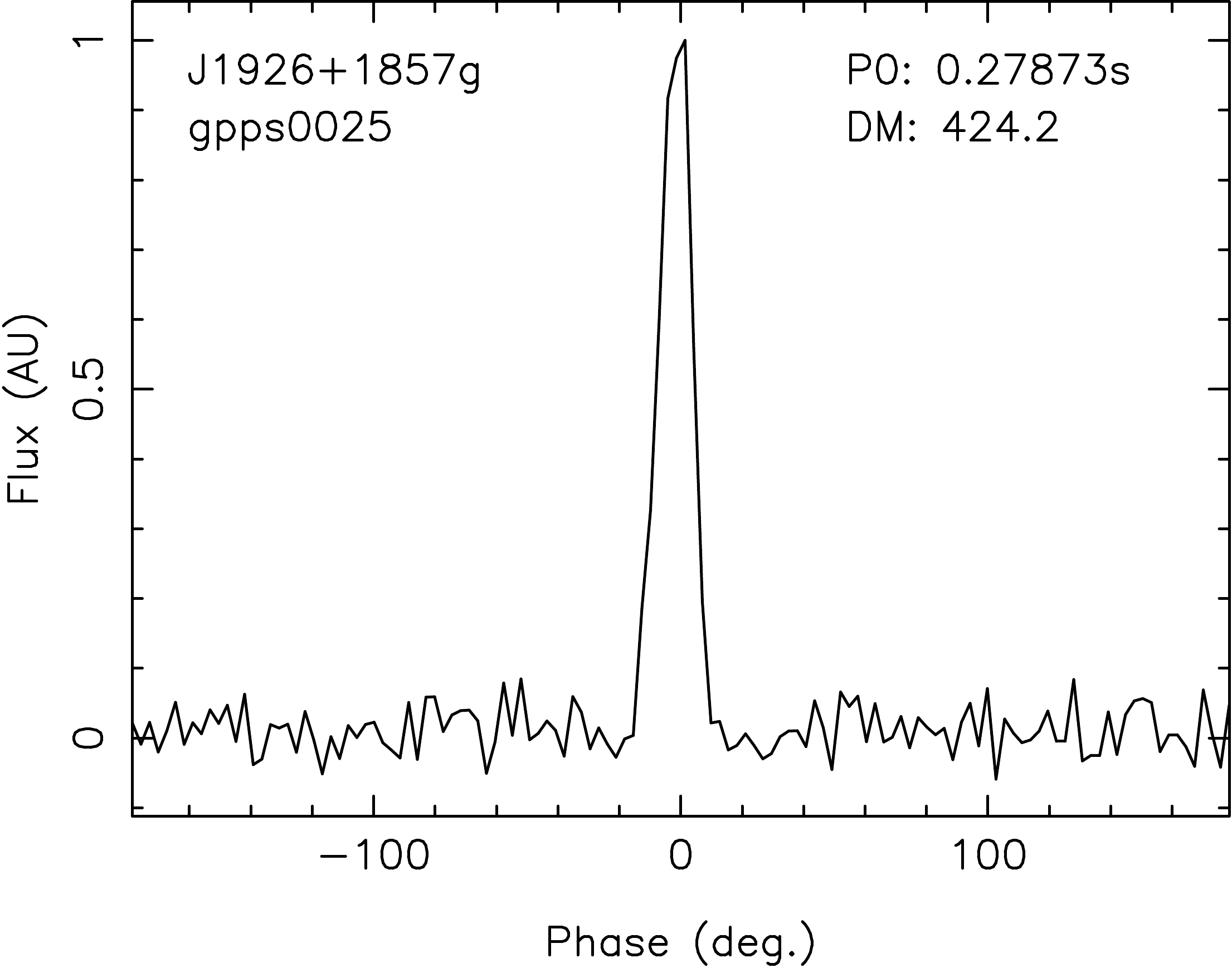}&
\includegraphics[width=39mm]{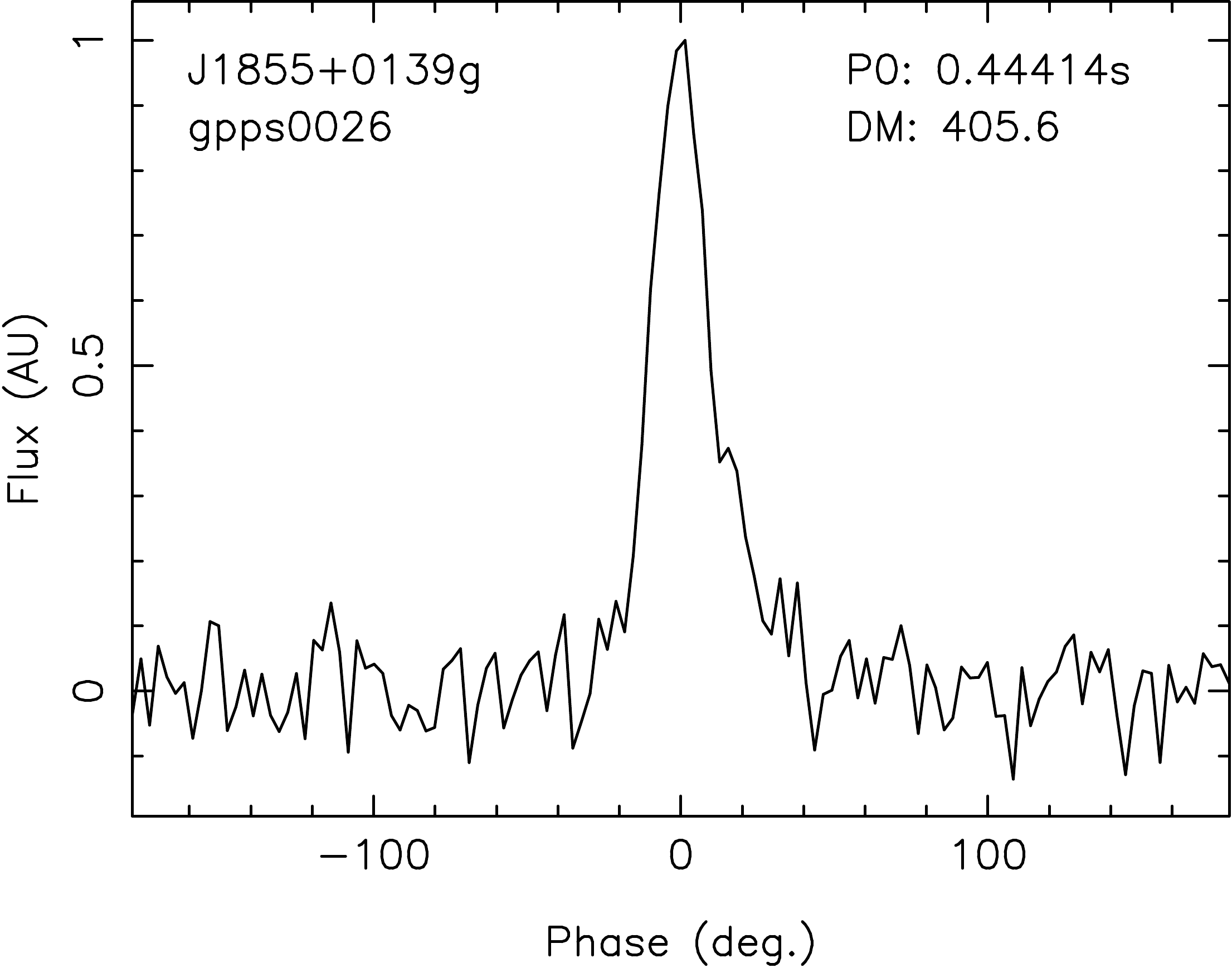}&
\includegraphics[width=39mm]{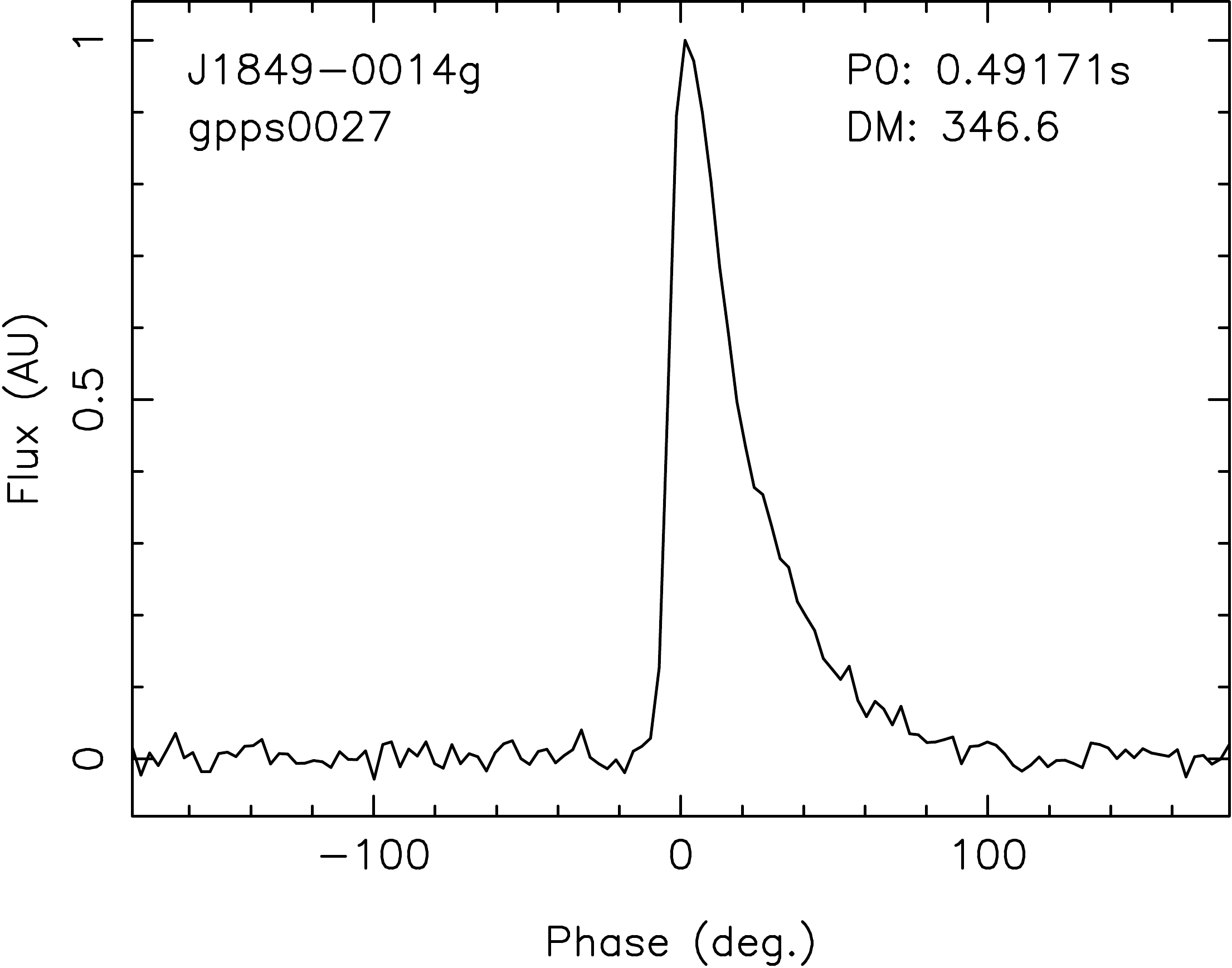}&
\includegraphics[width=39mm]{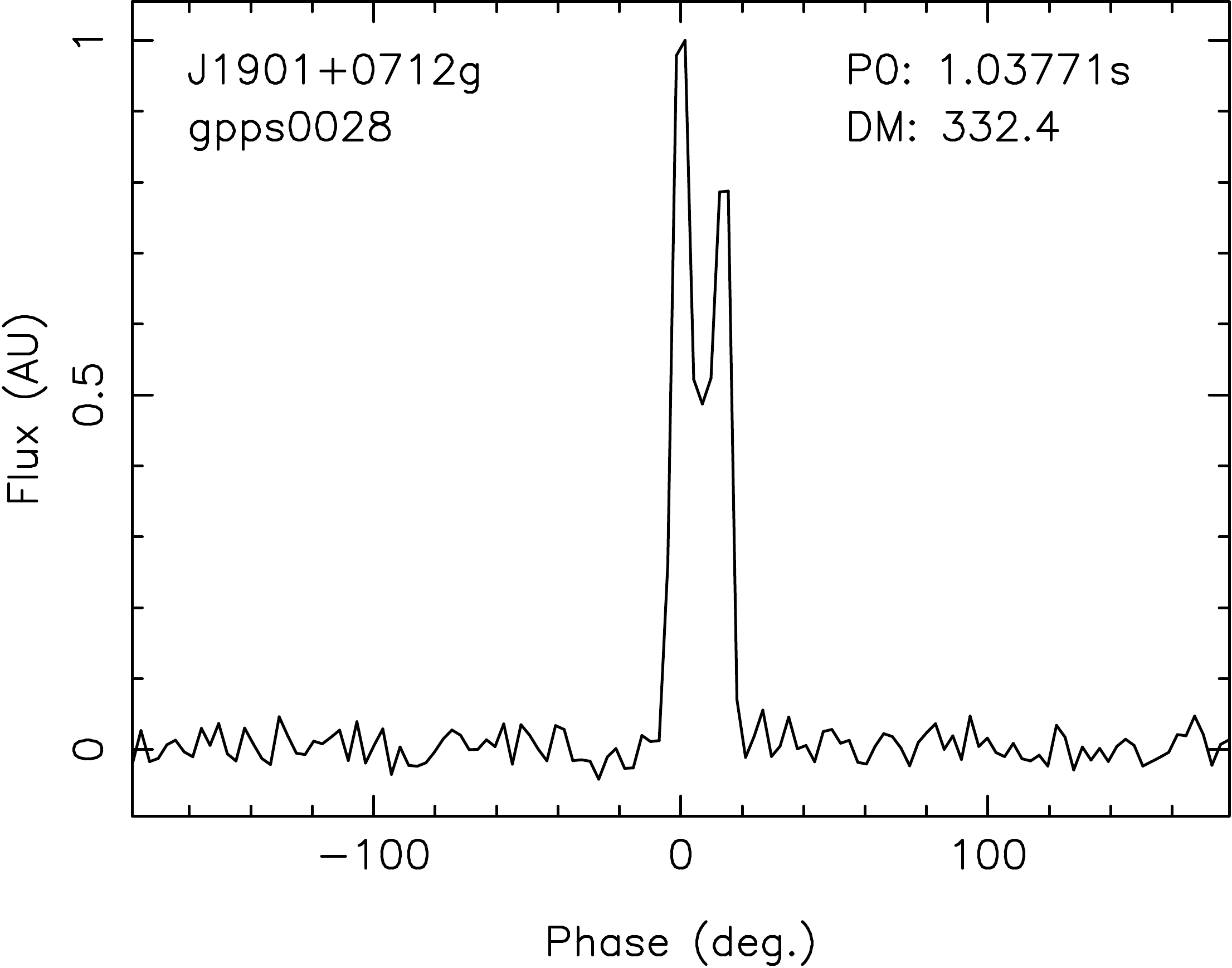}\\
\end{tabular}%
\caption[]{\baselineskip 3.8mm Integrated profiles of newly discovered pulsars, scaled to
  the peak and plotted in rotation phase of 360$^{\circ}$ for a full
  period. The pulsar name, gpps number, period and DM are noted in
  each panel. }
\label{fig11_gppsPSRprof}
\addtocounter{figure}{-1}
\end{figure*}%
\begin{figure*}
\centering
\begin{tabular}{rrrrrr}
\includegraphics[width=39mm]{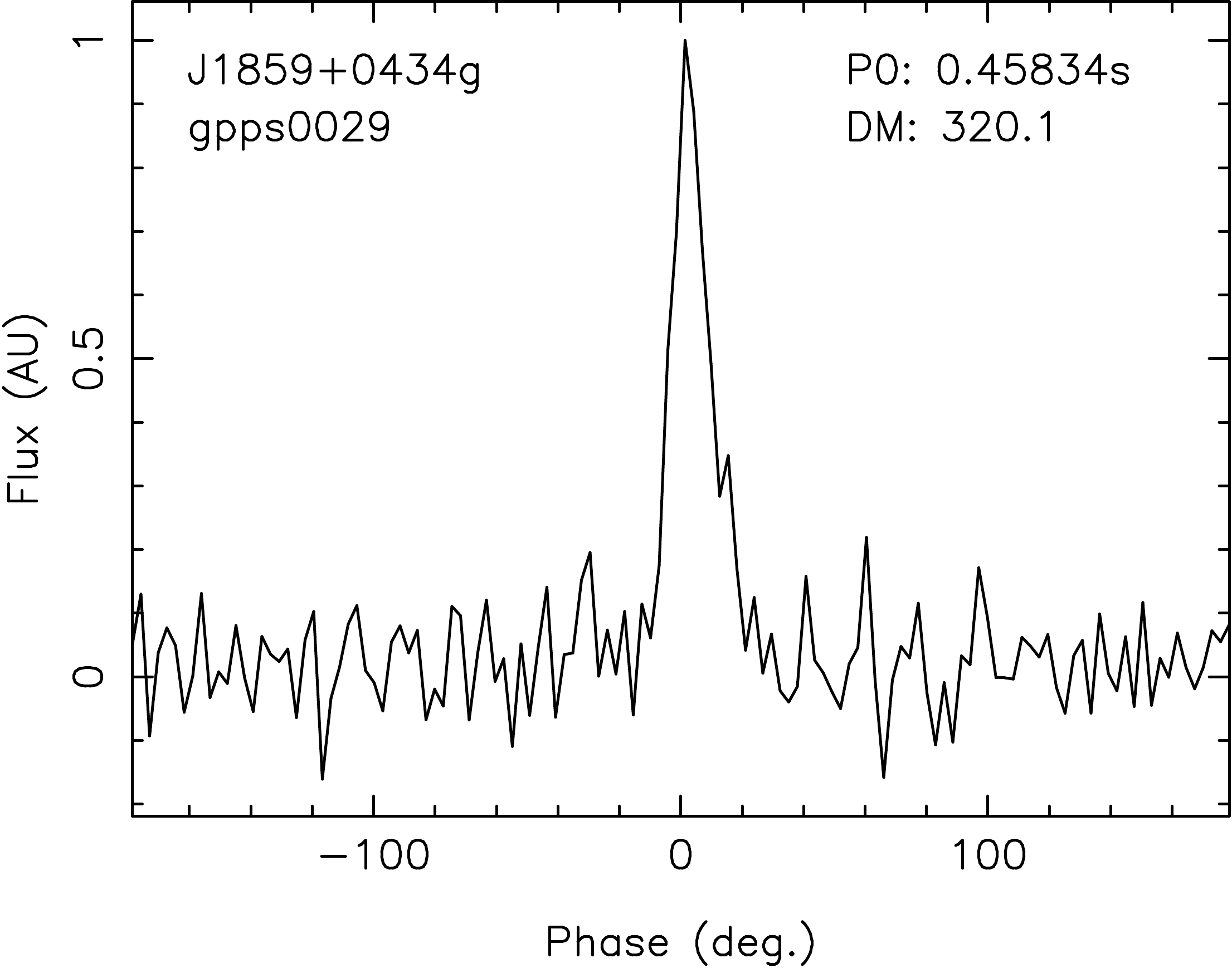}&
\includegraphics[width=39mm]{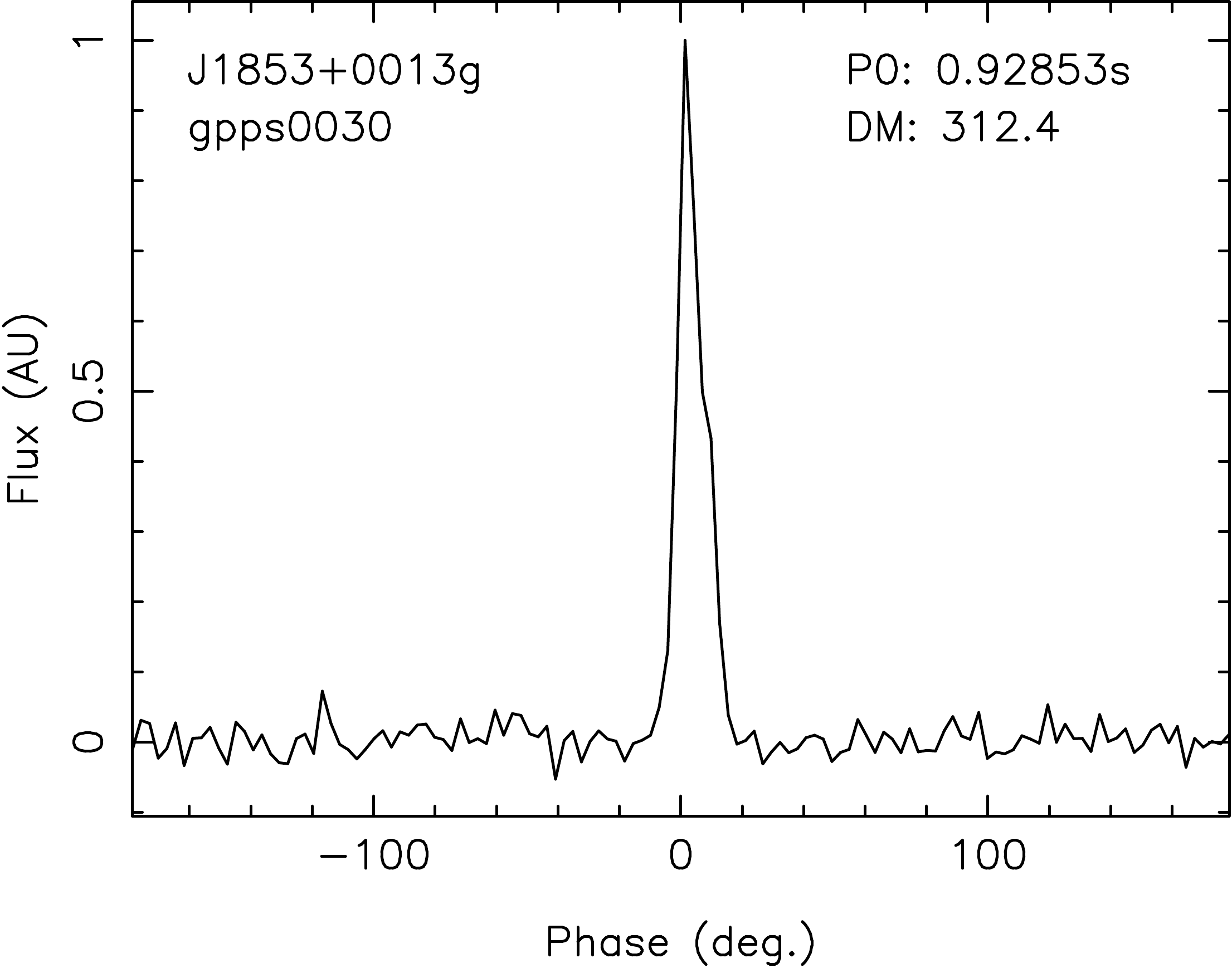}&
\includegraphics[width=39mm]{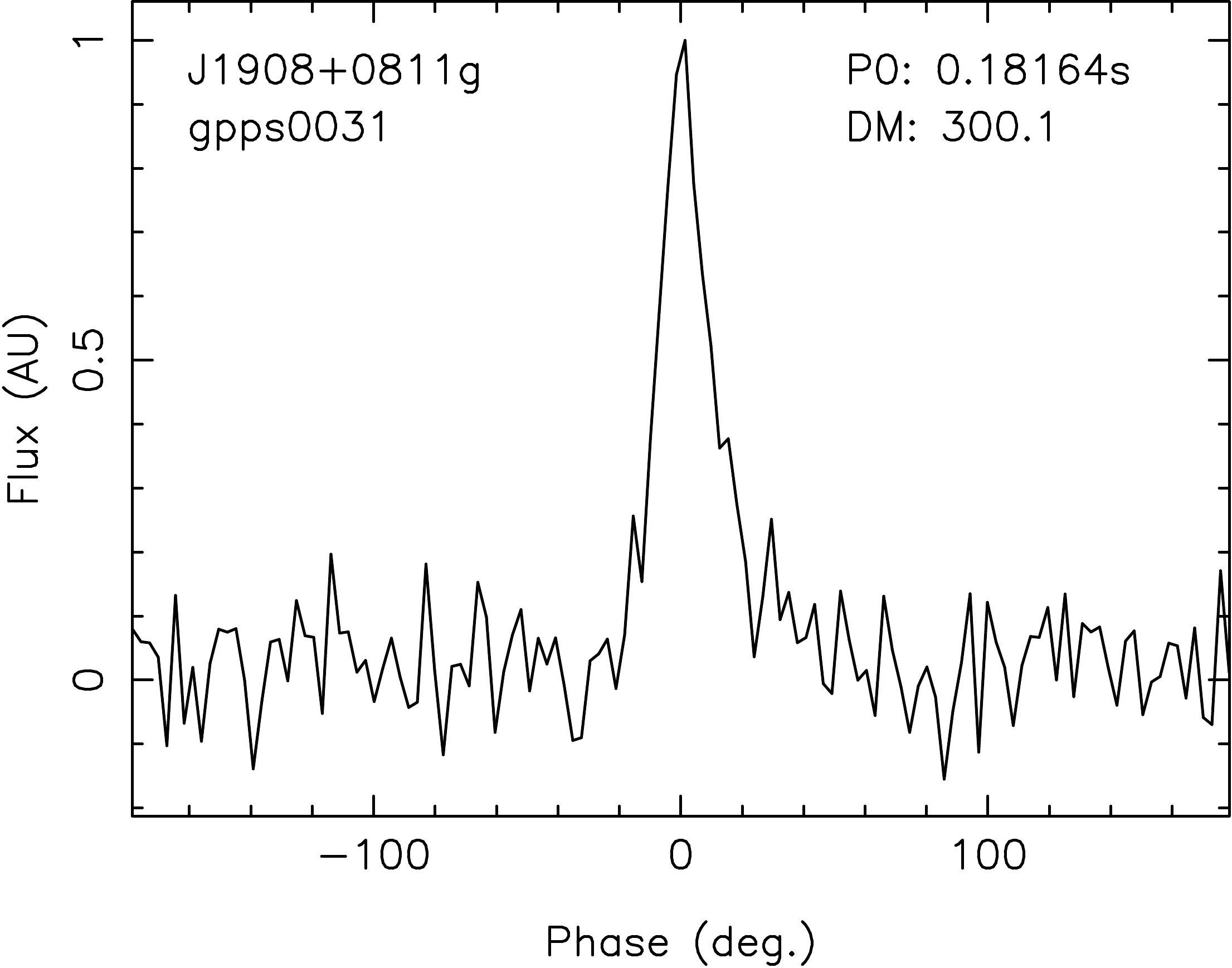}&
\includegraphics[width=39mm]{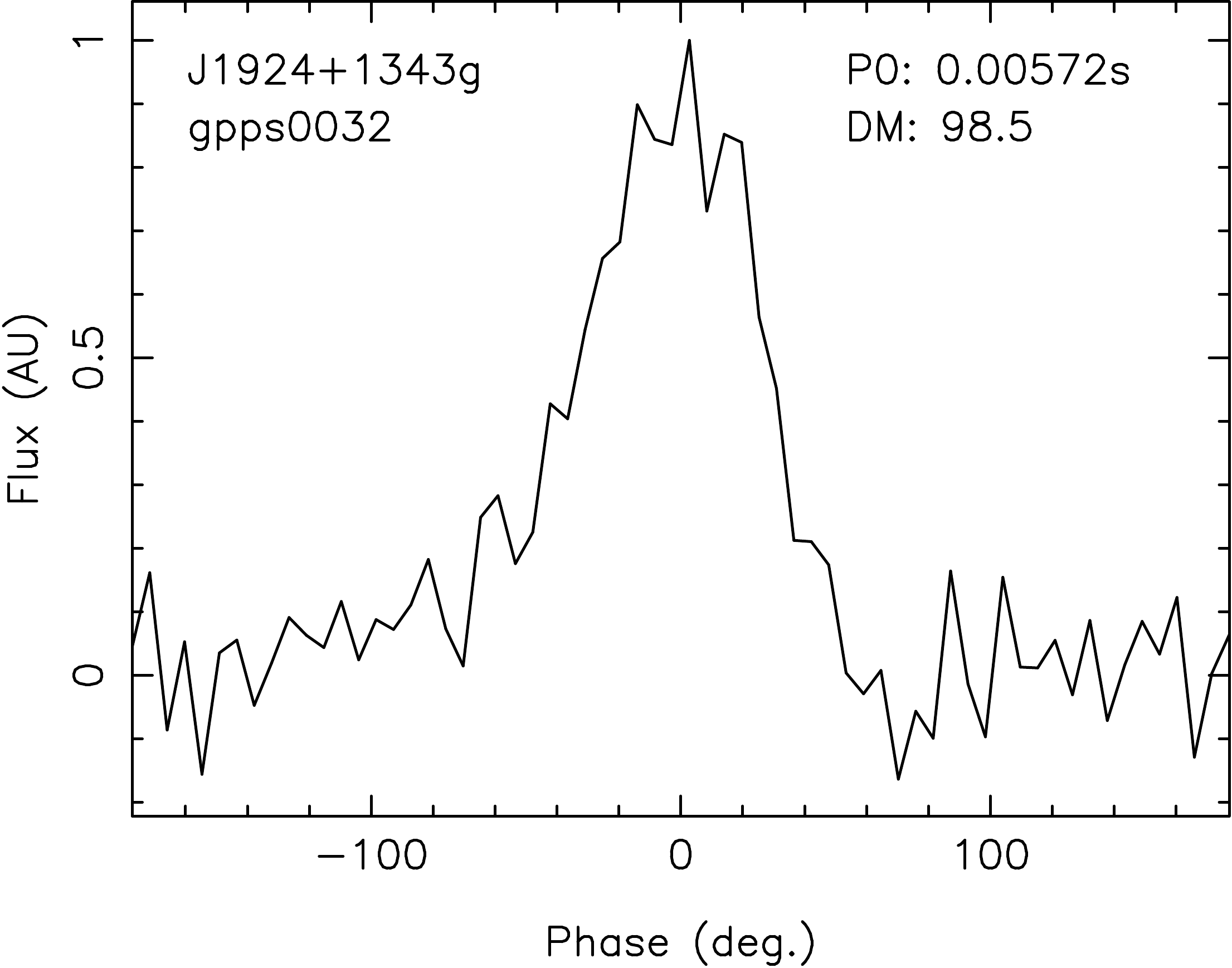}\\[2mm]
\includegraphics[width=39mm]{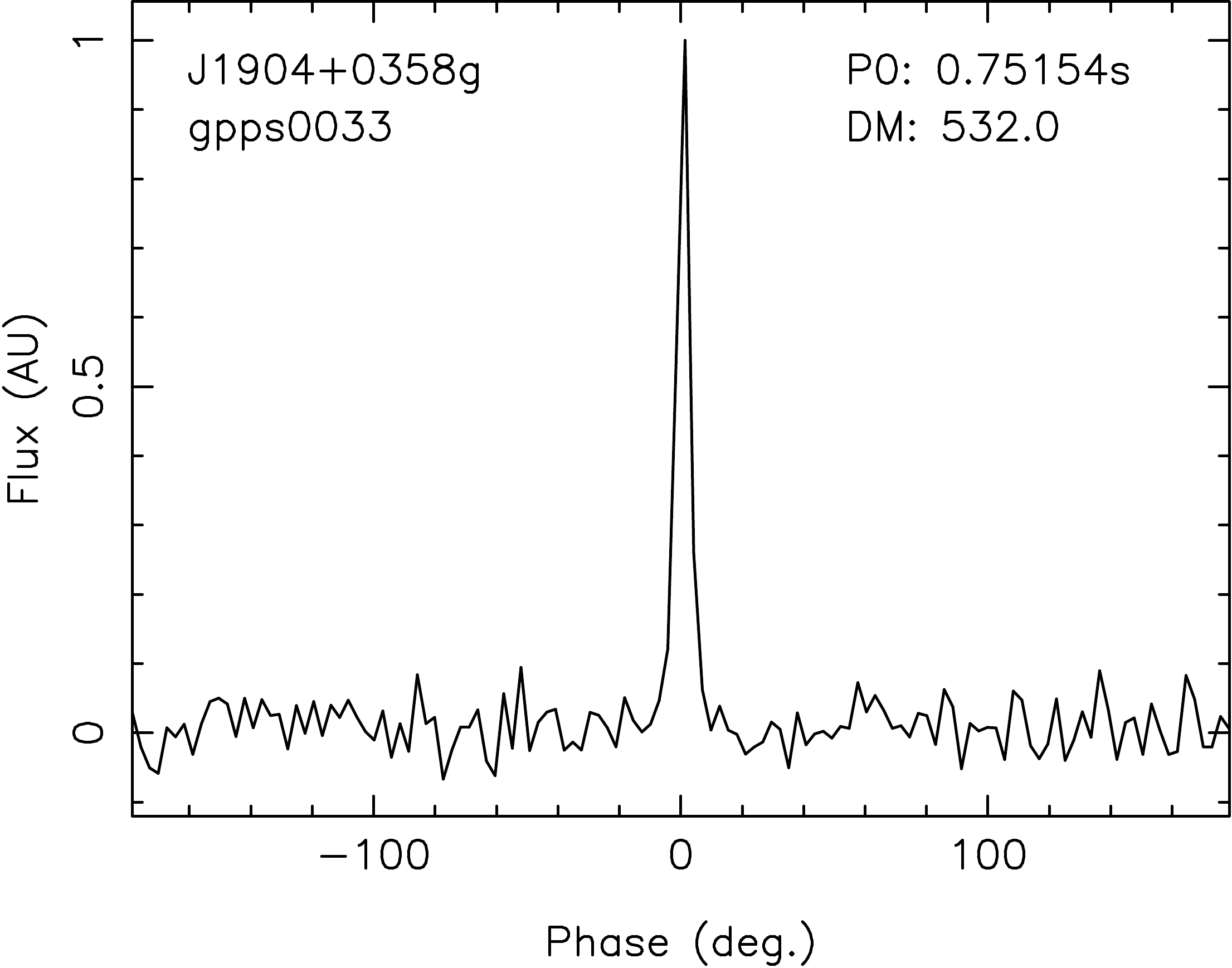}&
\includegraphics[width=39mm]{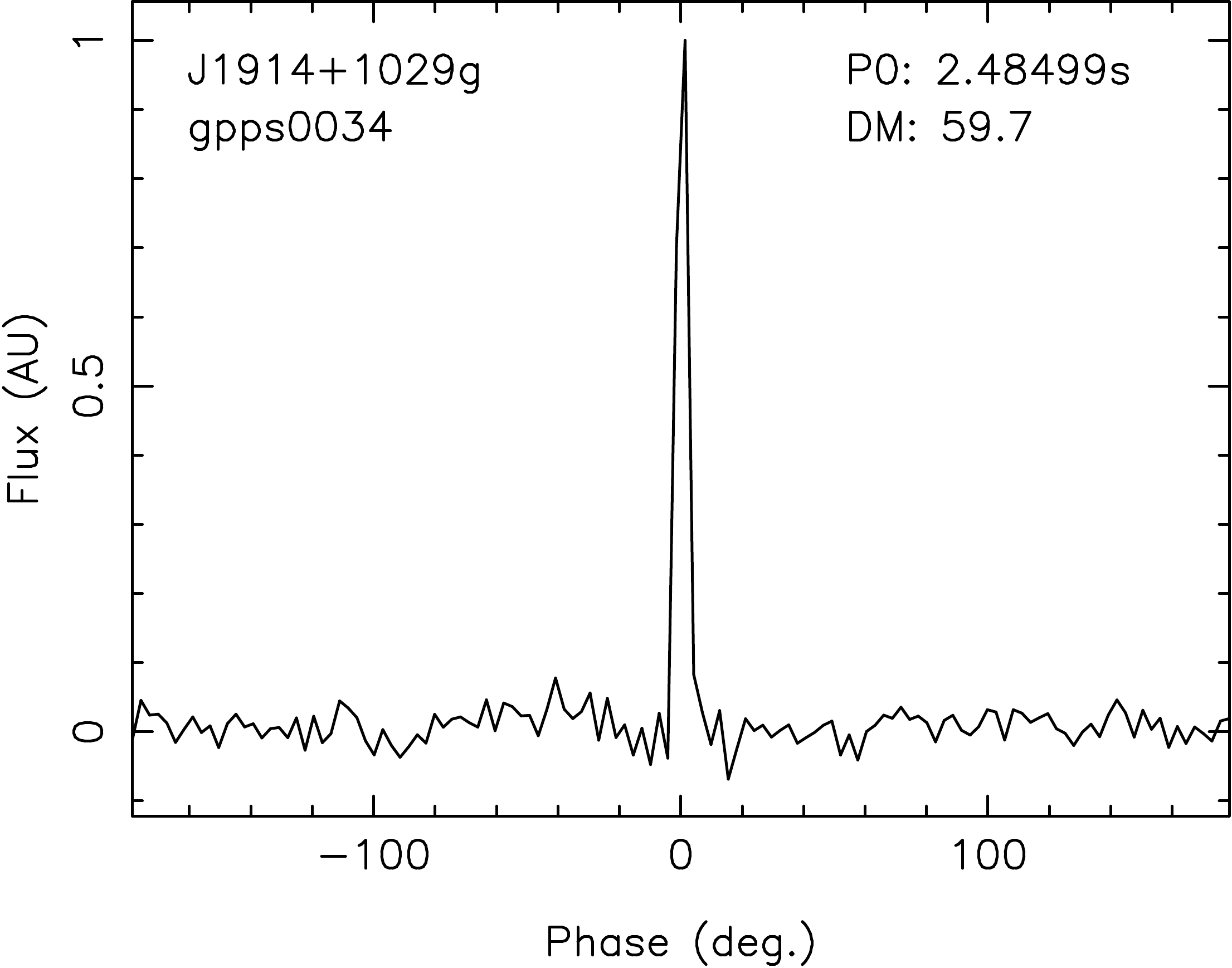}&
\includegraphics[width=39mm]{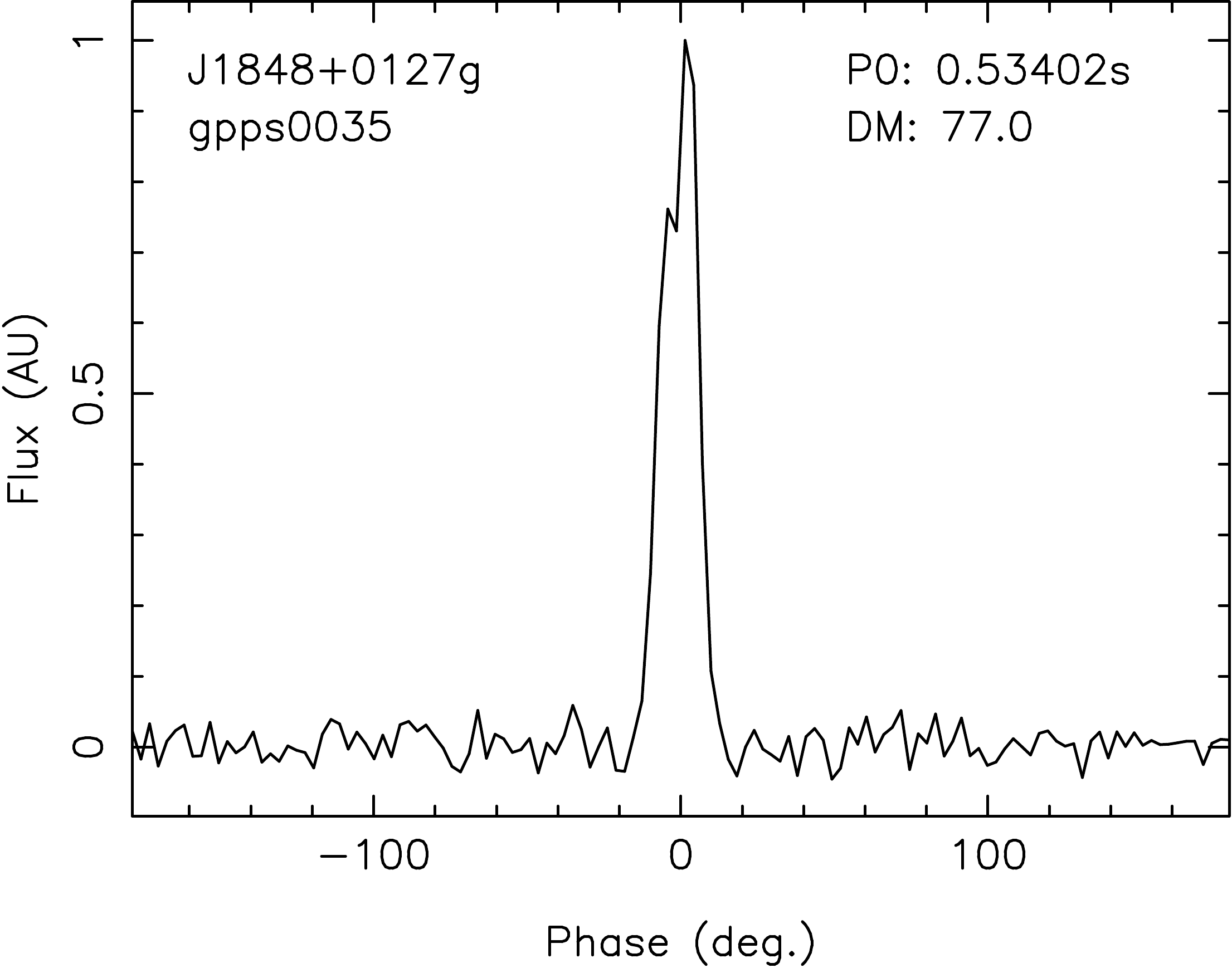}&
\includegraphics[width=39mm]{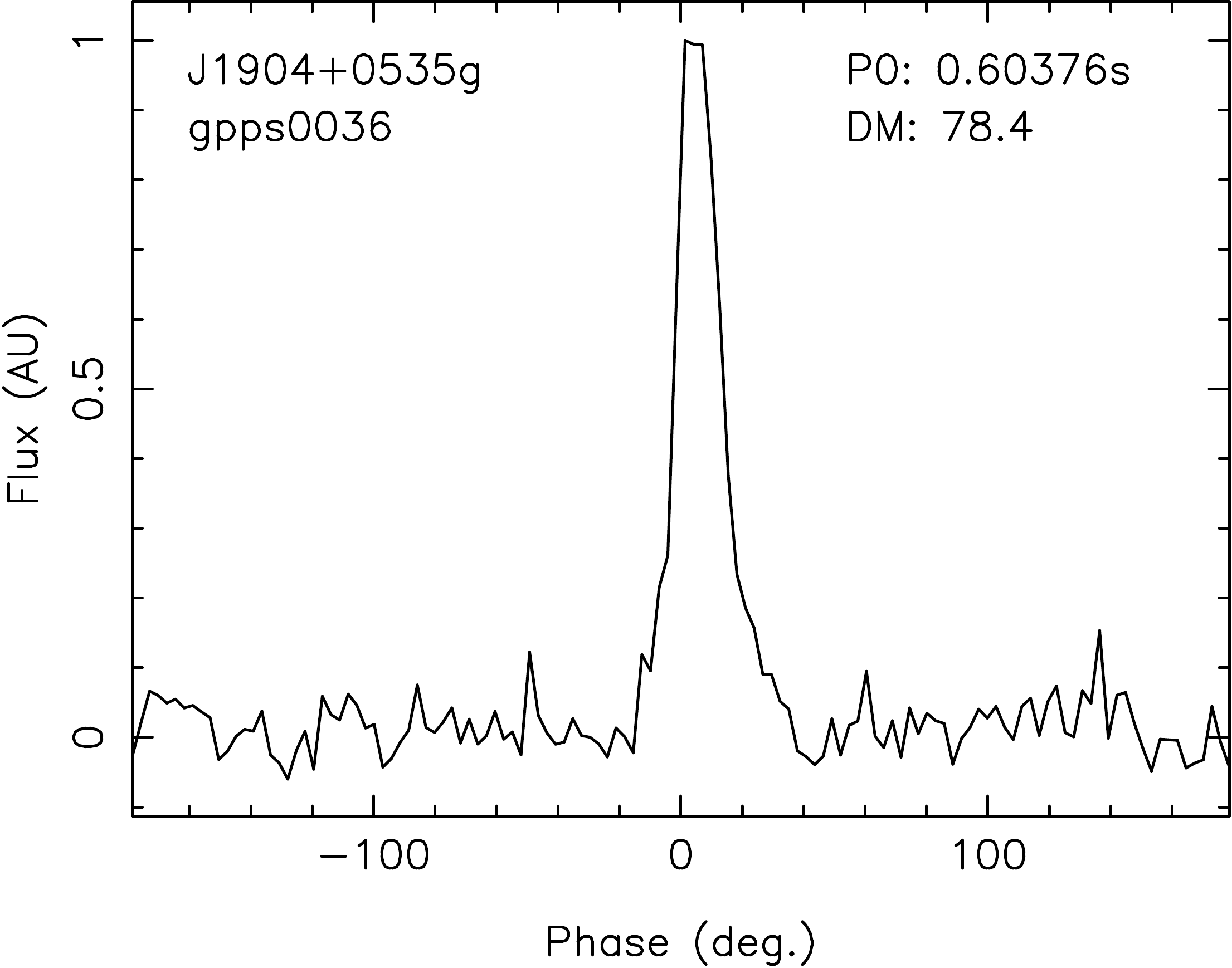}\\[2mm]
\includegraphics[width=39mm]{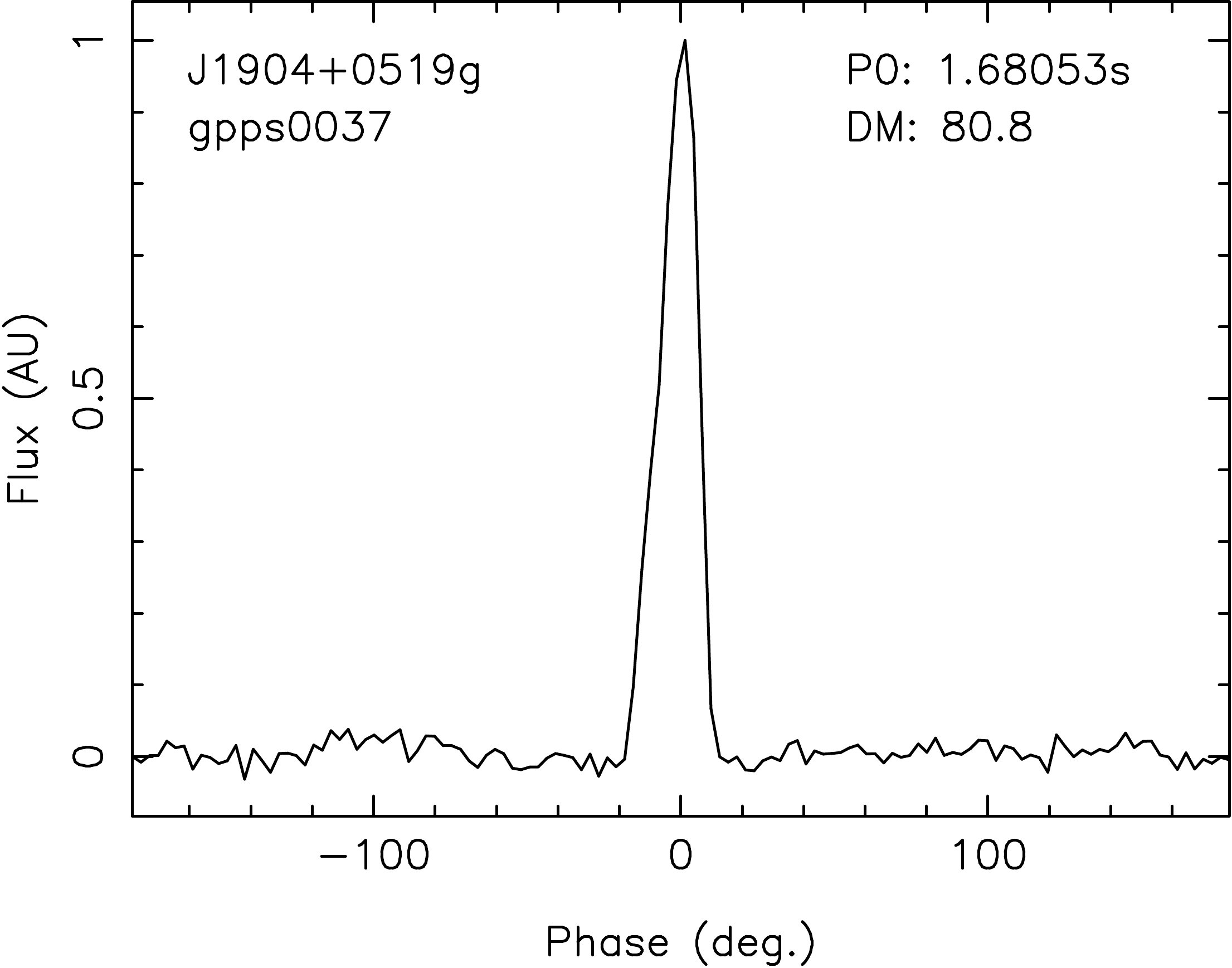}&
\includegraphics[width=39mm]{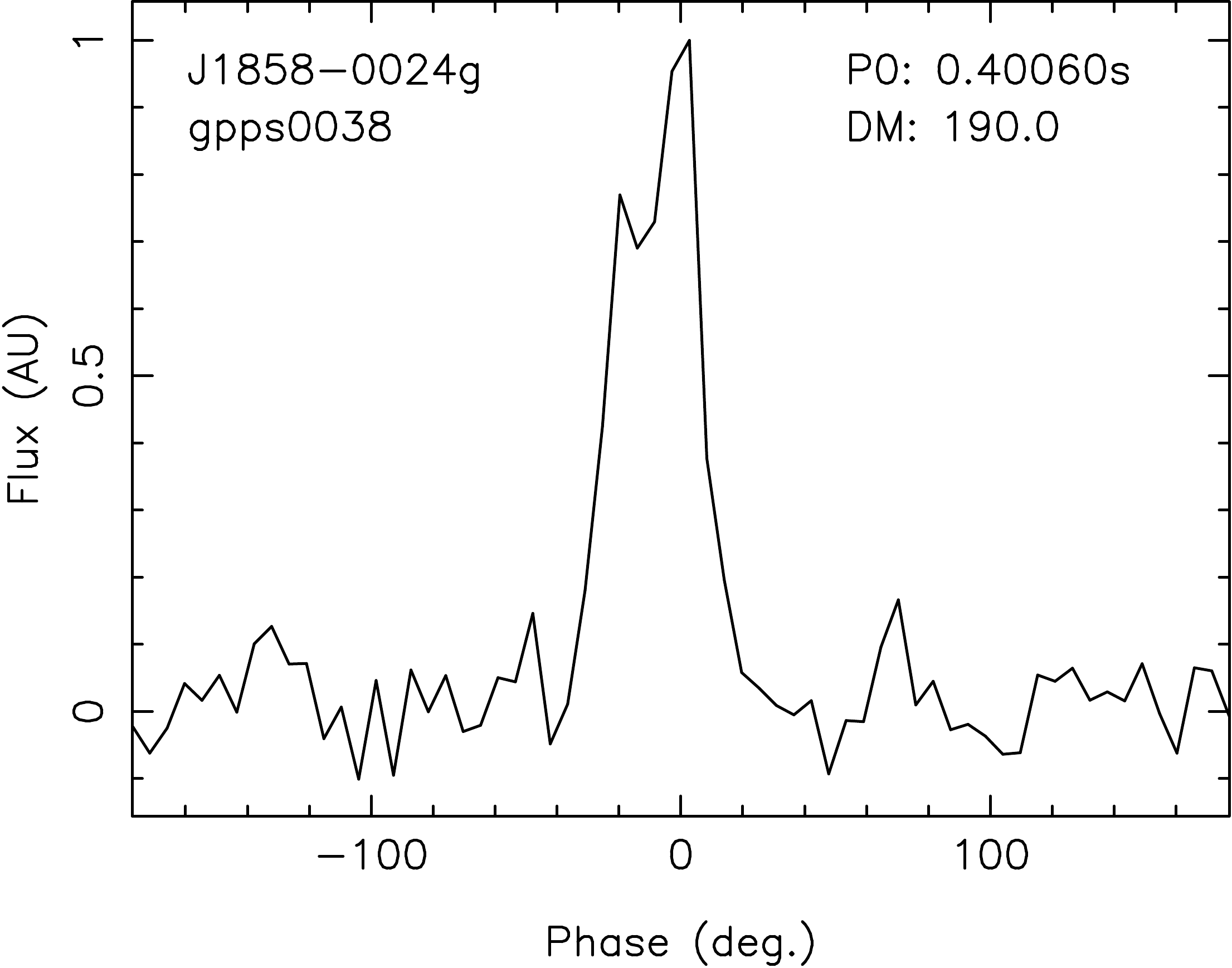}&
\includegraphics[width=39mm]{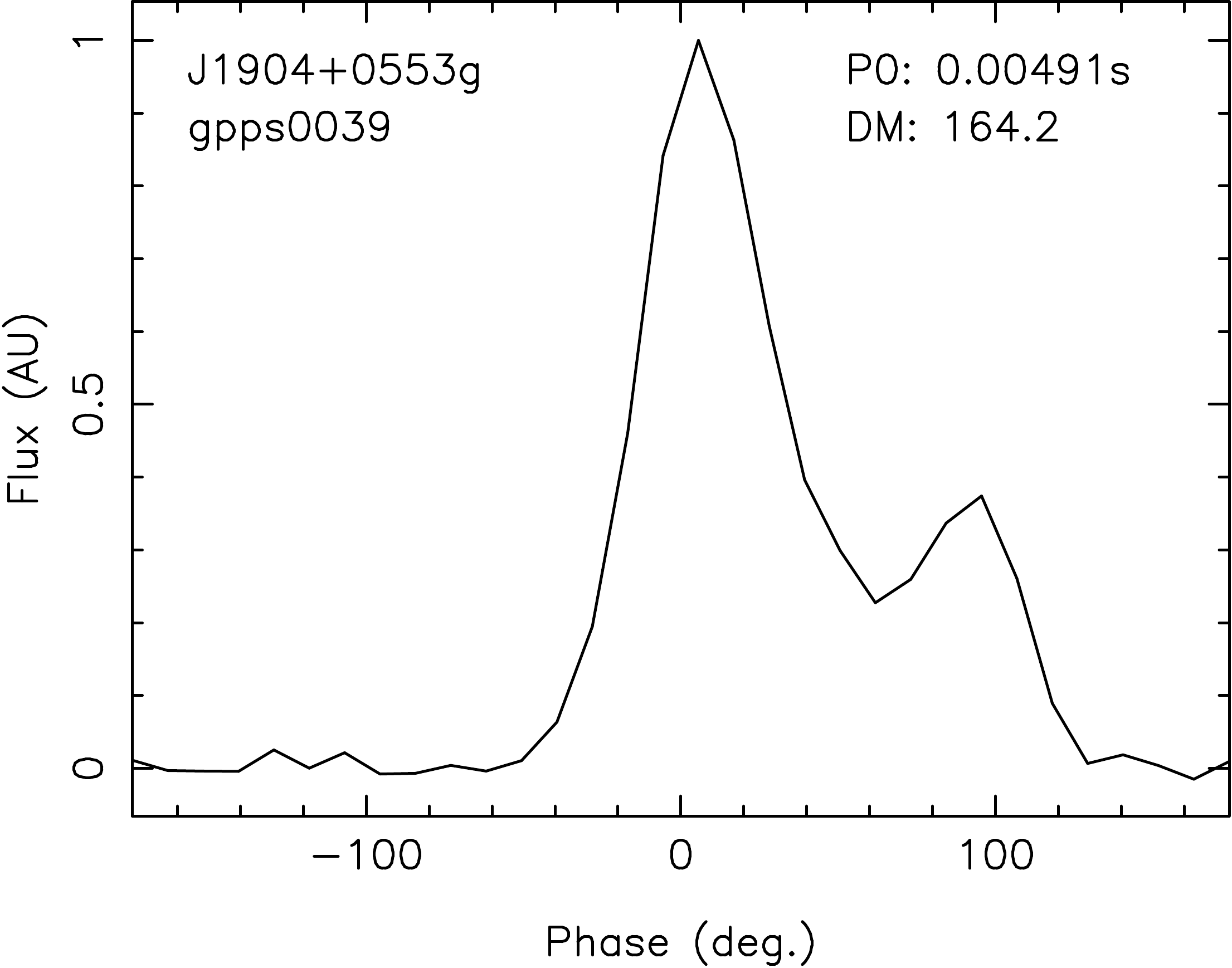}&
\includegraphics[width=39mm]{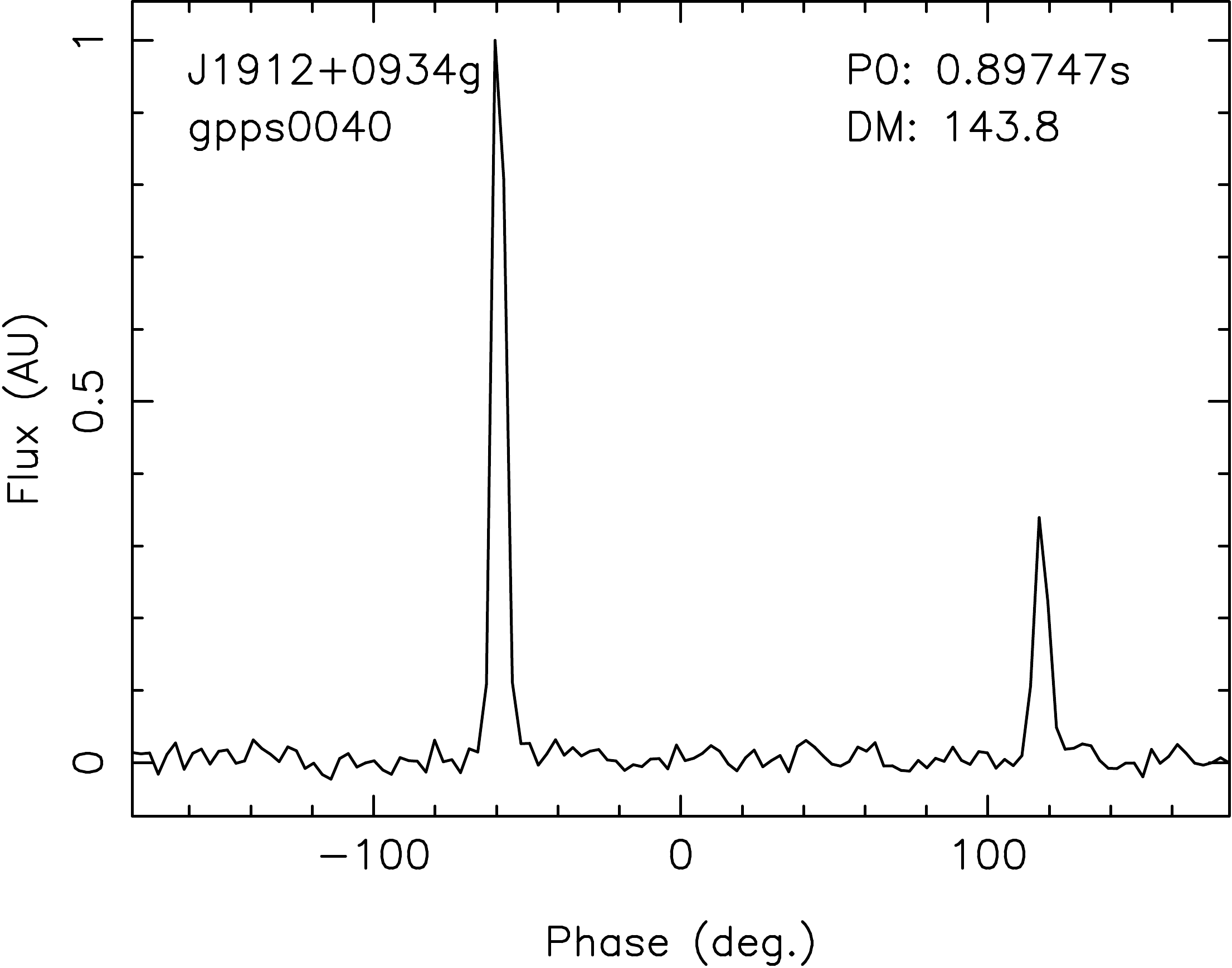}\\[2mm]
\includegraphics[width=39mm]{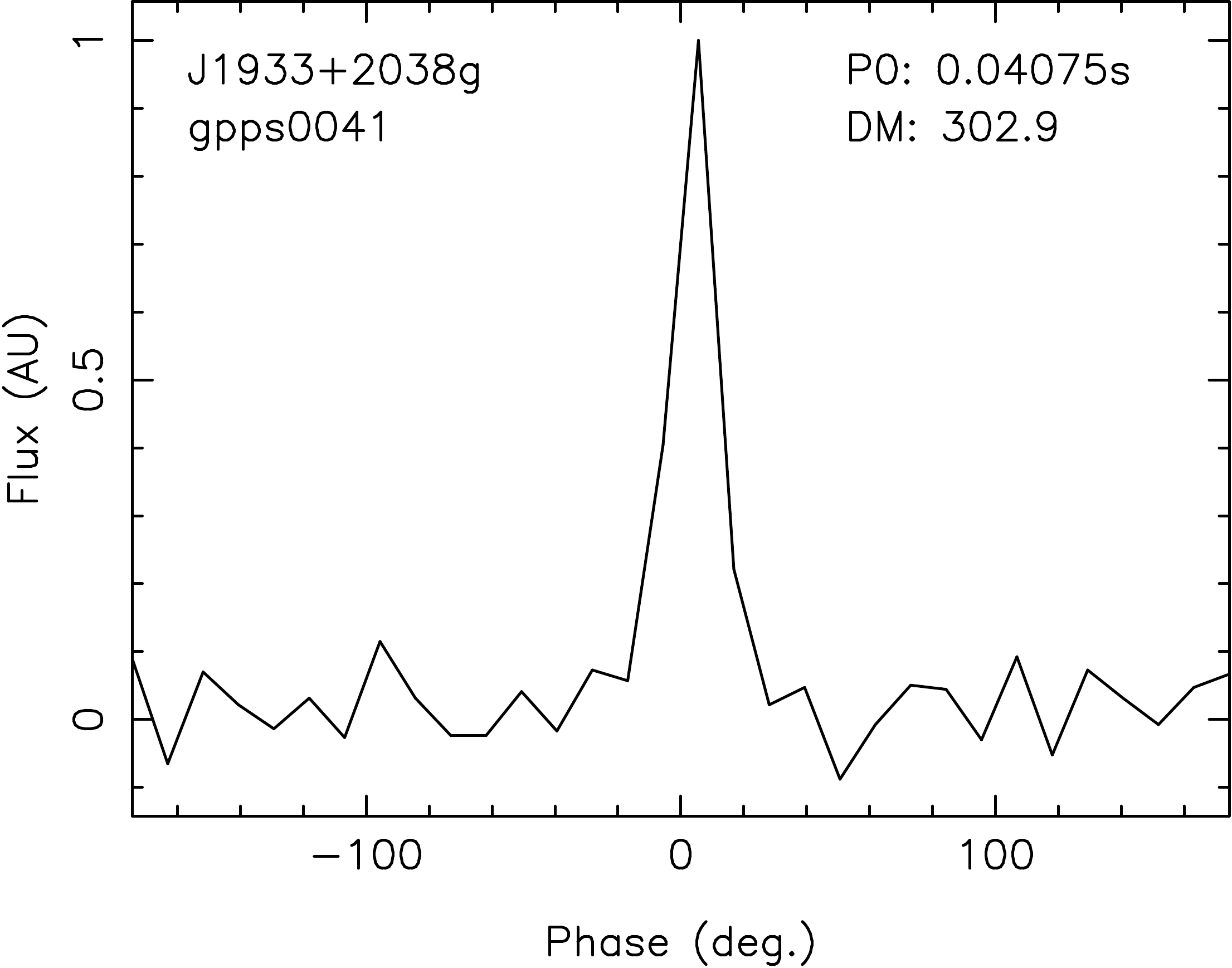}&
\includegraphics[width=39mm]{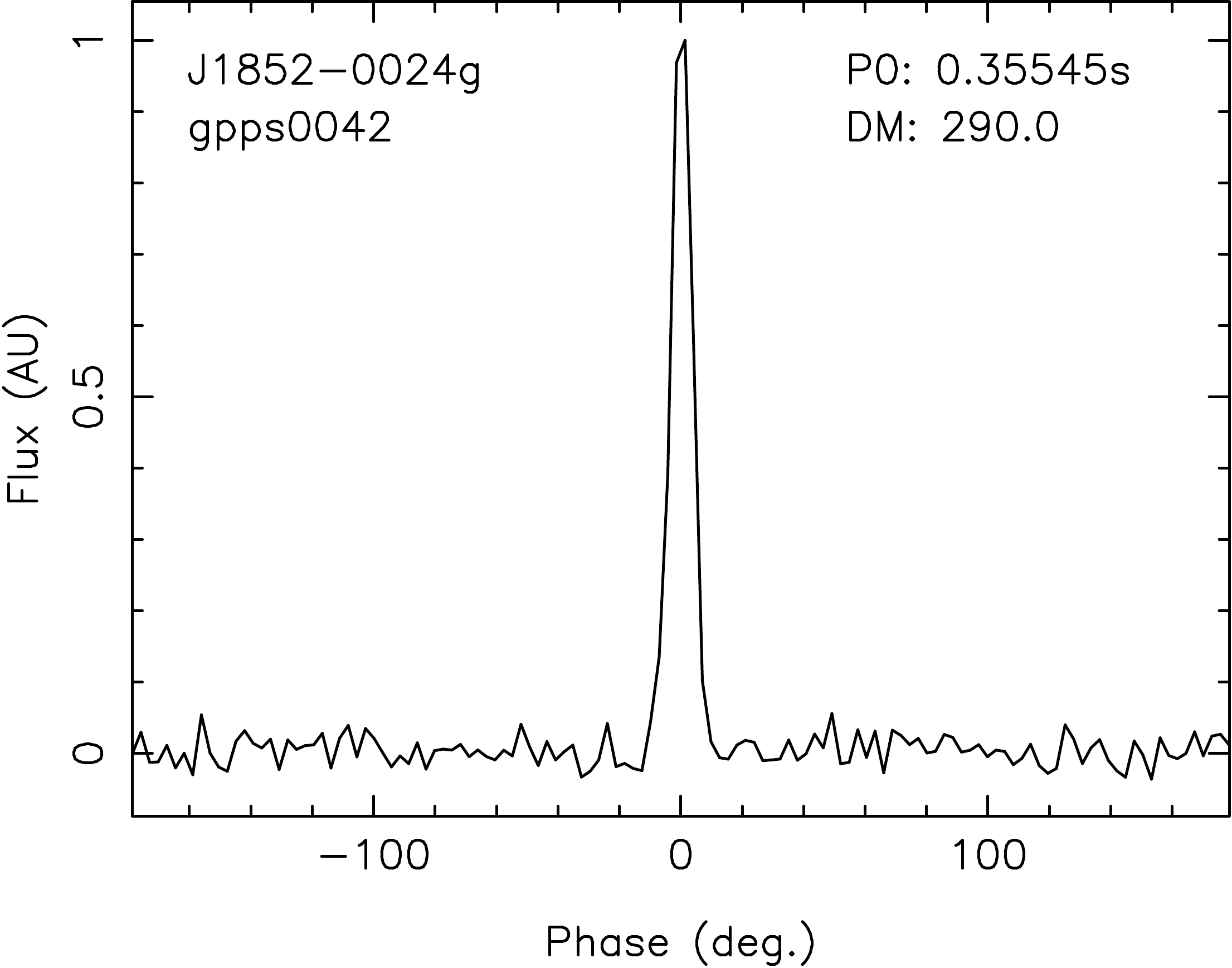}&
\includegraphics[width=39mm]{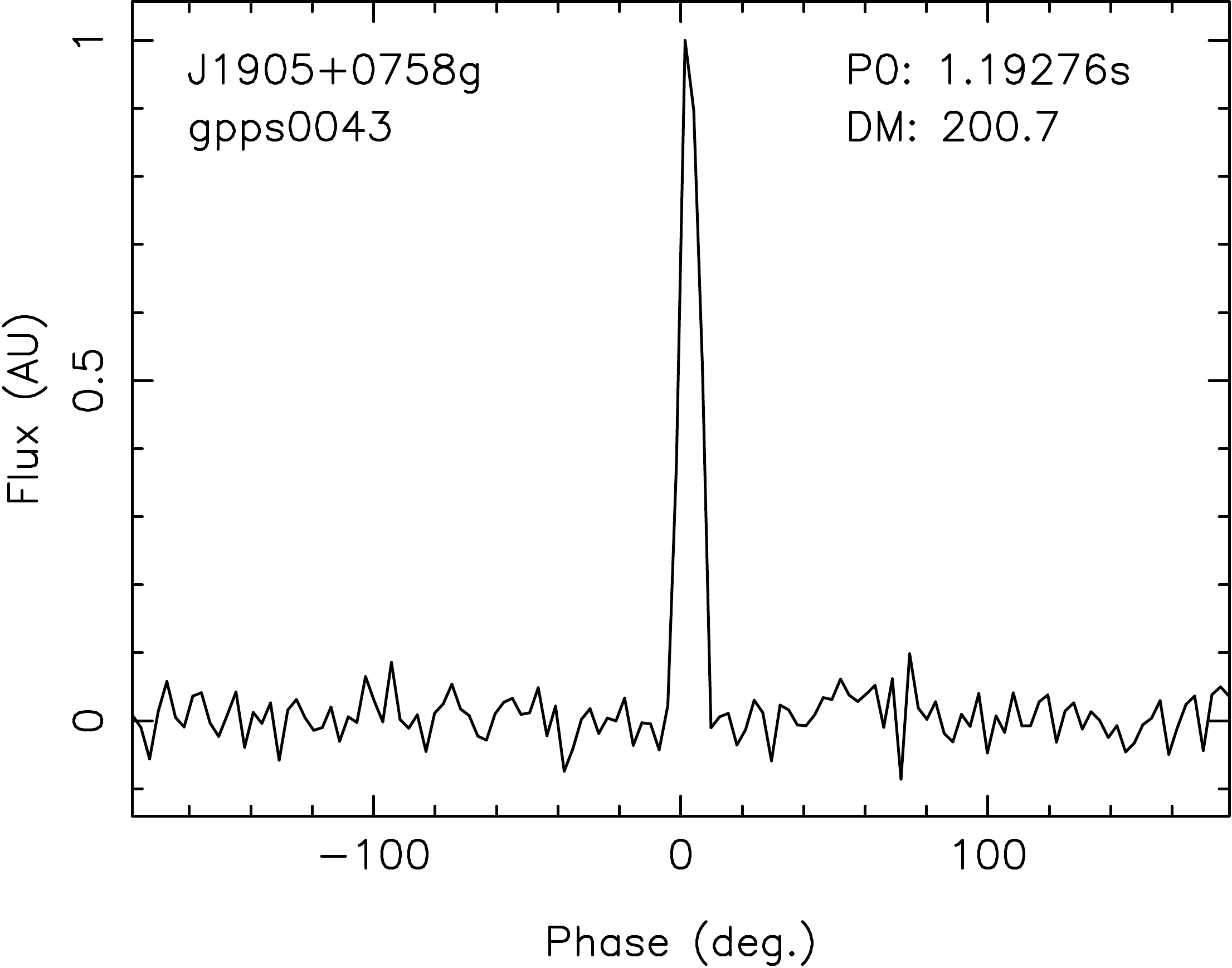}&
\includegraphics[width=39mm]{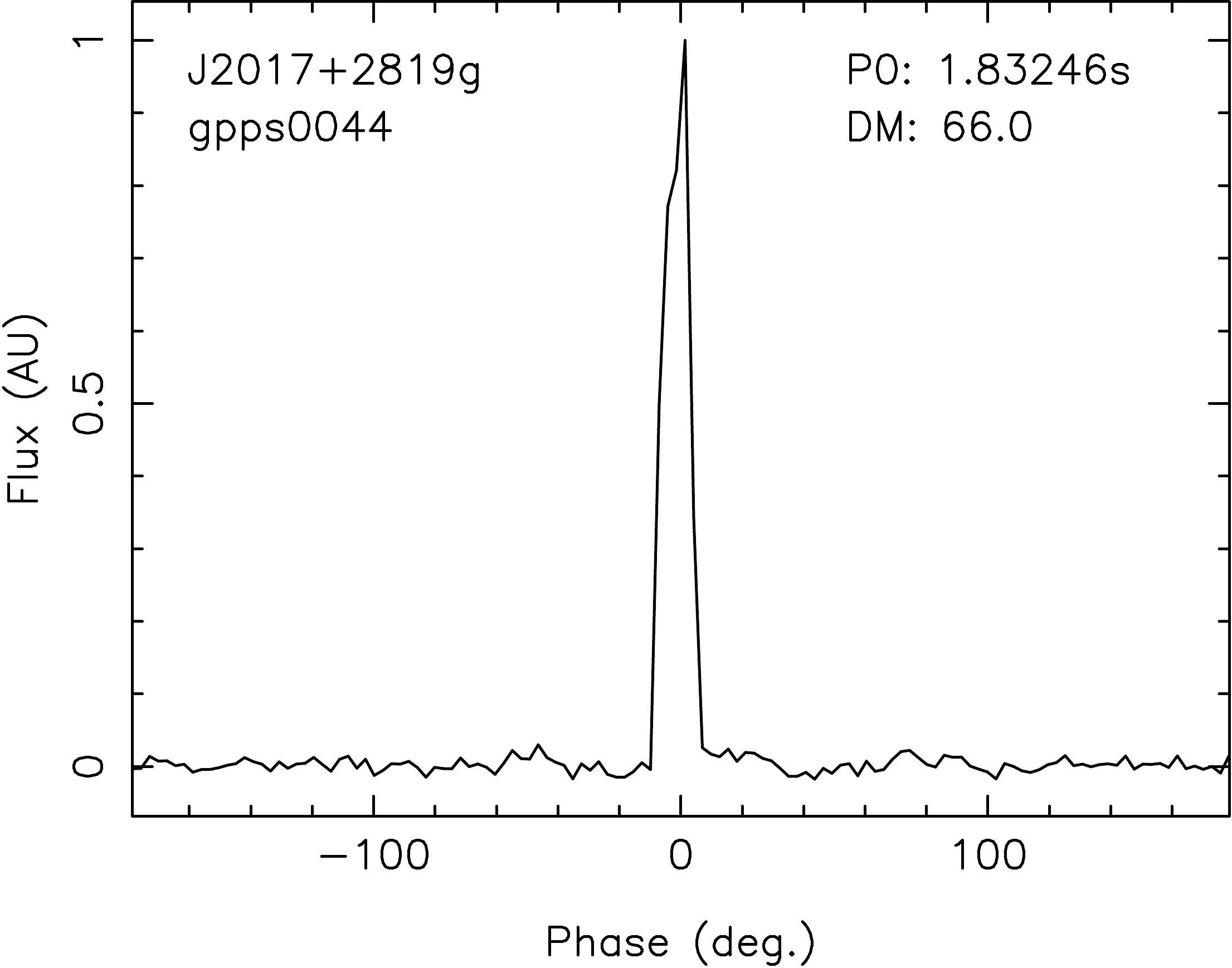}\\[2mm]
\includegraphics[width=39mm]{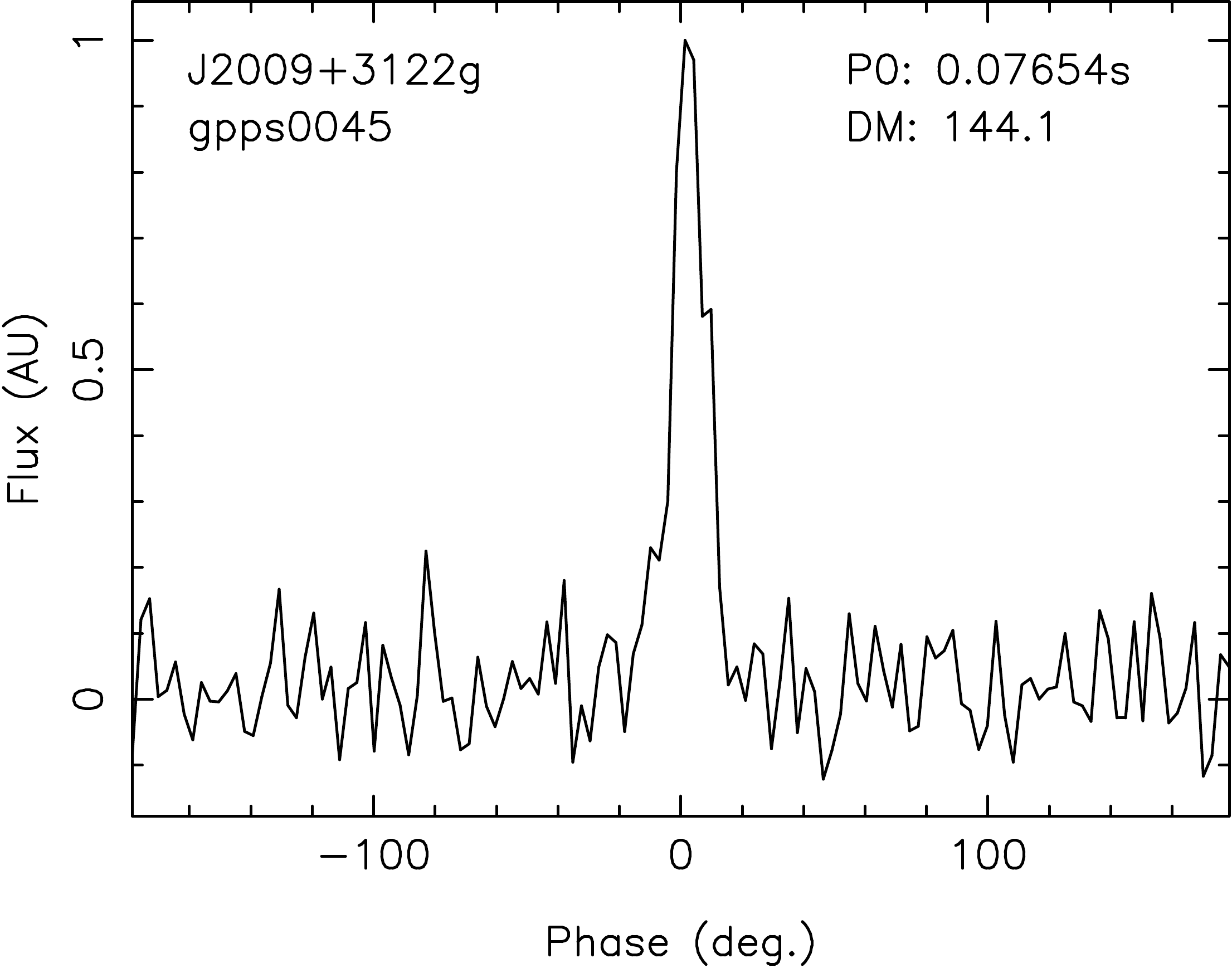}&
\includegraphics[width=39mm]{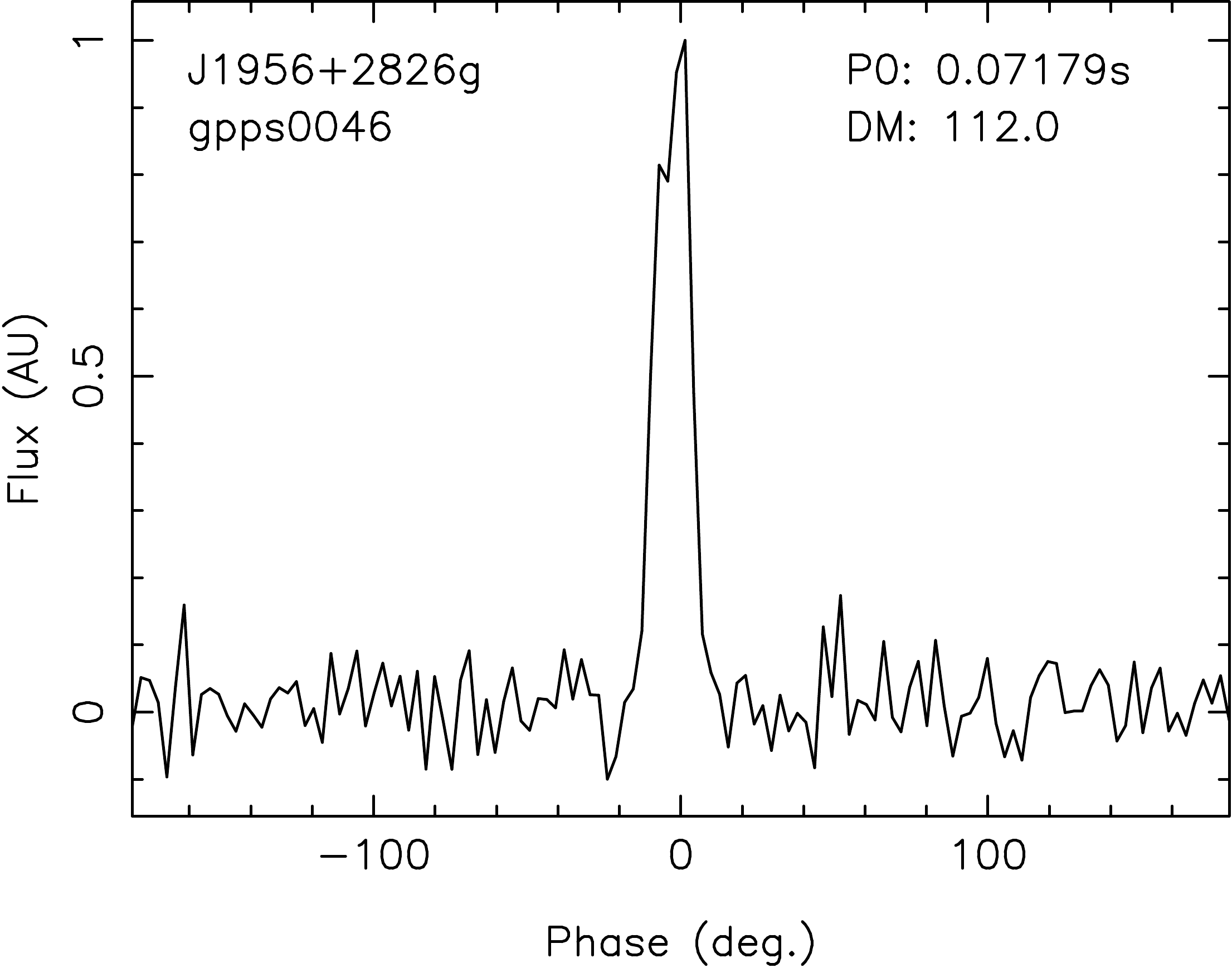}&
\includegraphics[width=39mm]{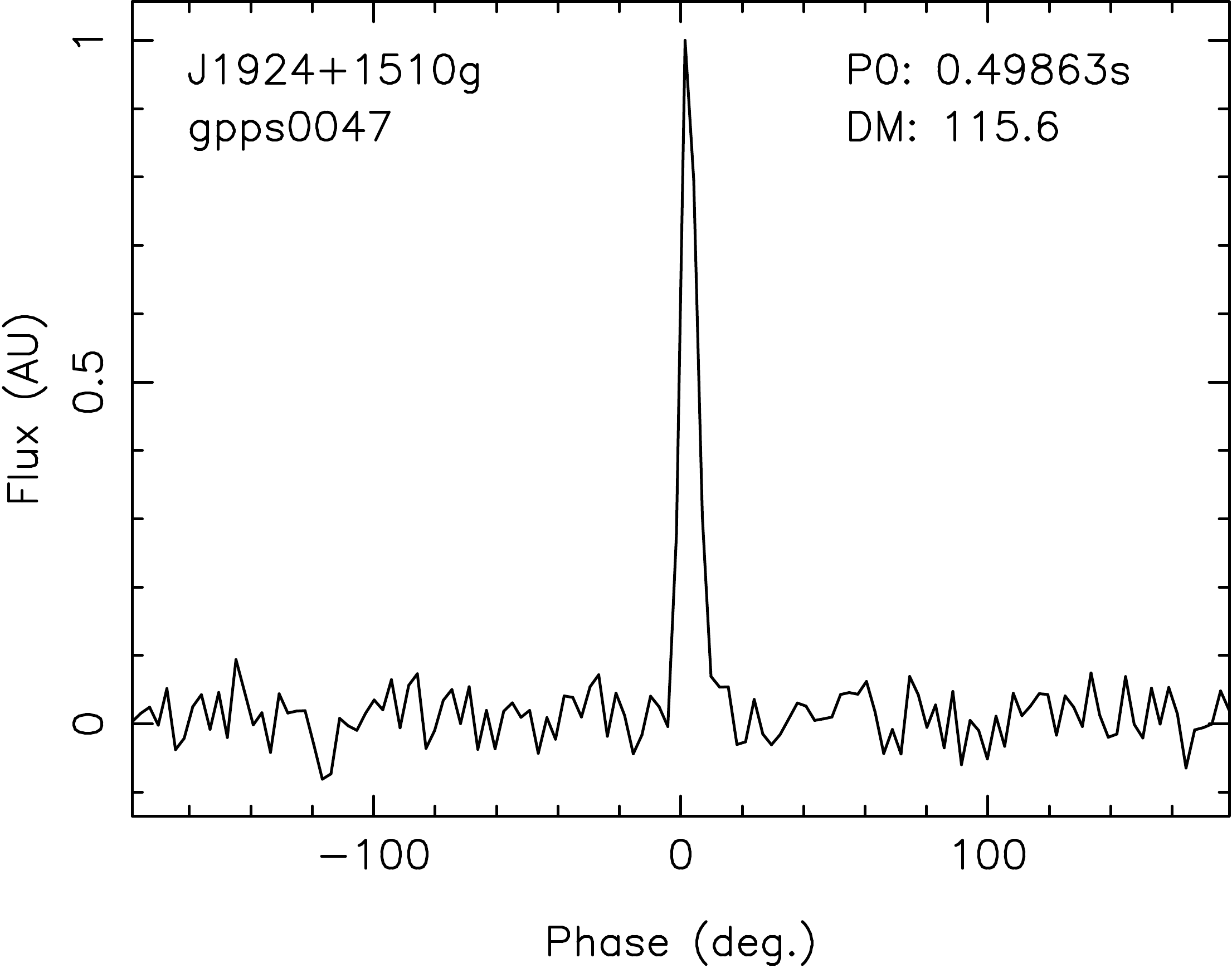}&
\includegraphics[width=39mm]{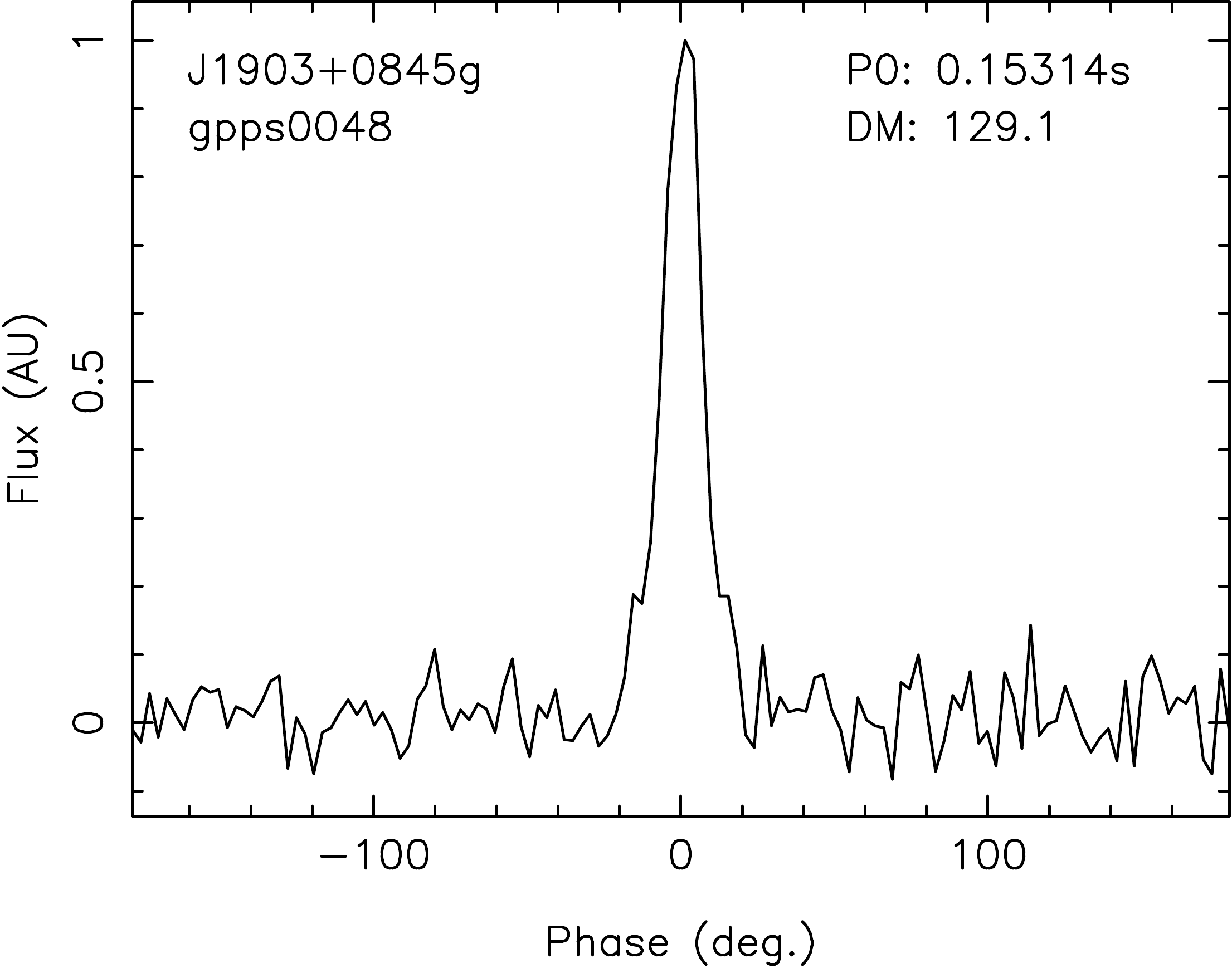}\\[2mm]
\includegraphics[width=39mm]{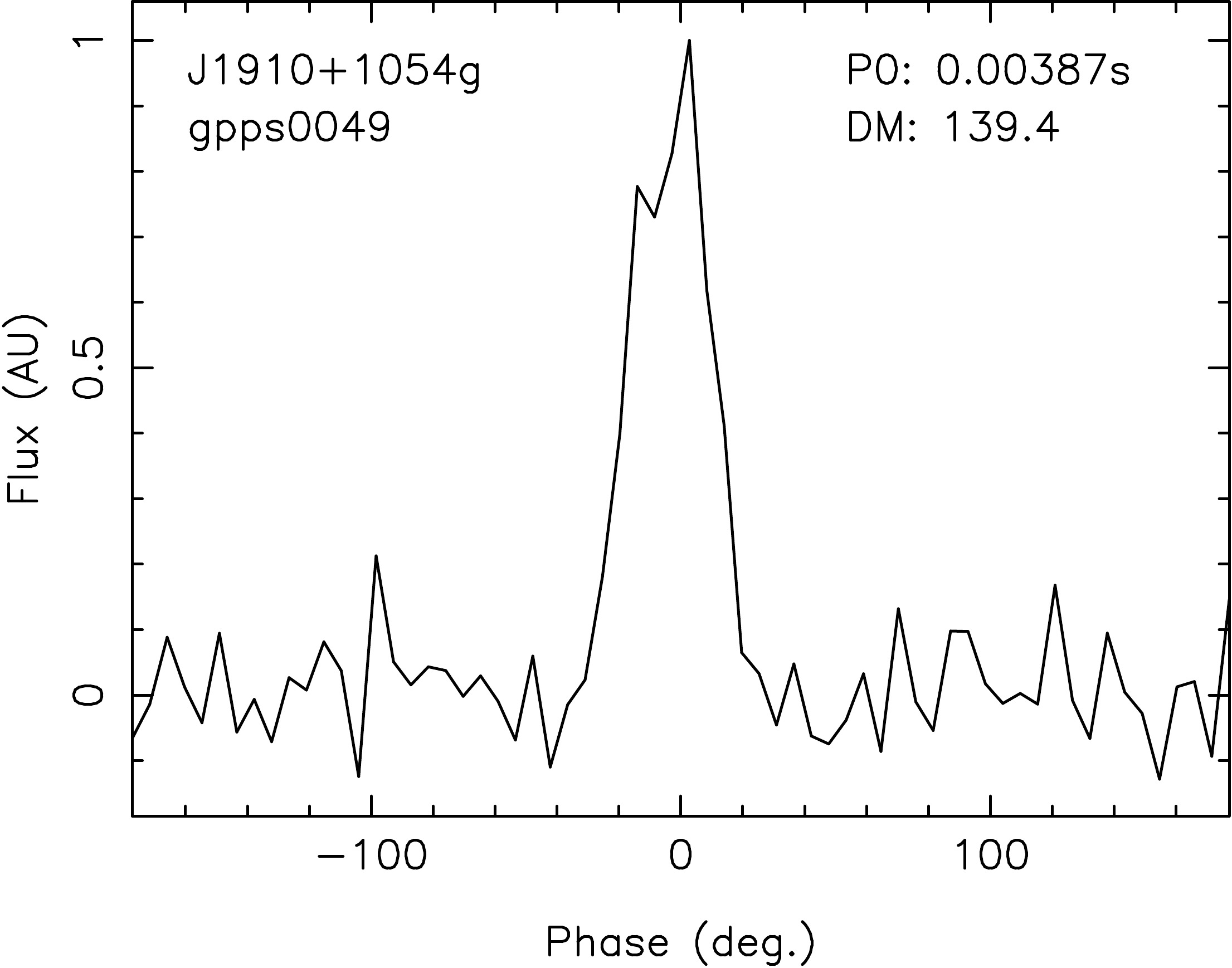}&
\includegraphics[width=39mm]{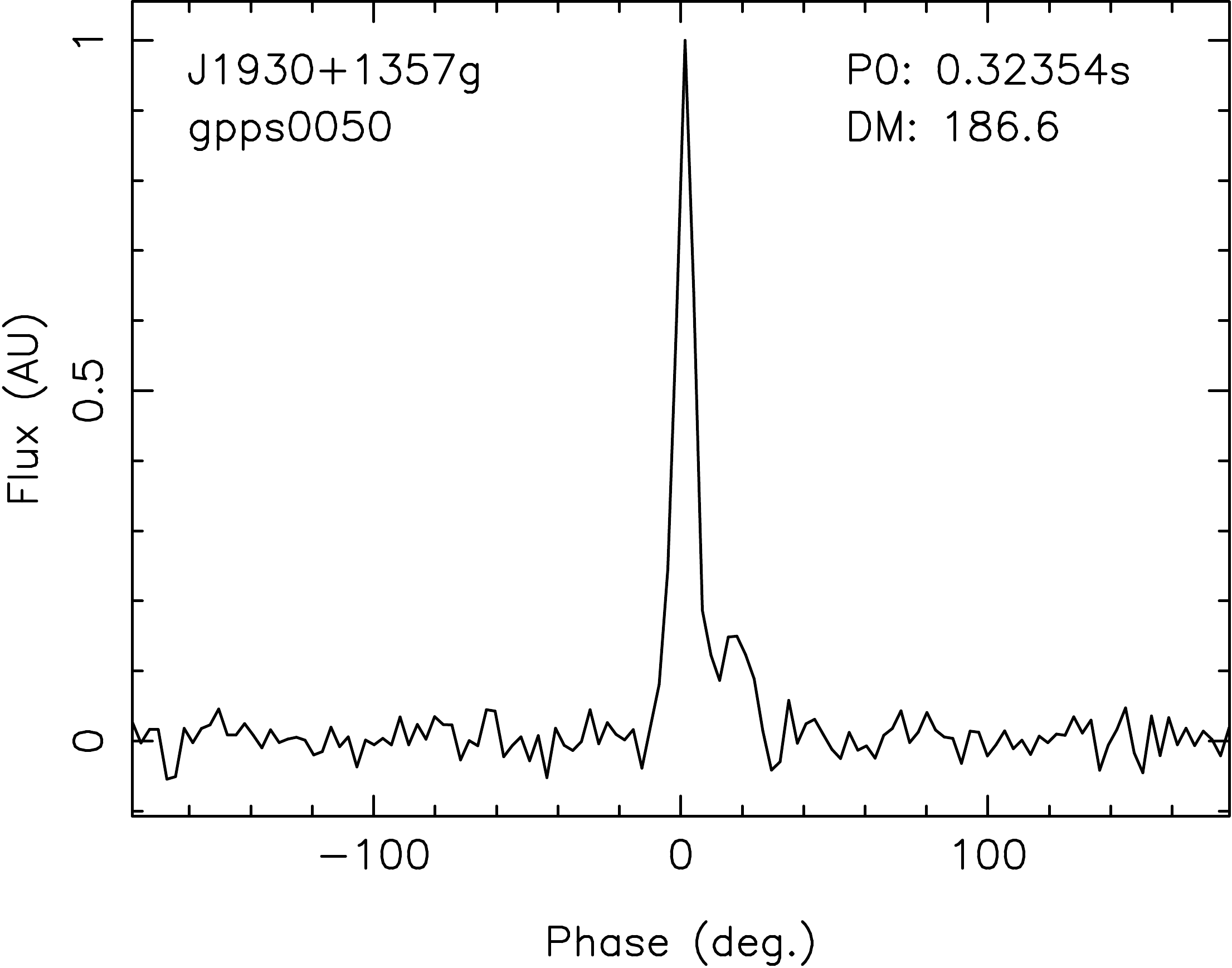}&
\includegraphics[width=39mm]{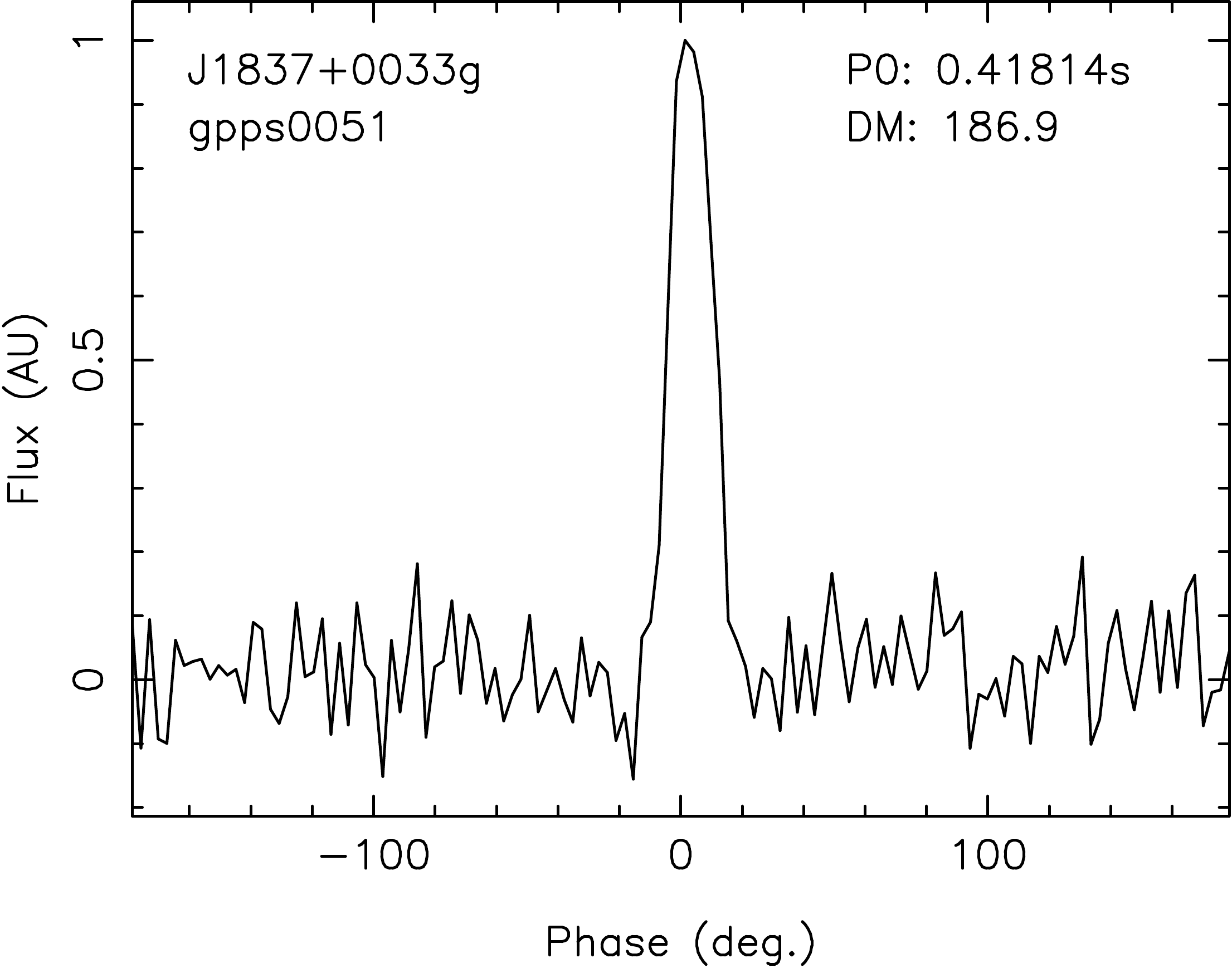}&
\includegraphics[width=39mm]{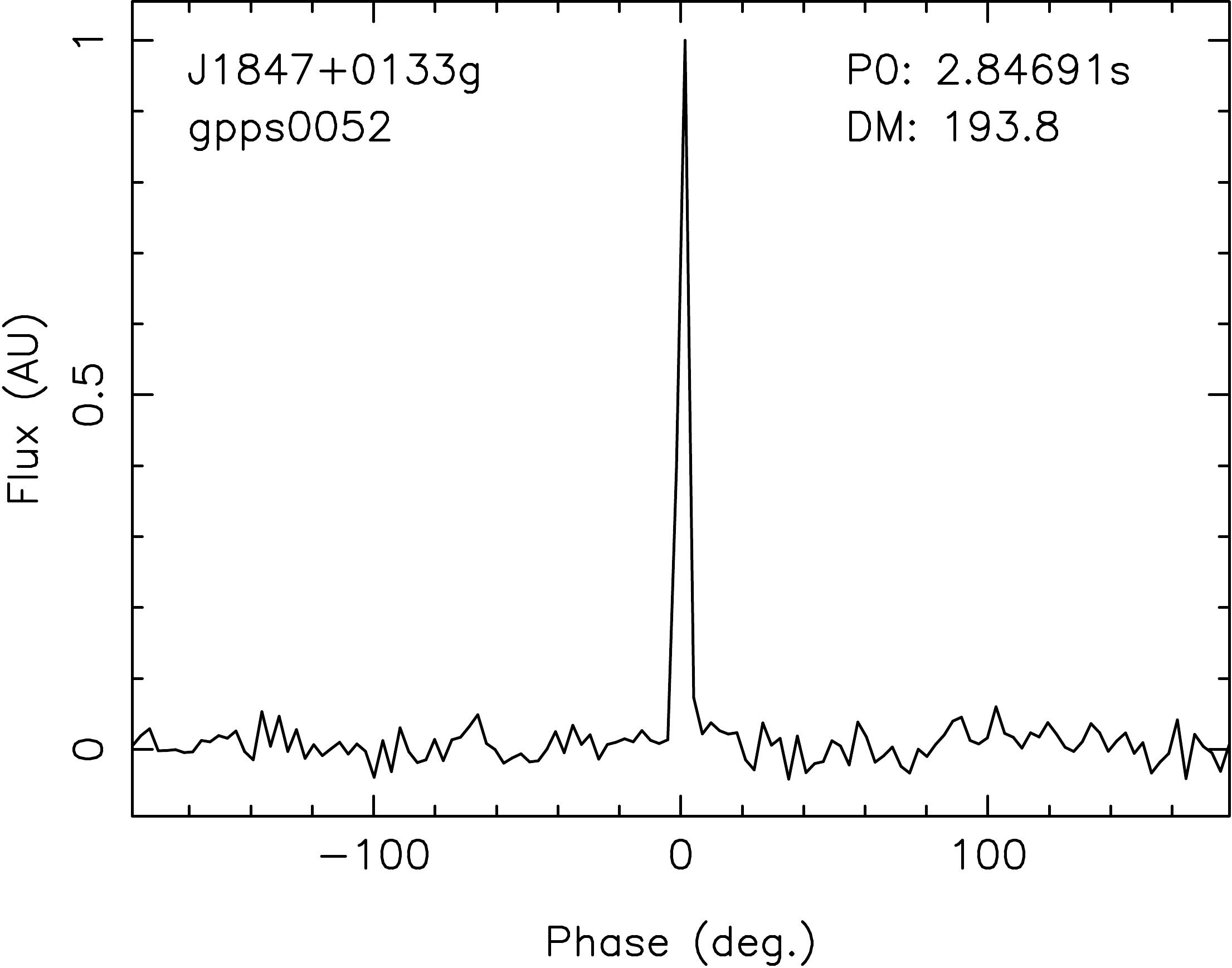}\\[2mm]
\includegraphics[width=39mm]{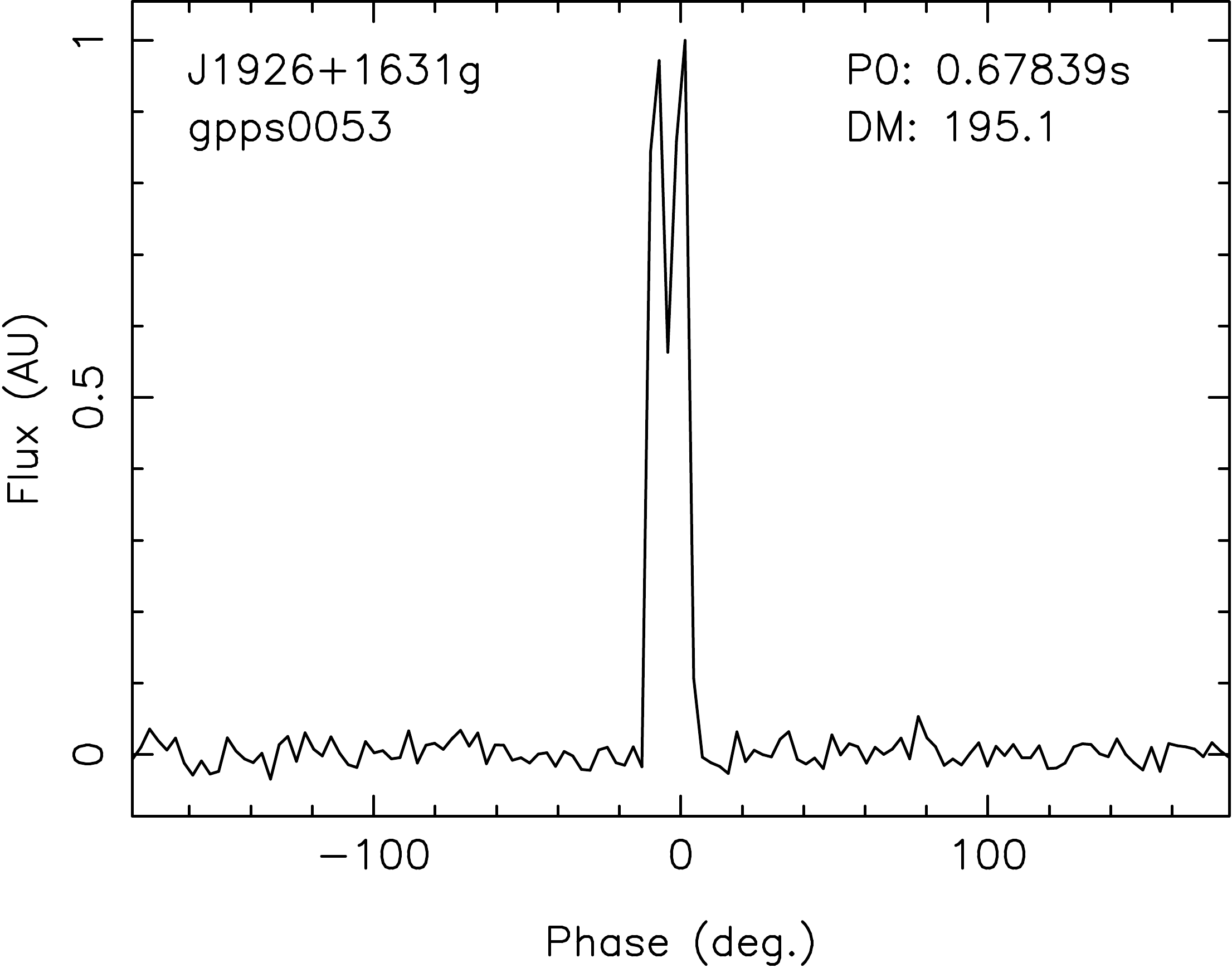}&
\includegraphics[width=39mm]{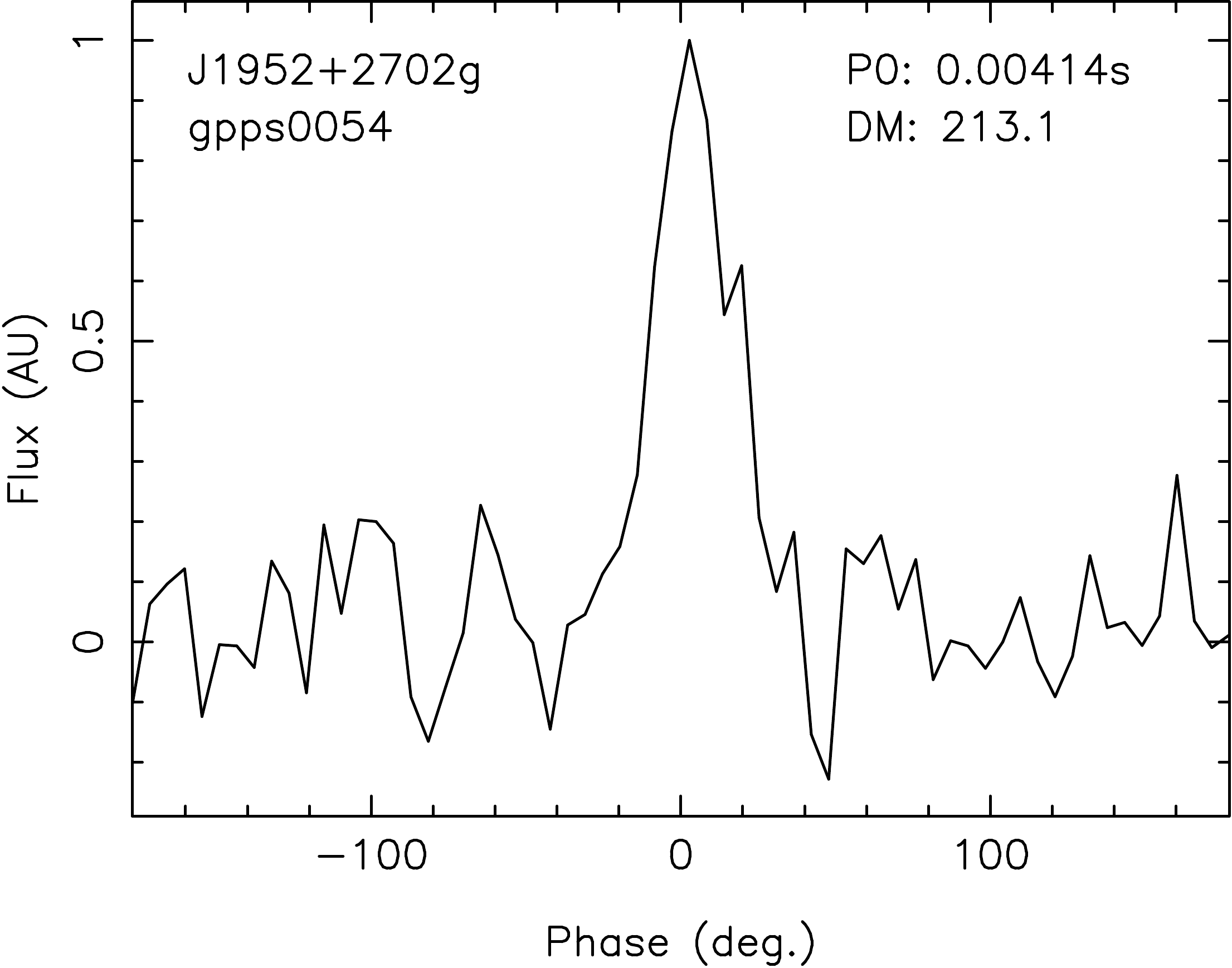}&
\includegraphics[width=39mm]{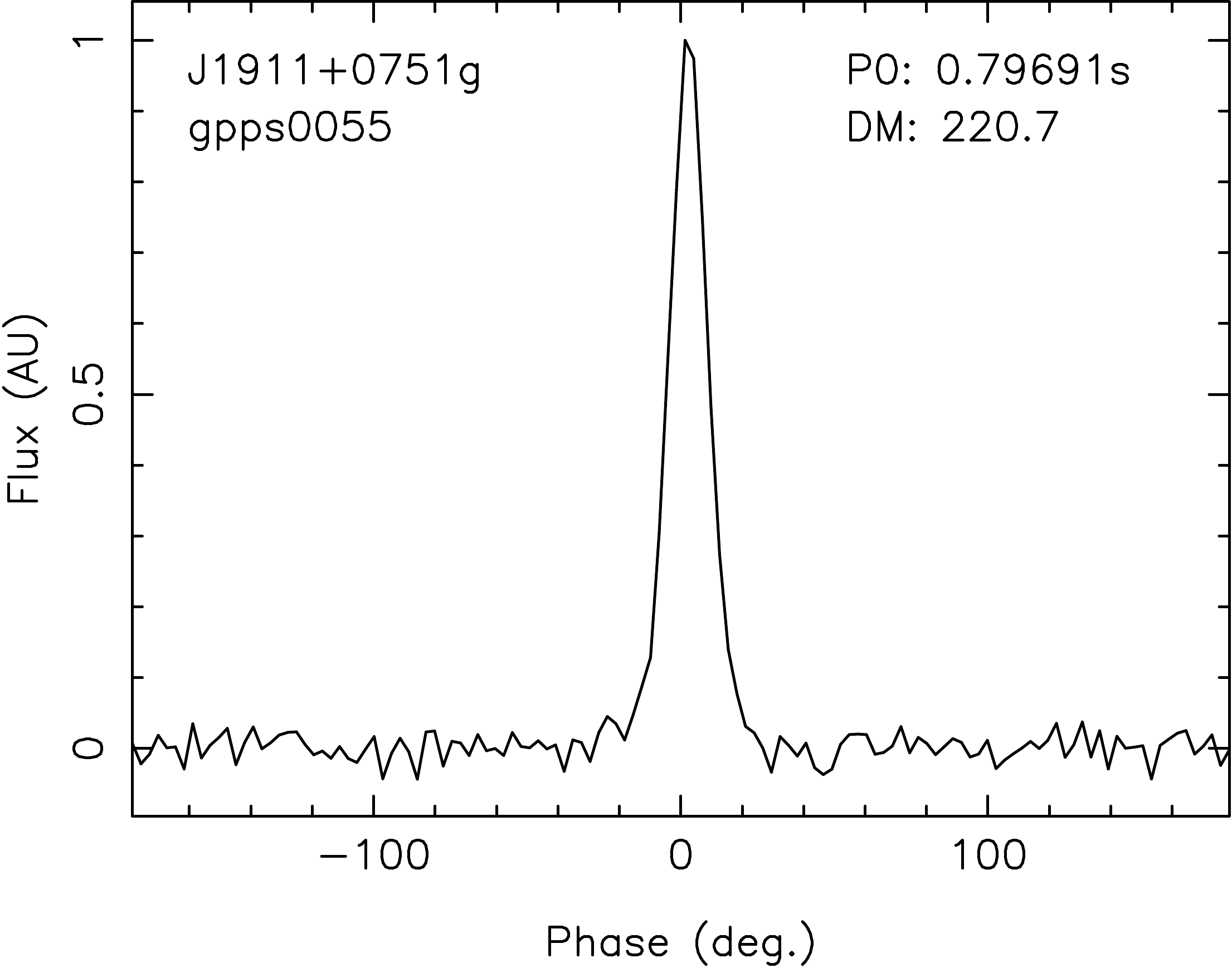}&
\includegraphics[width=39mm]{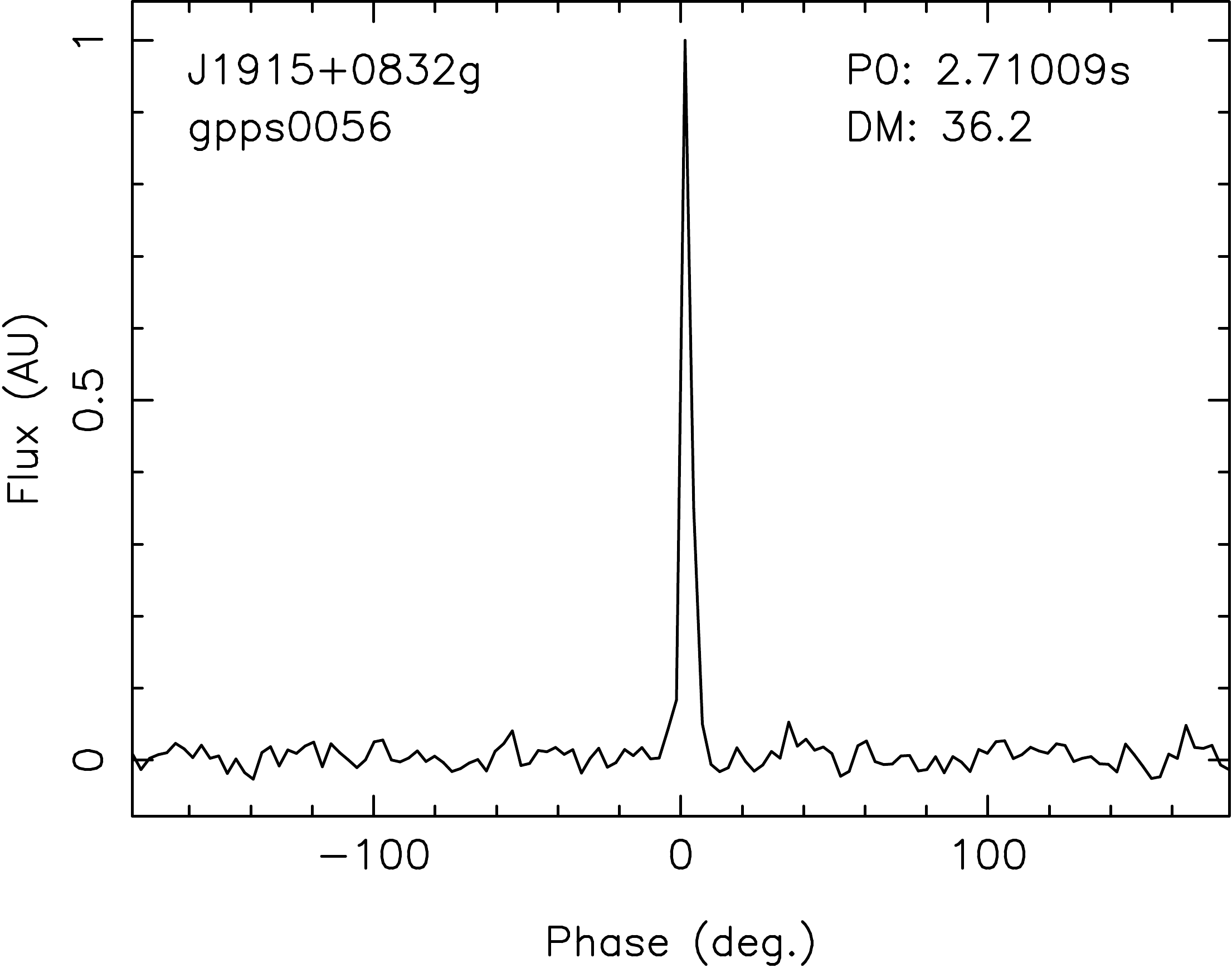}\\
\end{tabular}%

\begin{minipage}{3cm}
\caption[]{
-- {\it Continued}.}\end{minipage}
\addtocounter{figure}{-1}
\end{figure*}%
\begin{figure*}
\centering
\begin{tabular}{rrrrrr}
\includegraphics[width=39mm]{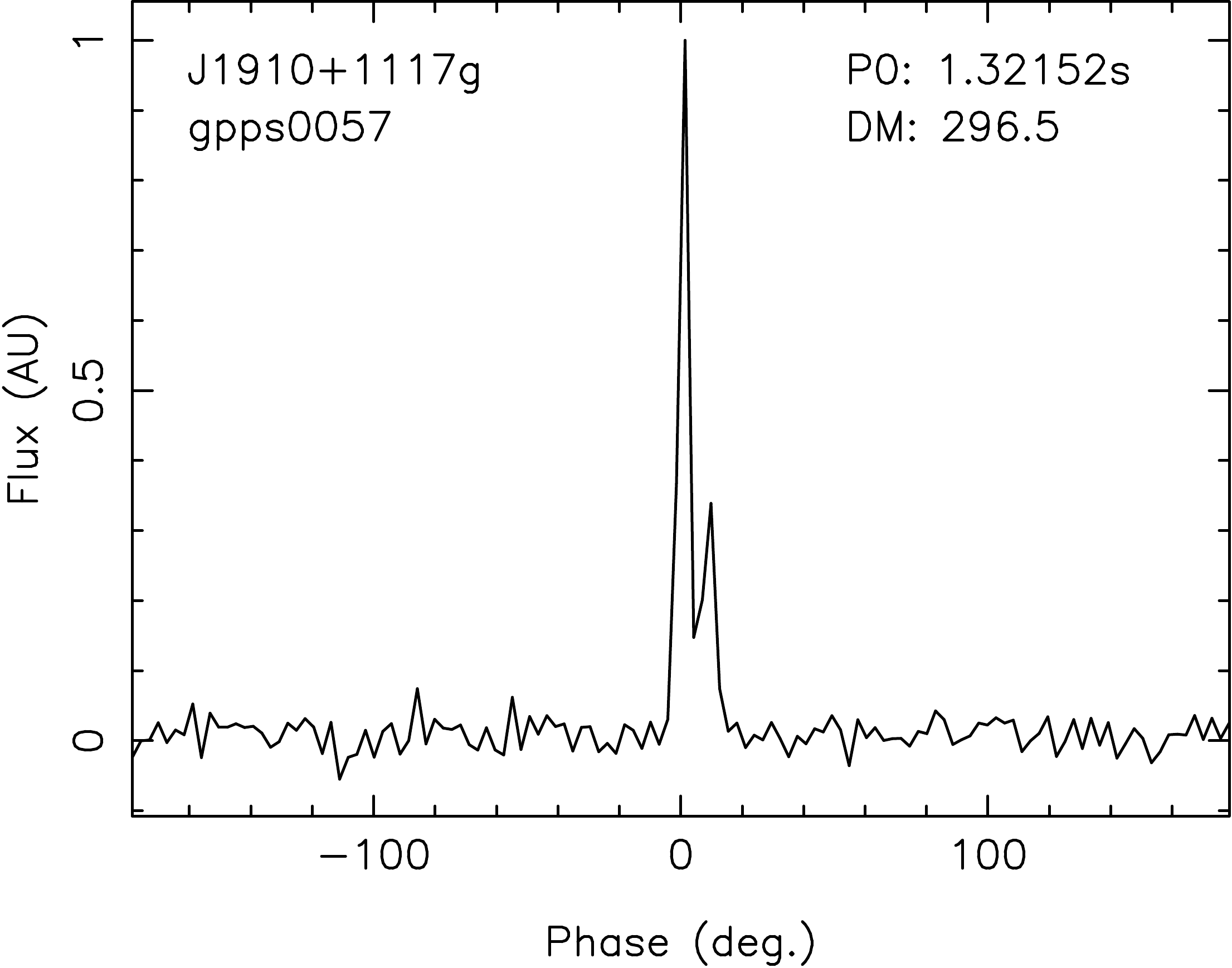}&
\includegraphics[width=39mm]{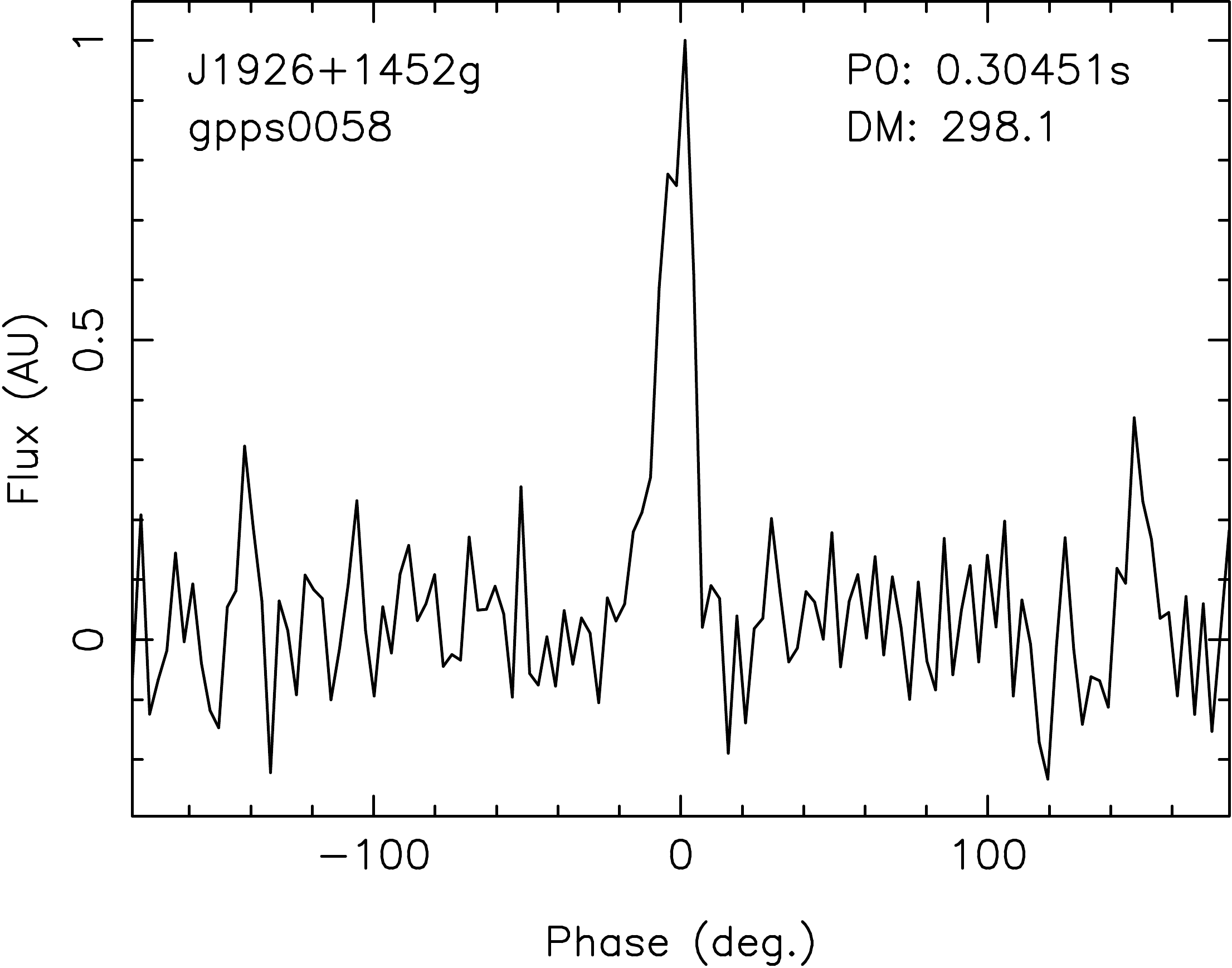}&
\includegraphics[width=39mm]{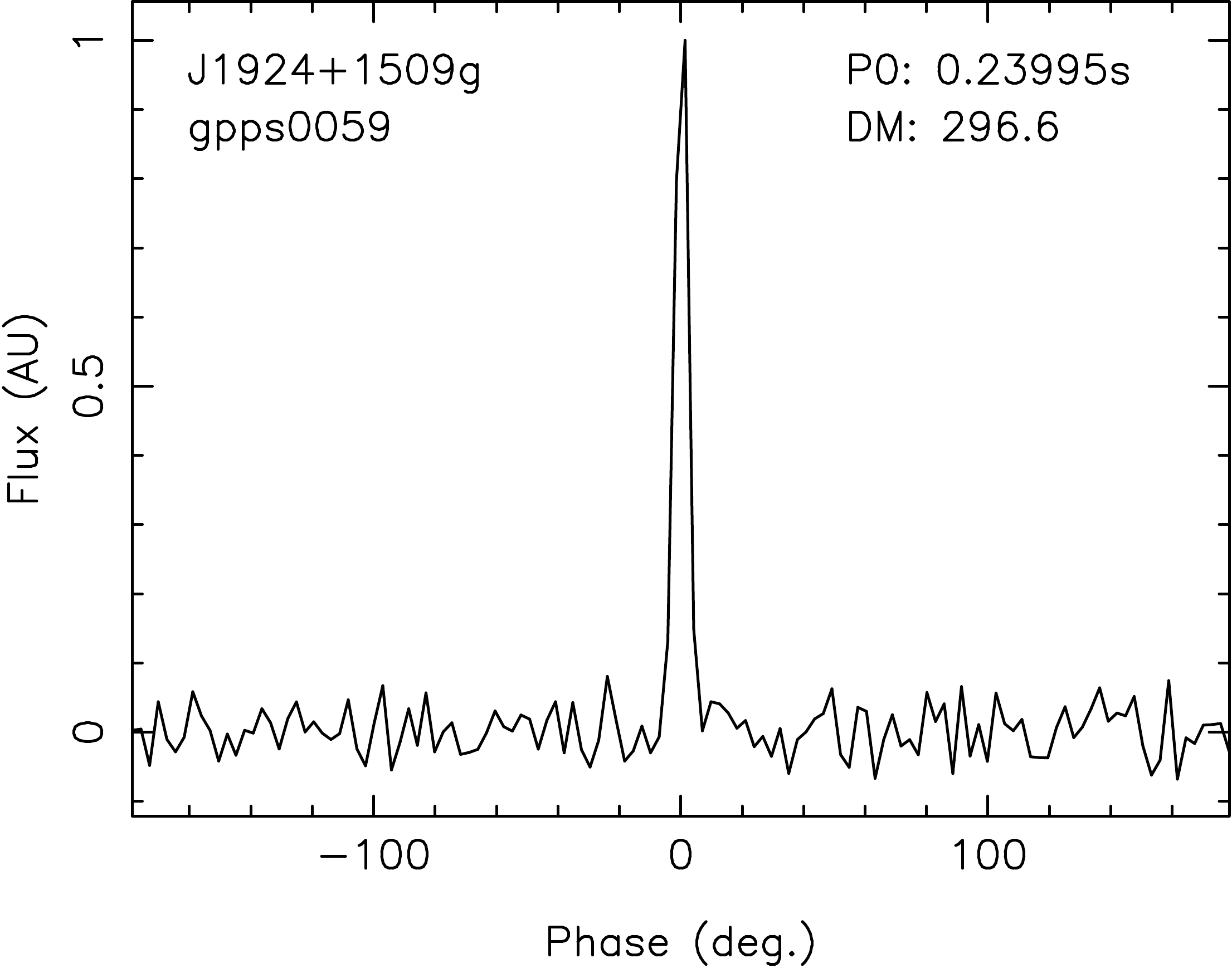}&
\includegraphics[width=39mm]{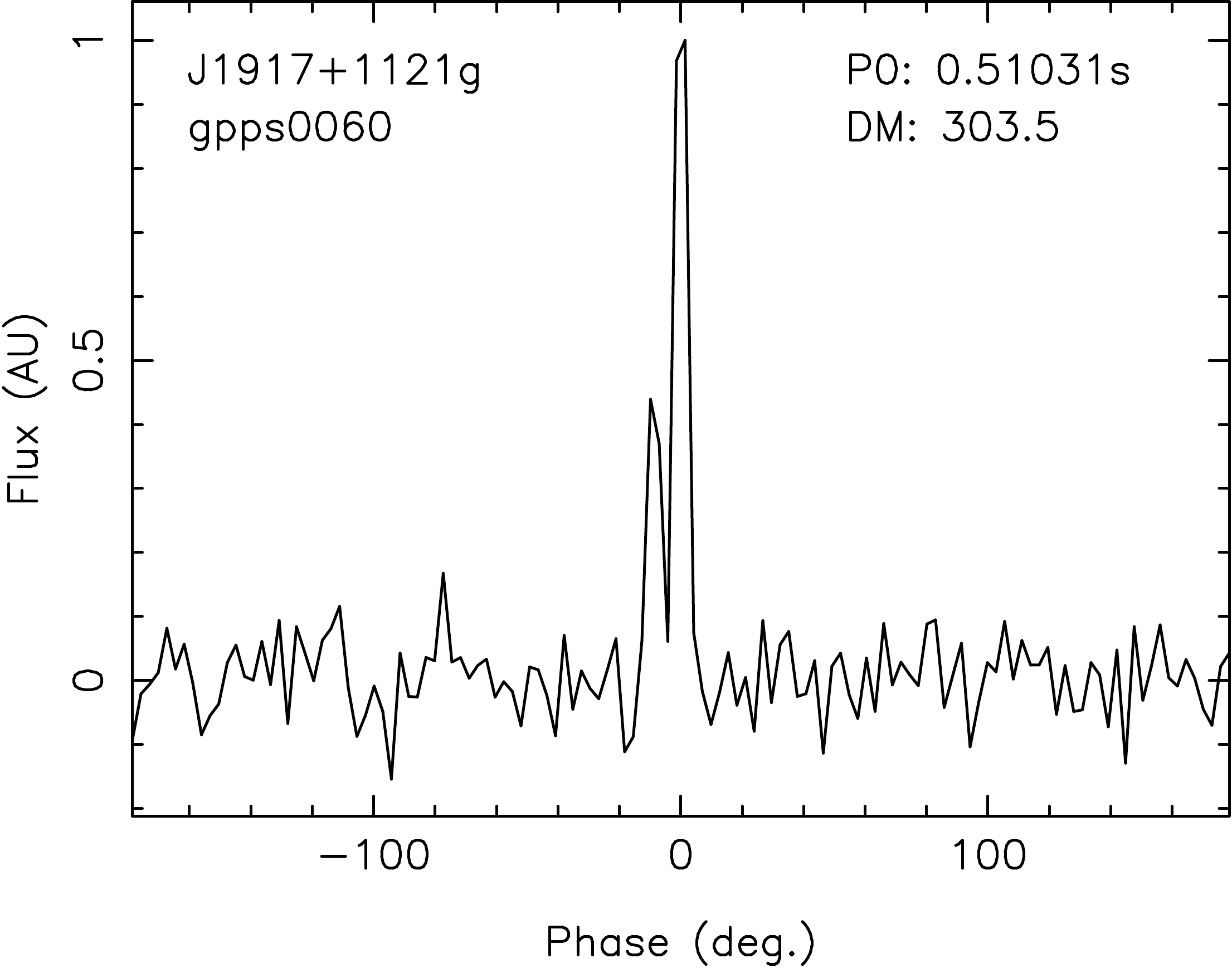}\\[2mm]
\includegraphics[width=39mm]{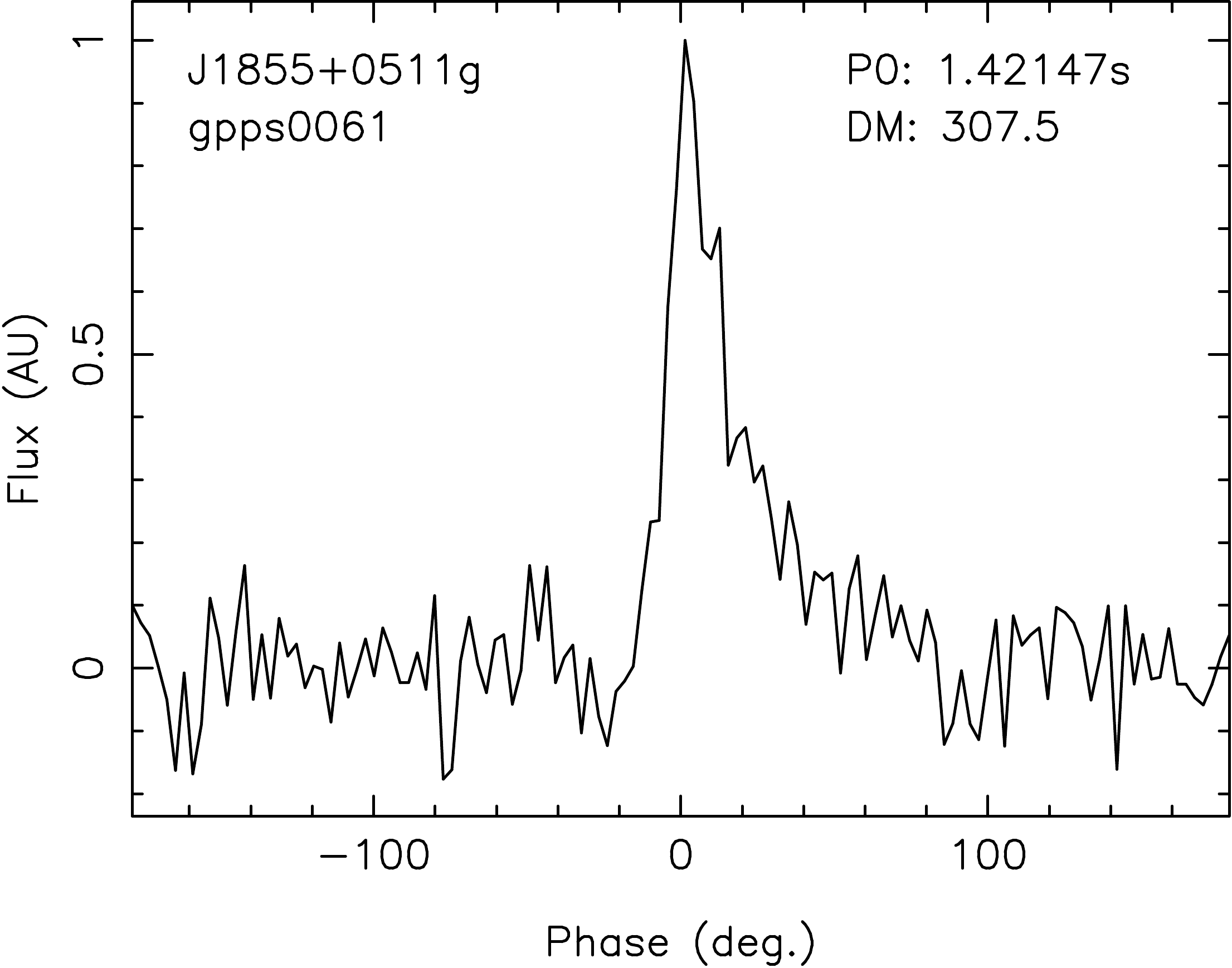}&
\includegraphics[width=39mm]{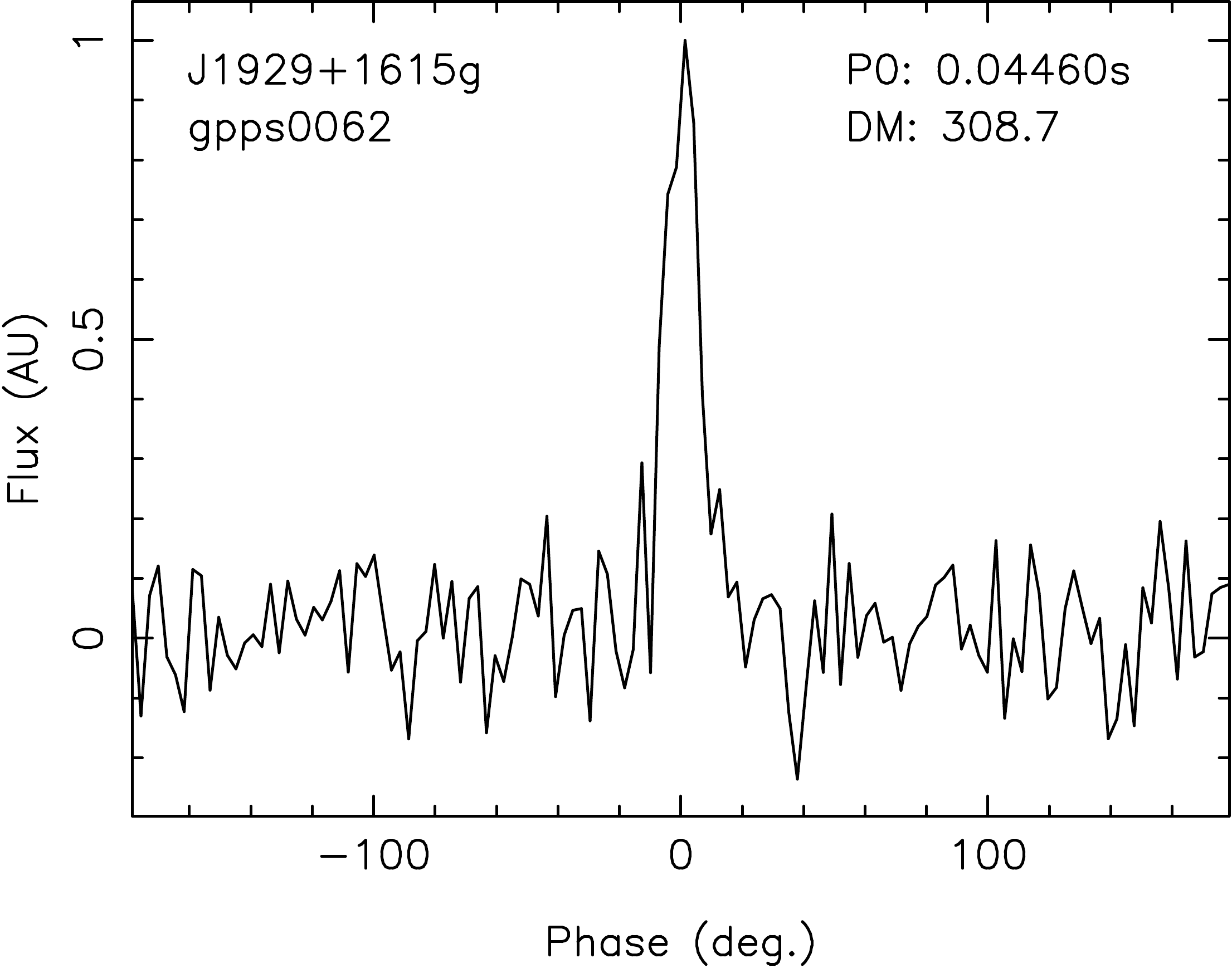}&
\includegraphics[width=39mm]{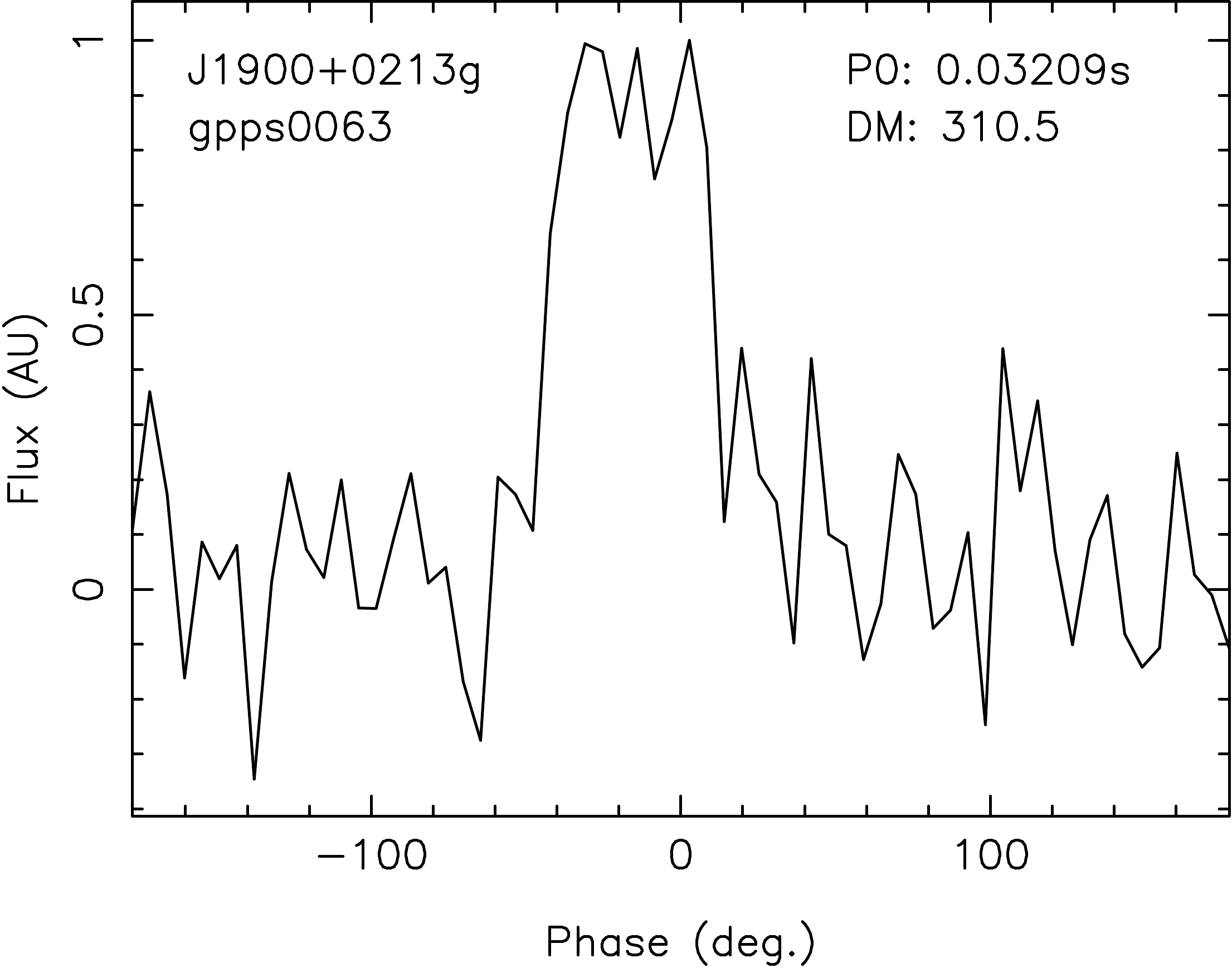}&
\includegraphics[width=39mm]{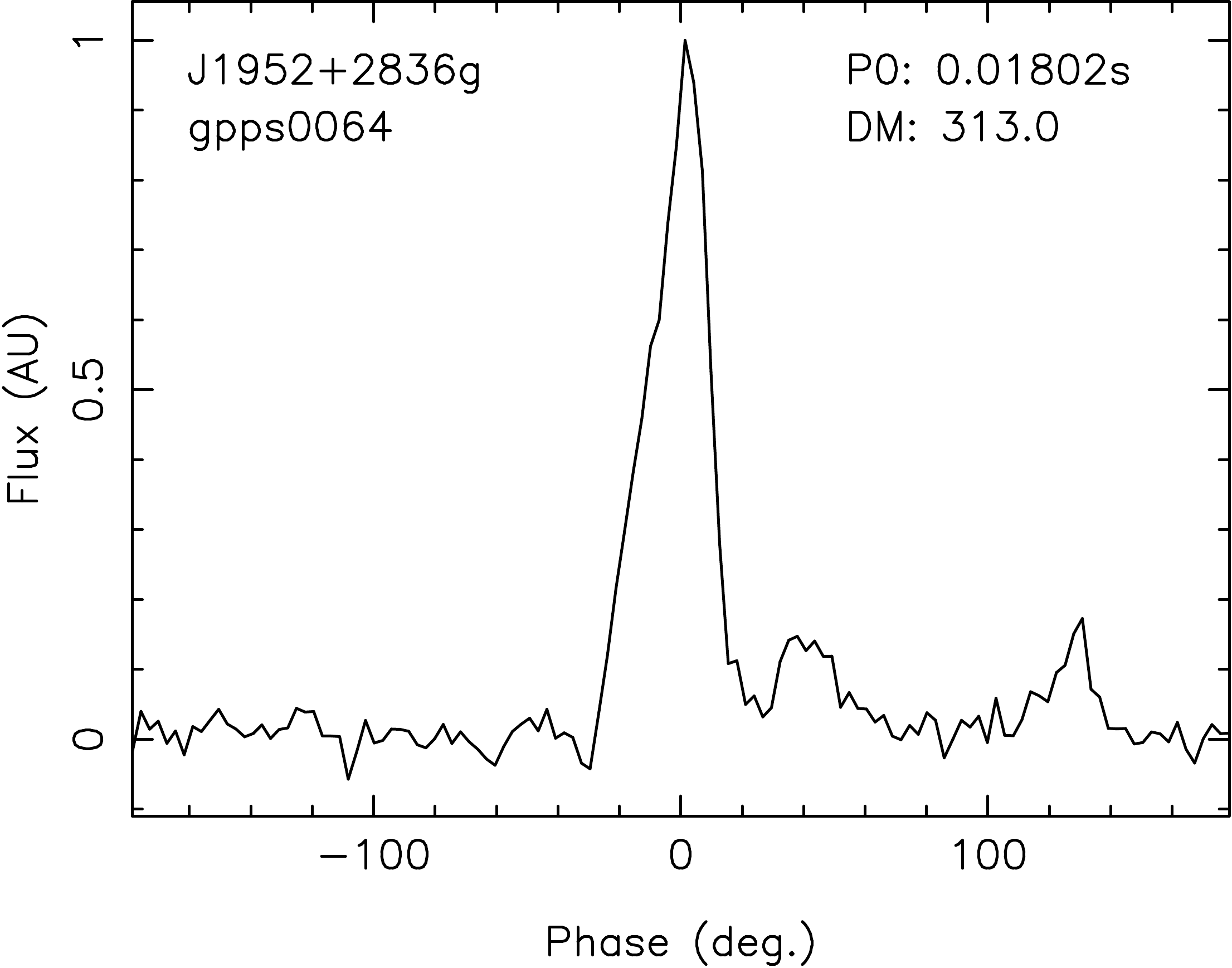}\\[2mm]
\includegraphics[width=39mm]{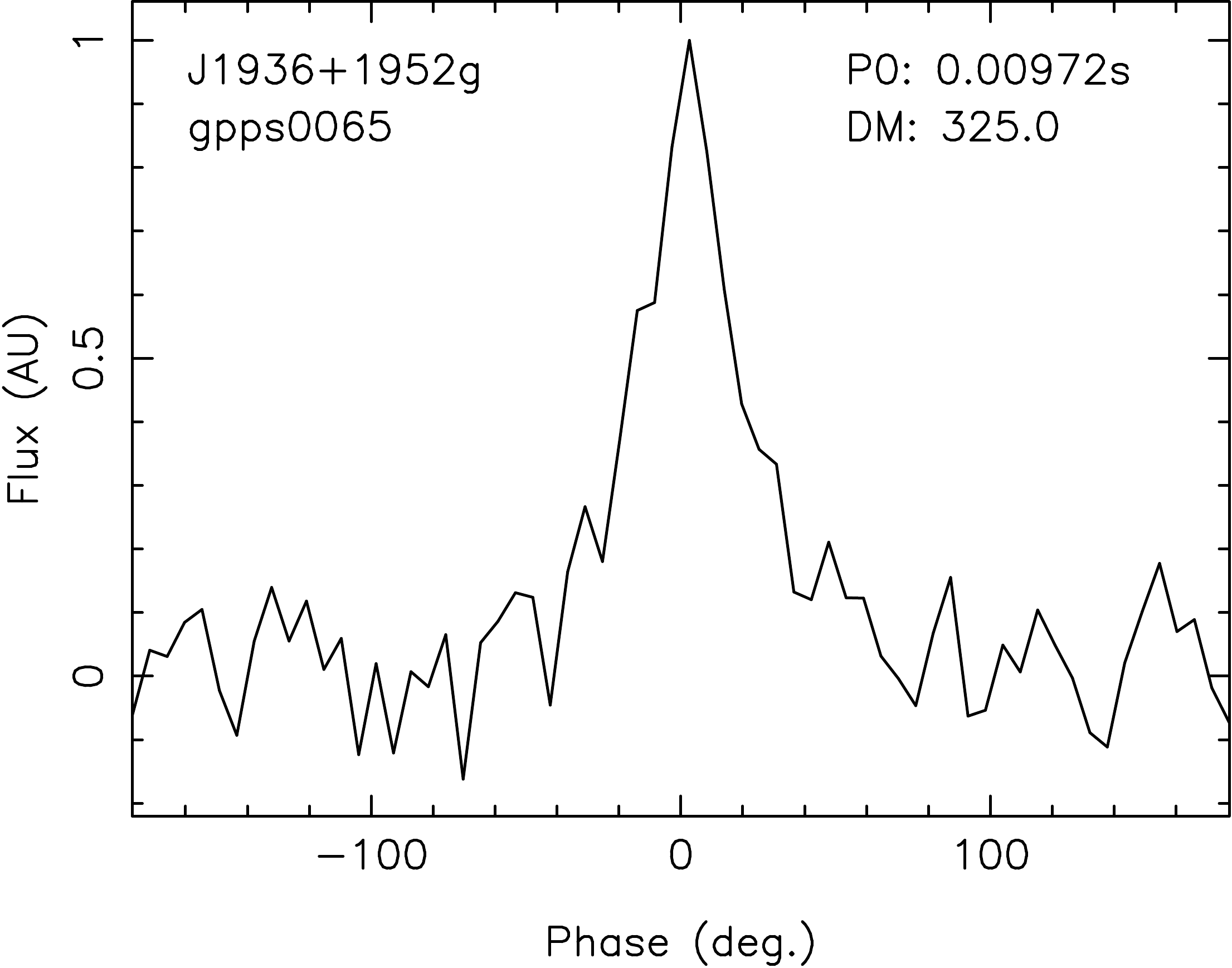}&
\includegraphics[width=39mm]{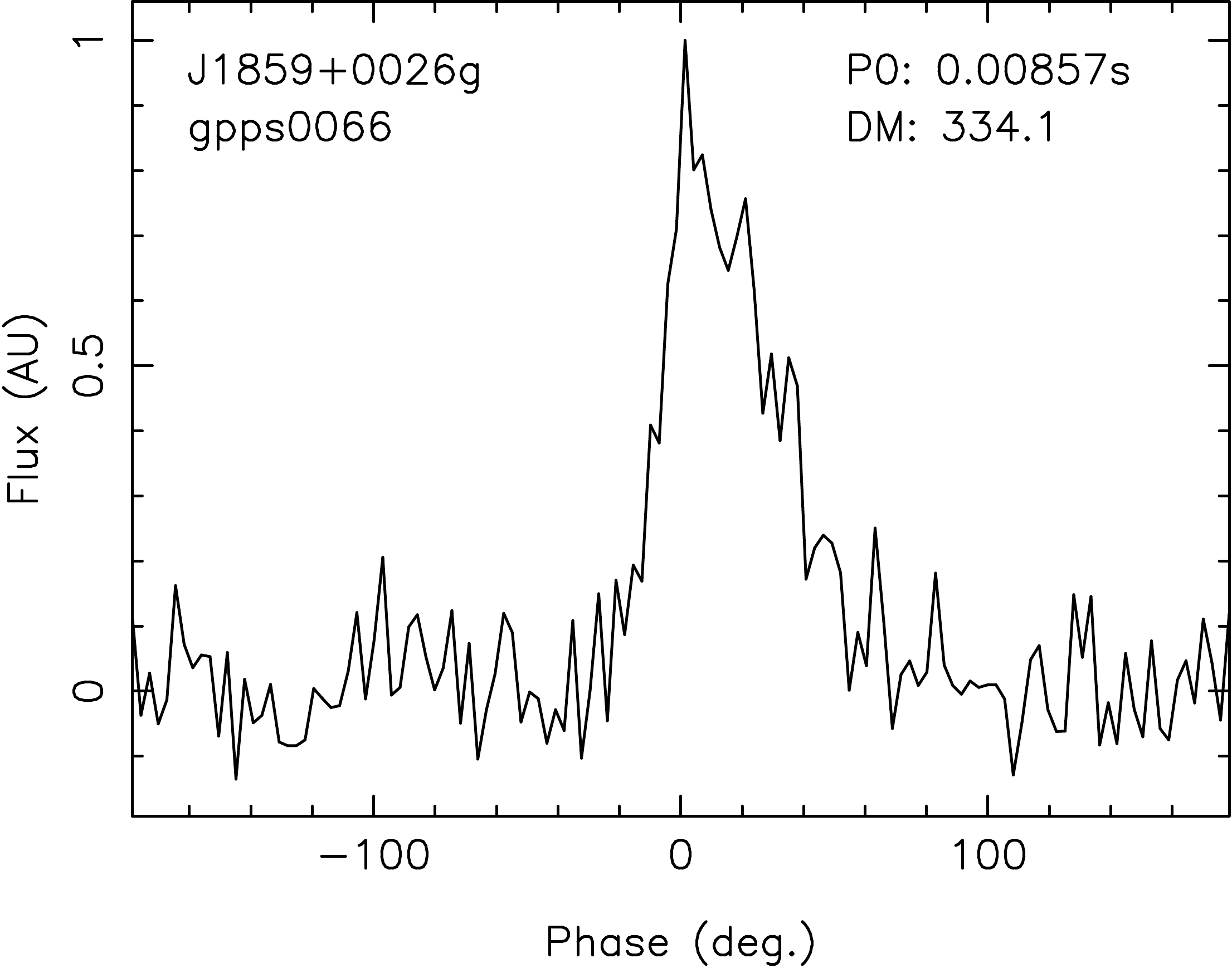}&
\includegraphics[width=39mm]{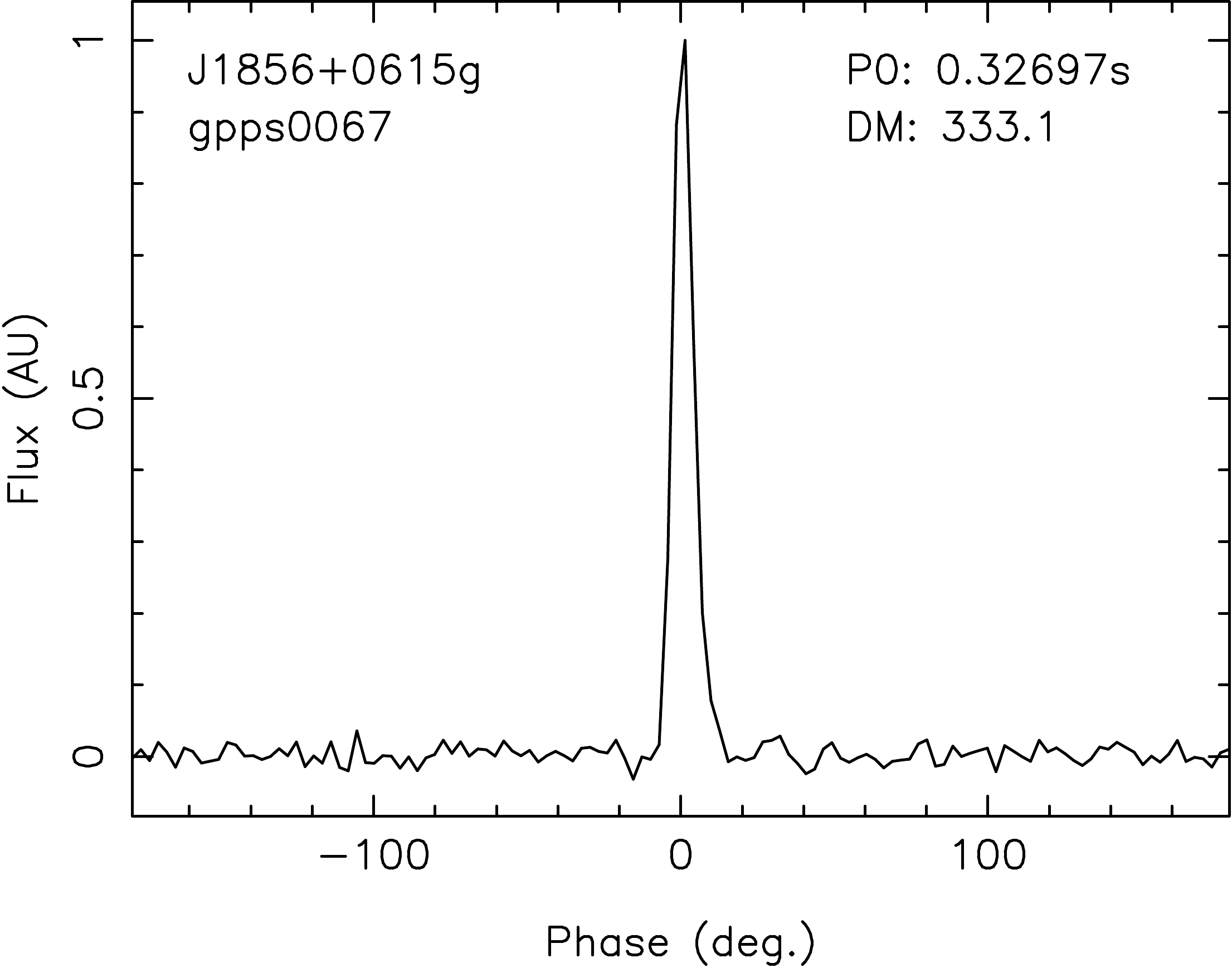}&
\includegraphics[width=39mm]{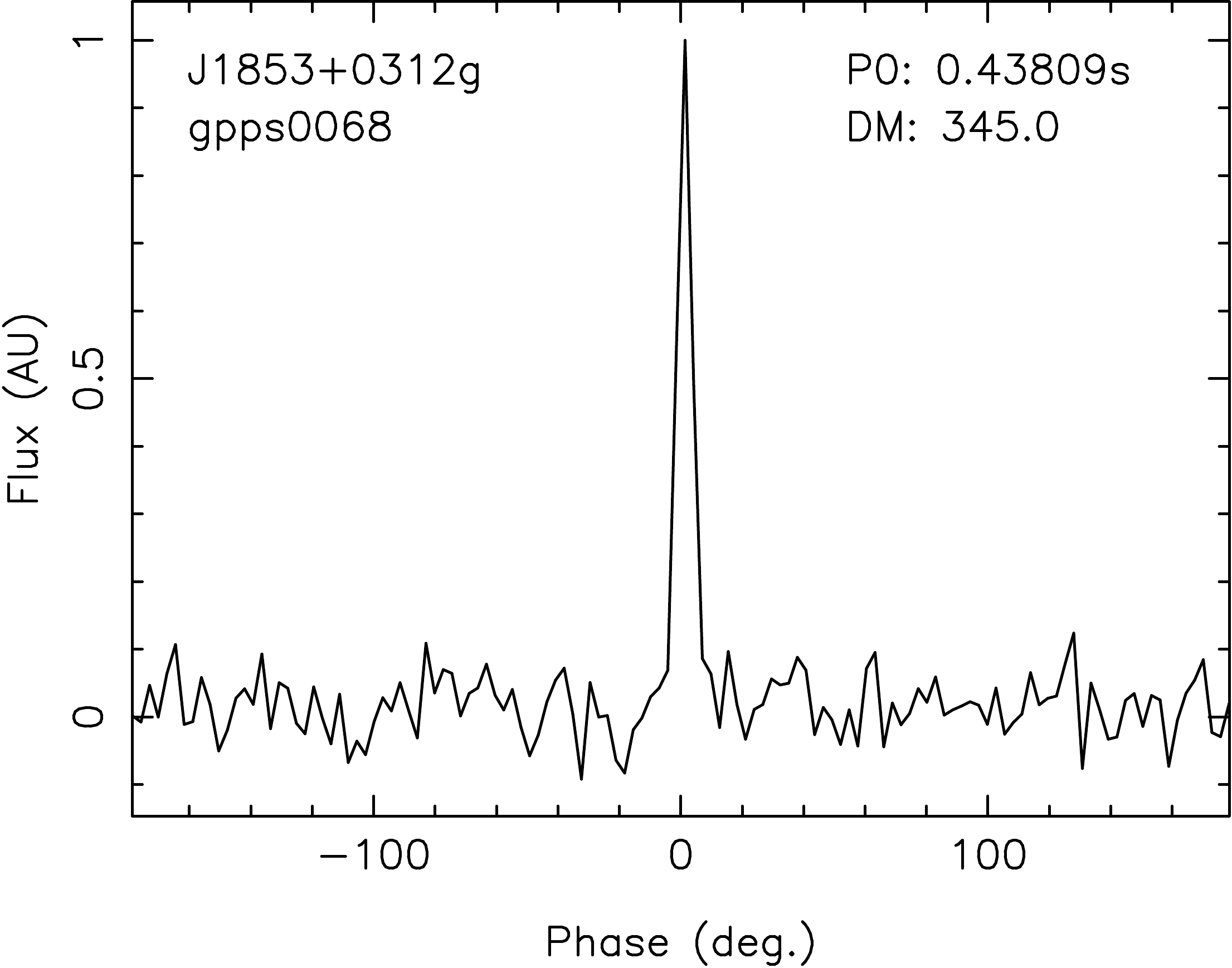}\\[2mm]
\includegraphics[width=39mm]{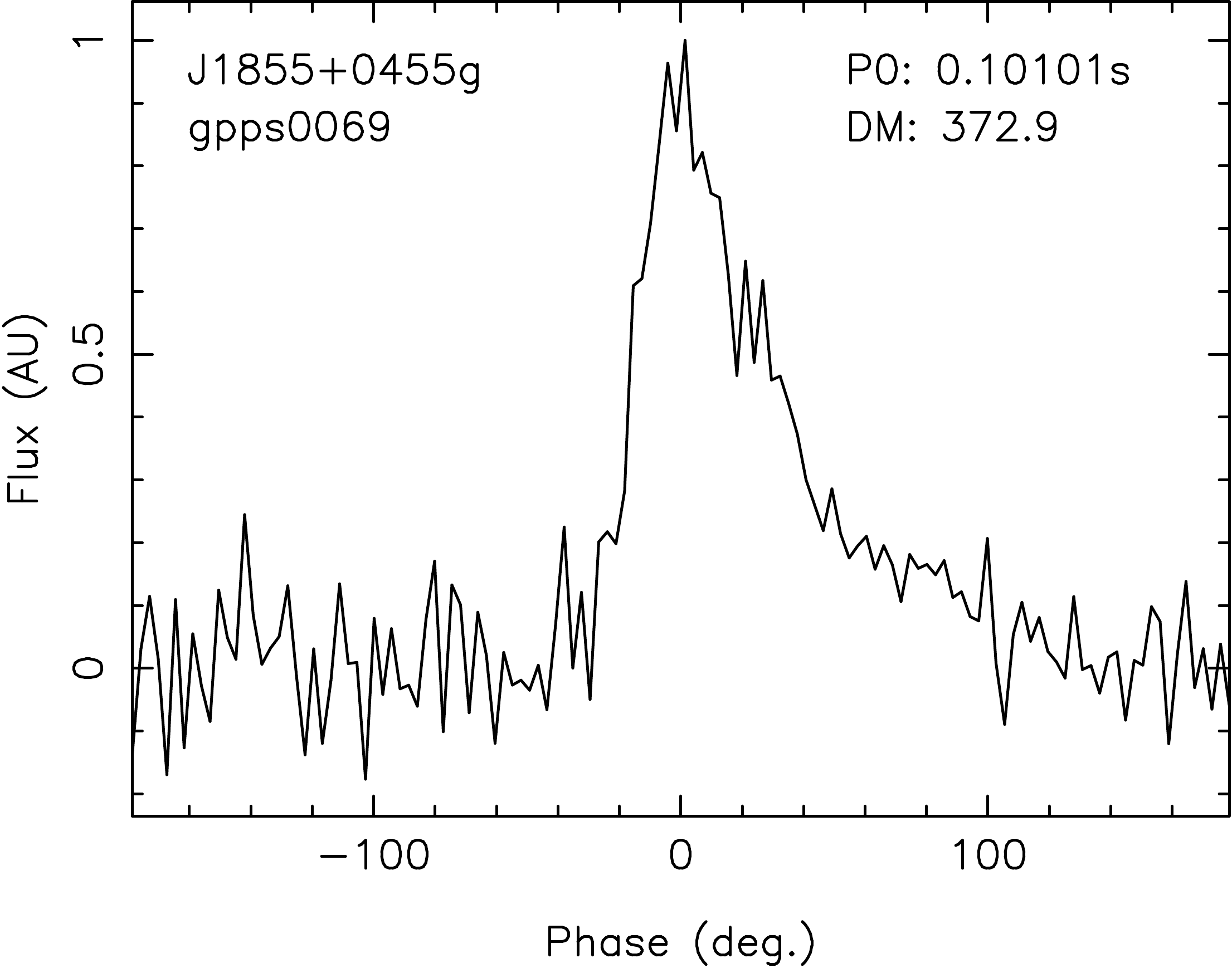}&
\includegraphics[width=39mm]{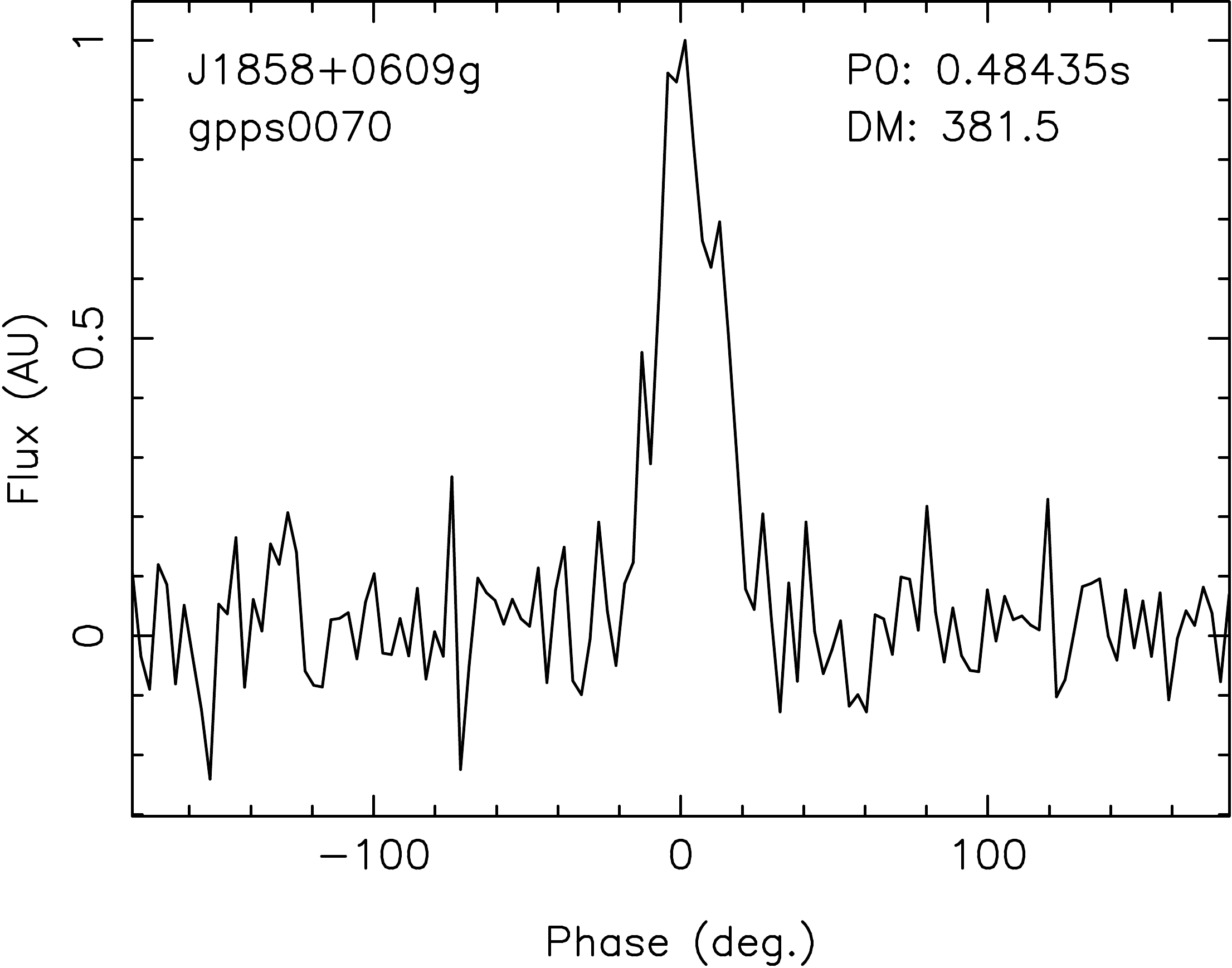}&
\includegraphics[width=39mm]{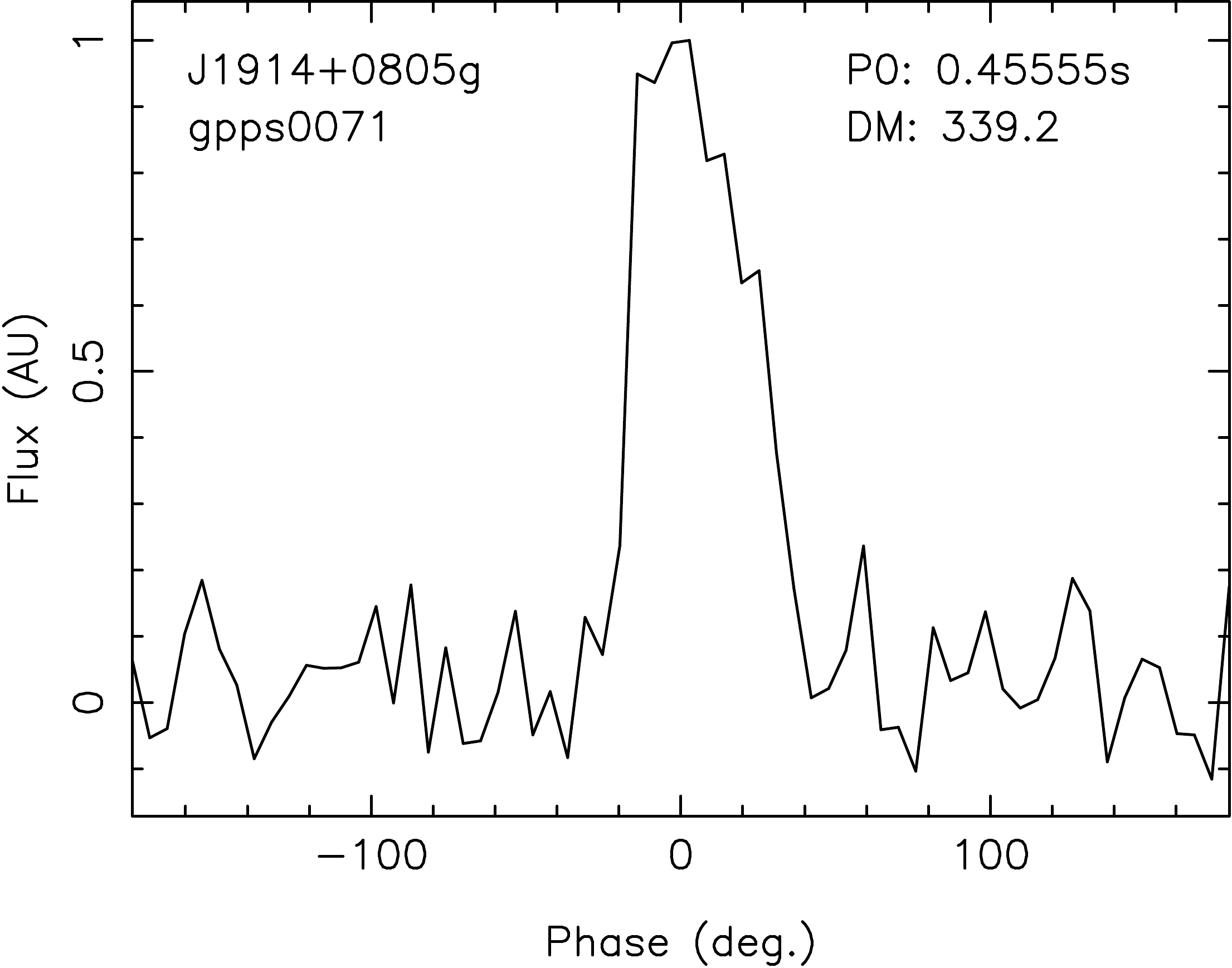}&
\includegraphics[width=39mm]{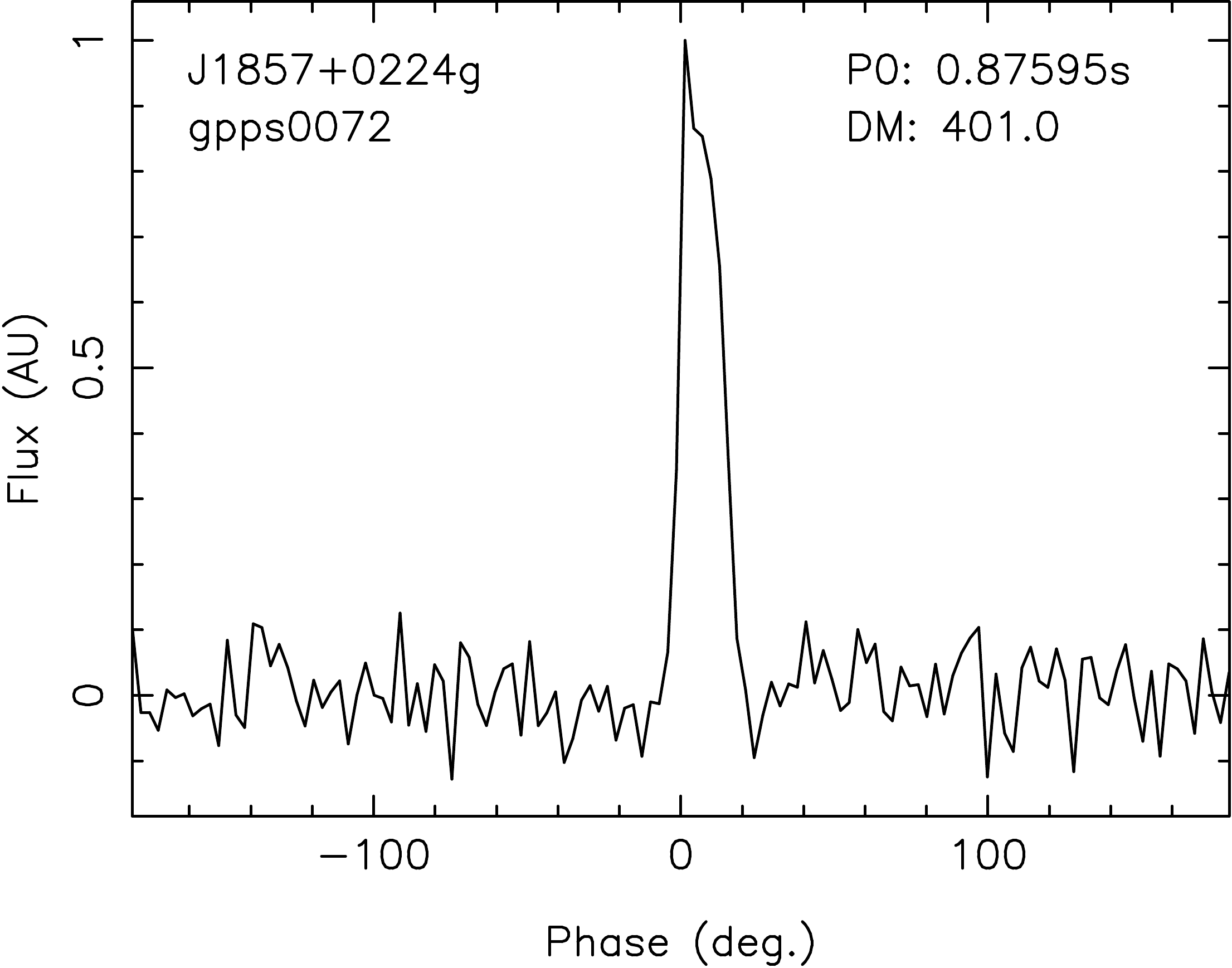}\\[2mm]
\includegraphics[width=39mm]{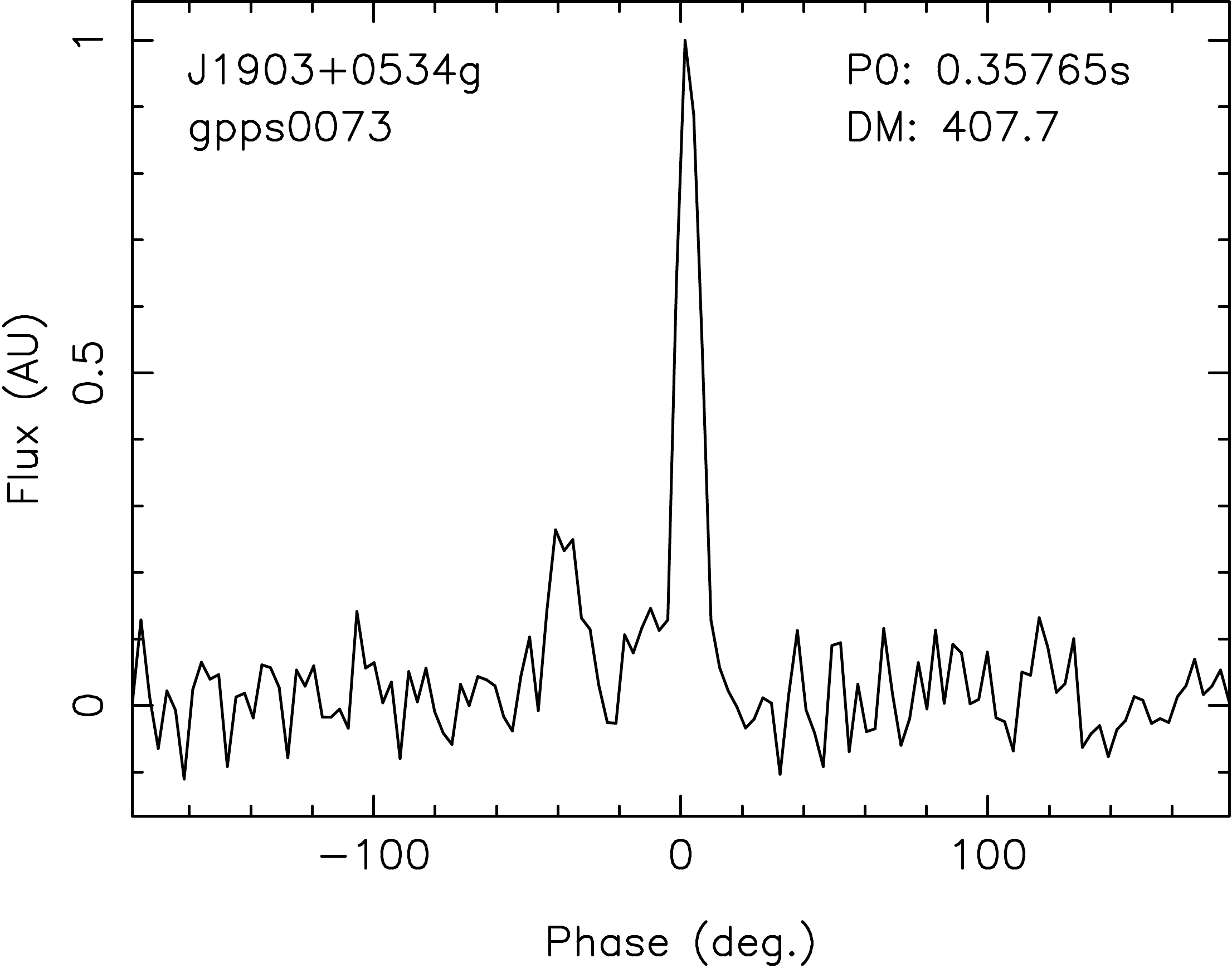}&
\includegraphics[width=39mm]{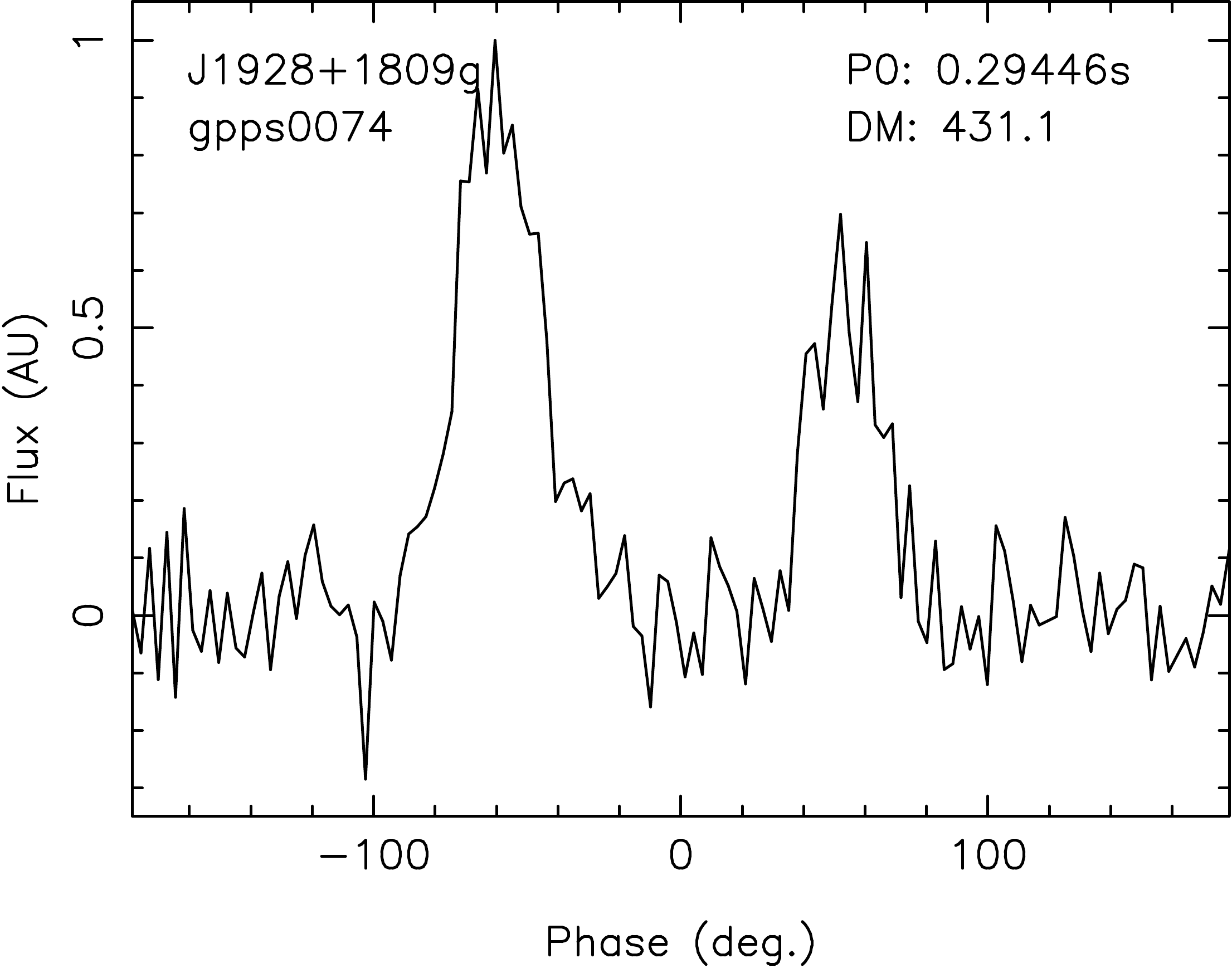}&
\includegraphics[width=39mm]{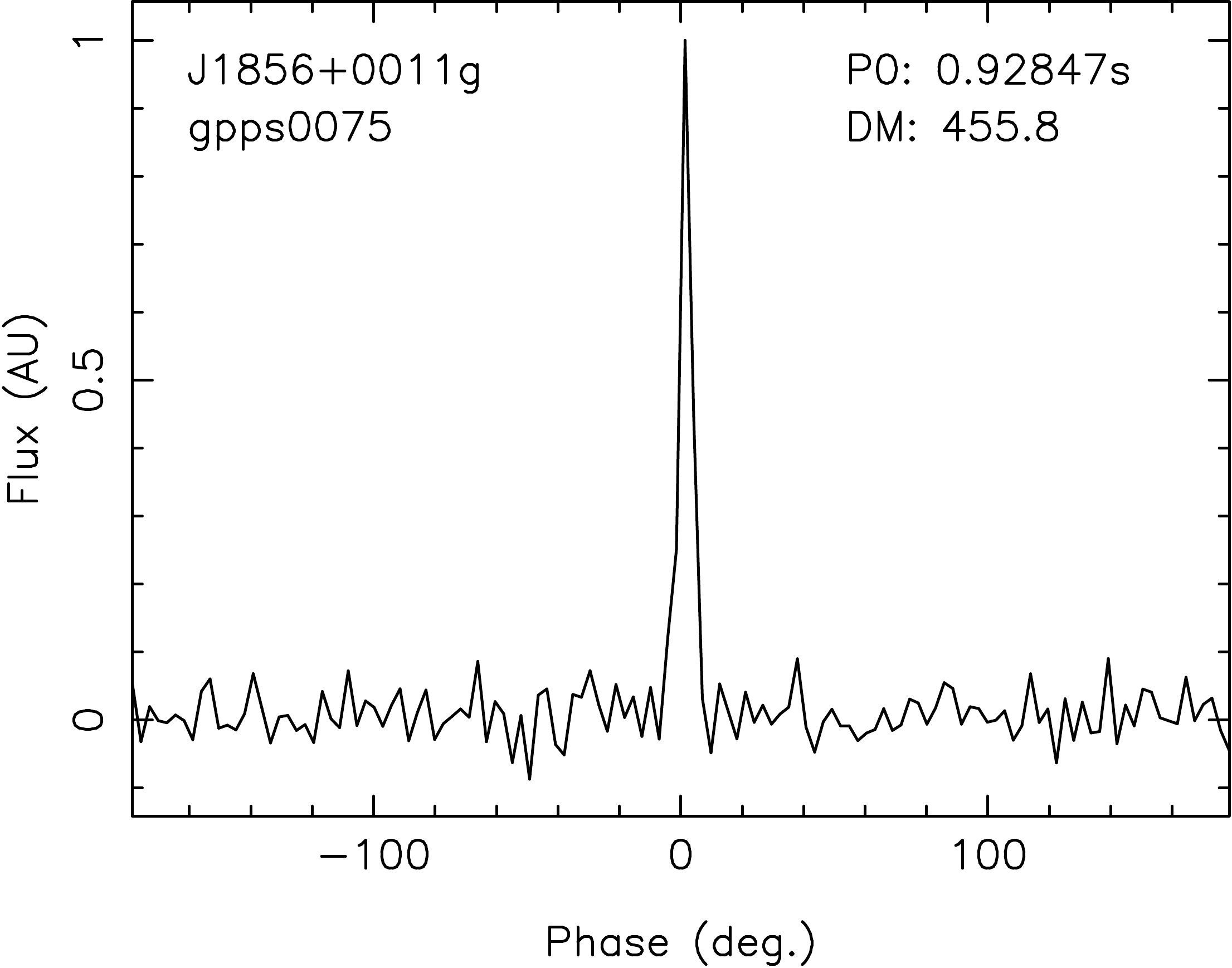}&
\includegraphics[width=39mm]{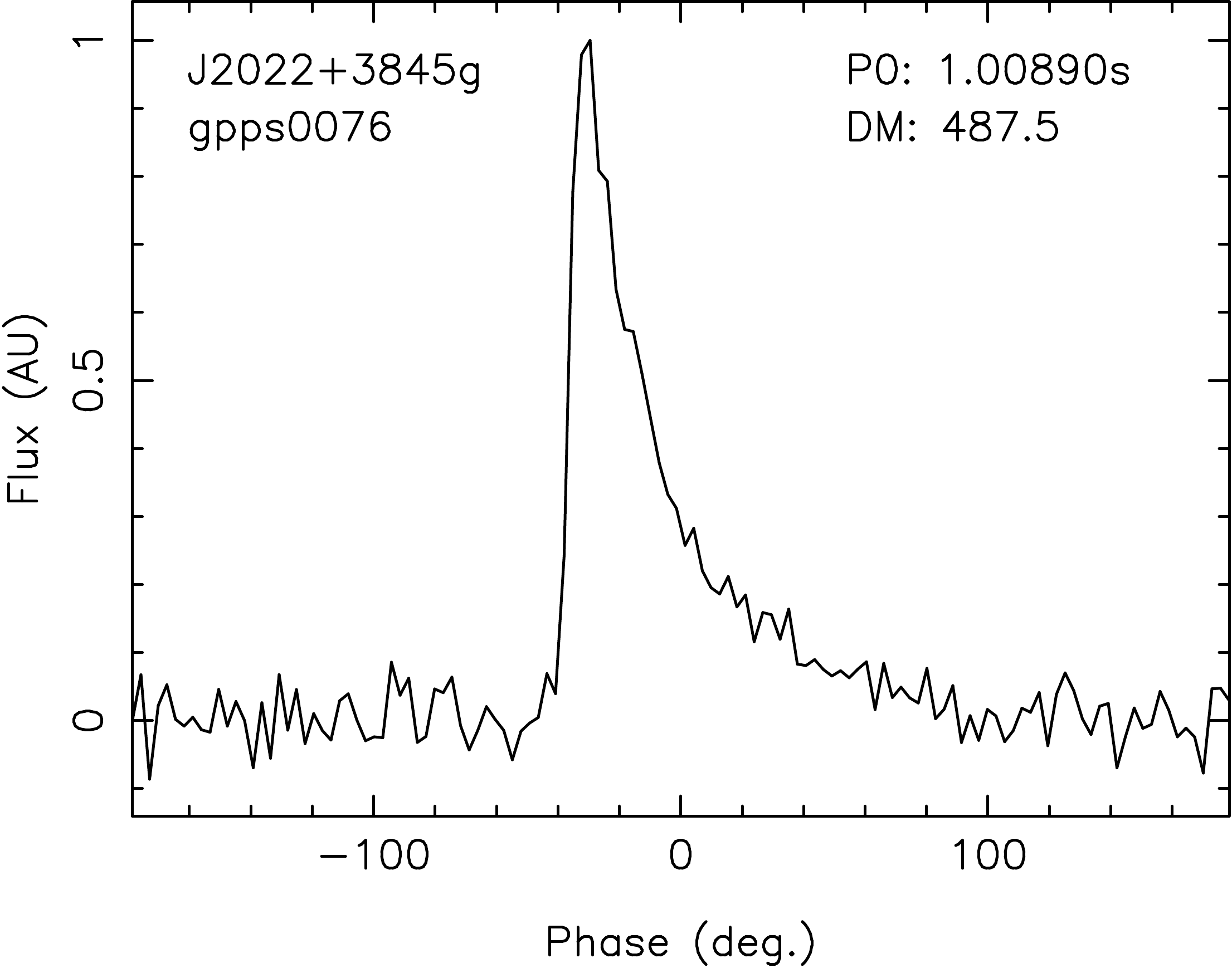}\\[2mm]
\includegraphics[width=39mm]{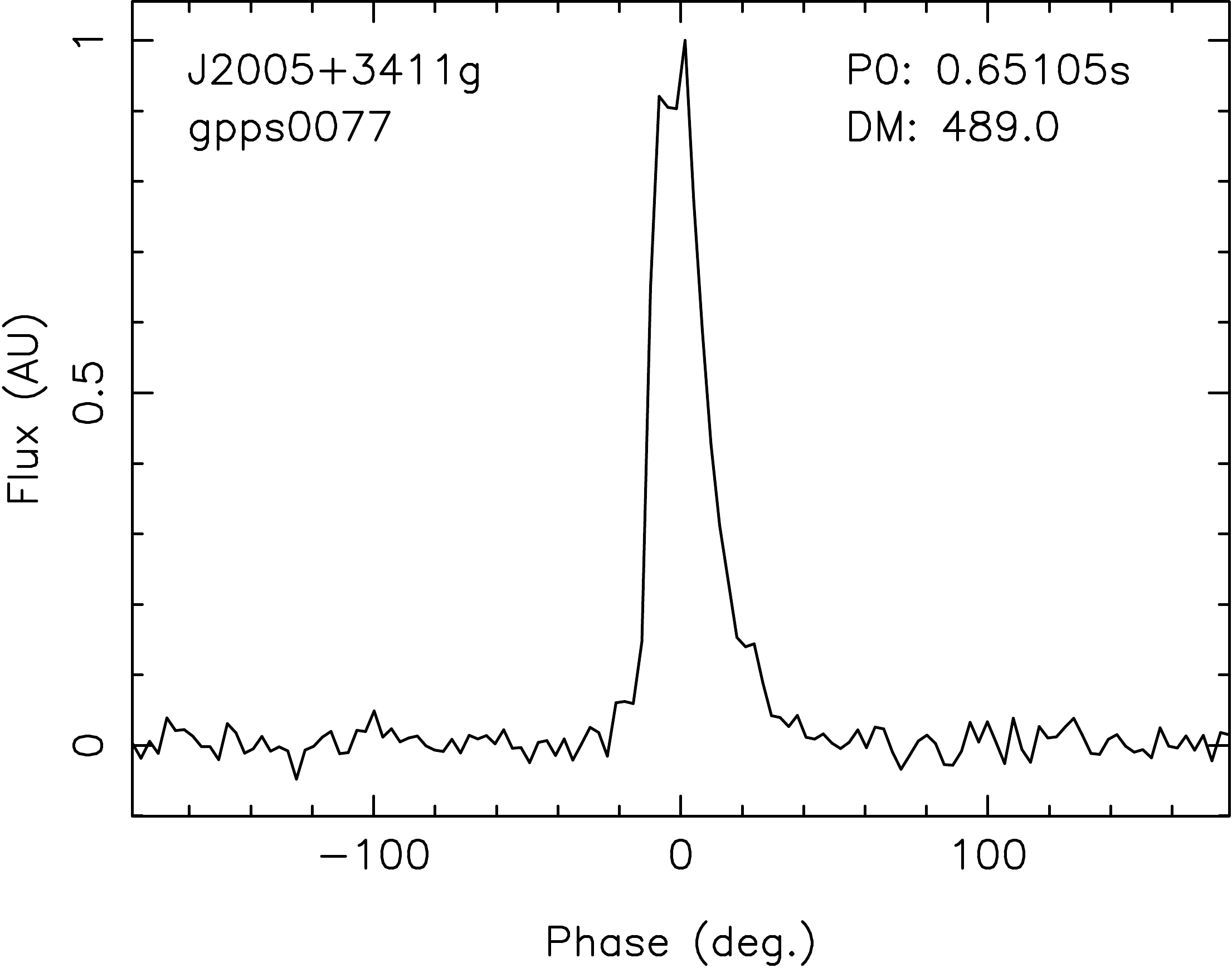}&
\includegraphics[width=39mm]{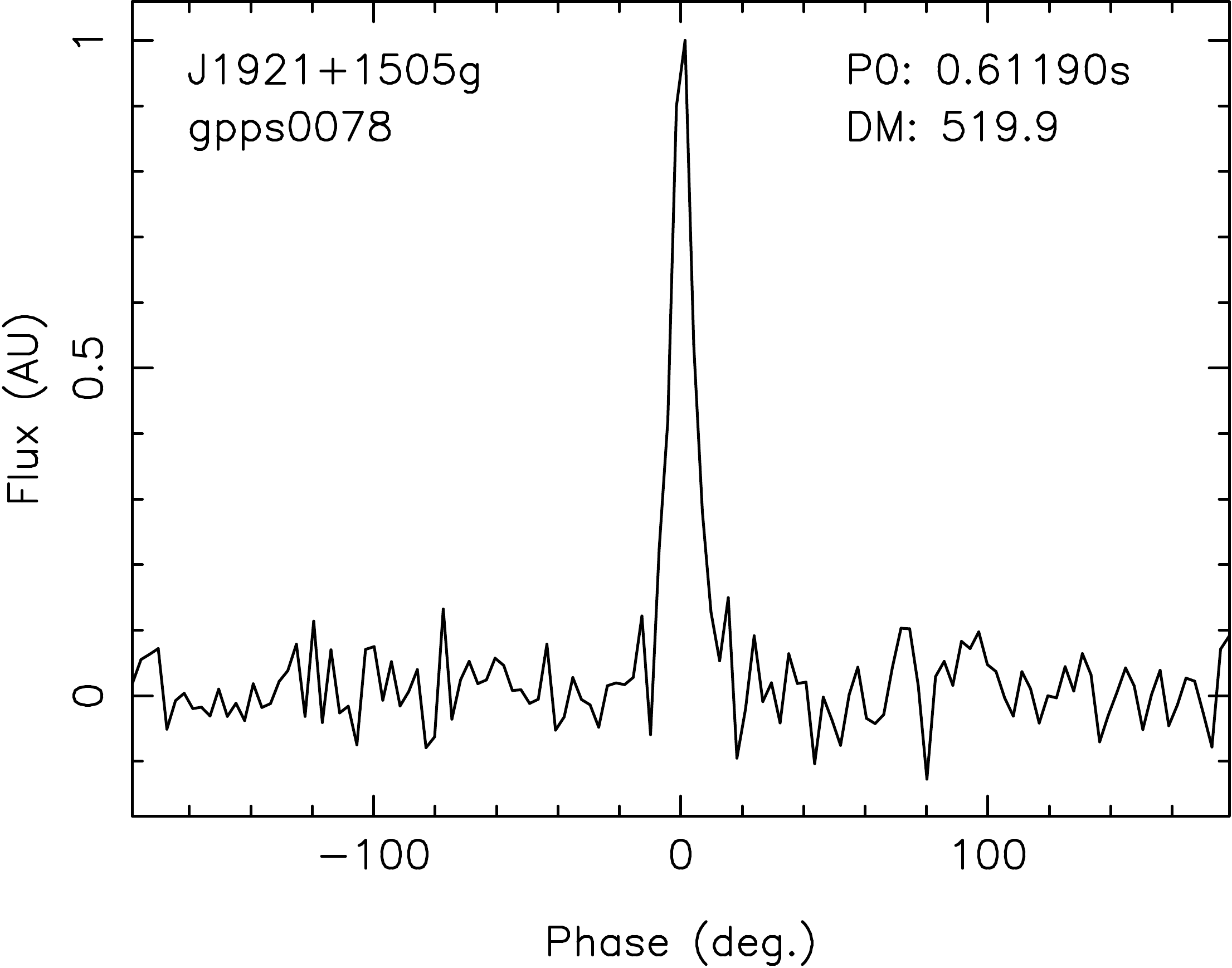}&
\includegraphics[width=39mm]{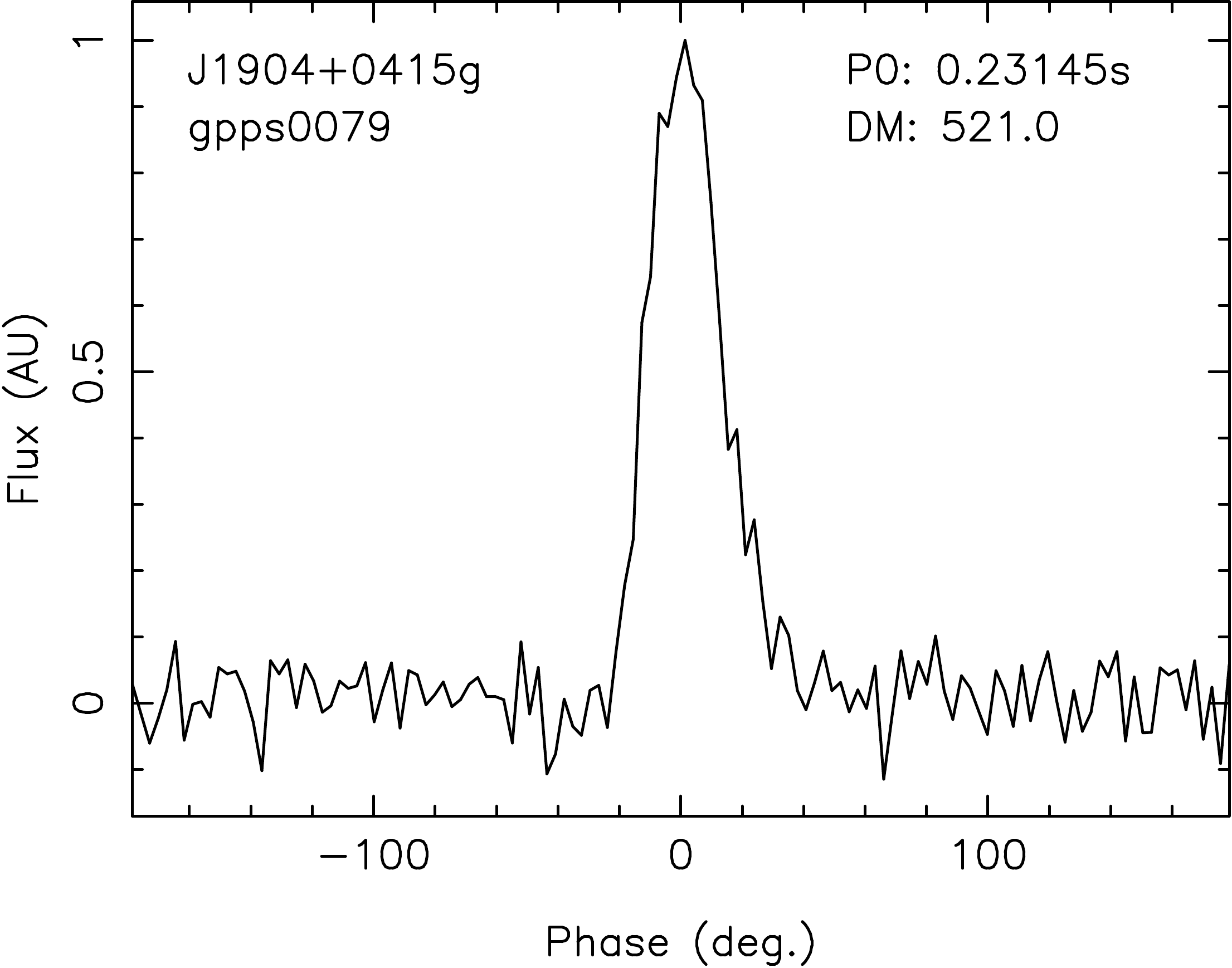}&
\includegraphics[width=39mm]{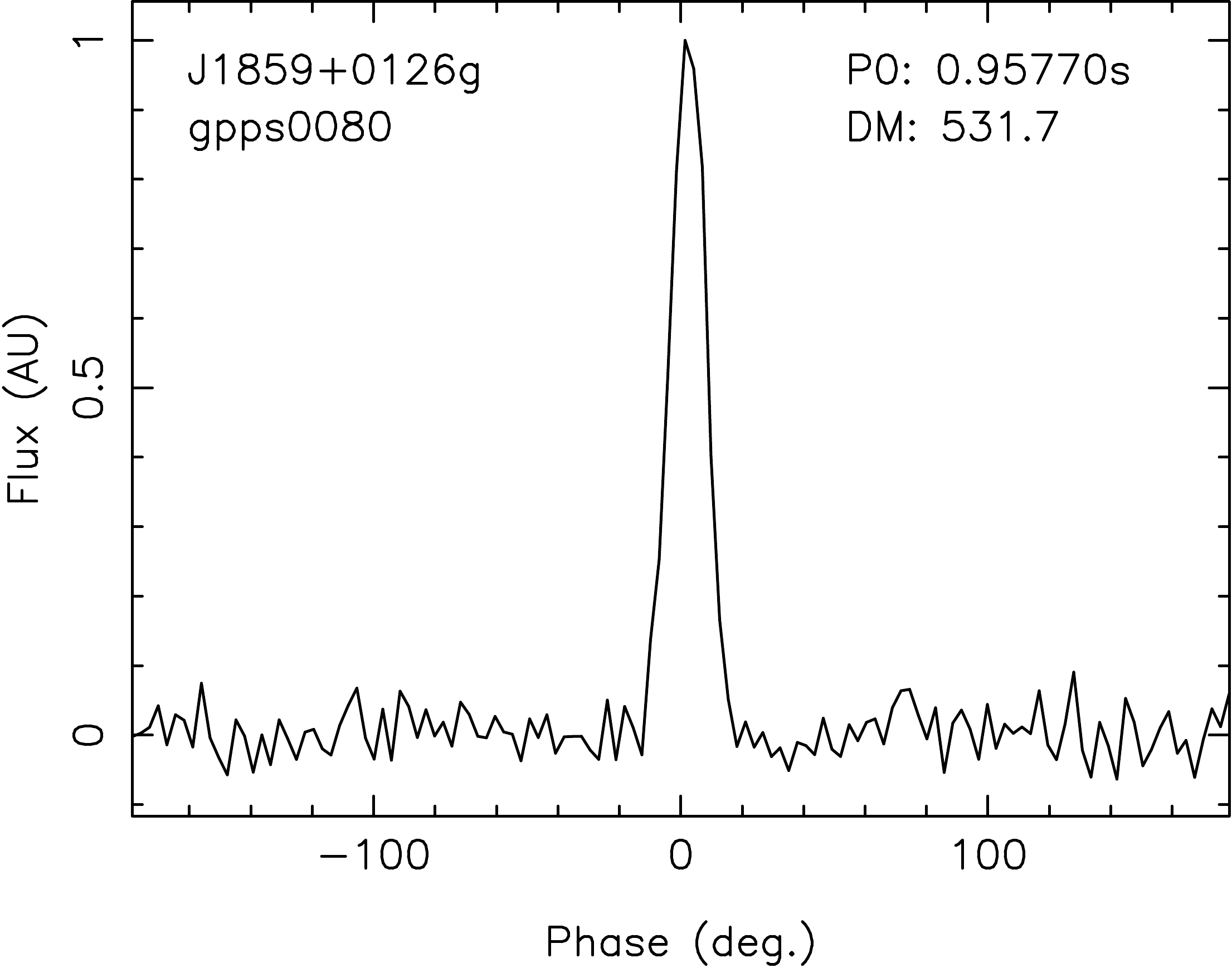}\\[2mm]
\includegraphics[width=39mm]{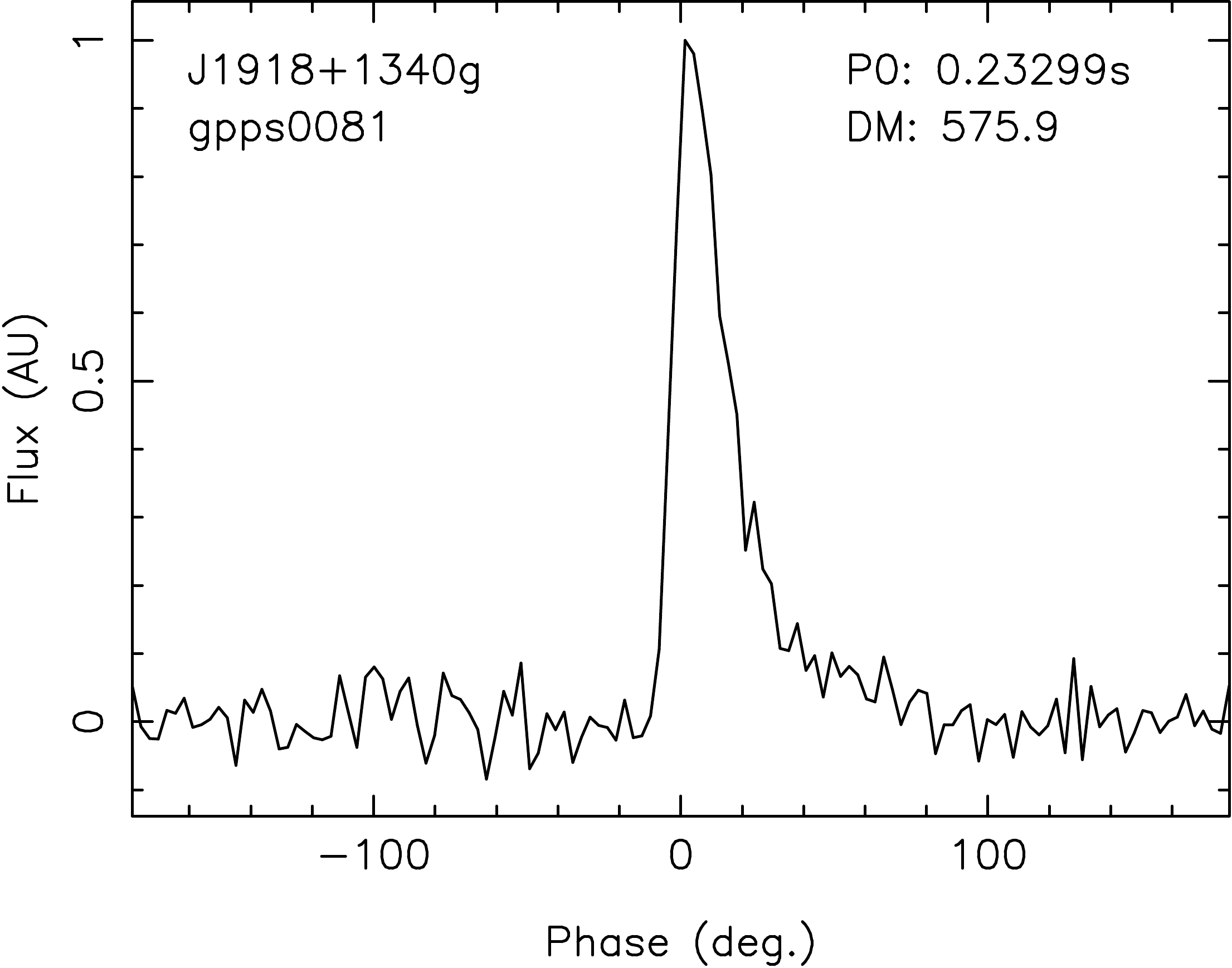}&
\includegraphics[width=39mm]{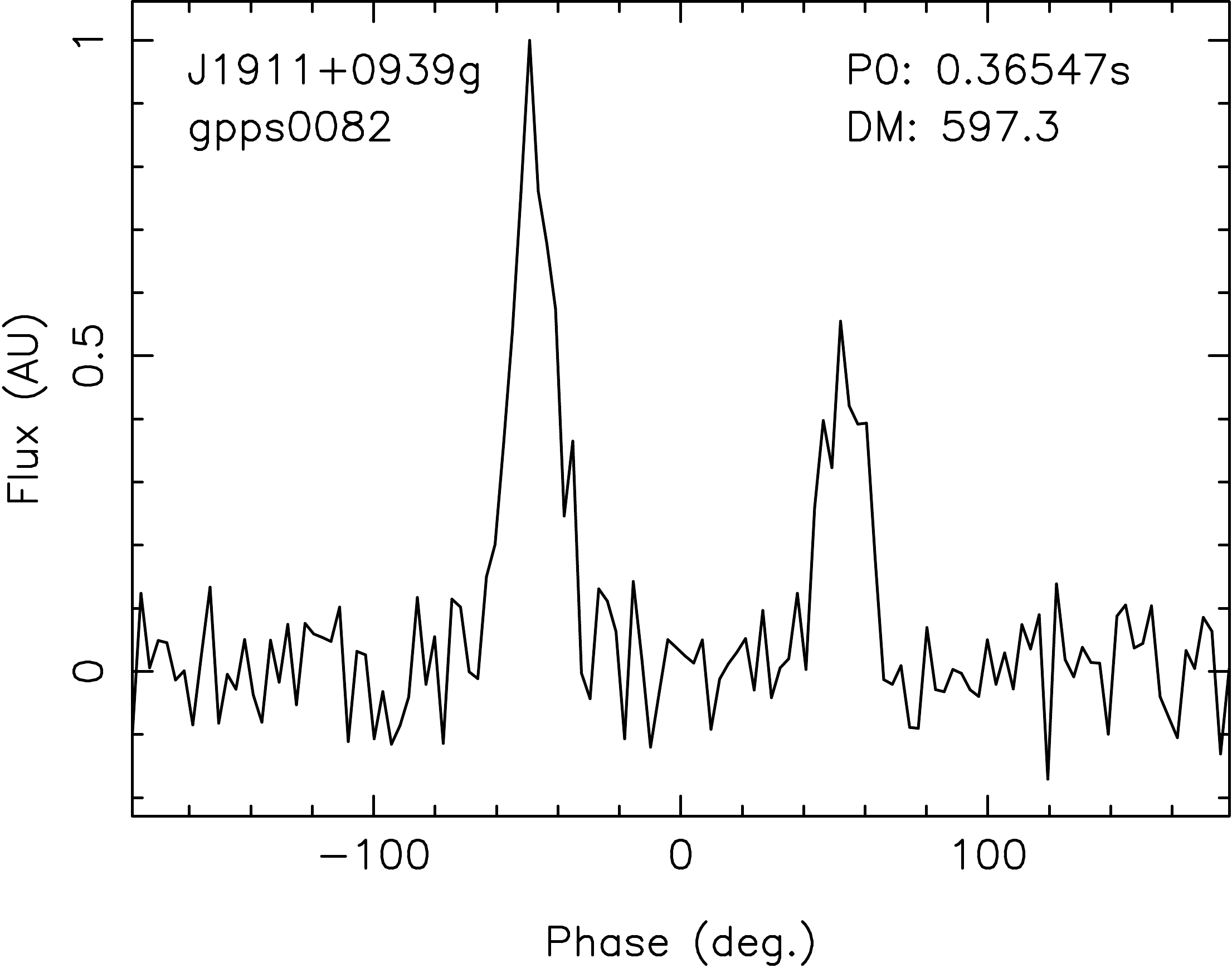}&
\includegraphics[width=39mm]{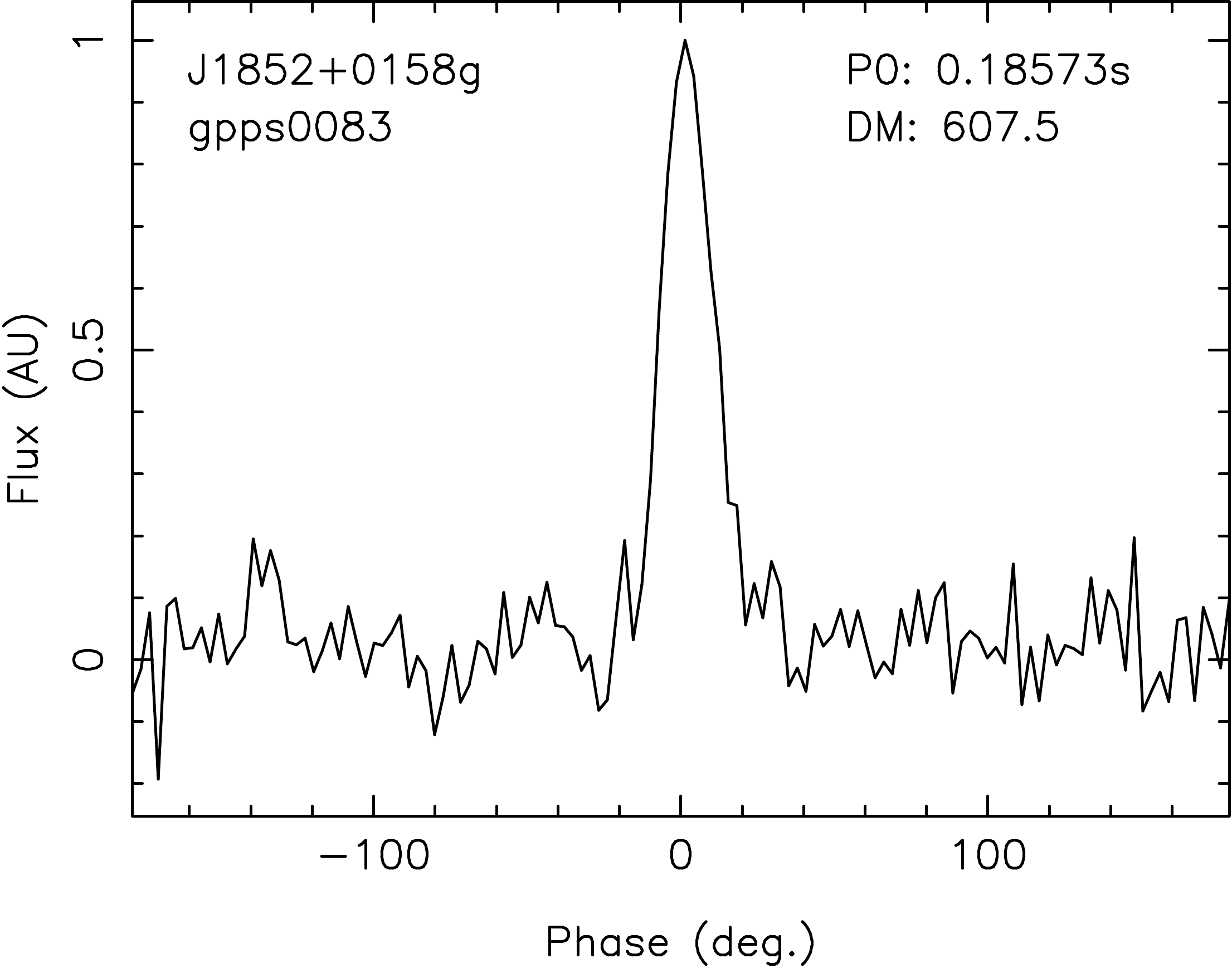}&
\includegraphics[width=39mm]{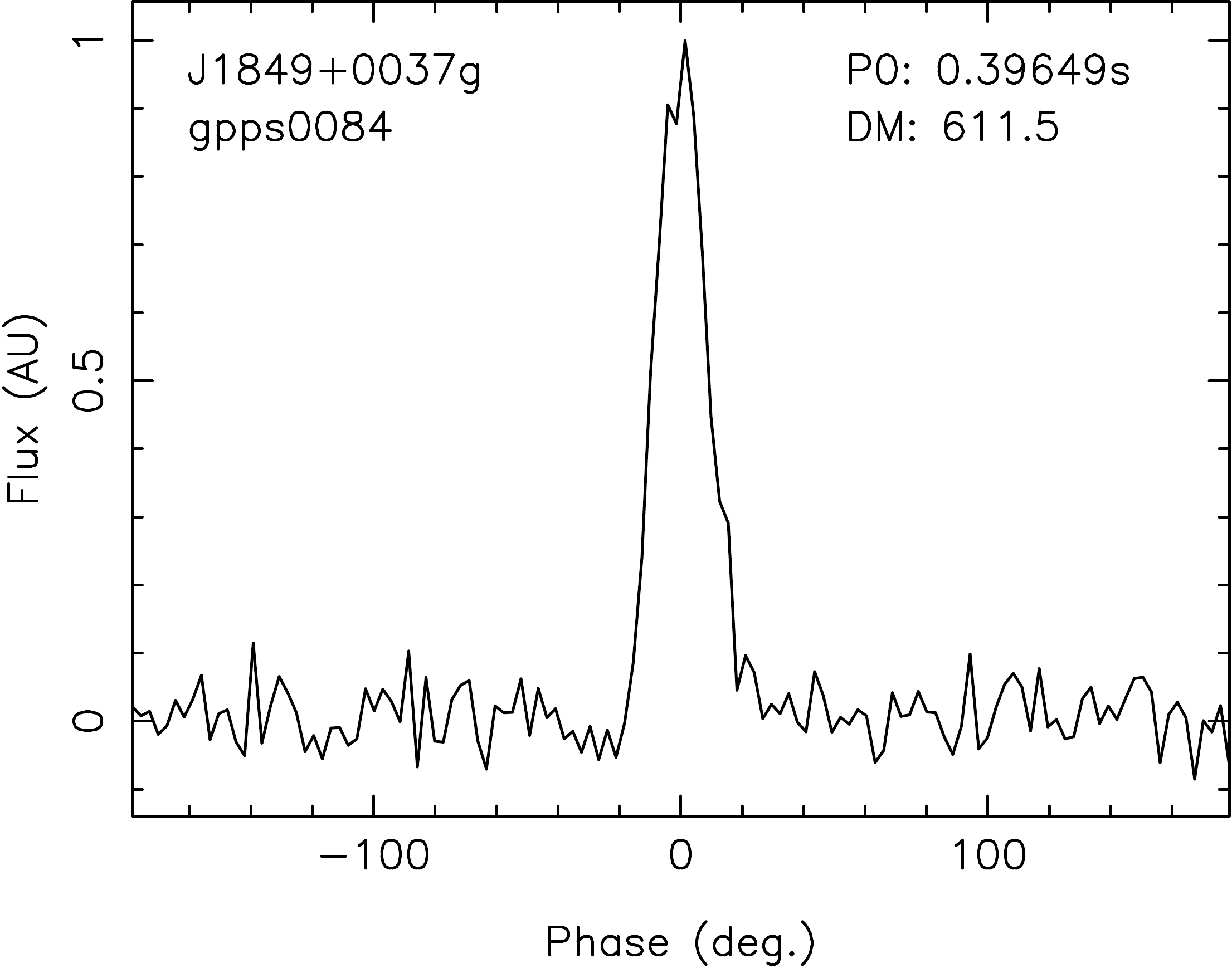}\\
\end{tabular}%

\begin{minipage}{3cm}
\caption[]{
-- {\it Continued}.}\end{minipage}
\addtocounter{figure}{-1}
\end{figure*}%
\begin{figure*}
\centering
\begin{tabular}{rrrrrr}
\includegraphics[width=39mm]{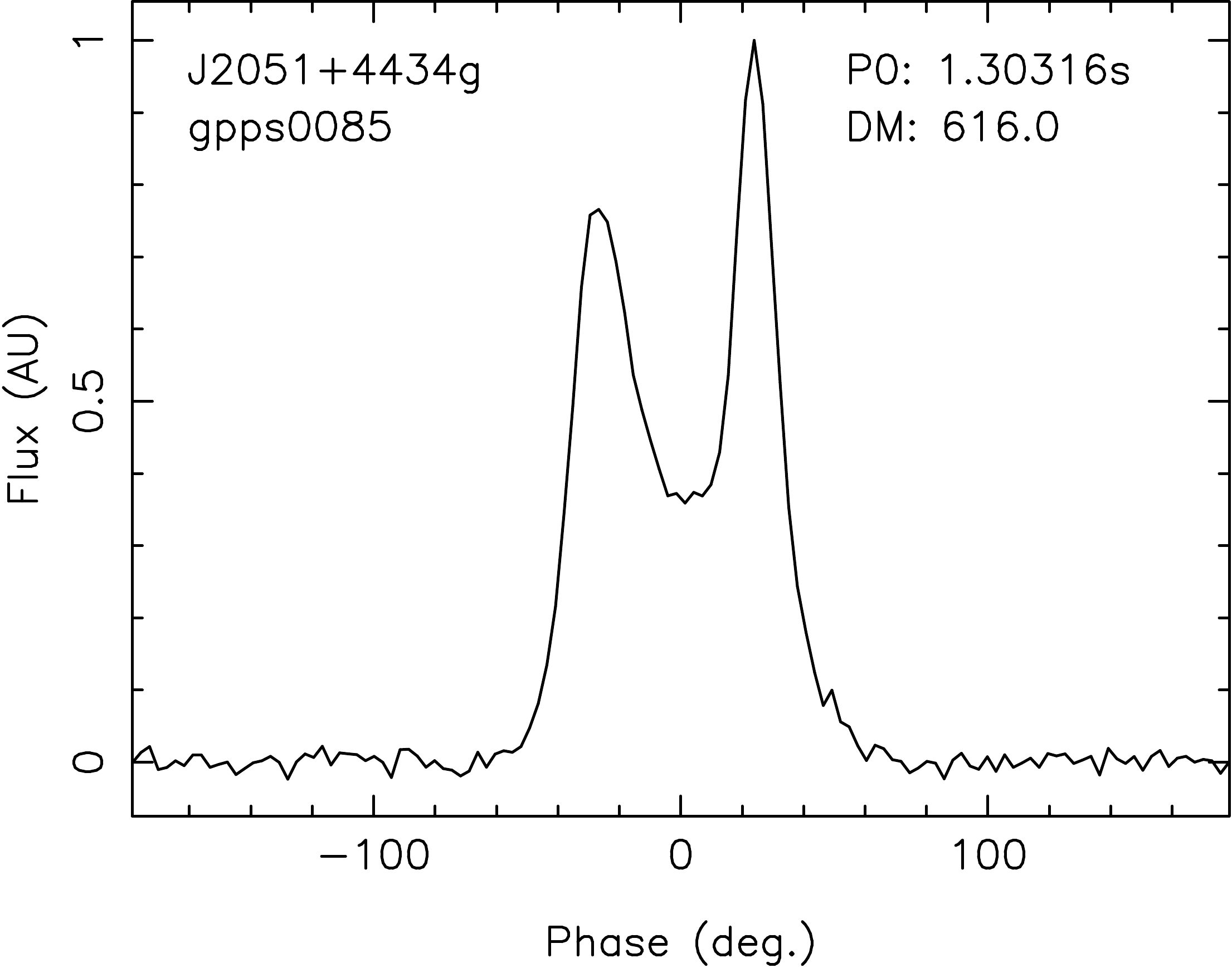}&
\includegraphics[width=39mm]{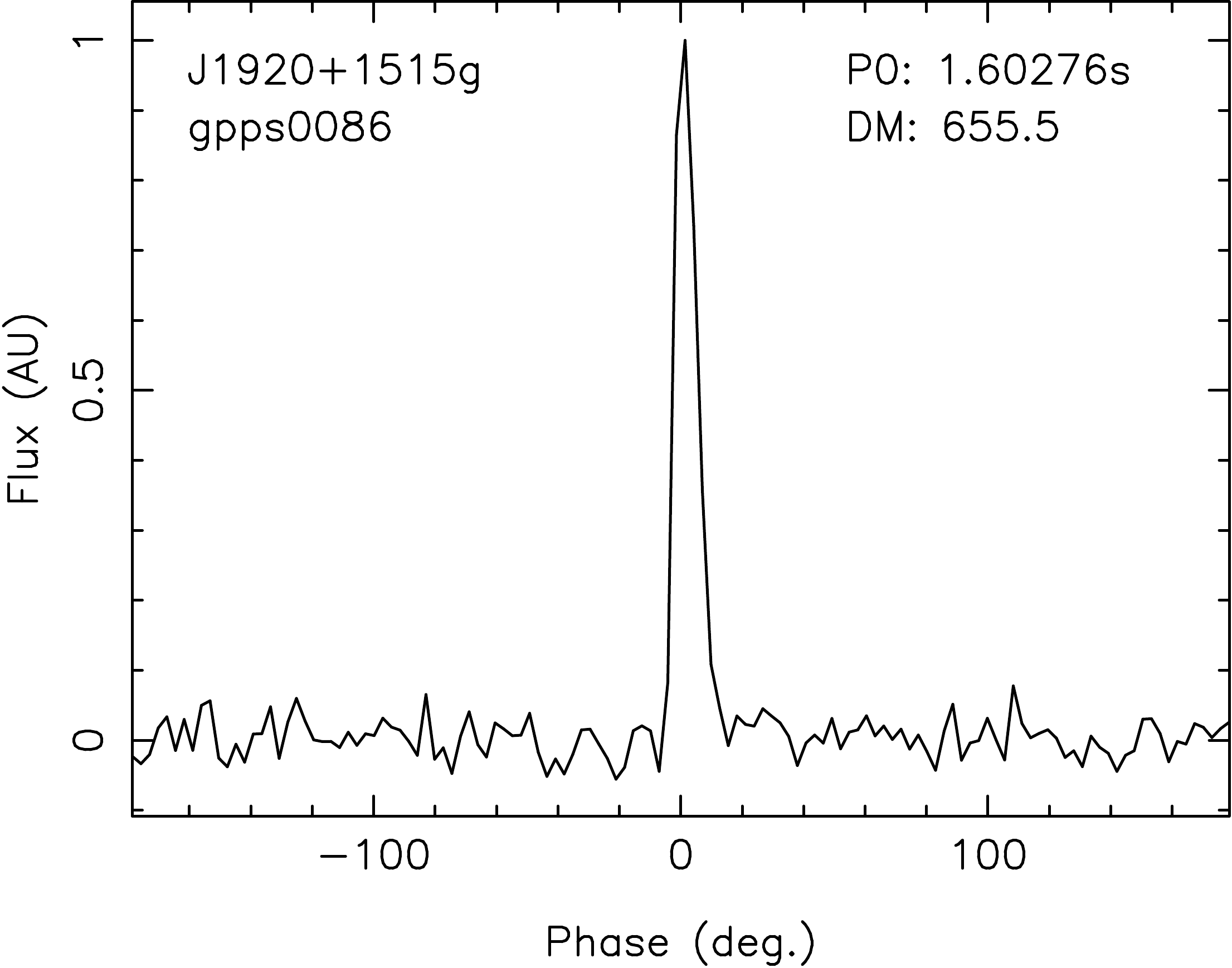}&
\includegraphics[width=39mm]{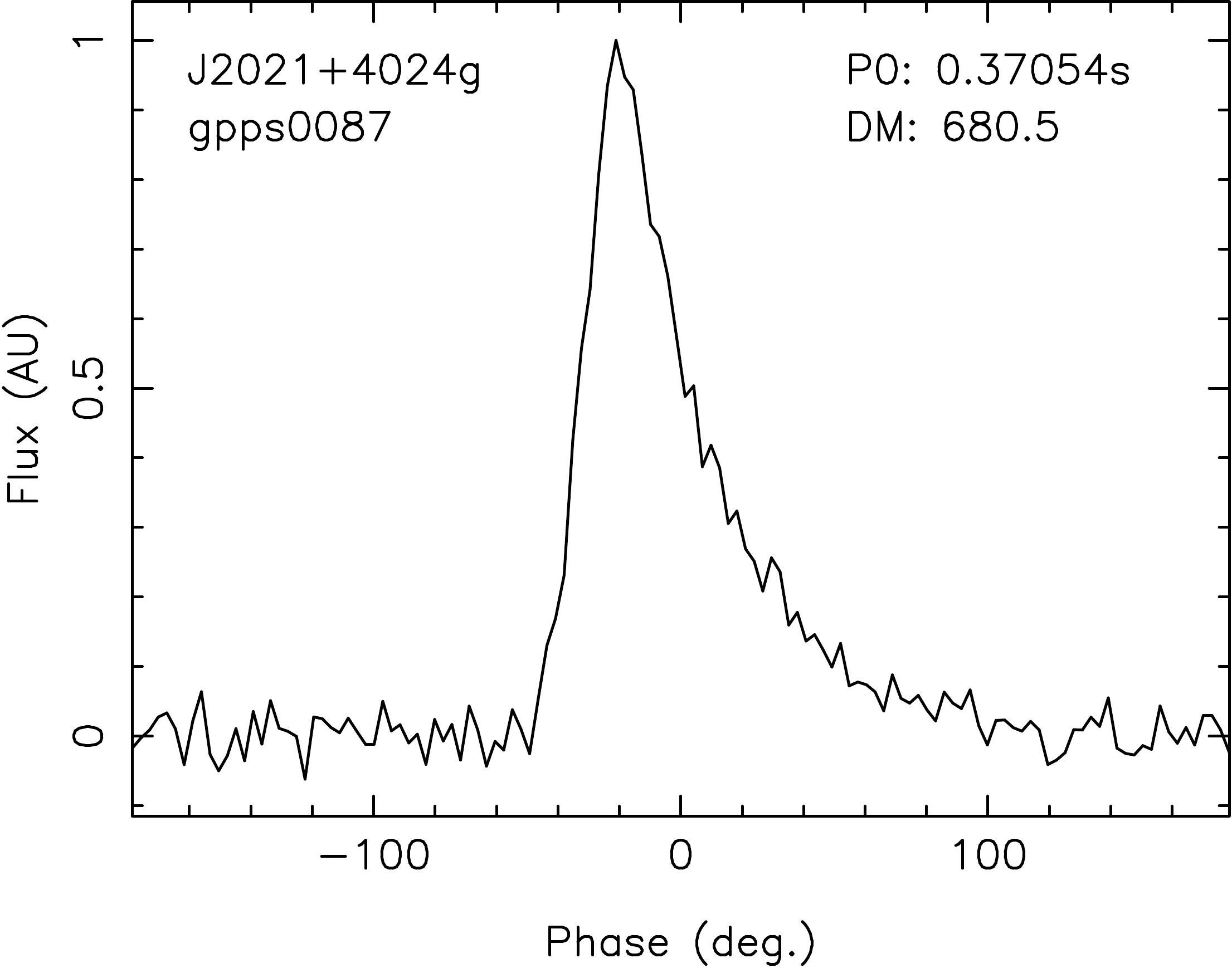}&
\includegraphics[width=39mm]{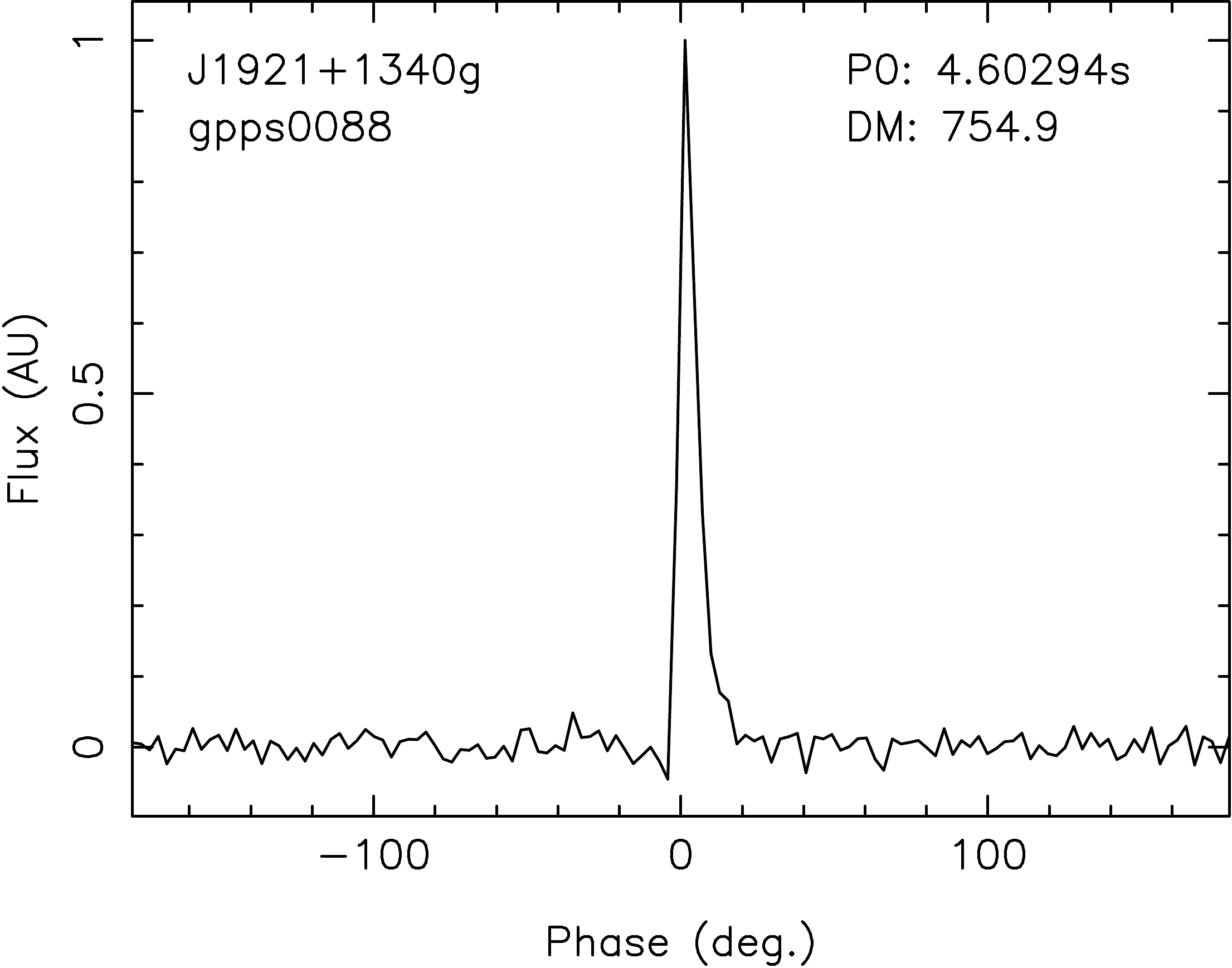}\\[2mm]
\includegraphics[width=39mm]{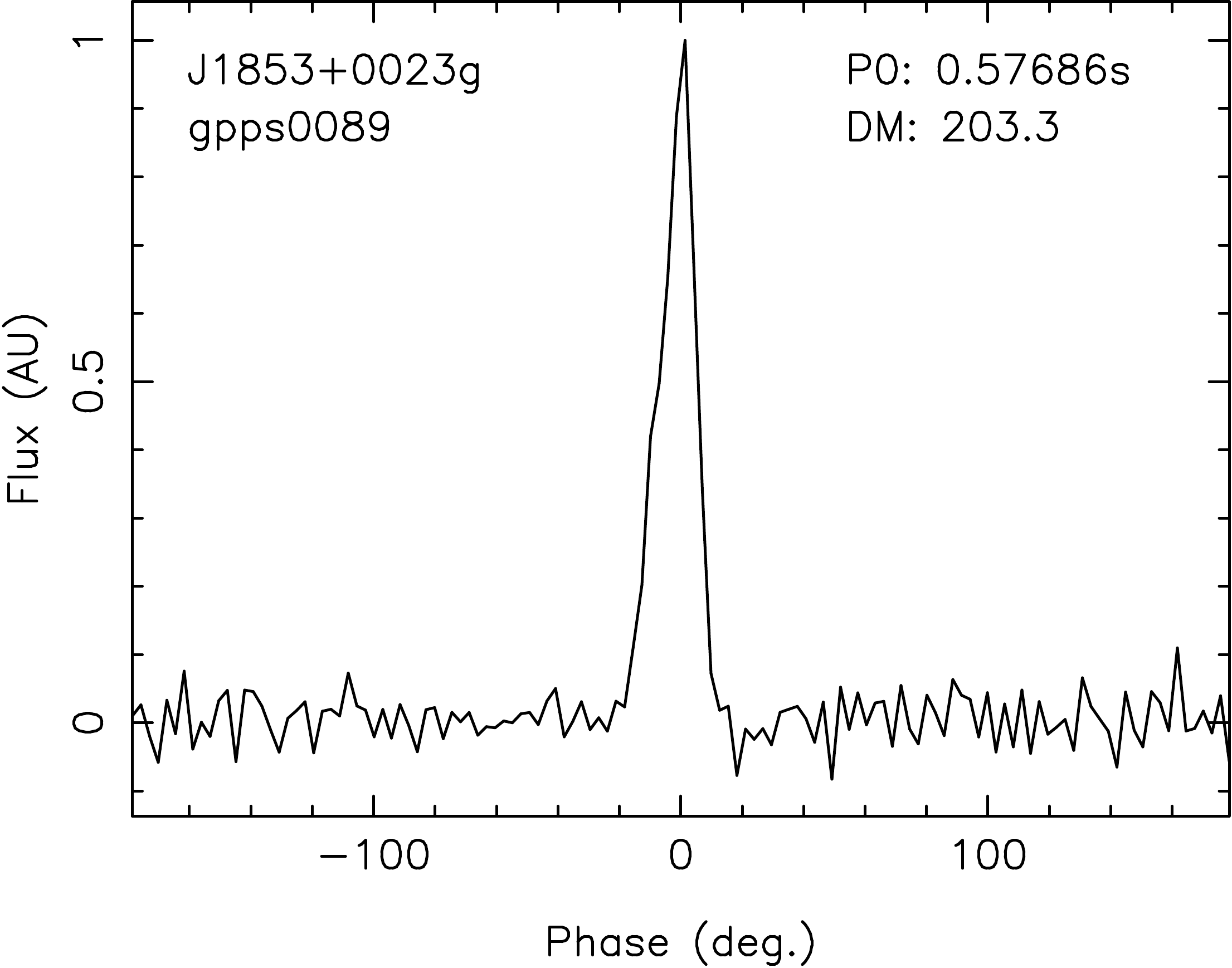}&
\includegraphics[width=39mm]{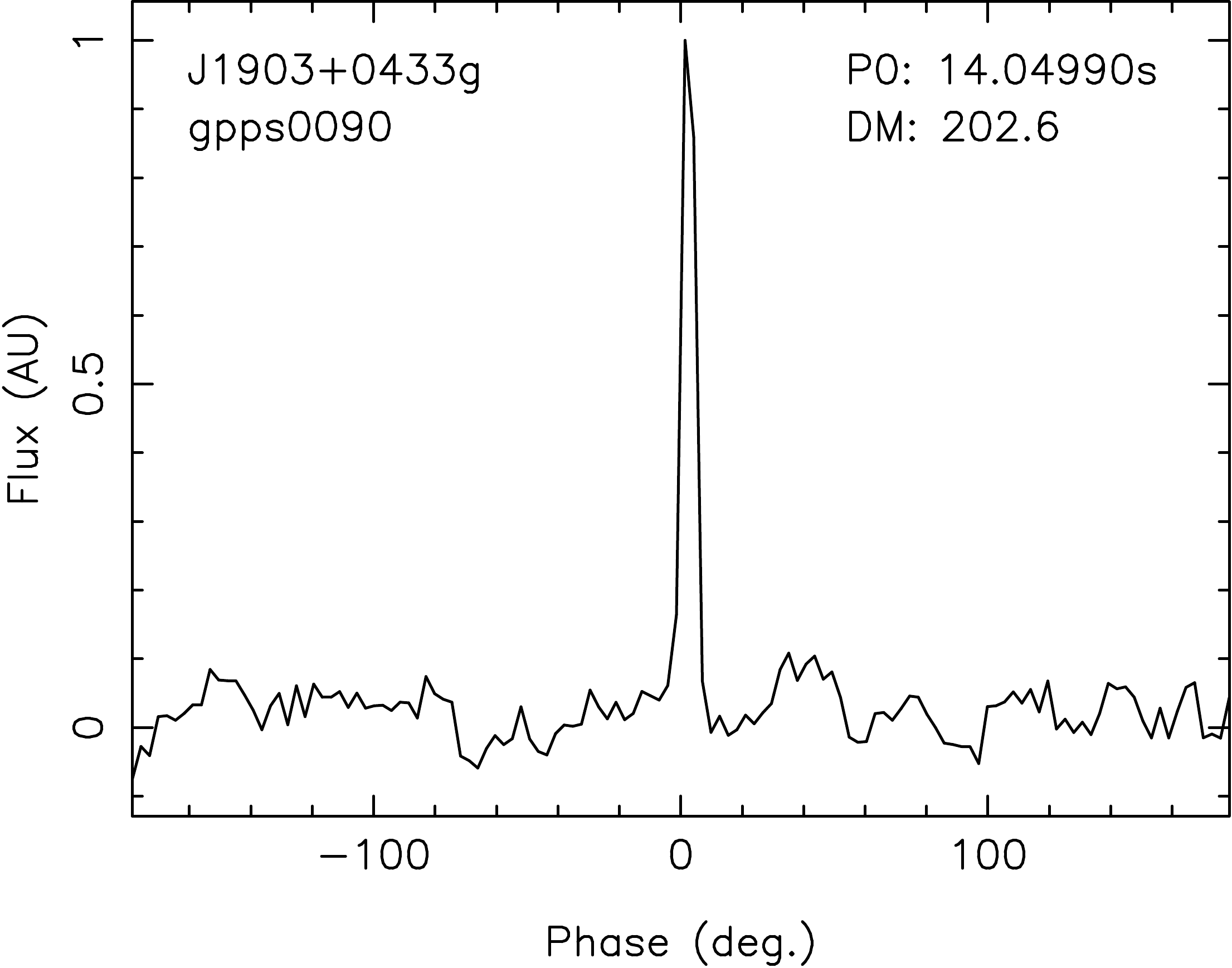}&
\includegraphics[width=39mm]{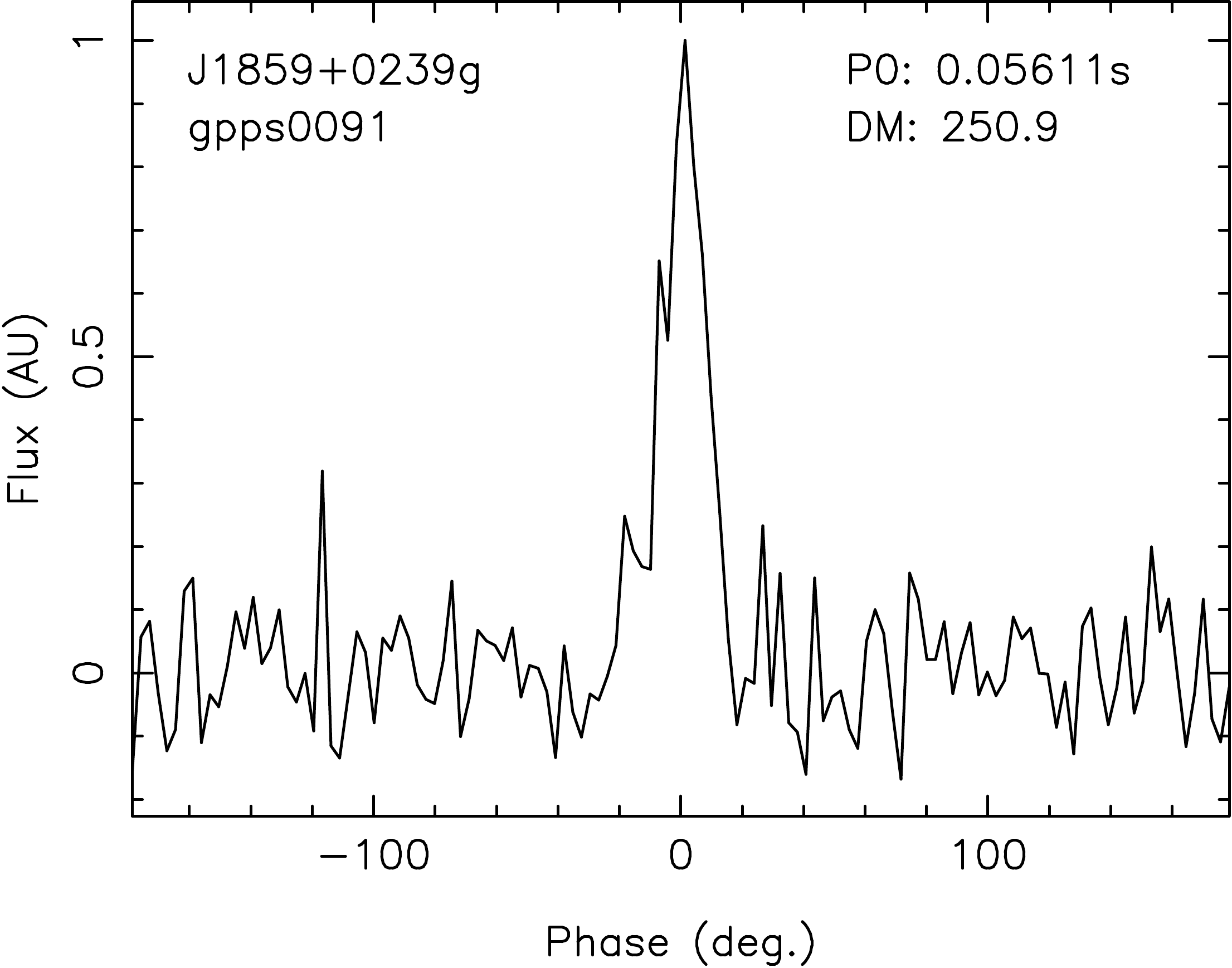}&
\includegraphics[width=39mm]{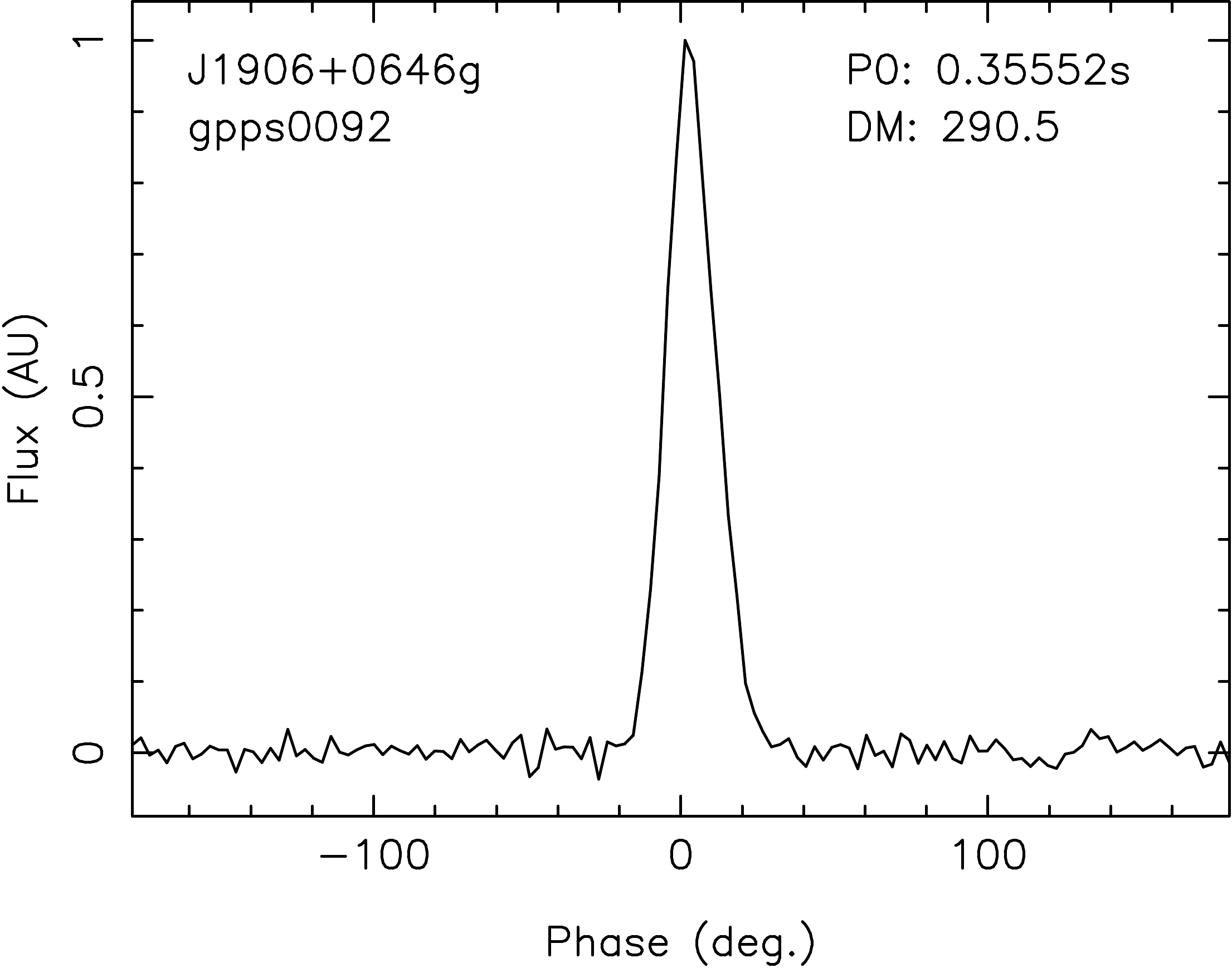}\\[2mm]
\includegraphics[width=39mm]{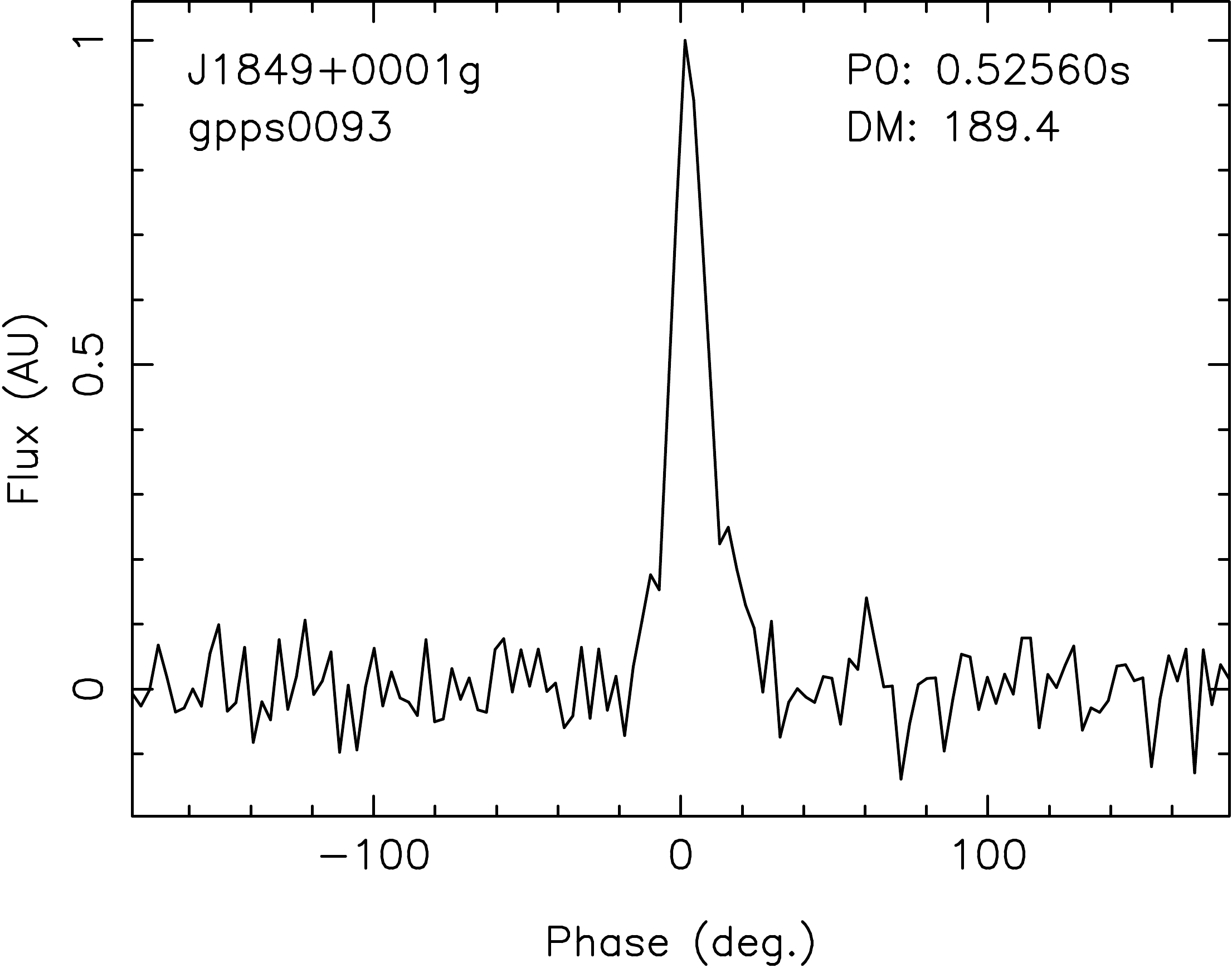}&
\includegraphics[width=39mm]{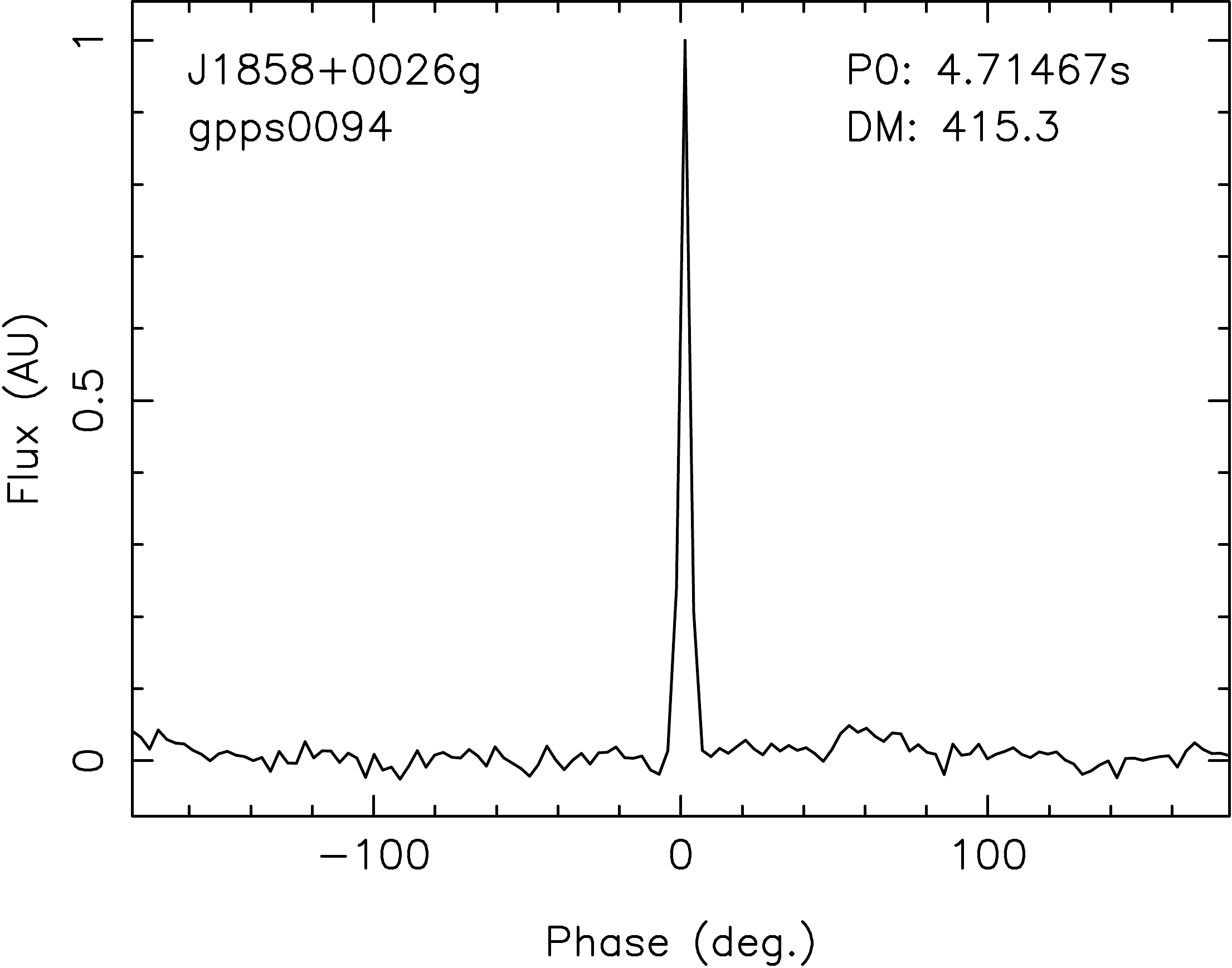}&
\includegraphics[width=39mm]{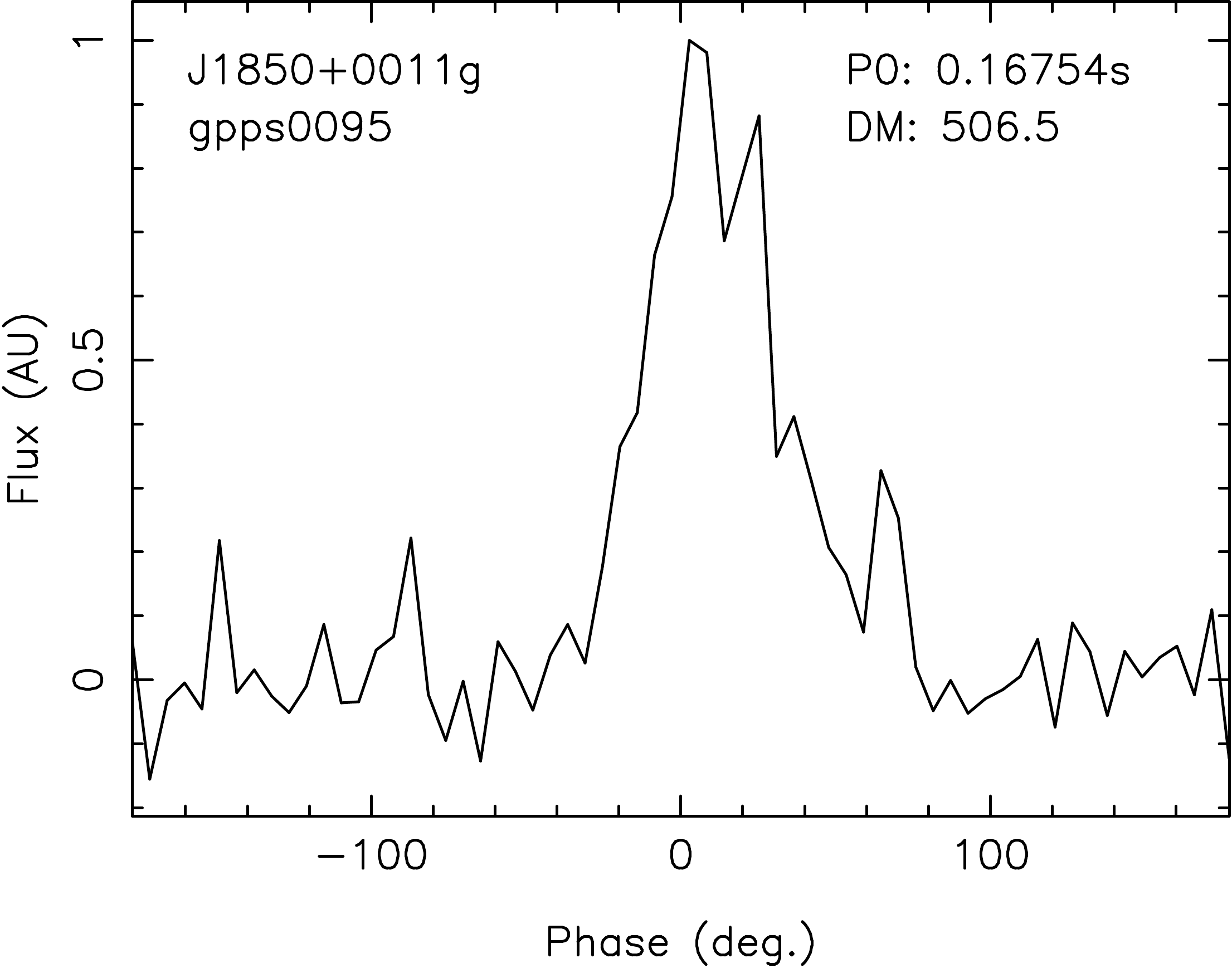}&
\includegraphics[width=39mm]{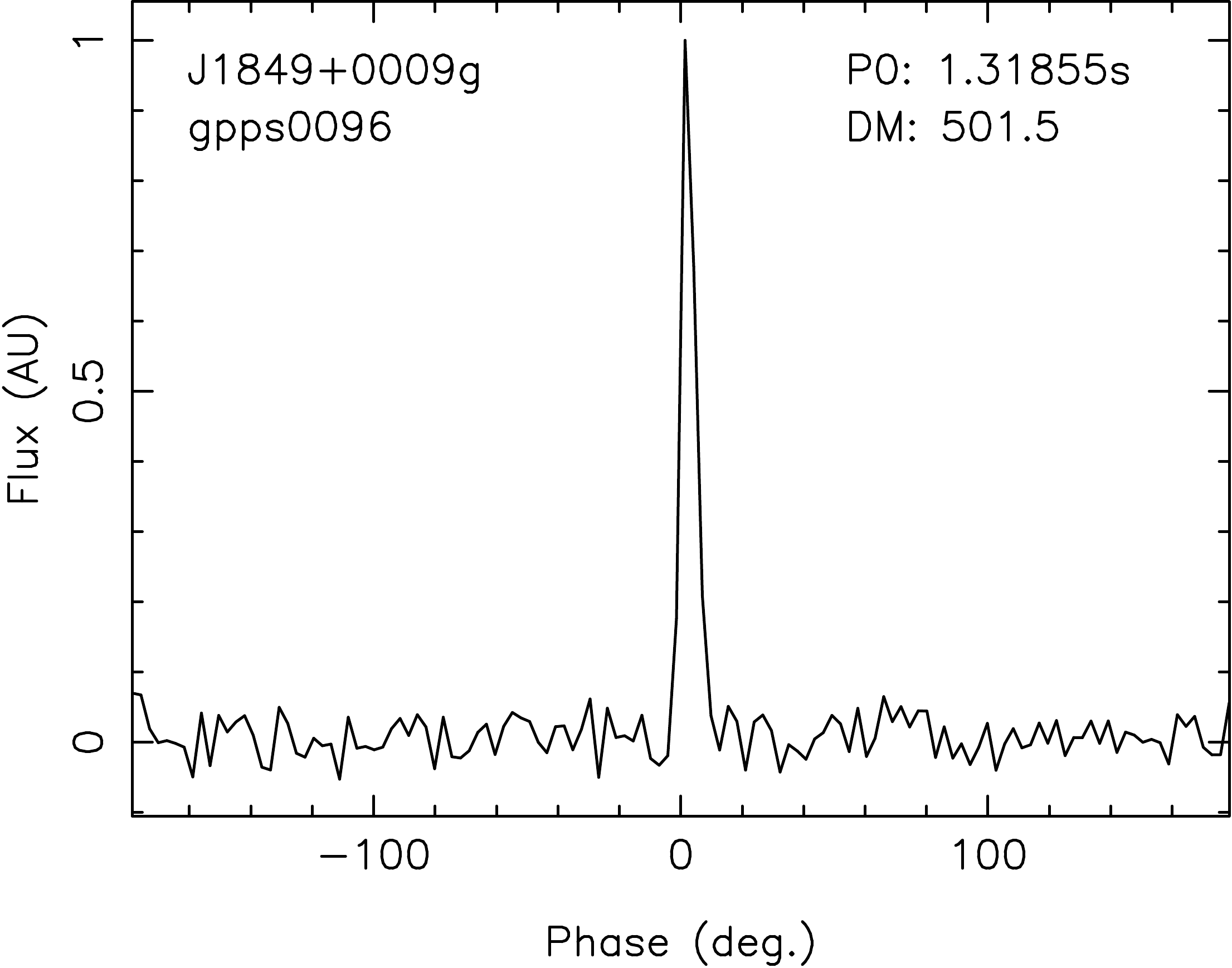}\\[2mm]
\includegraphics[width=39mm]{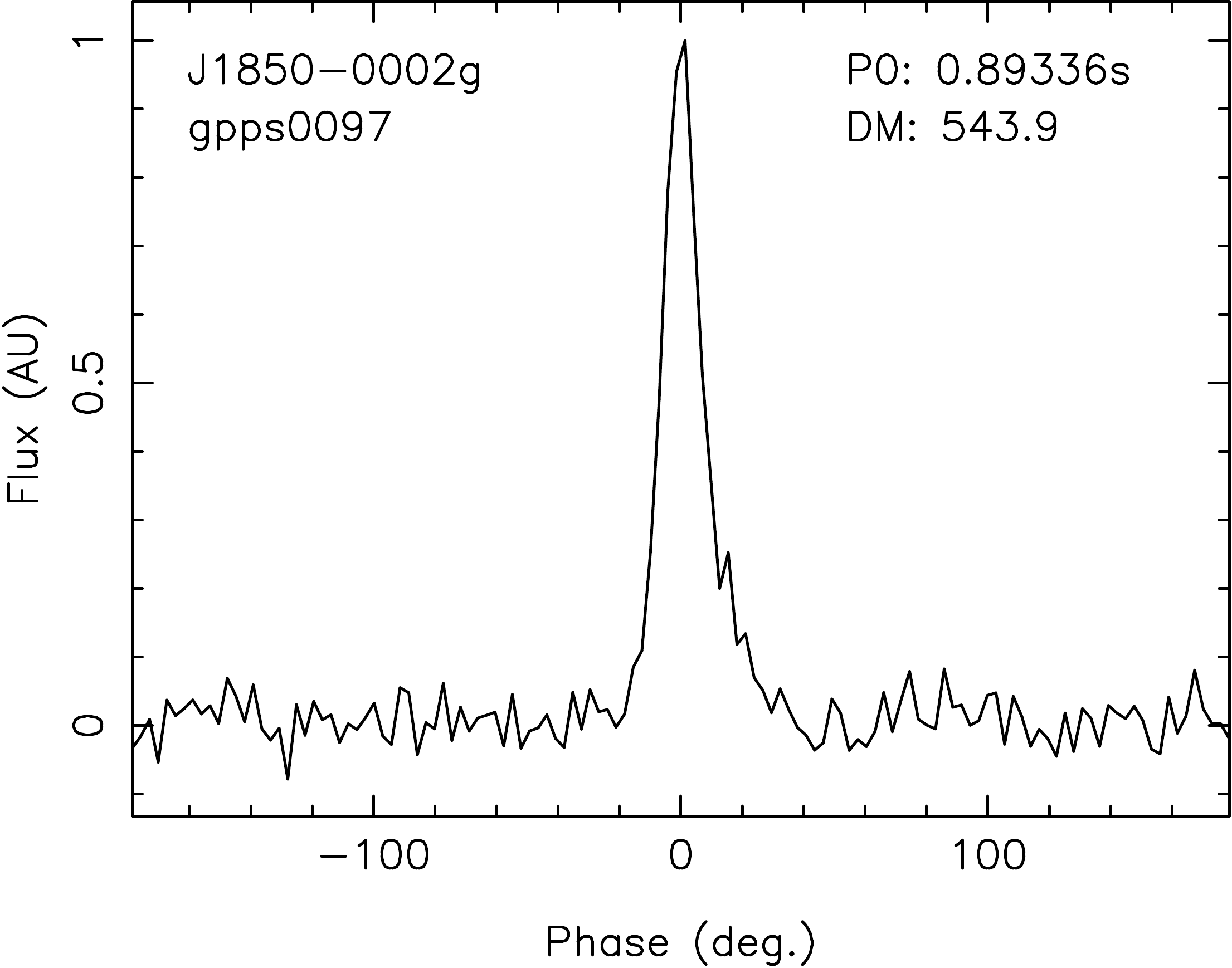}&
\includegraphics[width=39mm]{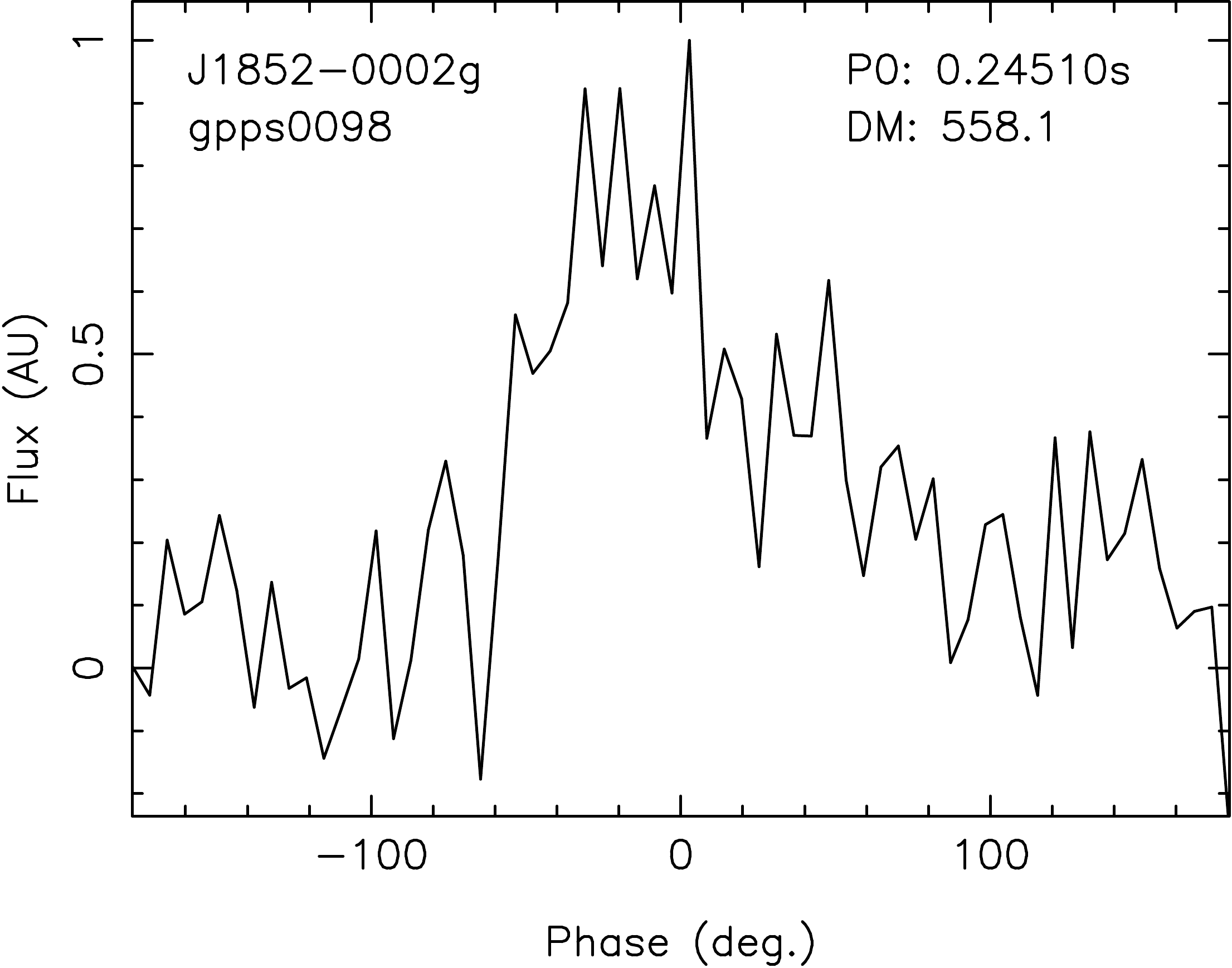}&
\includegraphics[width=39mm]{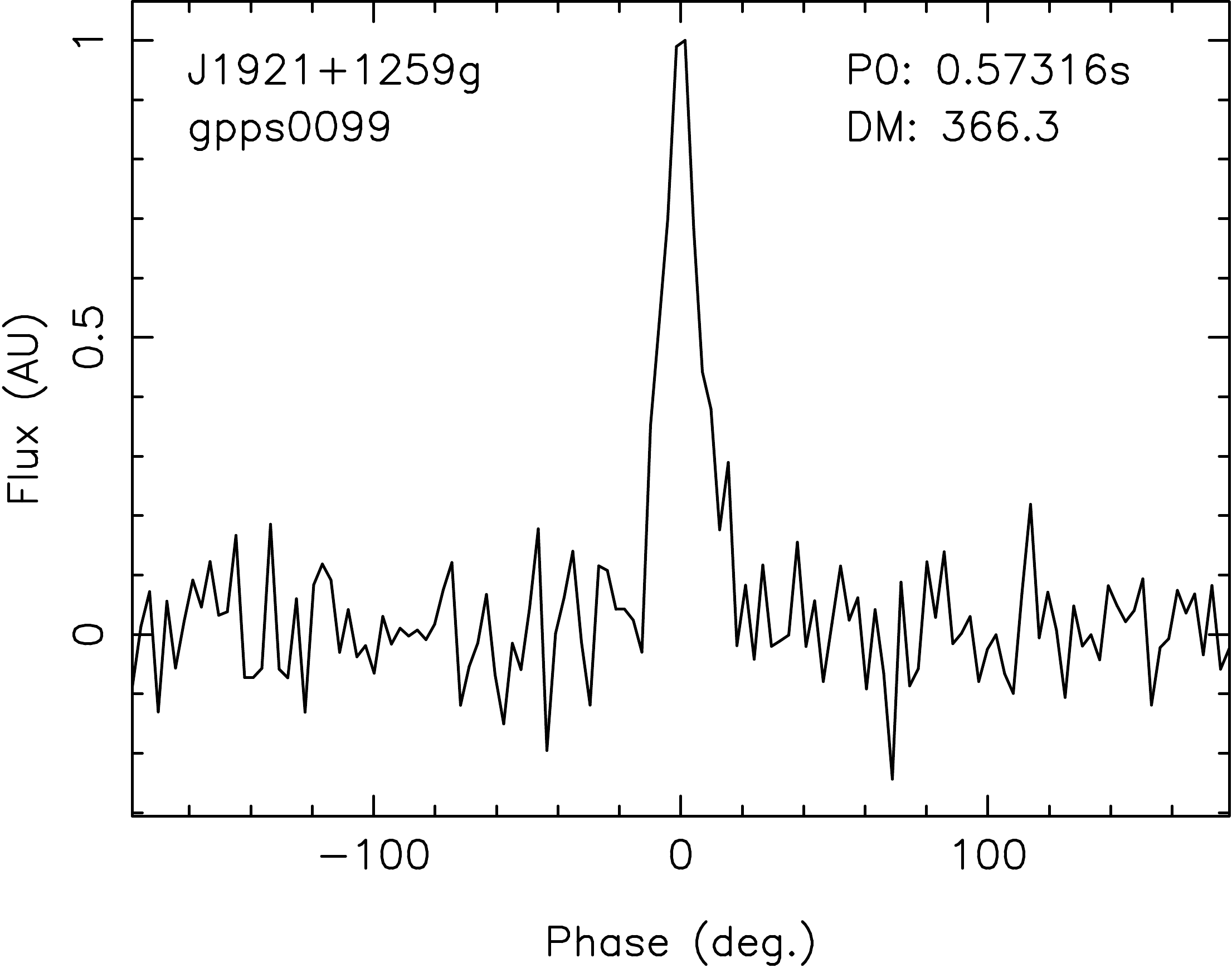}&
\includegraphics[width=39mm]{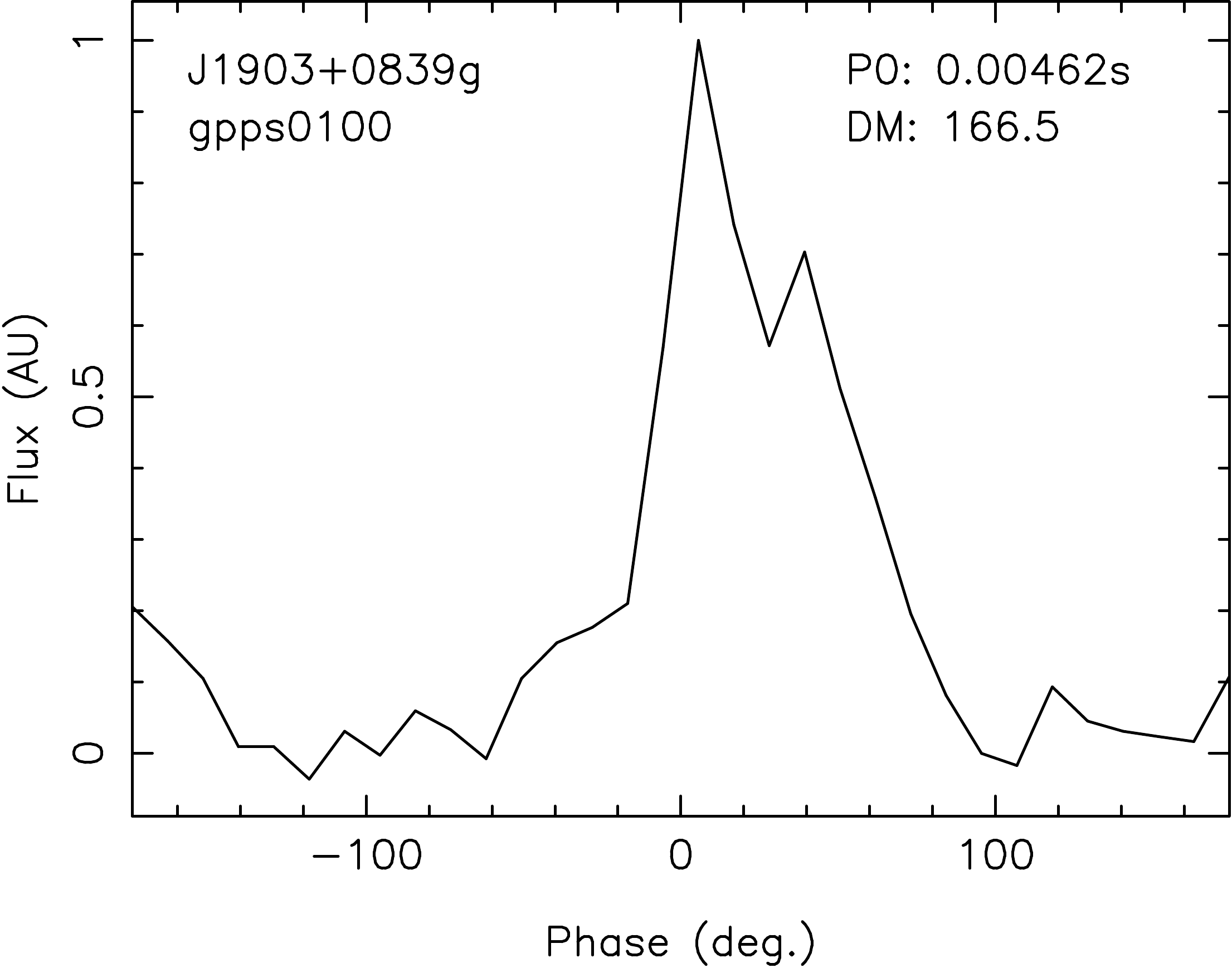}\\[2mm]
\includegraphics[width=39mm]{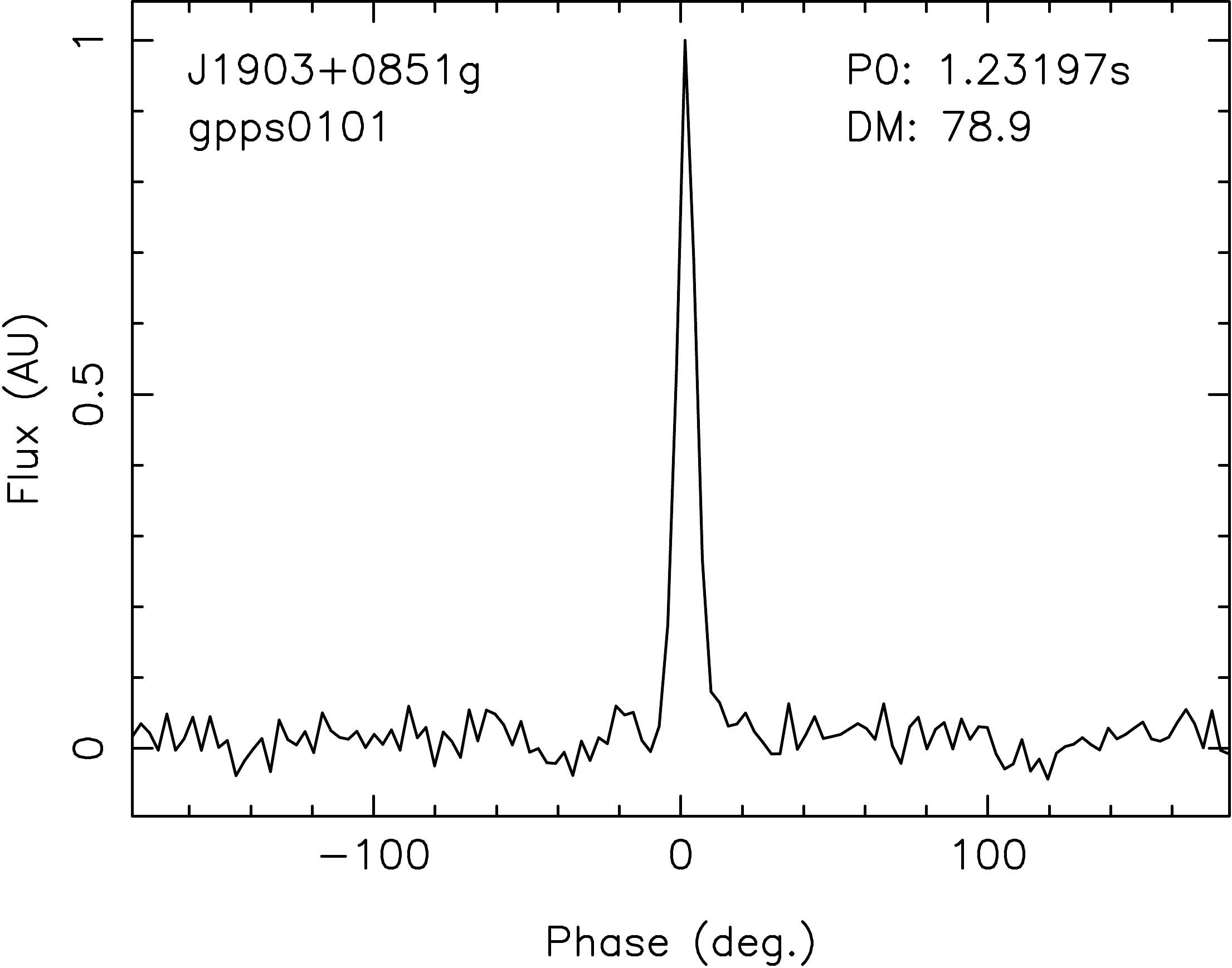}&
\includegraphics[width=39mm]{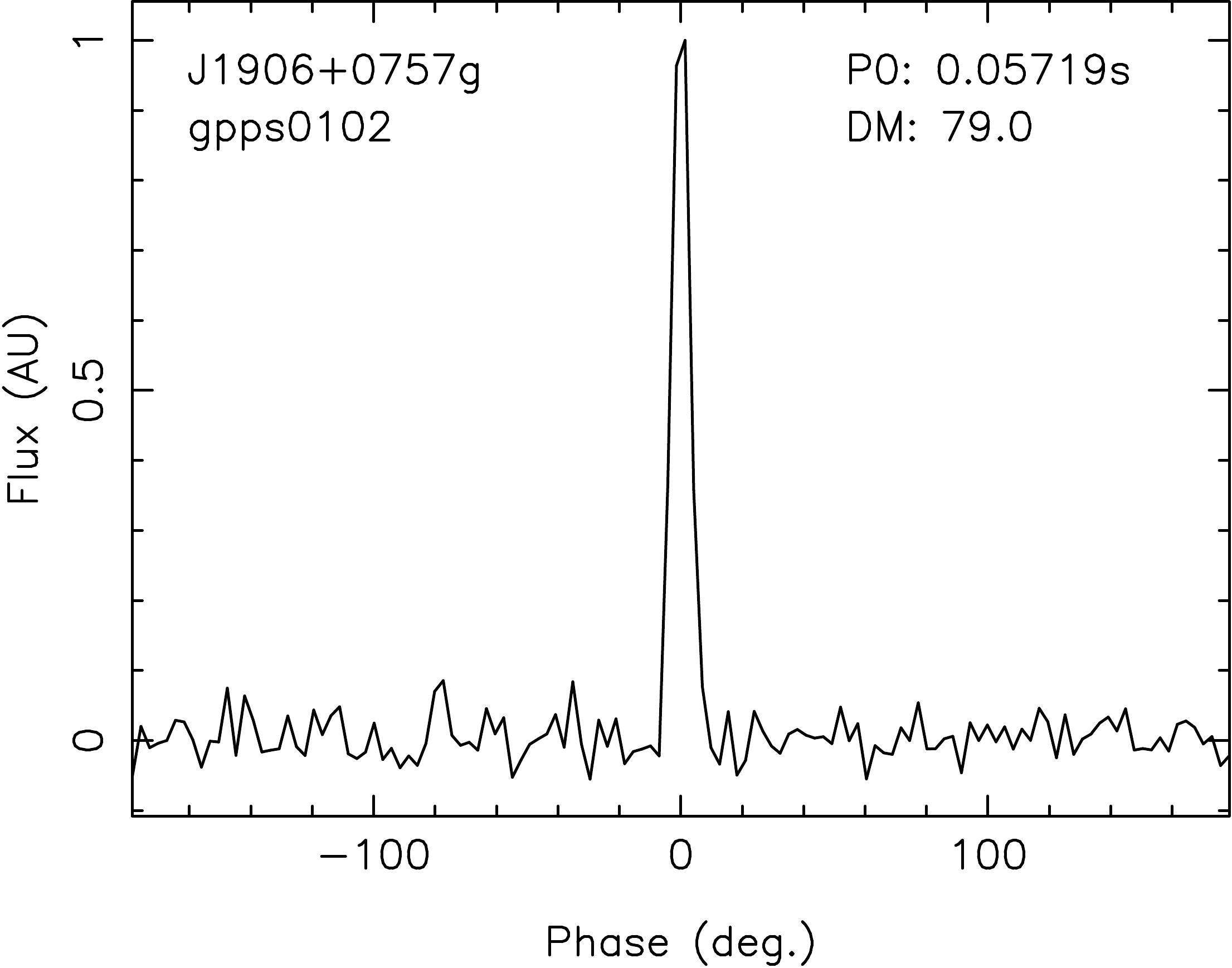}&
\includegraphics[width=39mm]{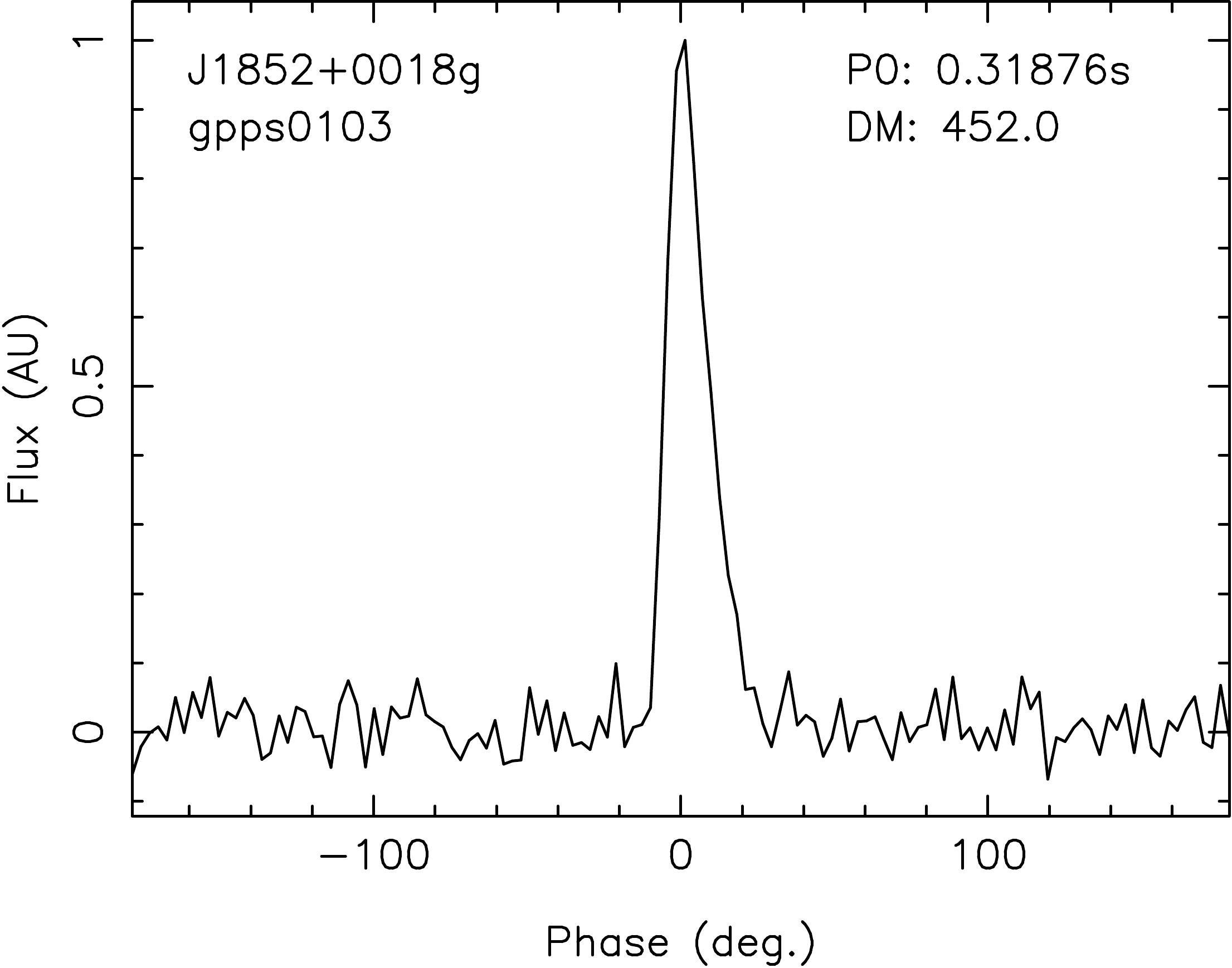}&
\includegraphics[width=39mm]{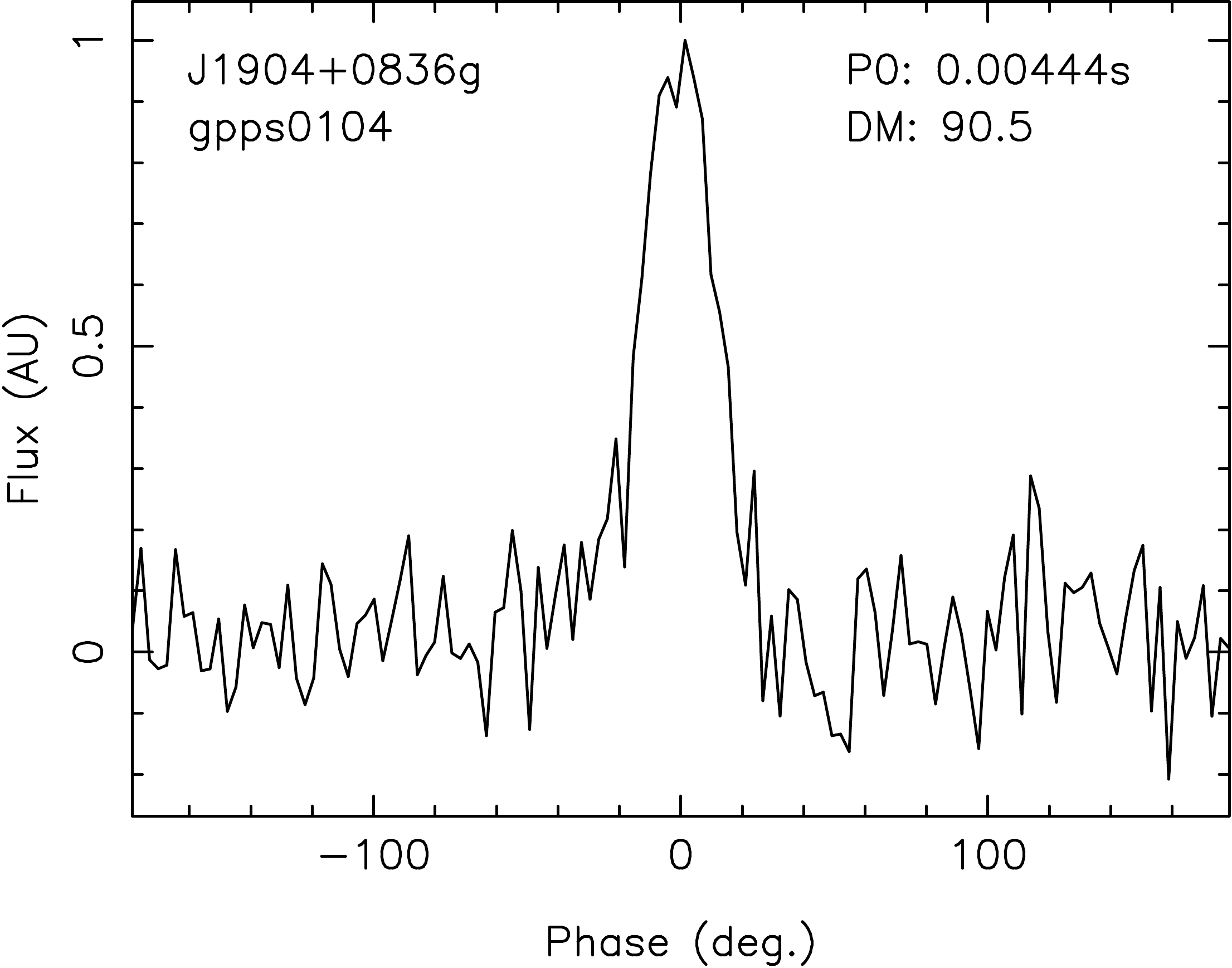}\\[2mm]
\includegraphics[width=39mm]{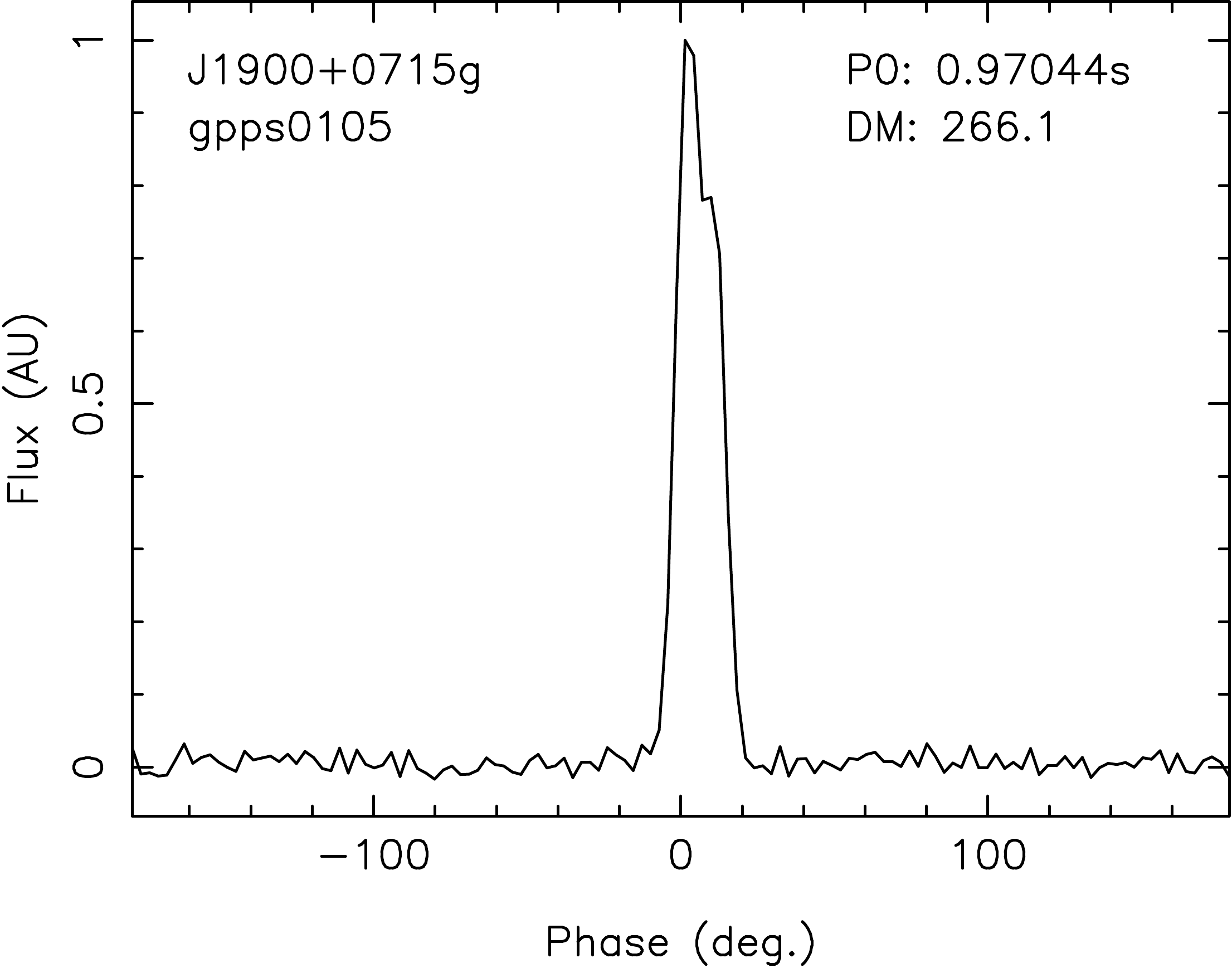}&
\includegraphics[width=39mm]{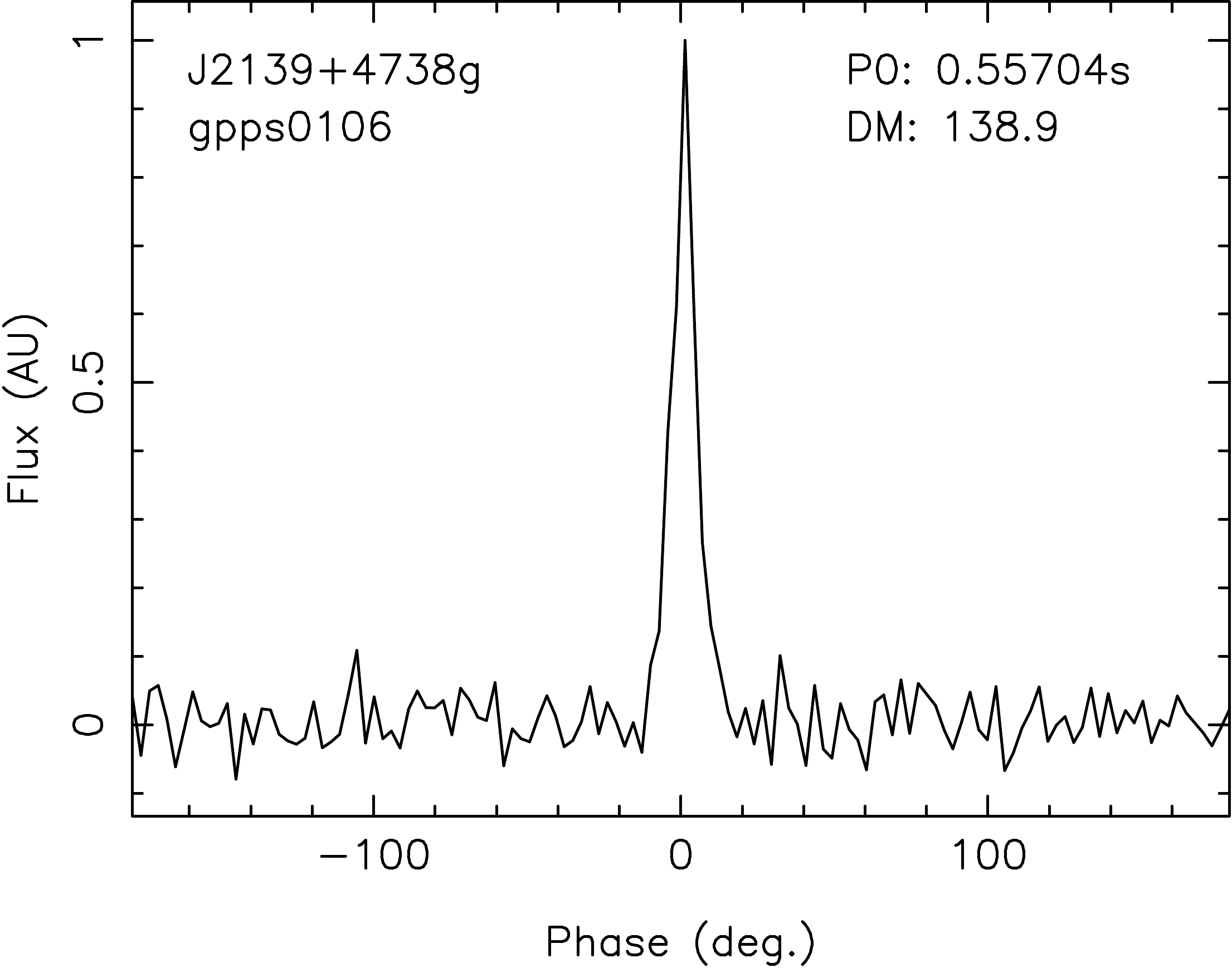}&
\includegraphics[width=39mm]{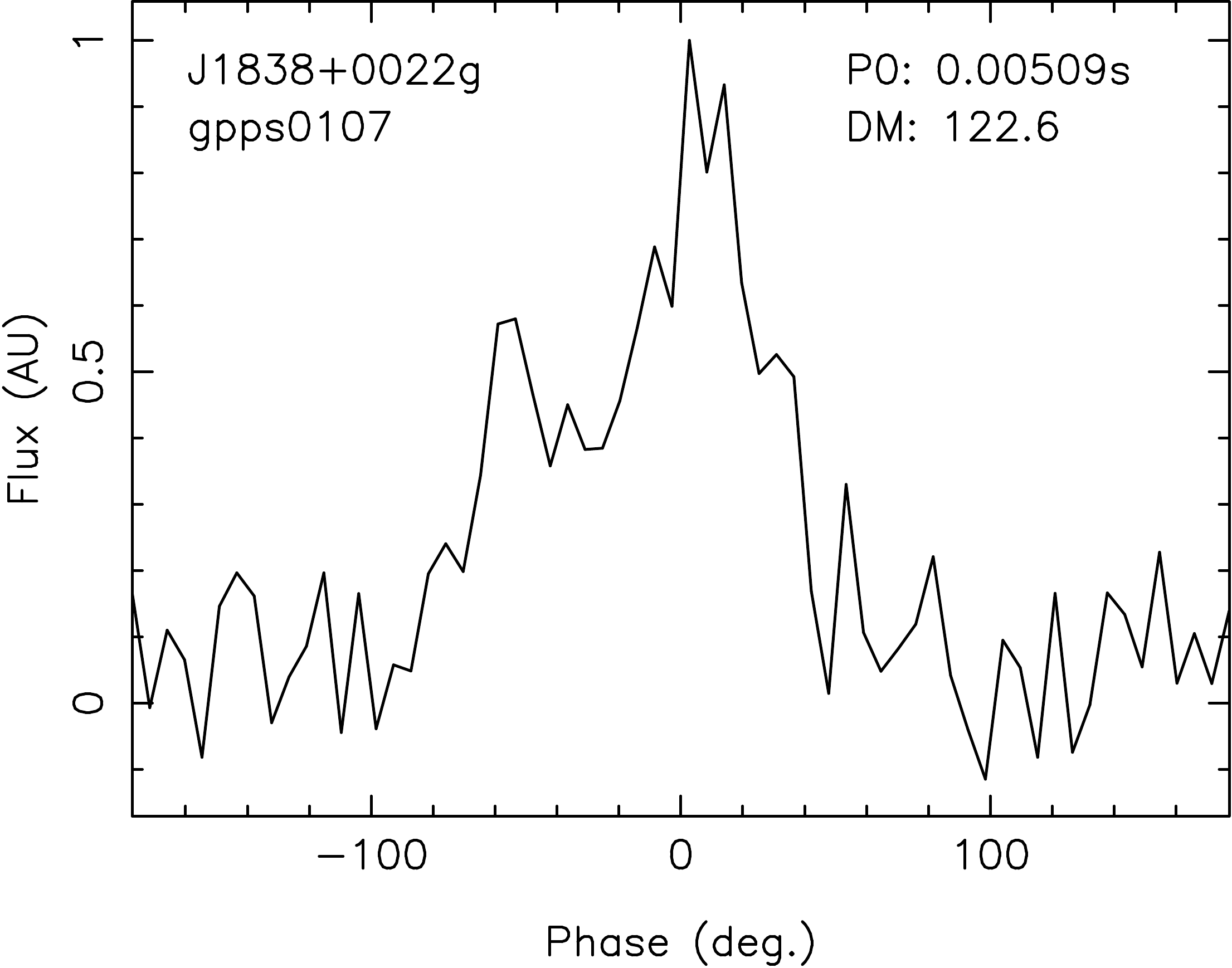}&
\includegraphics[width=39mm]{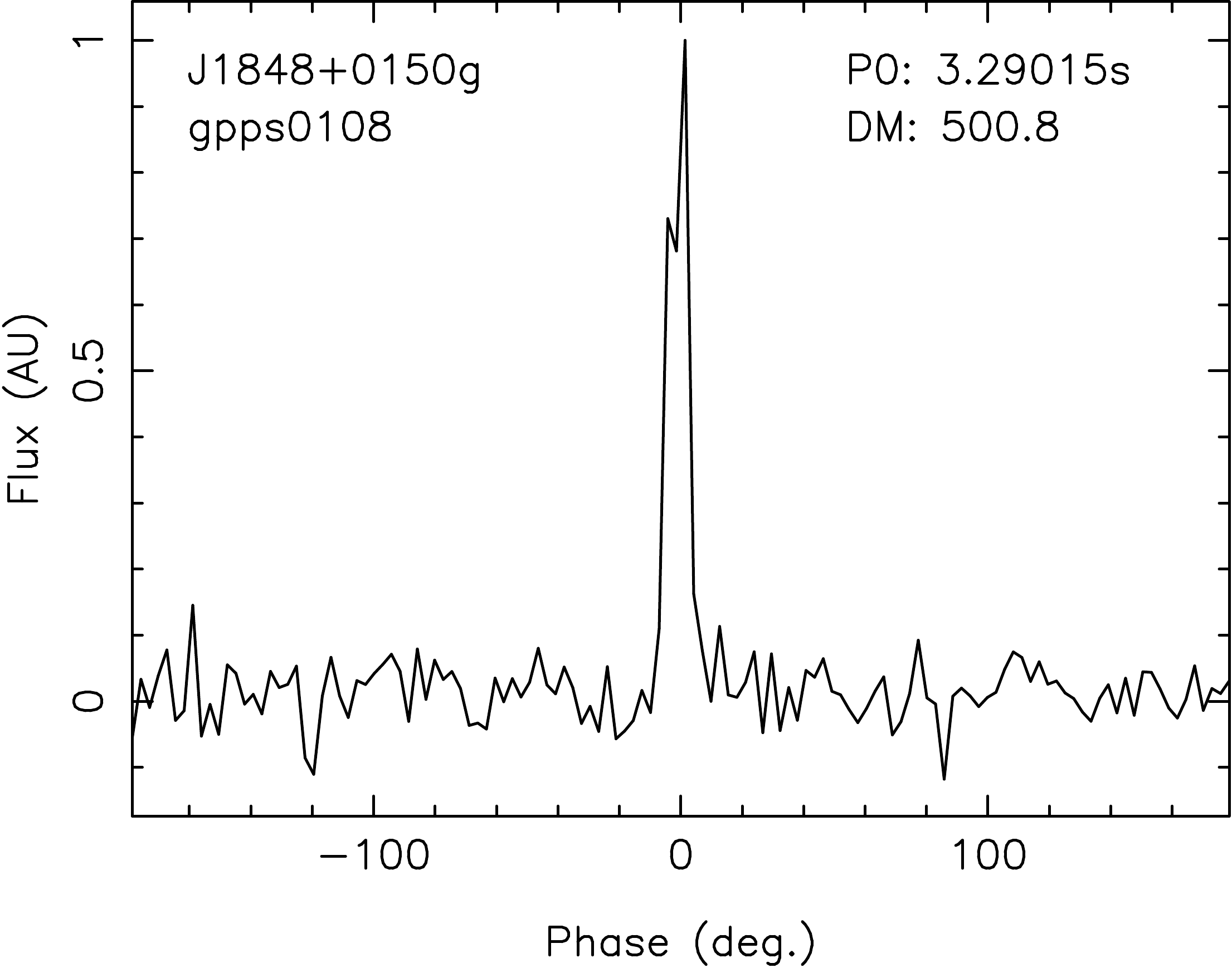}\\[2mm]
\includegraphics[width=39mm]{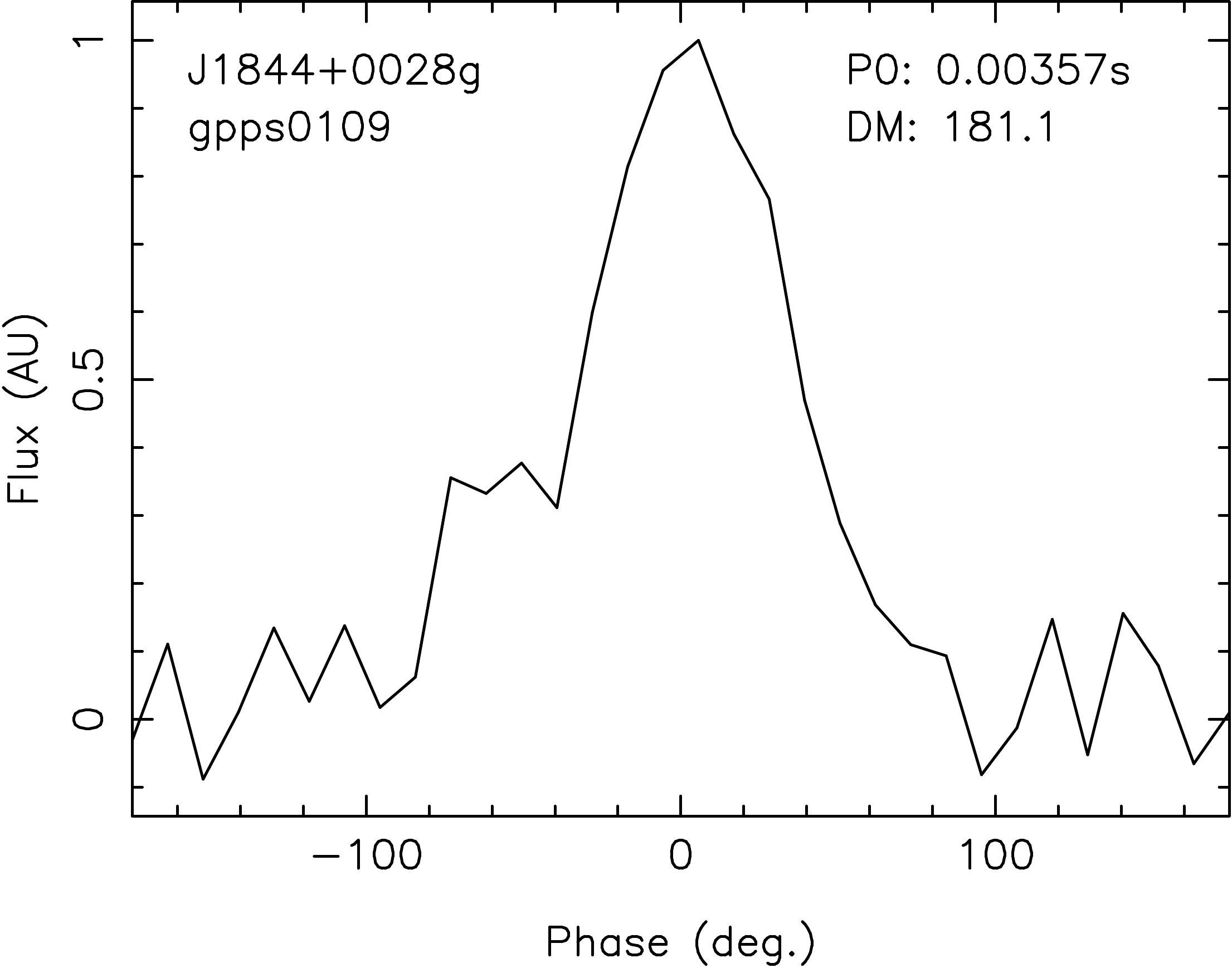}&
\includegraphics[width=39mm]{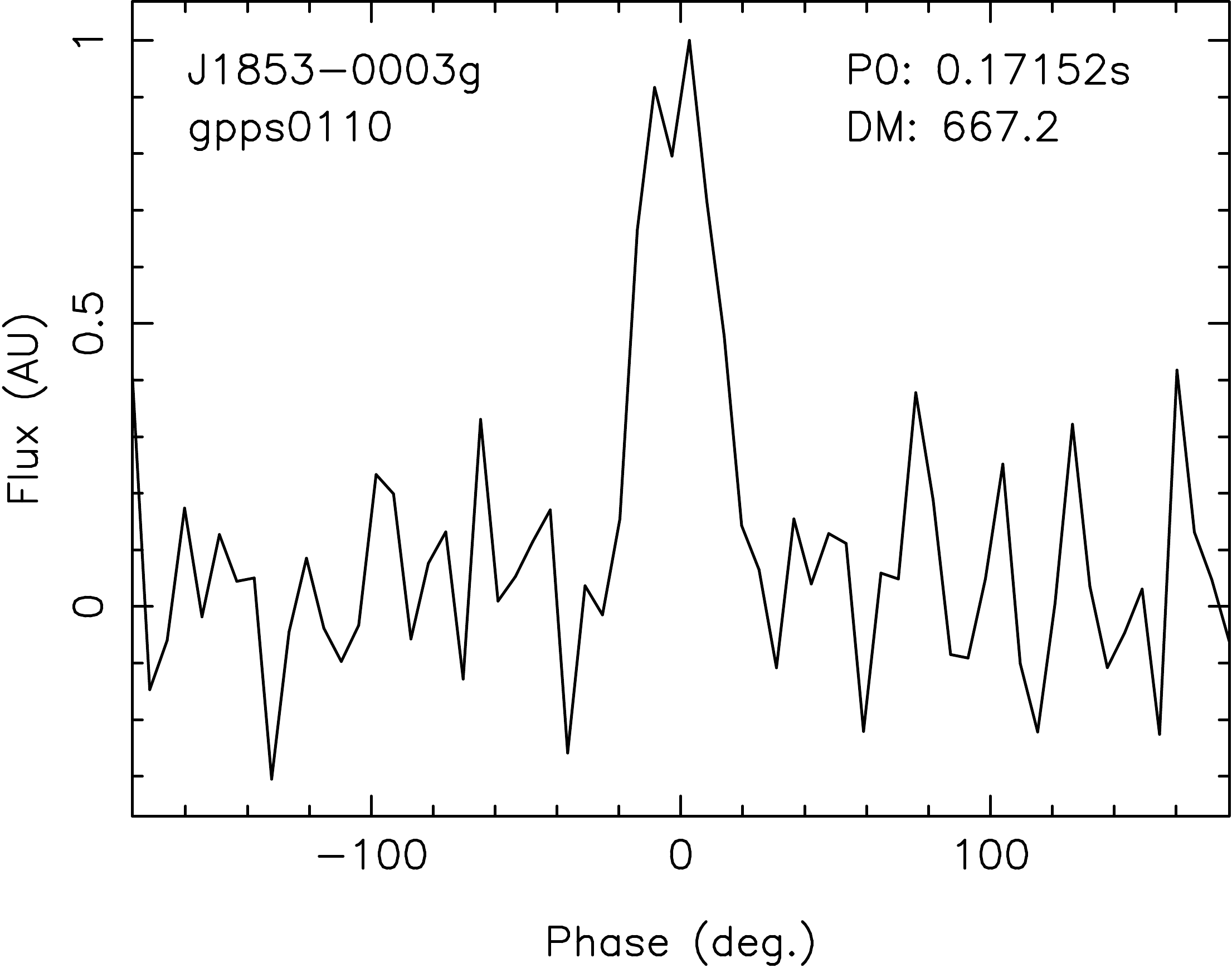}&
\includegraphics[width=39mm]{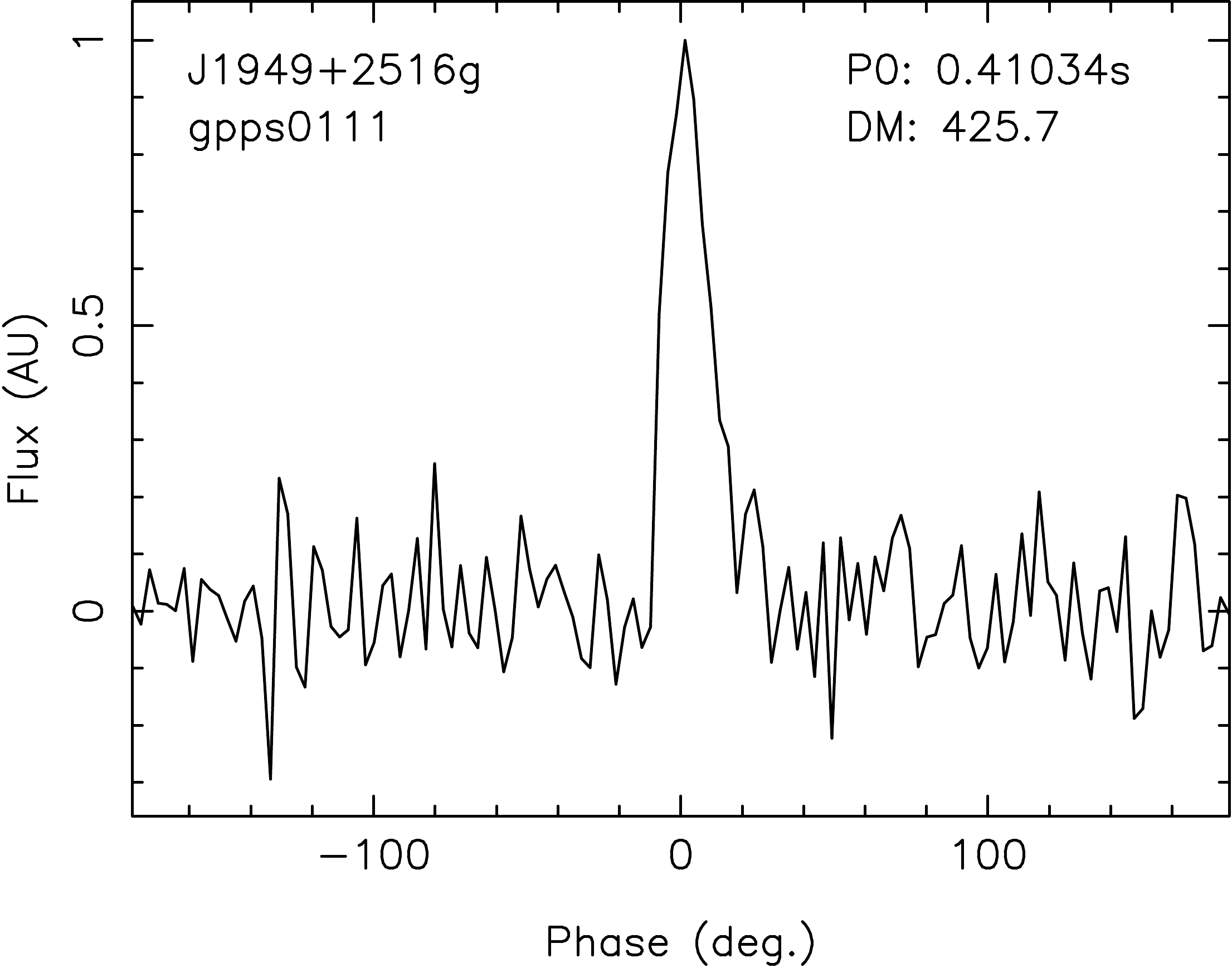}&
\includegraphics[width=39mm]{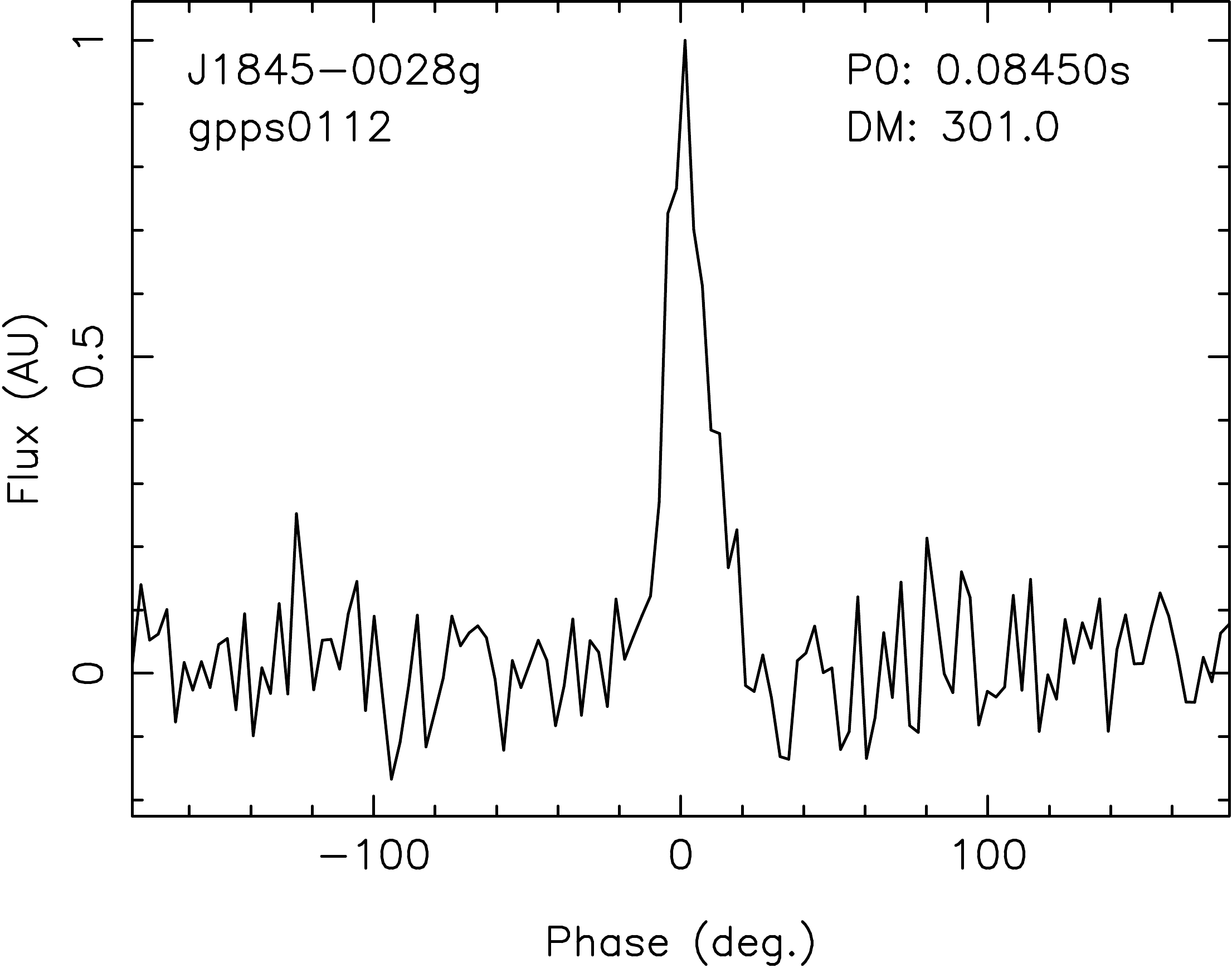}\\
 \end{tabular}%

\begin{minipage}{3cm}
\caption[]{
-- {\it Continued}.}\end{minipage}
\addtocounter{figure}{-1}
\end{figure*}%
\begin{figure*}
\centering
\begin{tabular}{rrrrrr}
\includegraphics[width=39mm]{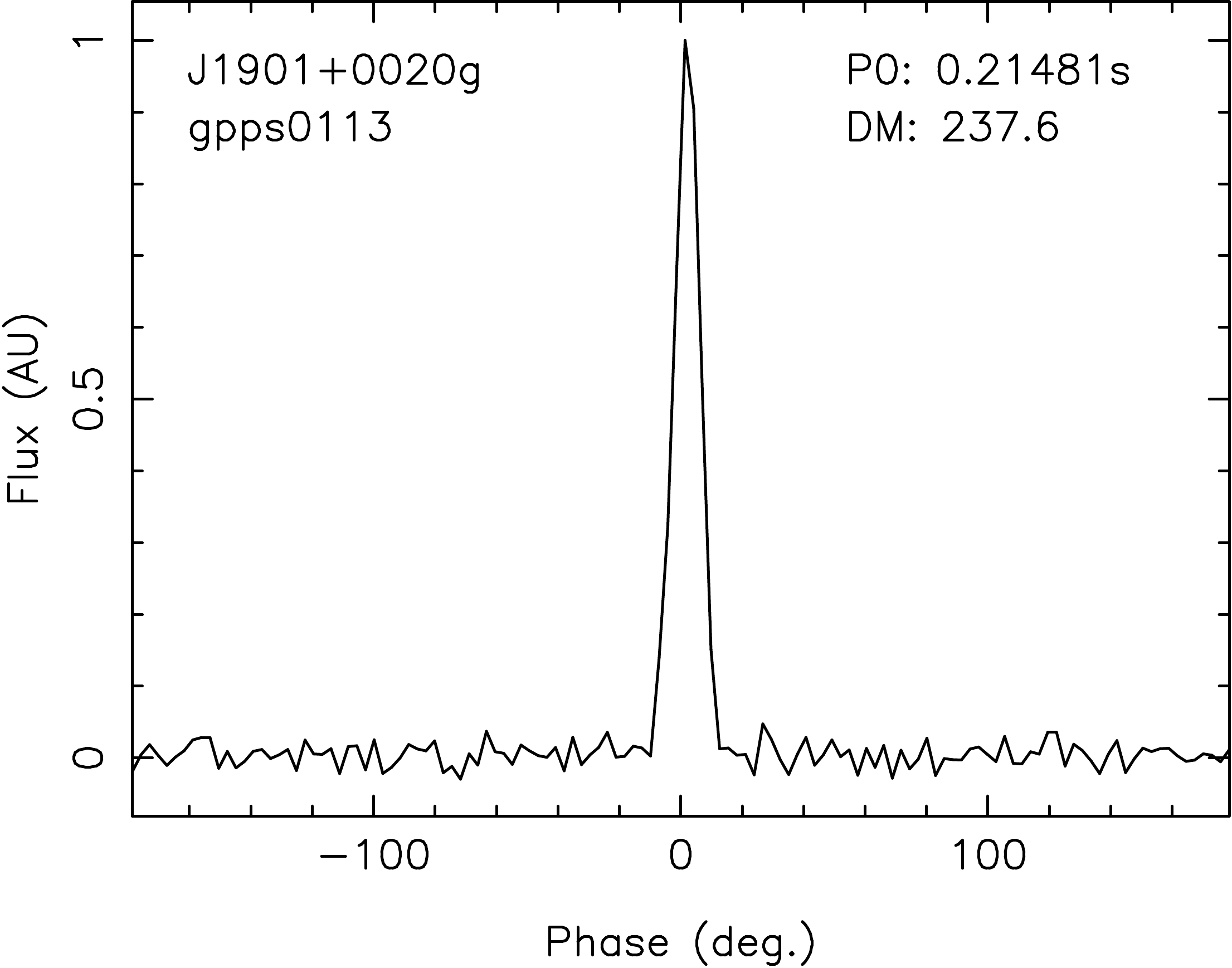}&
\includegraphics[width=39mm]{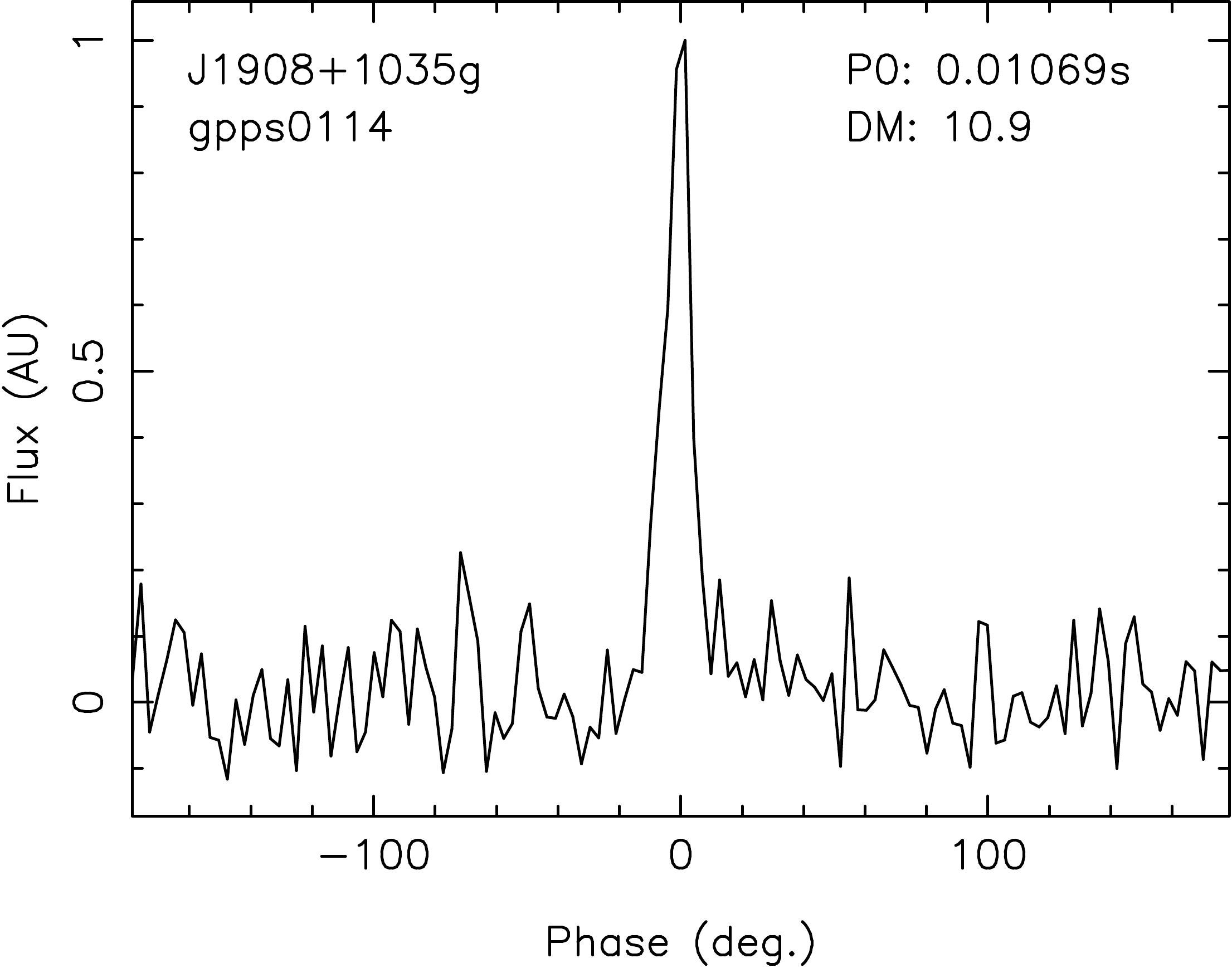}&
\includegraphics[width=39mm]{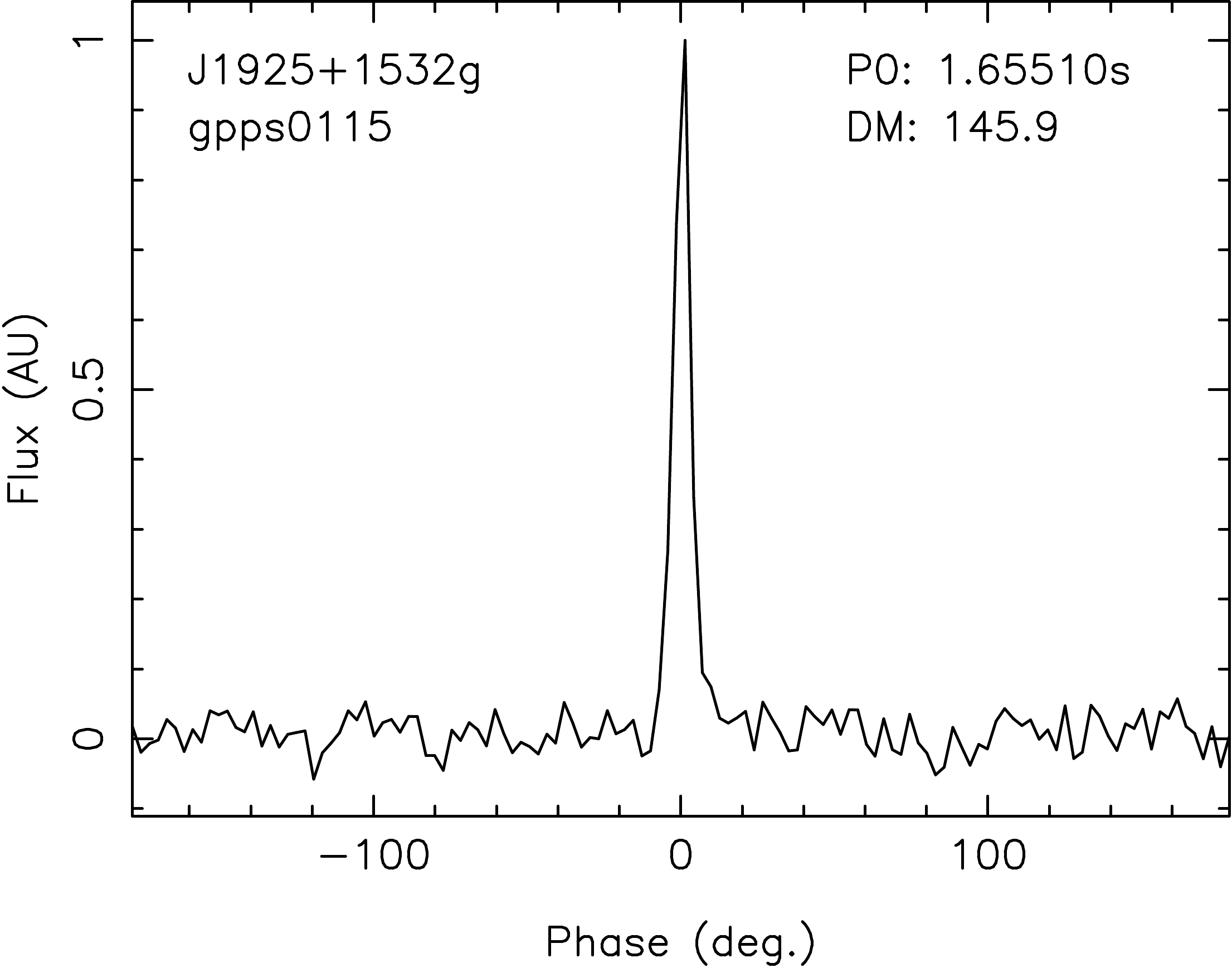}&
\includegraphics[width=39mm]{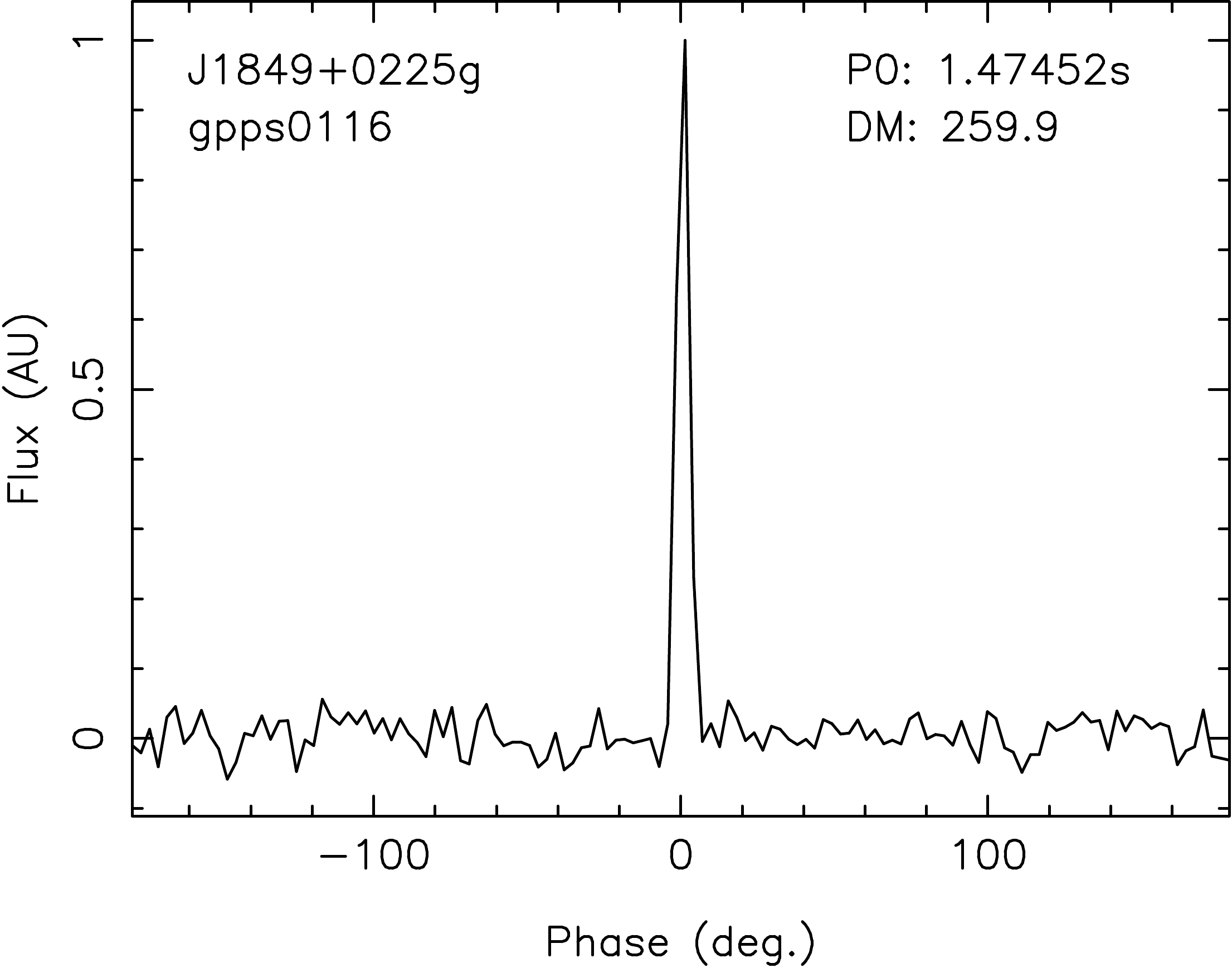}\\[2mm]
\includegraphics[width=39mm]{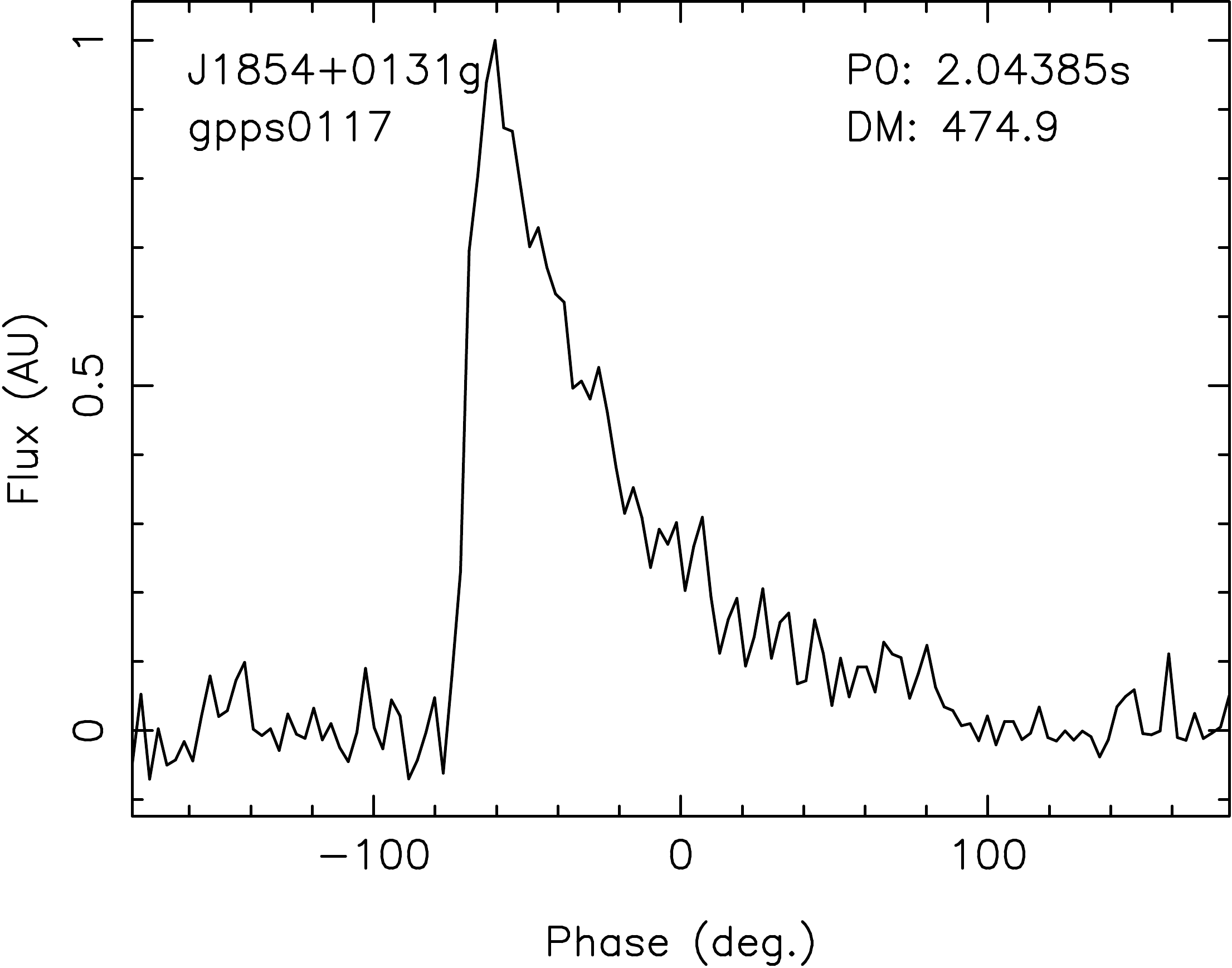}&
\includegraphics[width=39mm]{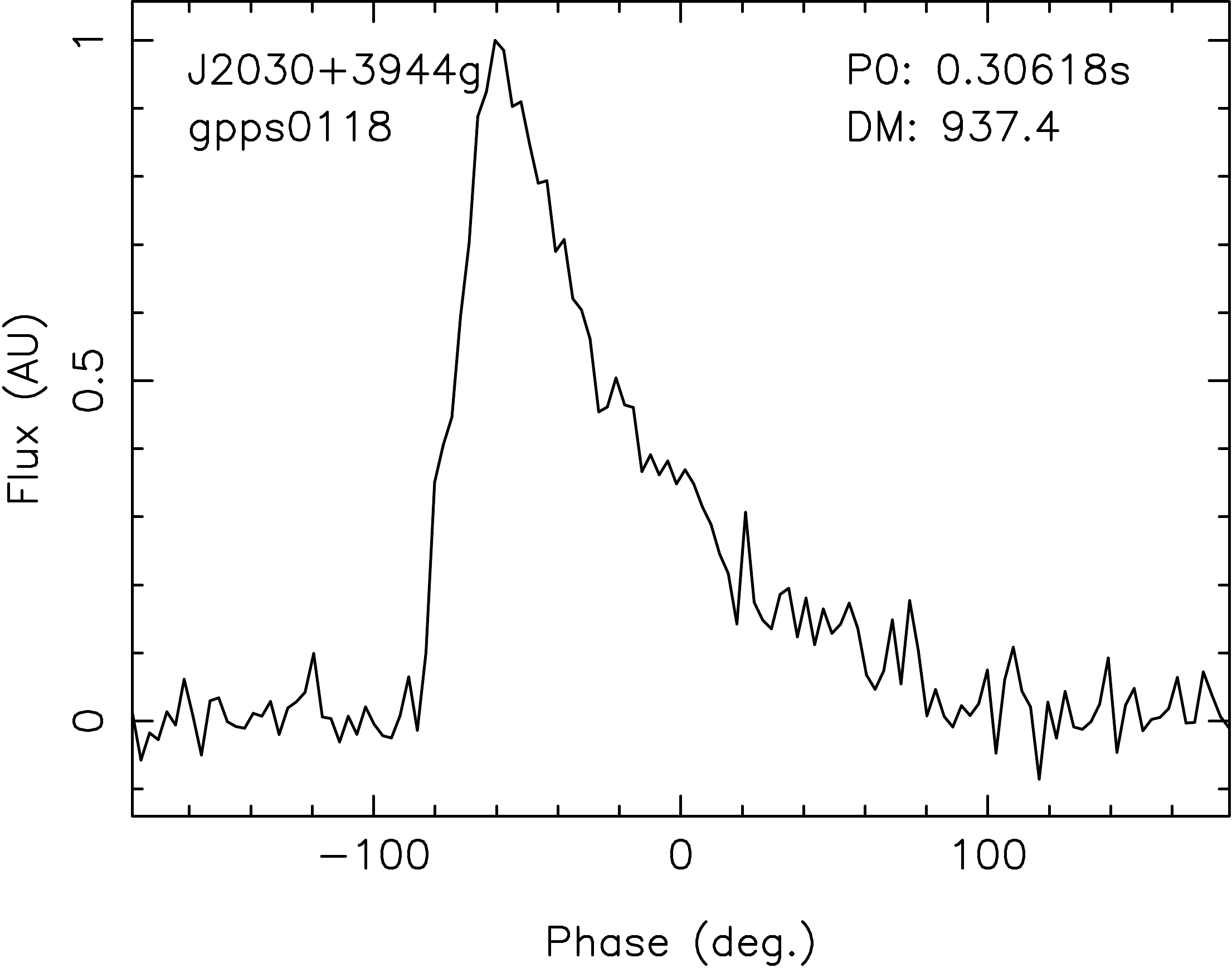}&
\includegraphics[width=39mm]{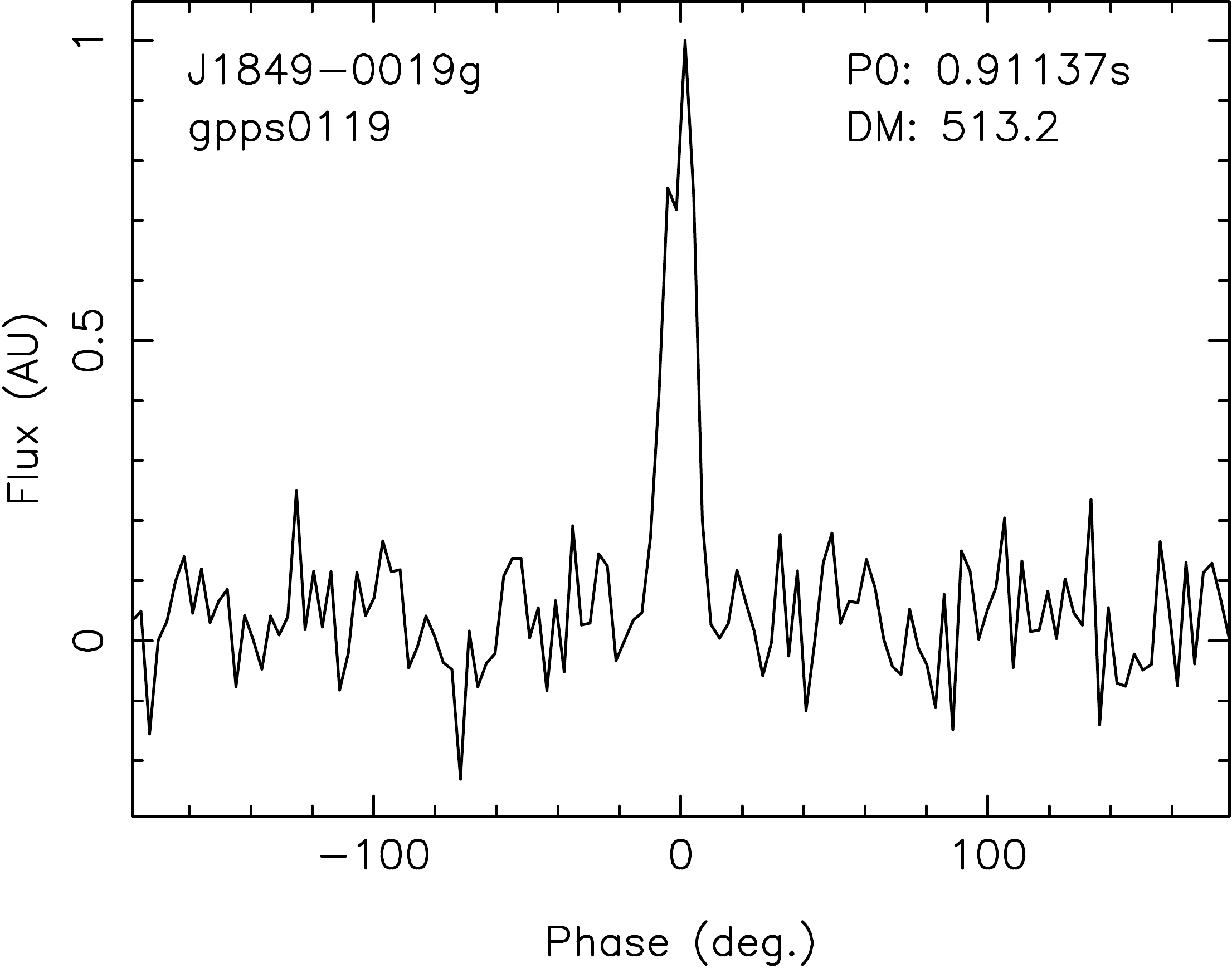}&
\includegraphics[width=39mm]{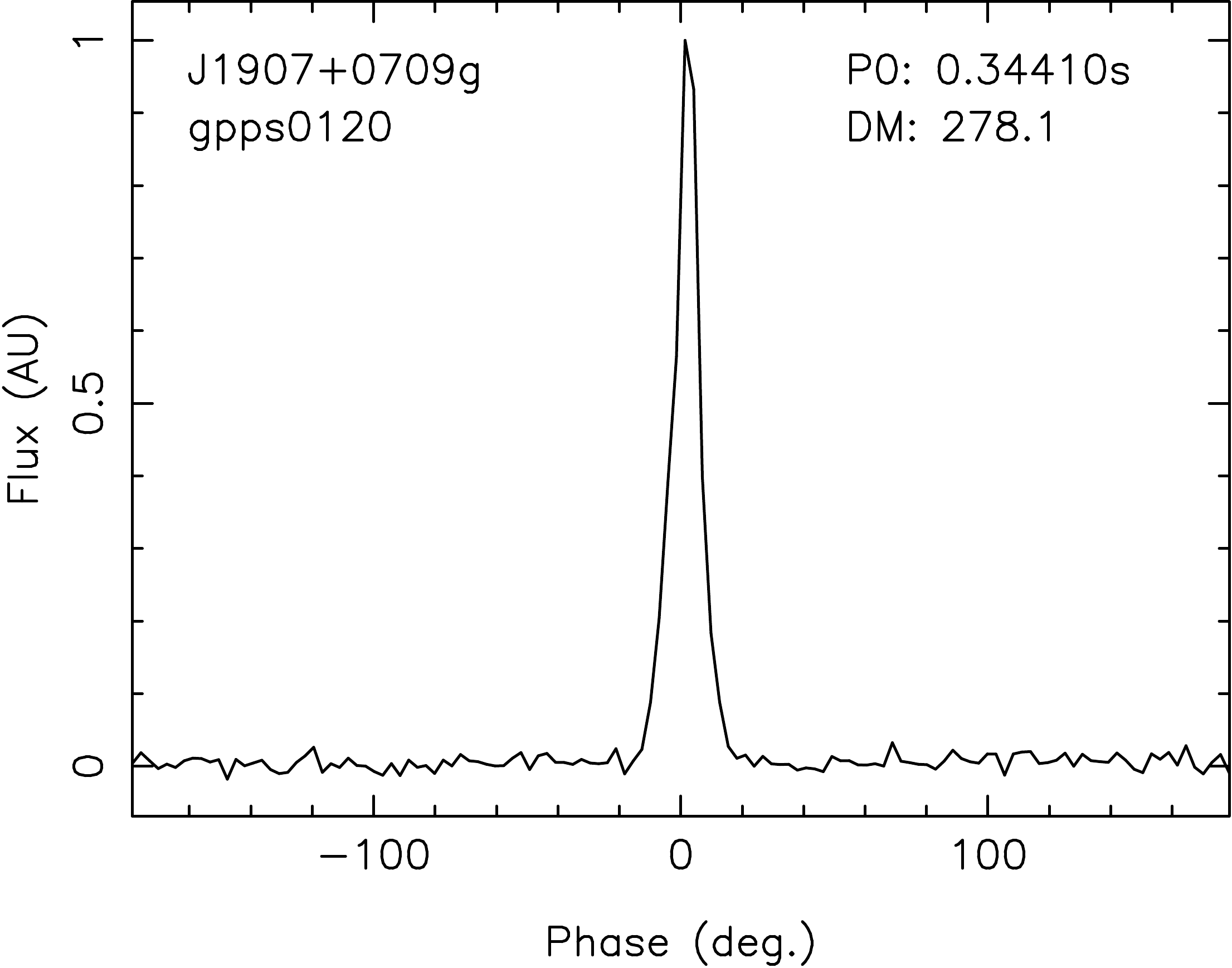}\\[2mm]
\includegraphics[width=39mm]{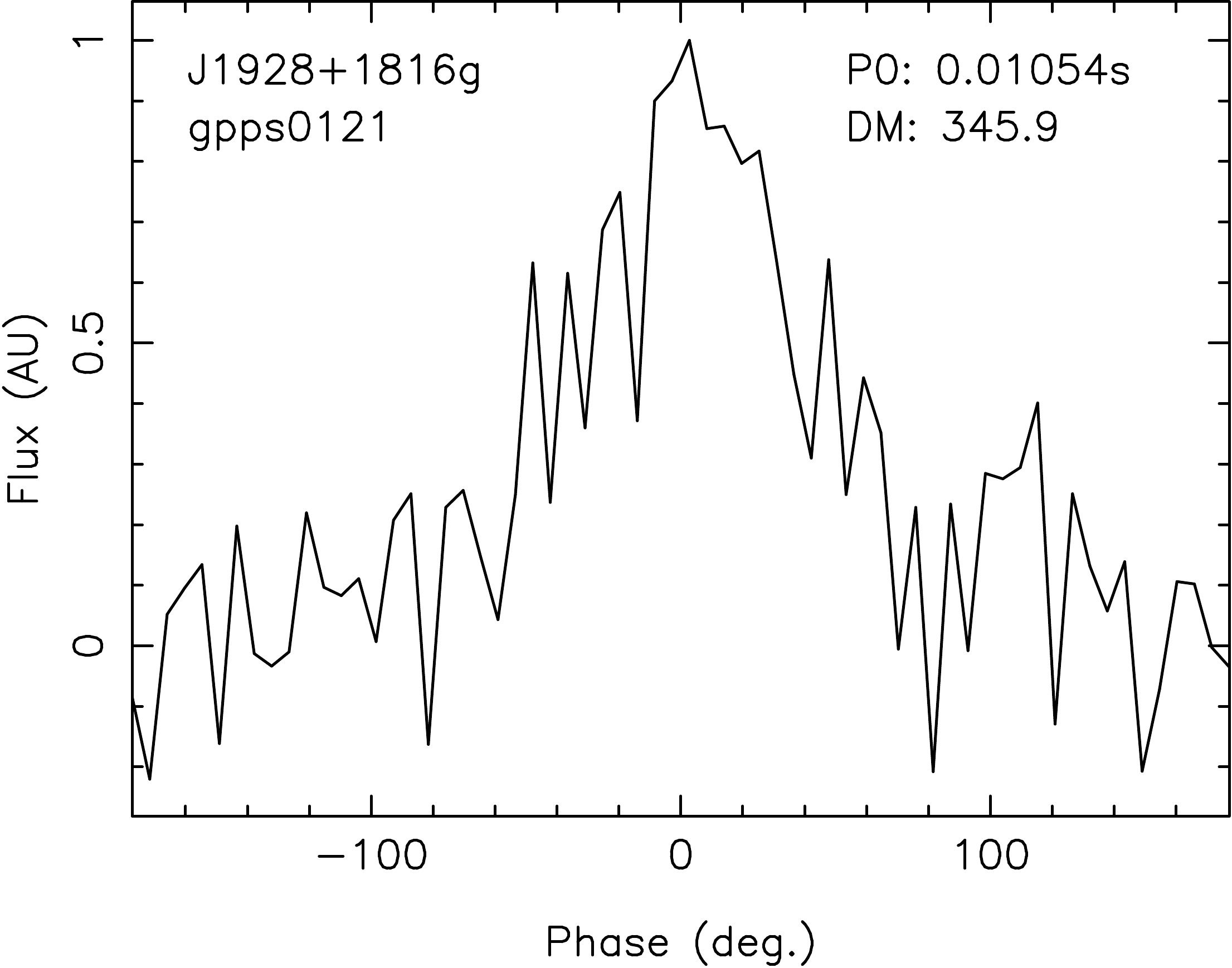}&
\includegraphics[width=39mm]{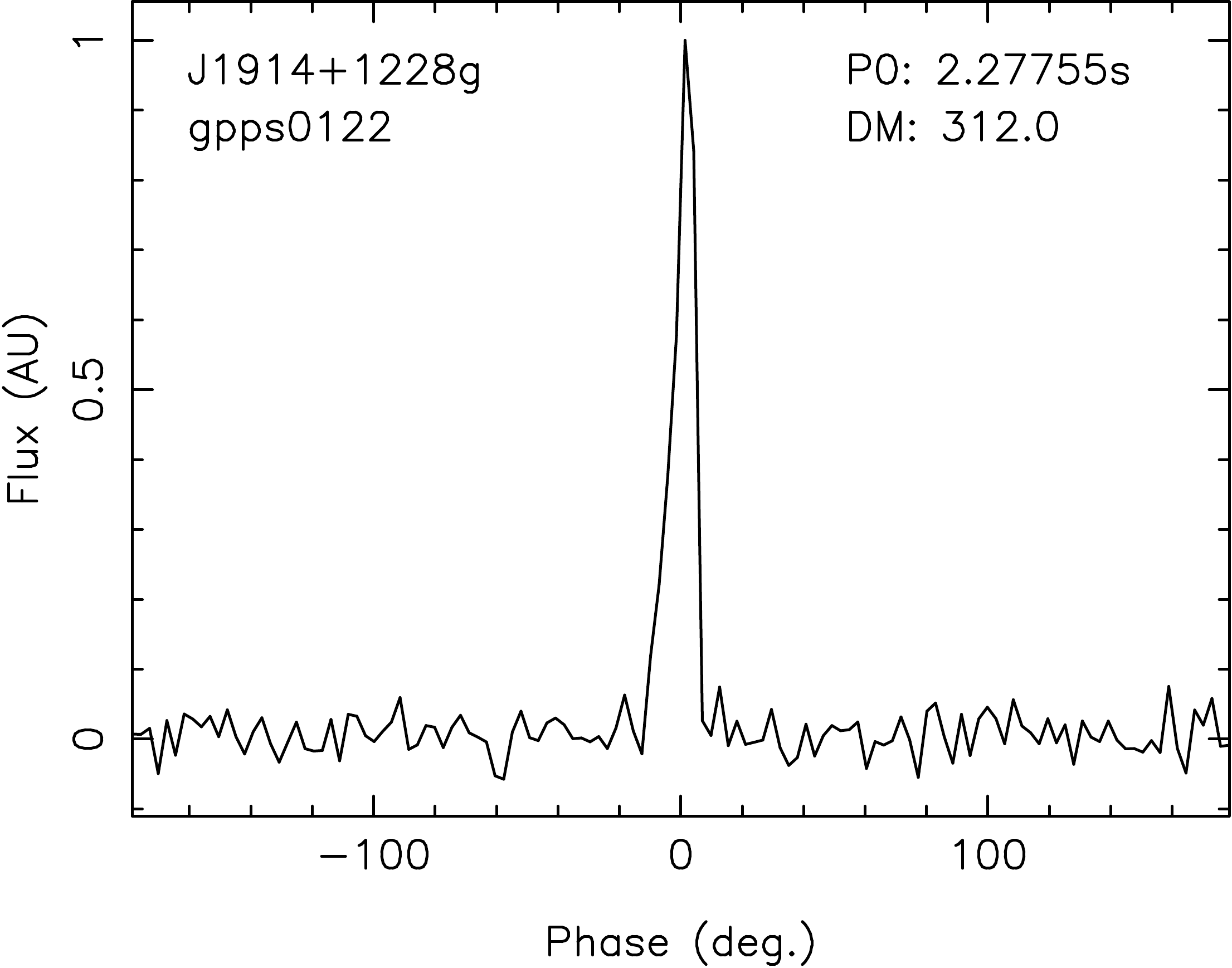}&
\includegraphics[width=39mm]{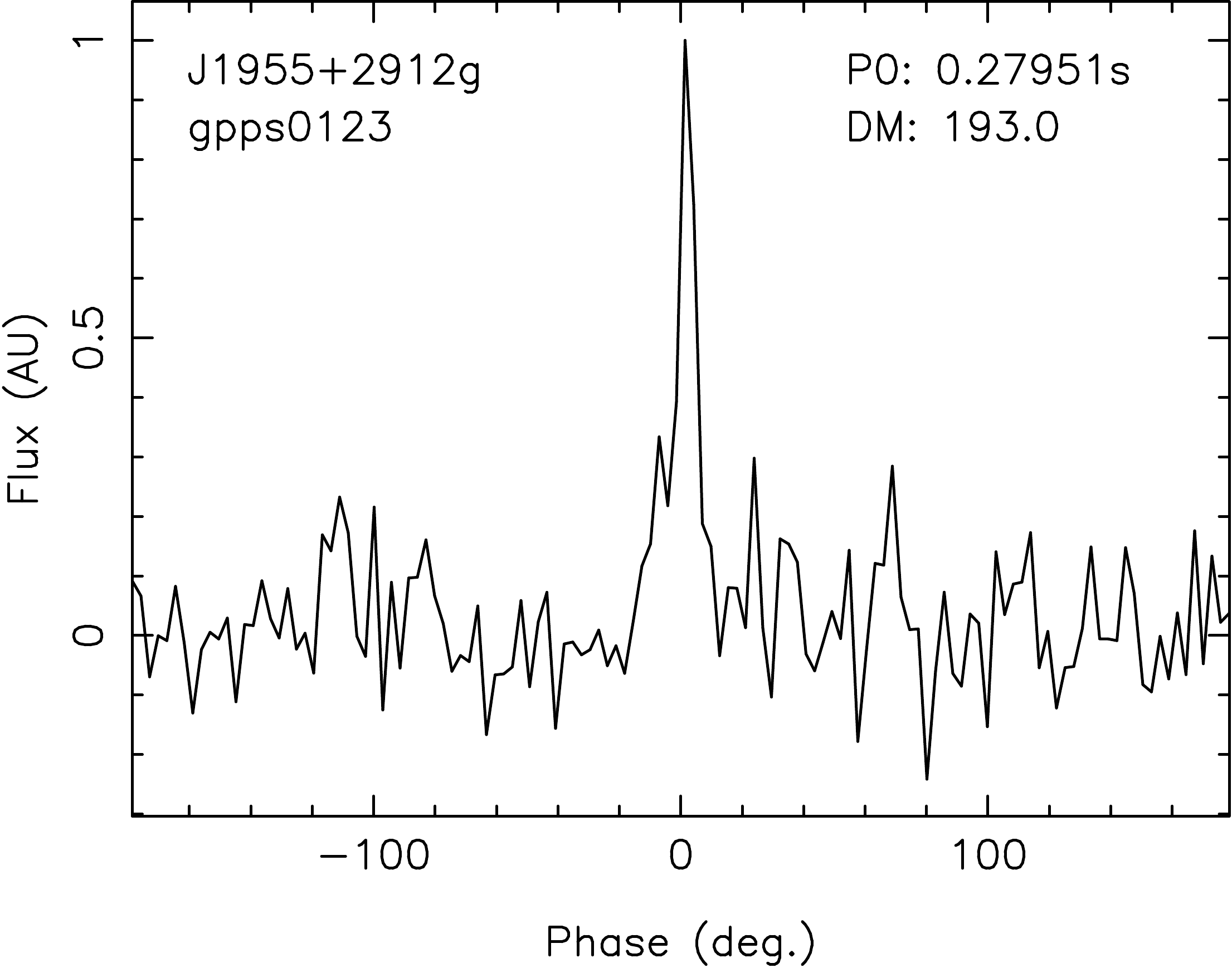}&
\includegraphics[width=39mm]{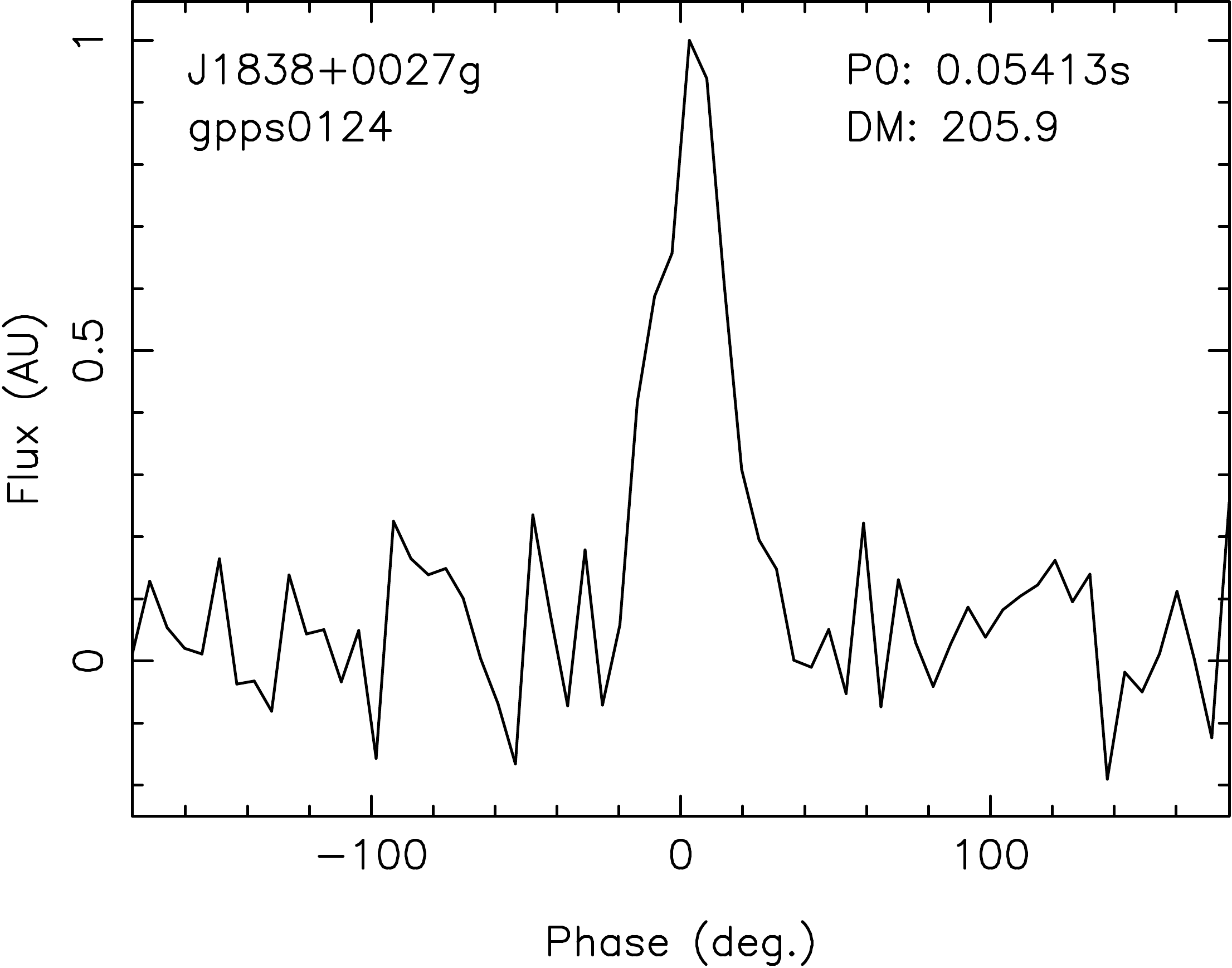}\\[2mm]
\includegraphics[width=39mm]{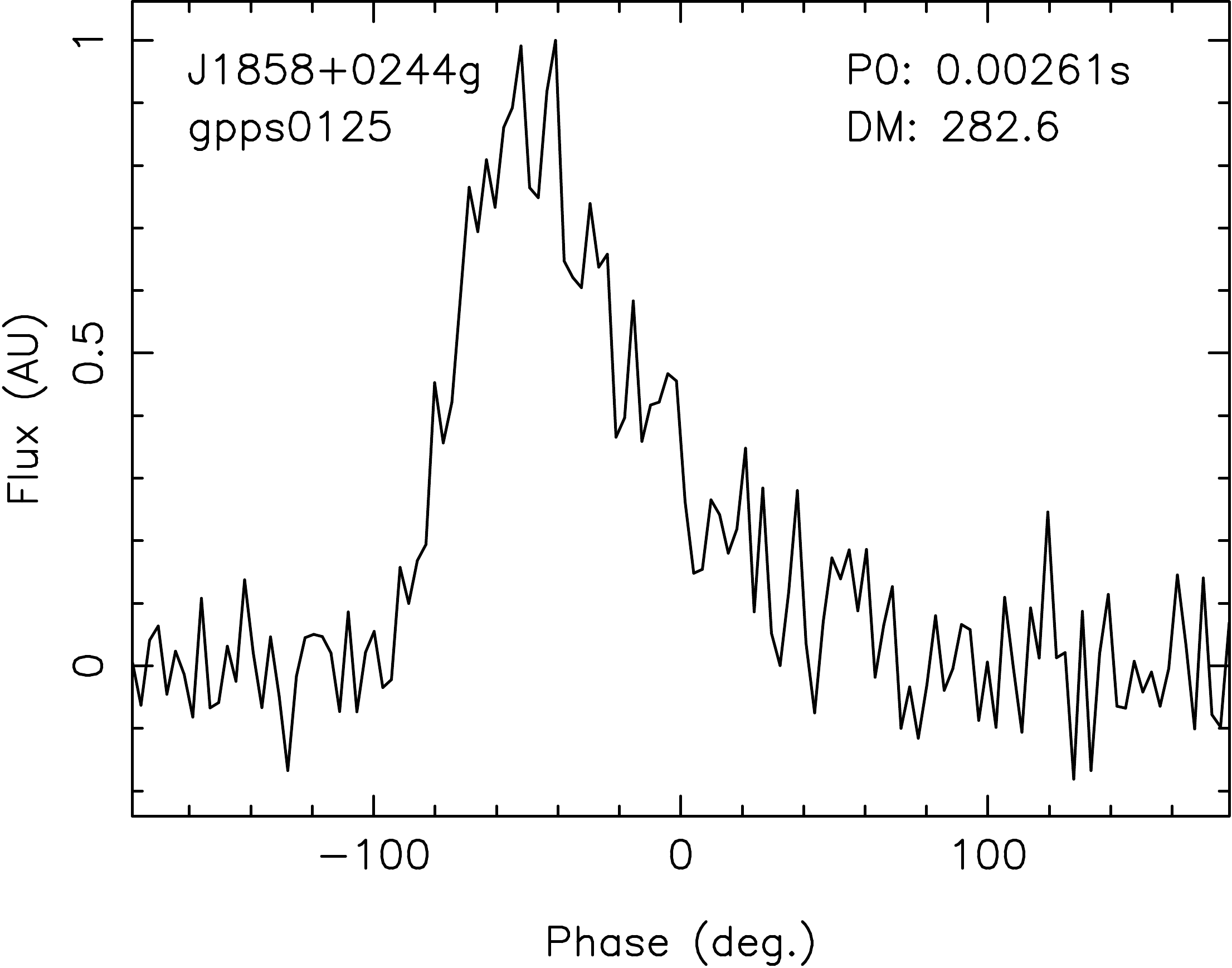}&
\includegraphics[width=39mm]{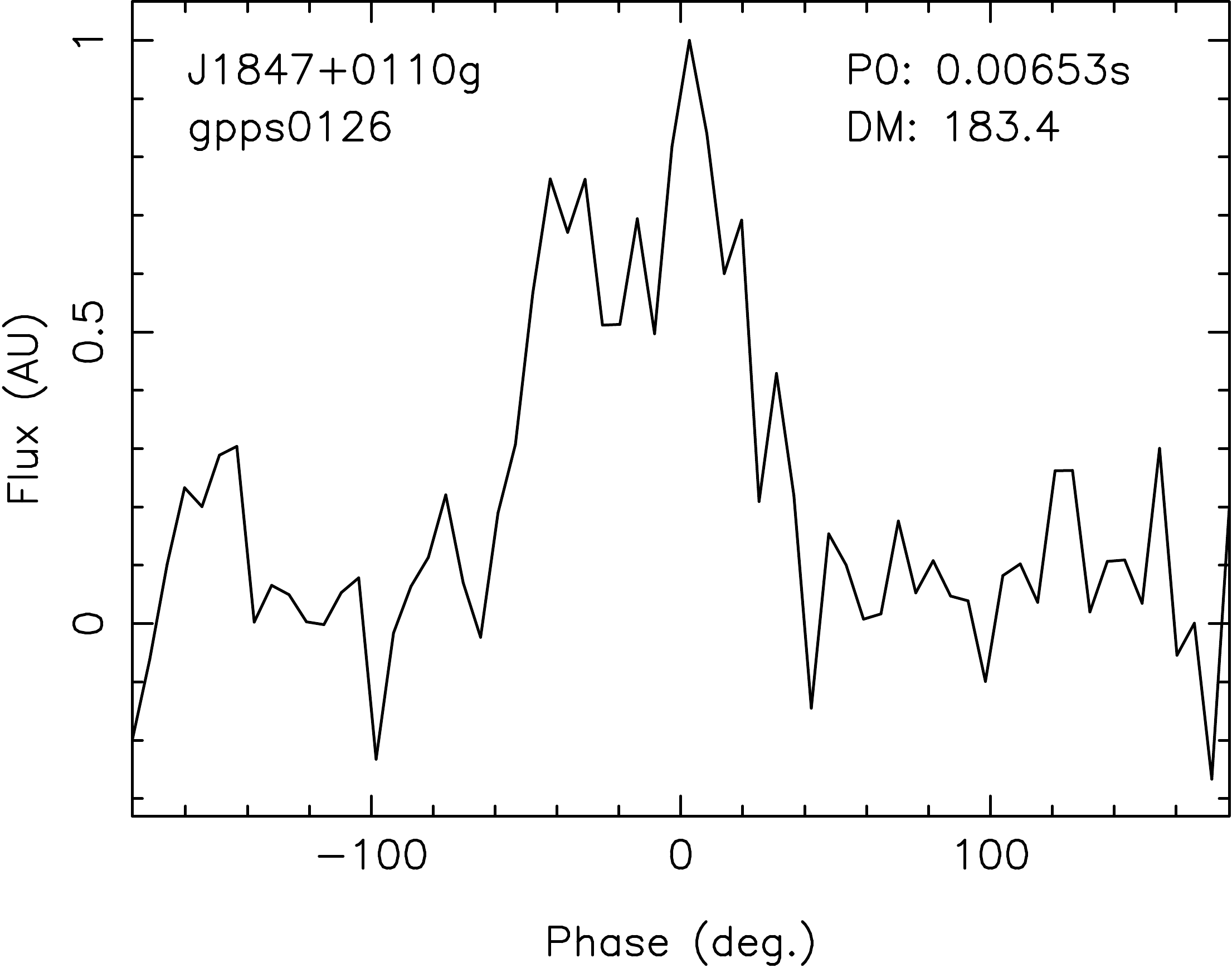}&
\includegraphics[width=39mm]{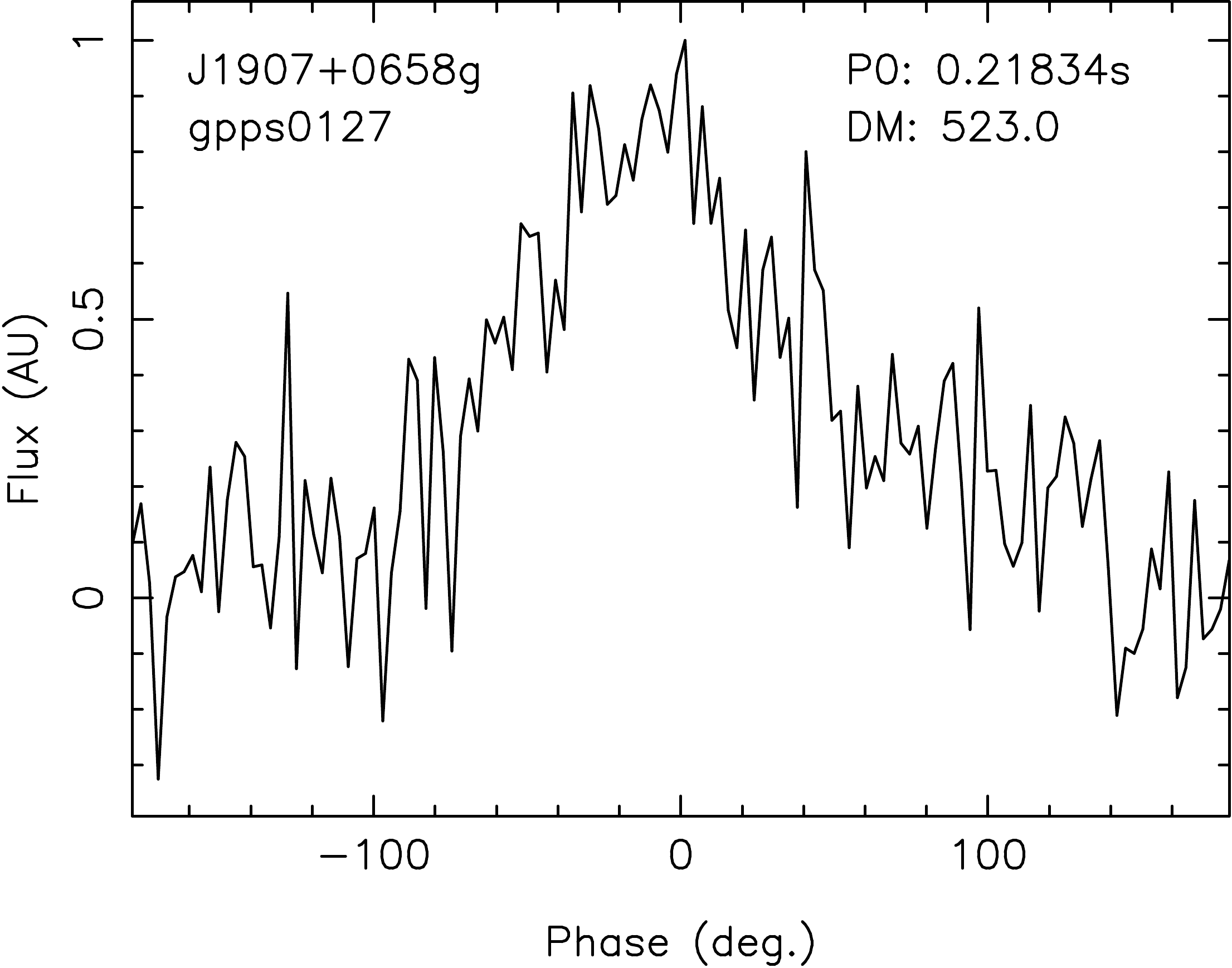}&
\includegraphics[width=39mm]{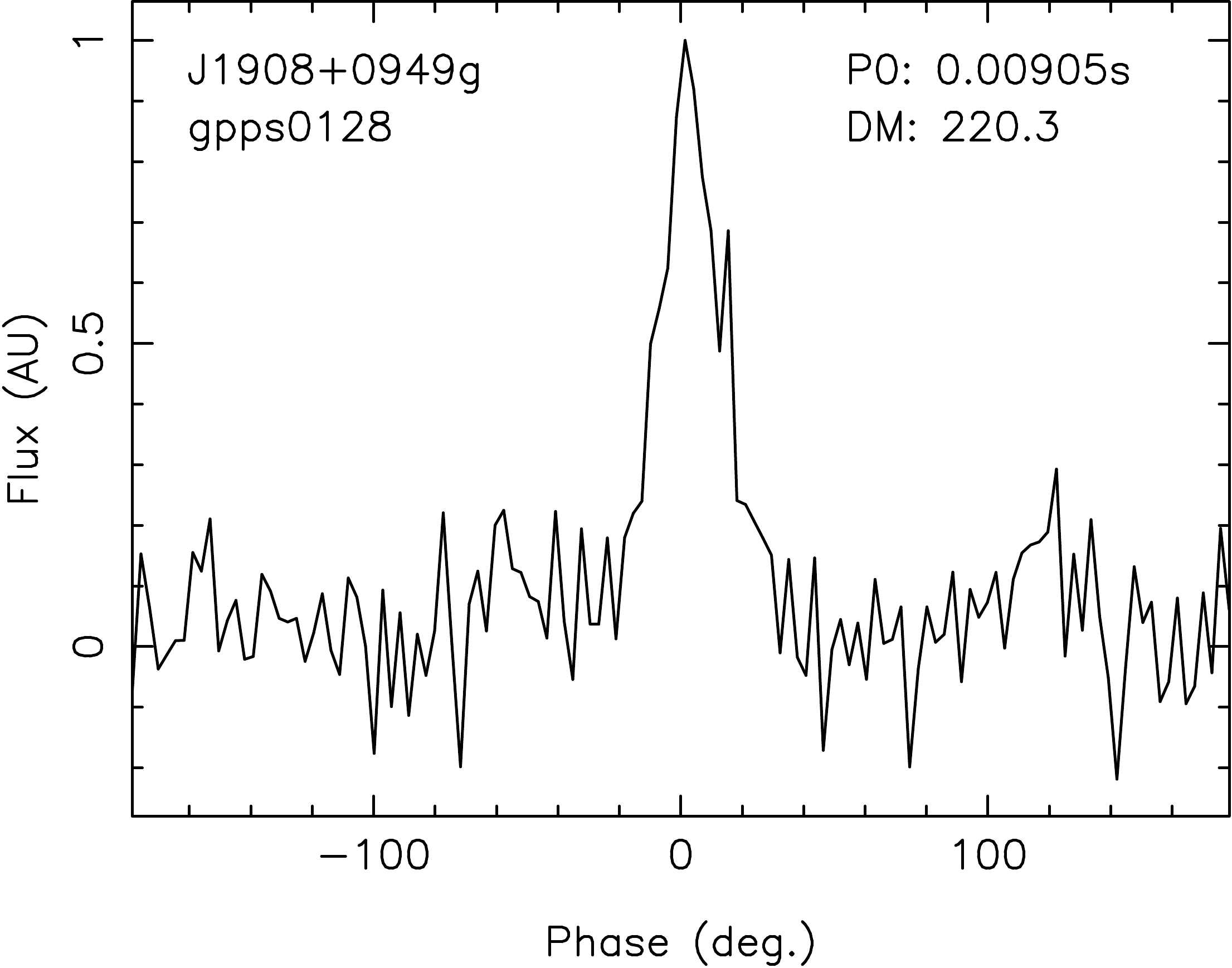}\\[2mm]
\includegraphics[width=39mm]{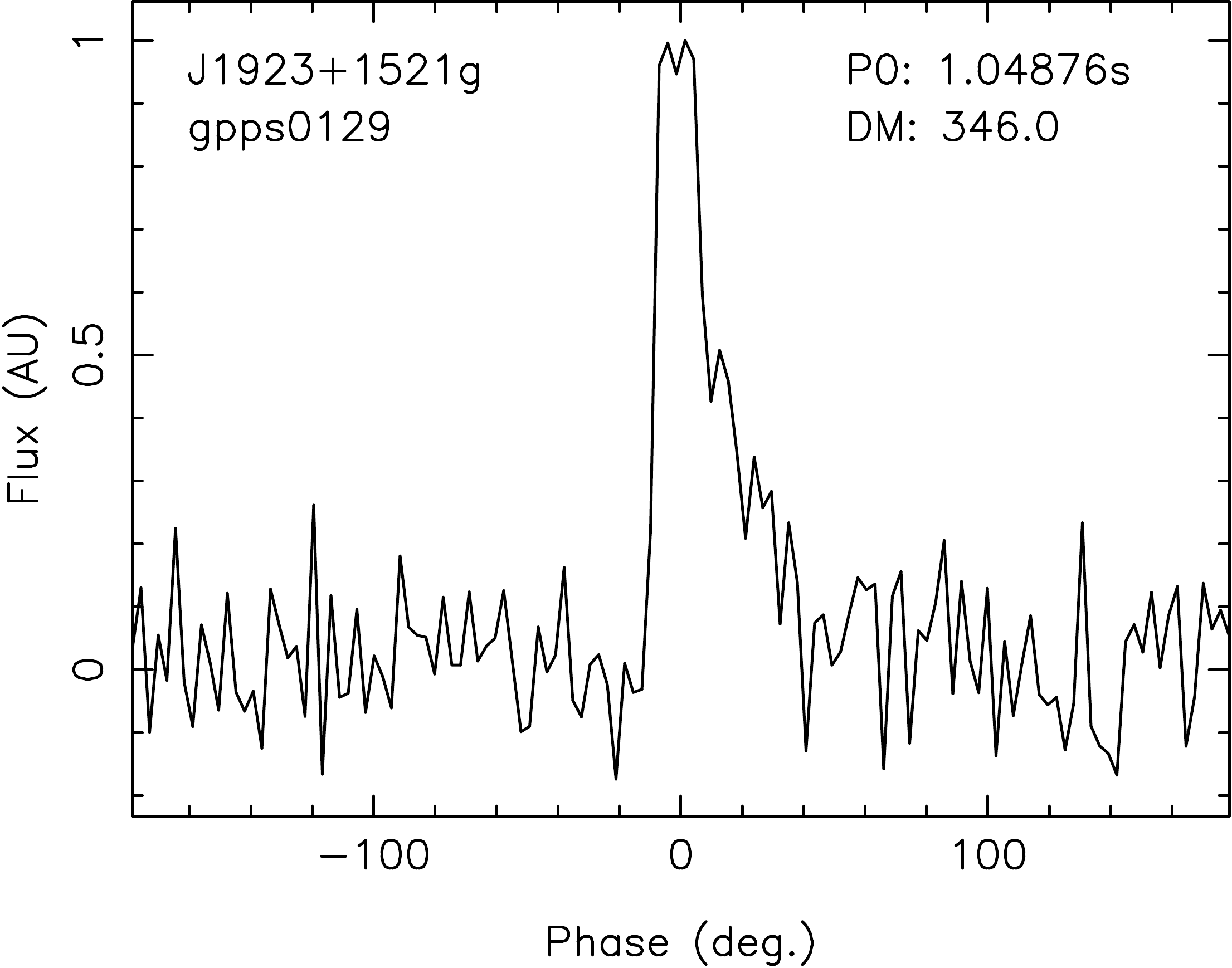}&
\includegraphics[width=39mm]{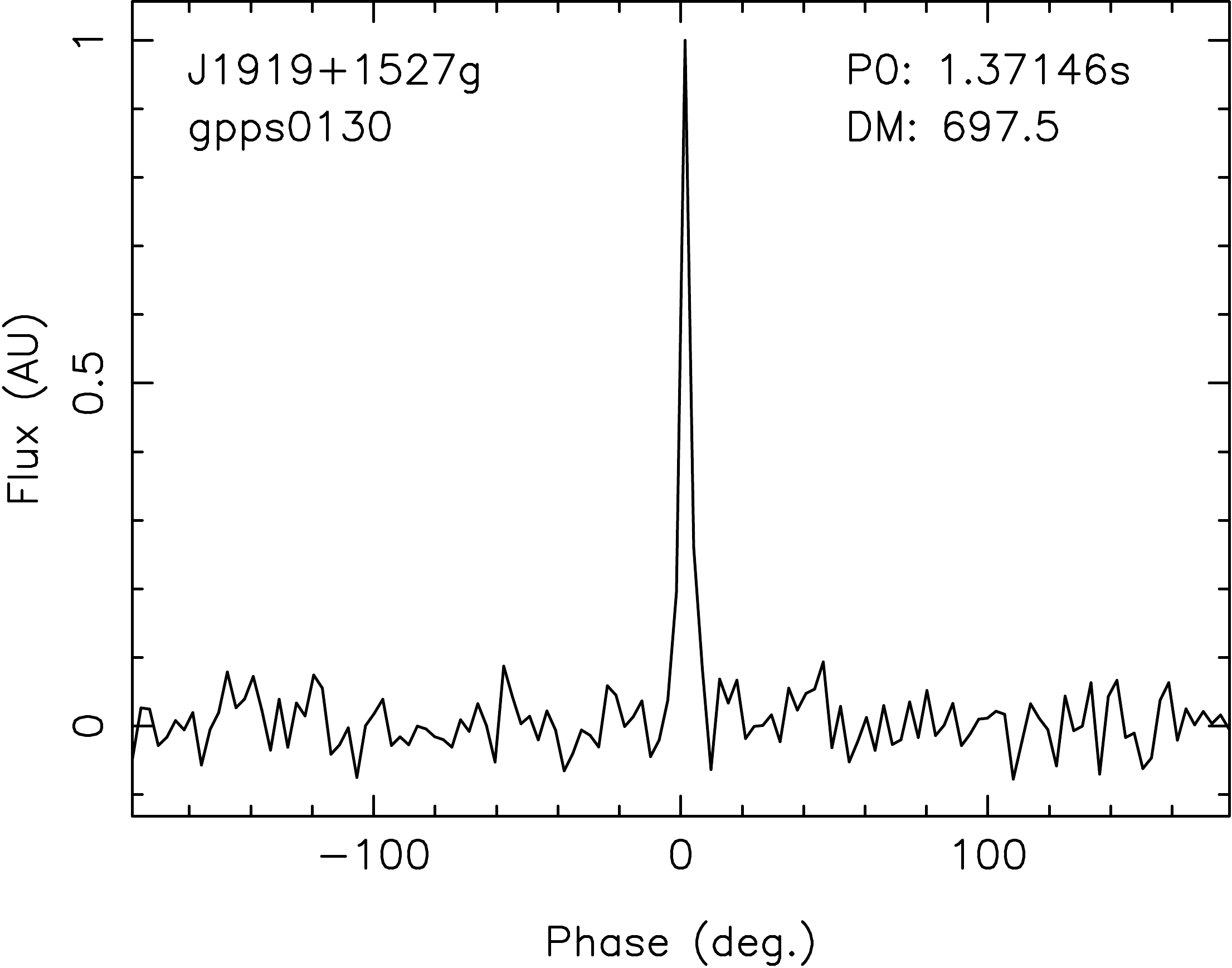}&
\includegraphics[width=39mm]{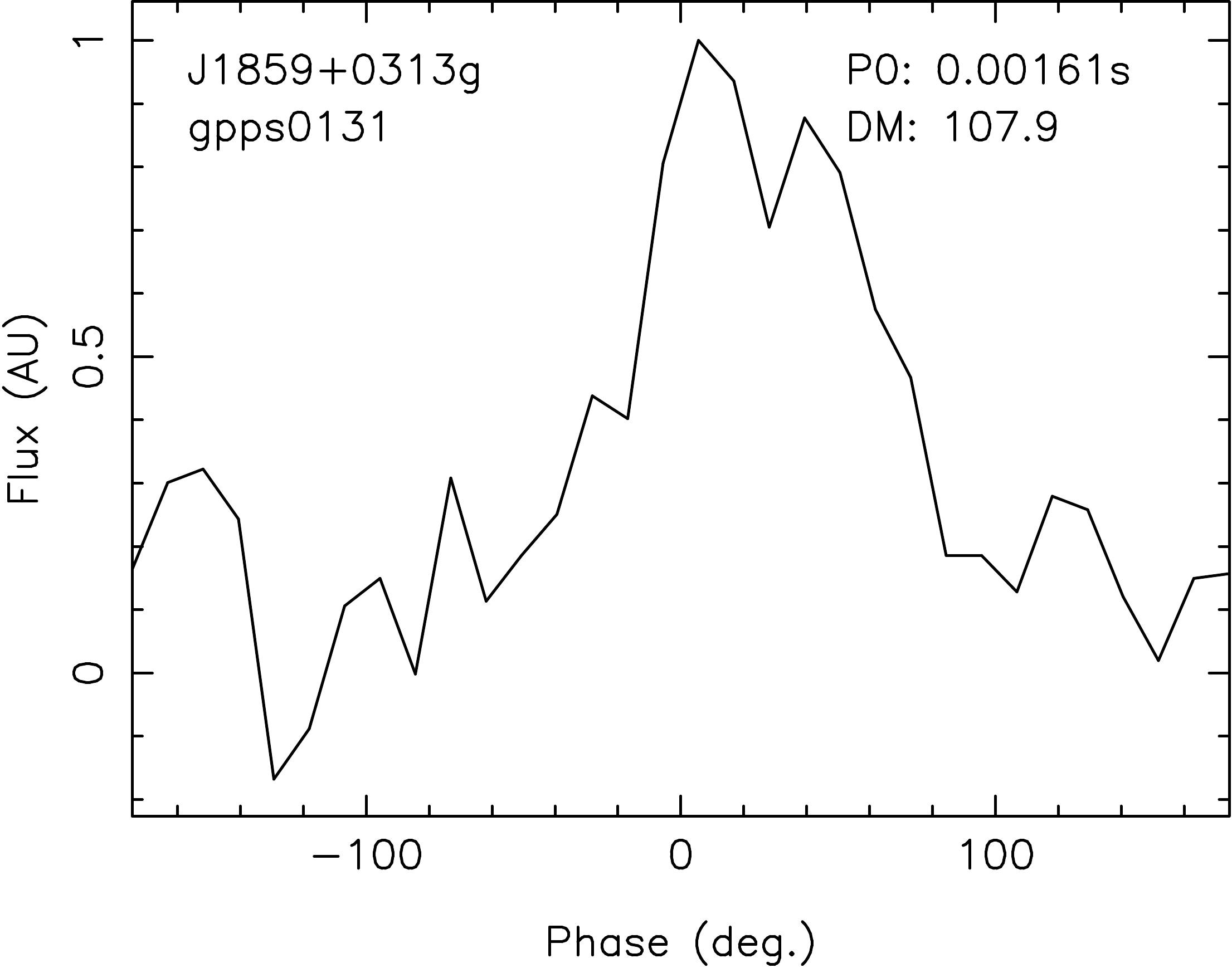}&
\includegraphics[width=39mm]{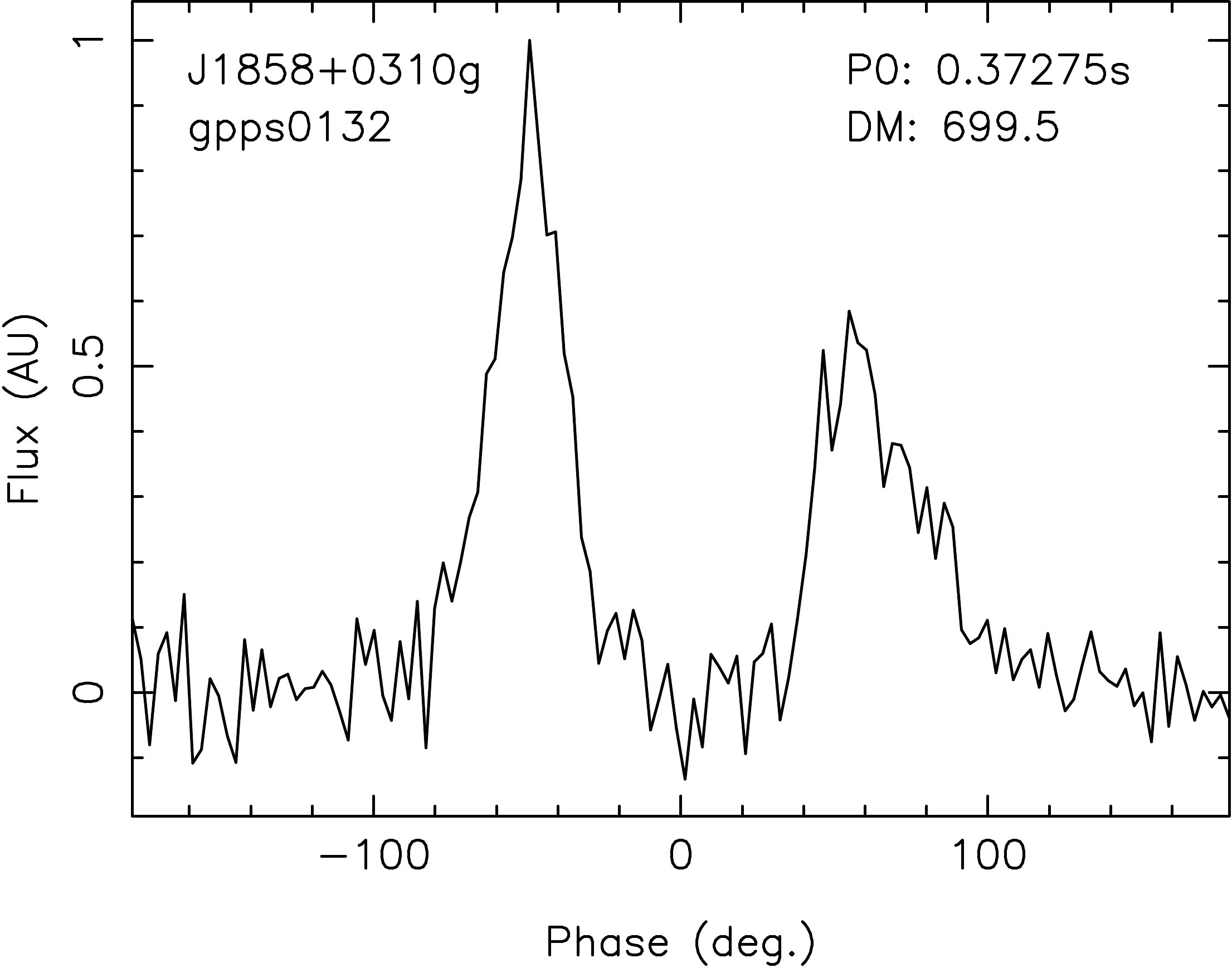}\\[2mm]
\includegraphics[width=39mm]{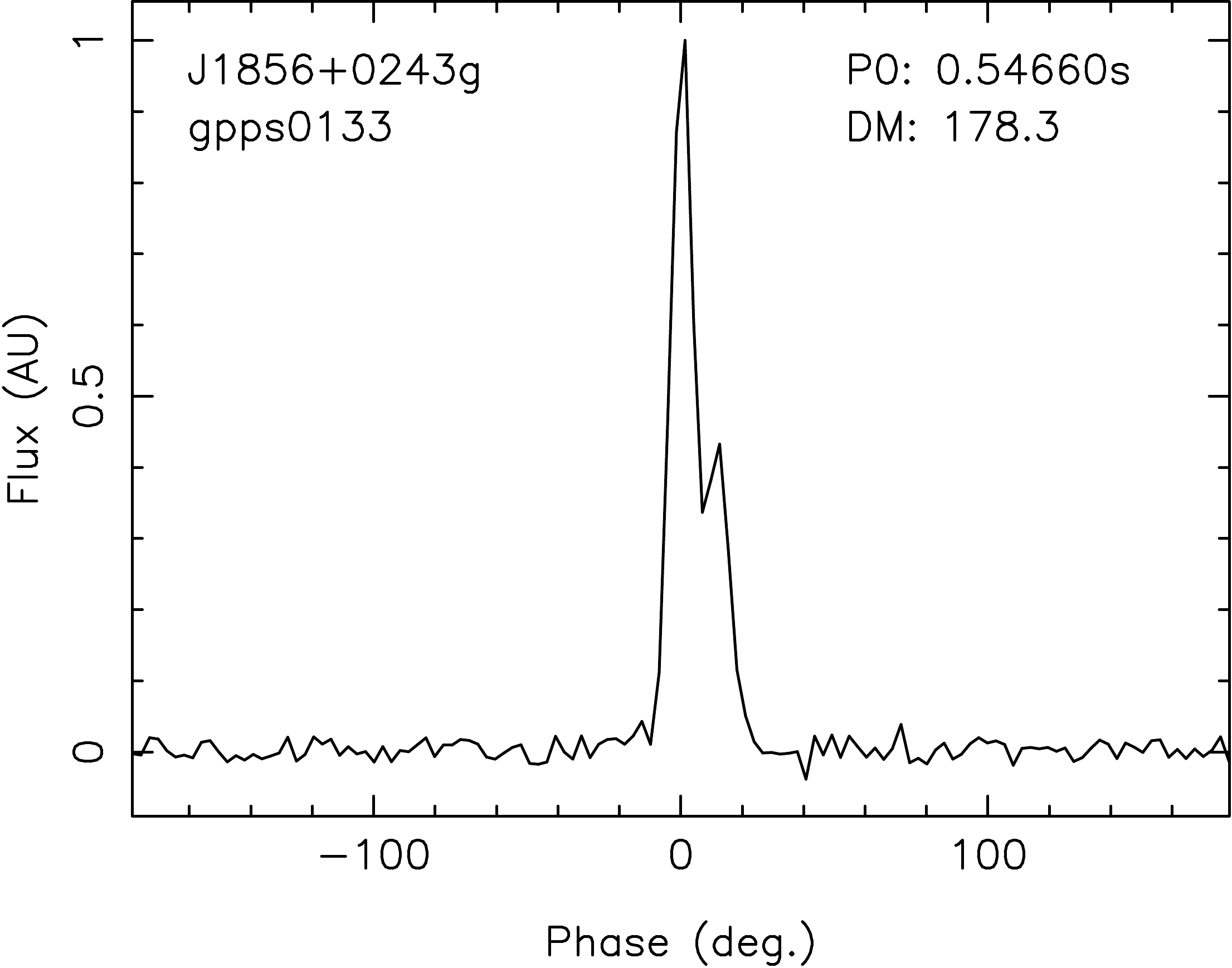}&
\includegraphics[width=39mm]{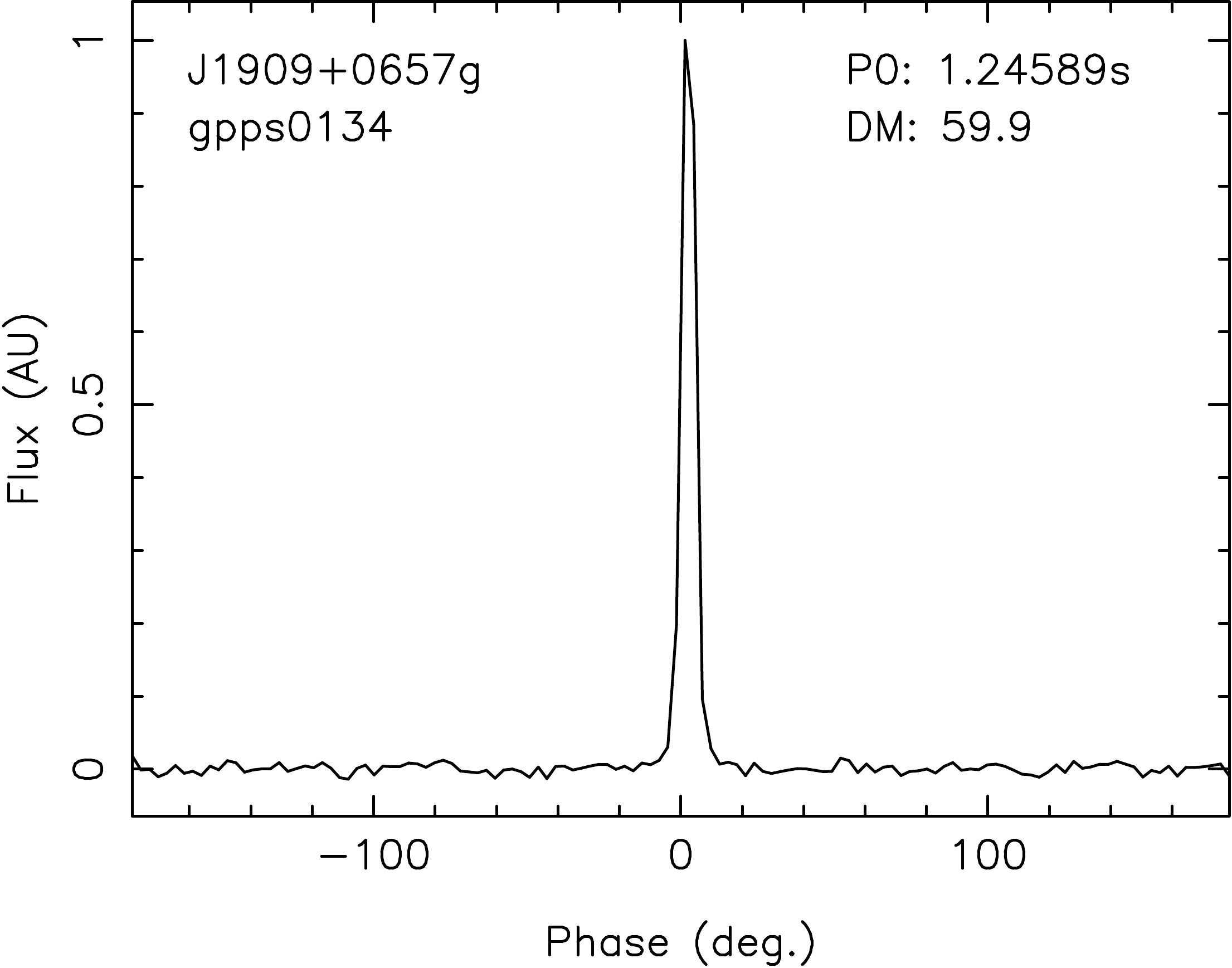}&
\includegraphics[width=39mm]{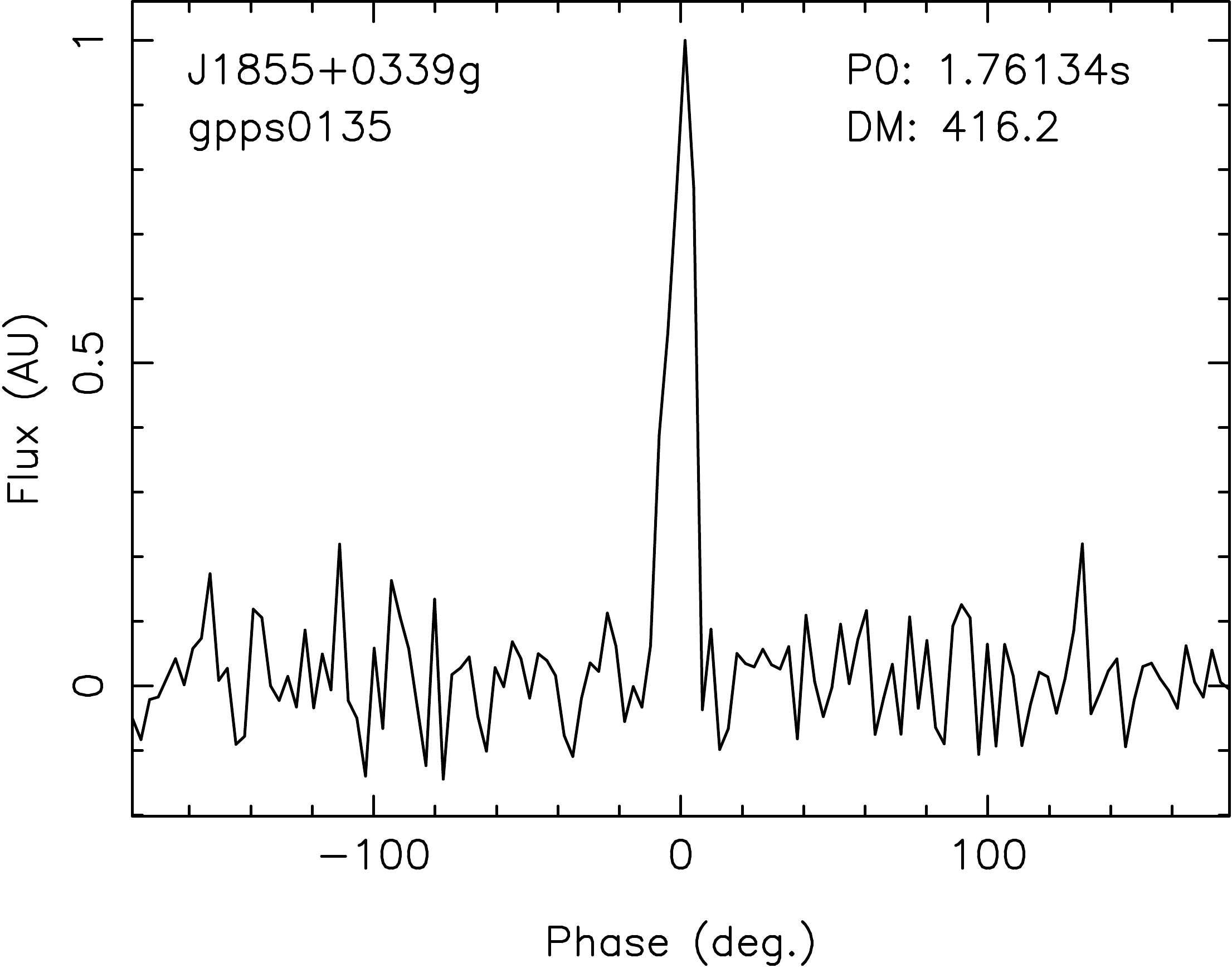}&
\includegraphics[width=39mm]{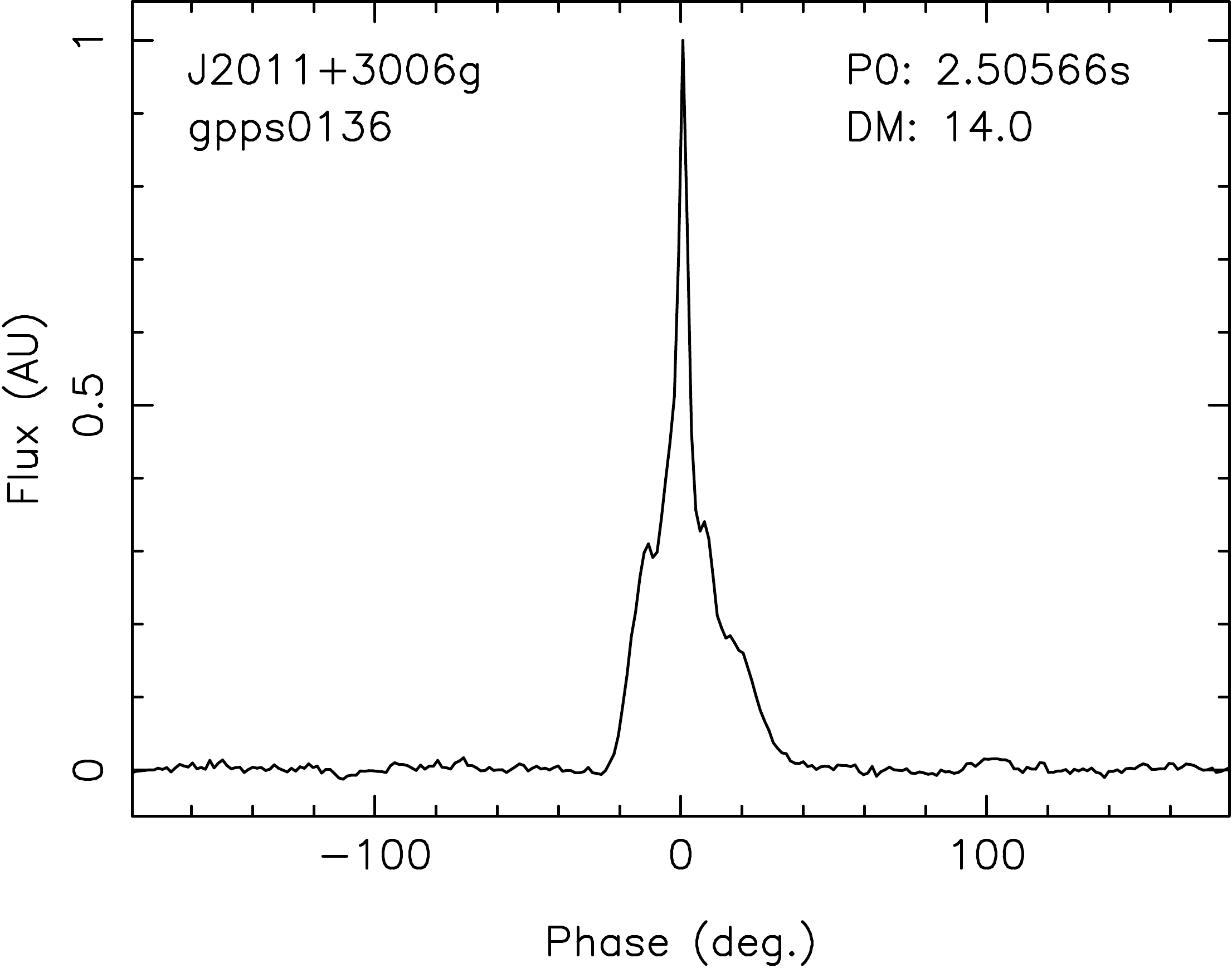}\\[2mm]
\includegraphics[width=39mm]{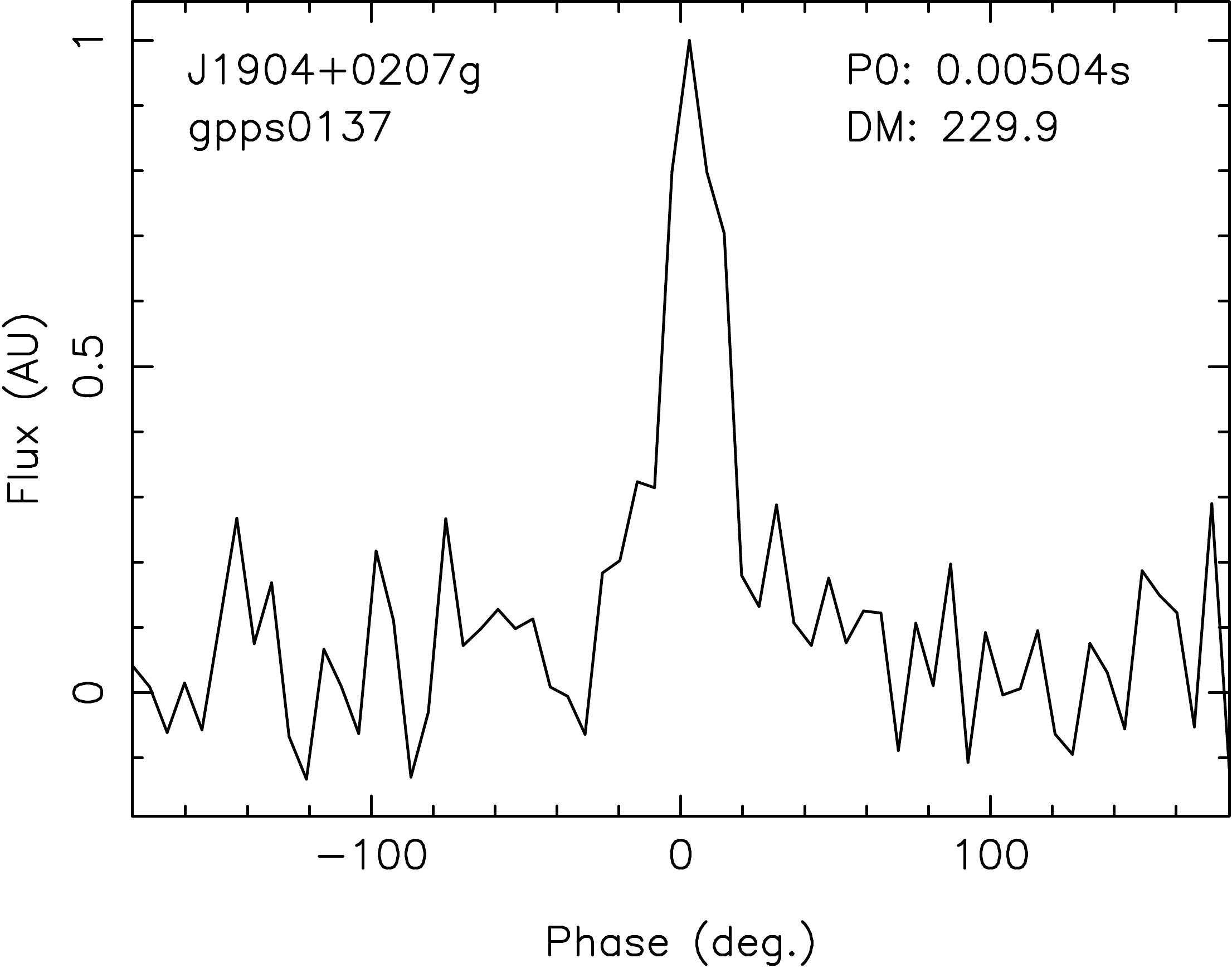}&
\includegraphics[width=39mm]{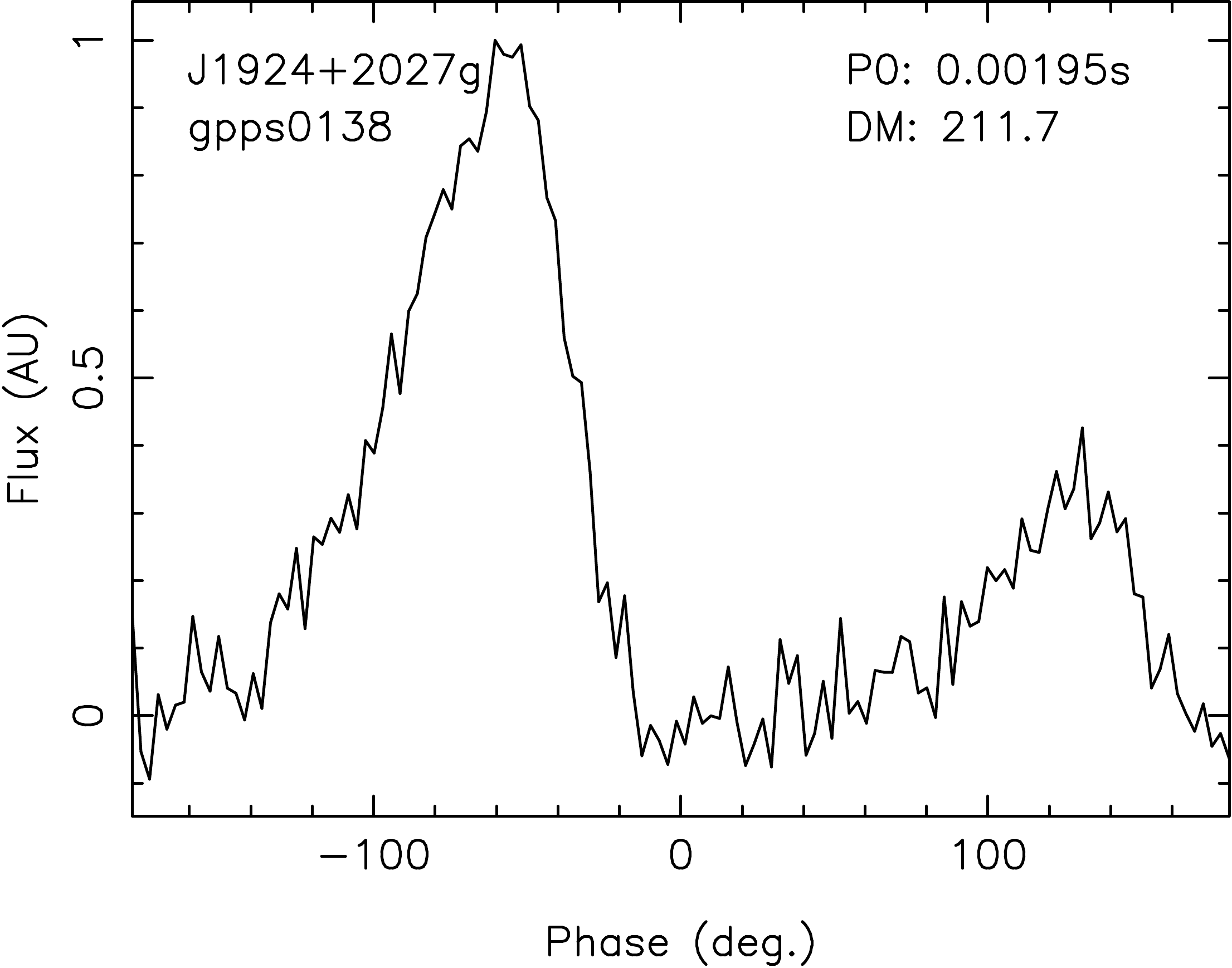}&
\includegraphics[width=39mm]{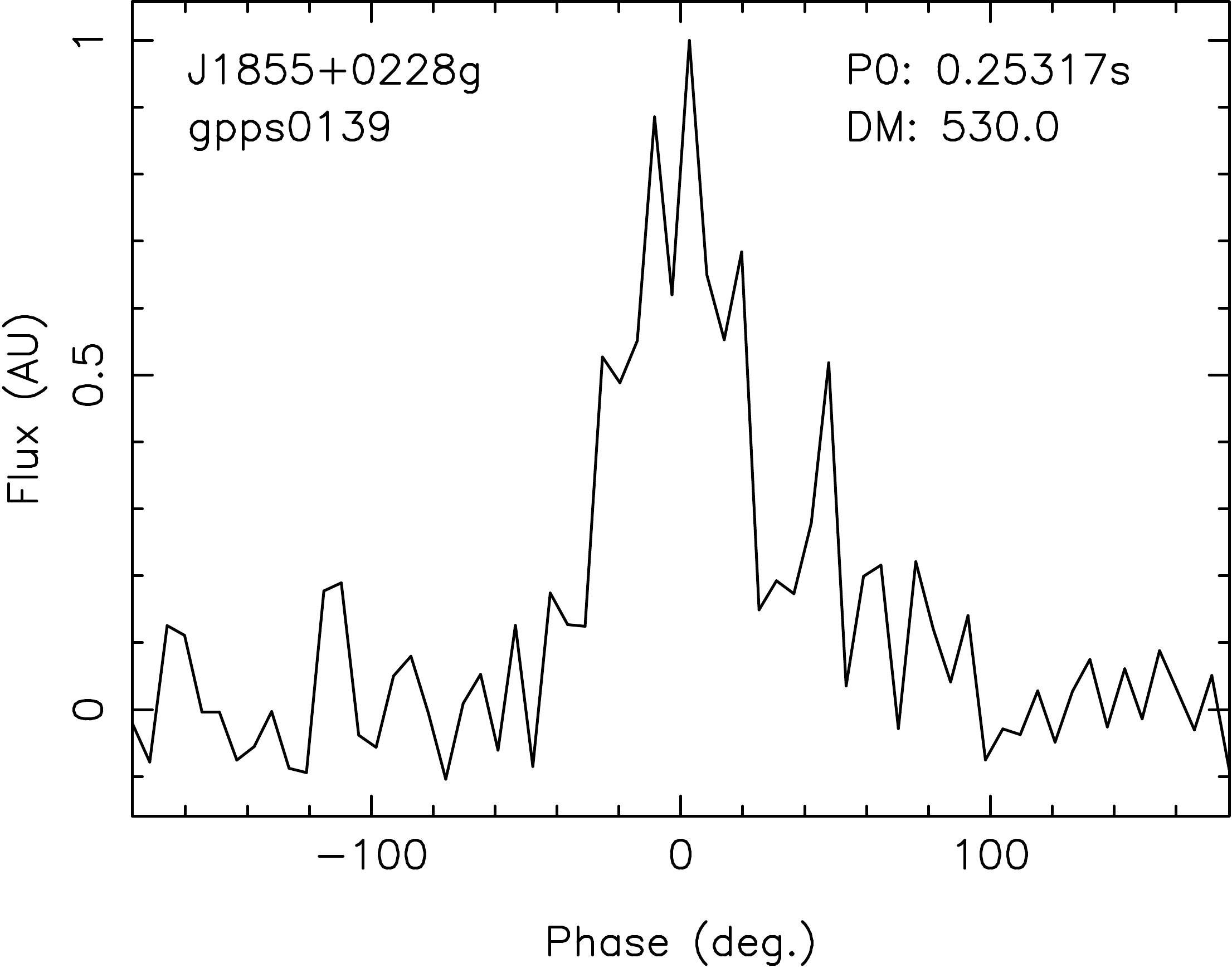}&
\includegraphics[width=39mm]{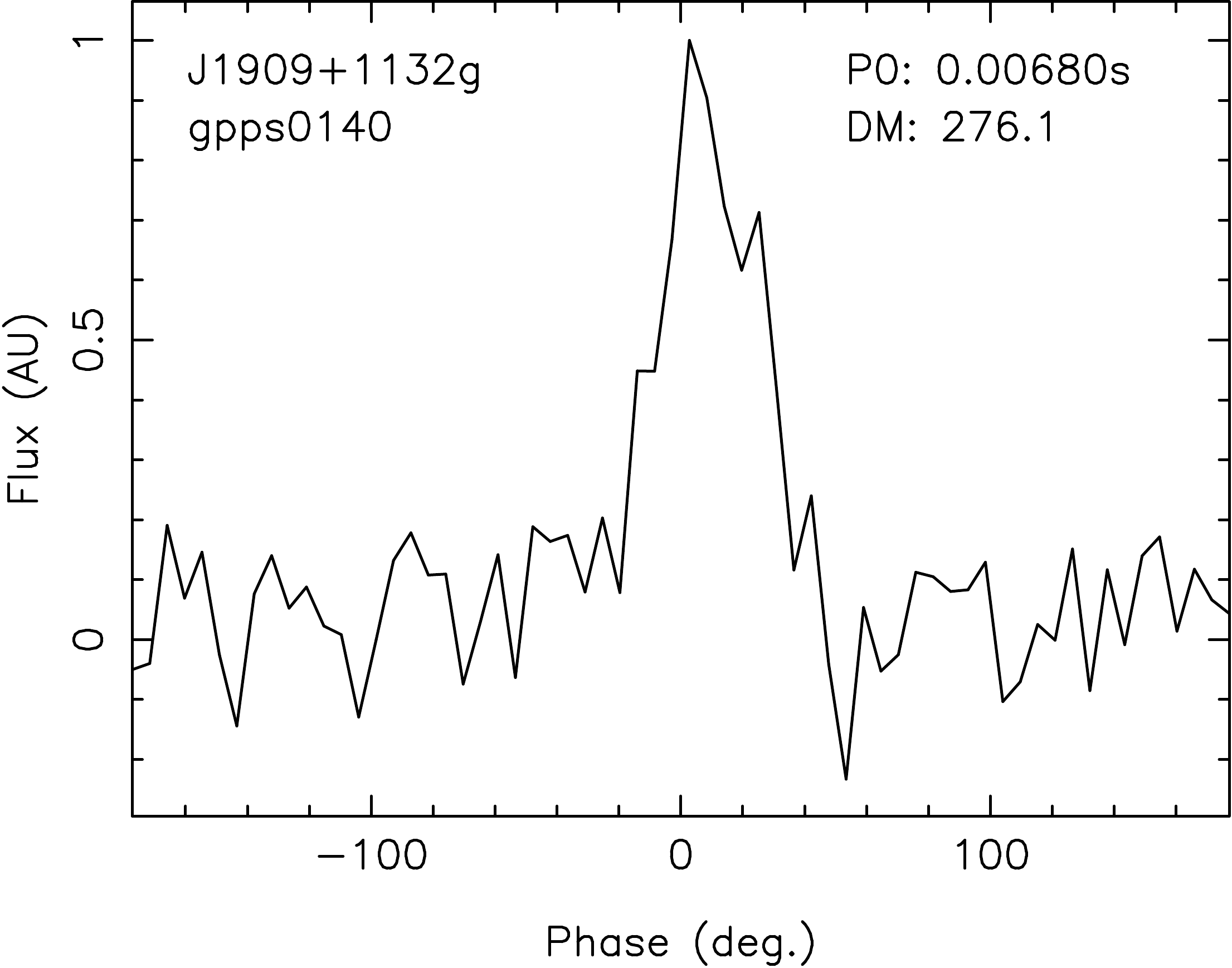}\\
\end{tabular}%

\begin{minipage}{3cm}
\caption[]{
-- {\it Continued}.}\end{minipage}
\addtocounter{figure}{-1}
\end{figure*}%
\begin{figure*}
\centering
\begin{tabular}{rrrr}
\includegraphics[width=39mm]{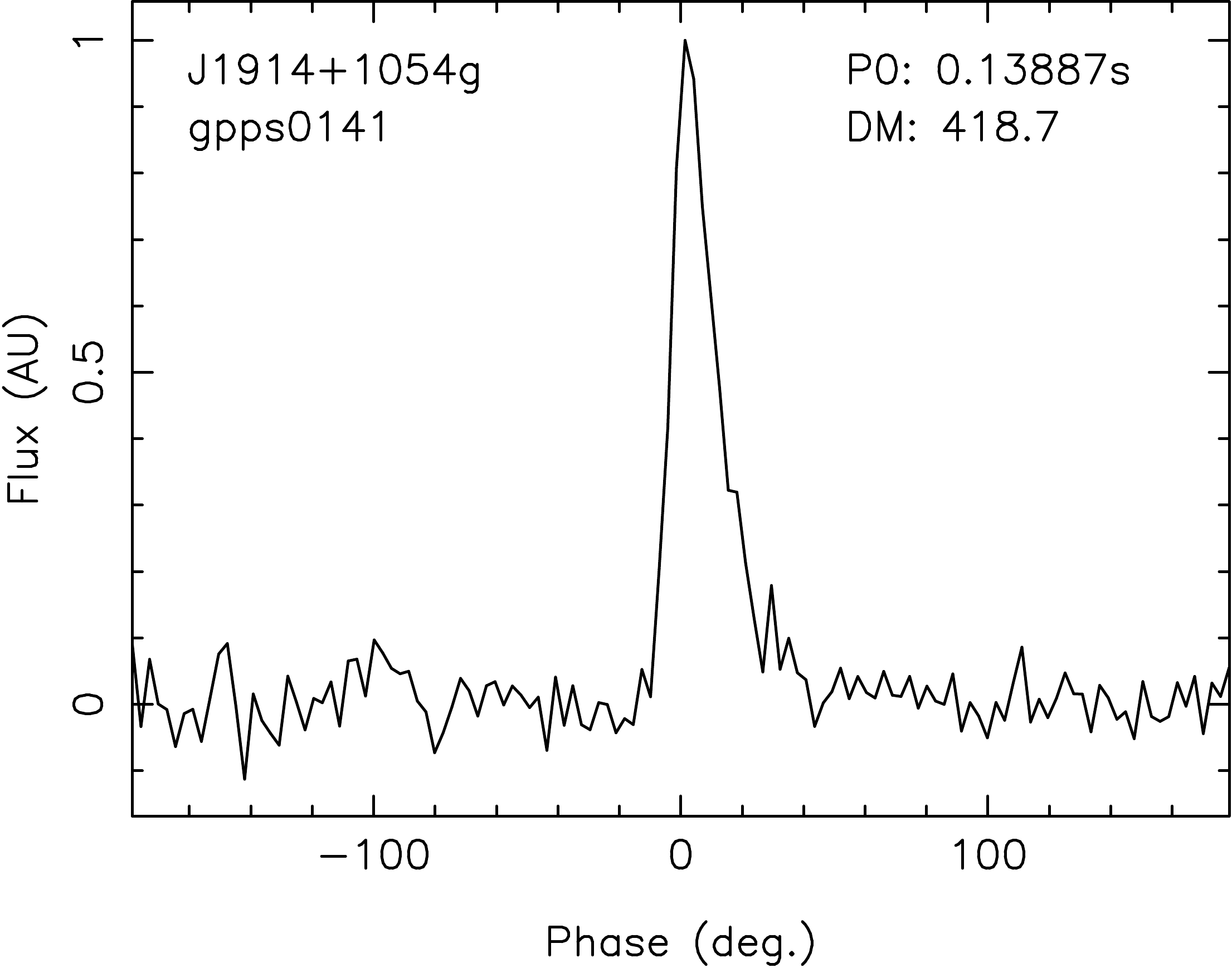}&
\includegraphics[width=39mm]{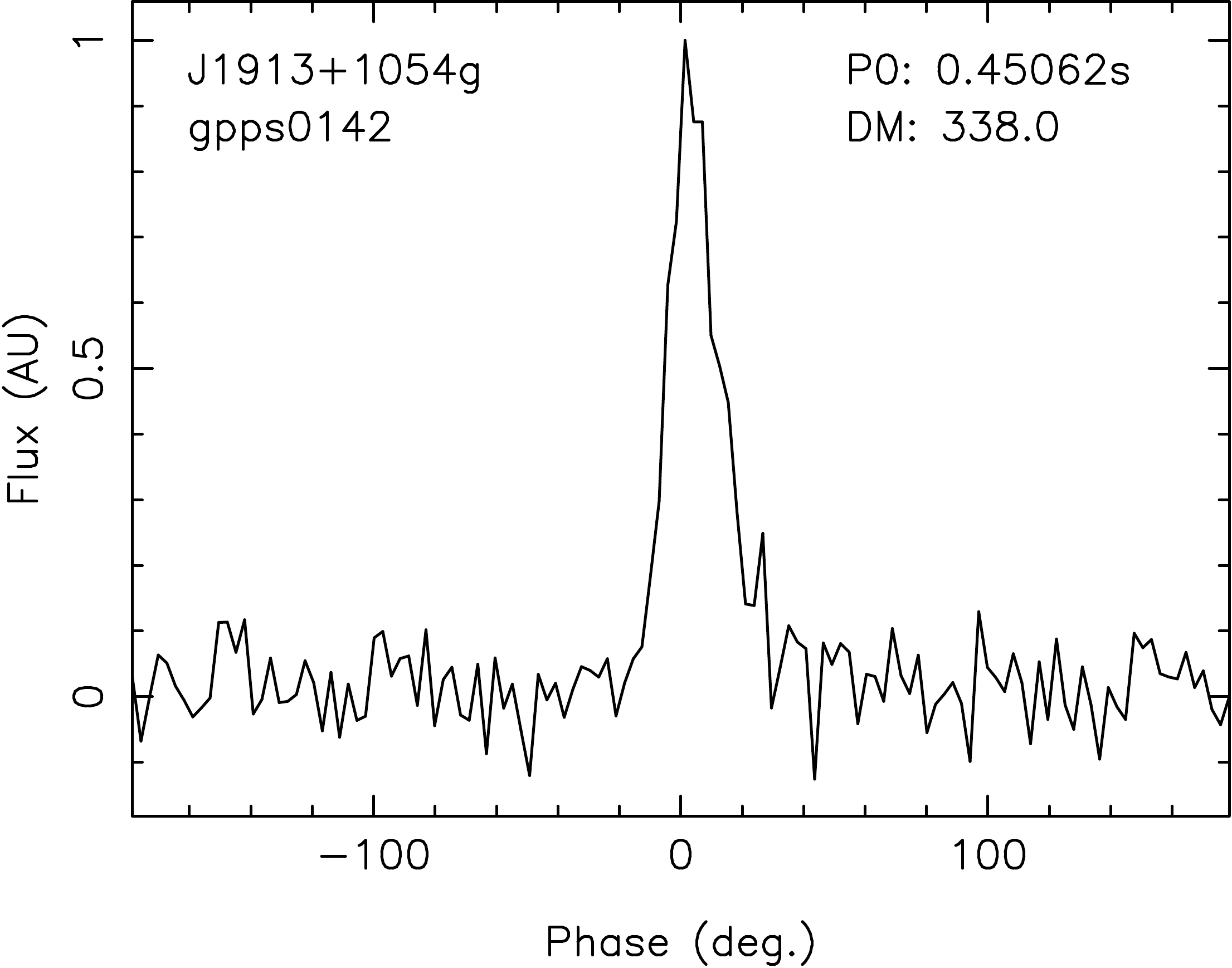}&
\includegraphics[width=39mm]{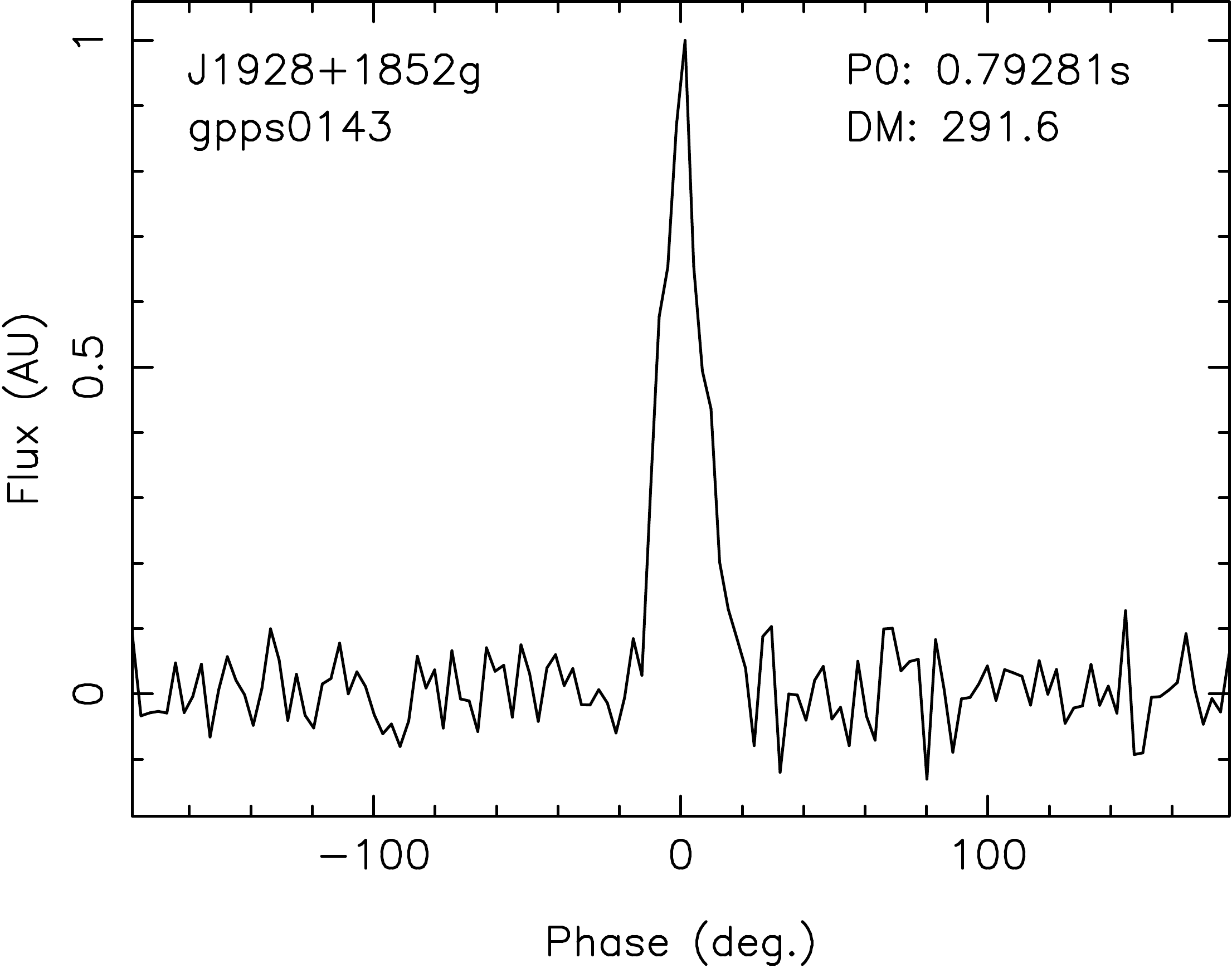}&
\includegraphics[width=39mm]{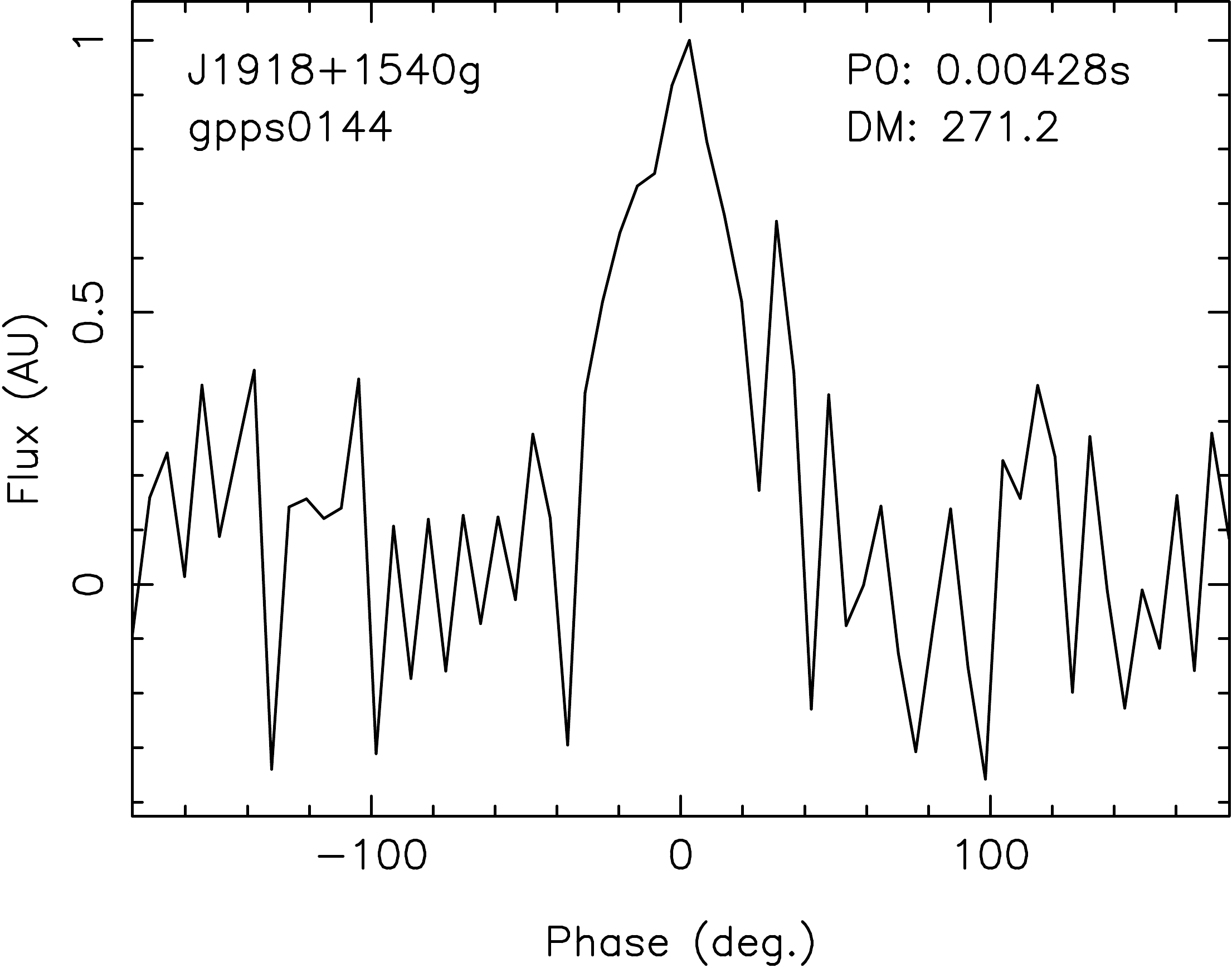} \\[2mm]
\includegraphics[width=39mm]{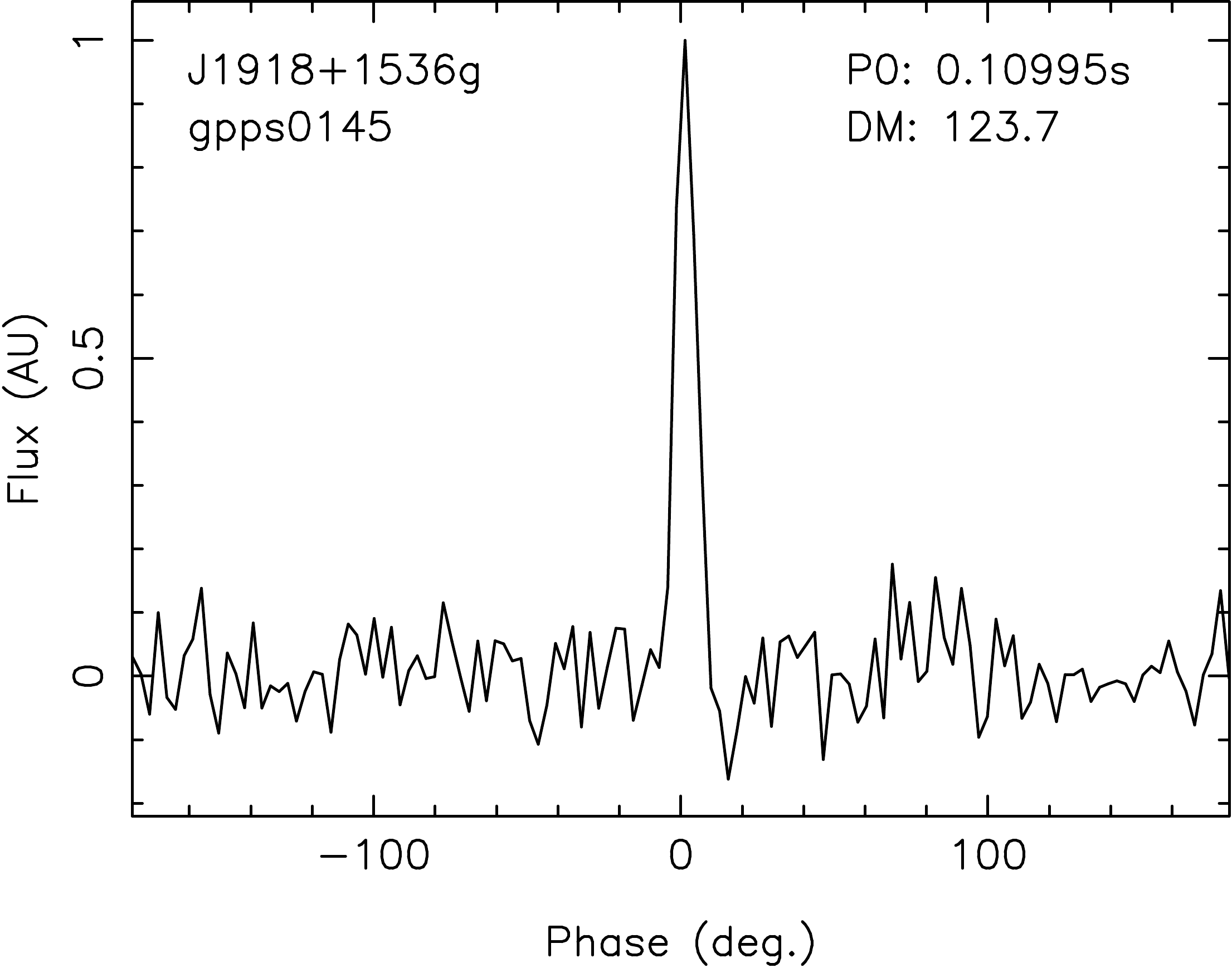} &
\includegraphics[width=39mm]{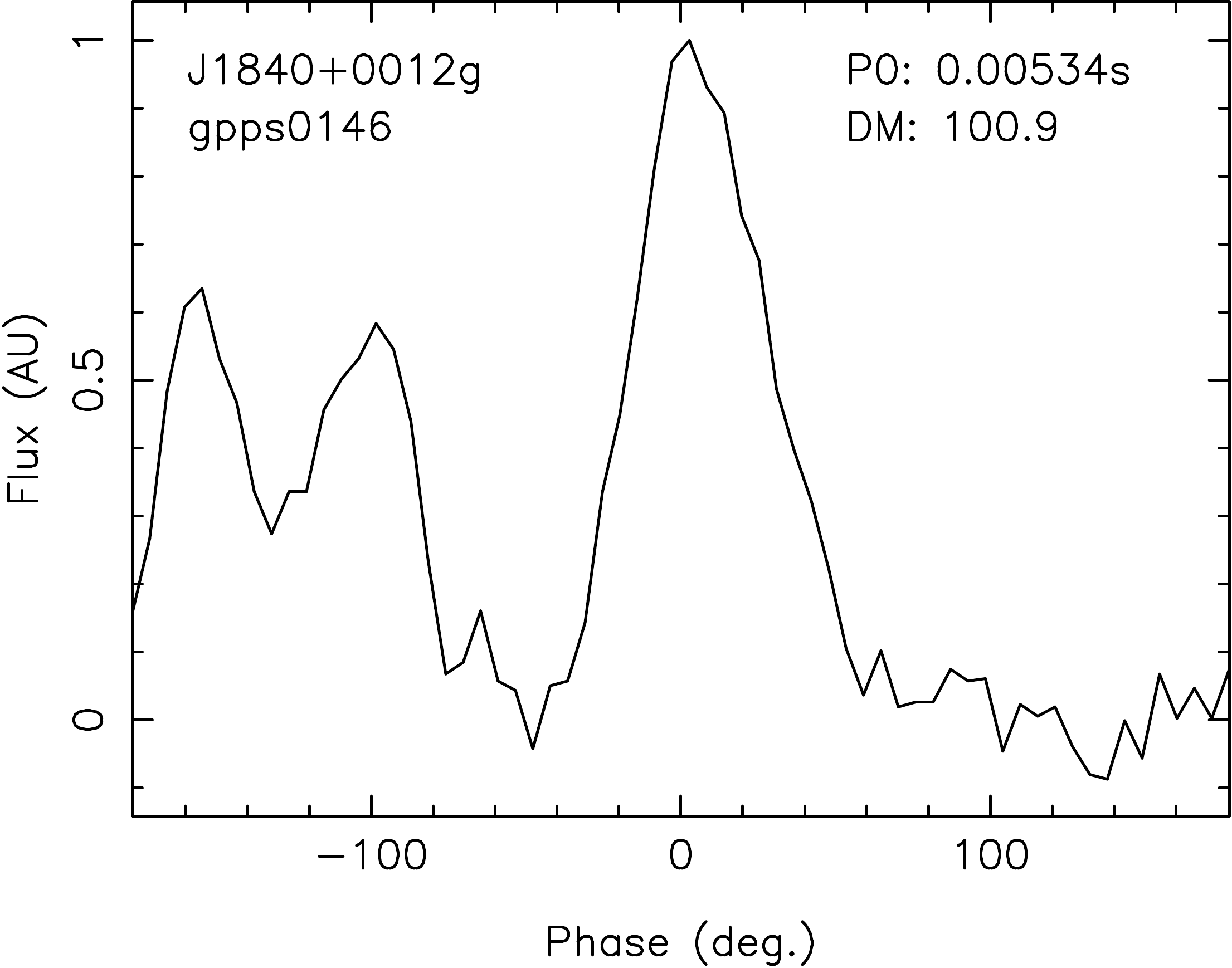} &
\includegraphics[width=39mm]{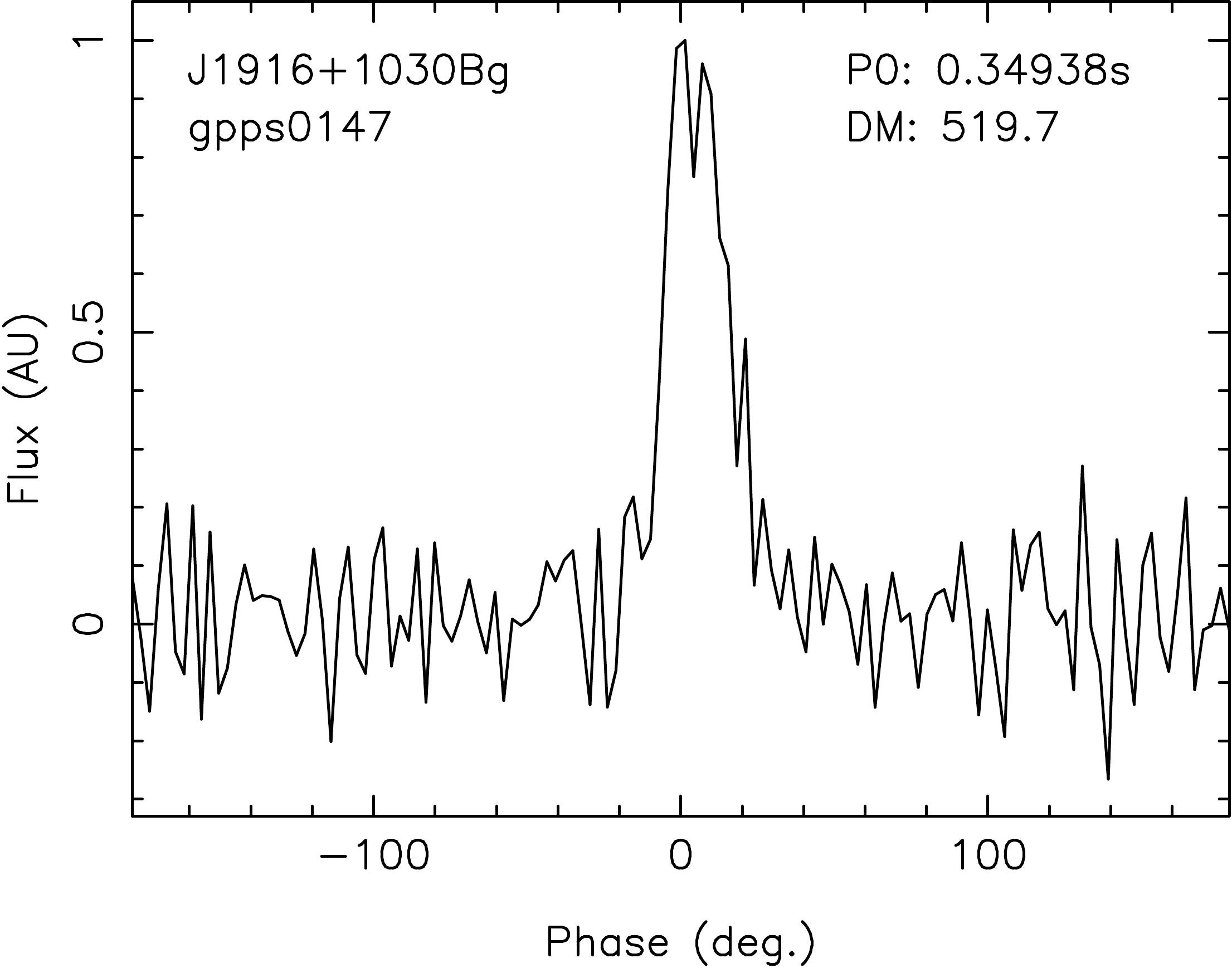}&
\includegraphics[width=39mm]{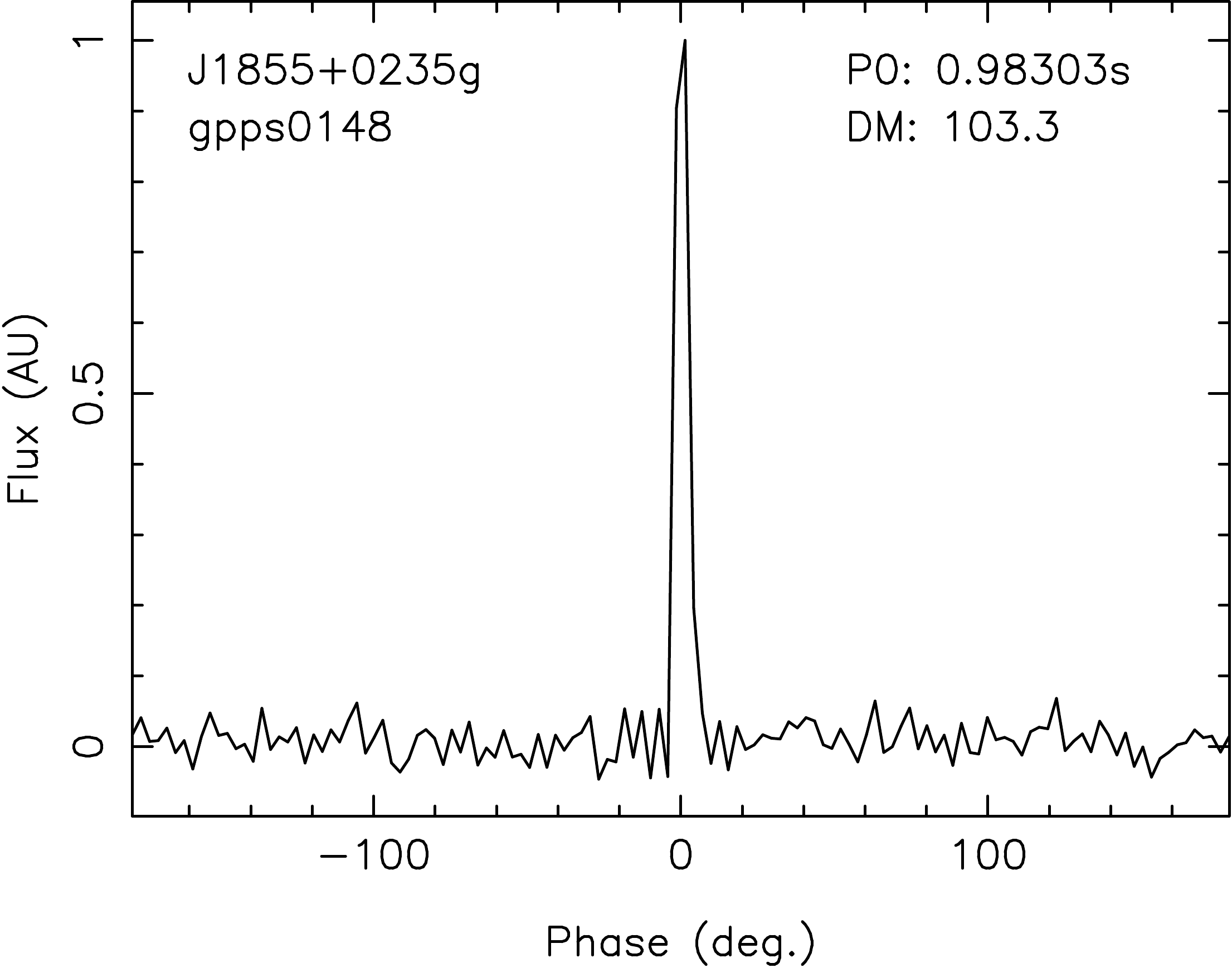}\\[2mm]
\includegraphics[width=39mm]{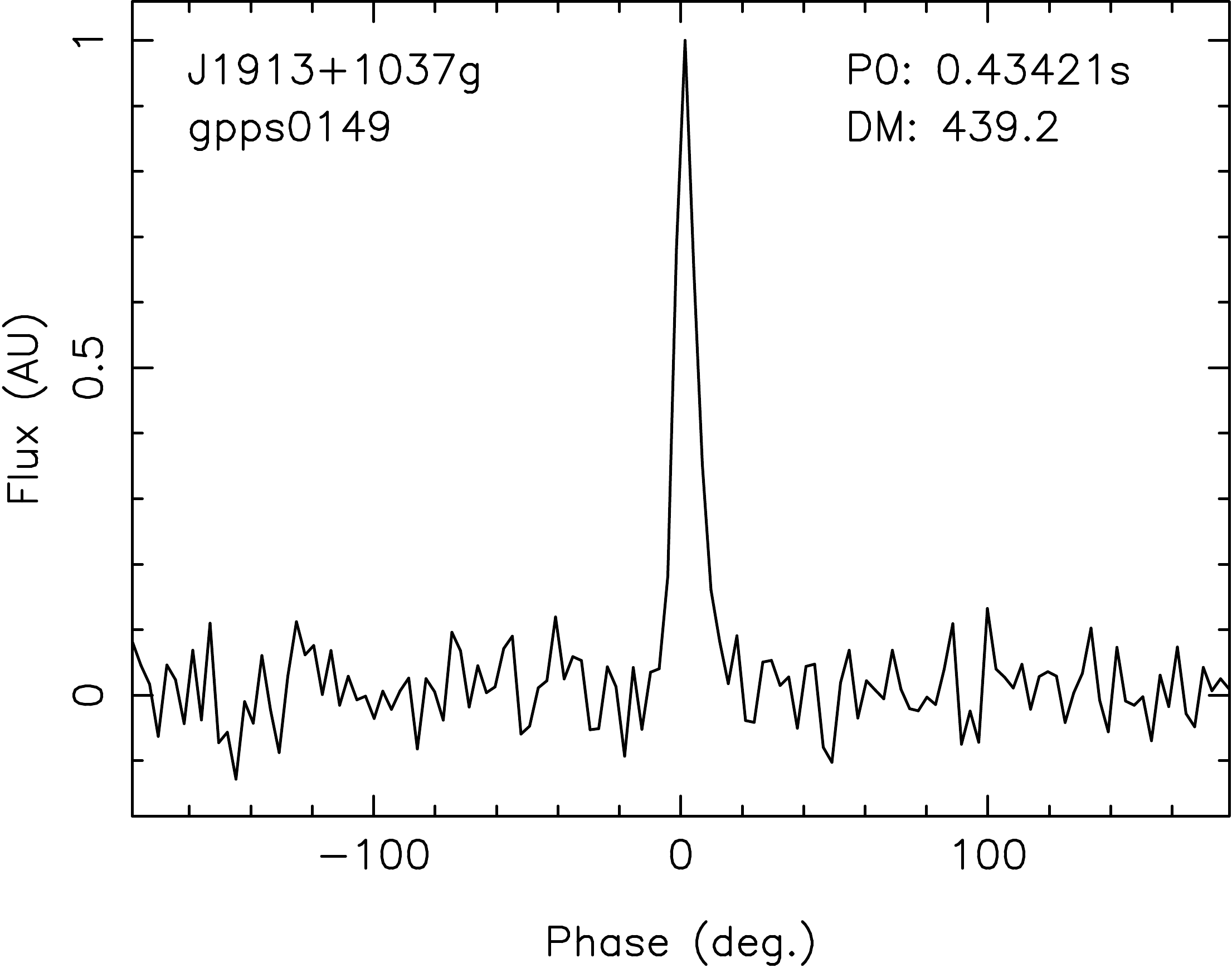}&
\includegraphics[width=39mm]{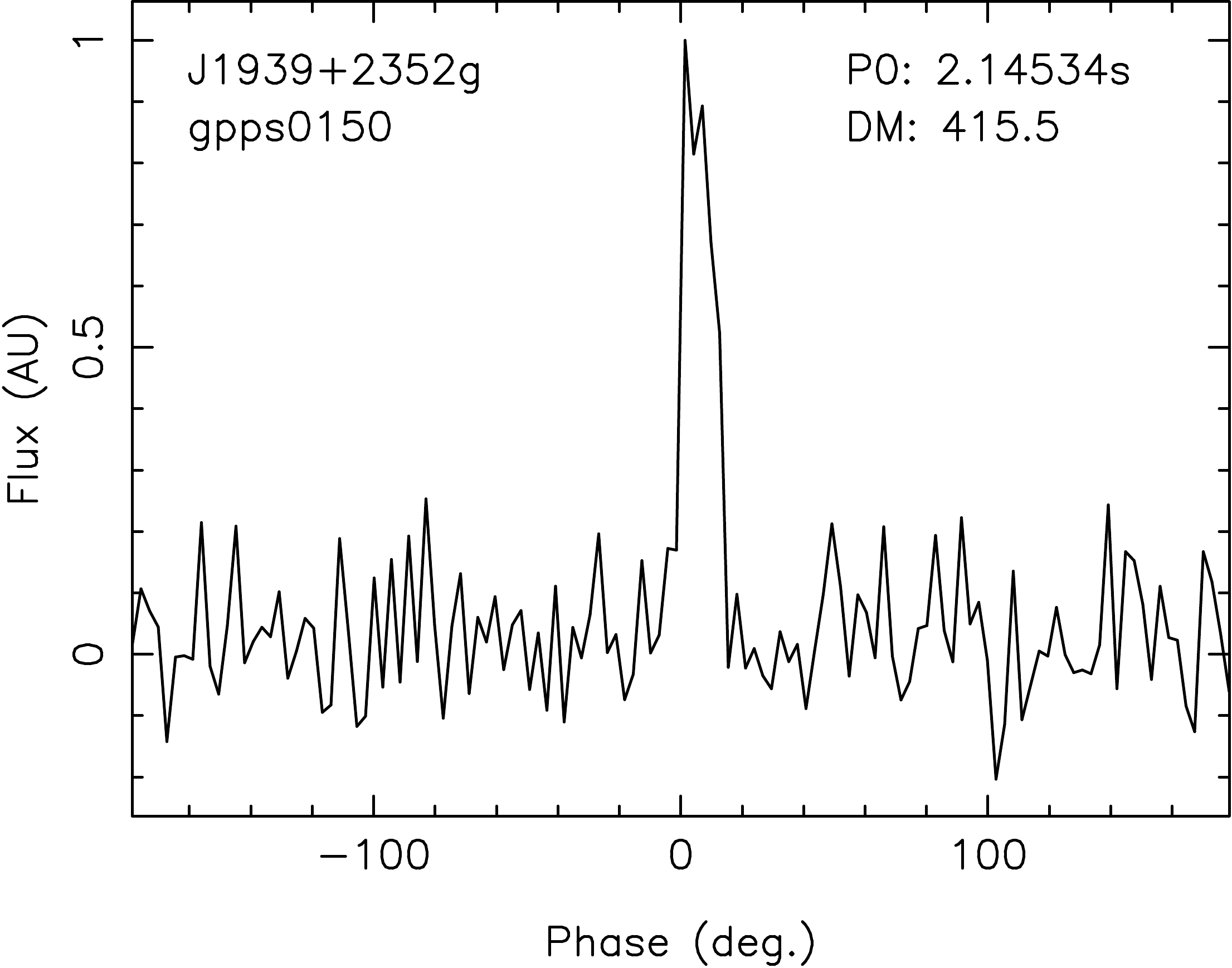}&
\includegraphics[width=39mm]{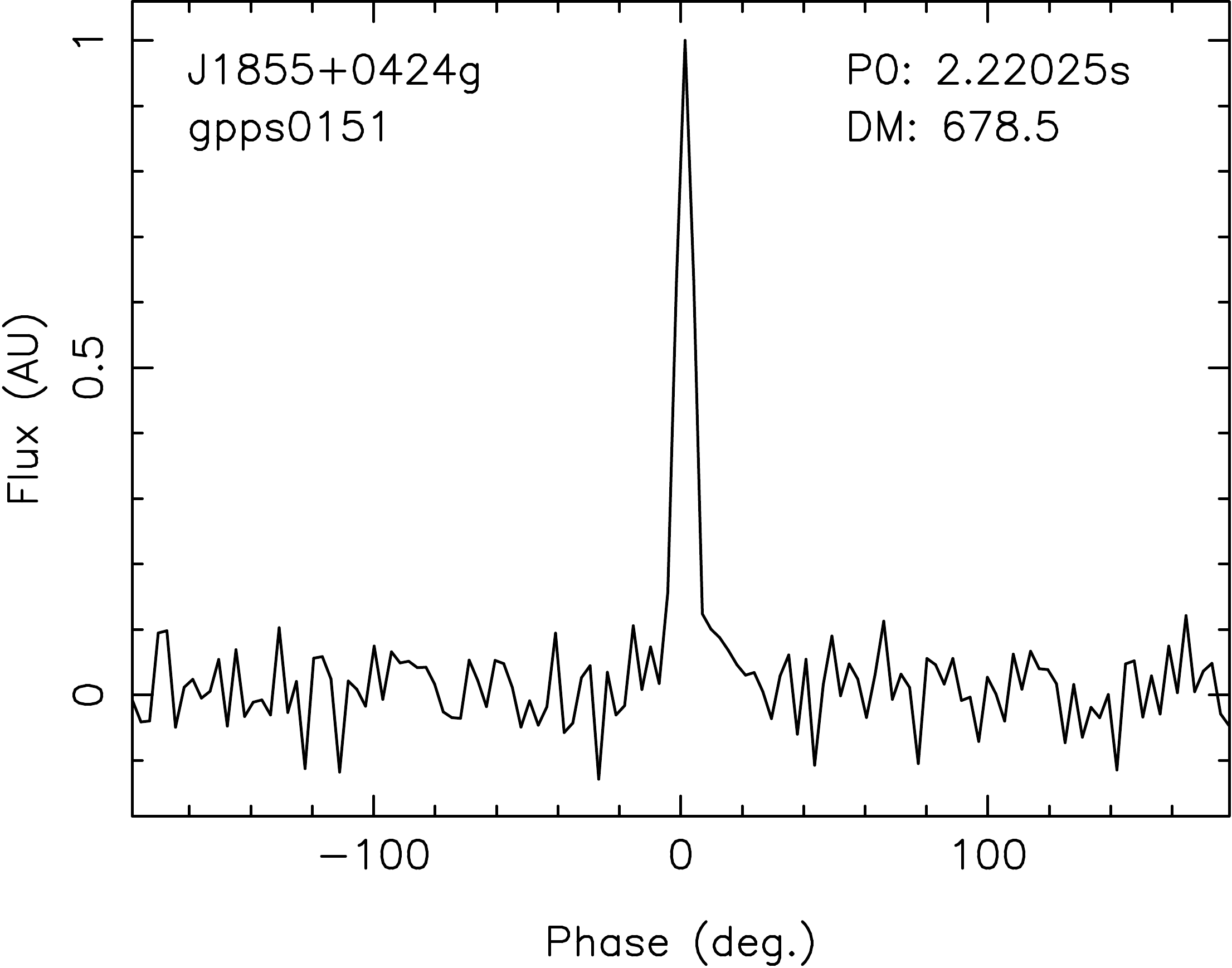}&	
\includegraphics[width=39mm]{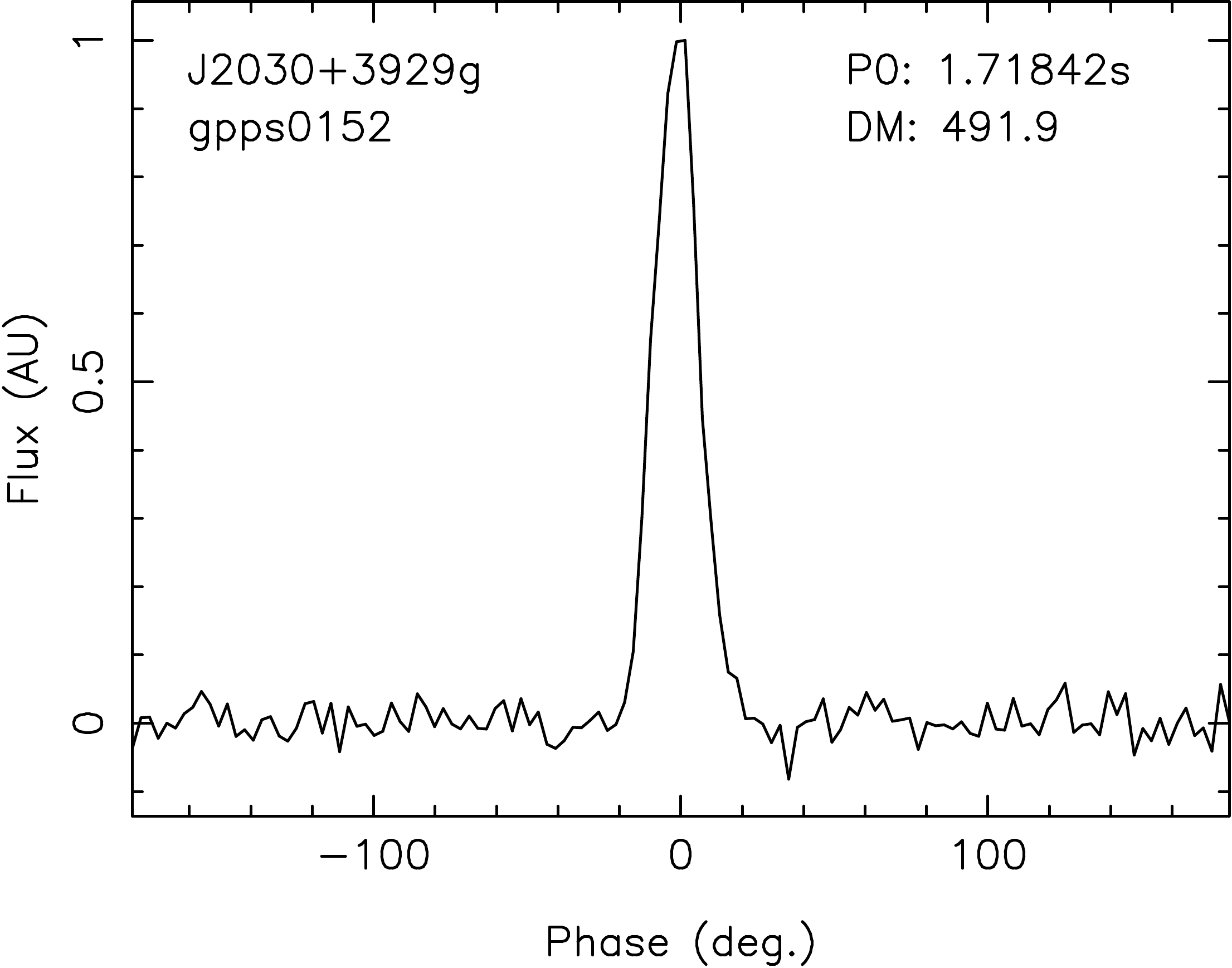}\\[2mm]
\includegraphics[width=39mm]{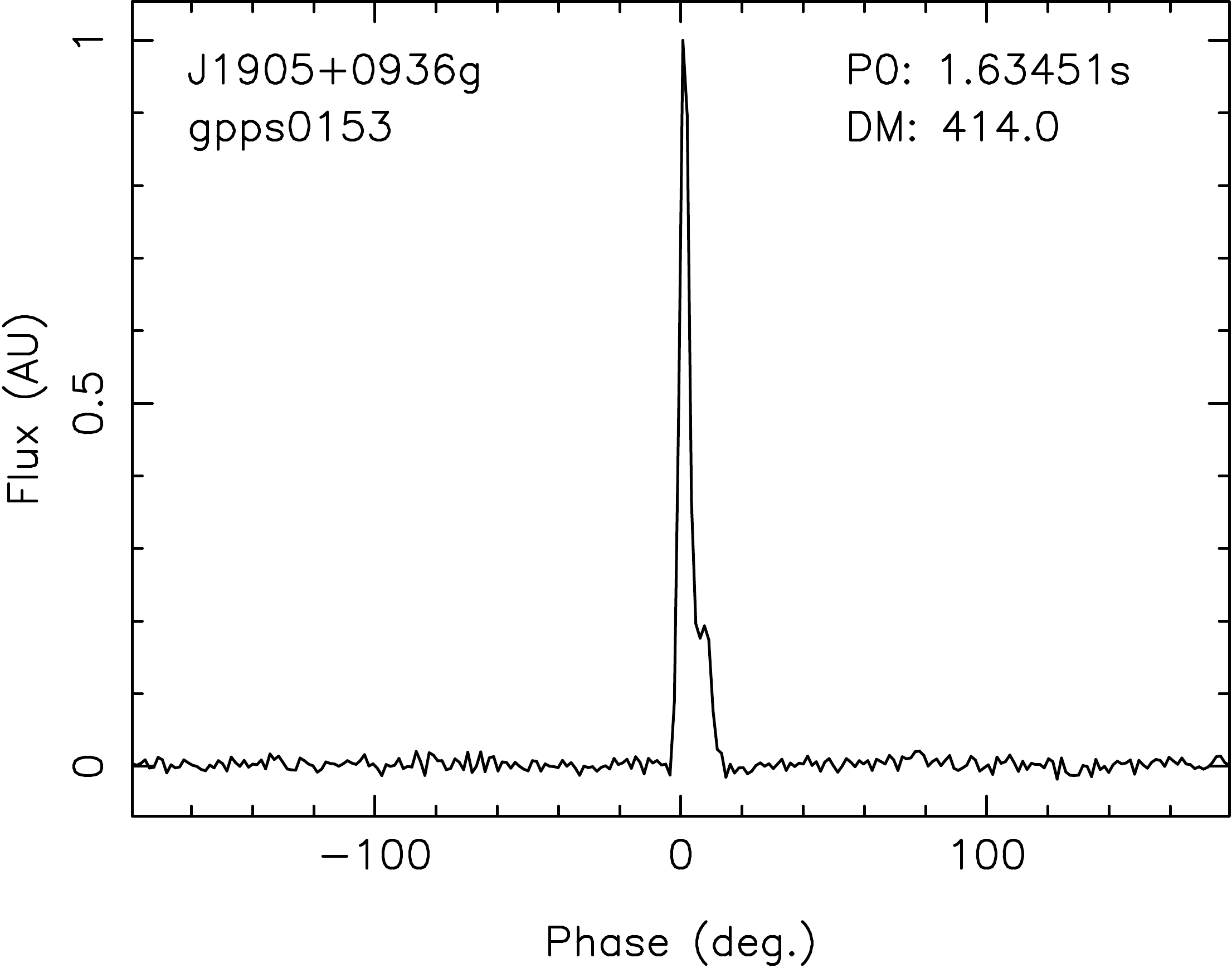}&	
\includegraphics[width=39mm]{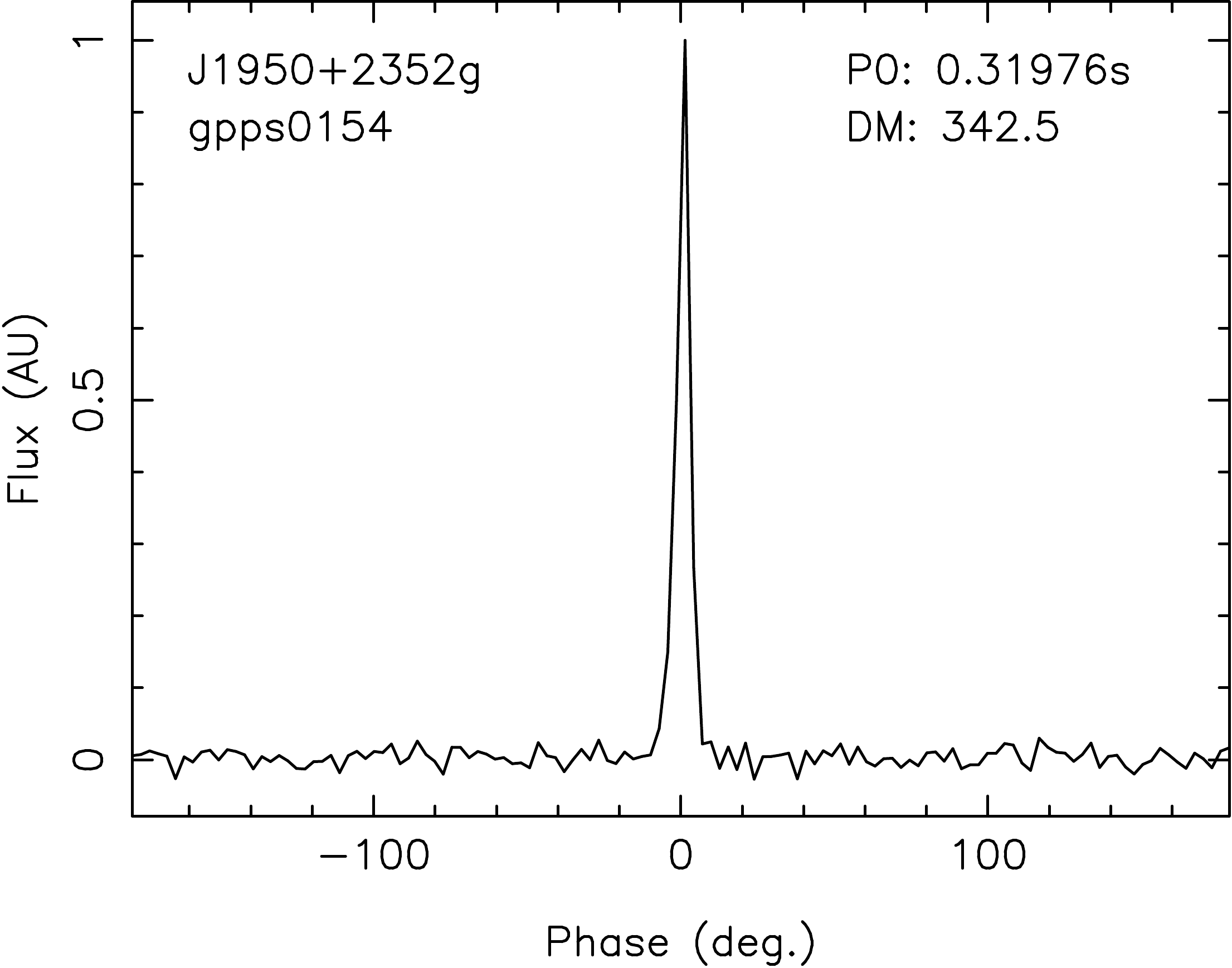}&	
\includegraphics[width=39mm]{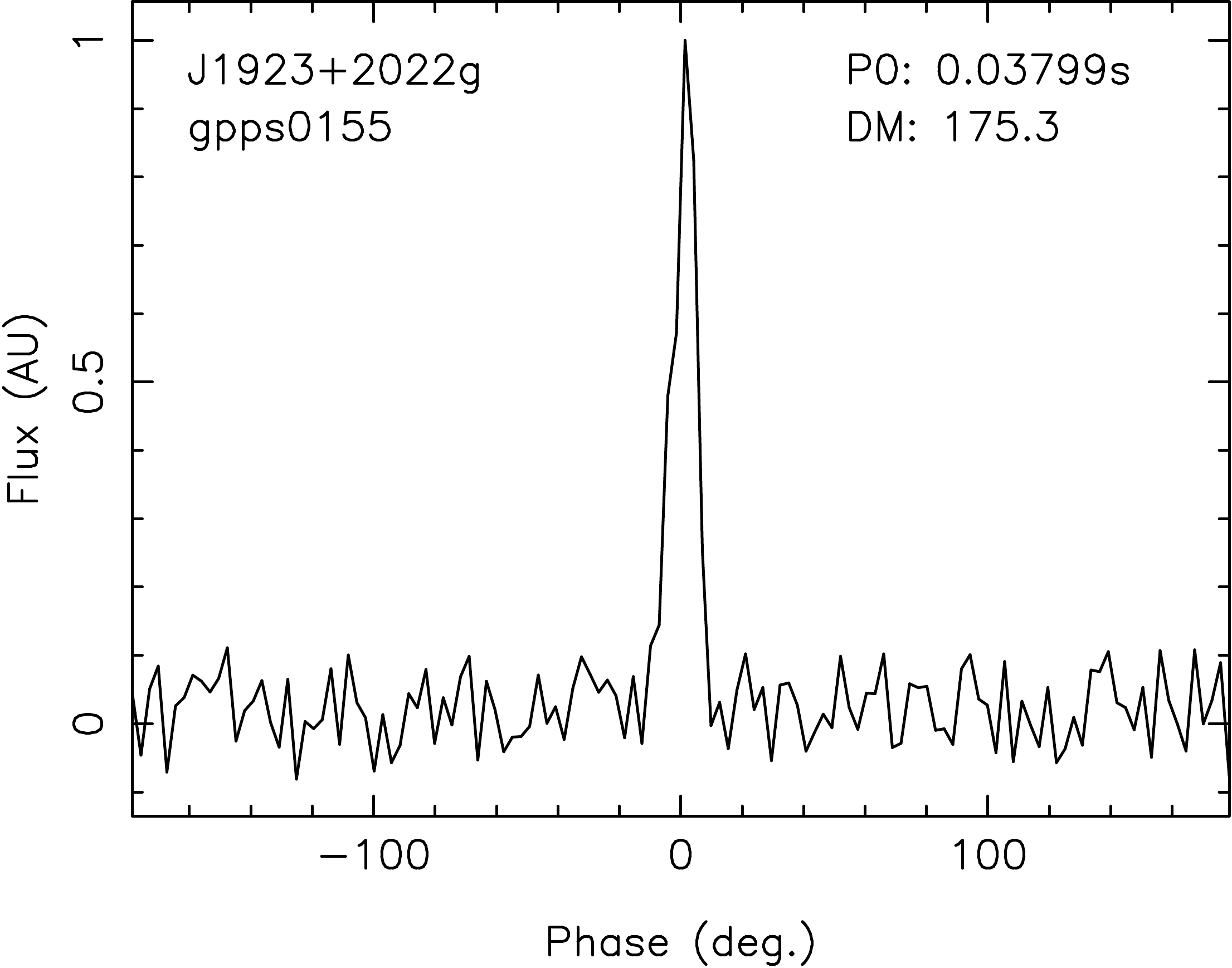}&	
\includegraphics[width=39mm]{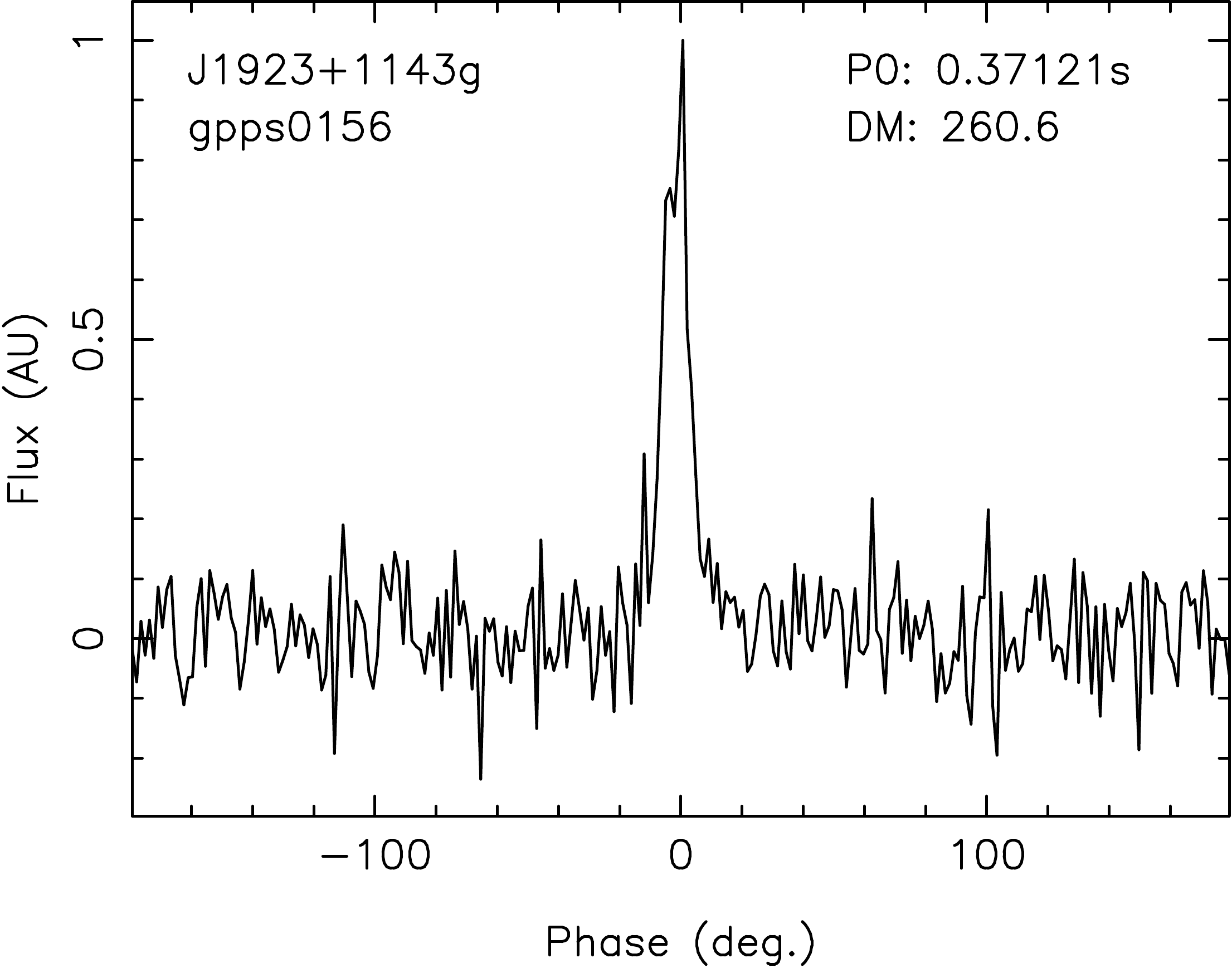}\\[2mm]
\includegraphics[width=39mm]{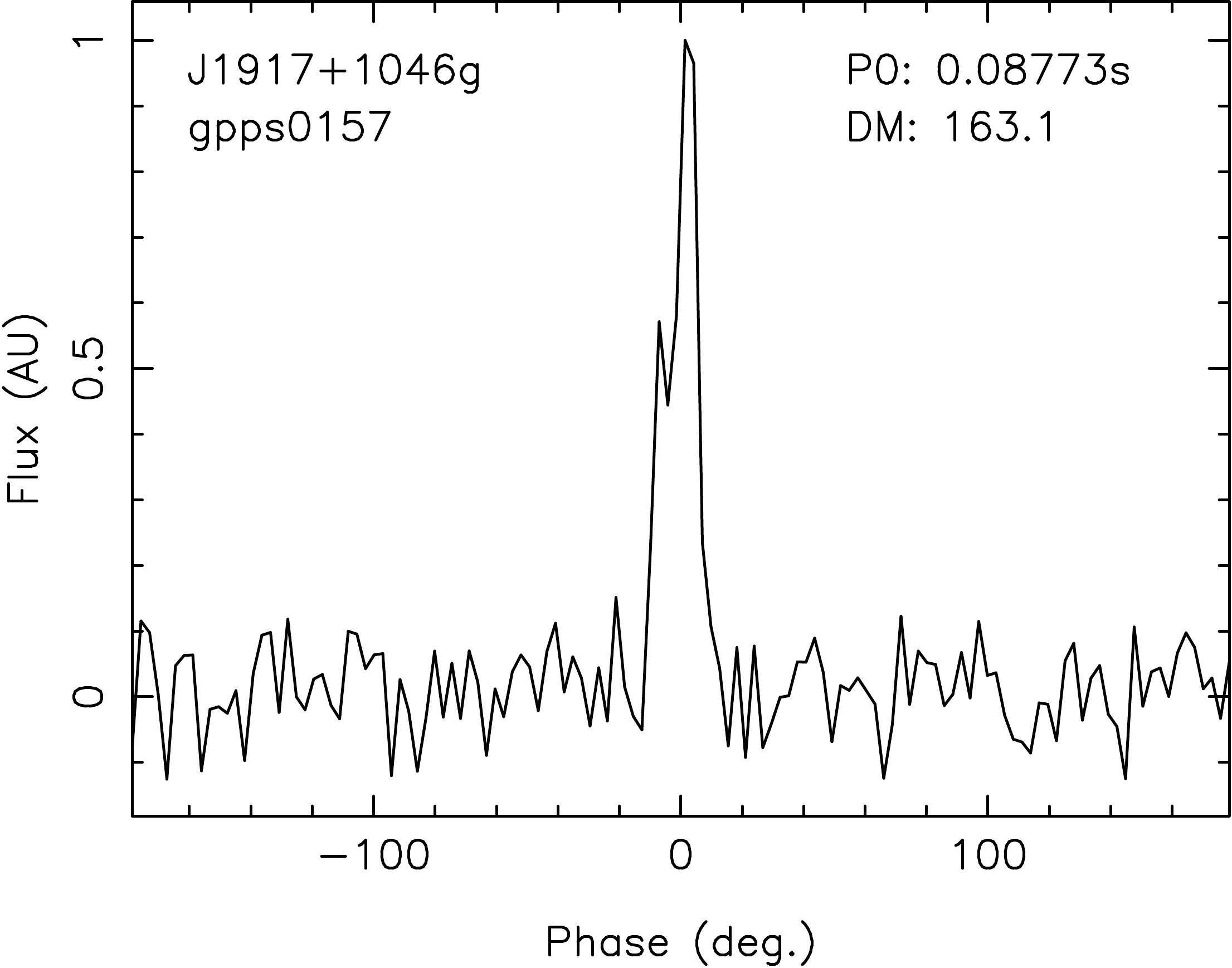}&	
\includegraphics[width=39mm]{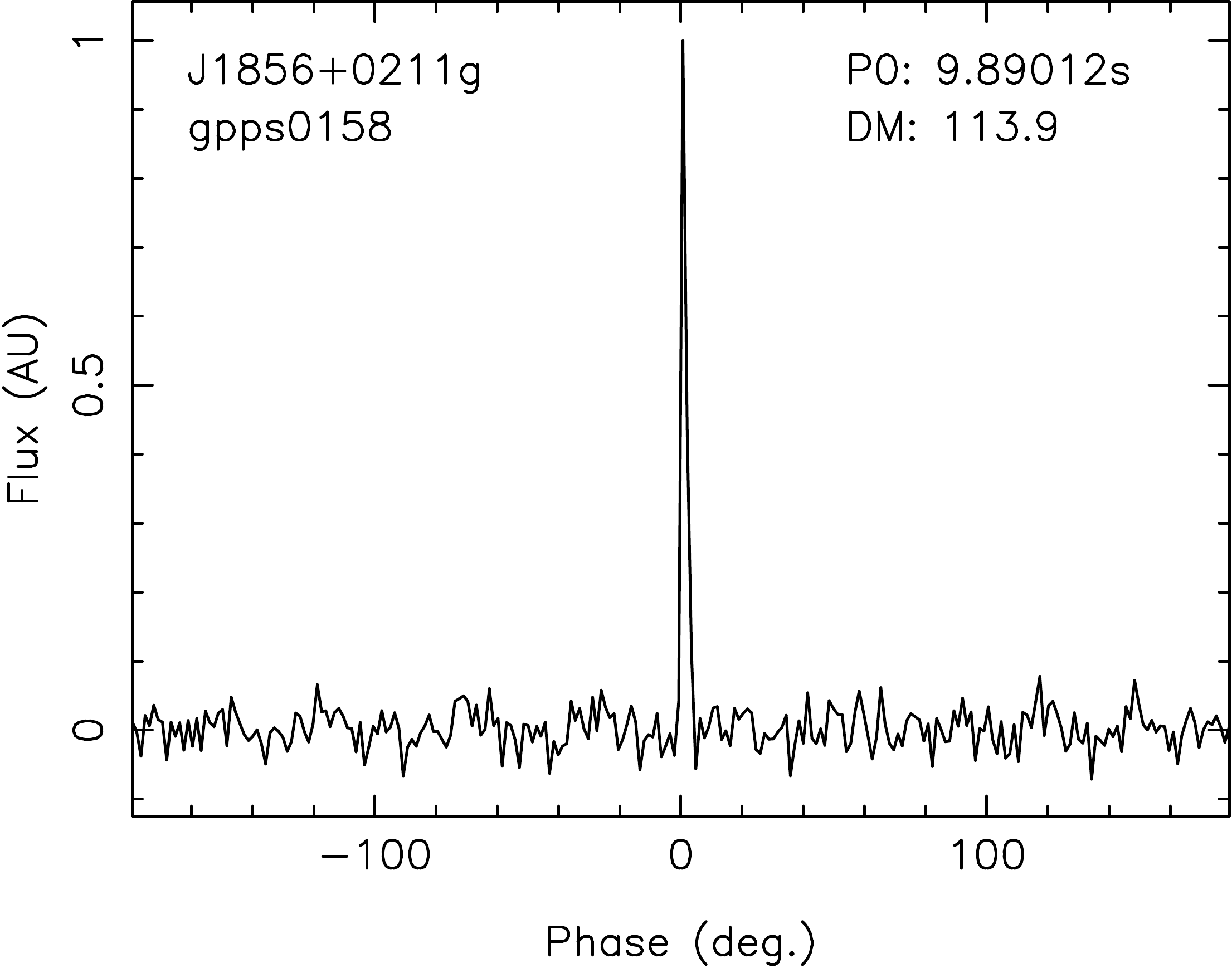}&	
\includegraphics[width=39mm]{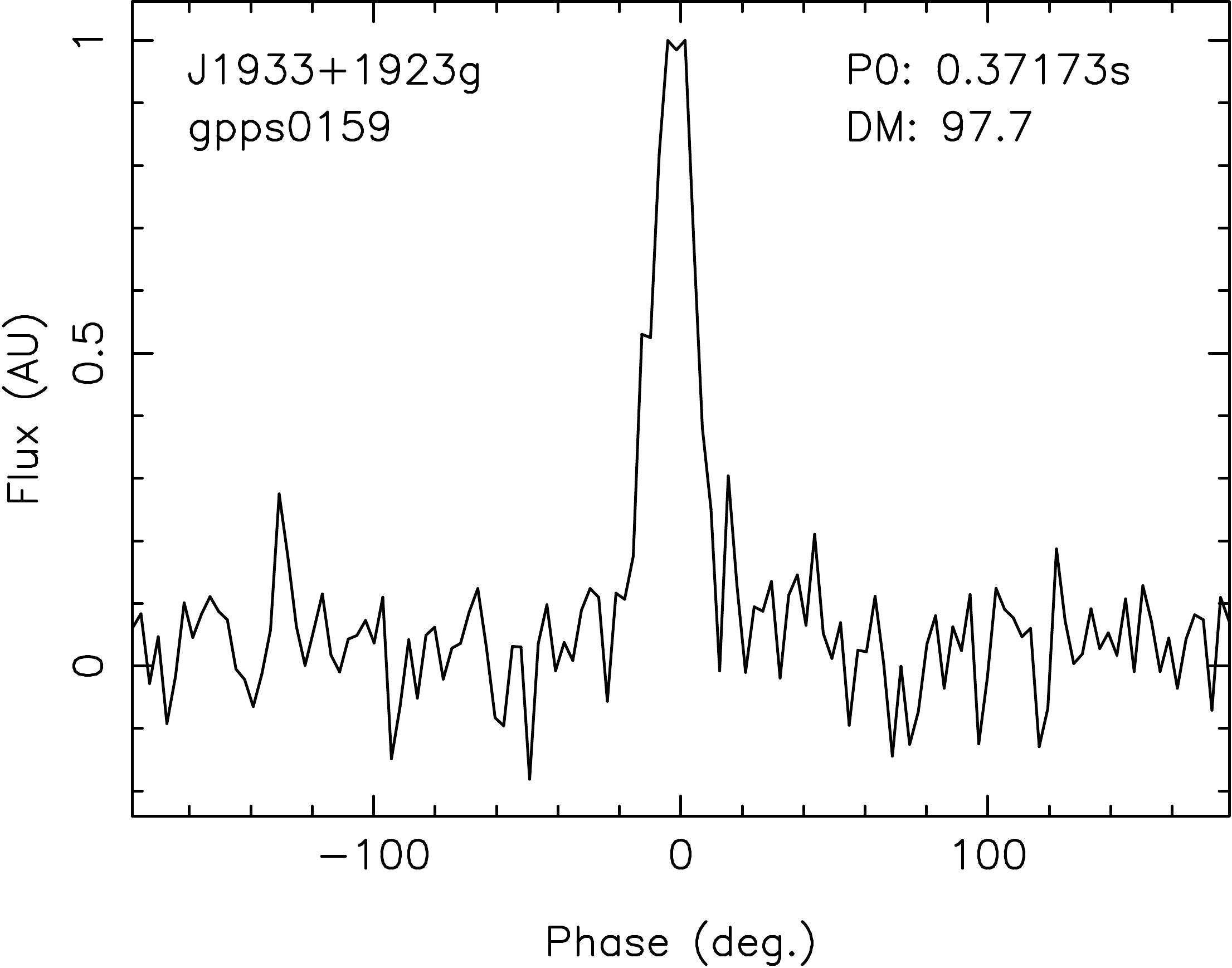}&	
\includegraphics[width=39mm]{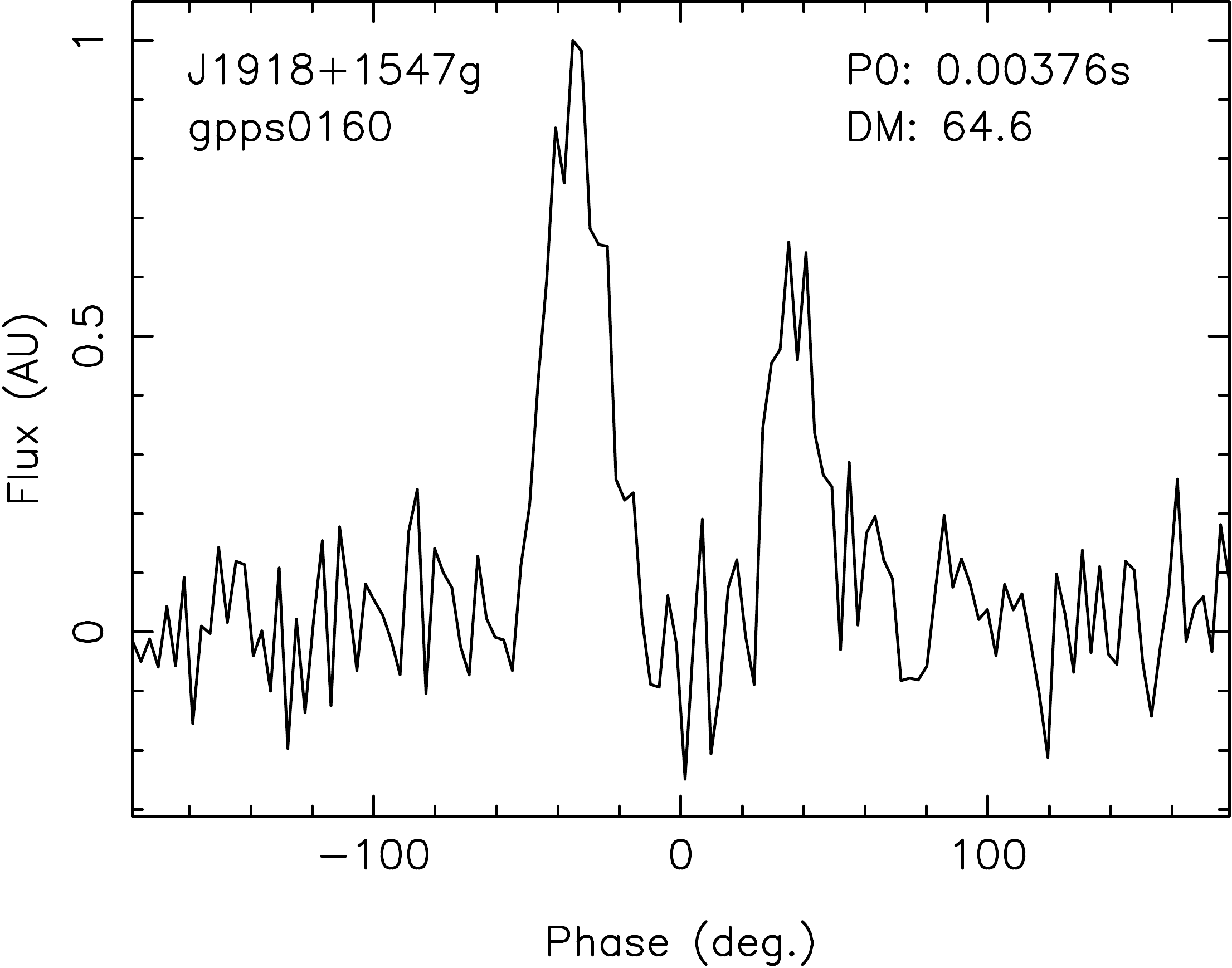}\\[2mm]
\includegraphics[width=39mm]{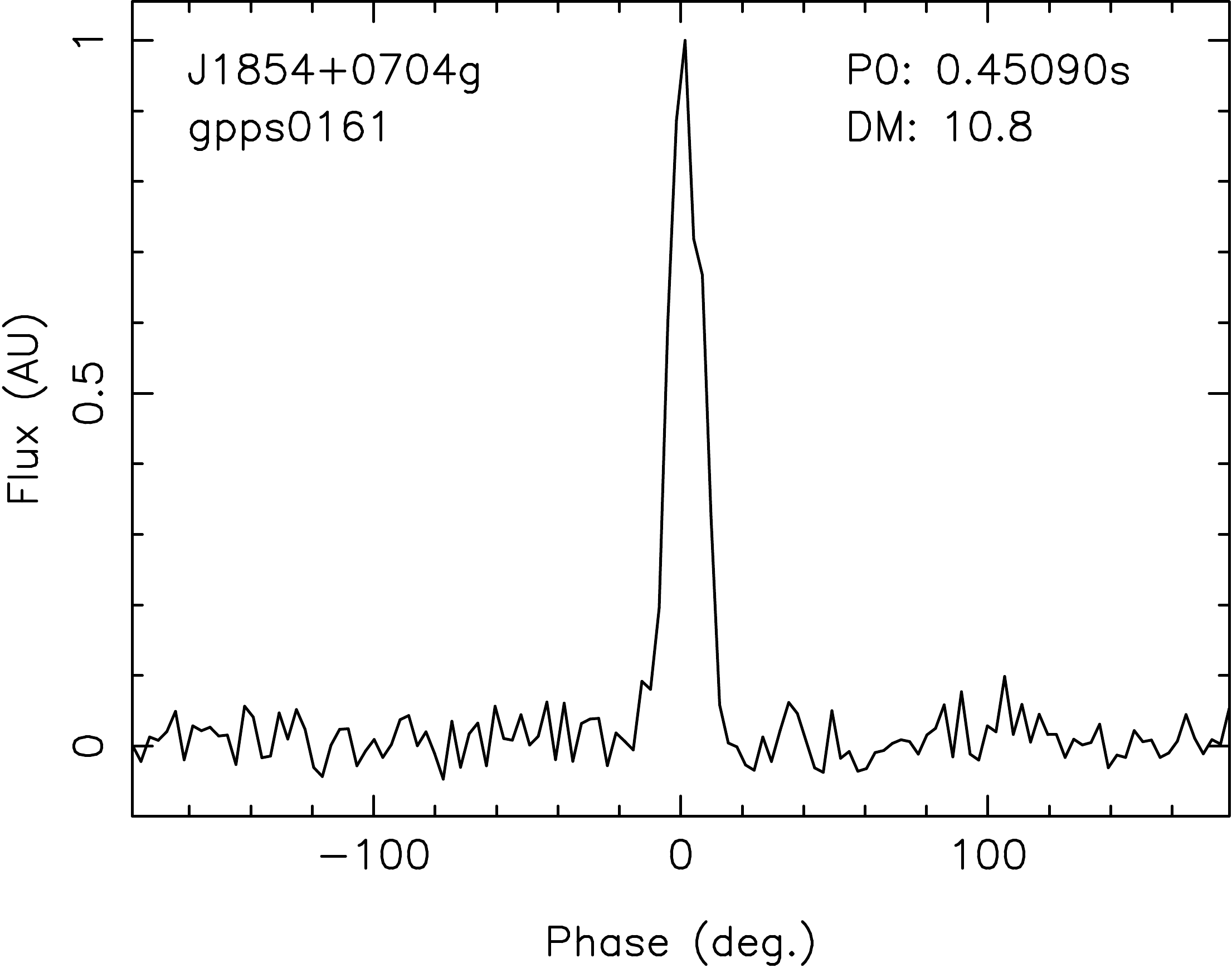}&	
\includegraphics[width=39mm]{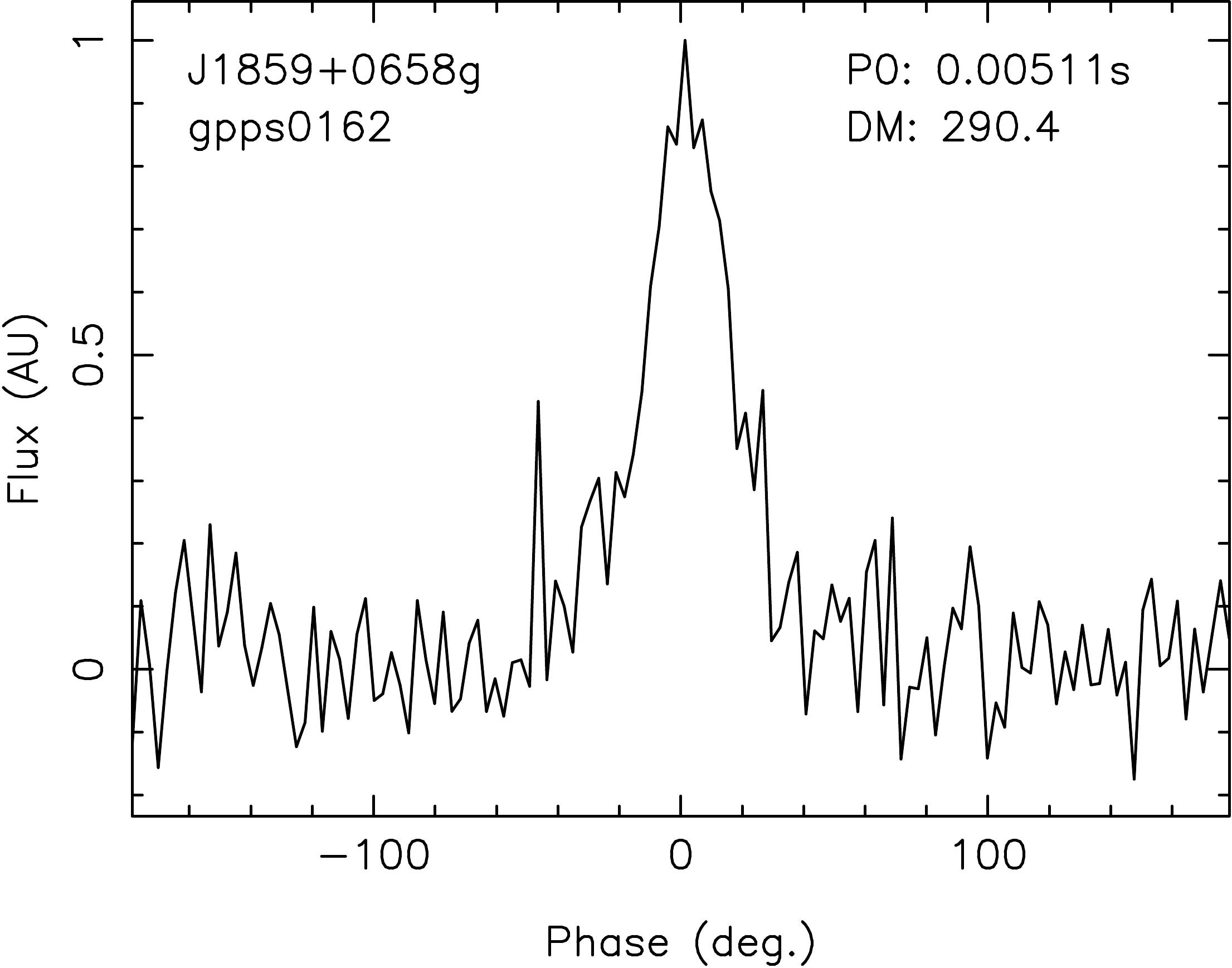}&
\includegraphics[width=39mm]{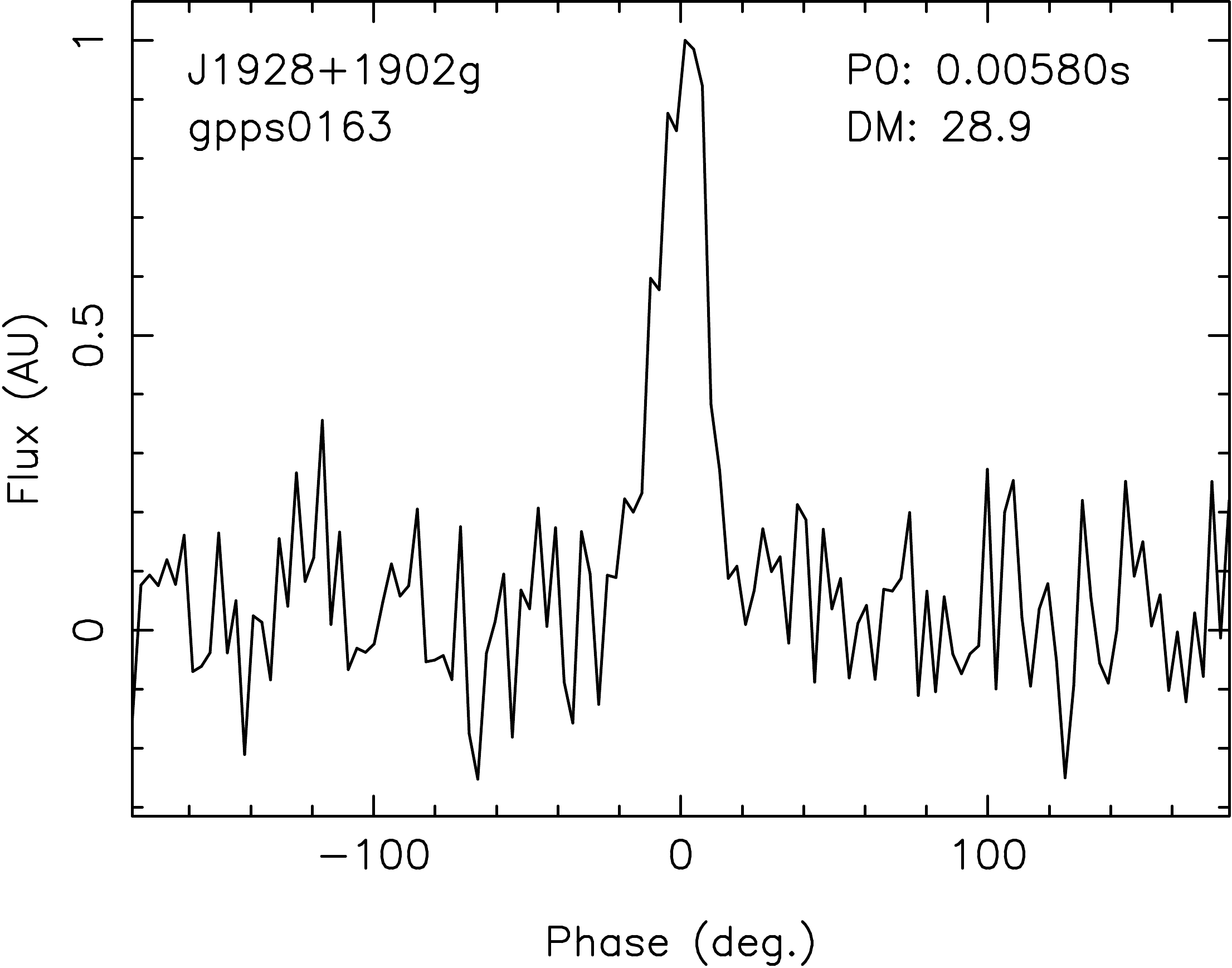}&	
\includegraphics[width=39mm]{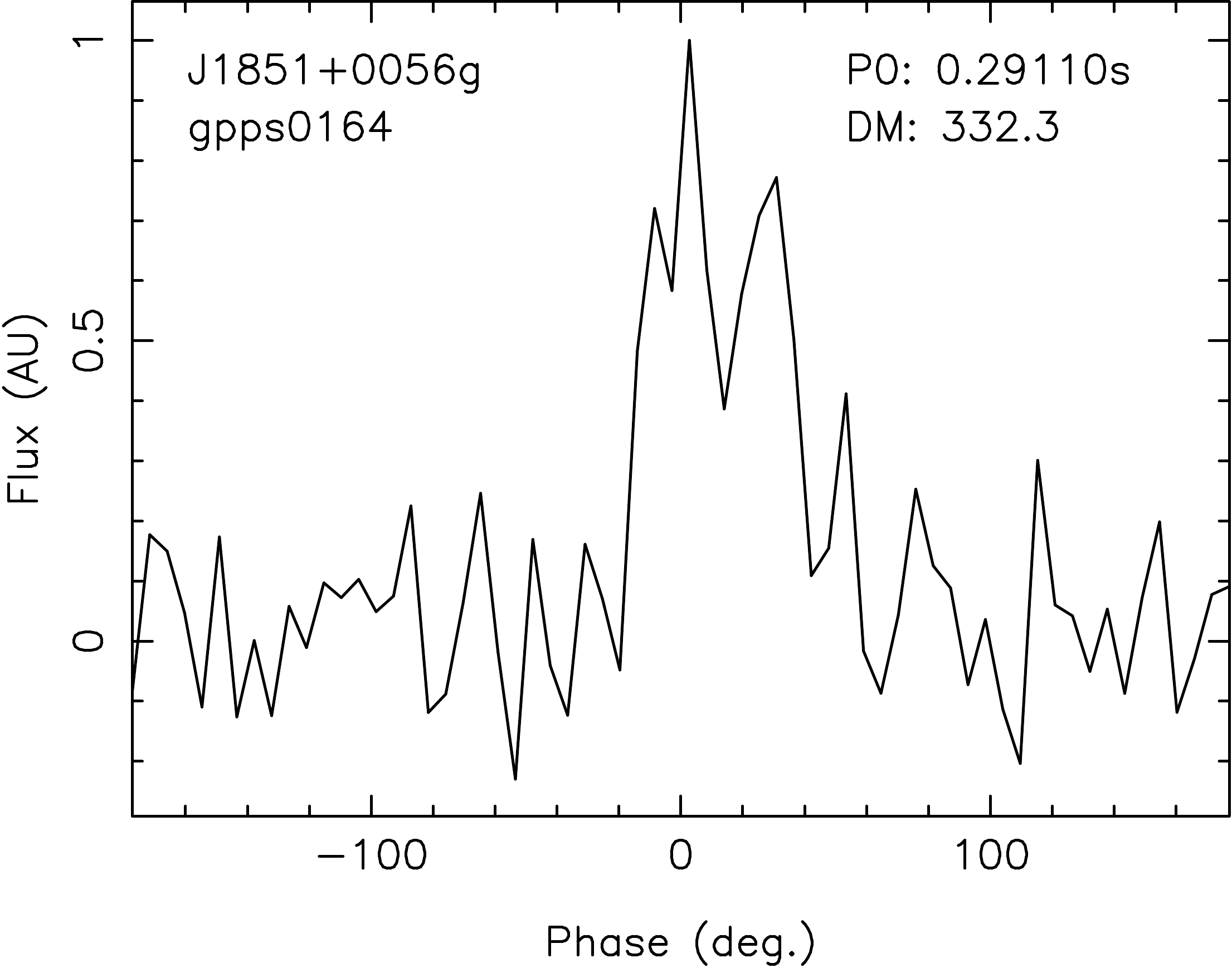}\\[2mm]
\includegraphics[width=39mm]{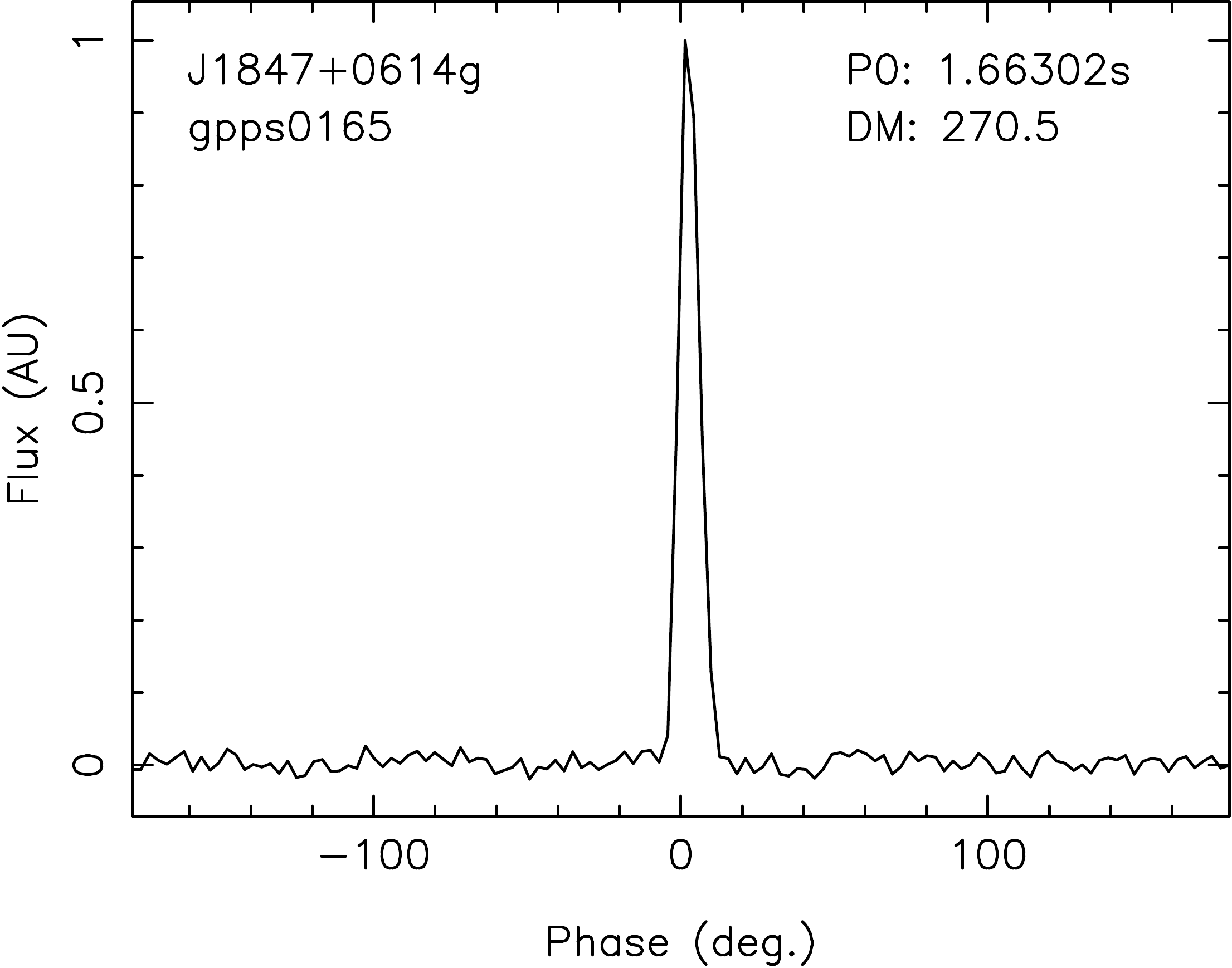}&	
\includegraphics[width=39mm]{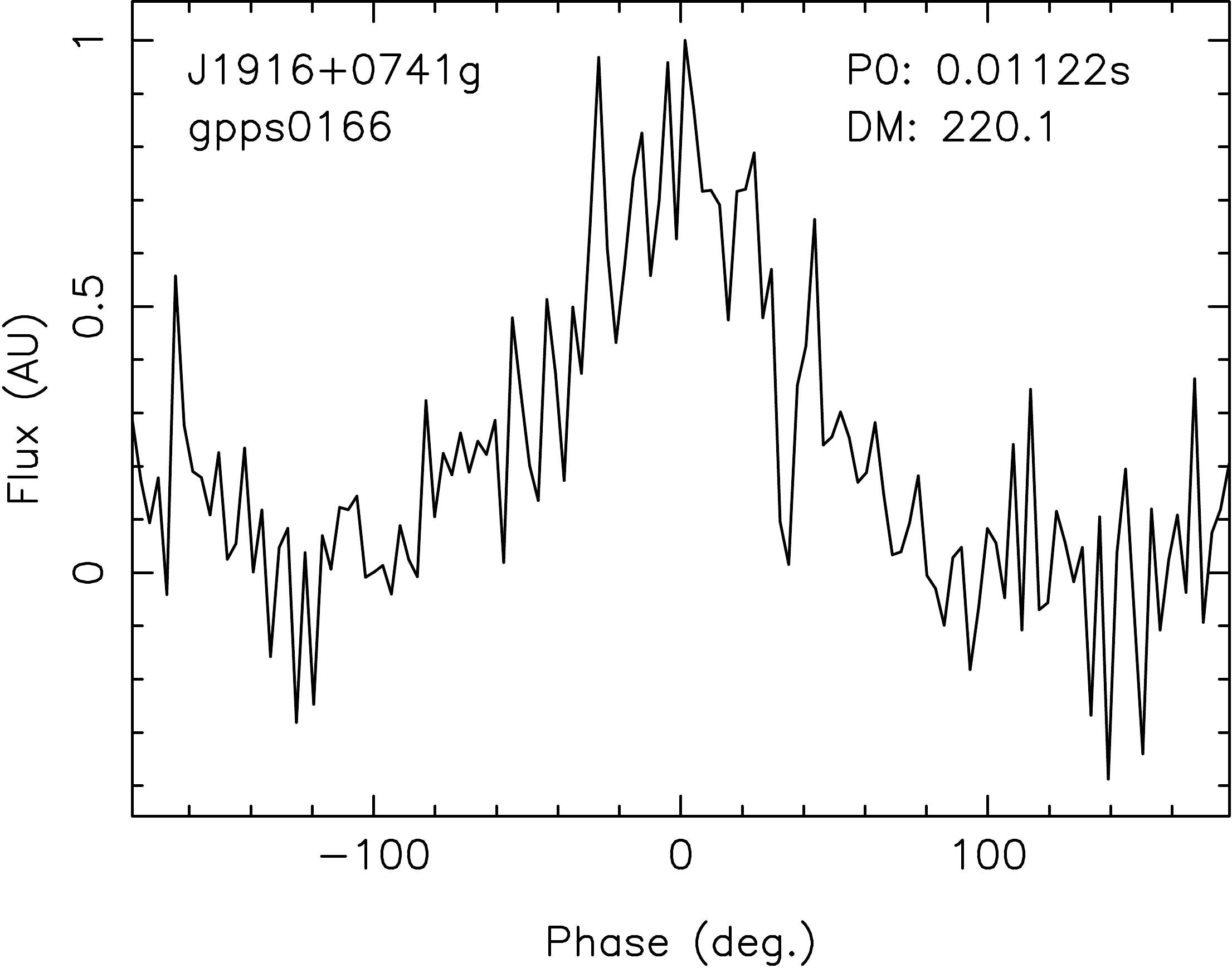}&	
\includegraphics[width=39mm]{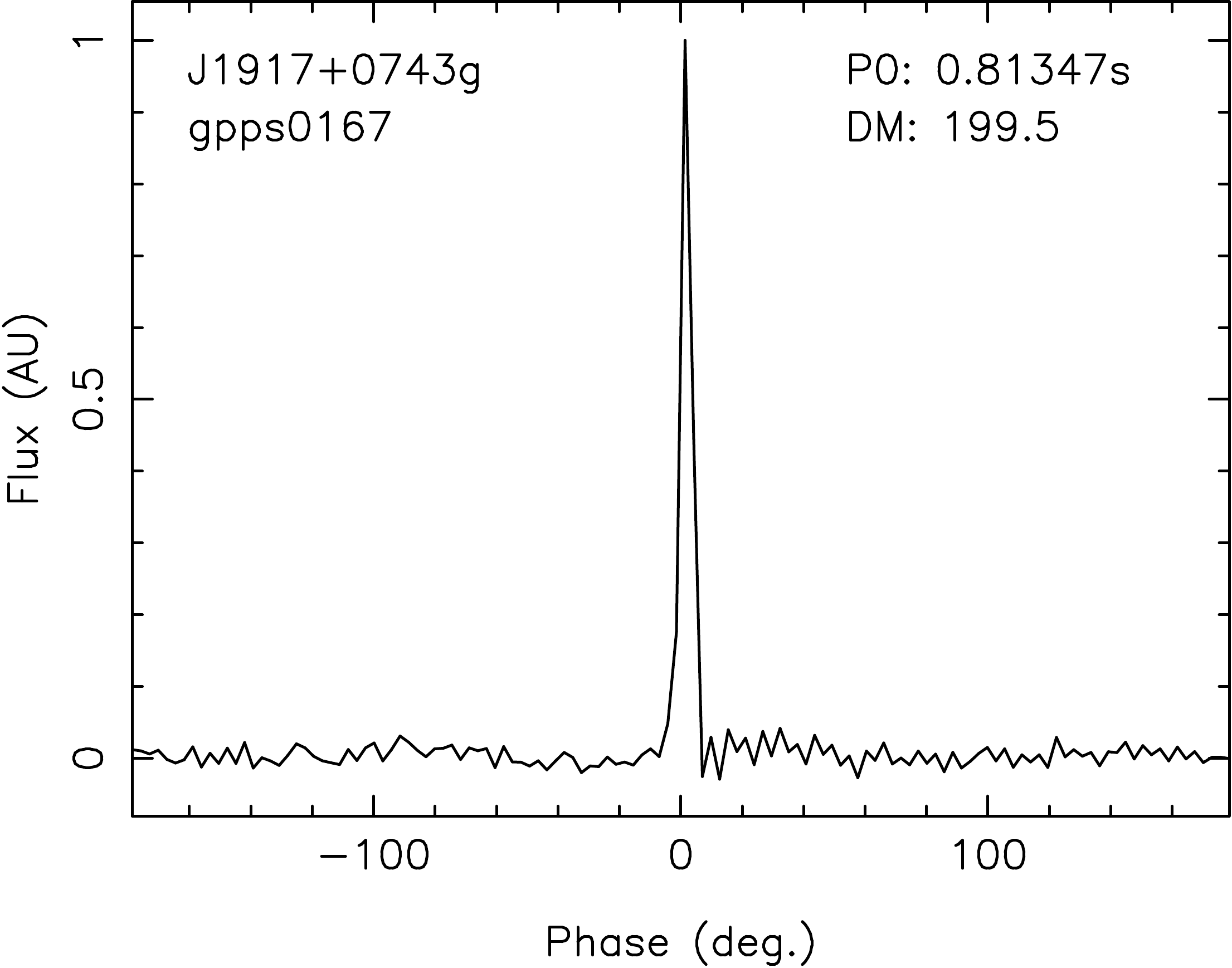}&	
\includegraphics[width=39mm]{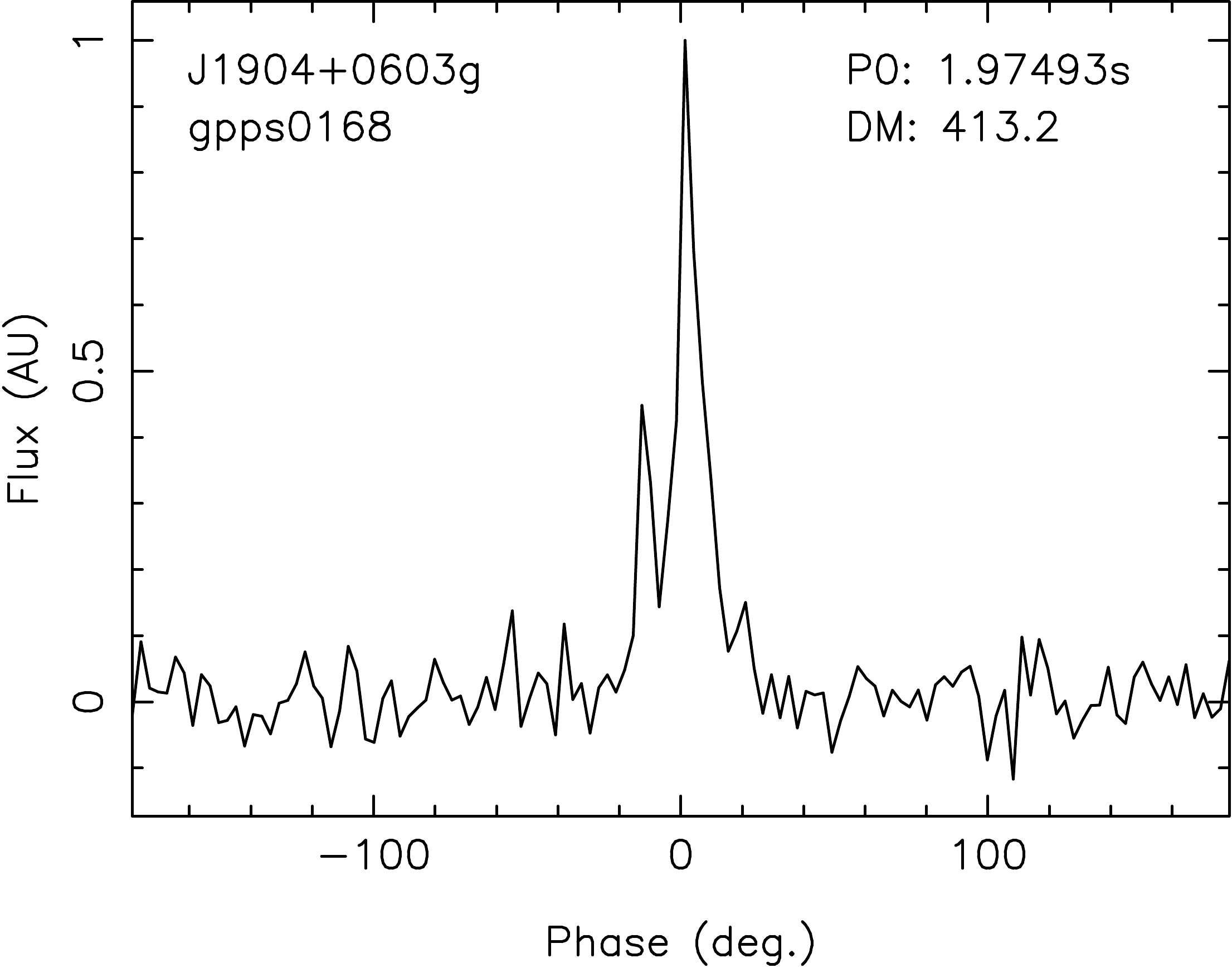}\\
\end{tabular}%

\begin{minipage}{3cm}
\caption[]{
-- {\it Continued}.}\end{minipage}
\addtocounter{figure}{-1}
\end{figure*}%
\begin{figure*}
\centering
\begin{tabular}{rrrr}
\includegraphics[width=39mm]{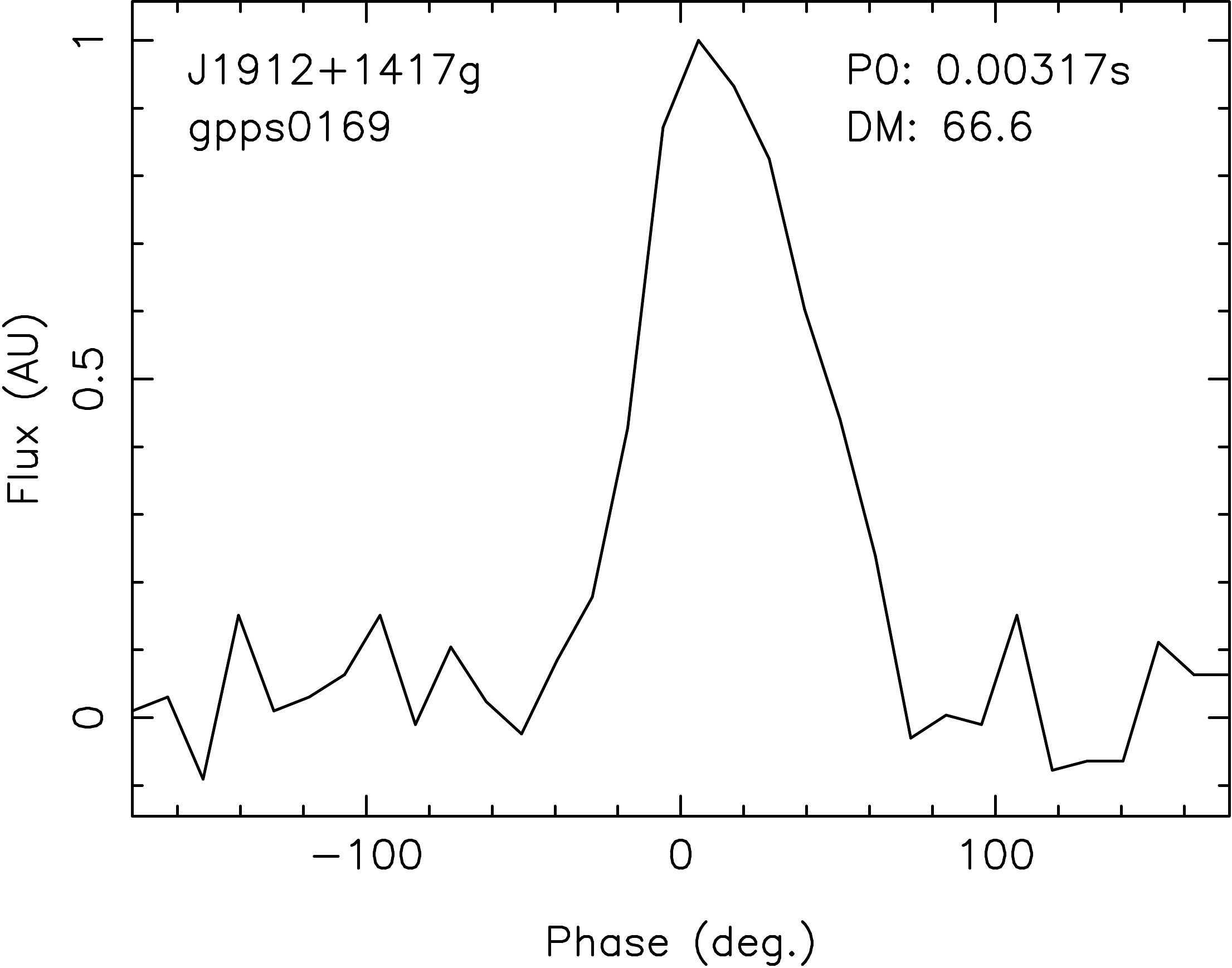}&	
\includegraphics[width=39mm]{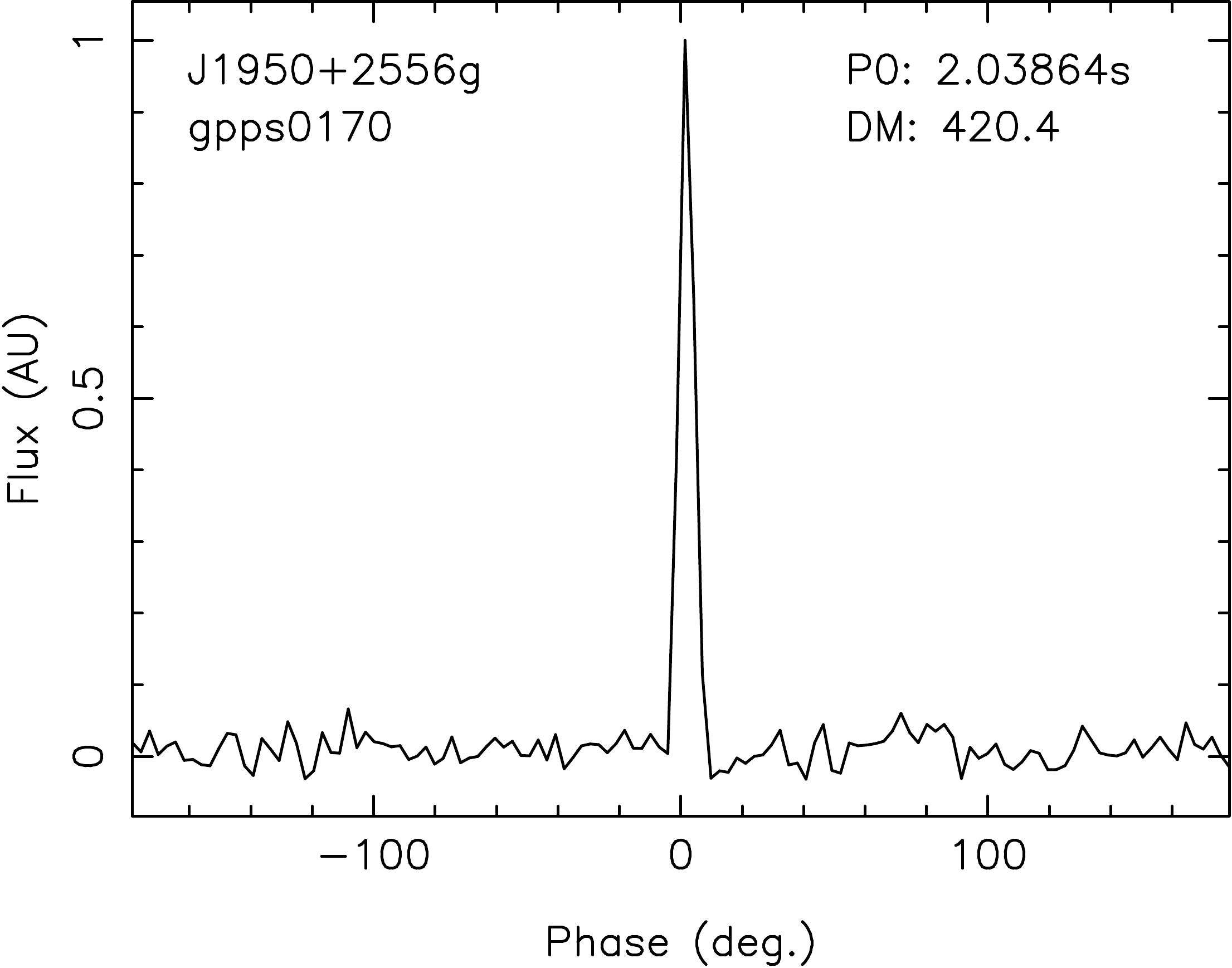}&
\includegraphics[width=39mm]{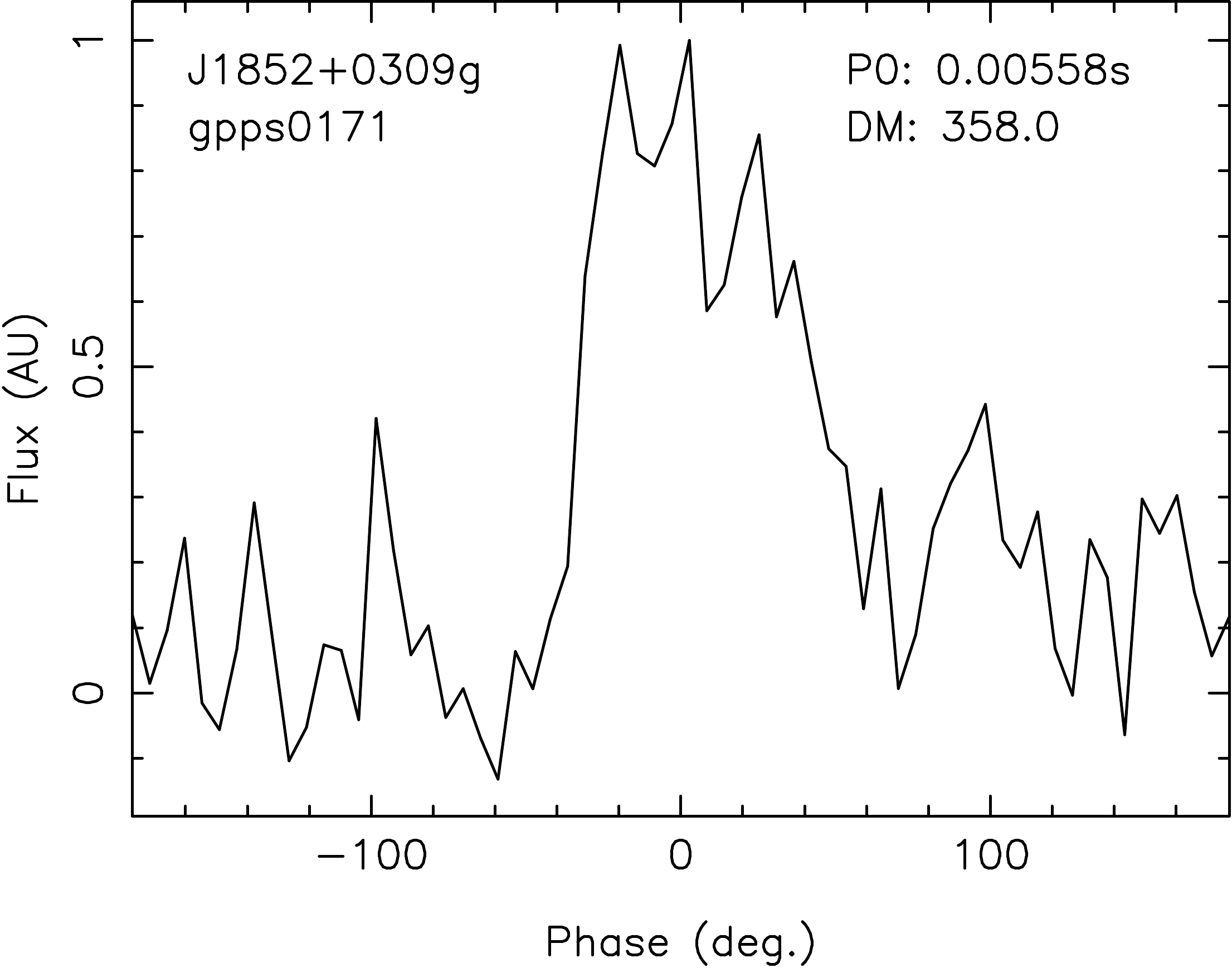}&
\includegraphics[width=39mm]{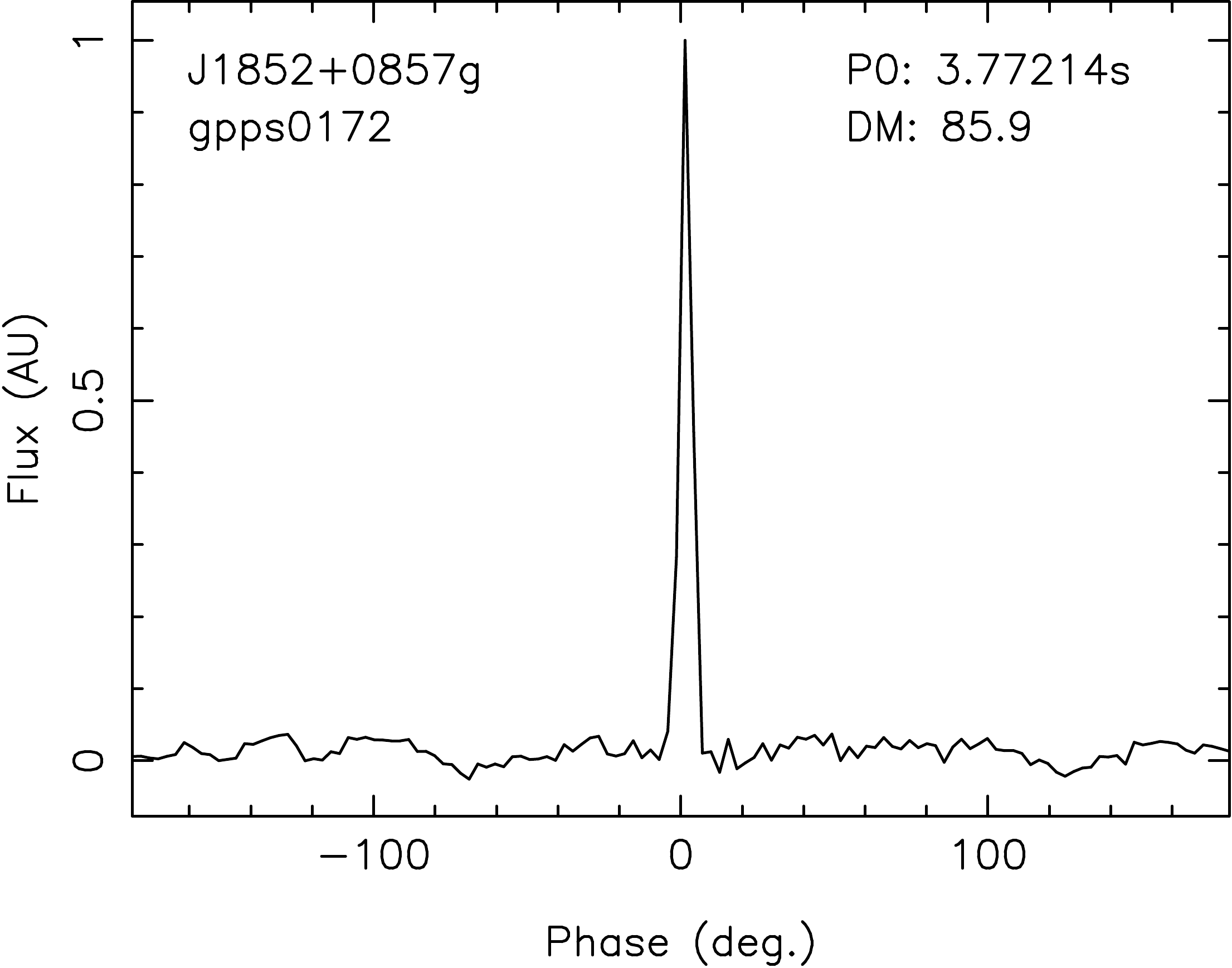}\\[2mm]
\includegraphics[width=39mm]{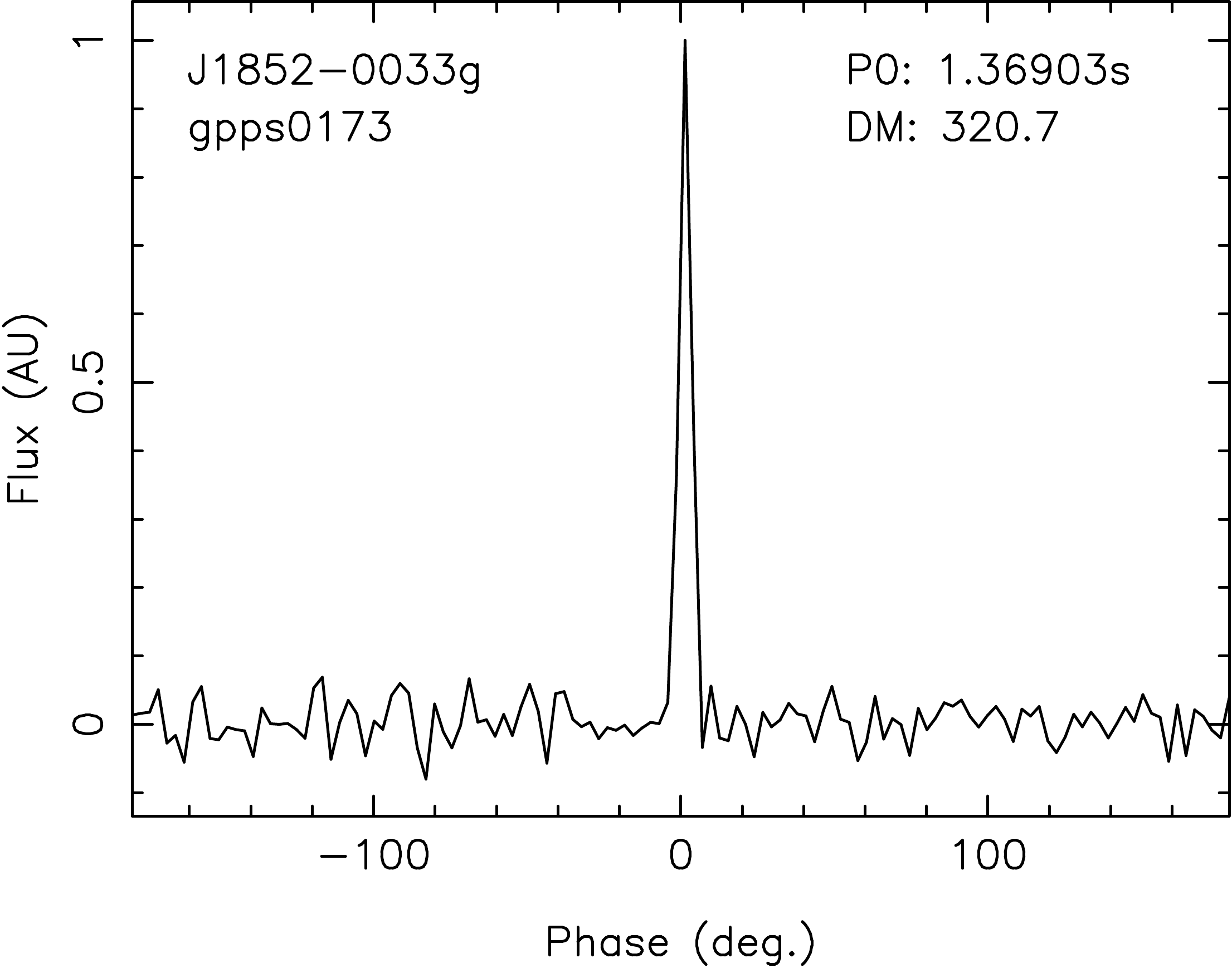}&
\includegraphics[width=39mm]{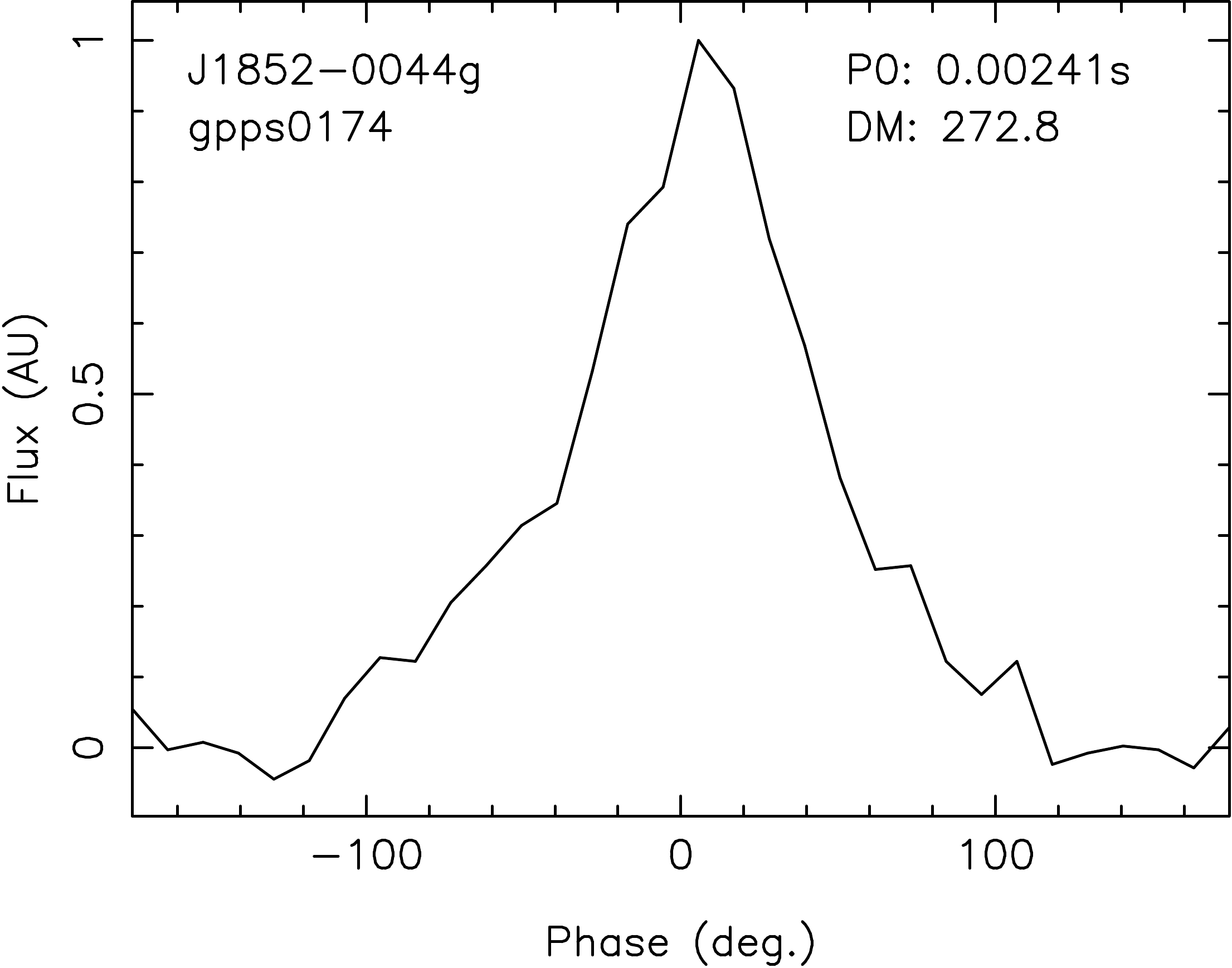}&
\includegraphics[width=39mm]{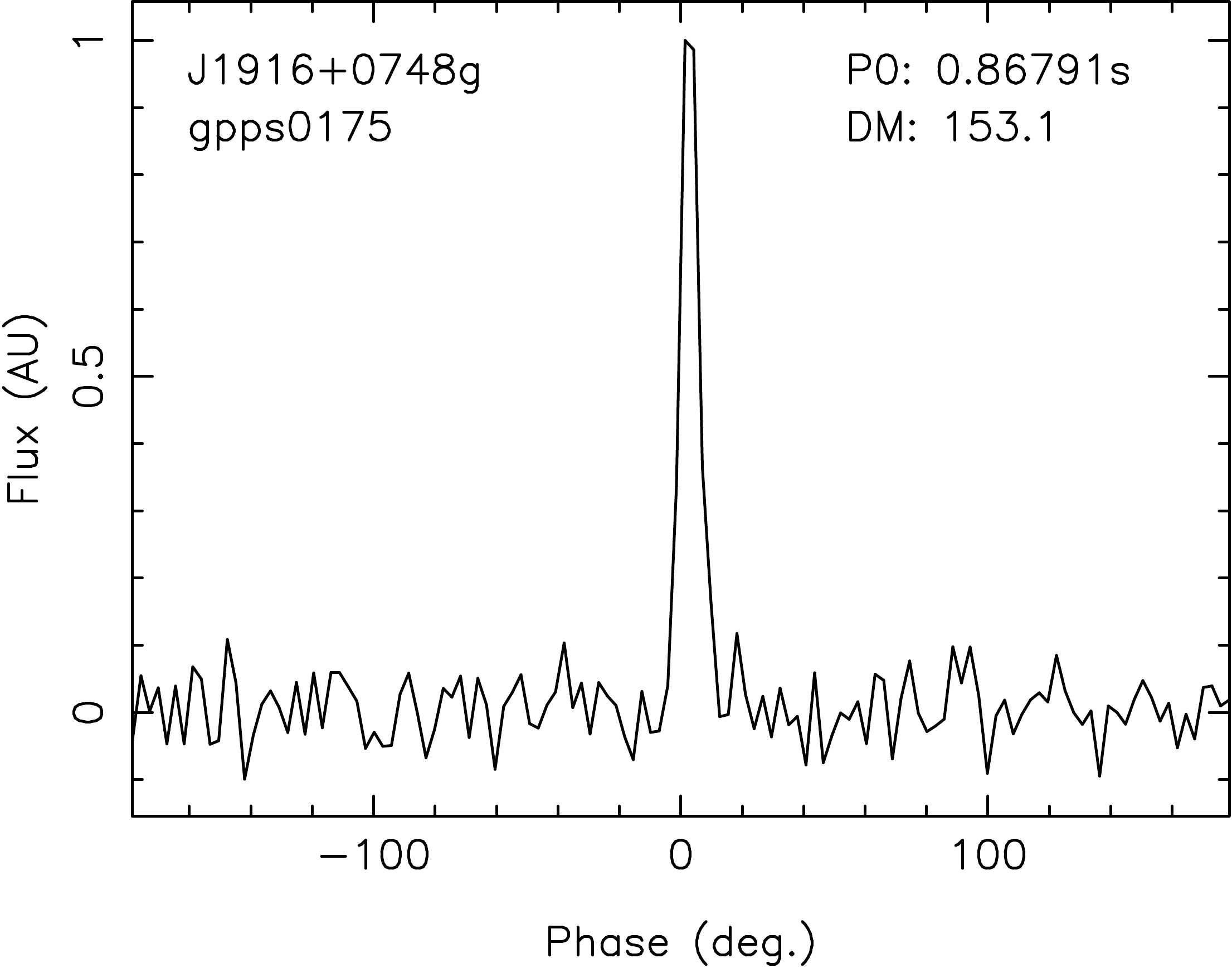}&
\includegraphics[width=39mm]{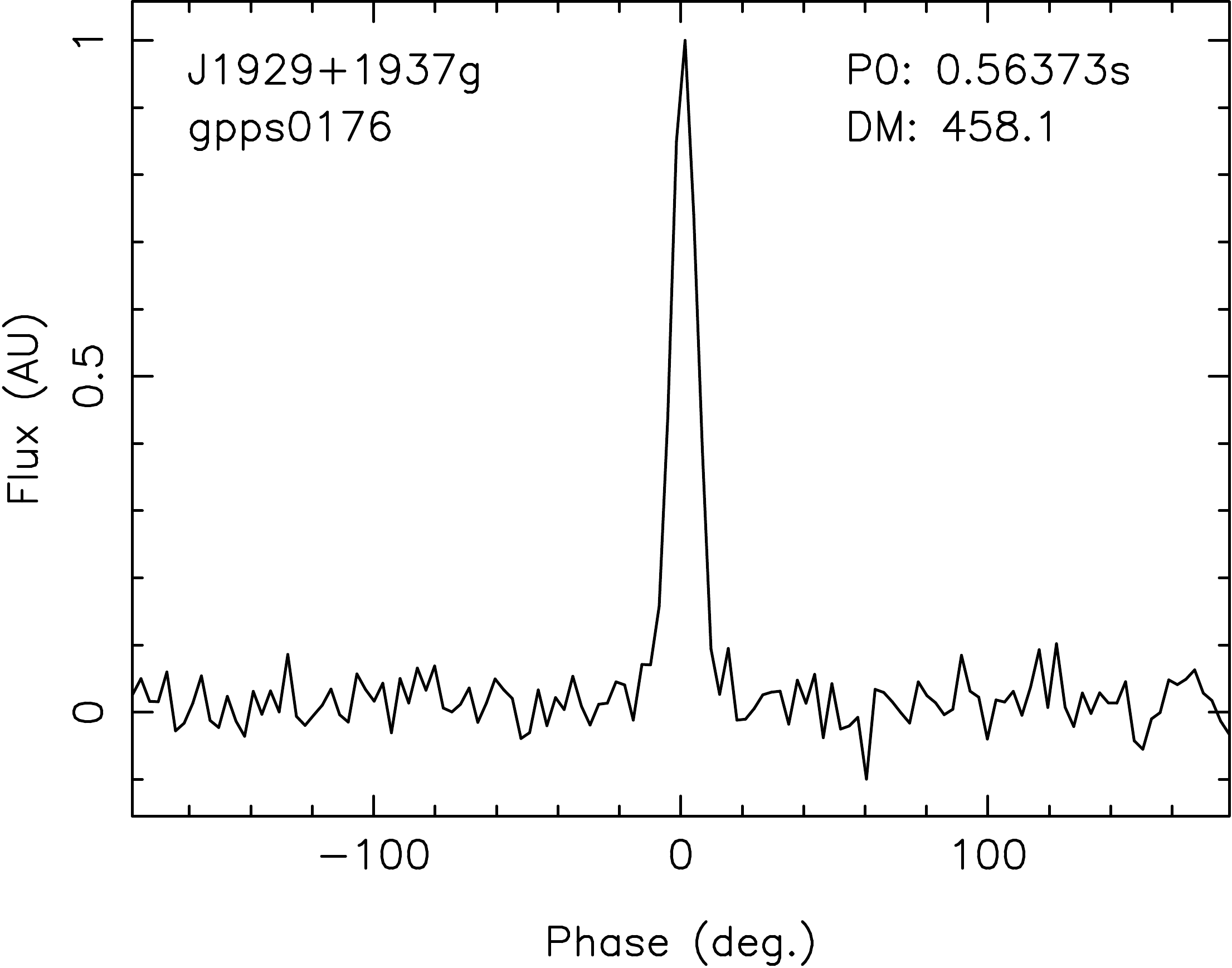}\\[2mm]
\includegraphics[width=39mm]{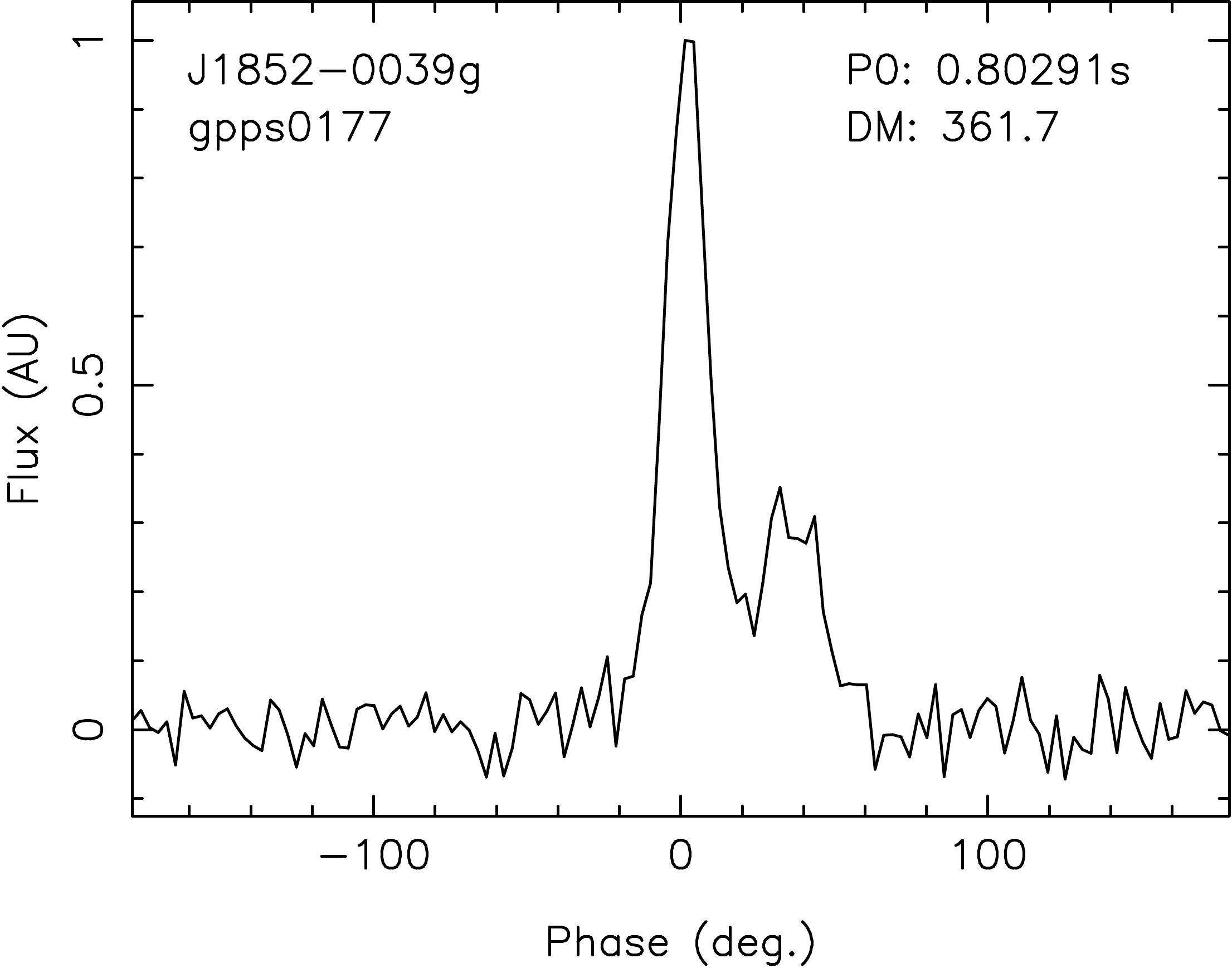}&
\includegraphics[width=39mm]{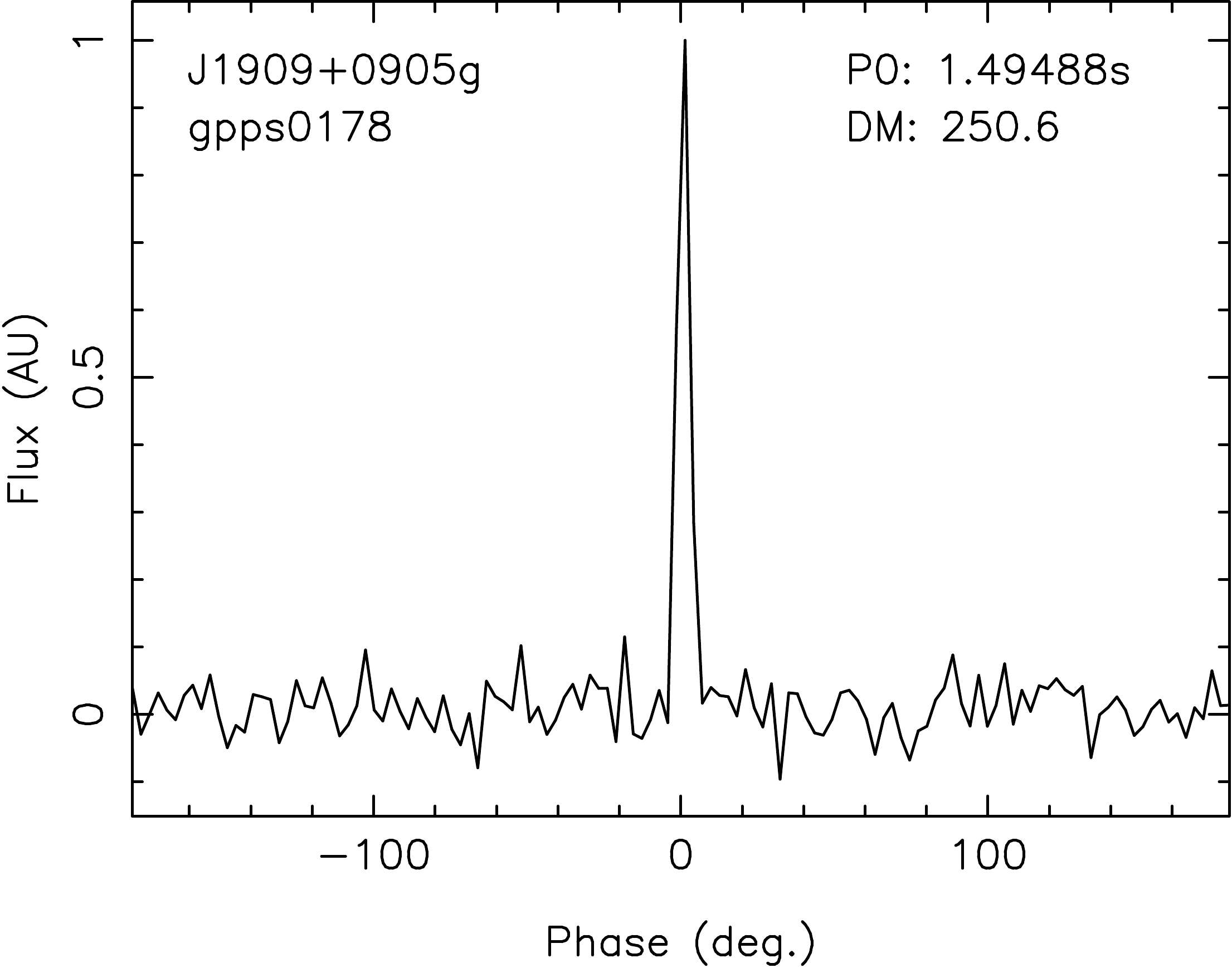}&
\includegraphics[width=39mm]{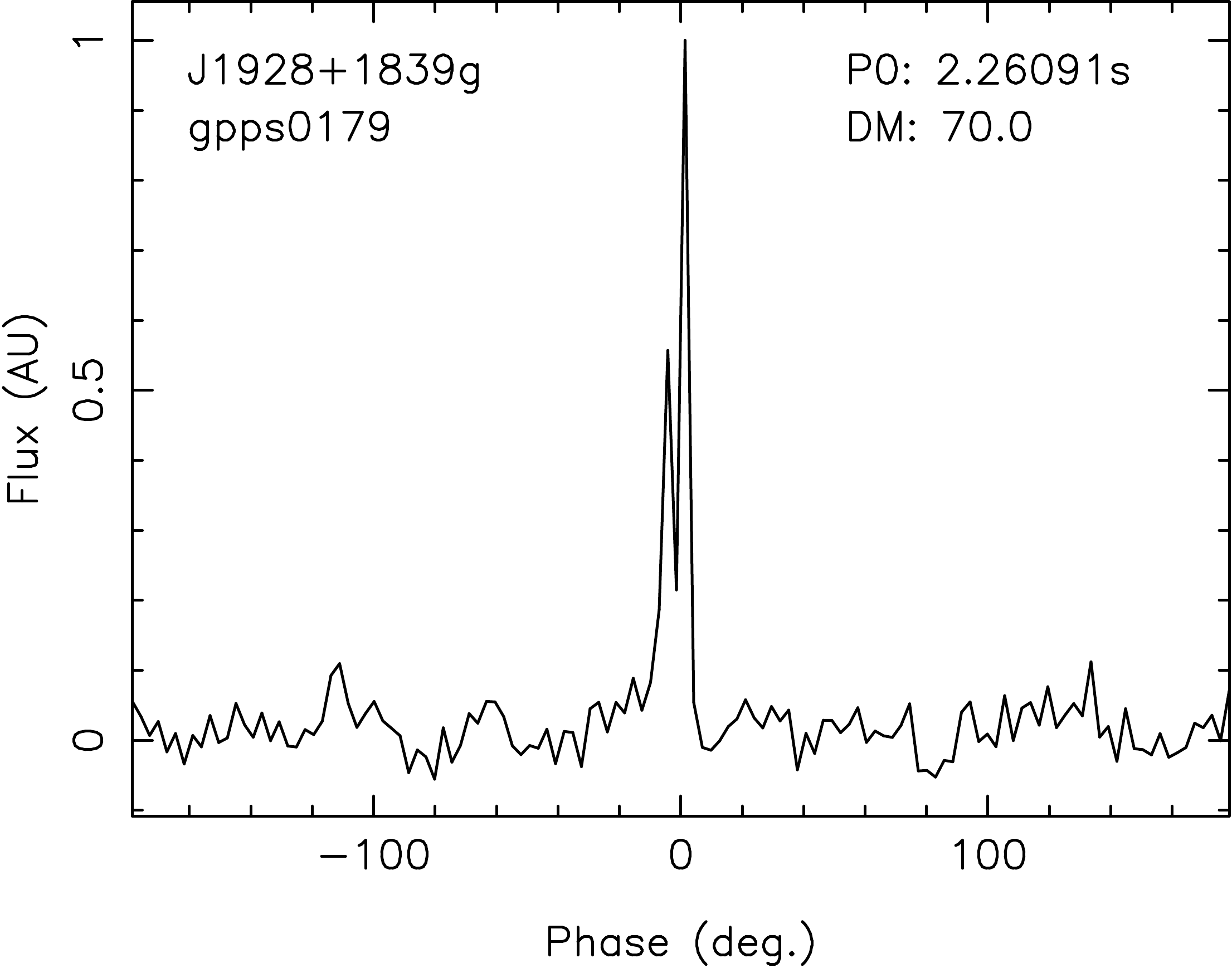}&
\includegraphics[width=39mm]{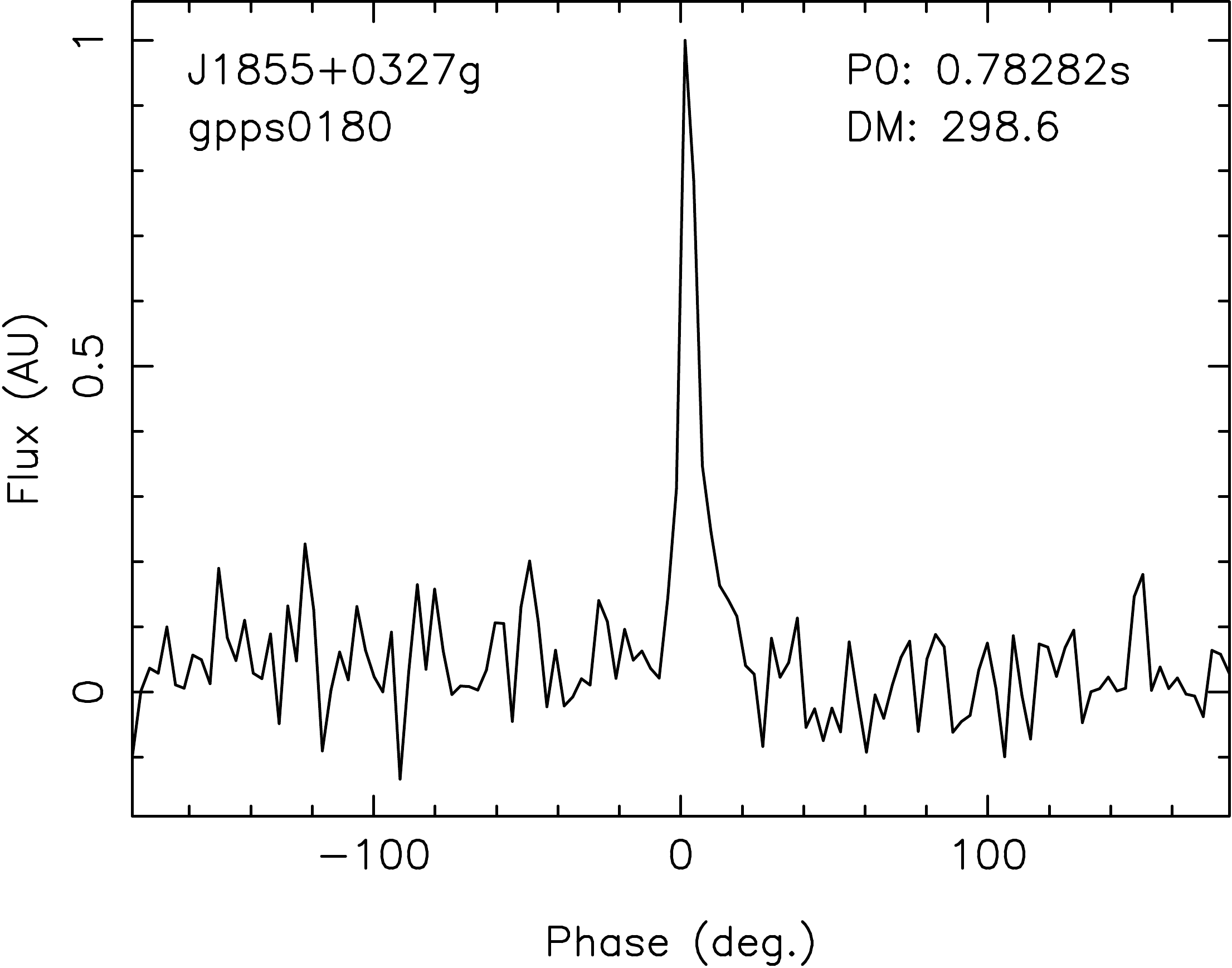}\\[2mm]
\includegraphics[width=39mm]{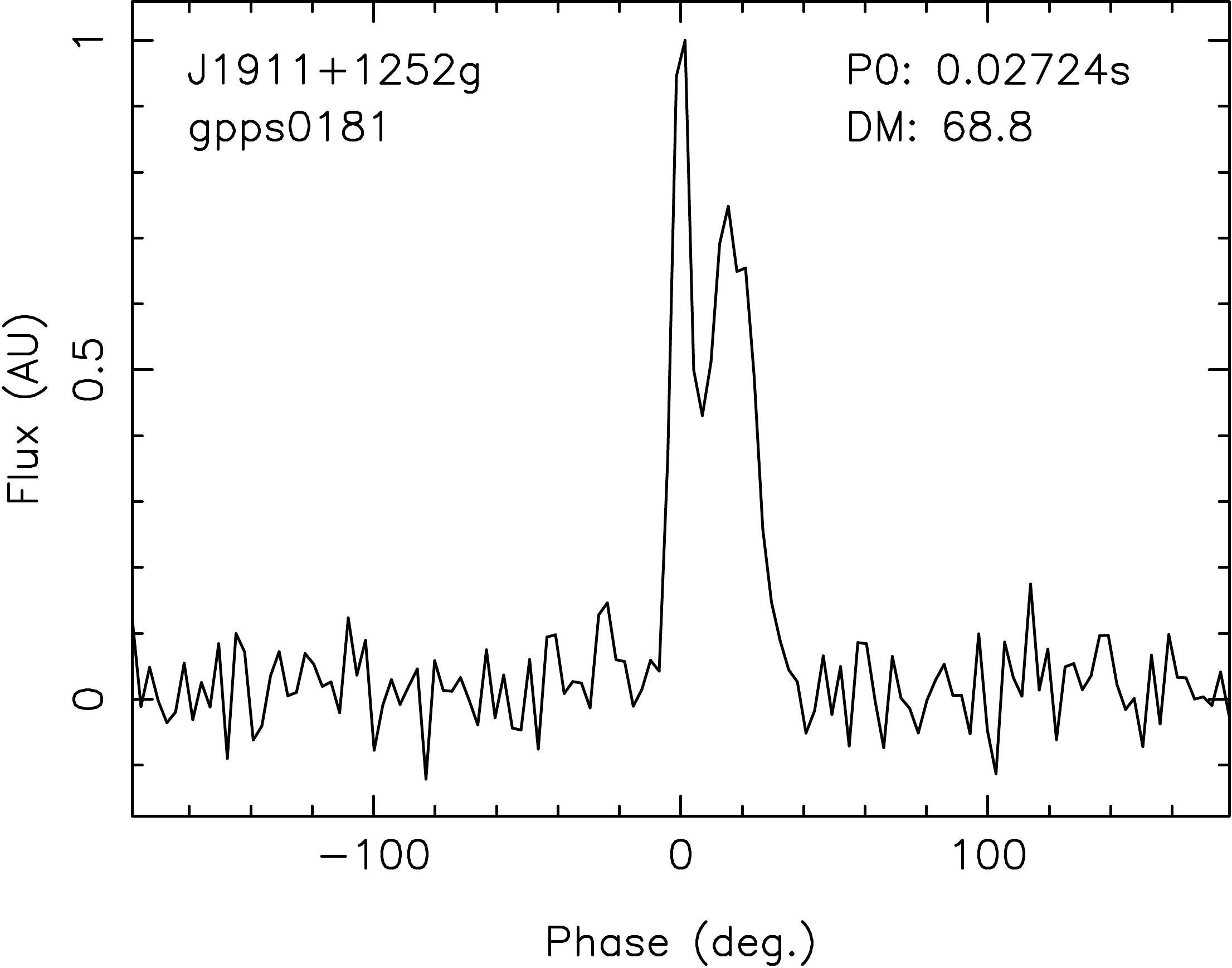}&
\includegraphics[width=39mm]{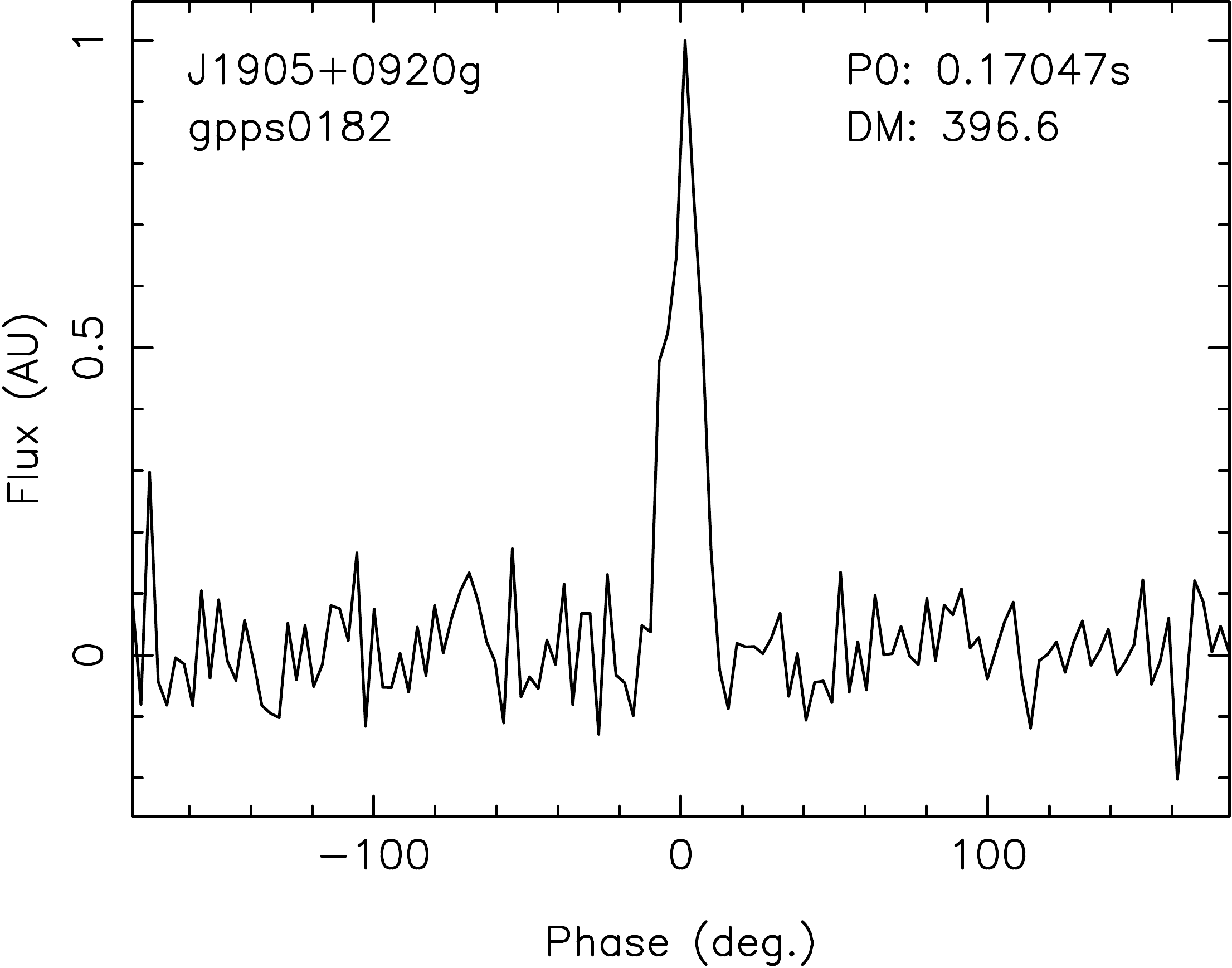}&
\includegraphics[width=39mm]{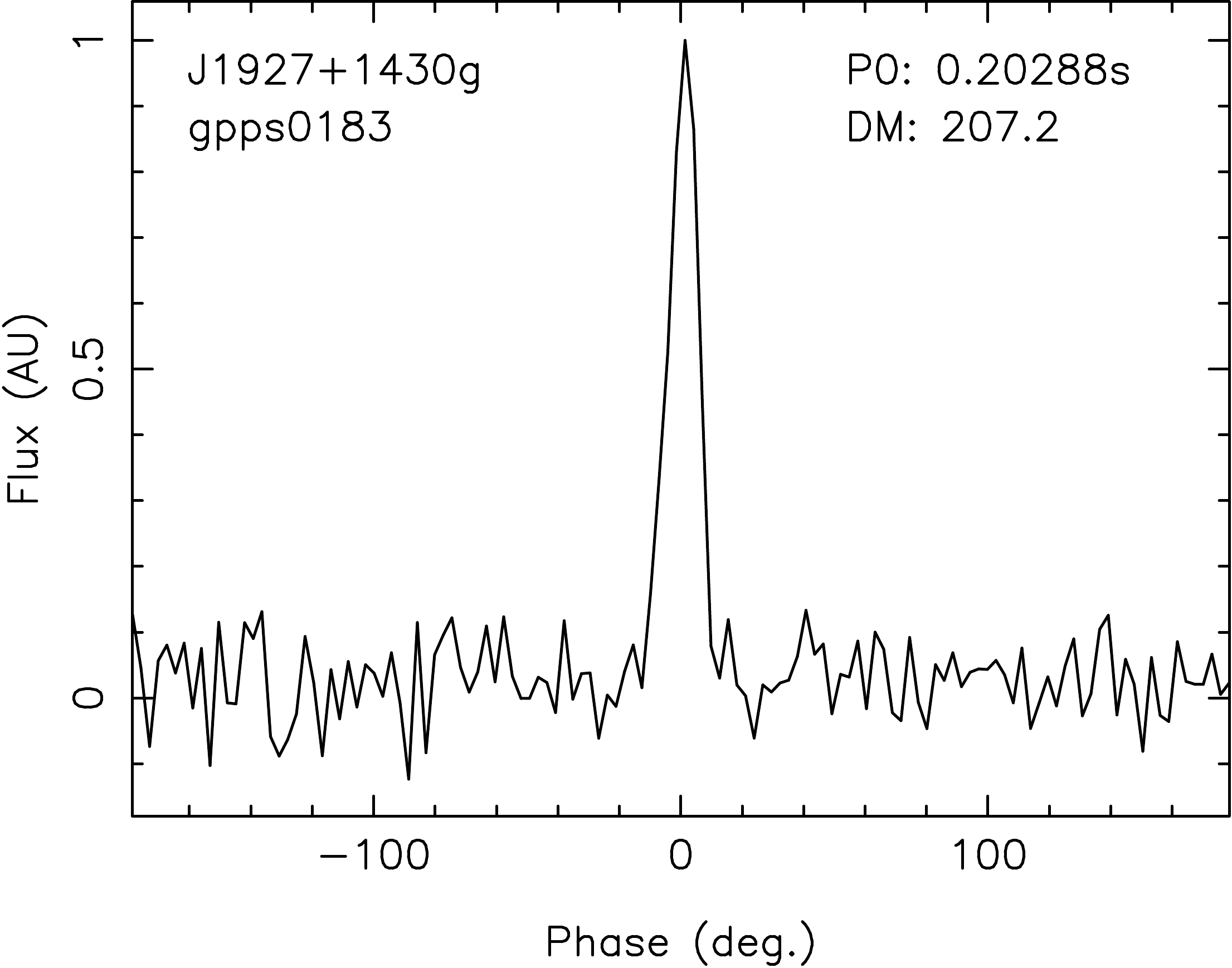}&
\includegraphics[width=39mm]{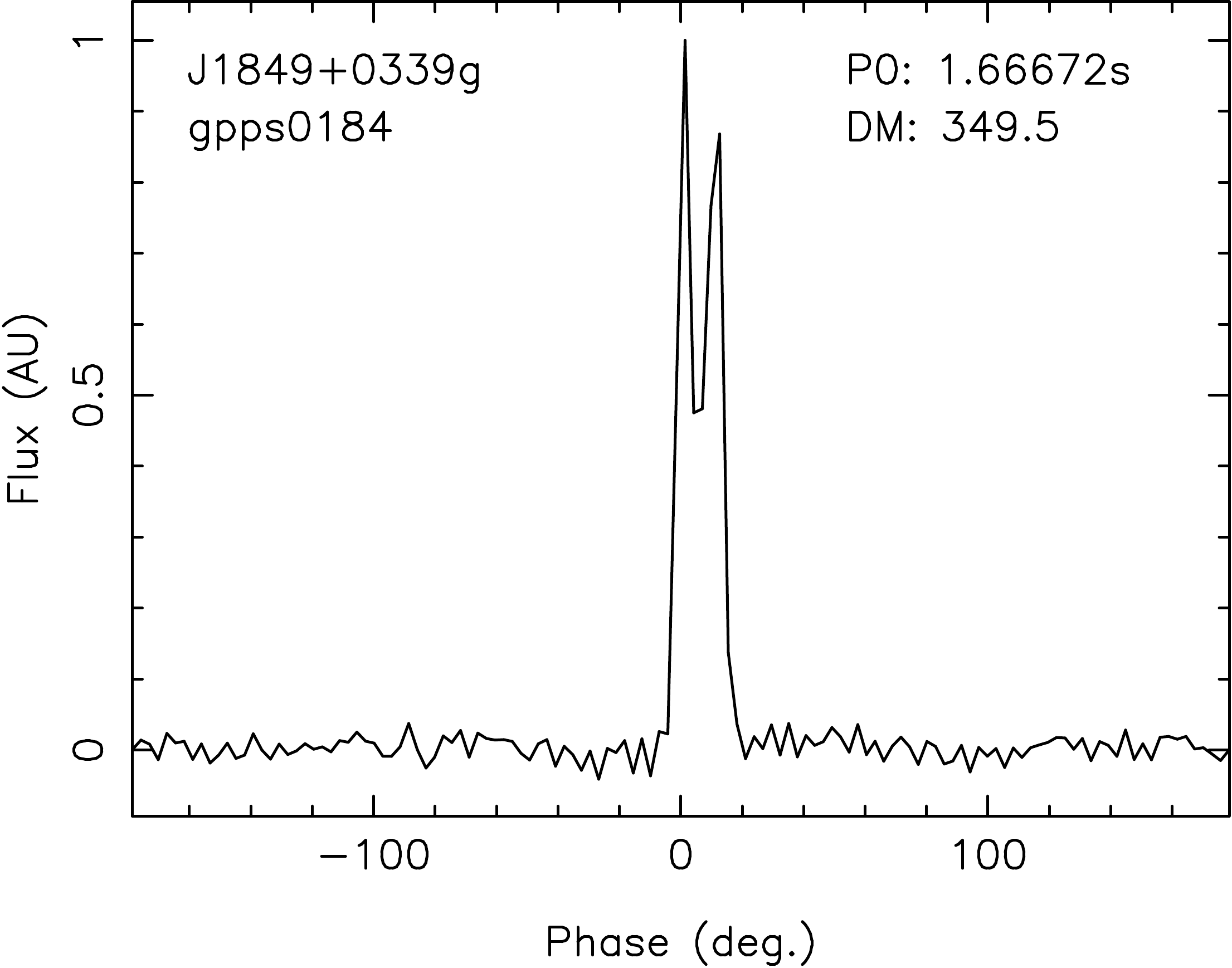}\\[2mm]
\includegraphics[width=39mm]{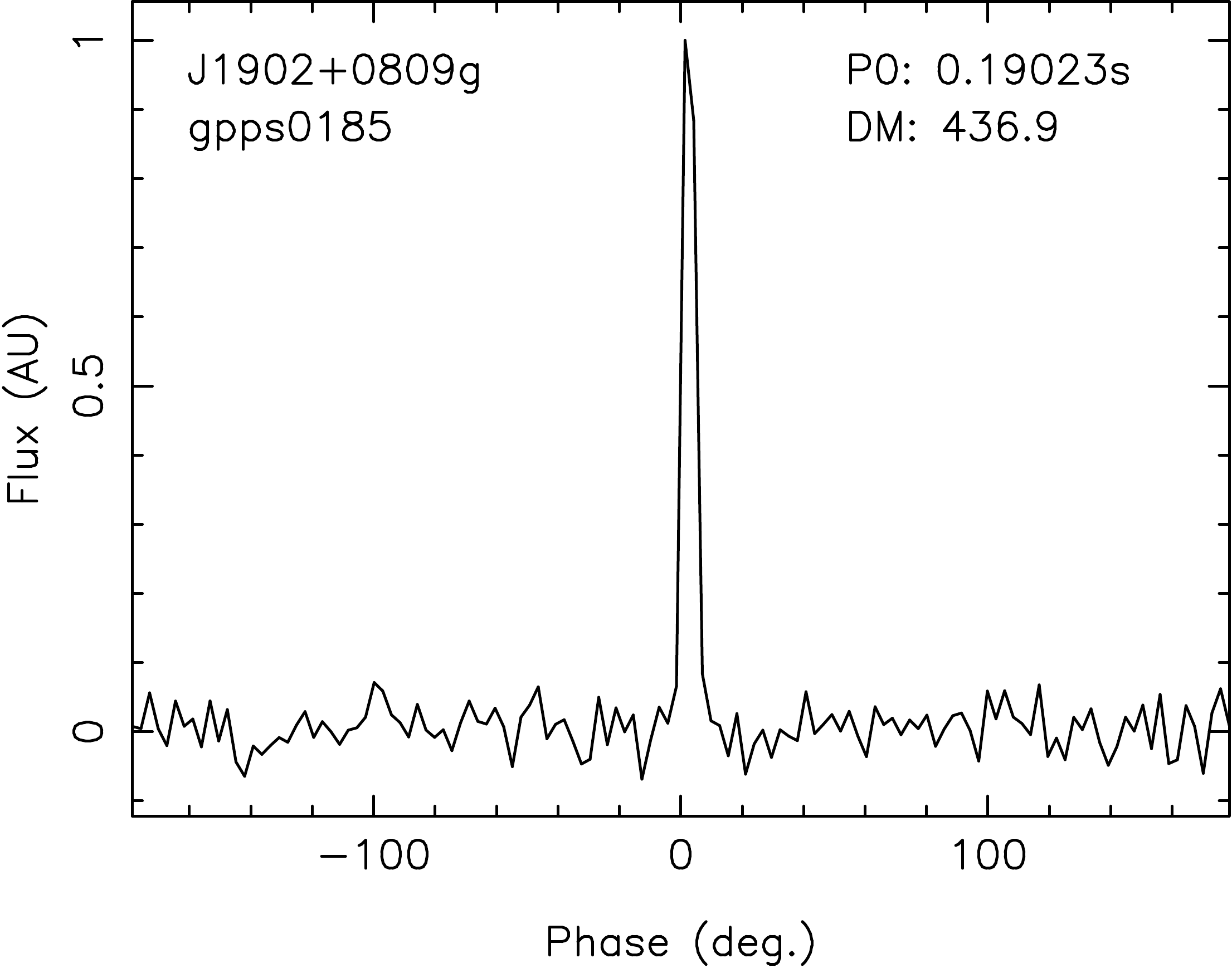}&
\includegraphics[width=39mm]{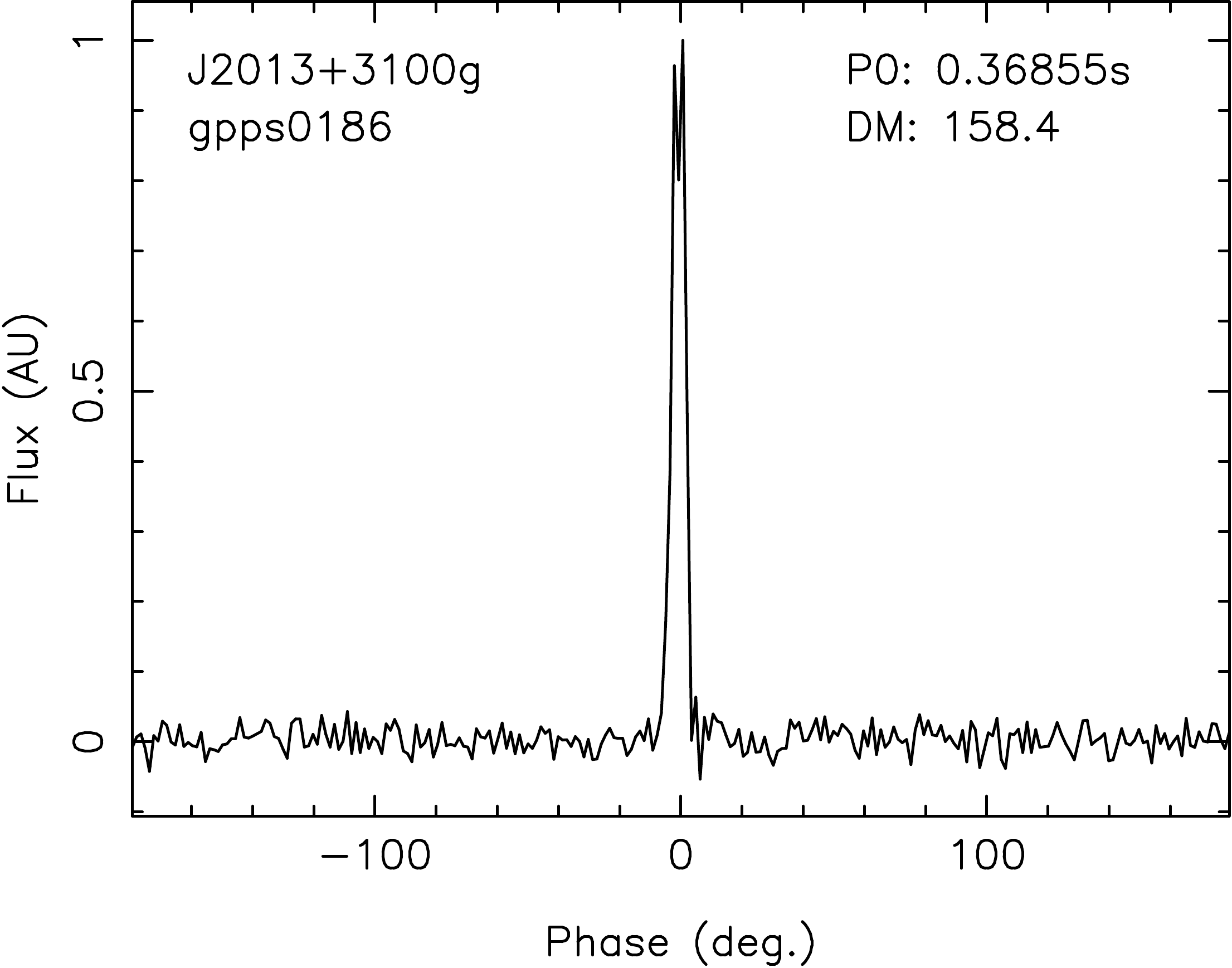}&
\includegraphics[width=39mm]{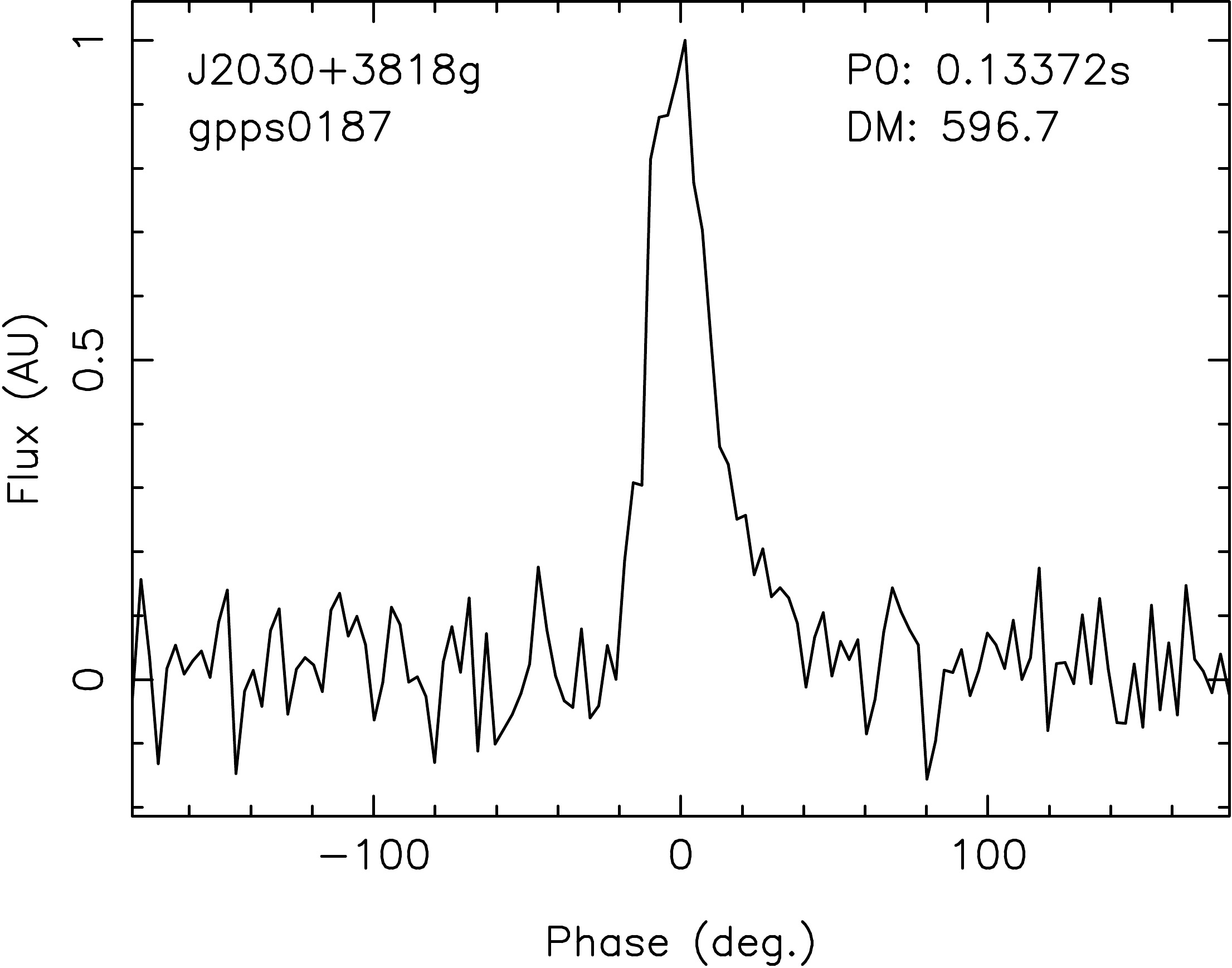}&
\includegraphics[width=39mm]{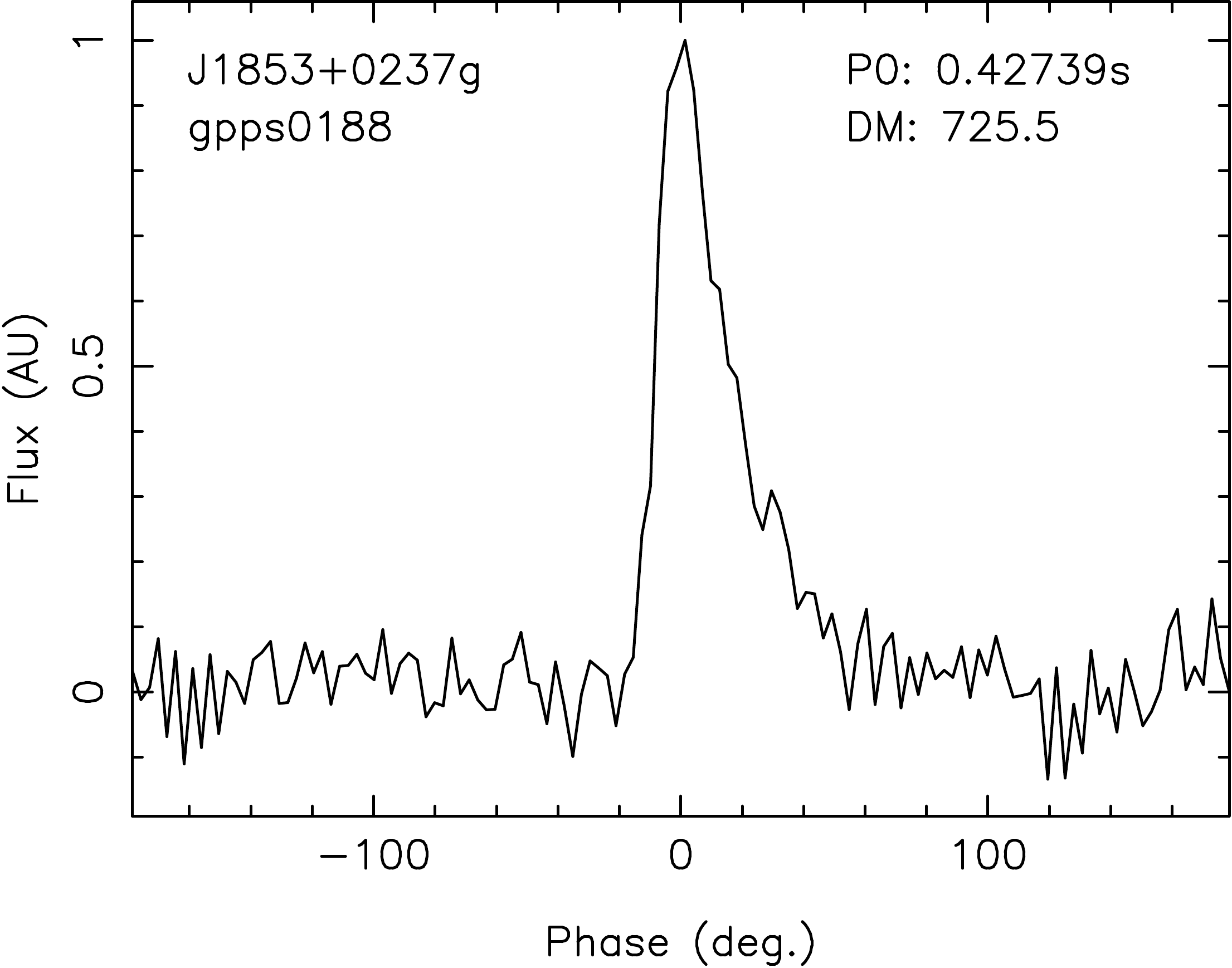}\\[2mm]
\includegraphics[width=39mm]{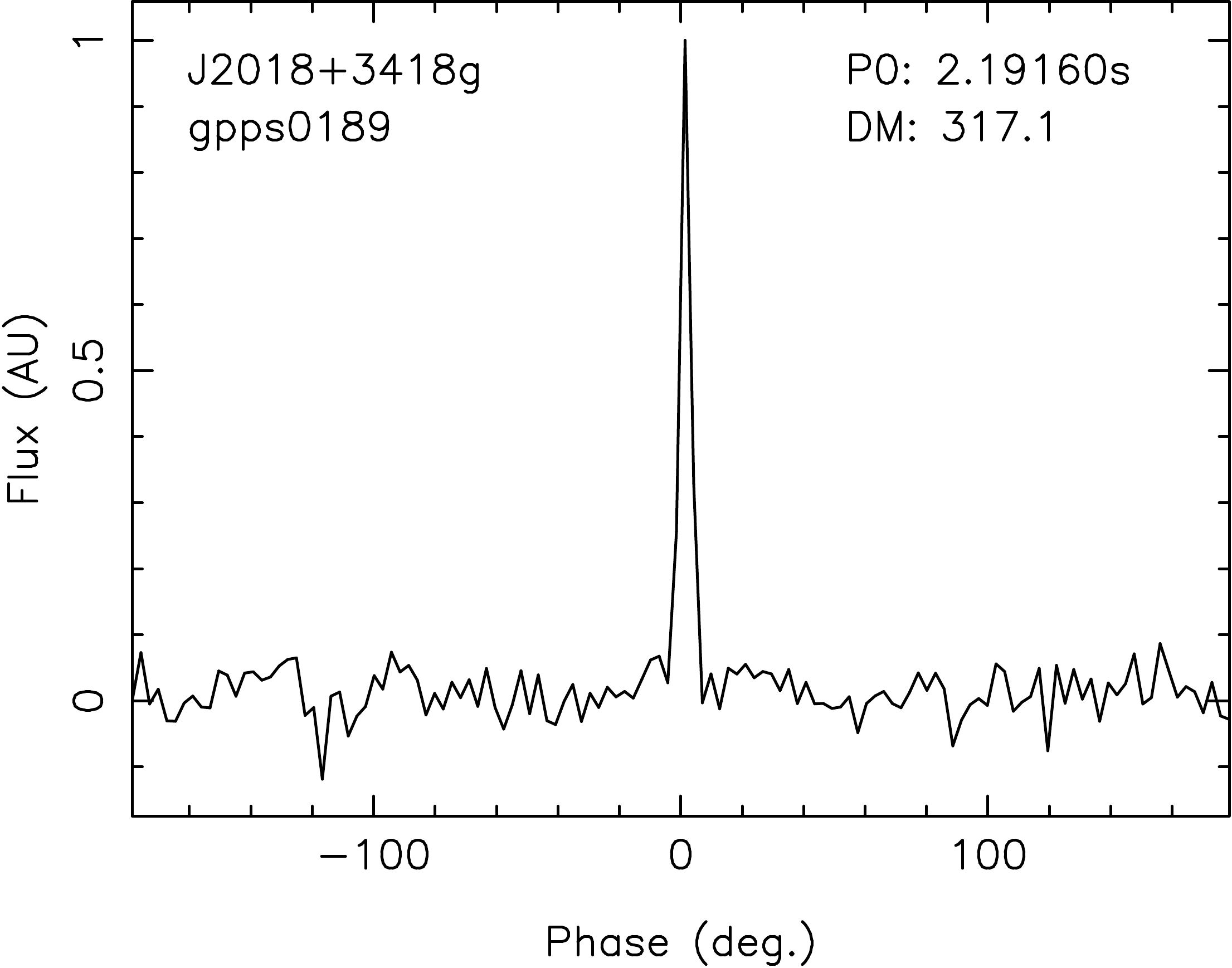}&
\includegraphics[width=39mm]{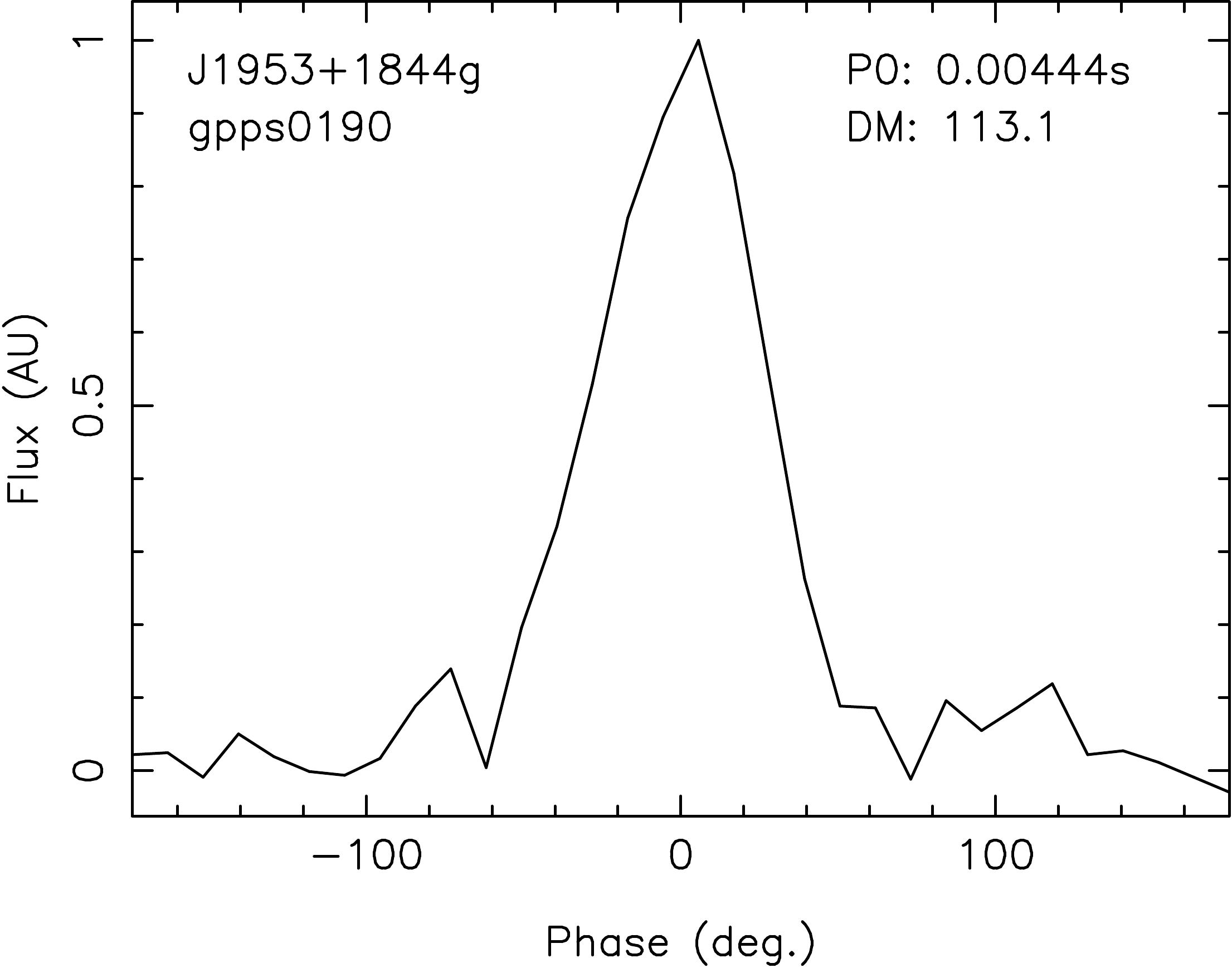}&
\includegraphics[width=39mm]{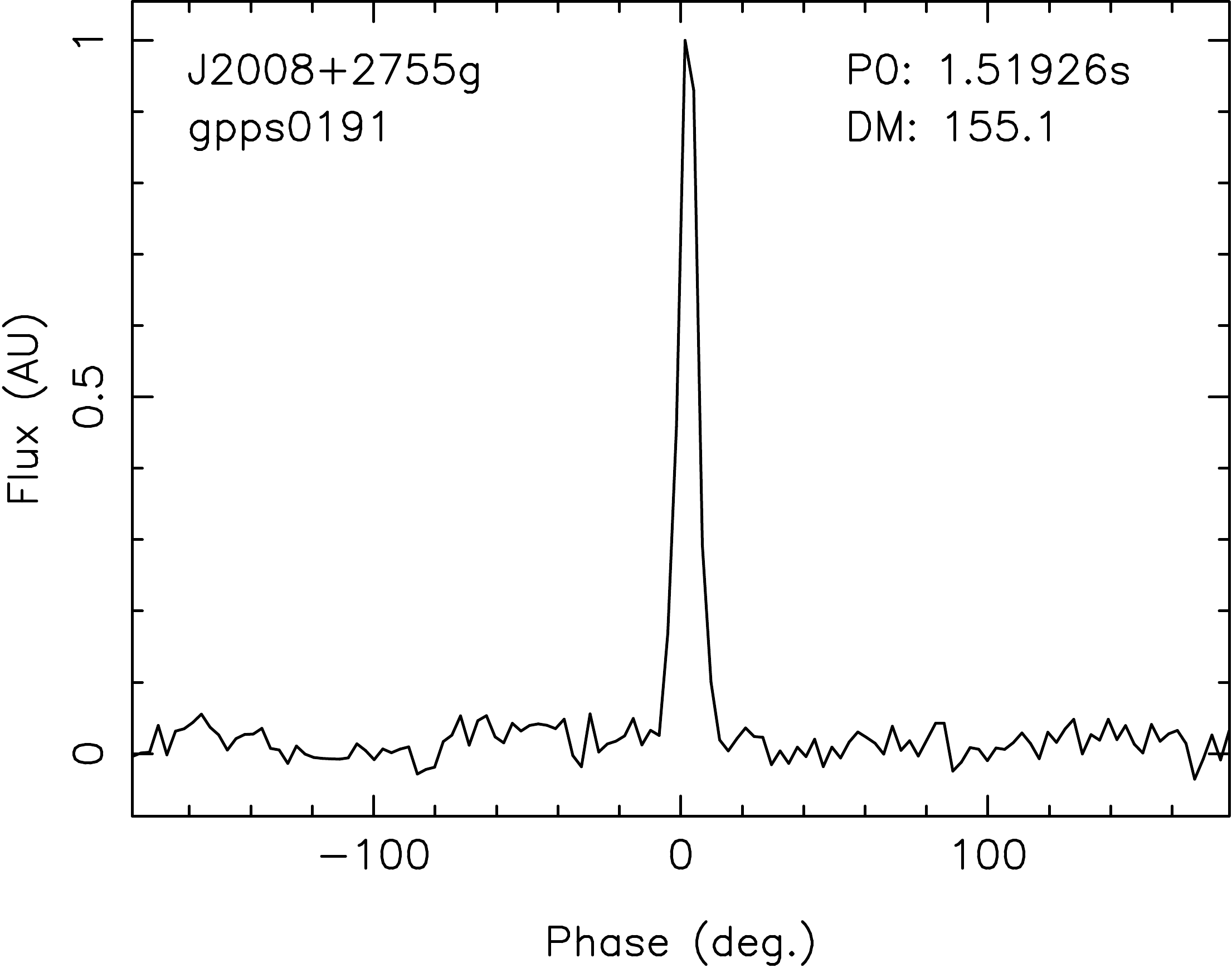}&
\includegraphics[width=39mm]{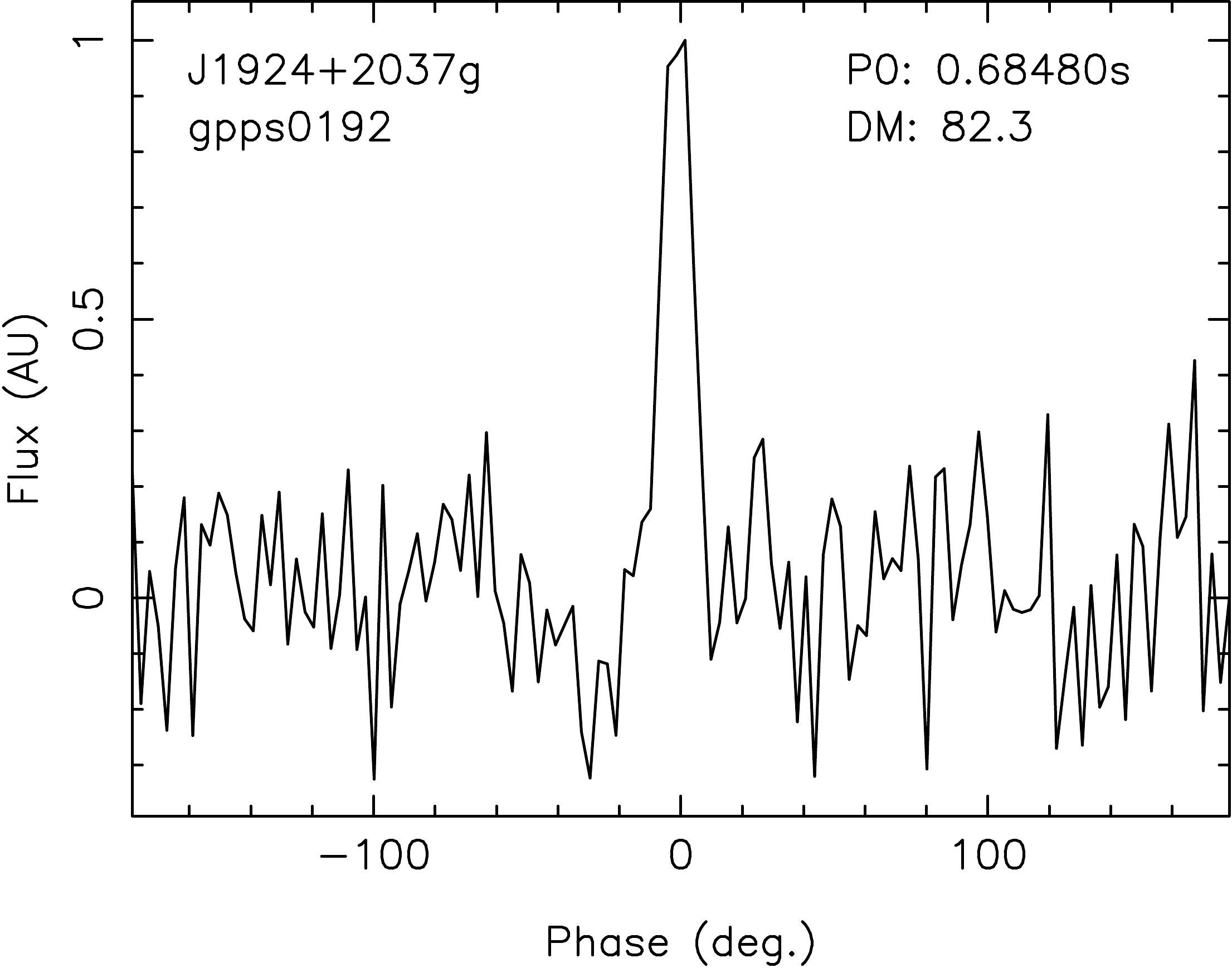}\\[2mm]
\includegraphics[width=39mm]{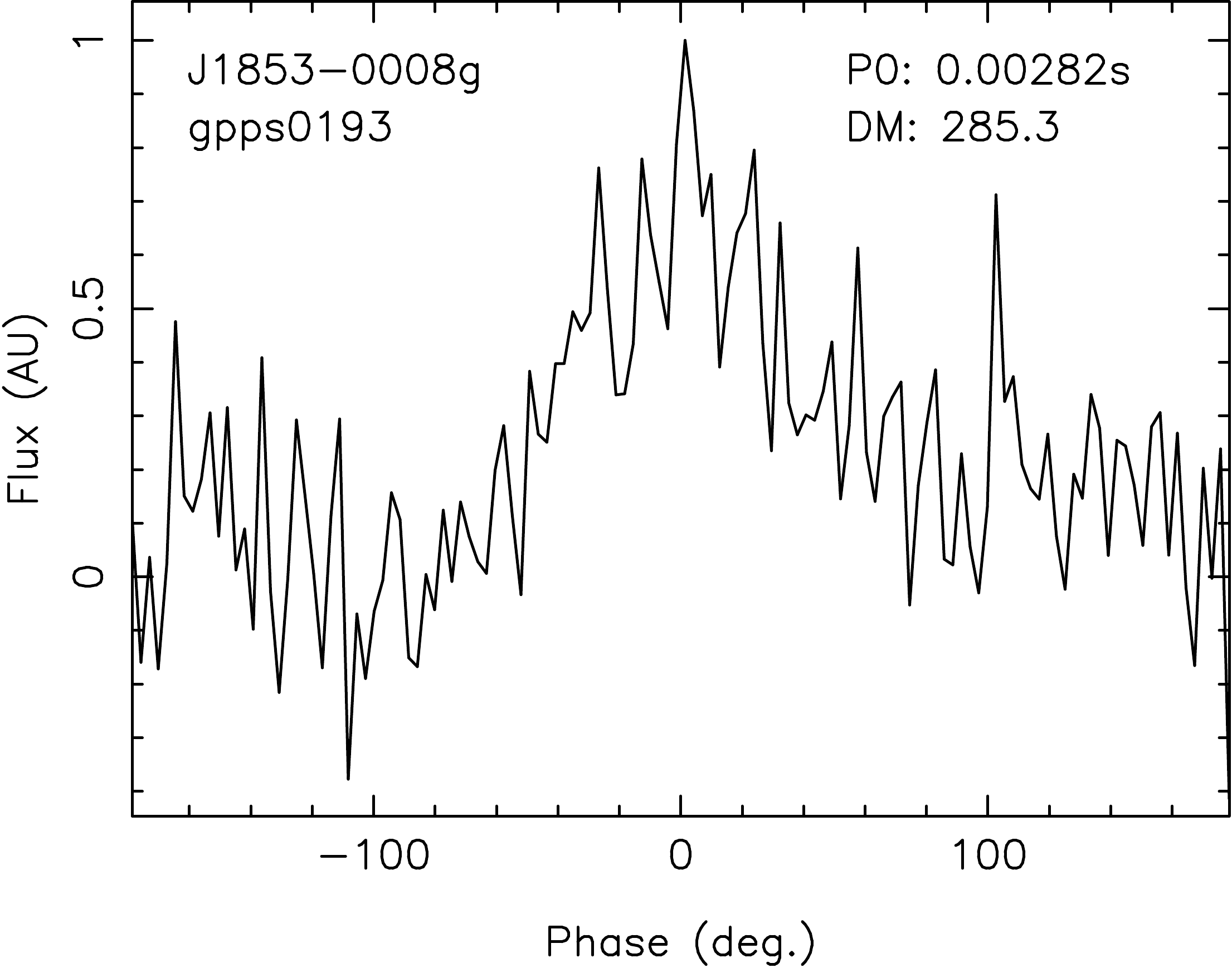}&
\includegraphics[width=39mm]{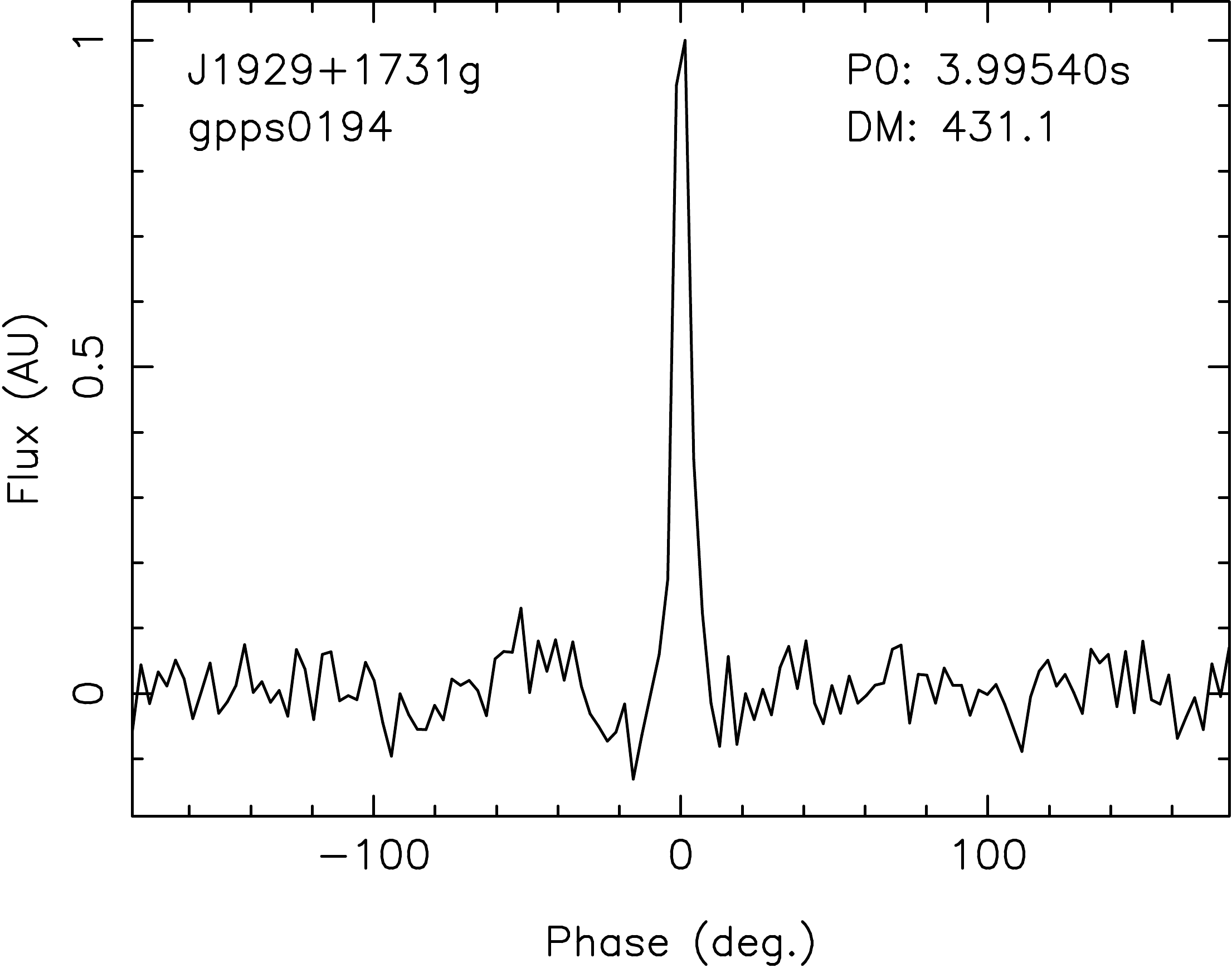}&
\includegraphics[width=39mm]{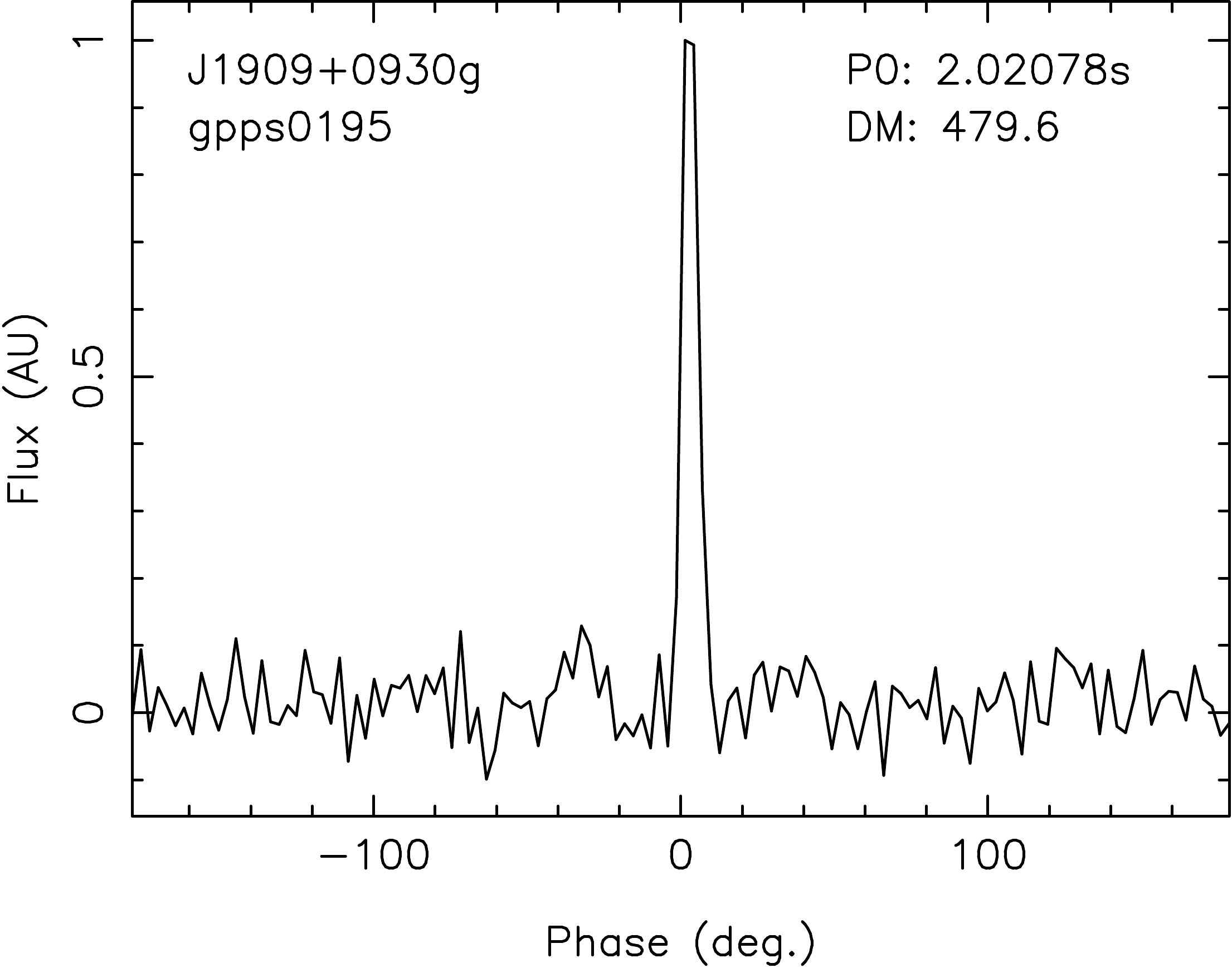}&
\includegraphics[width=39mm]{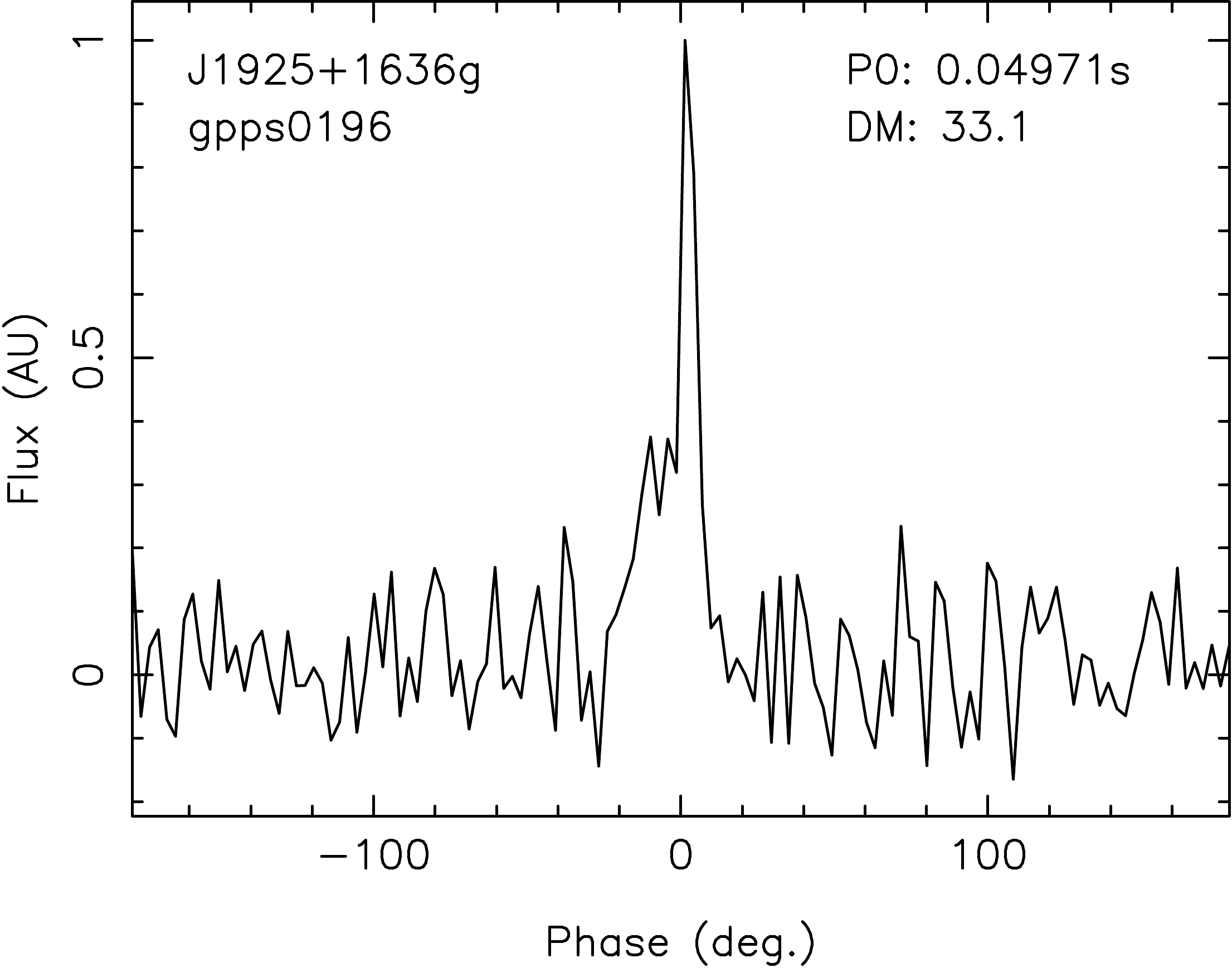}\\
\end{tabular}%

\begin{minipage}{3cm}
\caption[]{
-- {\it Continued}.}\end{minipage}
\addtocounter{figure}{-1}
\end{figure*}%
\begin{figure}
\centering
\begin{tabular}{rrrr}
\includegraphics[width=39mm]{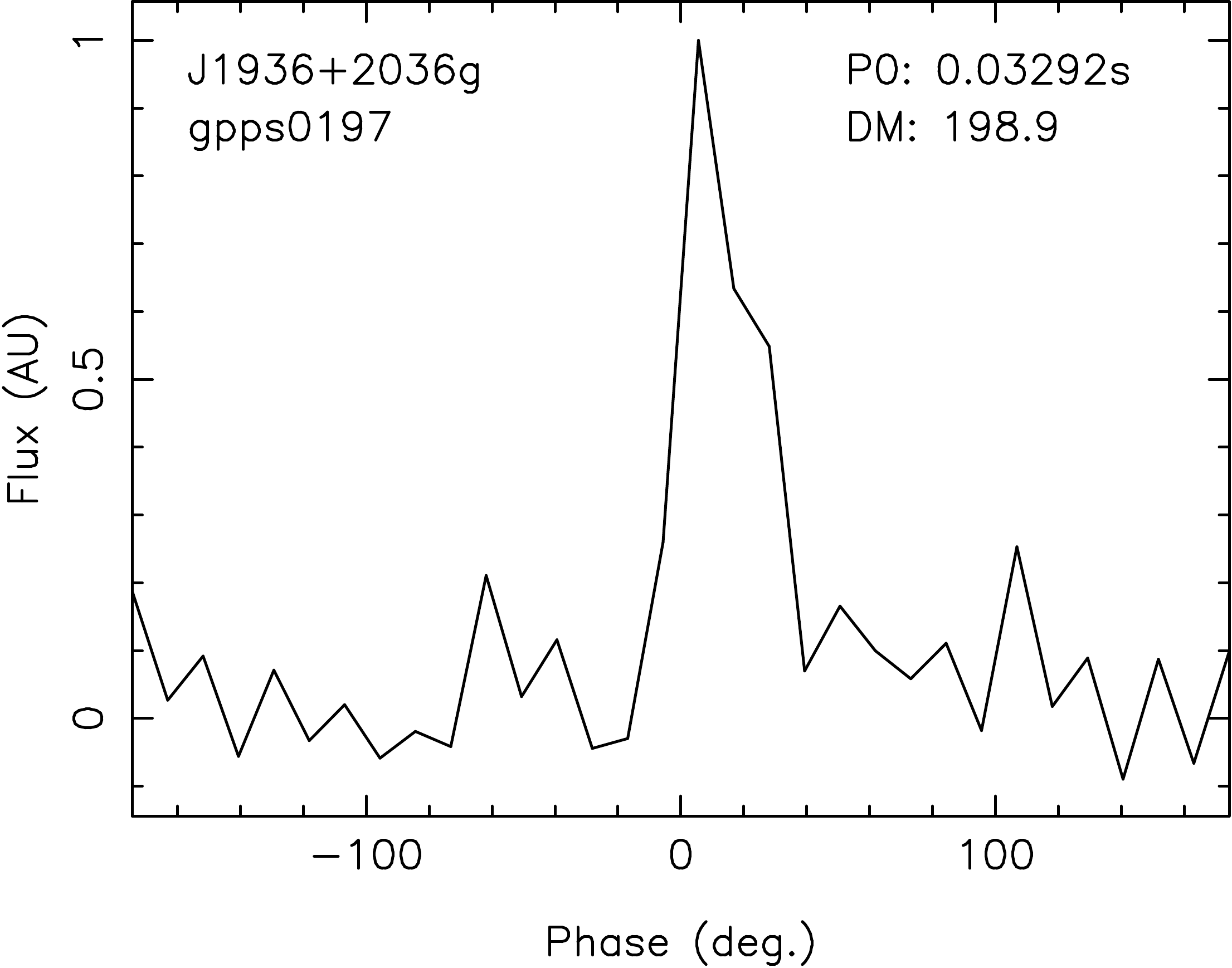} &
\includegraphics[width=39mm]{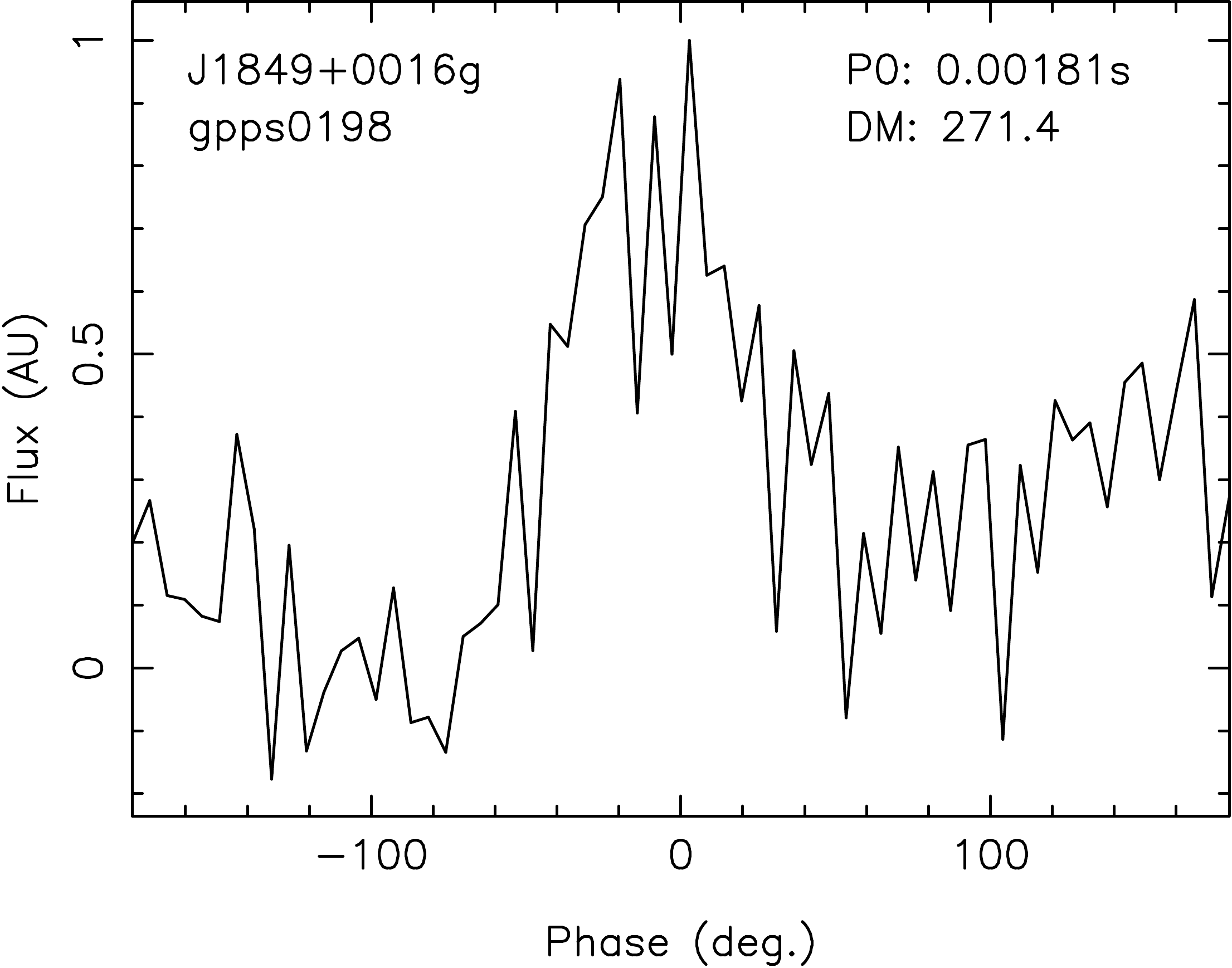} \\[2mm]
\includegraphics[width=39mm]{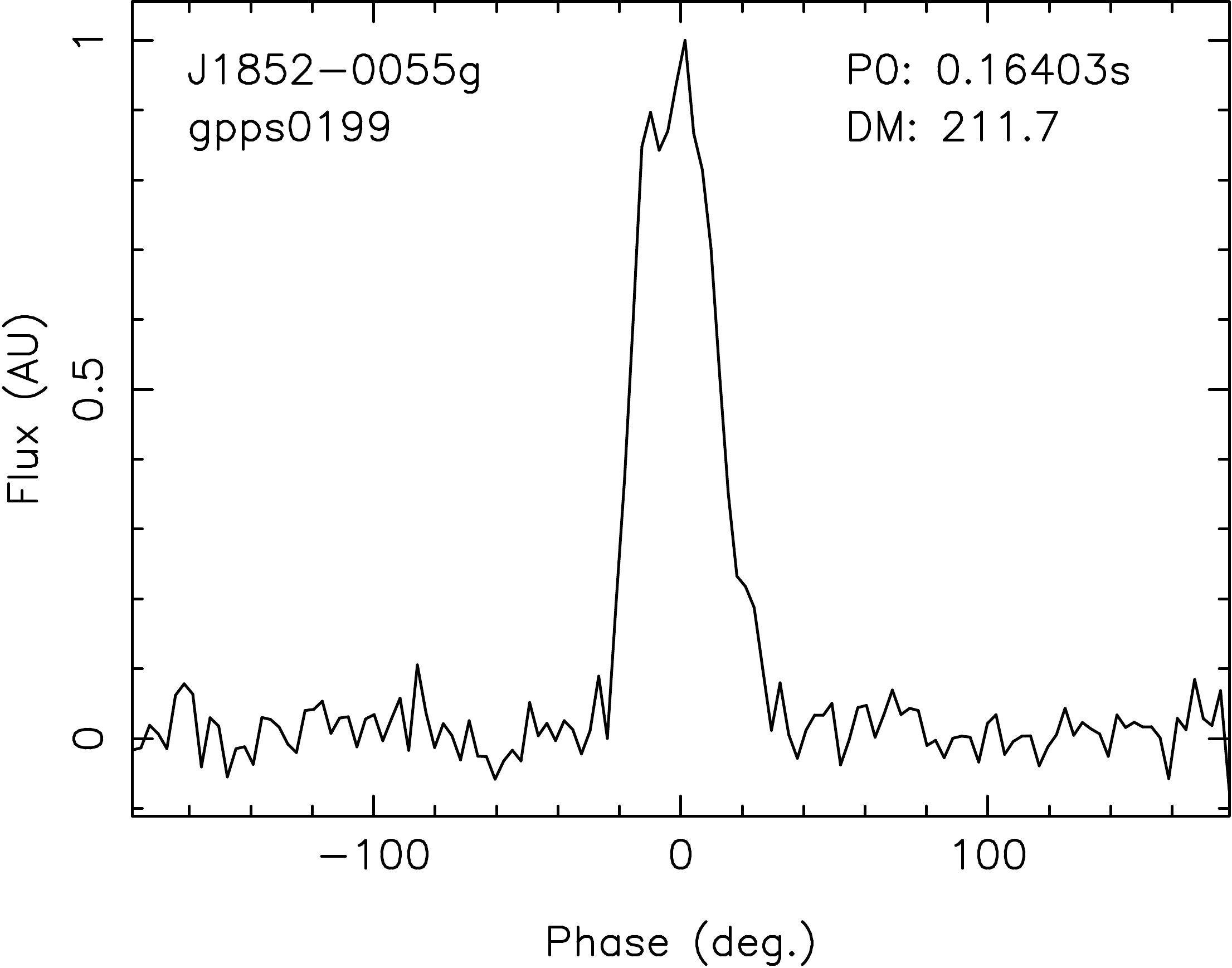} &
\includegraphics[width=39mm]{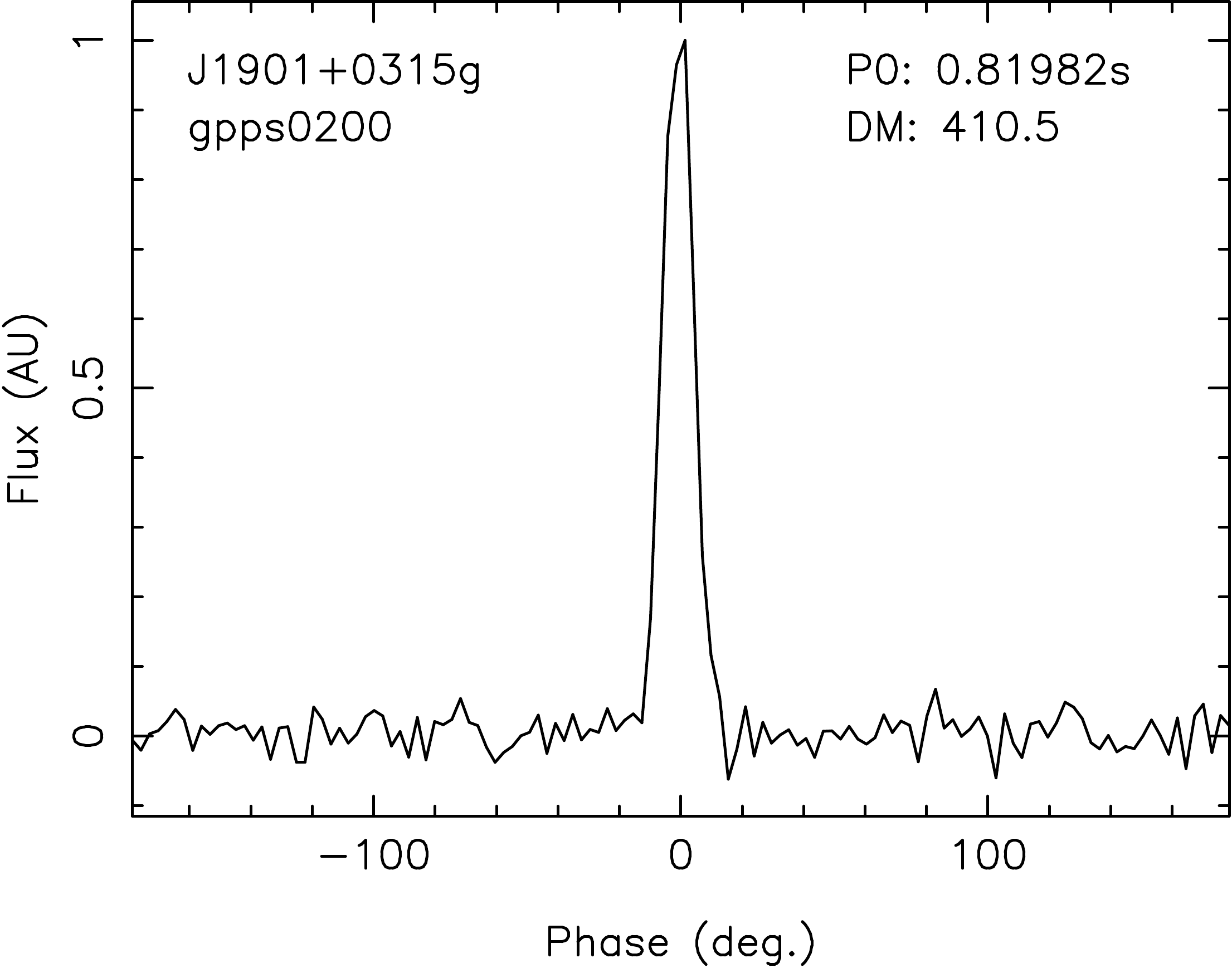} \\[2mm]
\includegraphics[width=39mm]{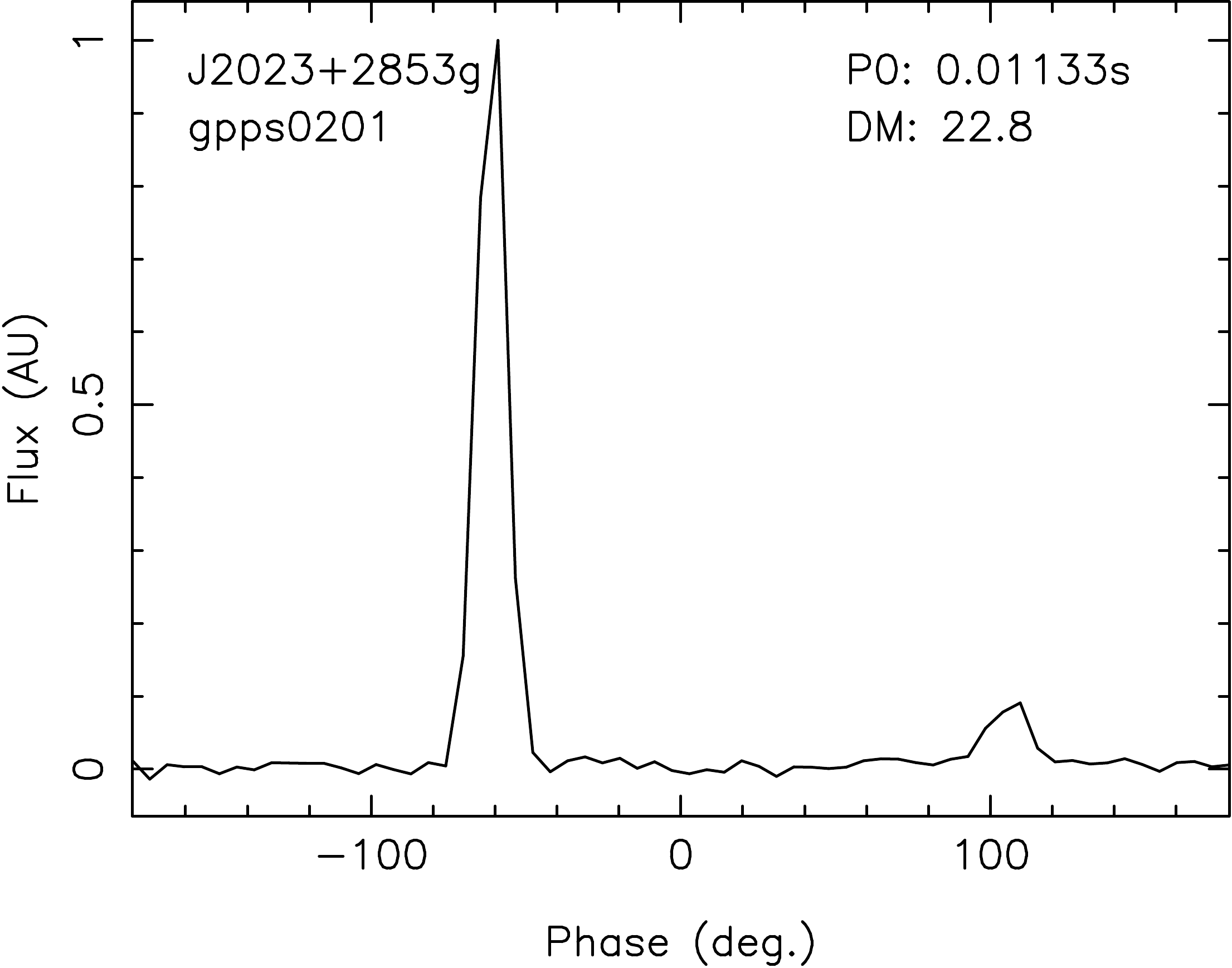} \\
\end{tabular}%

\begin{minipage}{3cm}
\caption[]{
-- {\it Continued}.}\end{minipage}
\end{figure}%
%

The next step is to add the data from $XX$ and $YY$ to form the total
power data series for pulsar searches. High-resolution data are good
for short-period pulsars so that 2048 channels and the default time
resolution of 49.152~$\upmu$s are taken as defaults. The so-combined
fits data are stored in the repository {\it data1j} for various
searches in the following steps.

When combining data from two polarization channels of $XX$ and $YY$,
their different band-passes, different signal to noise ratios (S/Ns),  and also
the RFI in different channels have to be considered. In order to
save computer disk space, we also cut off the 256 channels at both
sides of the band, which always have a very degraded gain, so that
there are only 1792 channels in the actual fits files in {\it data1j}.
All relevant parameters, such as the channel frequency resolution,
sampling time, MJD, etc., are accordingly stored in the
head of fits files.

\subsubsection{Searching for pulsars and pulses}

Three main approaches have been realized to search for pulsars and
individual pulses: PRESTO \citep{ran11}, SIGPROC \citep{lor11} and
our single pulse detection module developed in-house. Most of the
data processing is carried out via multi-jobs
\citep{tange_2020_4118697}.  Currently, the PRESTO module and the
single pulse module have been applied to searching for all covers,
and the SIGPROC module has been tested and will be applied for
re-searching for all GPPS survey data by utilizing a new computer
cluster that will be available in March 2021.

In the PRESTO module \citep{ran11}, a pipeline has been constructed
to perform the pulsar search with a few steps:\\
(1) read the fits file, and check the RFI and make RFI masks,
by using the {\it rfifind}; \\
(2) make the de-dispersion plan and de-disperse data by using {\it prepsubband}; \\
(3) search for periodical signals by using {\it accelsearch}; \\
(4) sift through the detected signals to find candidates and then fold data by using
{\it prepfold}. \\
Through this  processing, a number of candidates have been found, and
the {\it .pfd} and {\it .bestprof} files and also corresponding plots
are stored in the repository {\it data2j}.

In the SIGPROC module \citep{lor11}, the fits data file is converted
to filter-bank format first by utilizing the command {\it
filterbank}, then de-dispersion is done by running the command {\it
dedisperse\_all} and then periodical signals are searched by using
{\it seek}, and finally the results are combined and sorted to find
the best $P$ and $DM$ for pulsar candidates by utilizing {\it best}.
These candidates can be folded from the fits data by using {\it
prepfold} in the PRESTO module, and the results are also stored in
the repository {\it data2j}.

We have developed a  single pulse module, which has three steps:
de-dispersion, finding a single pulse from the image of data-array of
DM-time with speedy artificial intelligence (AI) recognition in GPU
clusters and the period finding for the picked pulses.

In addition, the pulsar acceleration search has been tested for a
few beams for binary candidates, and will be carried out for all
GPPS survey data when the new computer clusters are available.

\subsubsection{Evaluation of searching results}

After pulsar searching has been completed, a large number of pulsar candidates
have been found. Each candidate has a {\it .pfd}, {\it .bestprof} and {\it .ps} file in
the repository {\it data2j}. In fact, not only pulsar signals but also
some RFI which mimics pulsar signal features can be selected by
pulsar searching software. Therefore, evaluation of searching results
is desired.

\begin{figure}
   \centering
   \includegraphics[width=0.4\textwidth]{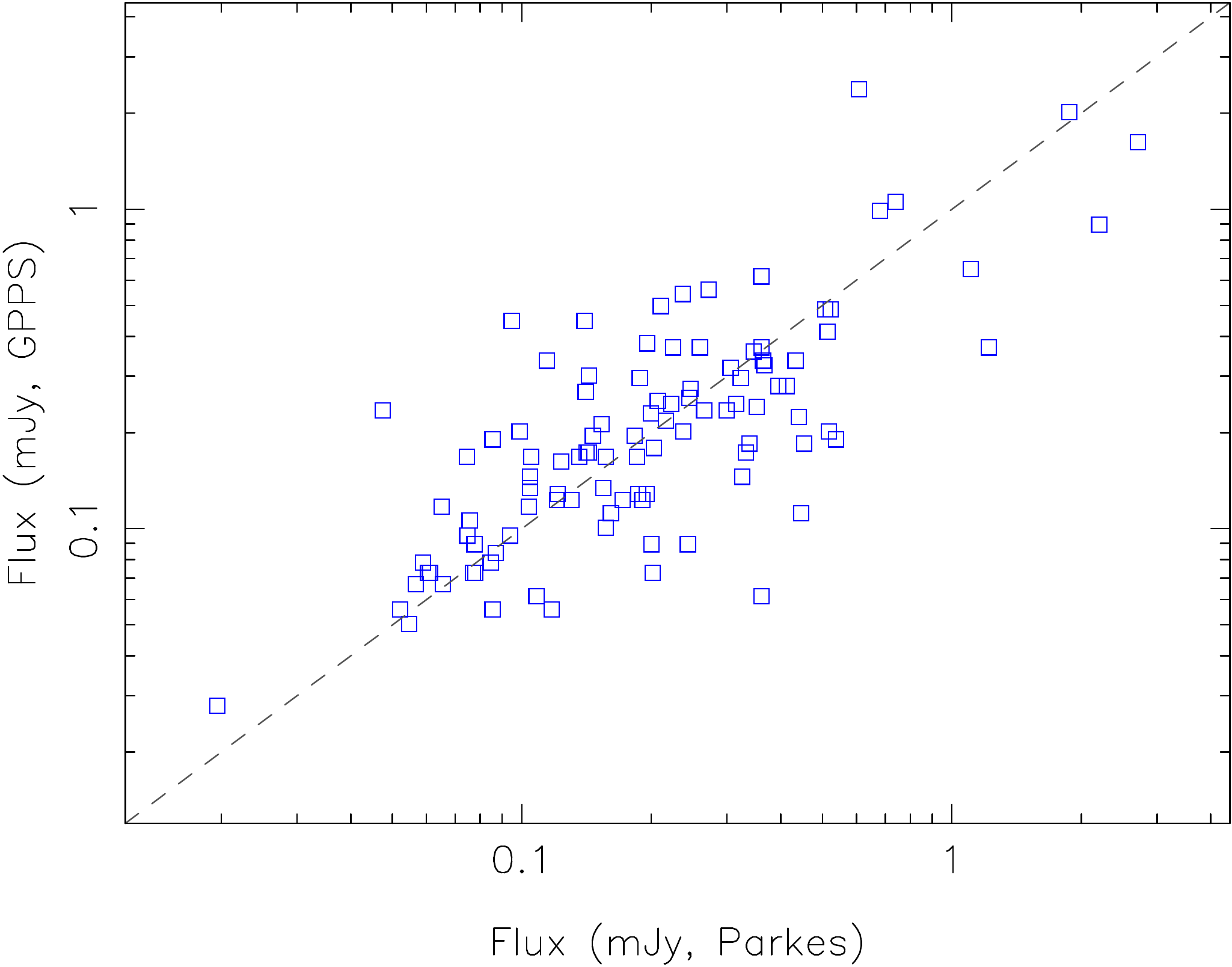}
   \caption{\baselineskip 3.8mm The estimated flux densities for Parkes pulsars detected
     in the GPPS survey are compared to the values in the ATNF pulsar
     catalog, which exhibits a remarkable consistence. }
   \label{fig12_fluxcomp}
\end{figure}

We found that the AI code developed by \citet{zbm+14} is very
efficient for discriminating a pulsar signal from RFI. A new AI module
has been developed and tested, and is applied in parallel.  After this
AI-sifting of candidates, only a small number of false candidates have
to be discarded during manual checking of pulsar candidates. The
relevant parameters can be extracted from {\it .bestprof} and
tabulated. These candidates from each survey observation are
cross-matched with known pulsars, and certainly only real candidates
for new pulsars will be further manually examined and undergo folding
by {\it pdmp}. If the result is good, the candidate will be observed
again for verification.

One side effect of high S/N pulses, which is seen more often with the
highly sensitivity FAST, is that there are too many candidates in a
wide range of cross-modulated period and DM values, as illuminated by
the peaks in Figure~\ref{fig10_DMt} caused by a strong pulsar. The
searching software can pick up these fake candidates, which have to be
carefully discarded by using a naive analysis above a high threshold.

\subsection{Verification Observations and Processing}

Thanks to the excellent pointing accuracy of FAST, the position
of any good candidate detected from one beam can be determined with an
accuracy better than $1.5'$ in radius. For strong pulsars detected in a few
nearby beams in the snapshot observations, more accurate coordinates
can be determined according to their beam center positions in the sky and
the S/N of a pulsar detected from these beams.

Once the position of a good candidate is determined, follow-up
tracking observation is carried out for 15 min, with the central beam
of the L-band 19-beam receiver pointed to the position. To investigate
the properties of the pulsar, data of four polarization channels
({$XX$}, {$YY$}, {$X^*Y$}, {$XY^*$}) from the verification
observations are always recorded. Before the beginning and after the
end of the observation session, the calibration signals are switched
on and off, for 1 second each, and the data are recorded for two
minutes and will be utilized for calibrations. In order to maximize
the FAST time efficiency for pulsar detection, data on the other 18
beams, in addition to the central beam, are also recorded for a deeper
pulsar search.

Data processing for the verification observations follows the same
searching procedure. In general, a new pulsar is detected with a
better S/N. In addition, the data are folded by running {\it pdmp}.  The
times of arrival (TOAs) of these observations will be used to derive
the period $P$ of a pulsar at a given epoch and the period derivative
$\dot{P}$ when several measurements are available.

\begin{figure}
   \centering
   \includegraphics[width=74mm]{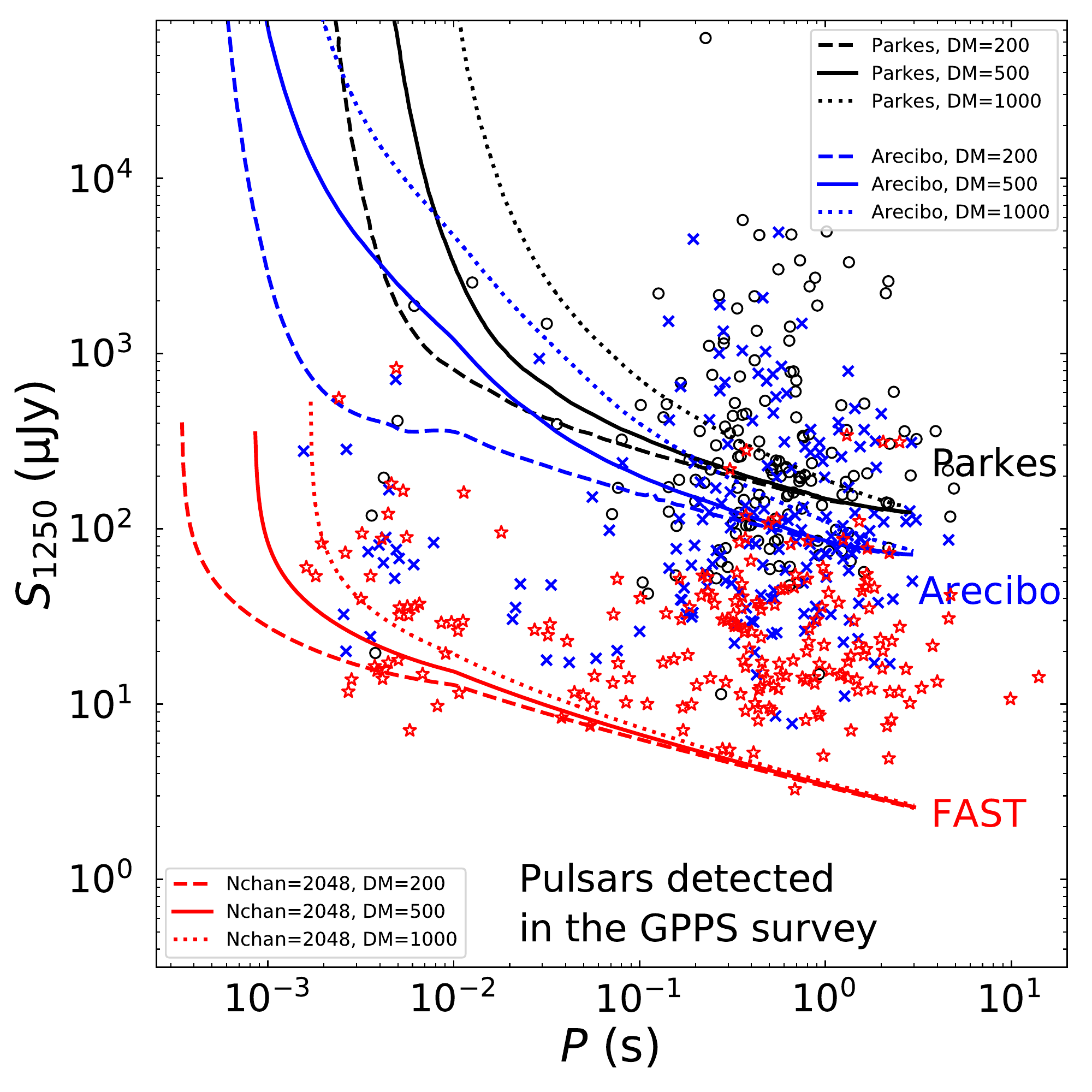}
   \includegraphics[width=72mm]{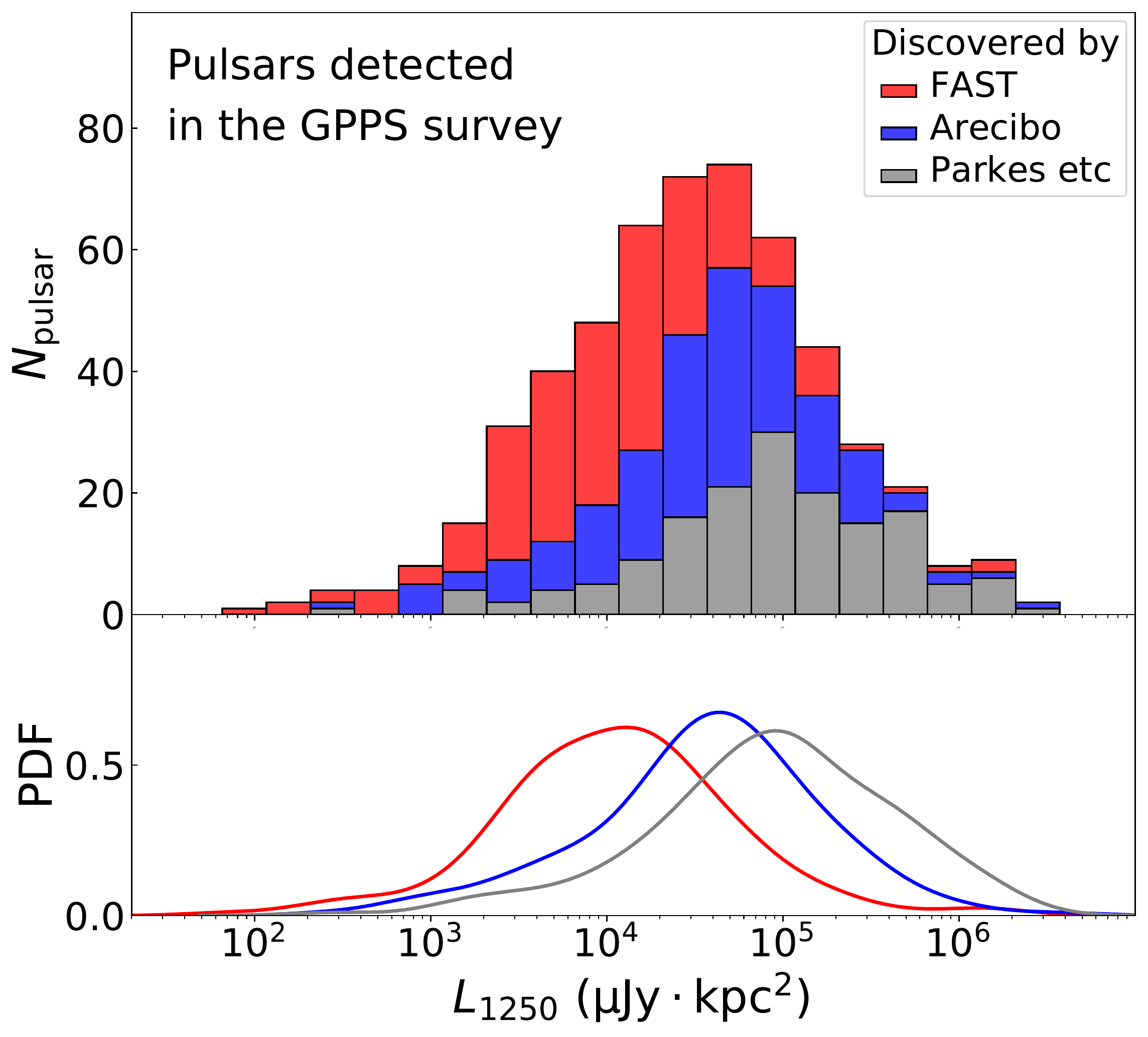}
   \caption{\baselineskip 3.8mm {\it Top}: the GPPS pulsars (201) always have a smaller
     flux, as indicated by red stars, compared to the pulsars
     discovered by, e.g., Arecibo (180), Parkes, etc. (156) radio
     telescopes. All pulsars {featured} here are detected in the GPPS
     survey. Some pulsars below the sensitivity curves are nulling
     pulsars, or pulsars with a narrower pulse than the assumed 5\% of
     the period, or pulsars first detected by verification
     observations. {\it Bottom}: comparison of the luminosity
     distribution of the GPPS pulsars with those of other surveys, in
     terms of the number distribution (the {\it upper panel}) and the
     probability density function (PDF, the {\it lower panel}).}
   \label{fig13_fluxP}
   \end{figure}

\begin{figure*}
   \centering
   \includegraphics[height=53mm]{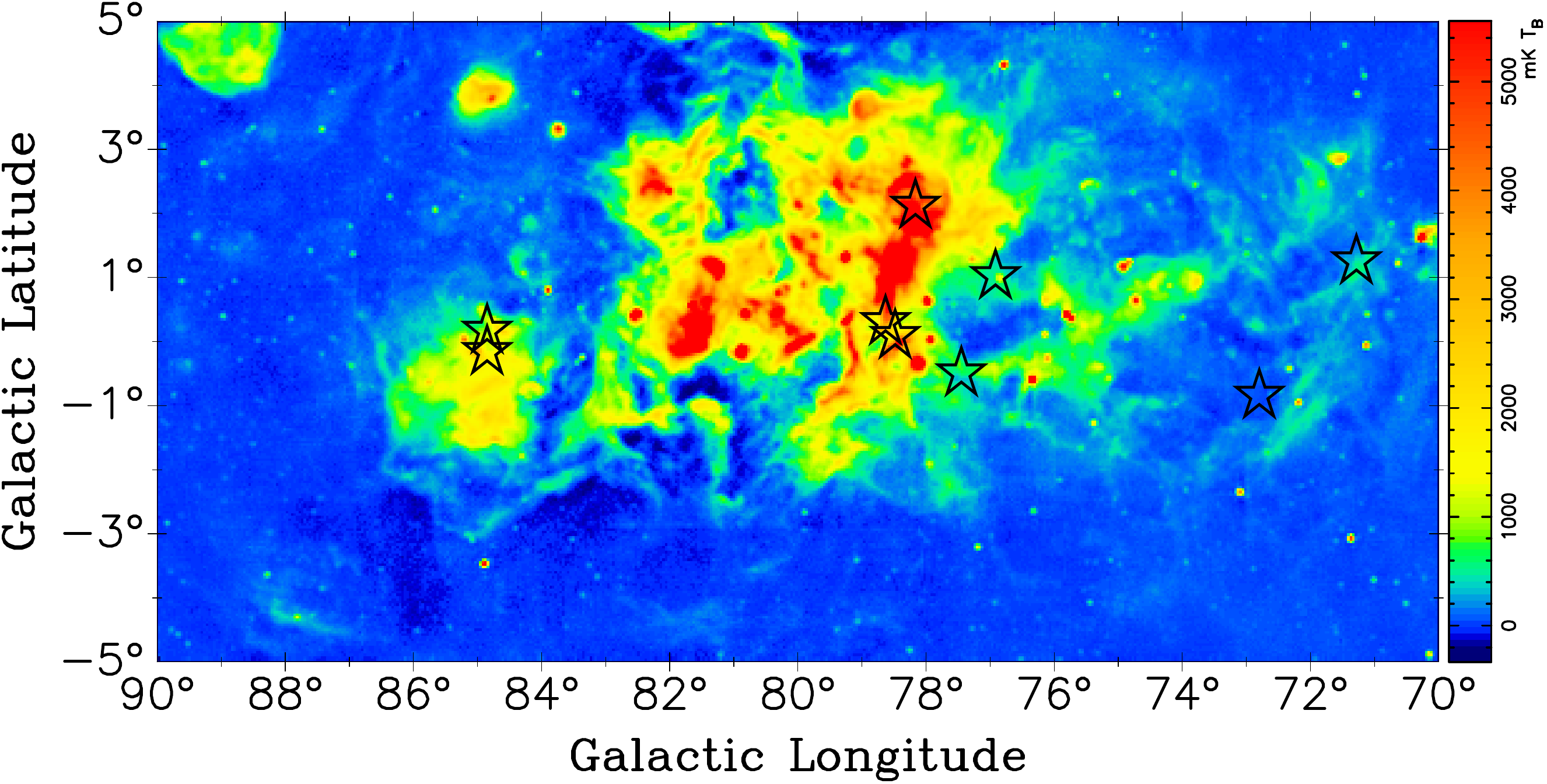}
   \includegraphics[height=52mm]{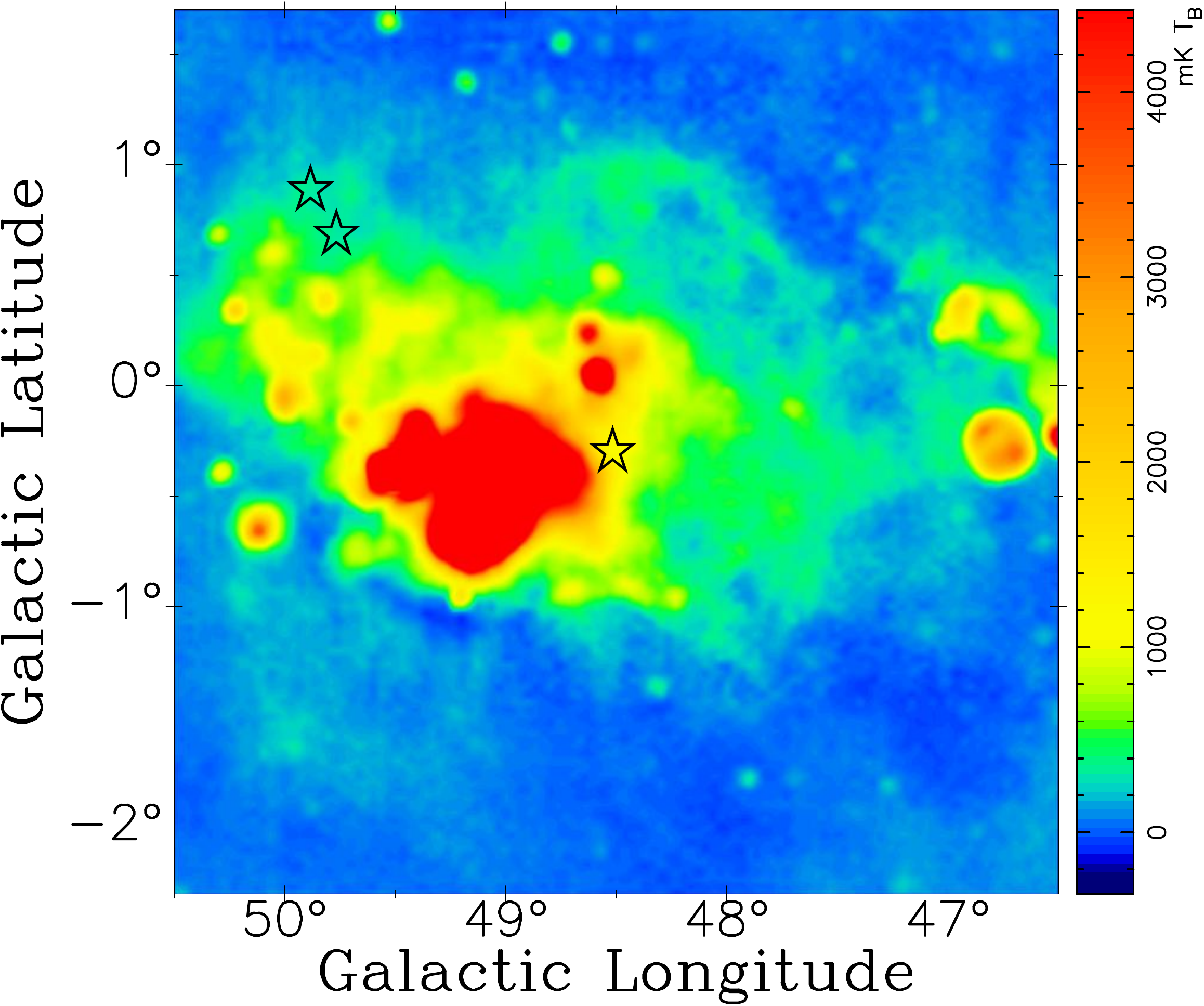}
   \caption{\baselineskip 3.8mm These pulsars have a large DM, marked as stars on the
     11\,cm radio map \citep{rfrr90,frrr90}, probably because they are
     located just in the tangential directions of the Local Arm ({\it the
     left panel}) and the Sagittarius arm ({\it the right panel}), where HII
     regions obviously contribute significant DMs.}
   \label{fig14_cygMap}
\end{figure*}

\section{Discoveries of new pulsars}
\label{sect:res}

The available GPPS survey data have been processed for only the first
round. About 330 previously known pulsars have been detected, some of
which will be discussed in the next section for updated
parameters. Here we report 201 newly discovered pulsars, including
several pulsars with polarization measurements and an RRAT found
through the single pulse module. The integrated profiles of newly
discovered pulsars are featured in Figure~\ref{fig11_gppsPSRprof}. In
Table~\ref{gppsPSRtab1}, we list the basic parameters for new
discoveries, including a temporary name with a suffix ``g'', period
(in second{s}), DM, coordinates in RA (J2000) and Dec (J2000) which
have an uncertainty of less than $1.5'$, and the Galactic coordinates
GL and GB, an averaged flux density at 1.25~GHz, two values of derived
distances by using the electron density models of NE2001 \citep{cl01}
and YMW16 \citep{ymw17}. Often the periods found in the search process
are harmonics, so that a correct period has to be found
(e.g. 2{$\times$}, 3{$\times$}, 5{$\times$}, 7{$\times$},
11{$\times$}, 13{$\times$} and 17{$\times$} the harmonic period). For
a bright pulsar detected in several beams, its position is determined
from a weighted average of the center positions of these beams; here
the S/Ns of profiles from these beams are taken as the weighting
factor. So-determined position has an accuracy of better than
$1'$. For a weak pulsar detected in just one beam, the beam center
is simply taken as the position, so that the position uncertainty is
about the beam size that is about $1.5'$ in radius. It is possible
that a pulsar is very offset from the center of a beam, so that it is
stronger but looks like a weaker pulsar due to the reduced gain of the
beam.

Pulsar flux densities are estimated from the integrated pulse energy
over the off-pulse deviations which are related to $T_{\rm sys} +
T_{\rm sky}$. We extract their values for the flux density estimations
since $T_{\rm sys}$ depends on the zenith angle \citep{jth+20} and
$T_{\rm sky}$ varies with the Galactic longitude and latitude
\citep{rfrr90,frrr90}. The so-estimated flux densities of Parkes
pulsars are compared to the values in the ATNF Pulsar Catalogue, and
we found excellent consistence (see Fig.~\ref{fig12_fluxcomp}).

Follow-up observations are going on for many newly discovered pulsars,
especially binaries. On the website for the FAST GPPS
survey\footnote{\it\url{http://zmtt.bao.ac.cn/GPPS/}} are releases of new
discoveries and updates of survey status and pulsar parameters.

In the following{,} we highlight some interesting pulsars, and the first are
$\upmu$Jy weak pulsars.

\begin{table*}
\centering
\begin{minipage}{13cm}
\centering\caption{Distant Pulsars with Excess DMs Challenging the Currently Widely Used Models \label{farPSR}}\end{minipage}

\fns
 \begin{tabular}{ccccccccc}
\hline \noalign{\smallskip}
PSR name &  gpps No.&     GL  &    GB       &      DM      & D$_{\rm NE2001}$ & D$_{\rm YMW16}$ &DM$^{\rm max}_{\rm NE2001}$ & DM$^{\rm max}_{\rm YMW16}$\\
    &     &($^{\circ}$)&($^{\circ}$) & (cm$^{-3}${\,}pc) & (kpc)  & (kpc) & (cm$^{-3}${\,}pc)  & (cm$^{-3}${\,}pc)  \\
 \hline\noalign{\smallskip}
J2052+4421g & gpps0019 &  84.8417  & $-$0.1616  &   547.0   &  50.0&  25.0 &  365.6  & 428.5  \\
J2051+4434g & gpps0085 &  84.8479  & $+$0.1706  &   616.0   &  50.0&  25.0 &  365.5  & 443.2  \\
J2030+3944g & gpps0118 &  78.6299  & $+$0.2988  &   937.4   &  50.0&  25.0 &  398.1  & 466.3  \\
J2030+3929g & gpps0152 &  78.4765  & $+$0.0848  &   491.9   &  50.0&  25.0 &  400.4  & 456.8  \\
J2021+4024g & gpps0087 &  78.1680  & $+$2.1153  &   680.5   &  50.0&  25.0 &  349.2  & 500.2  \\
J2030+3818g & gpps0187 &  77.4492  & $-$0.5085  &   596.7   &  50.0&  25.0 &  401.9  & 427.3  \\
J2022+3845g & gpps0076 &  76.9110  & $+$1.0169  &   487.5   &  50.0&  17.2 &  392.9  & 495.1  \\
J2005+3411g & gpps0077 &  71.2811  & $+$1.2438  &   489.0   &  50.0&  17.3 &  415.6  & 501.8  \\
J1919+1527g & gpps0130 &  49.8846  & $+$0.8895  &   697.5   &  50.0&  16.9 &  612.7  & 771.5  \\
J1920+1515g & gpps0086 &  49.7623  & $+$0.6780  &   655.5   &  50.0&  15.7 &  630.7  & 777.6  \\
J1921+1340g & gpps0088 &  48.4946  & $-$0.3047  &   754.9   &  50.0&  25.0 &  674.9  & 754.3  \\
 \hline
\end{tabular}
\end{table*}

\subsection{Discovery of Faintest Pulsars}

As shown in Figure~\ref{fig13_fluxP} and also flux values in
Table~\ref{gppsPSRtab1}, the GPPS survey can detect weak pulsars.
There are 23 pulsars with a flux density less than 10\,$\upmu$Jy at
1.25~GHz, much weaker than those previously detected by Parkes pulsar
surveys or Arecibo pulsar surveys. Up to now, the weakest known pulsar
{was} discovered in the GPPS survey, that is
PSR J1924+2037g (gpps0192, S$_{\rm   1250 MHz}$ = $3.3\,\upmu$Jy),
a very nulling pulsar discovered first via
the single-pulse module and later sorted by{ the} {PRESTO} module.
A number of weakest pulsars around $5\,\upmu$Jy are
PSR J2018+3418g (gpps0189, 4.9\,$\upmu$Jy),
PSR J1928+1915g (gpps0004, 5.1\,$\upmu$Jy),
PSR J1949+2516g (gpps0111, 5.3\,$\upmu$Jy),
PSR J1926+1452g (gpps0058, 5.5\,$\upmu$Jy) and
PSR J1955+2912g (gpps0123, 5.5\,$\upmu$Jy).

On the other hand, the intrinsically faintest pulsars should have
both flux density and distance considered.  Currently, three very nearby pulsars,
PSR J1908+1035g (gpps0114, 11.6\,$\upmu$Jy, DM = 10.9\,cm$^{-3}$pc, Dist$_{\rm YMW16} = 0.7$~kpc, L = 5.7 $\upmu$Jy~kpc$^2$){,}
PSR J1854+0704g (gpps0161, 34.6\,$\upmu$Jy, DM = 10.8\,cm$^{-3}$pc, Dist$_{\rm YMW16}$ = 0.6 kpc, L = 12.5 $\upmu$Jy~kpc$^2$){ and}
PSR J1928+1902g (gpps0163, 7.1\,$\upmu$Jy, DM = 29.8\,cm$^{-3}$pc, Dist$_{\rm YMW16}$ = 1.5 kpc, L = 16.0\,$\upmu$Jy~kpc$^2$)
have a luminosity of less than 20 $\upmu$Jy~kpc$^2$, which are the
faintest known pulsars discovered by the GPPS survey. The luminosity
distribution given in the lower panel of Figure~\ref{fig13_fluxP} clearly
affirms that the GPPS survey has significantly improved the
determination of the faint end of the pulsar luminosity function, which
has to be included in many relevant simulations
\citep[][]{lpr+19,hh20}.

\begin{figure}
   \centering
   \includegraphics[height=75mm]{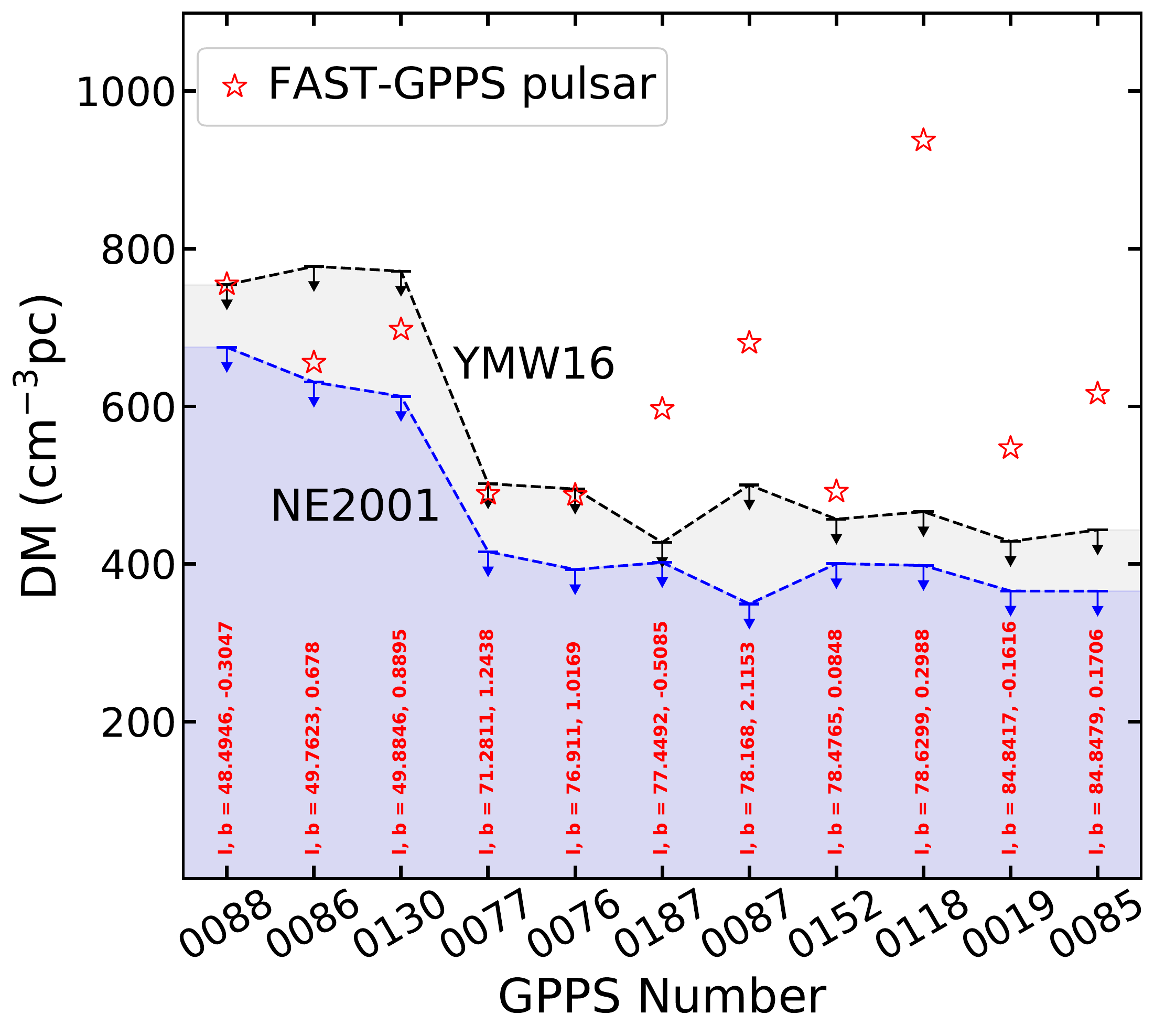}
   \caption{\baselineskip 3.8mm DMs of the newly discovered GPPS
     pulsars in Table~\ref{farPSR} are larger than the maximum given
     by YMW16 model \citep{ymw17} and/or NE2001 model \citep{cl01}.}
   \label{figAdd}
\end{figure}

\subsection{Discovery of Pulsars with Excess DMs Not Modeled}

By looking at Table~\ref{gppsPSRtab1}, one may notice that some
pulsars have large DM{s}, and hence their distances estimated based on
the electron distribution models \citep{cl01,ymw17} are very large,
i.e., $>25$~kpc in the YMW16 model \citep{ymw17} or $>50$~kpc in the
NE2001 model \citep{cl01}, see Fig.~\ref{figAdd}, which are an
indication of their possible locations outside the Milky Way. We
extract their relevant parameters and put them in
Table~\ref{farPSR}. Notice that these pulsars are in the direction of
the Local Arm \citep{xrd+16} or the tangential direction of spiral
arms \citep{hh14}, therefore the large DMs are not surprising at these
low Galactic latitudes since lines of sight will intersect spiral arms
that are wider than those in the models.  HII regions in the lines of
sight as shown in Figure~\ref{fig14_cygMap} should contribute a large
amount of thermal electrons.

\begin{table*}
\centering
\begin{minipage}{8cm}
  \caption{Pulsars Coincident with Supernova Remnants }
  \label{snrPSR}\end{minipage}

  \fns
\setlength{\tabcolsep}{1.5pt}
 \begin{tabular}{ccccccccccccc}
\hline\noalign{\smallskip}
SNR  &  RA(2000)   & Dec(2000) & Size &  D$_{\rm SNR}$ &  gpps & RA(2000) & Dec (2000)  & offset& P   &DM            &  D$_{\rm YMW16}$ & D$_{\rm NE2001}$ \\
     &  (hh:mm:ss) & (dd:mm)   & ($'$) &  (kps)      &   No. &  (hh:mm:ss) & (dd:mm)  & ($'$) & (s) &(cm$^{-3}${\,}pc)  &   (kpc)  &  (kpc)     \\
\hline\noalign{\smallskip}
G33.2$-$0.6 & 18:53:50 & $-$00:02 & 18          &     & gpps0110 & 18:53:24 & $-$00:03 &  $6.6'$ & 0.1715 & 667.2 & 9.1 & 6.3 \\
G65.1$+$0.6 & 19:54:40 & $+$28:35 & 90$\times$50 & 8.7--10.1 & gpps0046 & 19:56:48 & $+$28:26 & $29.4'$ & 0.0718 & 112.0 & 4.8 & 6.5 \\
            &          &          &              &    & gpps0064 & 19:52:48 & $+$28:36 & $24.5'$ & 0.0180 & 313.0 & 9.7 & 10.5 \\
            &          &          &              &    & gpps0123 & 19:55:11 & $+$29:12 & $39.7'$ & 0.2795 & 193.0 & 7.0 & 7.4 \\
G78.2$+$2.1 & 20:20:50 & $+$40:26 &  60          & 1.9& gpps0087 & 20:21:13 & $+$40:24 & $4.3'$  & 0.3705 & 680.5 & 25.0& 50.0 \\
 \hline
\end{tabular}
\end{table*}

How much DM can an HII region contribute to a pulsar? A quick answer
comes from the DMs of pulsars in the outer Galaxy
($90^{\circ}<$GL$<270^{\circ}$) where mostly HII regions in the
Perseus arm contribute. In the ATNF Pulsar Catalogue of
\citet{mhth05}, these pulsars in general have DM values of several
tens, but 10 pulsars have DM between 200 to 300 cm$^{-3}$~pc. If
occasionally the line of sight to a pulsar passes through three such
HII regions, it is not impossible to have a DM of several hundred. The
probability is very small for a pulsar with such radio flux densities
located in a background galaxy just behind these spiral arms. We
therefore suspect that these pulsars in Table~\ref{farPSR} are located
in or just behind the spiral arms and that these HII regions in these
spiral arms are responsible for the observed large DMs. In other
words, the electron density from HII regions in these spiral arms is
probably underestimated and should be updated in the electron density
distribution models. In Table~\ref{farPSR}, the asymptotic maximum DM
values provided by the two models are also listed, which shows that
the observed DMs have exceeded the maximum DM in the model. The
largest DM excess is given by PSR J2030+3944g (gpps0118, DM=937.4
cm$^{-3}$~pc), while the modeled DM is around 400 -- 500
cm$^{-3}$~pc. Obviously, a better model is desired to account for the
excess DMs.

\begin{figure*}
  \centering
  \includegraphics[width=0.33\textwidth]{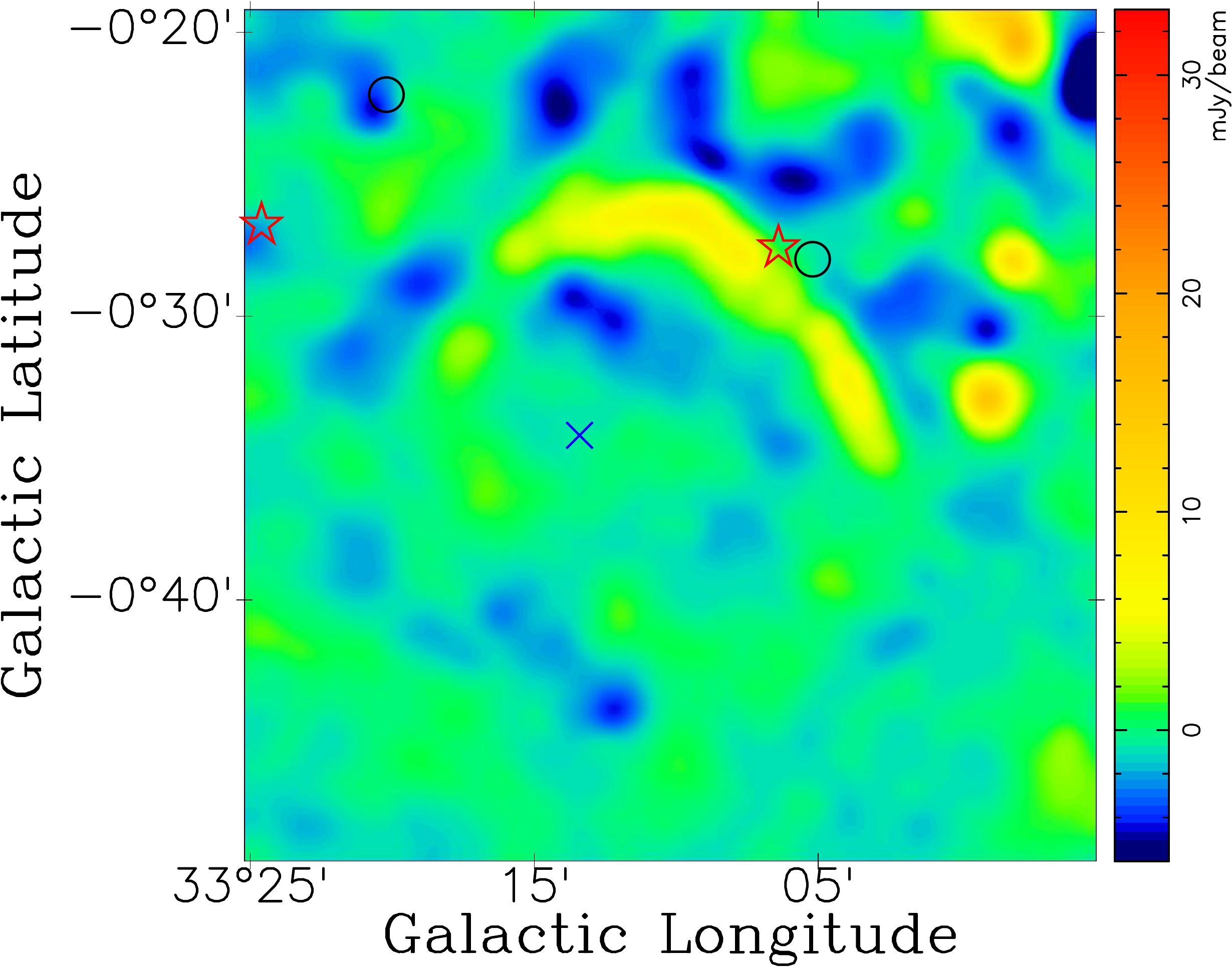}
  \includegraphics[width=0.33\textwidth]{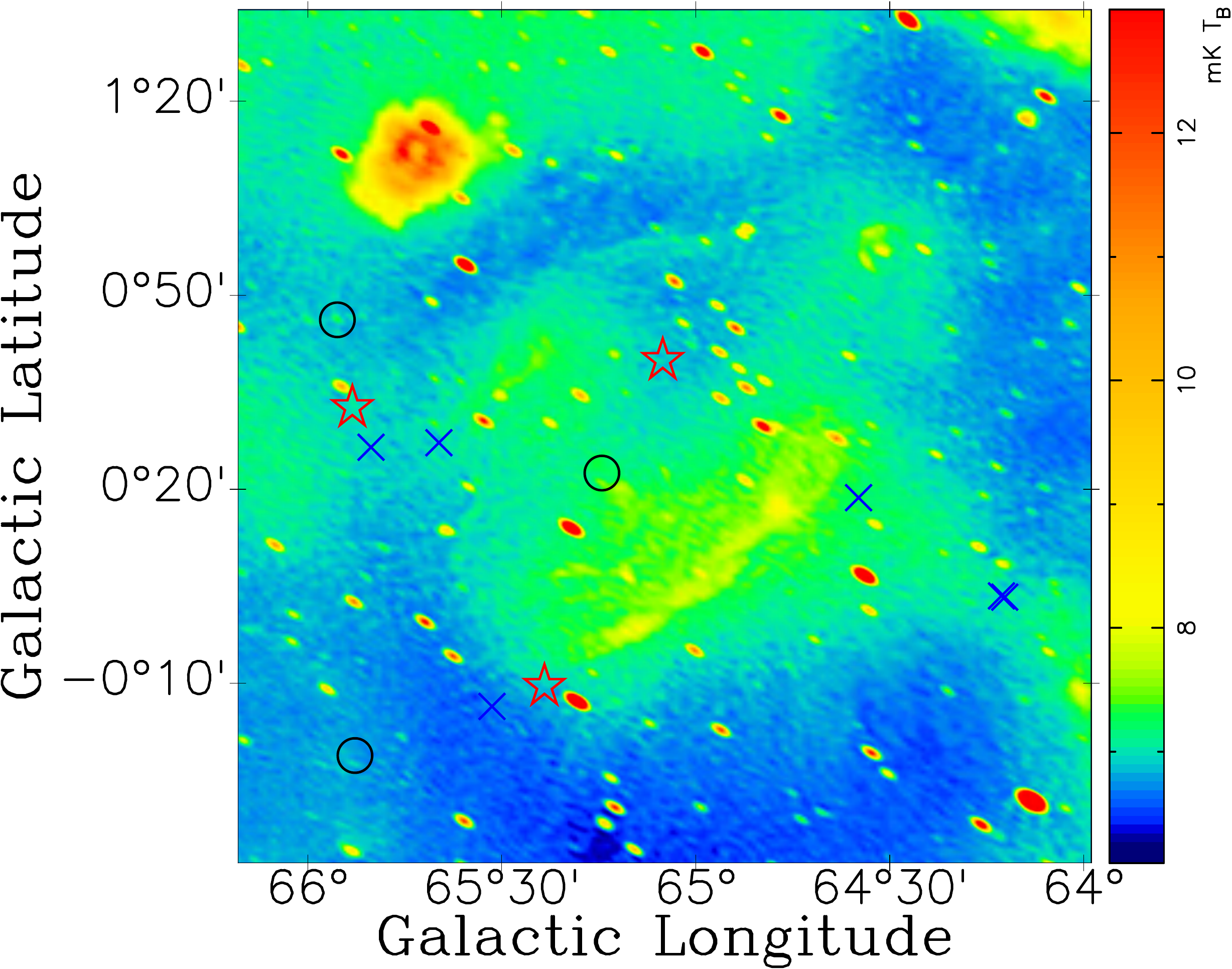}
  \includegraphics[width=0.33\textwidth]{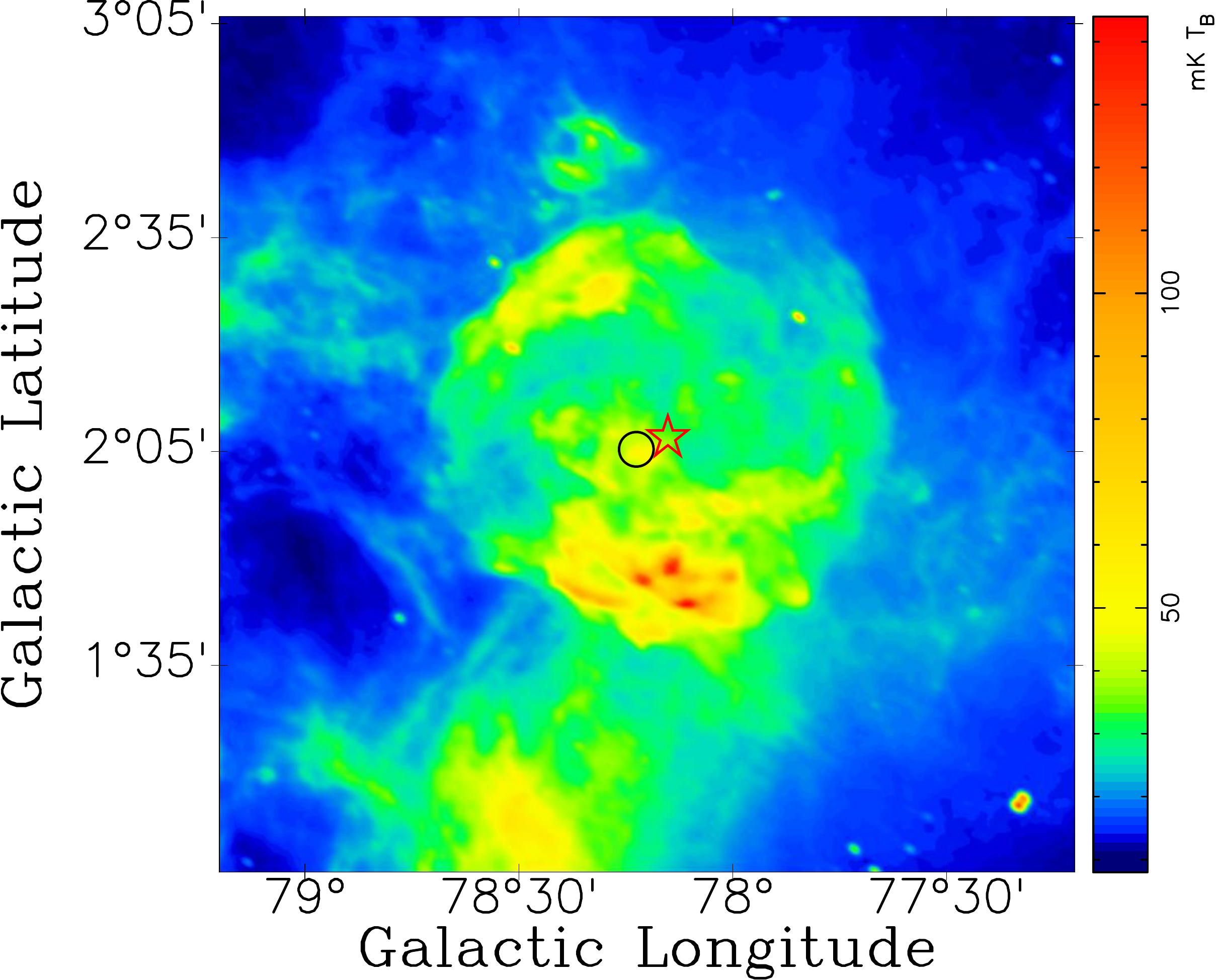}
  \caption{\baselineskip 3.8mm Newly discovered pulsars coincident with supernova
    remnants, signified as stars in these plots. Previously known pulsars
    are marked as crosses or circles (ref. Fig.~\ref{fig13_fluxP}). The
    background radio images are extracted from the NRAO VLA Sky Survey
    \citep{ccg+98} for SNR G33.2$-$0.6 and from the Canadian Galactic Plane
    Survey \citep{eti+98,tgp+03} for the others. }
   \label{fig15_psrSNR}
\end{figure*}

\subsection{Discovery of Pulsars in Supernova Remnants?}

Among the newly discovered pulsars, several are coincident with
known supernova remnants \citep[SNRs, see a catalog by][]{gre19}, as
seen in Figure~\ref{fig15_psrSNR} and listed in Table~\ref{snrPSR}. Then the
question arises: are they physically associated?

To ensure the association, a simple criterion must be satisfied that a
young pulsar must be inside the remnant in three-dimensional space. In
addition to coincidence on the sky, as displayed in
Figure~\ref{fig15_psrSNR}, accurate measurements for the distance to
the pulsar and the remnant are desired, which are in general difficult
to achieve. As seen in Table~\ref{snrPSR}, the distance estimate for
SNR G65.1+0.6 is in the range of 8.7$-$10.1~kpc \citep{tl06},
%
%
and that for G78.2+2.1 is 1.9~kpc \citep{lhh+13,szt+18}. The distance
of pulsars can be estimated from DMs. Given large uncertainties in the
estimated distances, other supplementary arguments for the association
could be that the pulsar is young, which has to be determined by the
period derivatives deduced from more TOA data which are being measured
now.

On the top-right of G33.2$-$0.6 (see Fig.~\ref{fig15_psrSNR}), we discovered
a pulsar PSR J1853$-$0003g (gpps0110) with a period of {$P$} = 0.17152~s and
DM = 667.2~cm$^{-3}$~pc, very nearby but different from a known pulsar
PSR J1853$-$0004 discovered by the Parkes multibeam survey \citep{hfs+04}
with a period of {$P$} = 0.101436~s and DM = 437.5~cm$^{-3}$~pc at a distance
of 5.3~kpc estimated by the YMW16 model \citep{ymw17}.

By looking at the period and estimated distances of these pulsars in
the field of SNR G65.1+0.6, one might guess that PSR J1956+2826g
(gpps0046) has a period similar to the values of young pulsars, but
the distance of PSR J1952+2836g (gpps0064) is closer to that of SNR
G65.1+0.6. \citet{tl06} argued previously that PSR J1957+2831
({$P$} = 0.307683~s, DM = 139.0~cm$^{-3}$~ps, GL = $65.5240^{\circ}$,
GB = $-0.2249^{\circ}$) is associated with the remnant, which in fact
is further away from the remnant than PSR J1956+2826g. Another new
pulsar, PSR J1955+2912g (gpps0123), is also further away and
probably not associated with the SNR.

\begin{figure*}
  \centering
  \includegraphics[width=40mm]{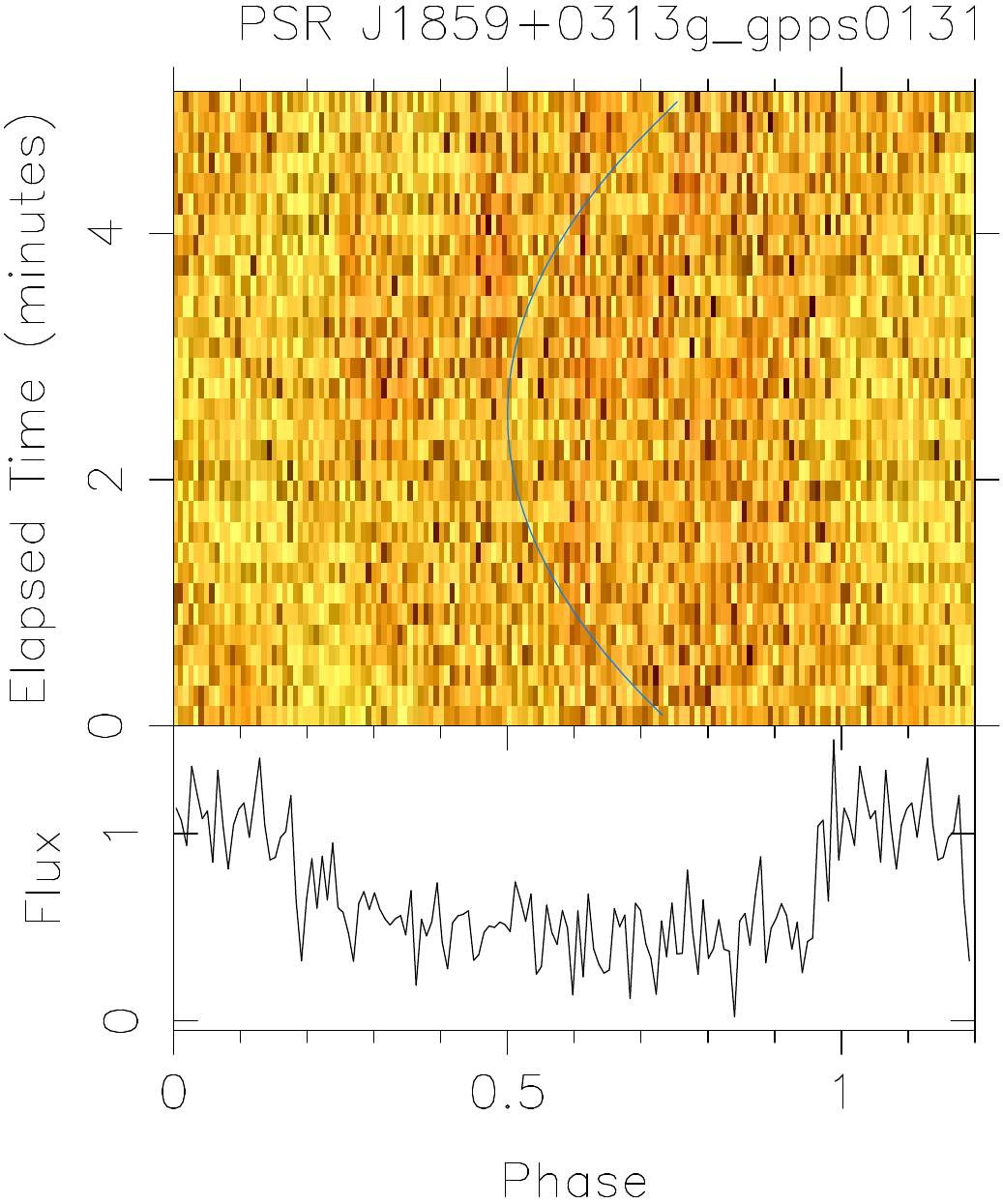}
  \includegraphics[width=40mm]{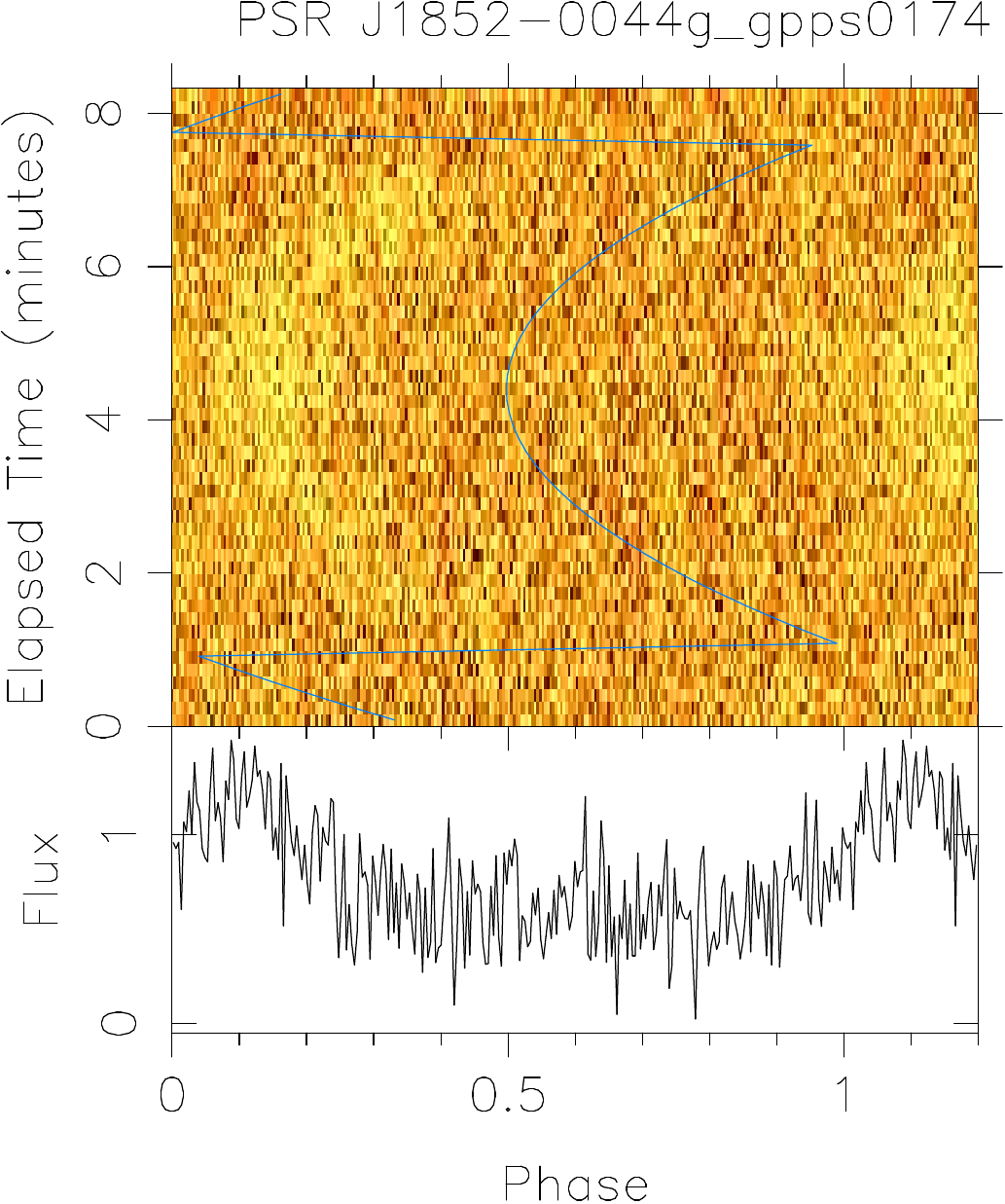}
  \includegraphics[width=40mm]{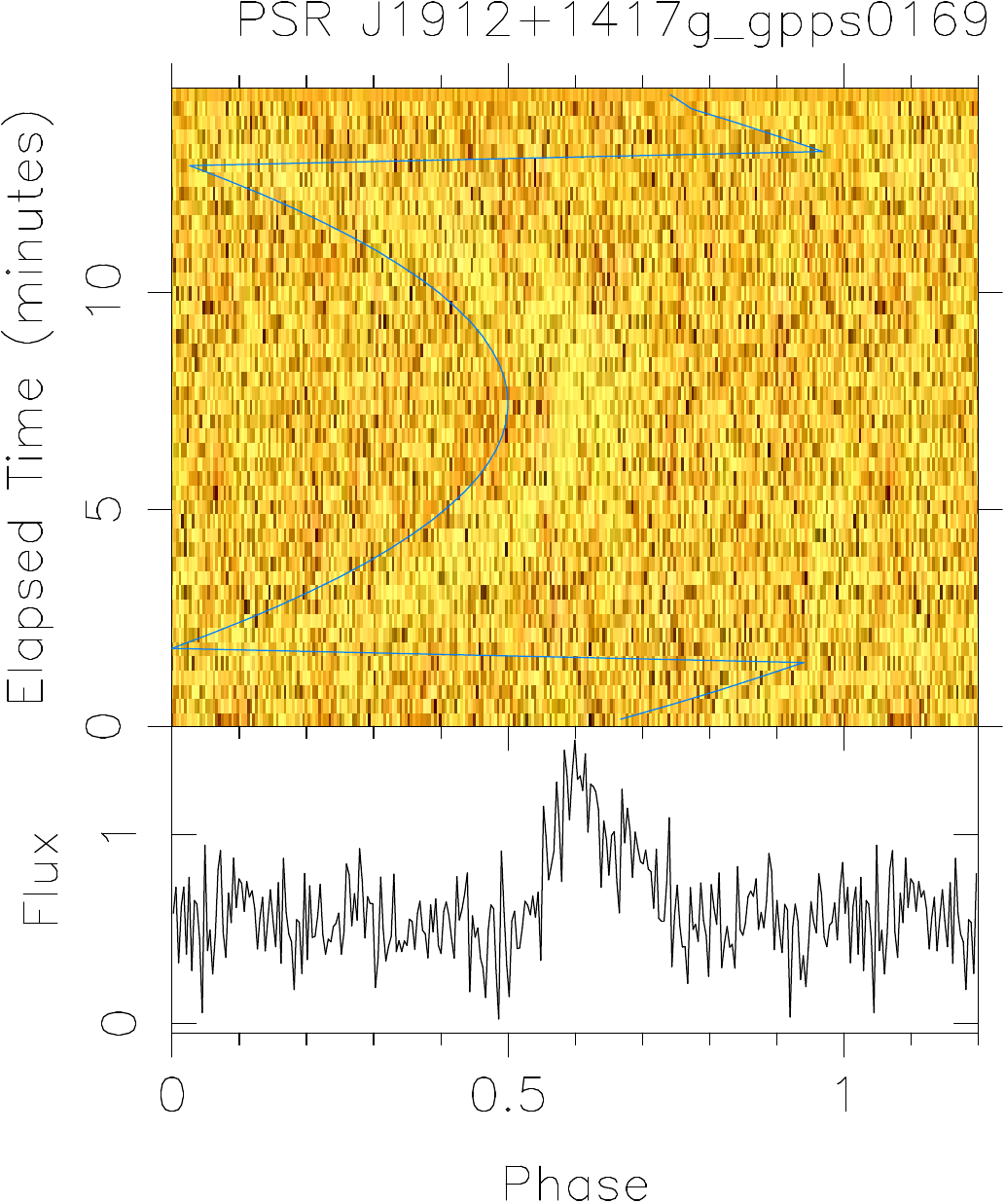}
  \includegraphics[width=40mm]{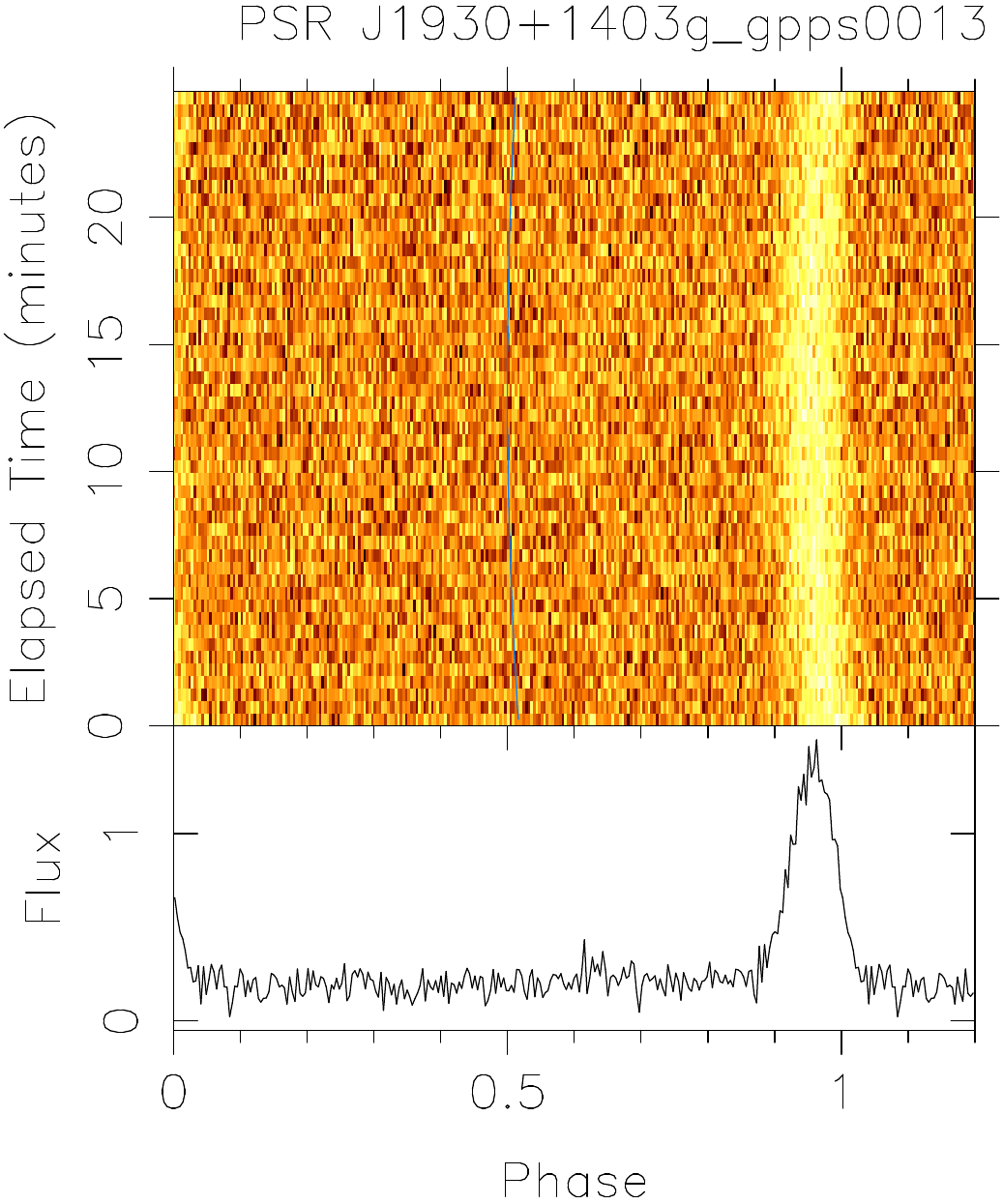}\\[1mm]
  \includegraphics[width=40mm]{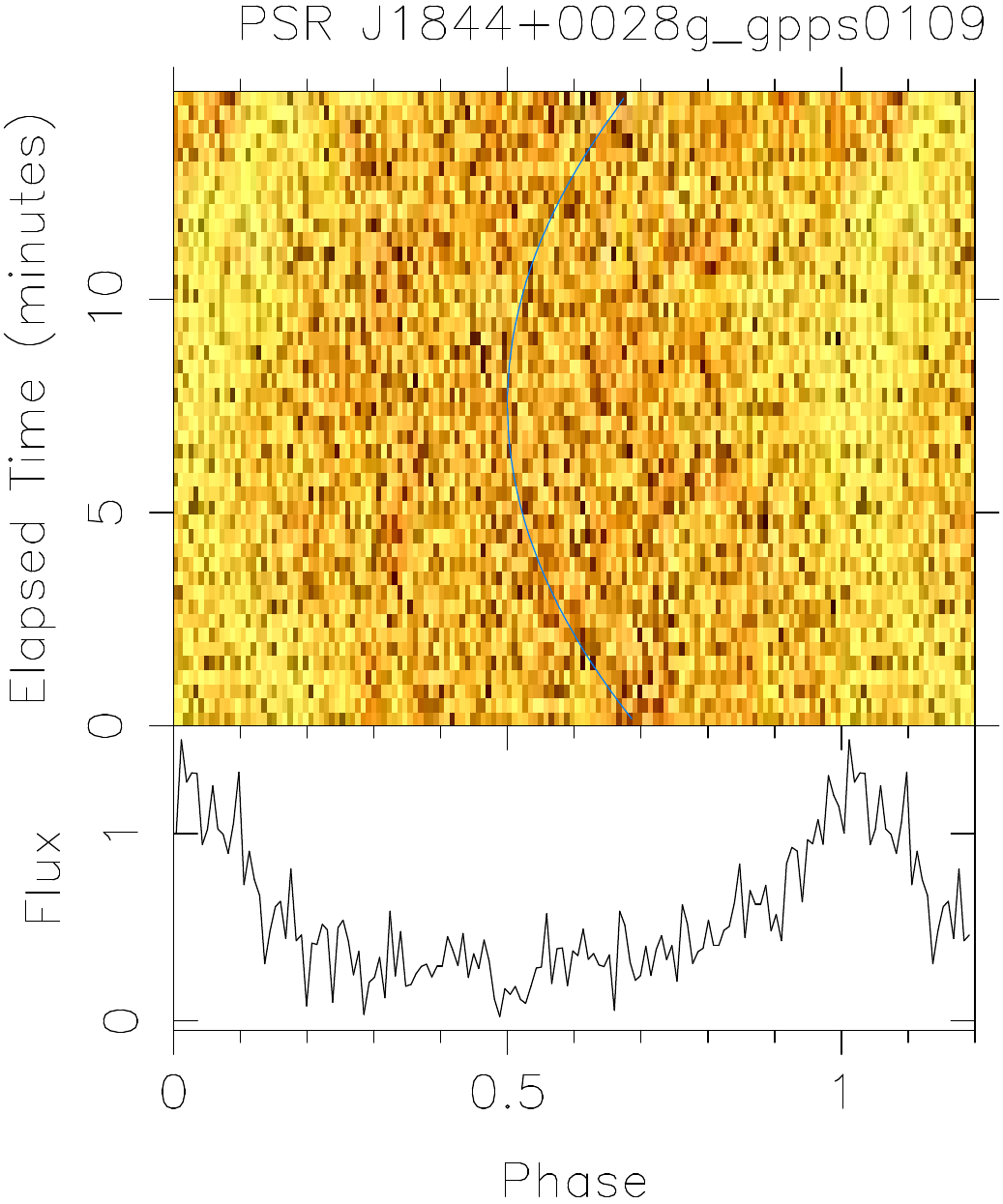}
  \includegraphics[width=40mm]{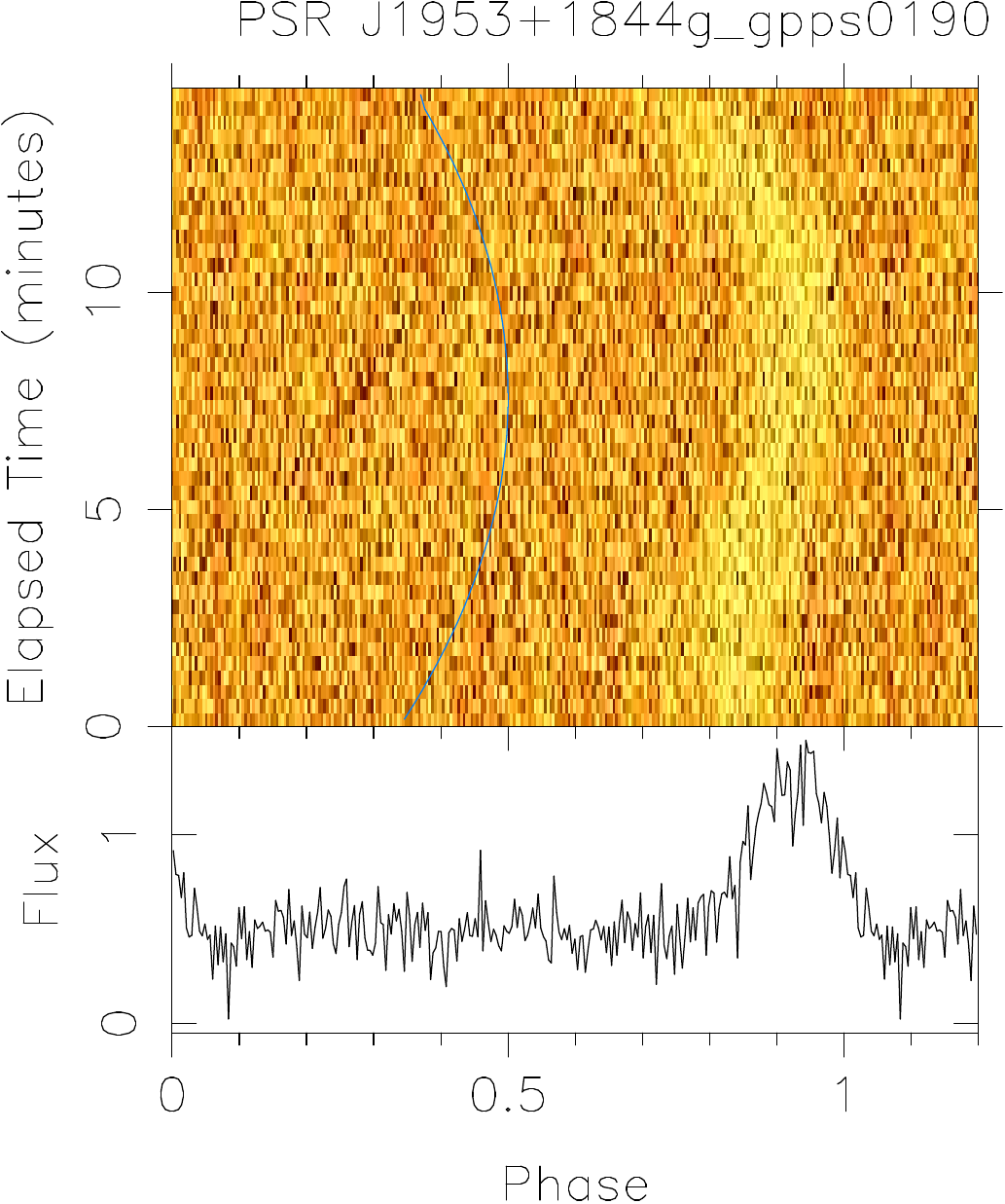}
  \includegraphics[width=40mm]{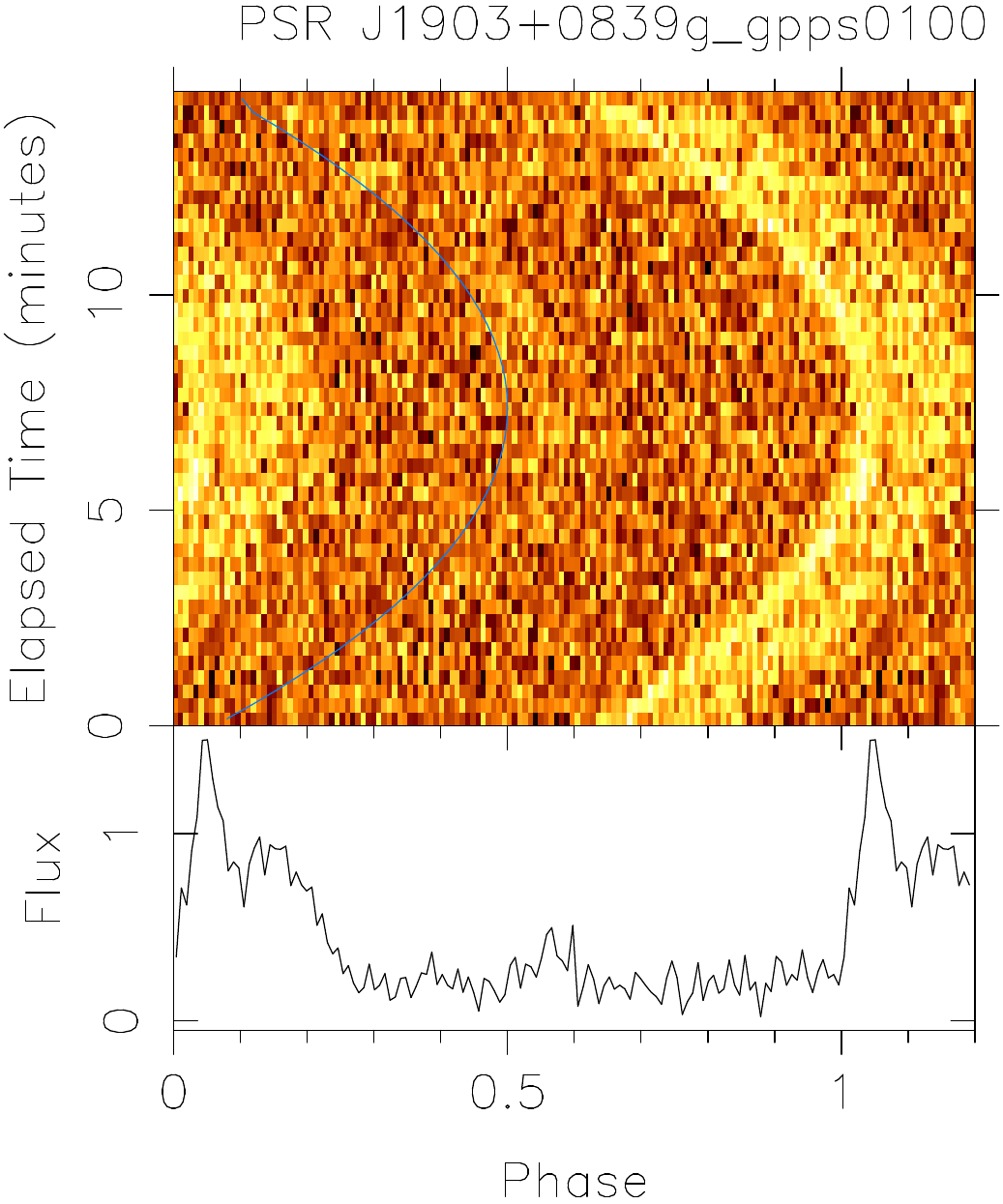}
  \includegraphics[width=40mm]{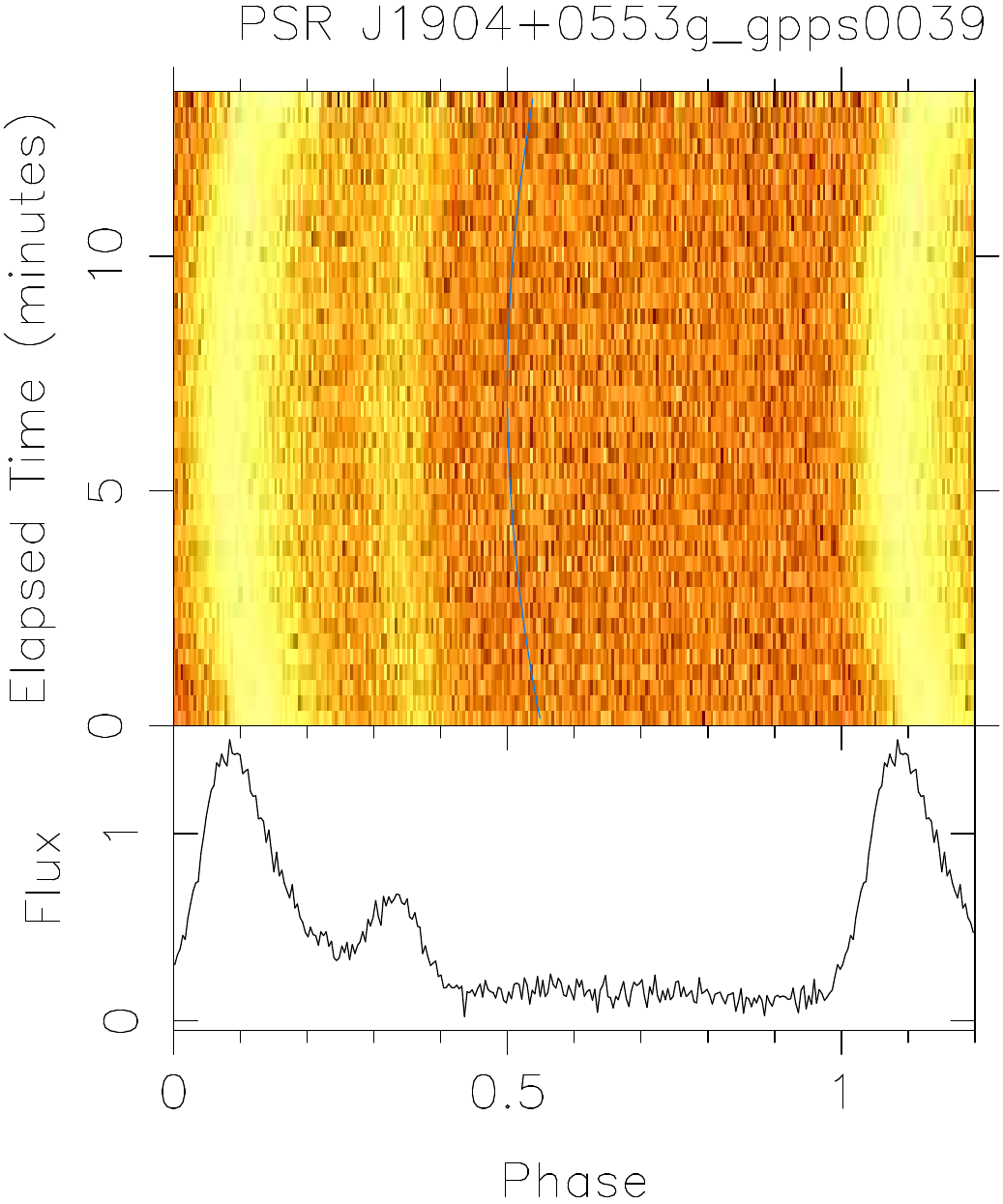}\\[1mm]
  \includegraphics[width=40mm]{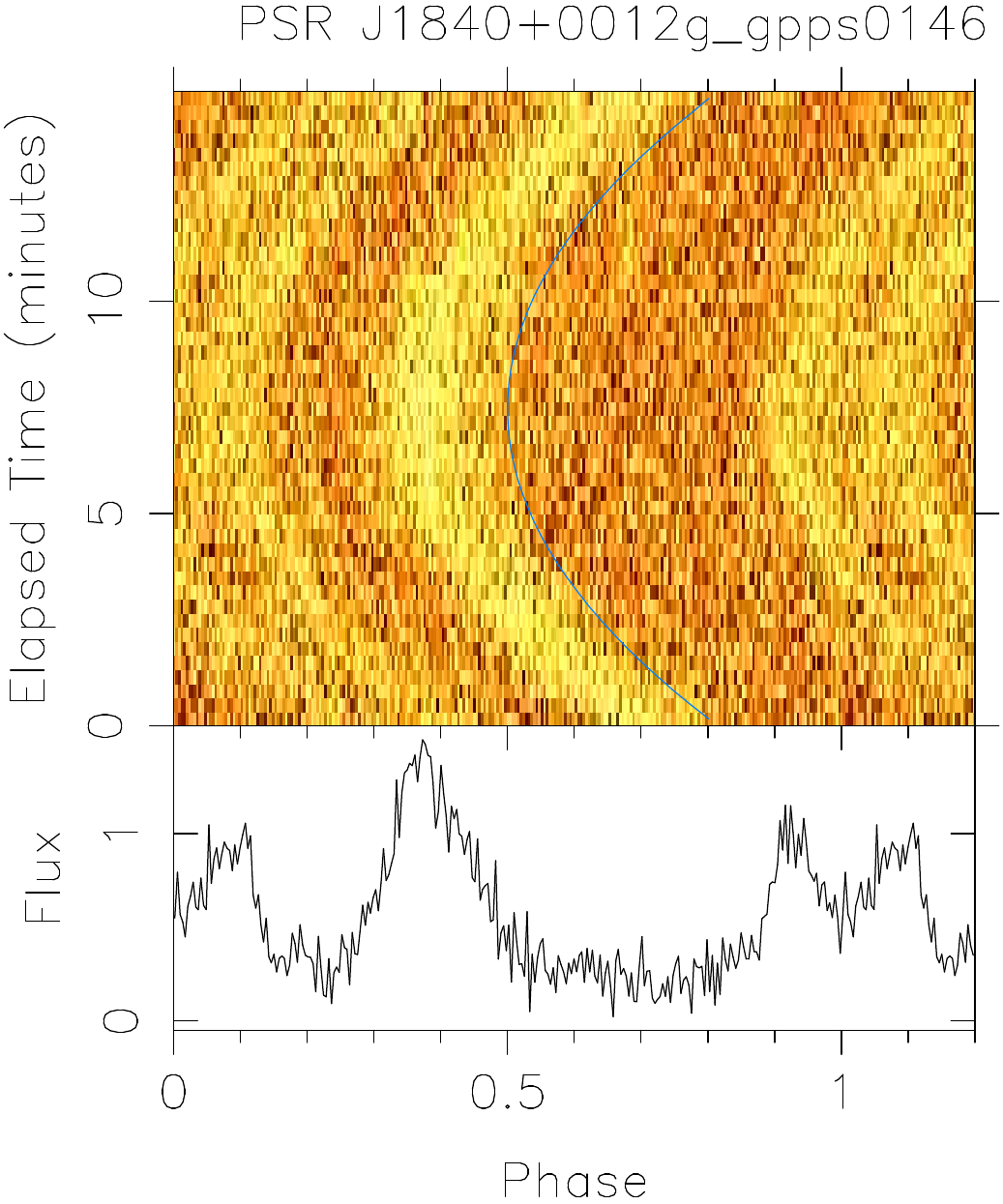}
  \includegraphics[width=40mm]{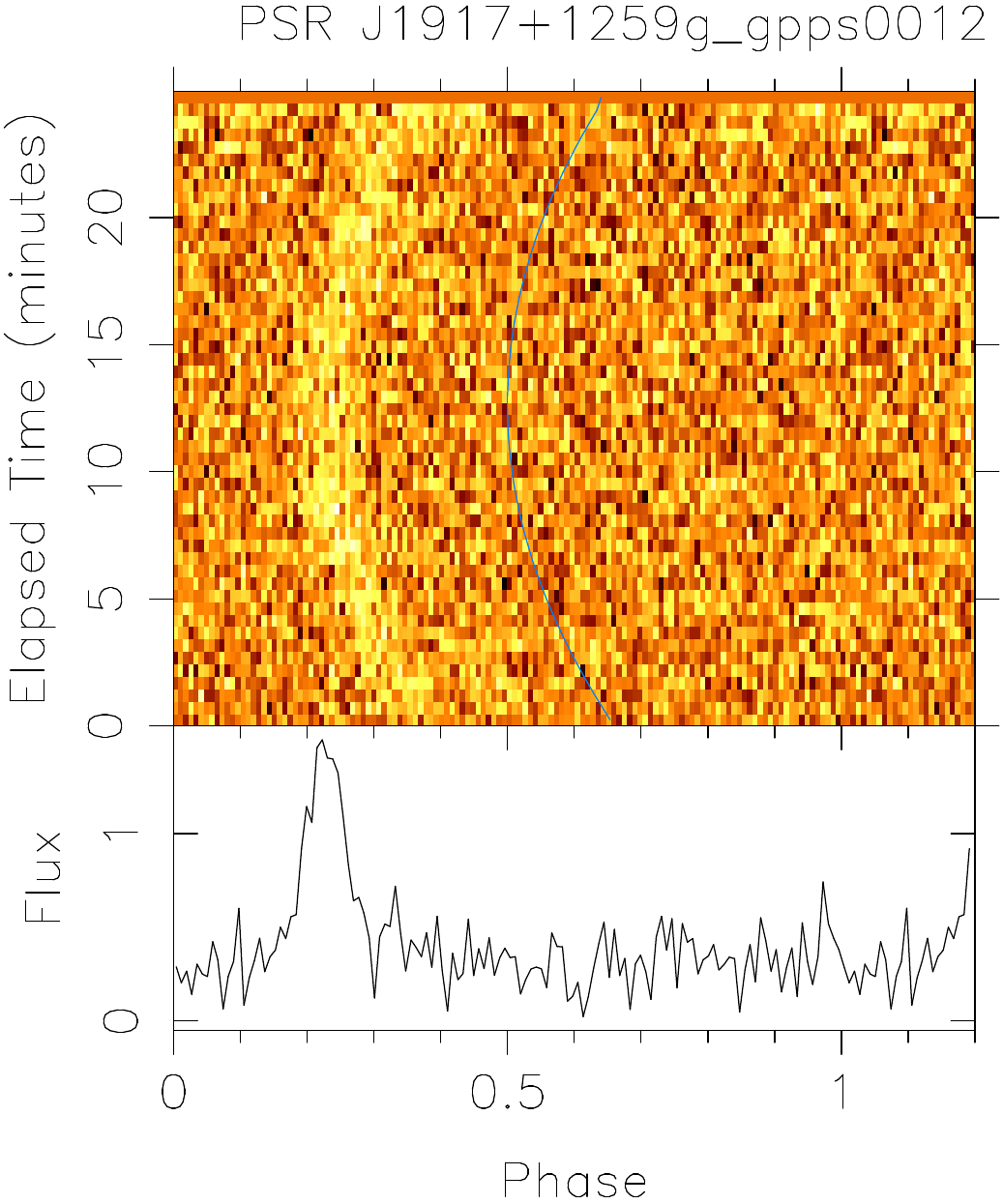}
  \includegraphics[width=40mm]{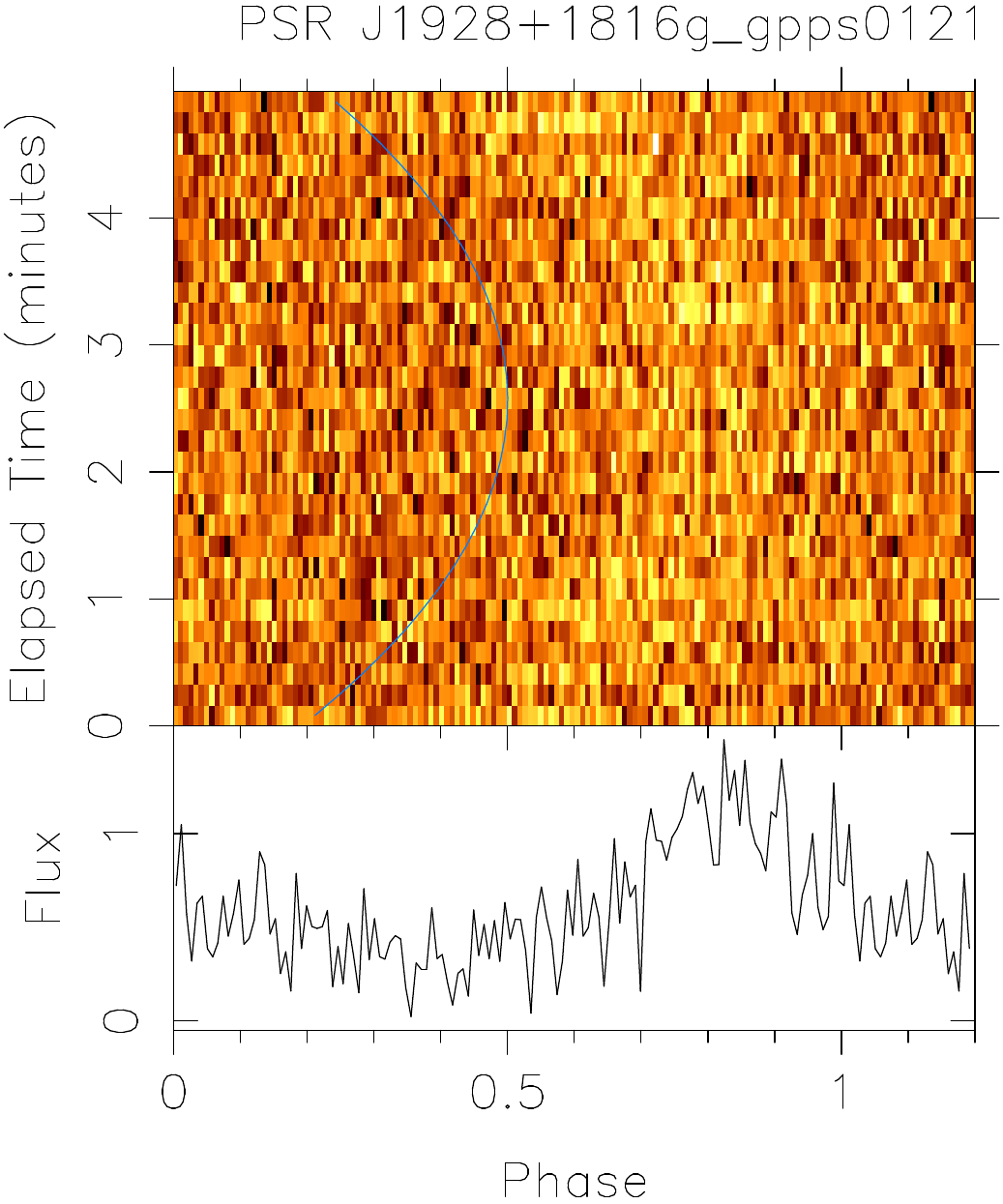}
  \includegraphics[width=40mm]{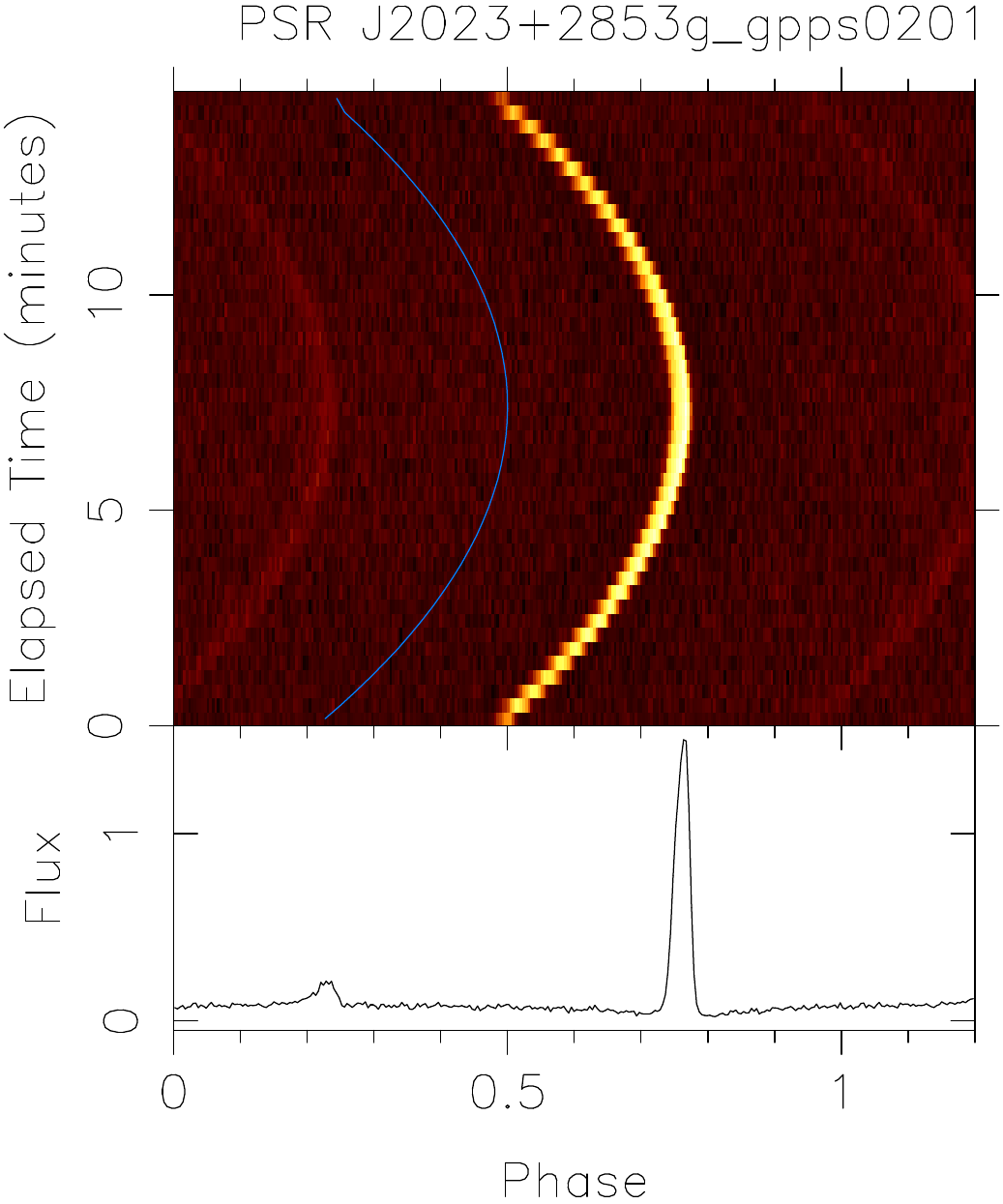}\\[1mm]
  \includegraphics[width=40mm]{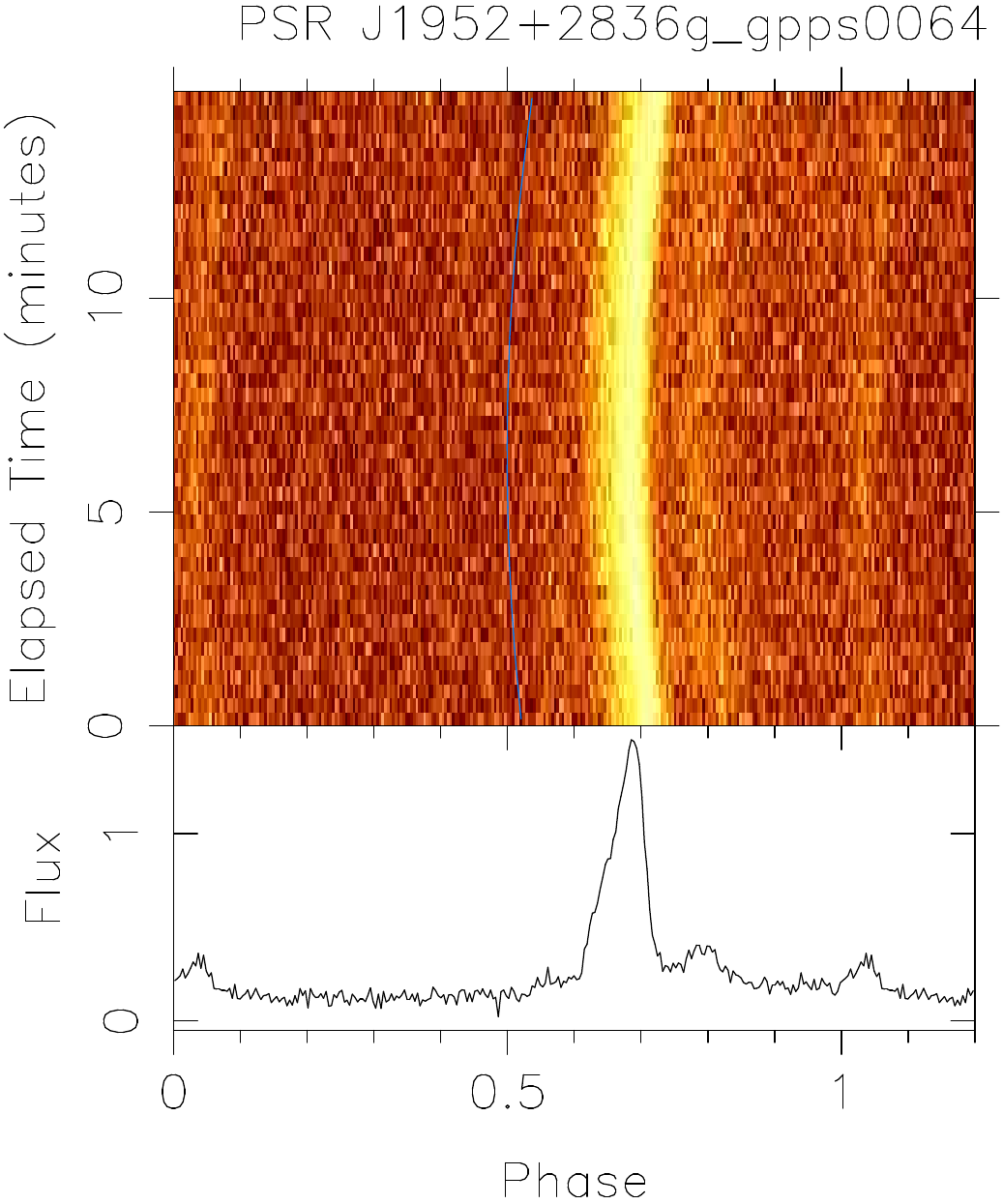}
  \includegraphics[width=40mm]{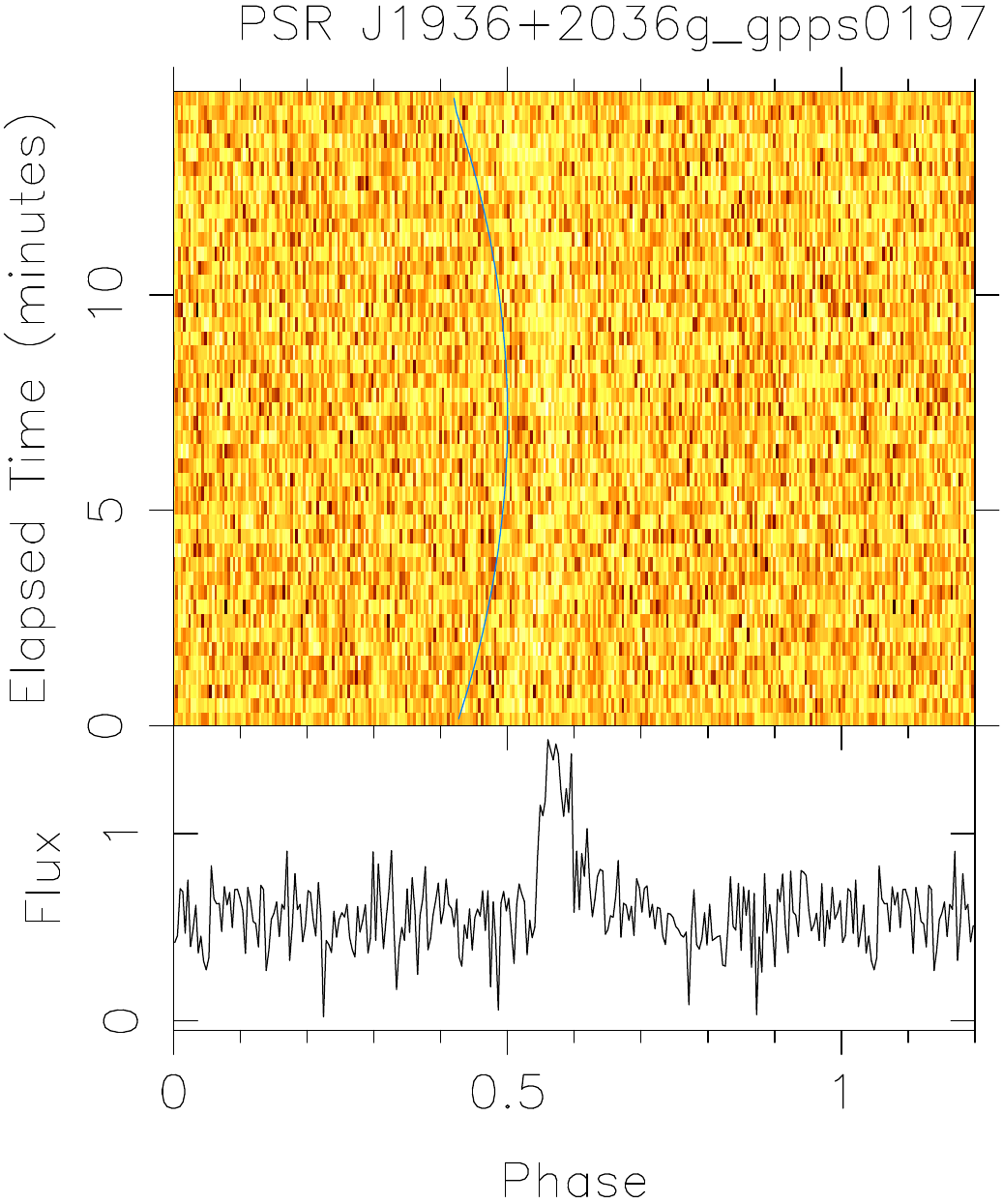}
  \includegraphics[width=40mm]{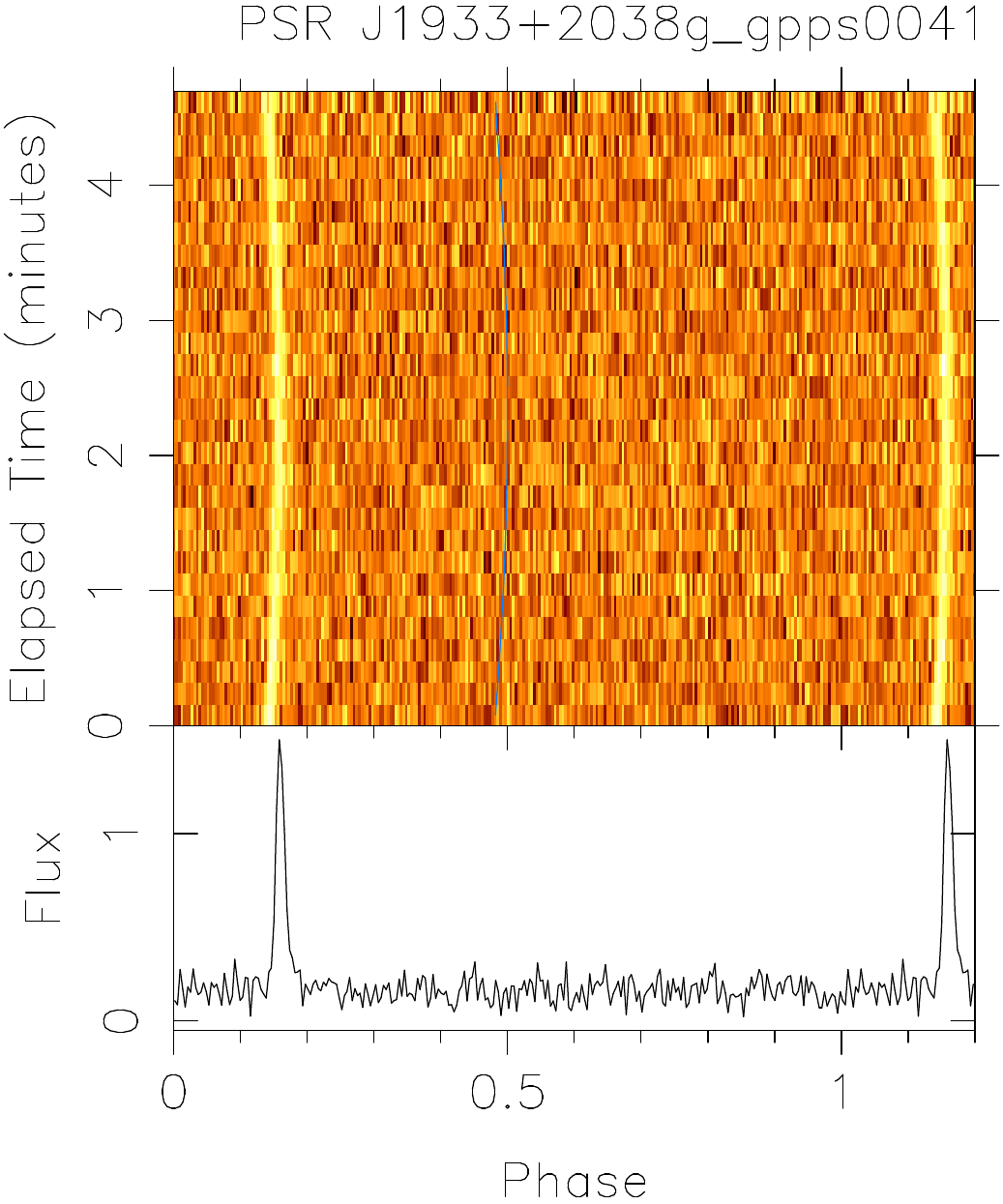}
  \includegraphics[width=40mm]{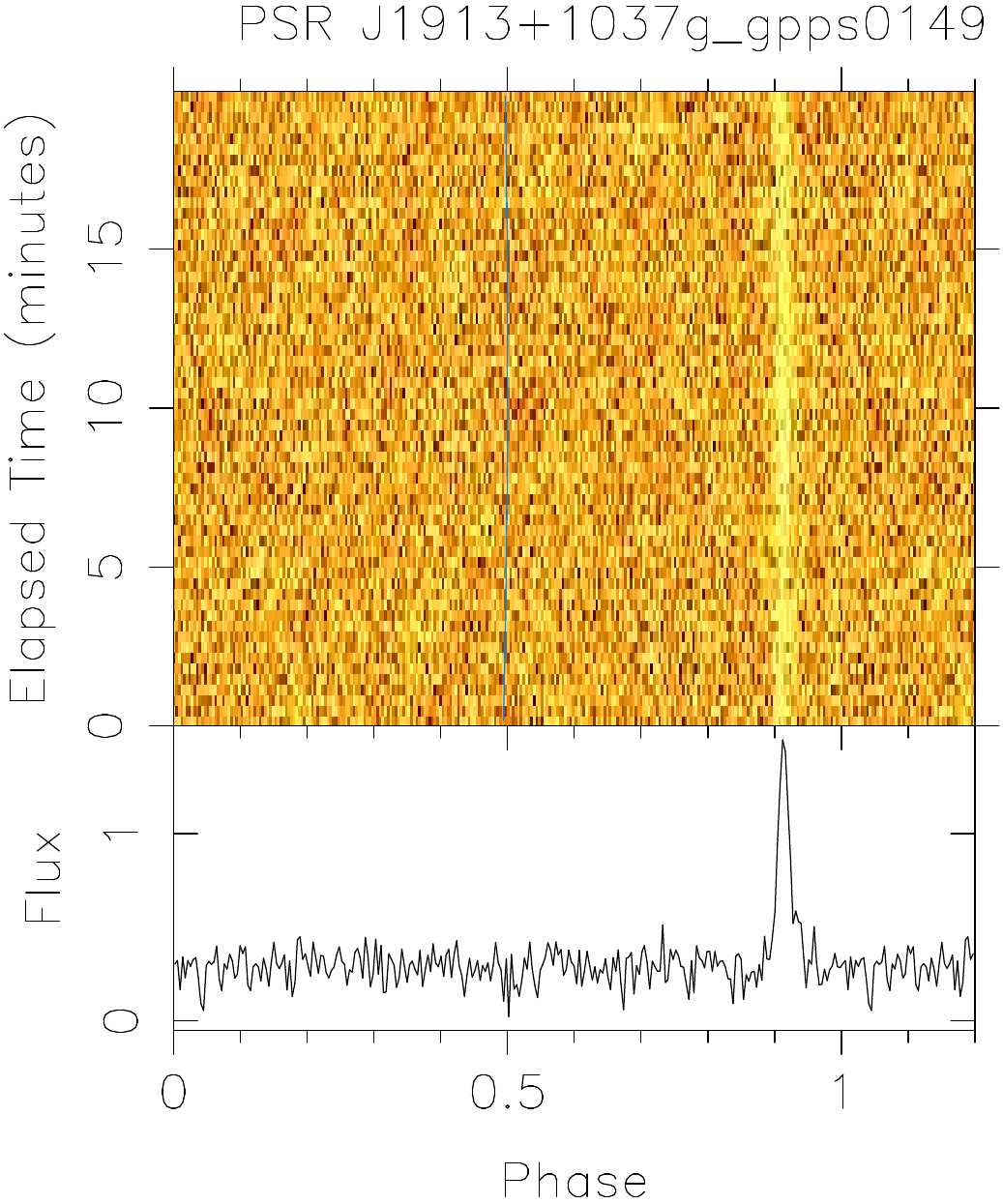}\\[1mm]
  \caption{\baselineskip 3.8mm Phase changes of pulses during a
    session of observations of a few newly discovered pulsars in the
    GPPS survey indicating their nature in a binary system. For each
    pulsar, the upper panel is the phase-time waterfall plot, with the
    phase defined in the unit of the spin period of a pulsar; the
    lower panel is the integrated profile after the phase shifts are
    corrected.}
   \label{fig16_binPSR}
\end{figure*}

Near the center of adiabatically expanding shell-type SNR
G78.2$+$2.1, a radio quiet X-ray and gamma-ray pulsar, PSR
J2021+4026, was located and has been claimed to be associated with
the SNR \citep{lhh+13}. The X-ray and gamma-ray timings show that it
is a young pulsar with a period of $P$ = 0.265318~s \citep{hsl+15},
$\dot{P}$ =$5.48\times10^{-14}$ s~s$^{-1}$ and hence a spin-down age
of $\sim$77 kyr \citep{aaa+09}. The newly discovered pulsar from the
GPPS survey, PSR J2020+4024g (gpps0087), which has a larger period
of 0.37054~s (about 7/5 period of PSR J2021+4026) and is also
located at the very center of G78.2+2.1. We suspected PSR
J2020+4024g is the radio counterpart of PSR J2021+4026. However, we
cannot get the folded profile around the period of PSR J2021+4026.
Considering that 1) the newly discovered pulsar has a very different
period; 2) it has a very large DM (680.5~cm$^{-3}$~ps) and hence a
much greater distance (though probably overestimated as mentioned
above) than the estimated distance of less than 2~kpc for the
remnant \citep{szt+18}; 3) its profile has an obviously scattered
tail, we therefore conclude the new pulsar PSR J2020+4024g is
distant and just coincident with the remnant in the direction of the
Local Arm.

In summary, no physical association for any newly discovered pulsars
with these supernova remnants can be concluded from available data.

\begin{table}
 \centering
\caption{Binary Pulsars Discovered in the GPPS
  Survey (Sorted by Pulsar Period)}
  \label{binTab}

  \fns
\setlength{\tabcolsep}{2.5pt}
 \begin{tabular}{crccr}
\hline 
PSR Name     & \multicolumn{1}{c}{P} &            MJD &  DM        &   Acc*  \\
             & \multicolumn{1}{c}{ (ms) } &  & (cm$^{-3}${ }pc)      &   (ms$^{-2}$)  \\
\hline\noalign{\smallskip}
J1859$+$0313g  &    1.6133516   &  58770.464631 &  107.7&   9.0  \\
J1852$-$0044g  &    2.4117532   &  59281.036276 &  272.9&   16.0  \\
J1912$+$1417g  &    3.1661676   &  59270.015017 &   66.6&   $-$8.7  \\
J1930$+$1403g  &    3.2092399   &  58885.155530 &  150.5&   0.1  \\
J1844$+$0028g  &    3.5705885   &  59126.433750 &  181.2&   2.0  \\
J1953$+$1844g  &    4.4441330   &  59293.074883 &  113.1&   $-$1.9  \\
J1903$+$0839g  &    4.6213782   &  59084.555017 &  166.5&   $-$5.6  \\
J1904$+$0553g  &    4.9073156   &  58940.003865 &  164.3&   0.9  \\
J1840$+$0012g  &    5.3394411   &  59215.208372 &  100.8&   5.2  \\
J1917$+$1259g  &    5.6378496   &  58887.175359 &  117.0&   0.8  \\
J1928$+$1816g  &    10.543555   &  58999.880554 &  346.3&   $-$73.0  \\
J2023$+$2853g  &    11.331753   &  59080.693995 &   22.8&   $-$9.6  \\
J1952$+$2836g  &    18.023112   &  59135.498861 &  313.4&   1.5  \\
J1936$+$2036g  &    32.923094   &  59091.606022 &  198.8&   $-$7.5  \\
J1933$+$2038g  &    40.706716   &  59004.760866 &  302.8&   $-$18.0  \\
J1913$+$1037g  &    434.21498   &  59214.259211 &  437.0&   $-$4.0  \\
\hline\noalign{\smallskip}
\end{tabular}
\parbox{7cm}{*Acceleration parameter obtained from {\it {pdmp}}. }
\end{table}

\begin{figure*}
  \centering
  \includegraphics[width=41mm]{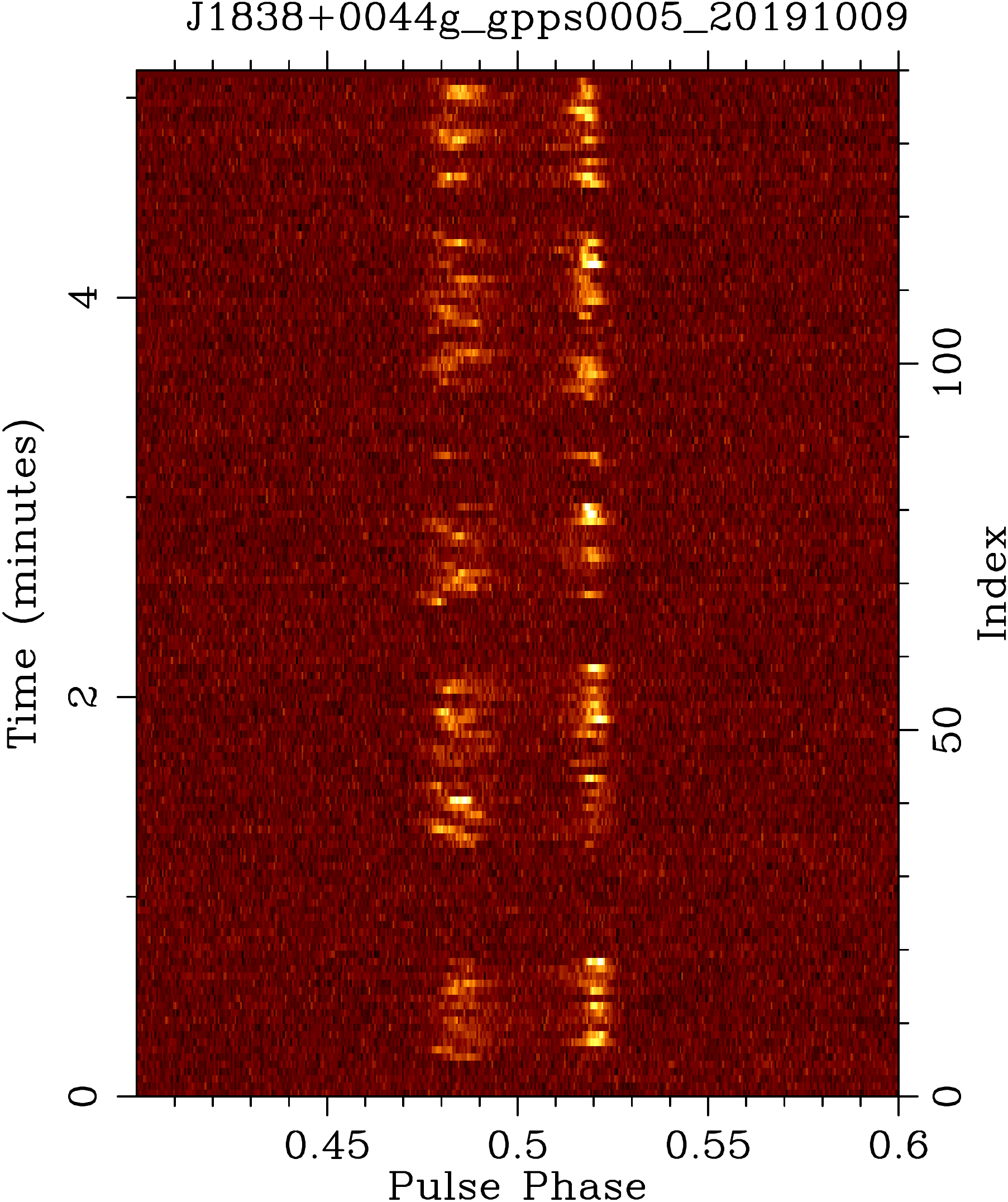}
  \includegraphics[width=41mm]{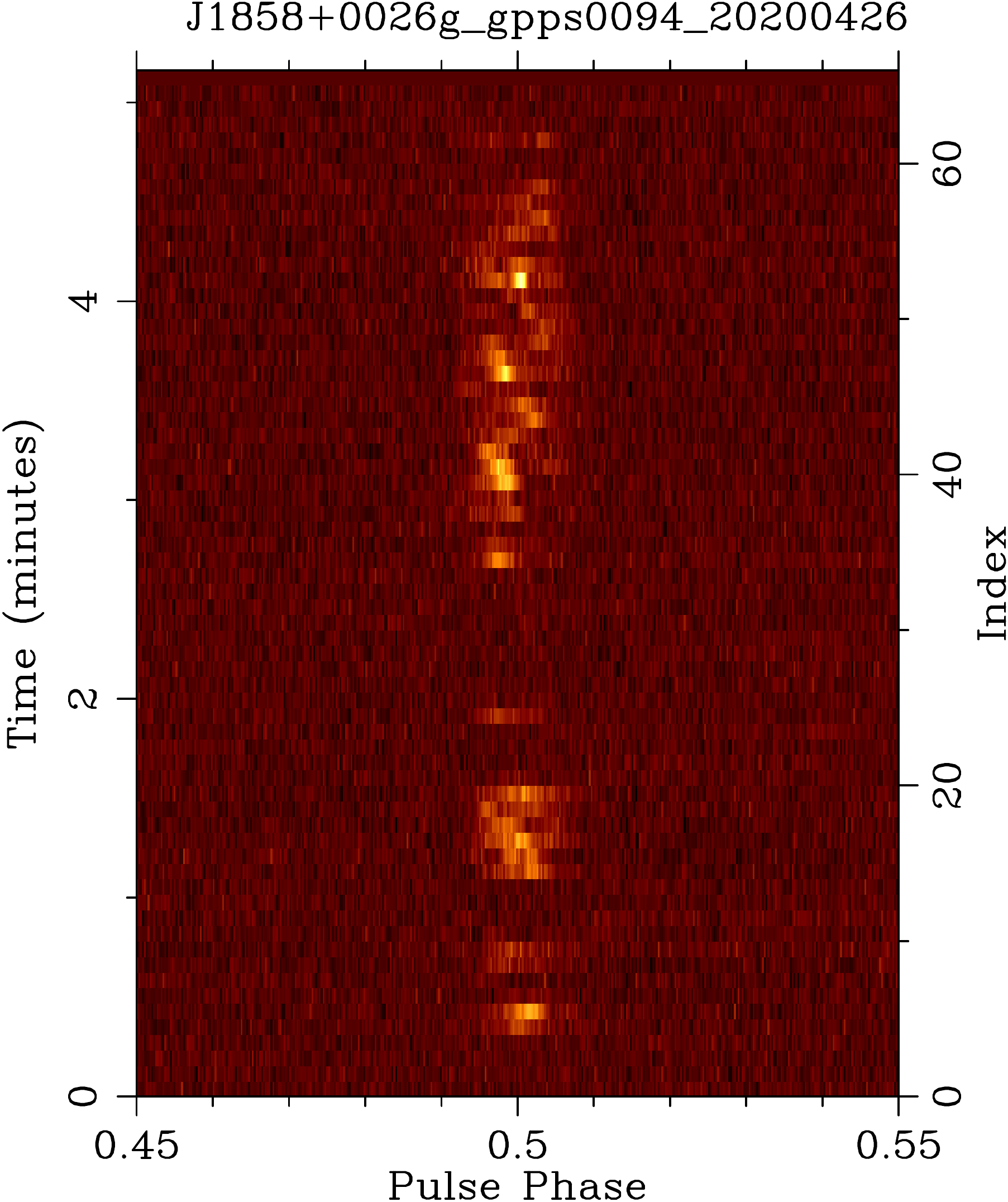}
  \includegraphics[width=41mm]{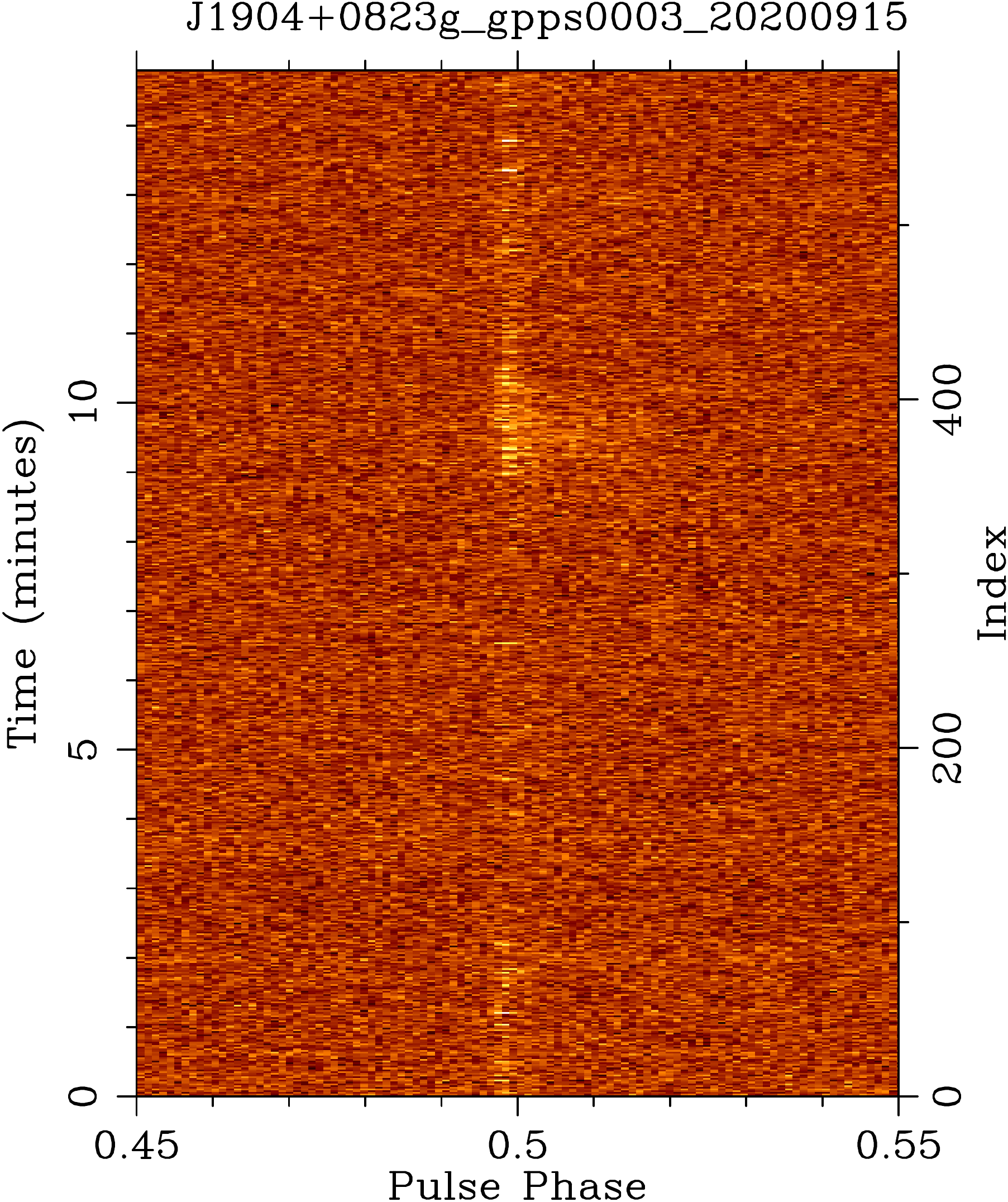}
  \includegraphics[width=41mm]{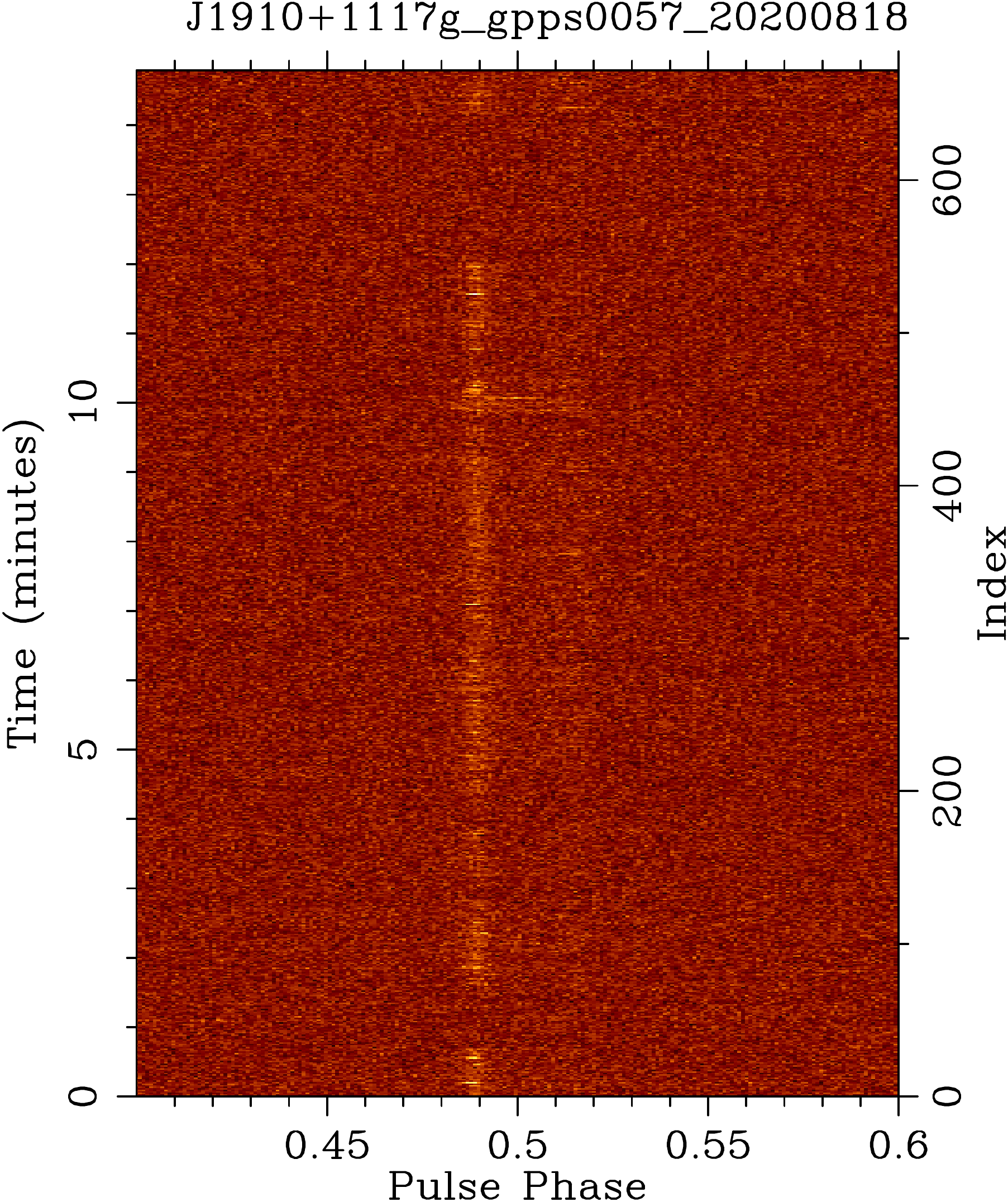} \\[2mm]
  \includegraphics[width=41mm]{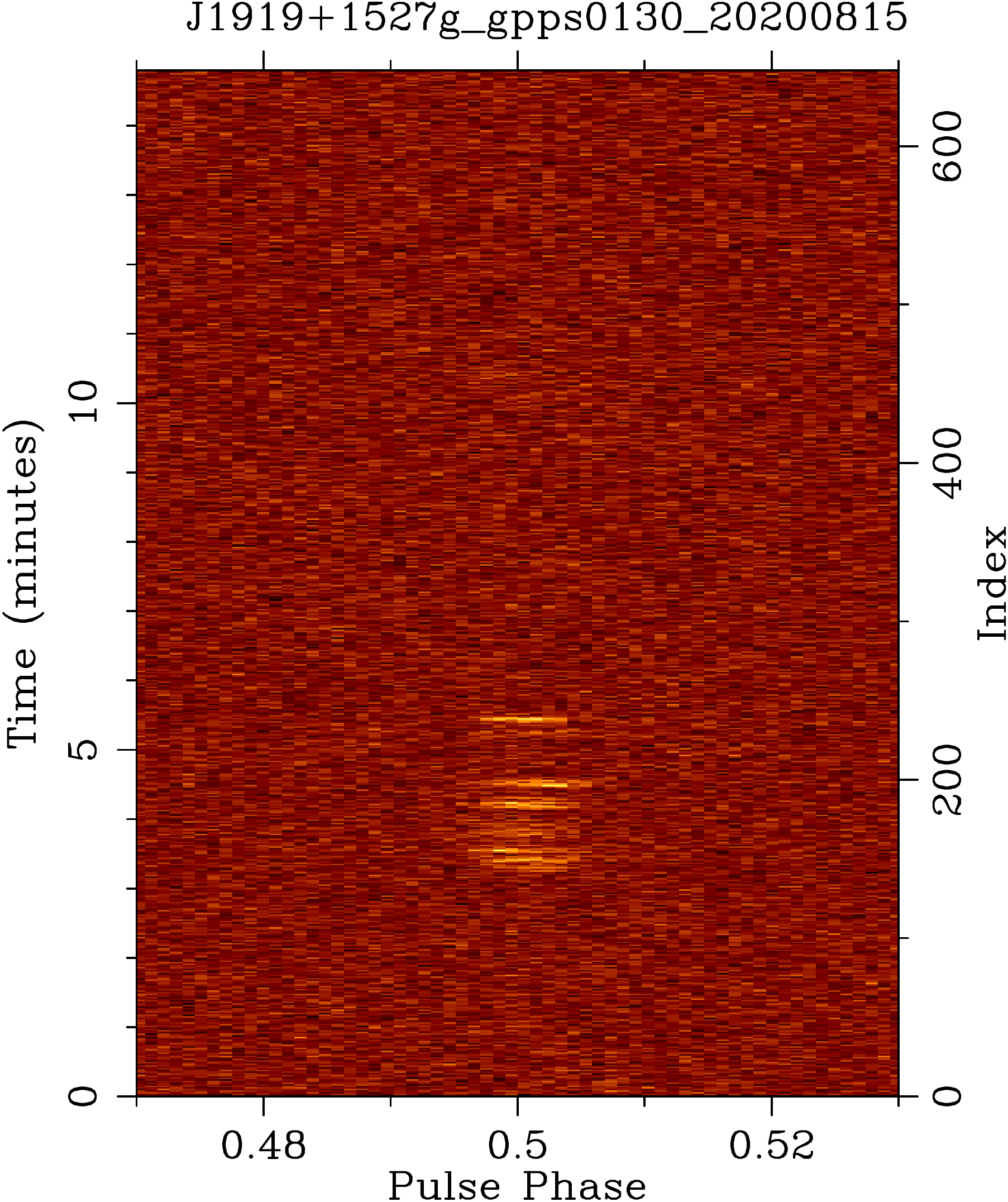}
  \includegraphics[width=41mm]{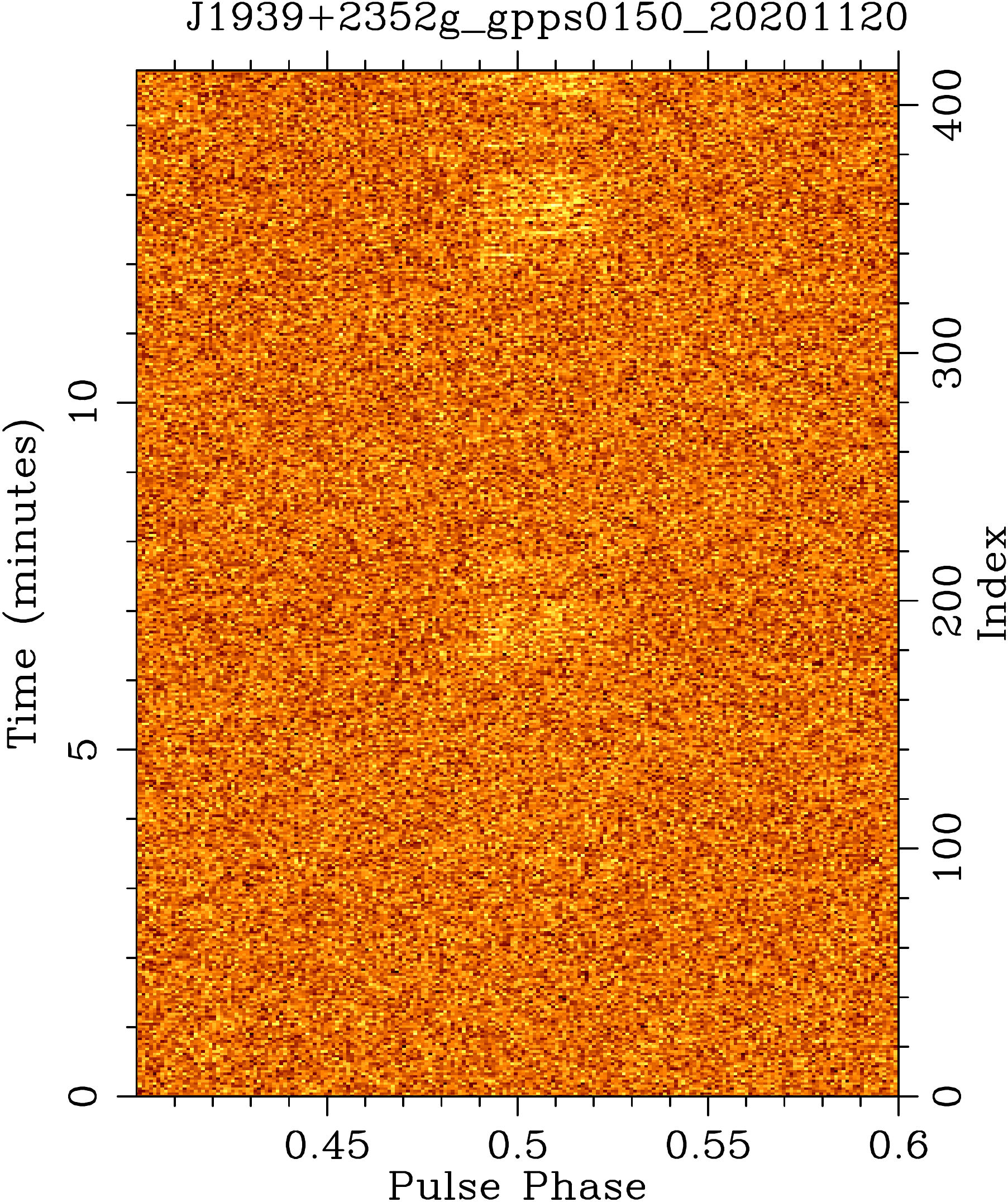}
  \includegraphics[width=41mm]{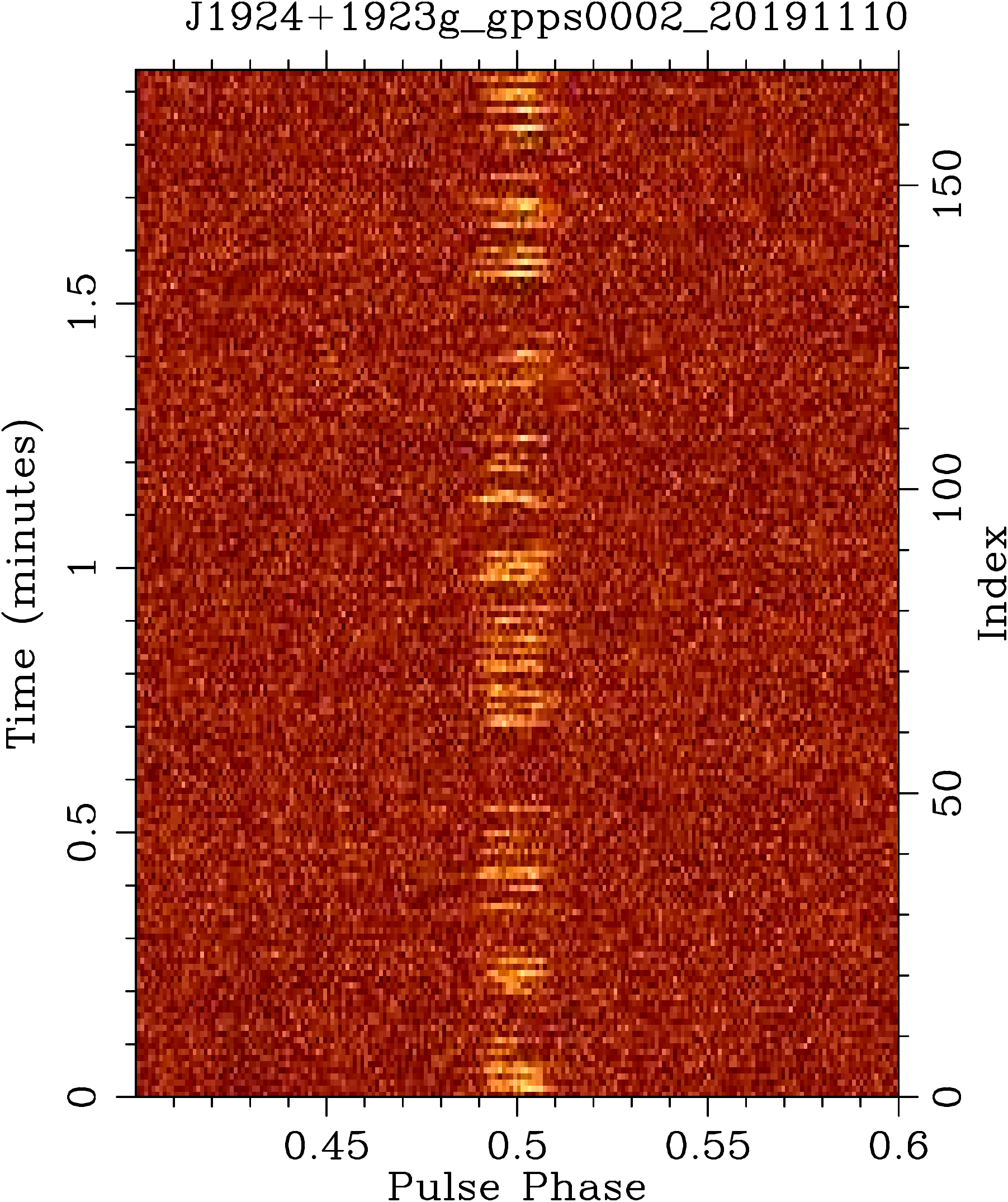}
  \includegraphics[width=41mm]{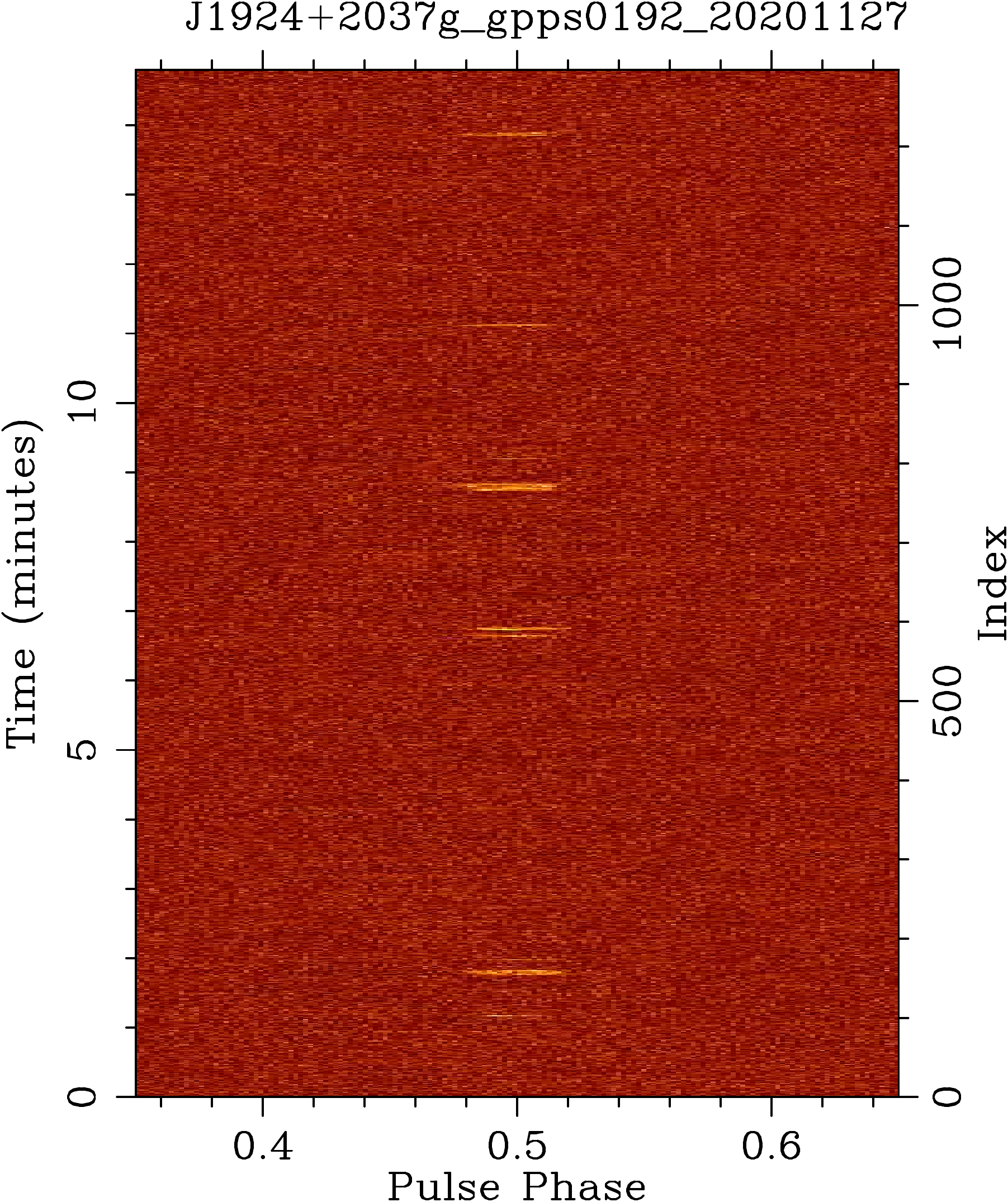}

  \begin{minipage}{15cm}
  \caption{\baselineskip 3.8mm Phase-time plots exhibit nulling
    phenomena for eight newly discovered pulsars in the GPPS survey.
  }
   \label{fig17_nullingPSR}\end{minipage}
\end{figure*}

\begin{figure*}
  \begin{minipage}[]{57mm}
  \centering
  \includegraphics[width=55mm]{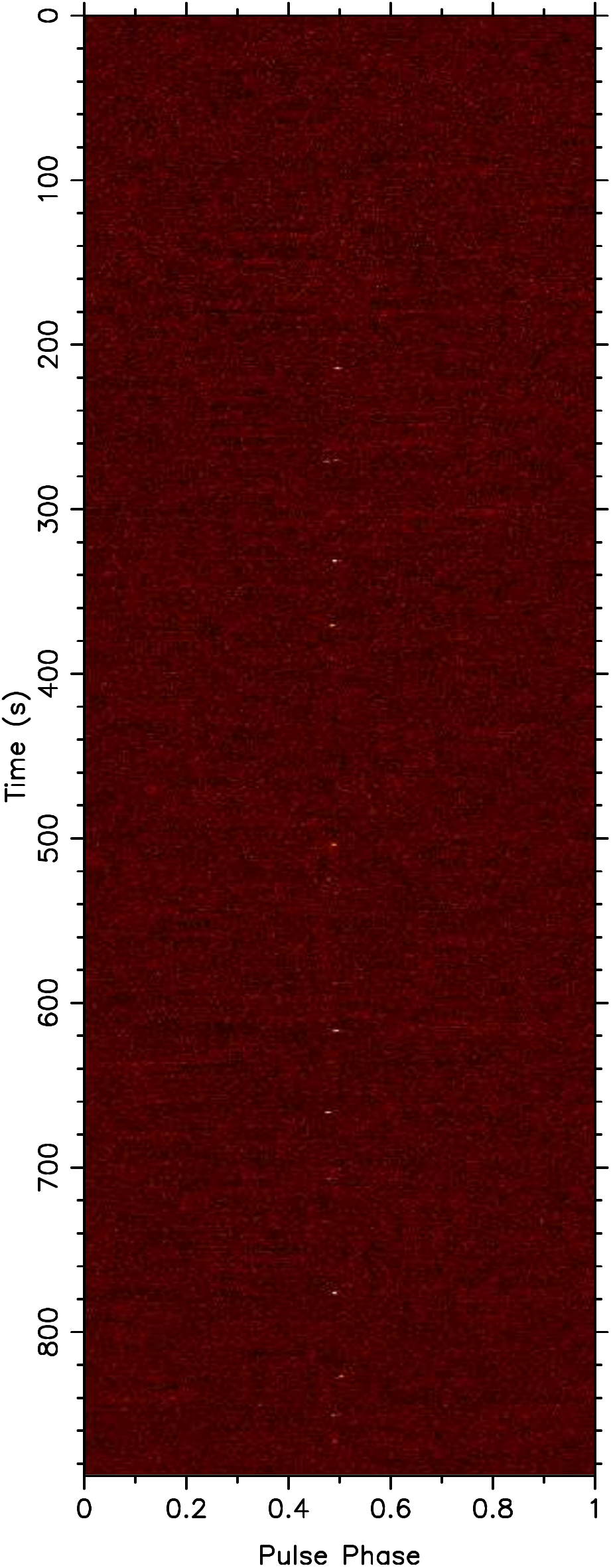}
\end{minipage}
\begin{minipage}[]{113mm}
  \centering
  \includegraphics[width=37mm]{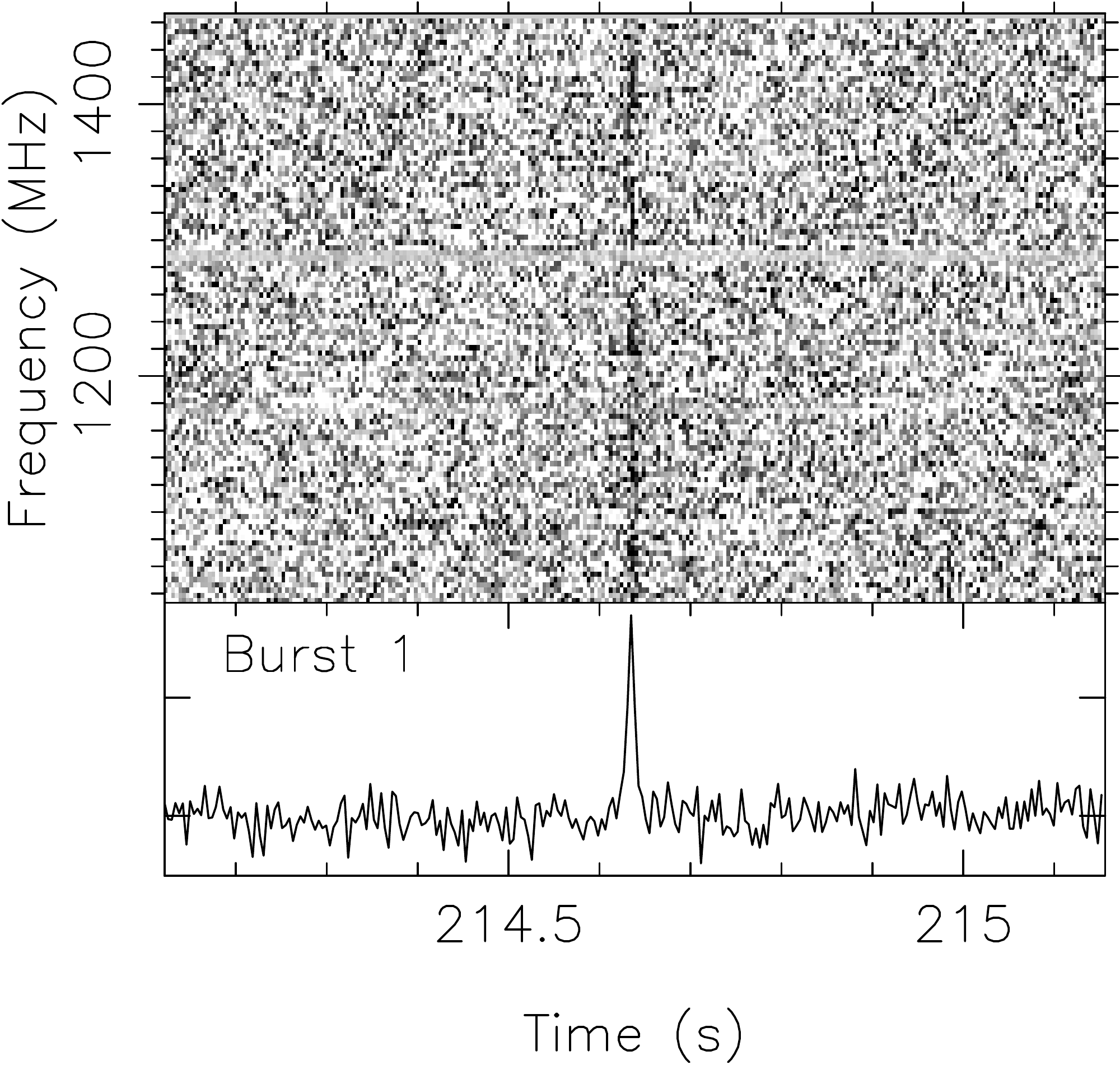}
  \includegraphics[width=37mm]{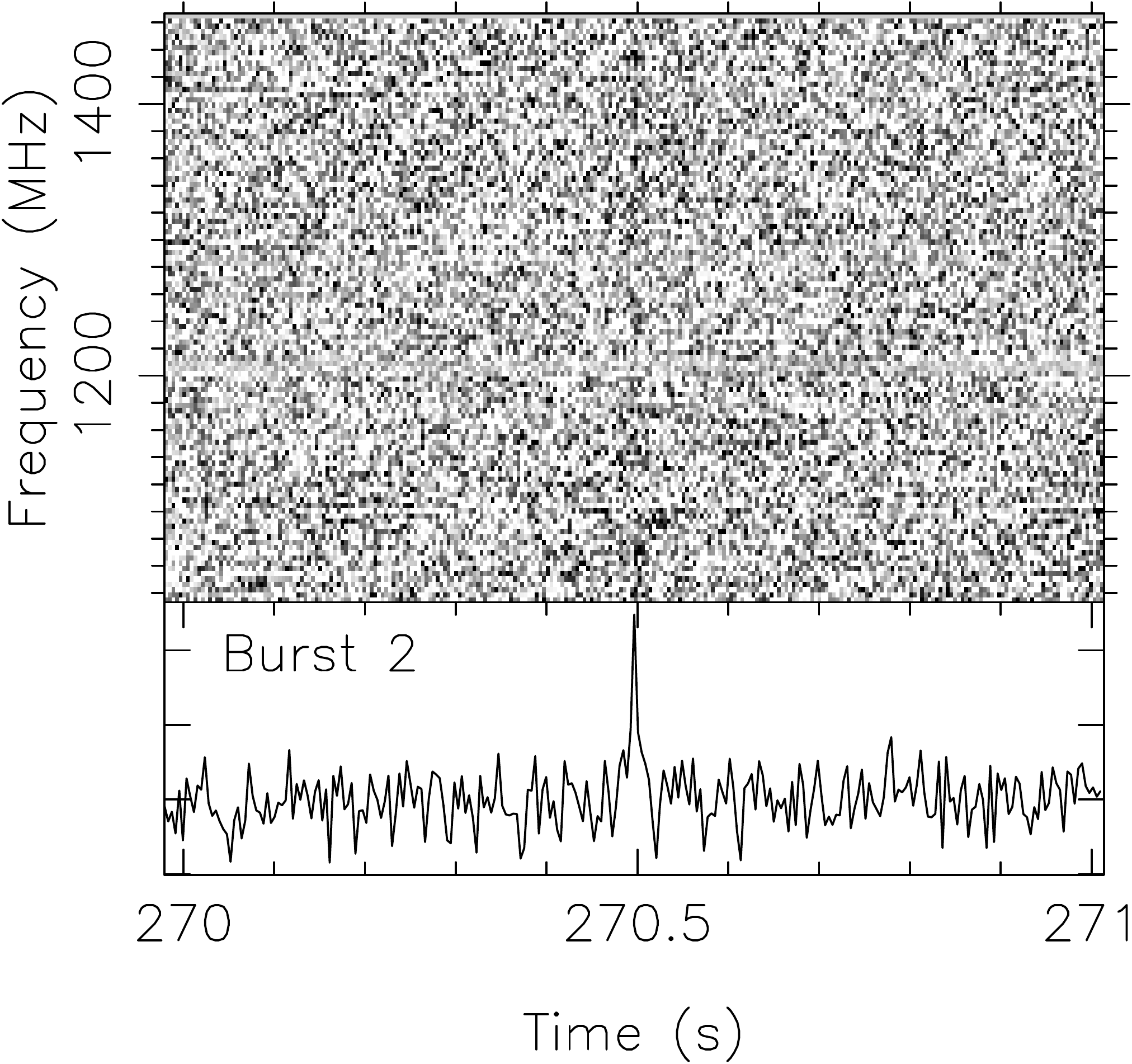}
  \includegraphics[width=37mm]{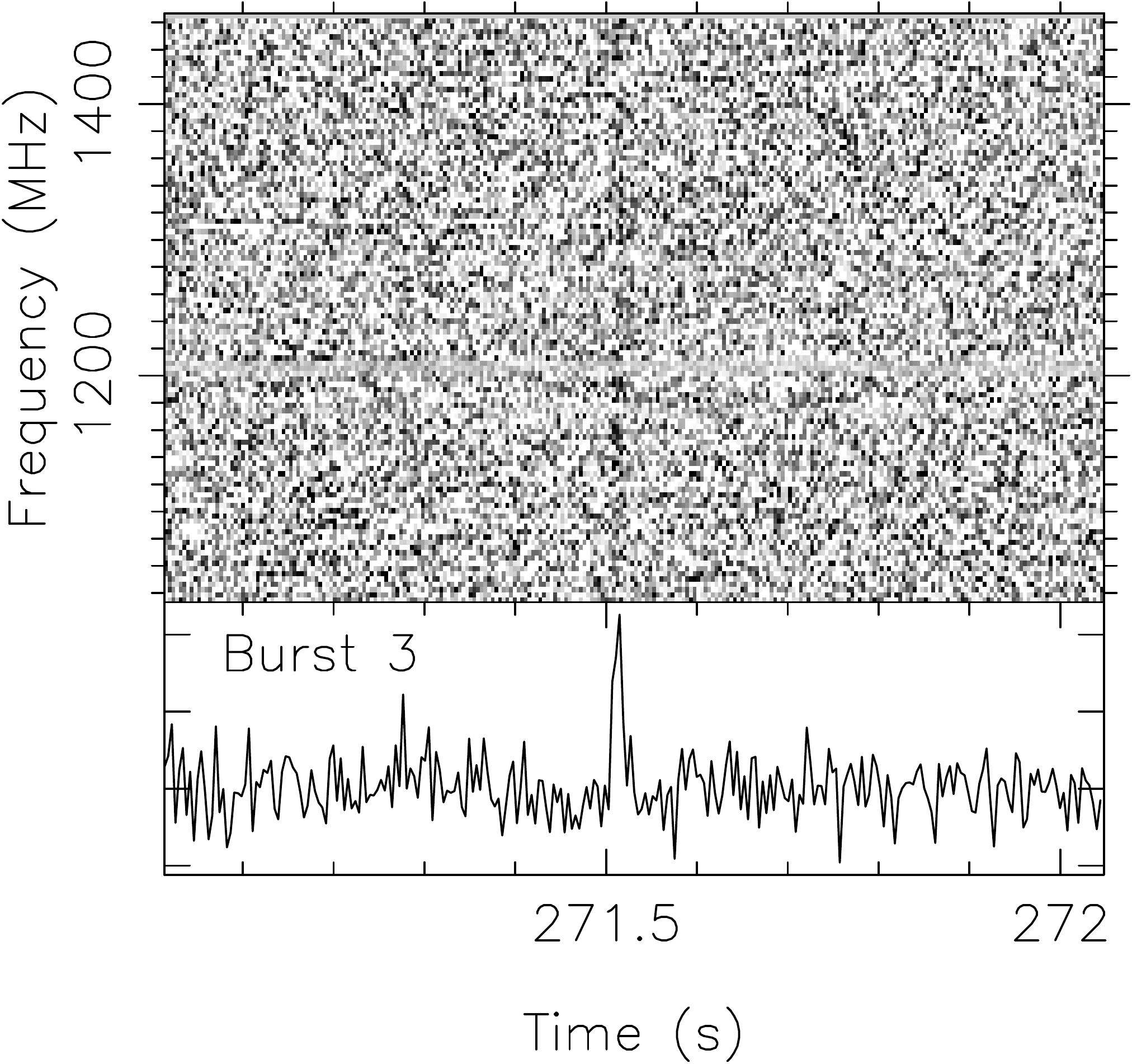}
  \includegraphics[width=37mm]{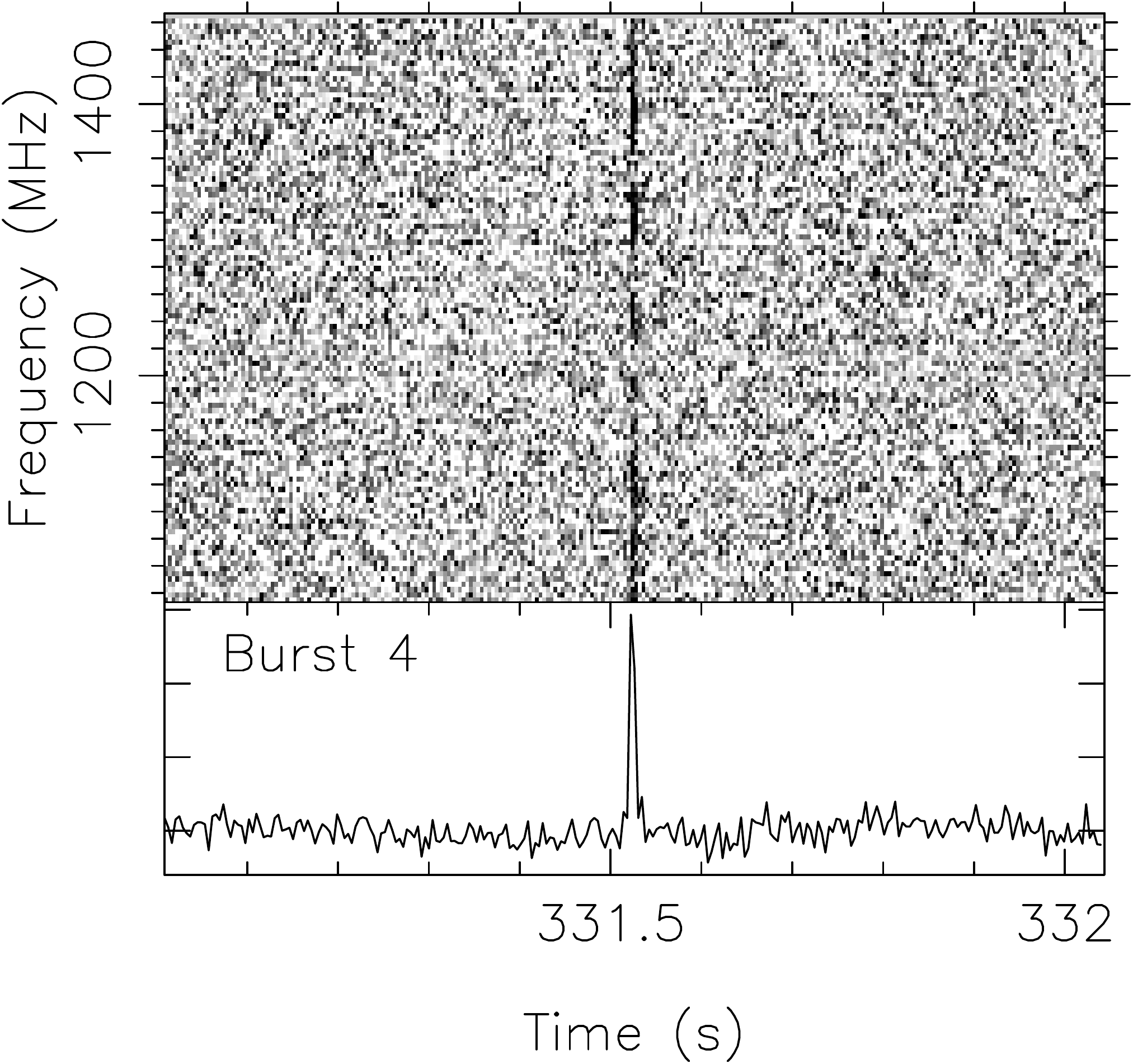}
  \includegraphics[width=37mm]{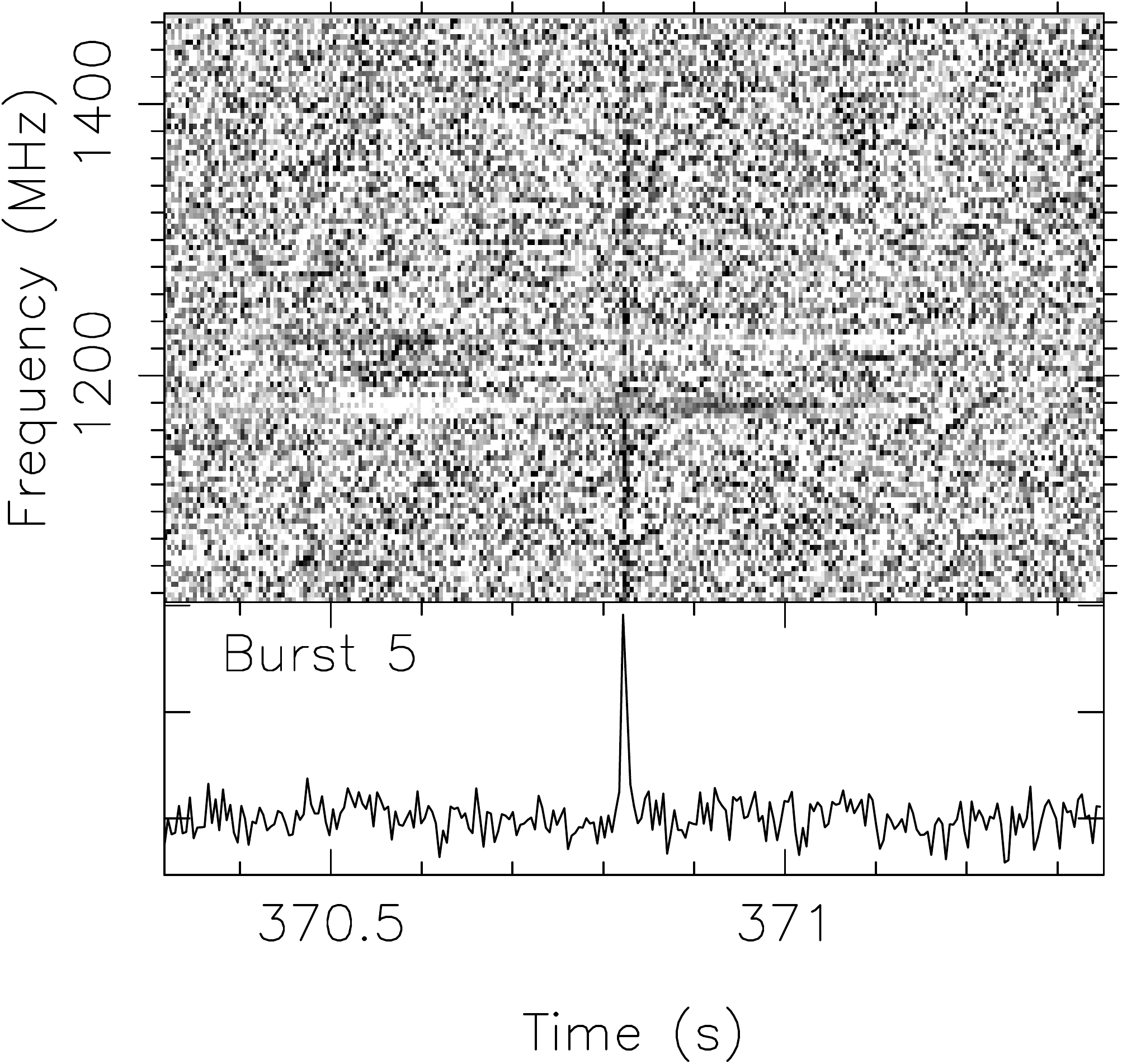}
  \includegraphics[width=37mm]{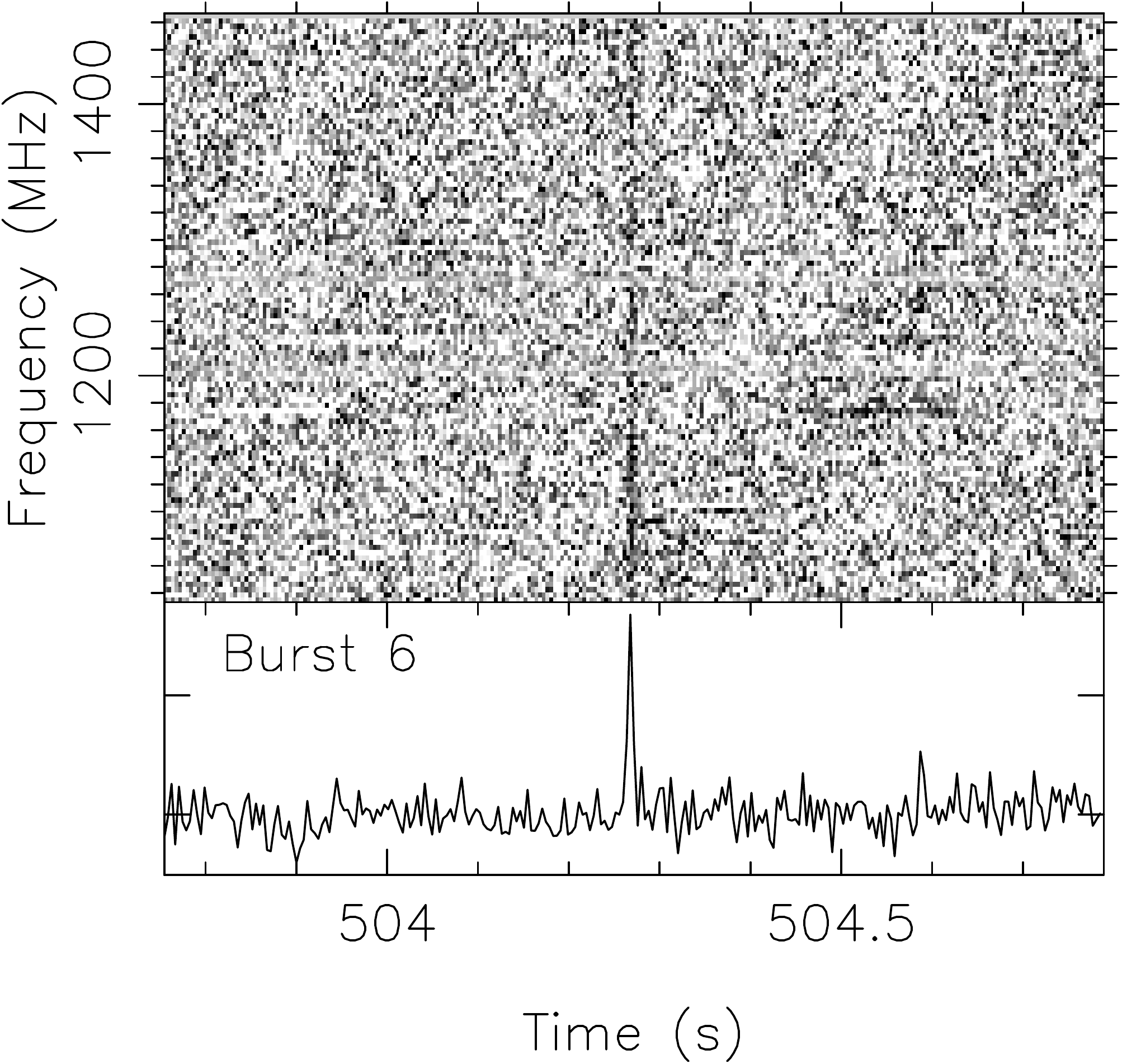}
  \includegraphics[width=37mm]{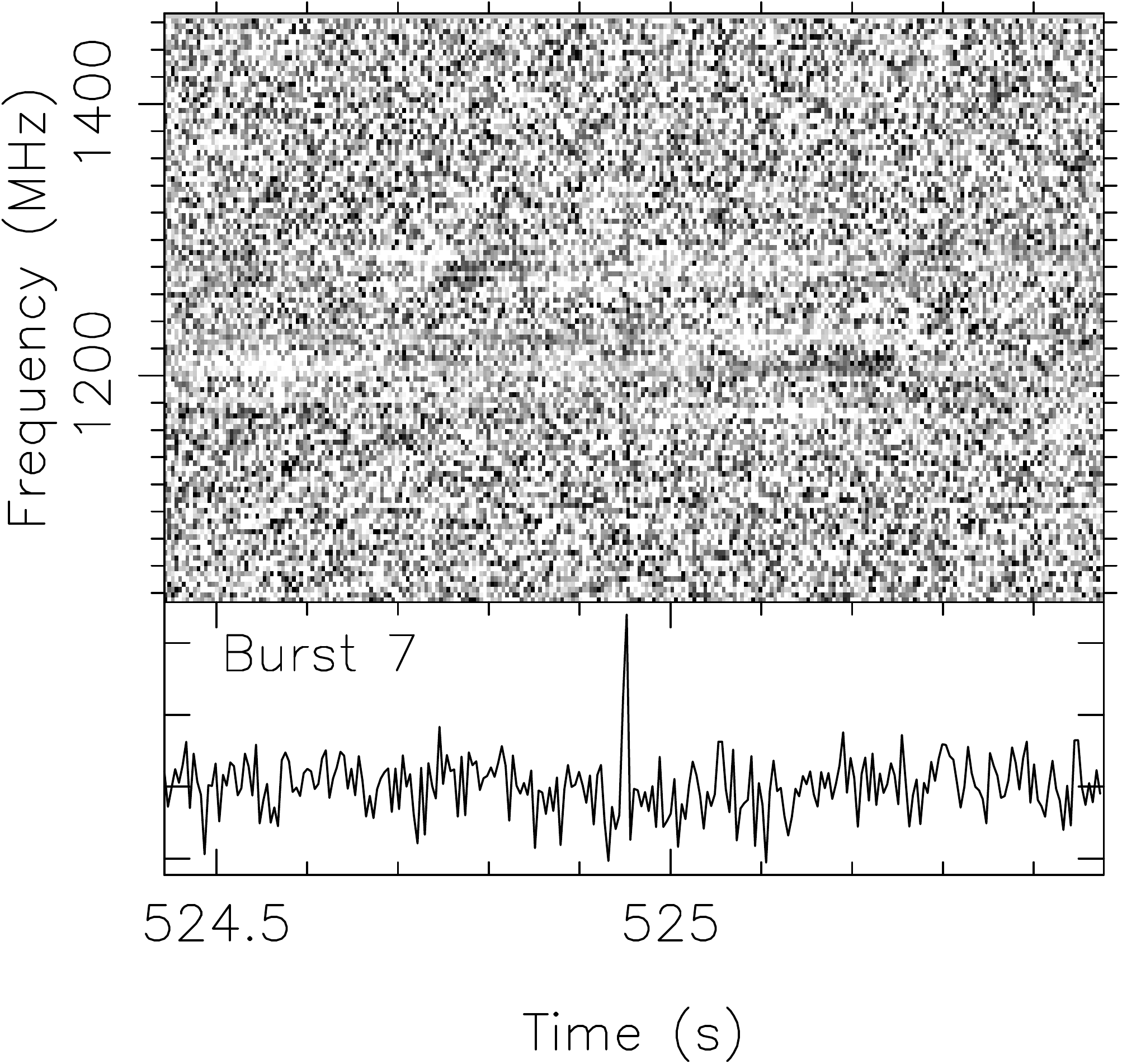}
  \includegraphics[width=37mm]{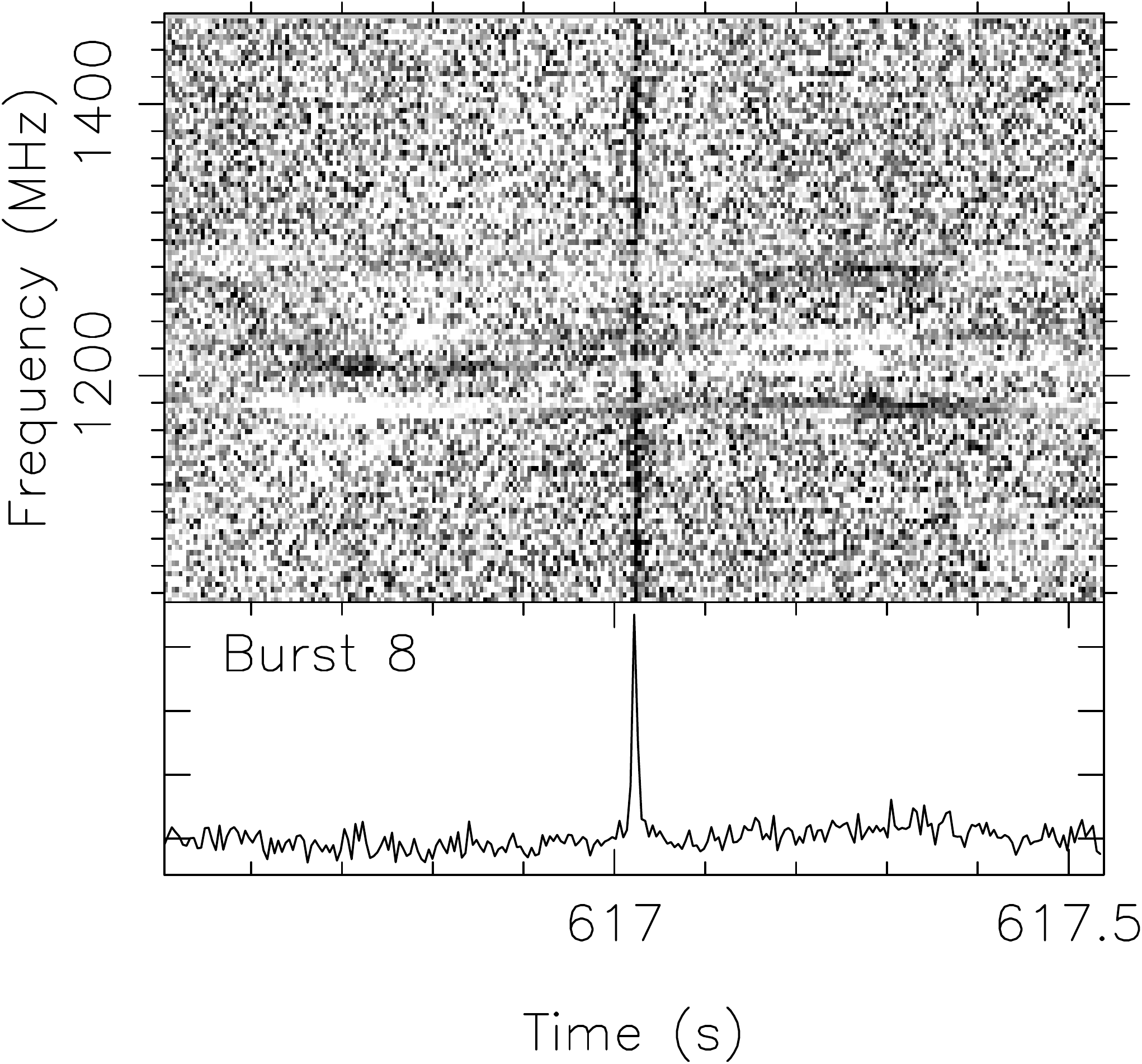}
  \includegraphics[width=37mm]{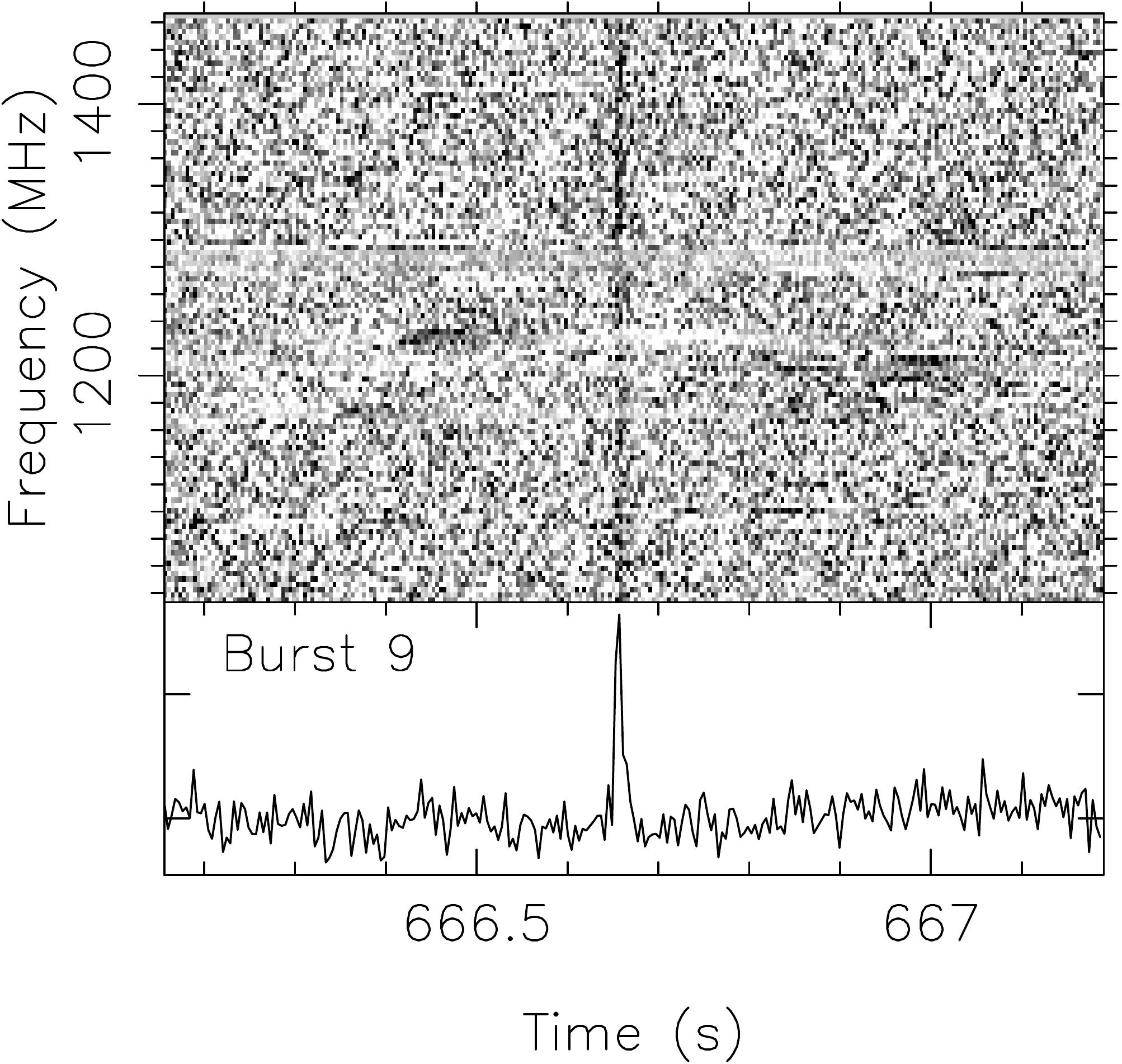}
  \includegraphics[width=37mm]{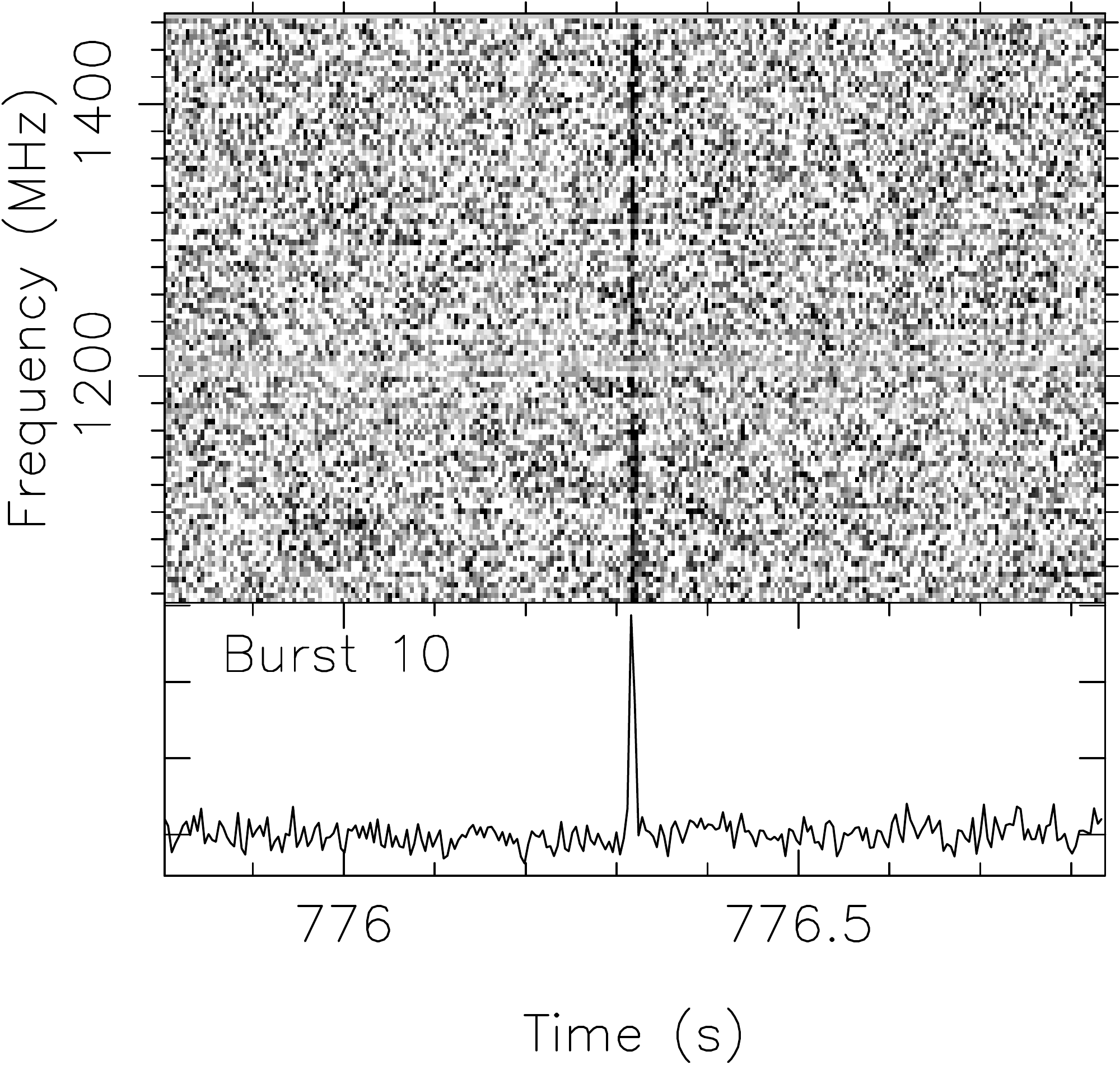}
  \includegraphics[width=37mm]{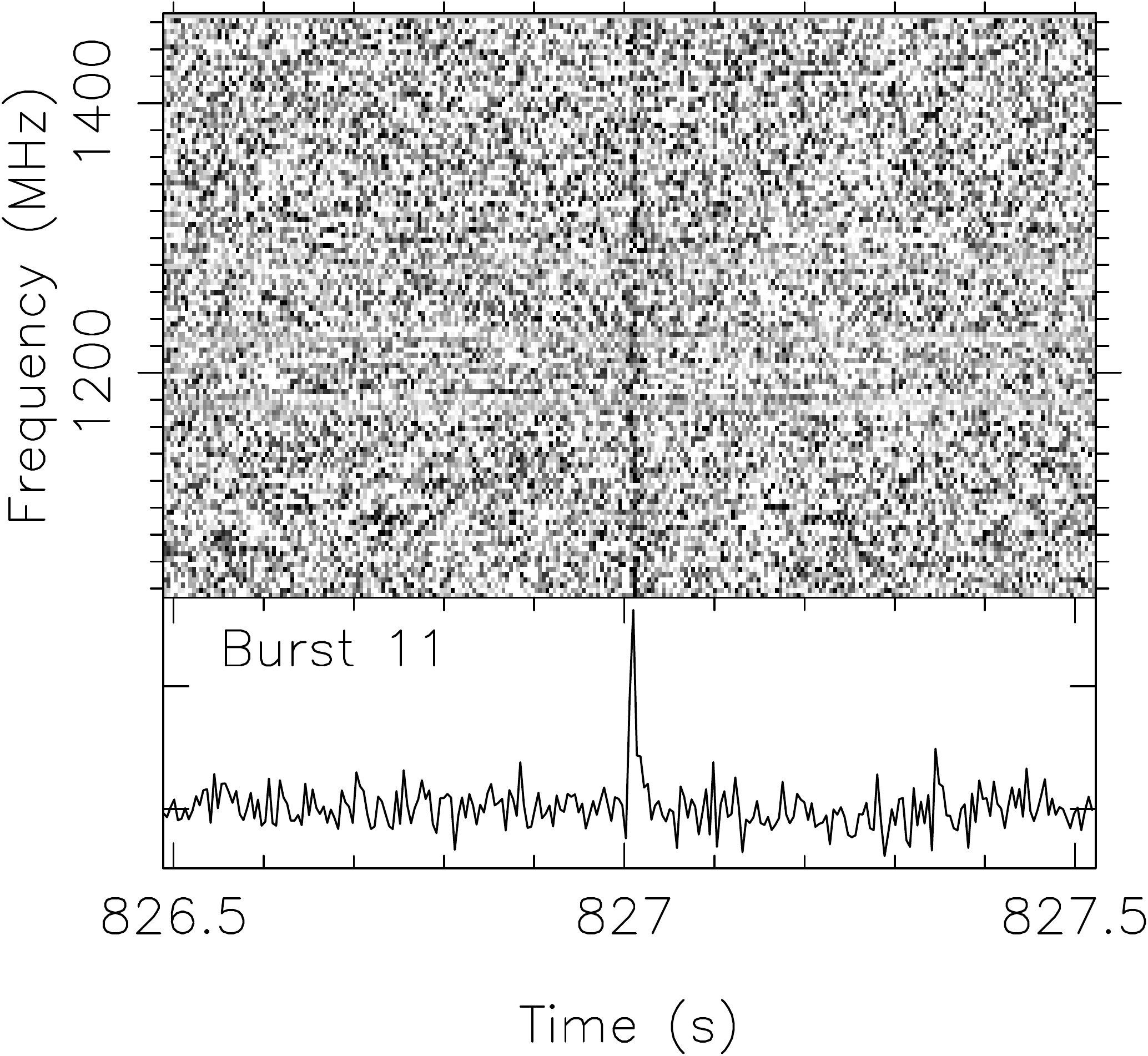}
  \includegraphics[width=37mm]{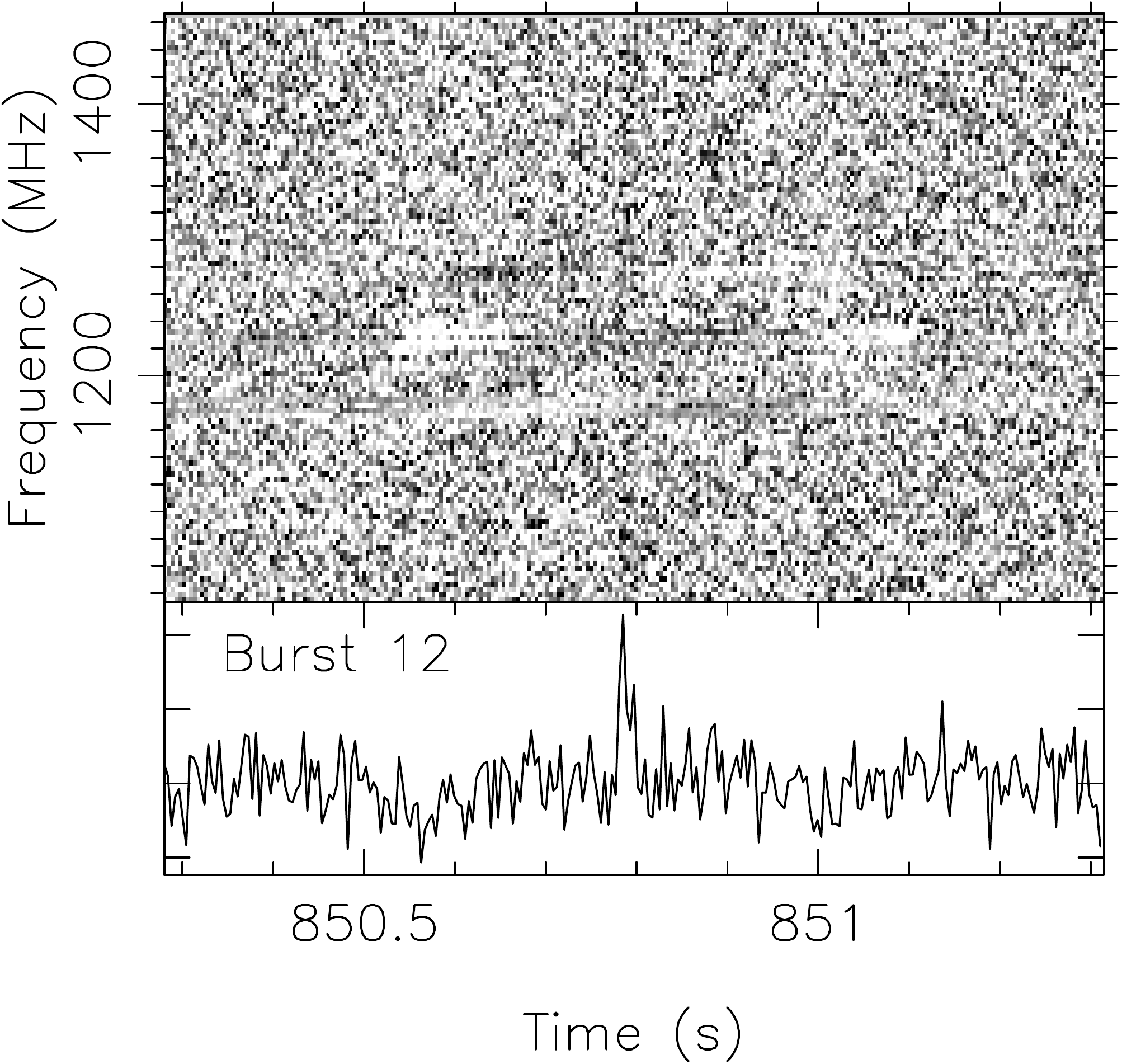}
\end{minipage}
\caption{\baselineskip 3.8mm Discovery of RRAT J1905$+$0849 in the
  GPPS survey, which has a period of $1.0343\pm0.0052$~s and a DM of
  $257.8\pm2.3$~cm$^{-3}$~pc. In the {\it left long panel}, the 12
  bright pulses were detected in the verification observations on 2020
  Nov. 21, and aligned with the folding period. On the {\it right
    panels}, the de-dispersed waterfall plots for the intensity on the
  frequency against time are depicted for 12 pulses, and with an
  integrated profile in the bottom panel for each pulse.}
   \label{fig18_rrat}
\end{figure*}

\subsection{Discovery of Millisecond Pulsars and Binary Pulsars}

In Table~\ref{gppsPSRtab1}, there are 40 pulsars with a period less
than 30~ms, which can be regarded as MSPs according to the $P-\dot{P}$
diagram for previously known pulsars, though the $\dot{P}$ values of
newly discovered pulsars are not available yet. The current young
pulsars all have a period greater than 30~ms.
Among these MSPs, 14 pulsars manifest their obvious pulse shift
(coming early or delayed from the best phase bin, as shown in
Fig.~\ref{fig16_binPSR}) in a short observation session, which
indicates their binary nature, as listed in Table~\ref{binTab}, as do
two longer period pulsars J1933+2038g (gpps0041, $P$ = 40.7~ms) and
PSR J1913+1037g (gpps0149, $P$ = 434.2~ms).

Two interesting binaries are remarkable. The binary pulsar,
J1953+1844g (see Fig.~\ref{fig16_binPSR}) was discovered in a snapshot
survey with $P$ = 4.44~ms and DM = 113.1~cm$^{-3}$~pc, and is probably
located in the global cluster M71, because it is only $2.5'$ away from
M71A/PSR J1953+1846 \citep[DM = 117.0~cm$^{-3}$~pc,][]{hrs+07} and has
a similar but smaller DM than the four previously known pulsars in
M71\footnote{\it\url{https://fast.bao.ac.cn/cms/article/65/}}.
The binary pulsar J2023+2853g (gpps0201,{$P$ = 11.33~ms, DM
=  22.8~cm$^{-3}$~pc) was found in the vicinity of a bright known pulsar,
J2022+2854 ($P$ = 343.402~ms, DM = 24.6~cm$^{-3}$~pc). Its signals were
easily misinterpreted as a harmonic of the known pulsar because they
have similar but slightly different DM values.

Parameters of all these millisecond and binary pulsars have to be
determined in more follow-up observations, which are going on by the
survey team members. Careful observations and detailed analyses of
these binary systems may reveal a number of relativistic effects and
lead to excellent tests of some fundamental properties of gravity.

\subsection{Discovery of Nulling and Mode-Changing Pulsars}

By viewing the phase-time plots of newly discovered pulsars, we
noticed that several pulsars display the mode-changing or nulling
phenomenon (see examples in Fig.~\ref{fig17_nullingPSR}) in the
duration of observations, i.e., they switch their emission modes or
even cease their emission for some periods. More interesting are
subpulses of PSR J1838+0046g, which drift and modulate both in the
leading and trailing profile components, in addition to the nullings
around pulses, e.g., Nos. 20, 60, 90 and 120. PSR J1858+0028g
undergoes nulling and subpulse-drifting. PSRs J1904+0823g and
J1910+1117g occasionally have a bright pulse, even during a long
nulling session (e.g., PSR J1904+0823g at the period index No.260)
similar to that for PSR B0826$-$34 reported by
\citet{eamn12}. Mode-changing and nulling are obvious for PSR
J1910+1117g.
Very nulling pulsars, PSR J1919+1527g (gpps0130), J1939+2352g
(gpps0150) and PSR J1924+2037g (gpps0192), have a short duration for
emission but a longer duration for nulling (see
Fig.~\ref{fig17_nullingPSR}). They {were} first detected as a few individual
pulses via {a} {self}-developed single pulse module, and later the period
emission was found from a track observation for 15~minutes via the
PRESTO search module.

Since the duration of the GPPS survey is only 300~s and that for
follow-up verification observations is only 15~minutes, it is hard to
get the nulling fraction or study the details of mode-changes or get
statistical properties of these nullings from available observation
sessions, though we do see the nulling or mode-changing of these
pulsars from available data. Longer observations with high sensitivity
are desired for this purpose \citep[e.g.][]{whh+20}. Presented here
are just the first results from short GPPS survey observations. More
results on newly detected nulling, mode-changing and subpulse-drifting
phenomena for previously known pulsars and also the newly discovered
pulsars in the GPPS survey will be reported by Yan et al. (in
preparation).

\subsection{Discovery of Long Period Pulsars}
\label{longP}

The GPPS survey discovered a few pulsars with very long period{s}.
For example, PSR J1903+0433g (gpps0090) has a period of 14.05~s, which
is the second longest period known, and PSR J1856+0211g (gpps0158) has
a period of $P$ = 9.89012~s. Currently in the ATNF Pulsar Catalogue
\citep{mhth05} the longest period is $P$ = 23.535378~s for PSR
J0250+5854, which was discovered by LOFAR \citep{tbc+18}.

PSR J1903+0433g (gpps0090) was discovered in the 15th harmonic ({$P$}
=0.93671~s) from a snapshot survey on 2020 Mar. 30 with a very good
S/N = 21.2. It was rediscovered in the 17th harmonic ($P$ =0.82684~s)
in an another survey cover on 2020 Apr. 22. Within the uncertainty
they have a similar DM value. In the verification observations on
2020 Aug.~29, the 6th harmonic of $P$ =2.3417~s was detected, and
on 2020 Sep.~11 the 5th harmonic of $P$ =2.8099~s. From these
harmonics, the right period was found to be $P$ =14.0508~s and
the pulse width is 123.5~ms.

Discovery of PSR J1856+0211g (gpps0158) has a different story. Its
single pulses were first detected via the single pulse module with
marginal significance in a snapshot survey for the cover of
G35.58+0.00\_20201003. In the follow-up observations that lasted for
15 minutes on 2021 Jan.~13, the pulsar was detected in its 8th
harmonic ($P$ = 1.23627~s). In further verification observations on
2021 Jan. 26, the pulsar was detected by its 4th harmonic ($P$ =
2.47253~s). The procedure for harmonic-checking led to the discovery
of the proper period of $P$ =  9.89012~s. This pulsar also has a
relatively very narrow pulse, with a pulse width of about 38~ms.


\begin{table}
  \centering
\caption{Parameters of RRAT J1905$+$0849}
  \label{rratTab}

  \fns
\setlength{\tabcolsep}{2.5pt}
 \begin{tabular}{ll}
\hline
Parameter (unit)  & {V}alue  \\
\hline
First discovery date / MJD & 20190917 / 58743  \\
Confirmed date / MJD  & 20201121 / 59174 \\
Burst period, {$P$} (s)  &  1.034 \\
Dispersion measure, DM (pc~cm$^3$) & 257.8 (23) \\
Distance (kpc, YMW2016) & 7.73 \\
Right ascension (J2000, hh:mm:ss) & 19:05:03 \\
Declination (J2000, dd:mm)  &  08:49 \\
Galactic longitude (deg.)  & 42.327 \\
Galactic latitude (deg.)   & 1.017 \\
Max burst rate (n hr$^{-1}$) &  48 \\
Mean width (ms) of pulses & 3.74 \\
S$_{\rm mean}$ (mJy) of pulses at 1.25 GHz &    14.67 \\
RM (rad~m$^{-2}$)  & 493.9(14)   \\
\hline
\end{tabular}
\end{table}

\begin{figure}
  \centering
  \includegraphics[width=4cm]{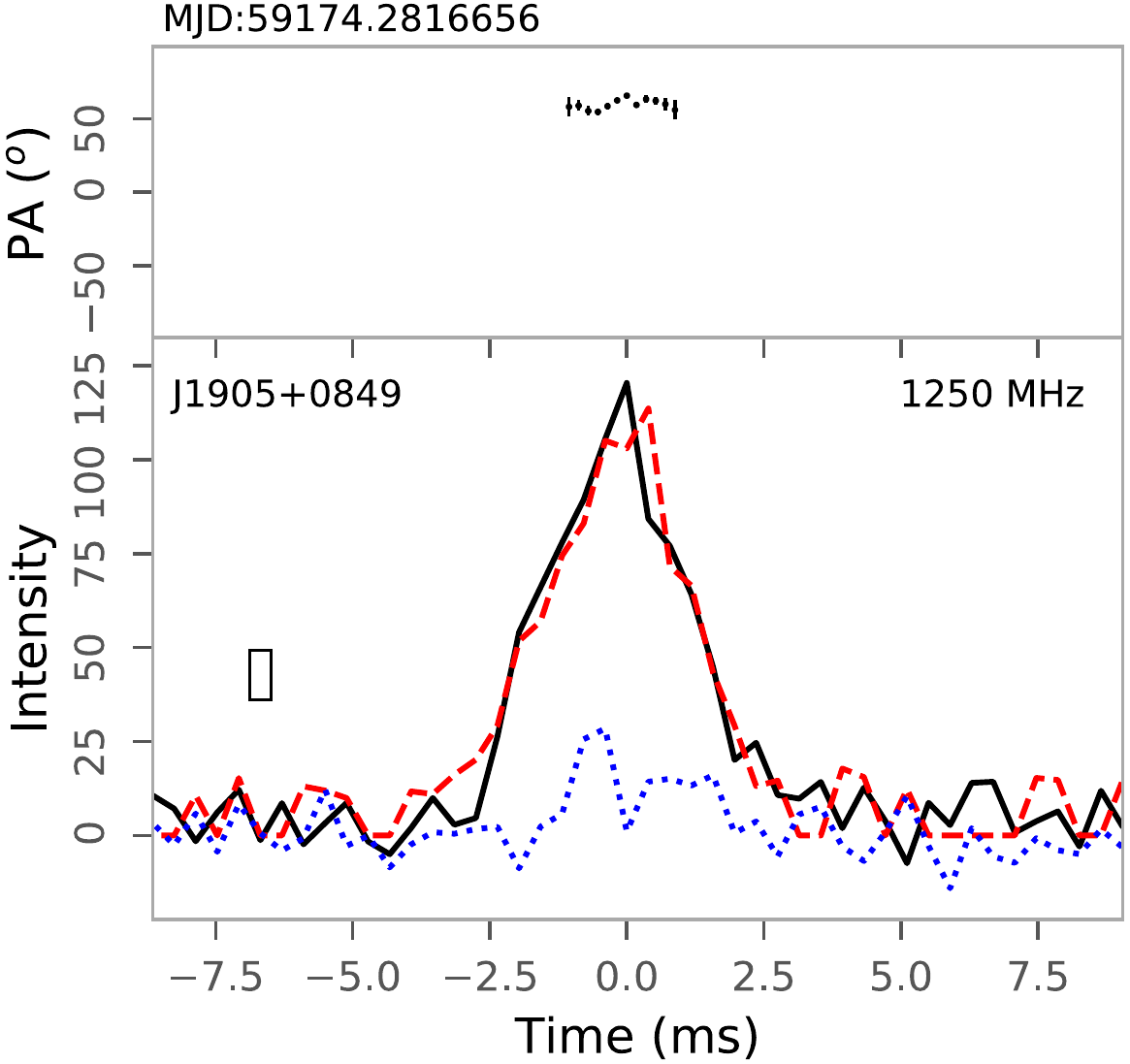}
  \includegraphics[width=4cm]{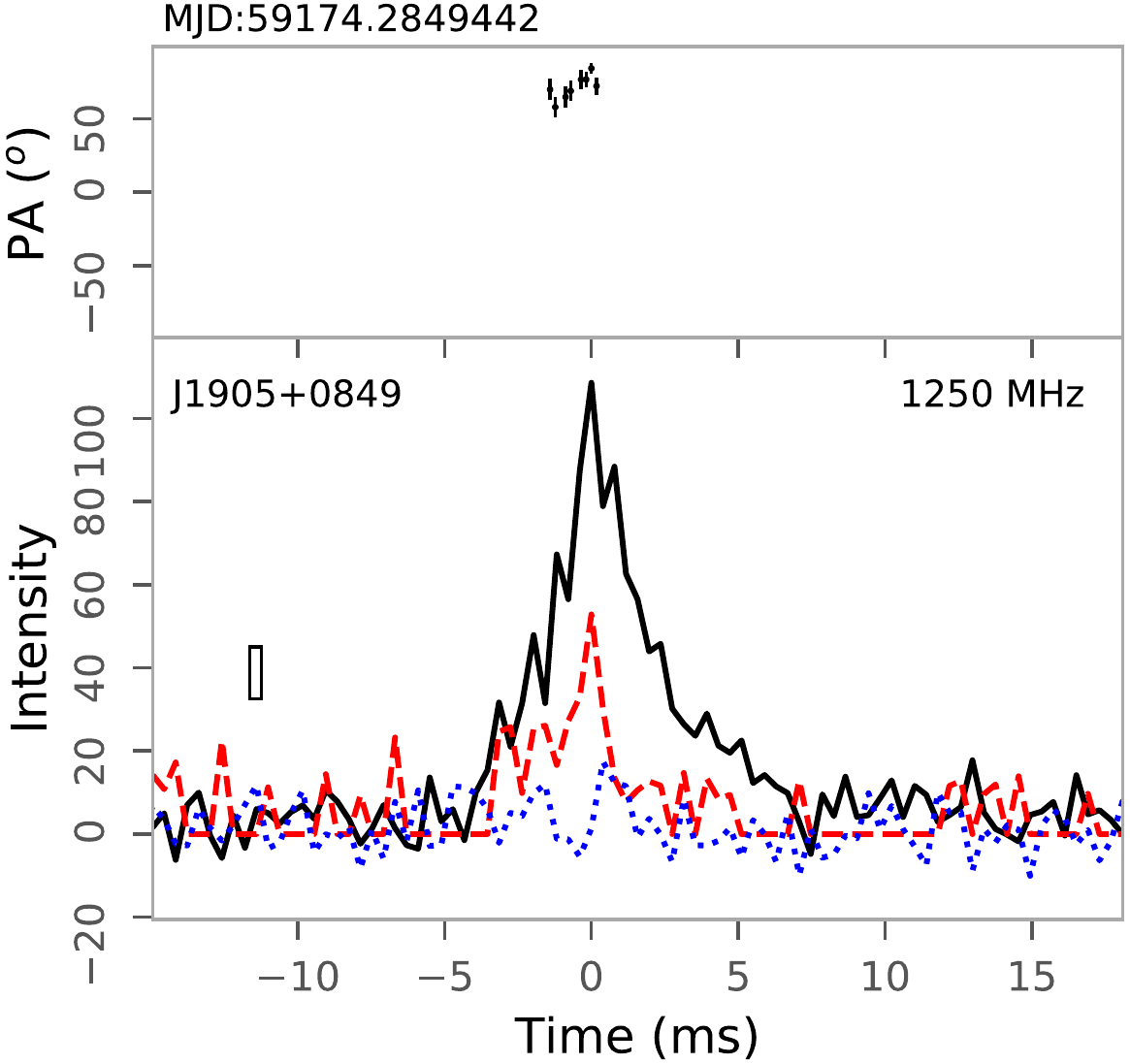}
  \caption{\baselineskip 3.8mm Polarization profiles of two pulses of RRAT J1905+0849,
    No.4 and No.8. The polarization angle variations are plotted in the
    upper panels, and the total intensity ({\it thick line}), linear
    ({\it dashed line}) and circular ({\it dotted line}) polarization ($>0$
    for the left-hand sense) are featured in the bottom panels.}
  \label{fig19_RRATpoln}
\end{figure}

\subsection{Discovery of RRAT J1905$+$0849}

In addition to pulsars listed in Table~\ref{gppsPSRtab1}, the single
pulse search o{f the} GPPS survey data using the newly developed single
pulse module detect{ed} some RRATs. Here is the first case, RRAT
J1905$+$0849, {featured} in Figure \ref{fig18_rrat}. Four pulses around DM of
$257.8\pm2.3$~cm$^{-3}${ }pc of this RRAT were first detected from a
cover of the GPPS survey G42.38+0.93\_20190917, on the beam of
M06-P1. During the verification observation on 2020 Nov. 21, twelve
pulses were discovered, with the polarization signals
recorded. Further analyses of TOAs of 12 pulses \citep[see the method
  of][]{kle+10} yield a period of $1.0343\pm0.0052$~s. Basic parameters
of this RRAT are listed in Table~\ref{rratTab}. The DM is obtained by
maximizing S/N for pulses, and S$_{\rm mean}$ is the mean of the
averaged flux densities in the duration of detected pulses (not an
average over a period or even the whole session). Polarization
profiles of two pulses are plotted in Figure~\ref{fig19_RRATpoln}. The first is
pulse No.4, which is almost 100\% polarized with a flat
polarization angle curve from which the rotation measure (RM) is derived
(see Table~\ref{rratTab}). The other is for No.8, which is mildly
polarized with a sweep-up polarization angle curve. The pulses of this
RRAT have a diverse polarization feature.

More RRATs discovered in the GPPS survey will be presented by Zhou et
al. (2021, in preparation).

\begin{table}
  \centering
  \caption{Polarization Parameters of Eight Newly Discovered Pulsars}
  \label{table:poln}
  \fns
  \tabcolsep 1.0mm
  \begin{tabular}{crrrrr}
    \hline
PSR         & $W_{50}$ & $L/I$ & $V/I$  & $|V|/I$ &\multicolumn{1}{c}{RM}    \\
            & ($^{\circ}$) & (\%) & (\%) & (\%) &($\rm rad{\,} m^{-2}$) \\
\hline
J1848$+$0127g &   7.0 &  49.2 &  2.5 &  8.2 &$  -50.1(15)$   \\
J1849$-$0014g &  26.7 &  22.8 &$-1.6$&  3.4 &$ -110.3(16)$   \\
J1852$-$0023g &   8.4 &  27.2 &  1.2 &  9.1 &$   93.8(20)$   \\
J1852$+$0056g &  16.9 &  58.7 &$-29.6$&30.1 &$  -93.6(16)$   \\
J1856$+$0211g &   1.4 &  58.3 & 30.0 & 37.2 &$ -180.7(27)$   \\
J1903$+$0433g &   3.8 &  16.6 & -8.3 & 16.9 &$  213.2(31)$  \\
J1926$+$1857g &  16.9 &  78.7 & 13.6 & 15.2 &$  262.3(13)$   \\
J2052$+$4421g &  40.8 &  62.4 &  9.5 &  9.8 &$ -276.7(5)$   \\
\hline
  \end{tabular}
\end{table}

\begin{figure}
  \centering
  \includegraphics[width = 0.222\textwidth]{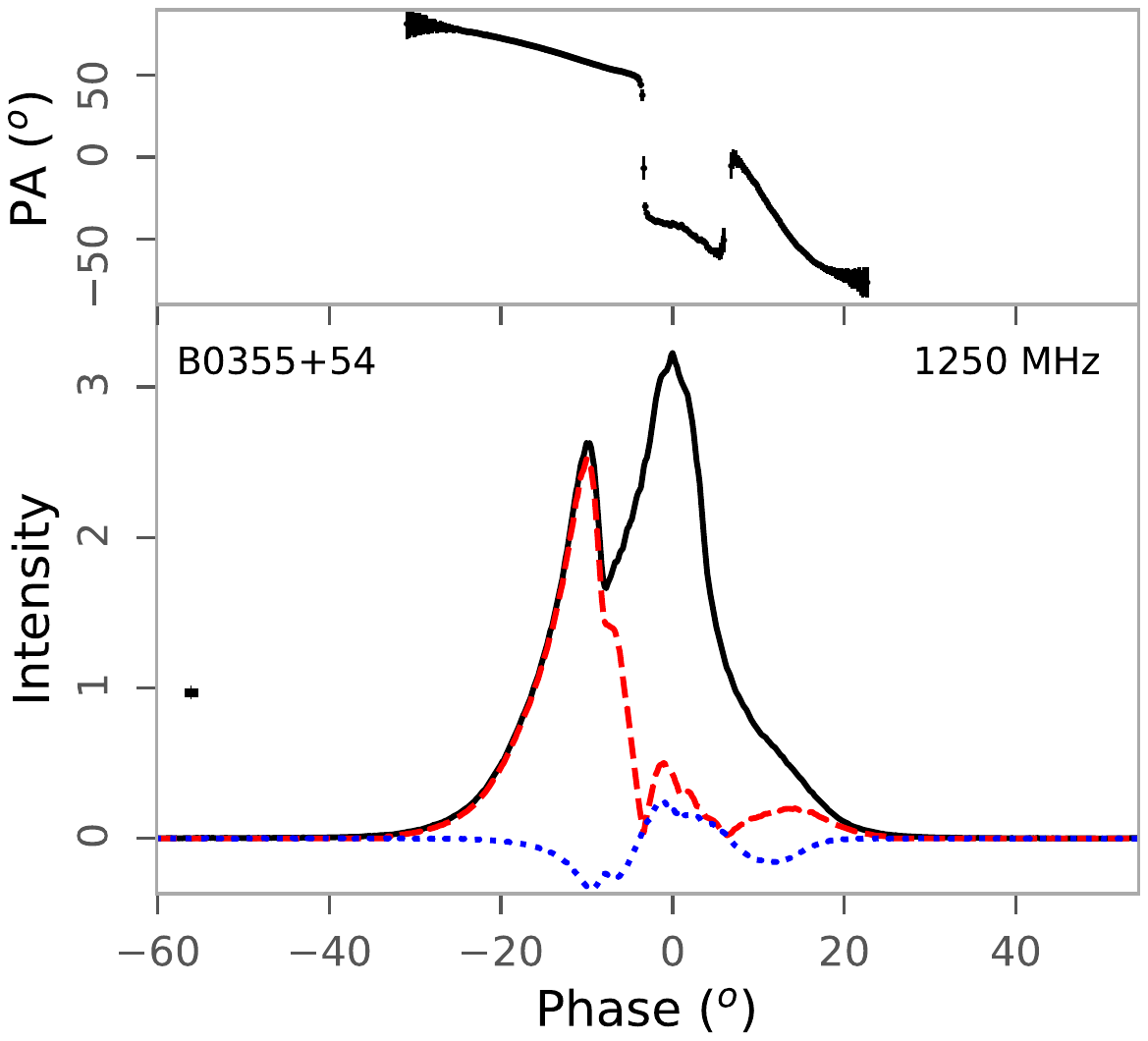}
  \includegraphics[width = 0.222\textwidth]{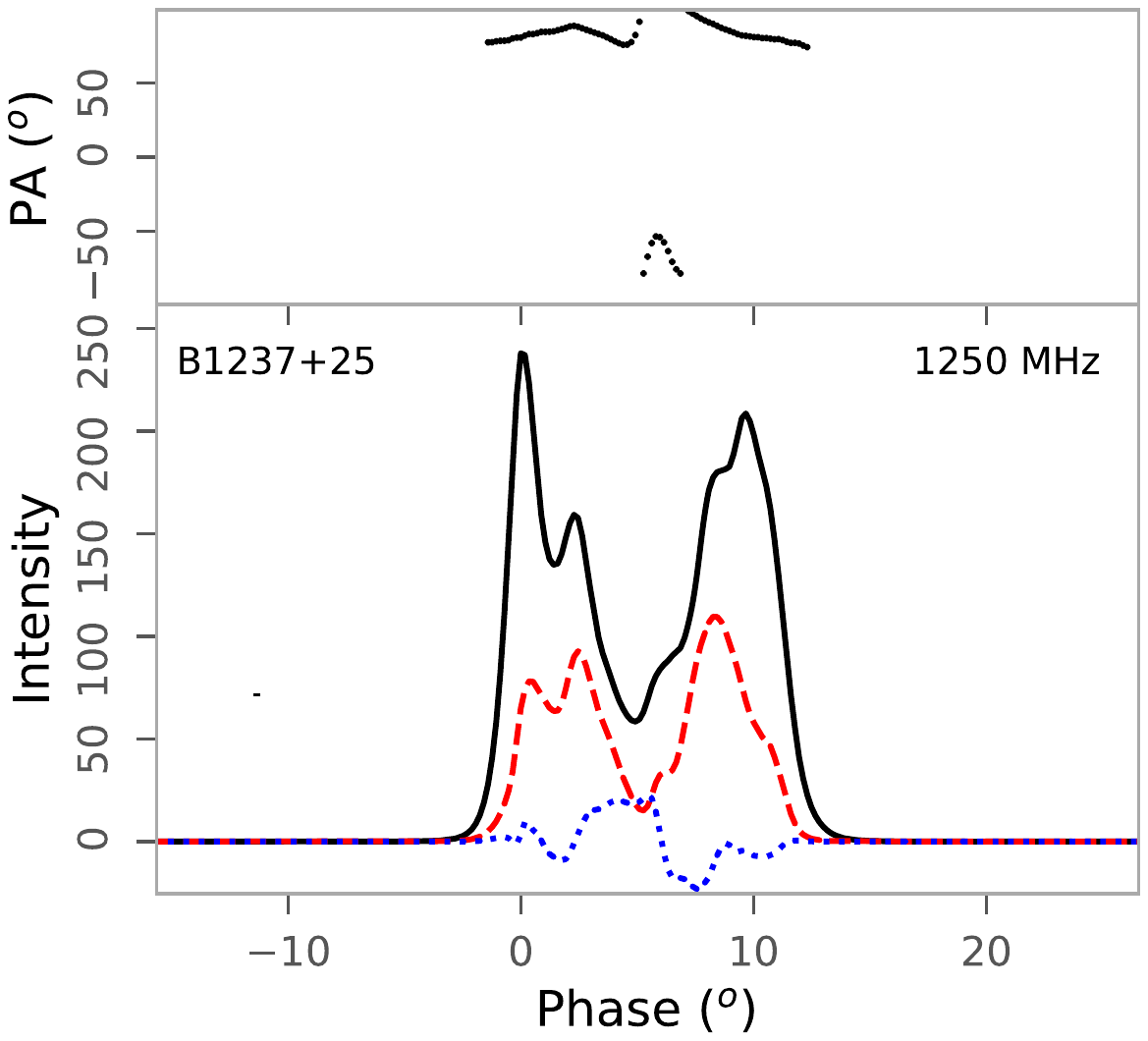}
  \includegraphics[width = 0.222\textwidth]{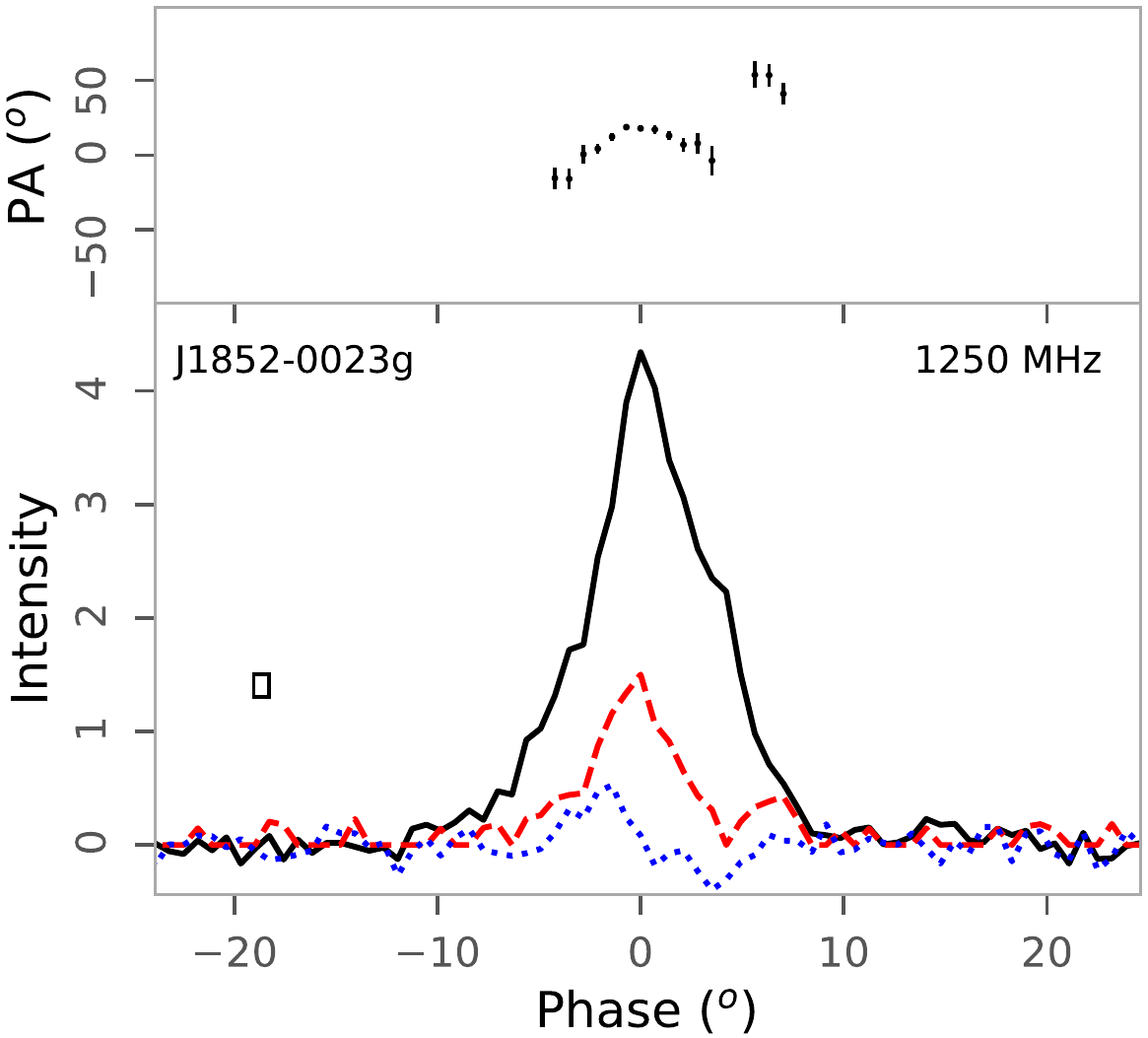}
  \includegraphics[width = 0.222\textwidth]{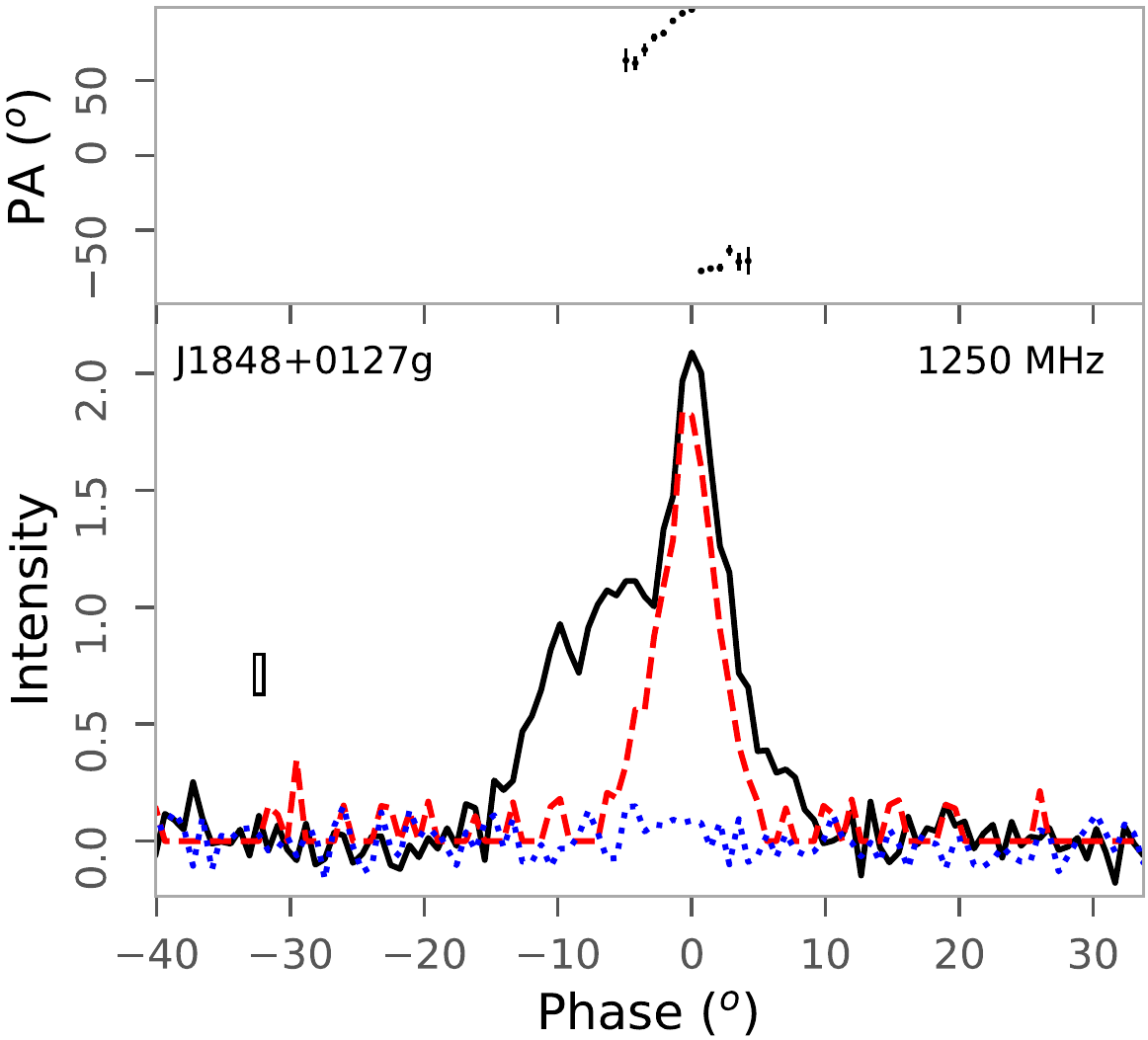}
  \includegraphics[width = 0.222\textwidth]{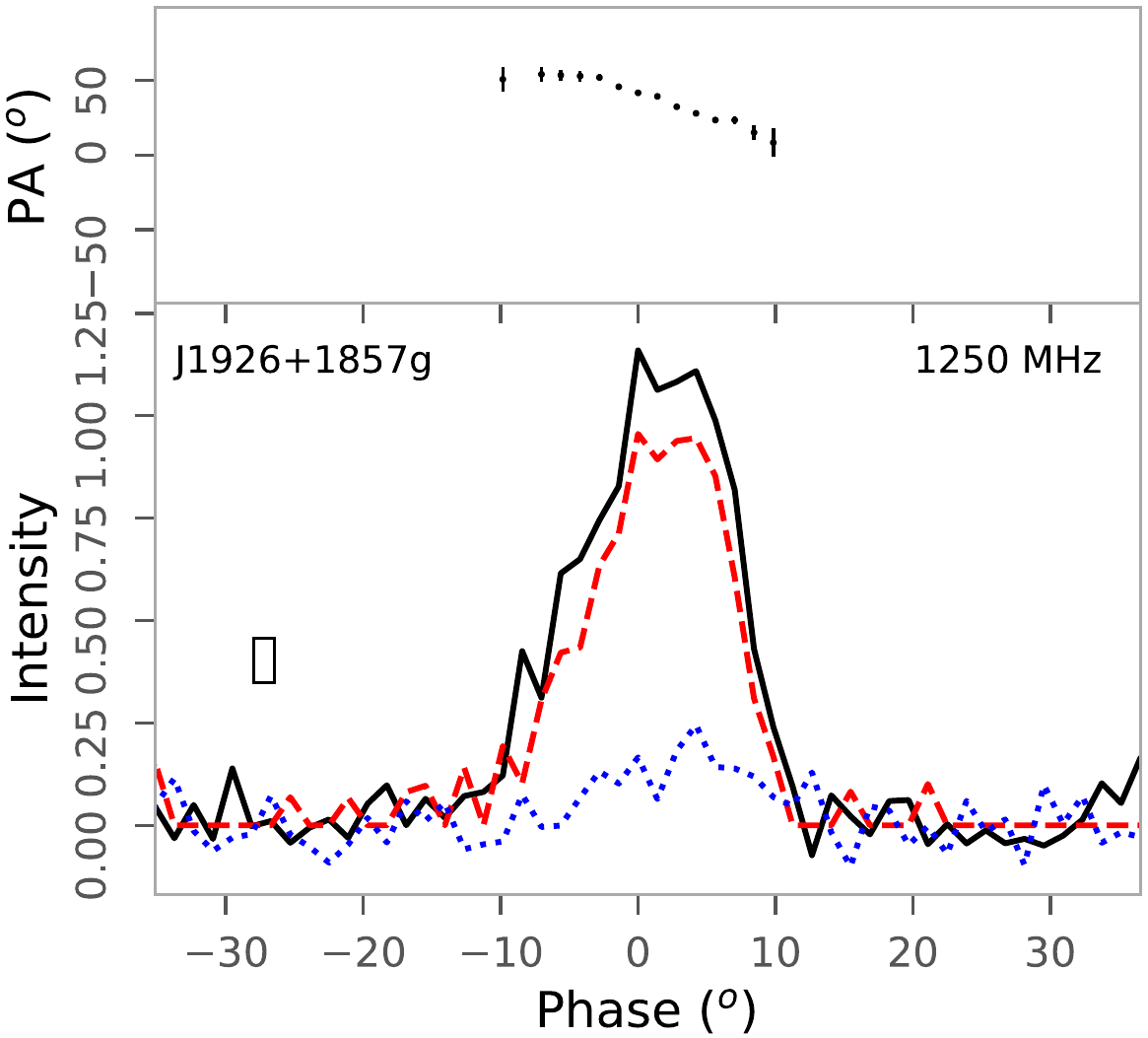}
  \includegraphics[width = 0.222\textwidth]{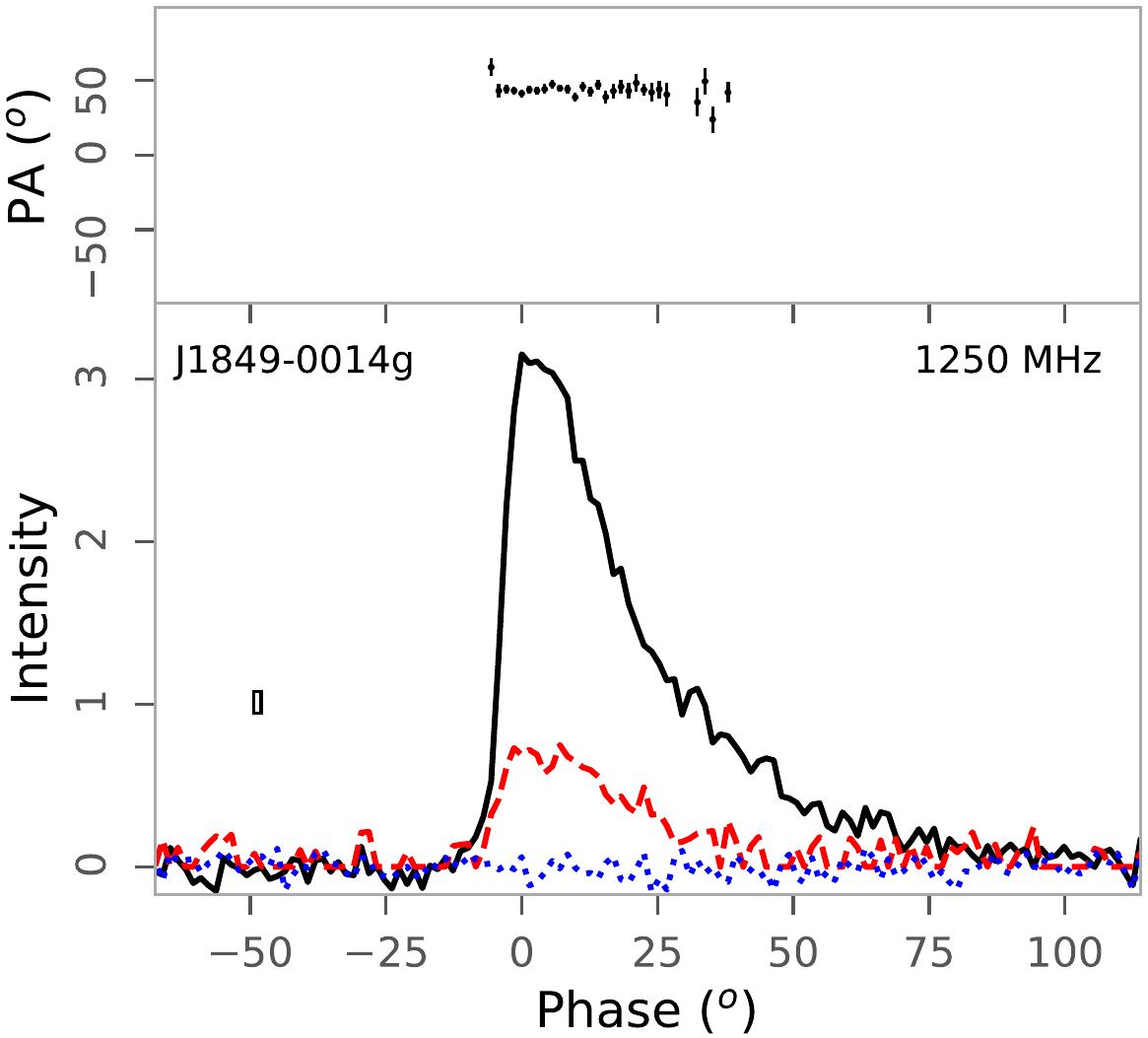}
  \includegraphics[width = 0.222\textwidth]{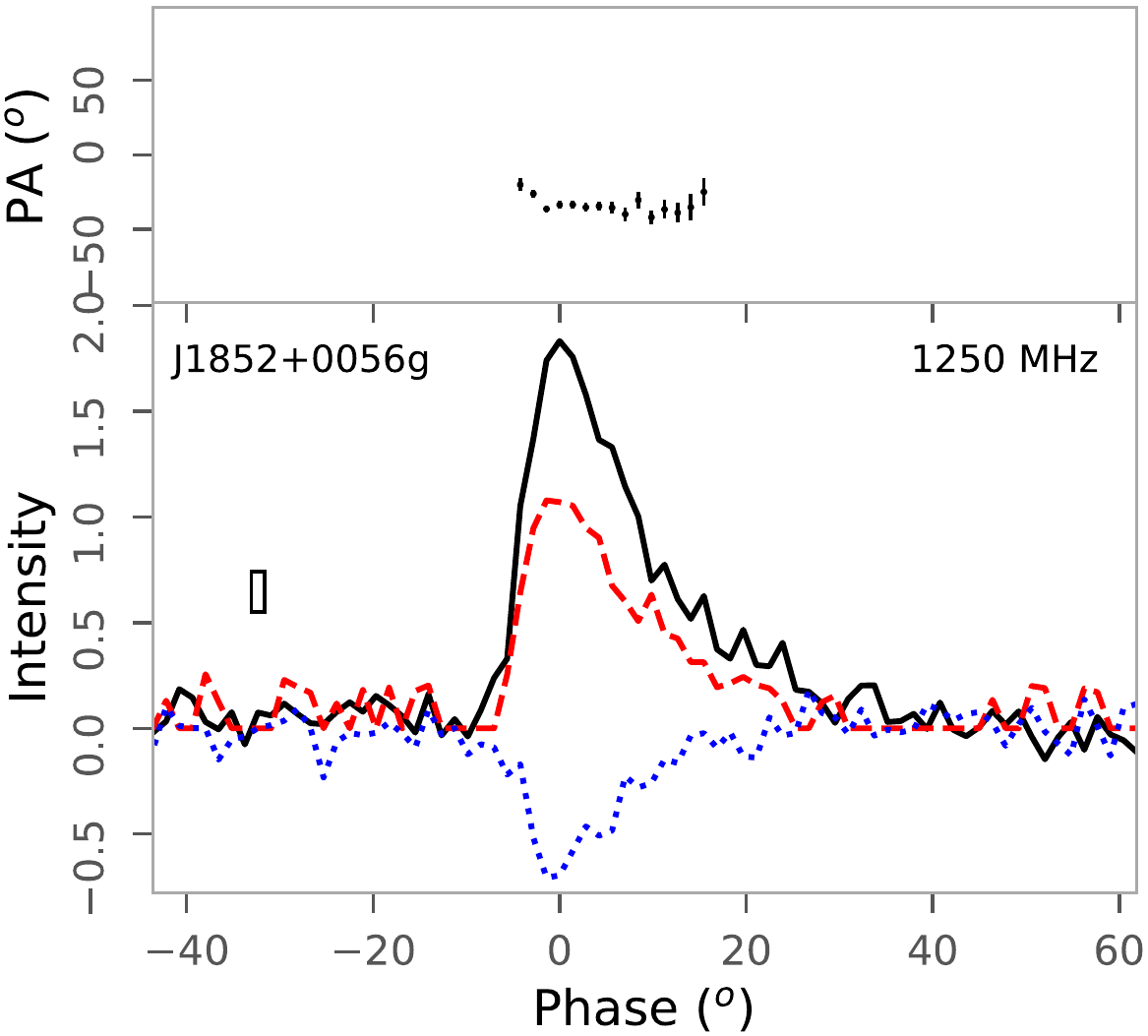}
  \includegraphics[width = 0.222\textwidth]{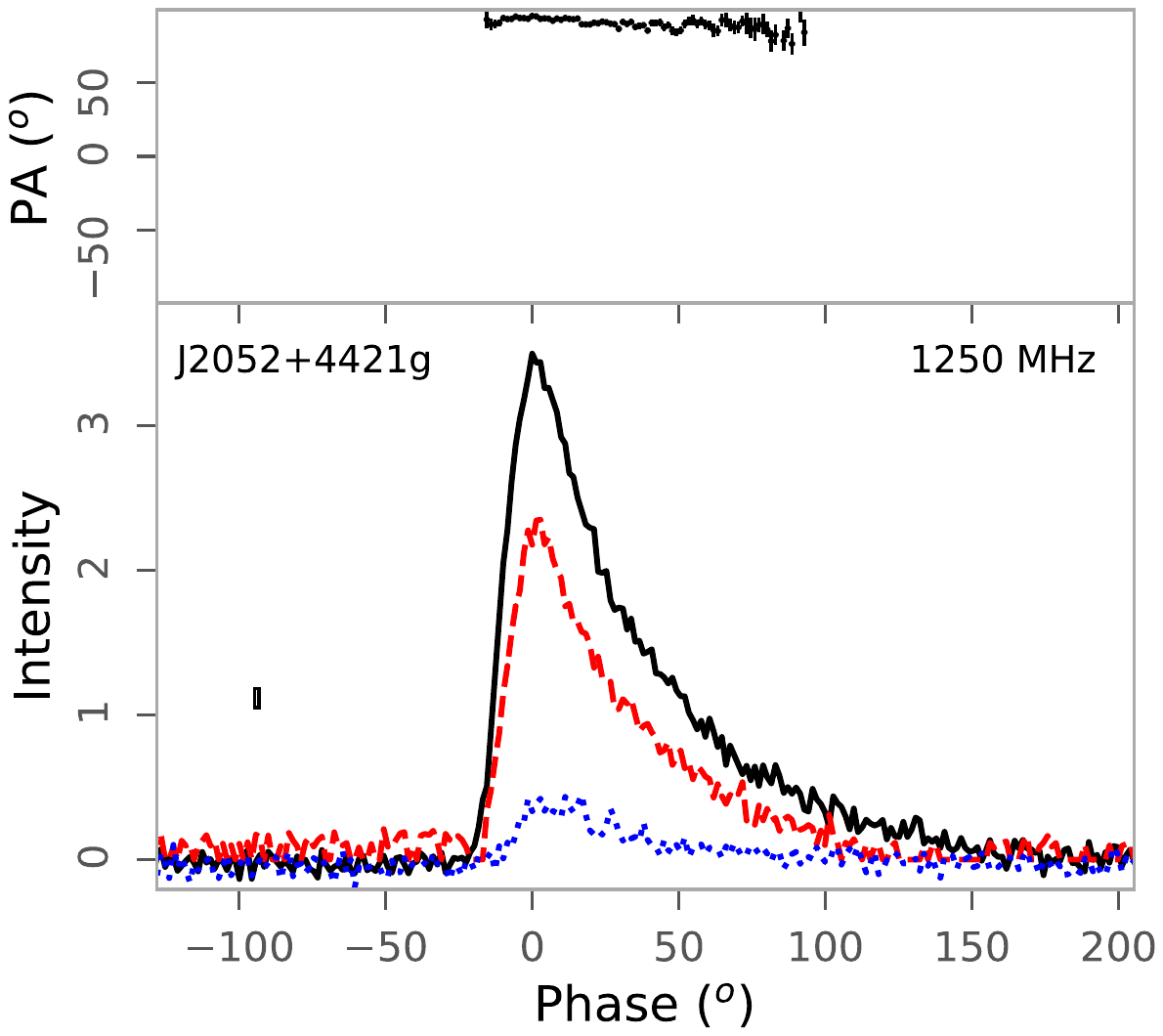}
  \includegraphics[width = 0.222\textwidth]{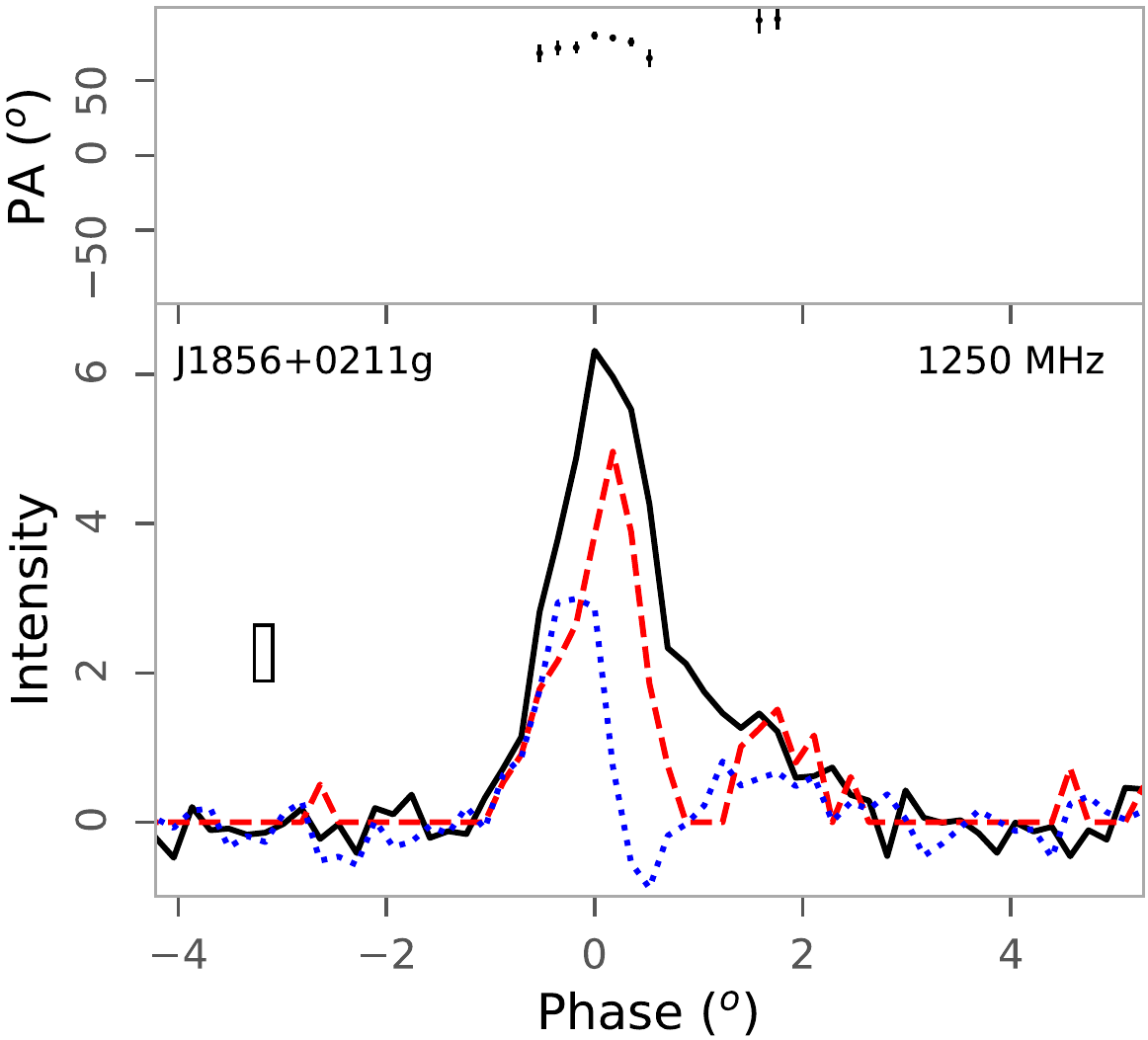}
  \includegraphics[width = 0.222\textwidth]{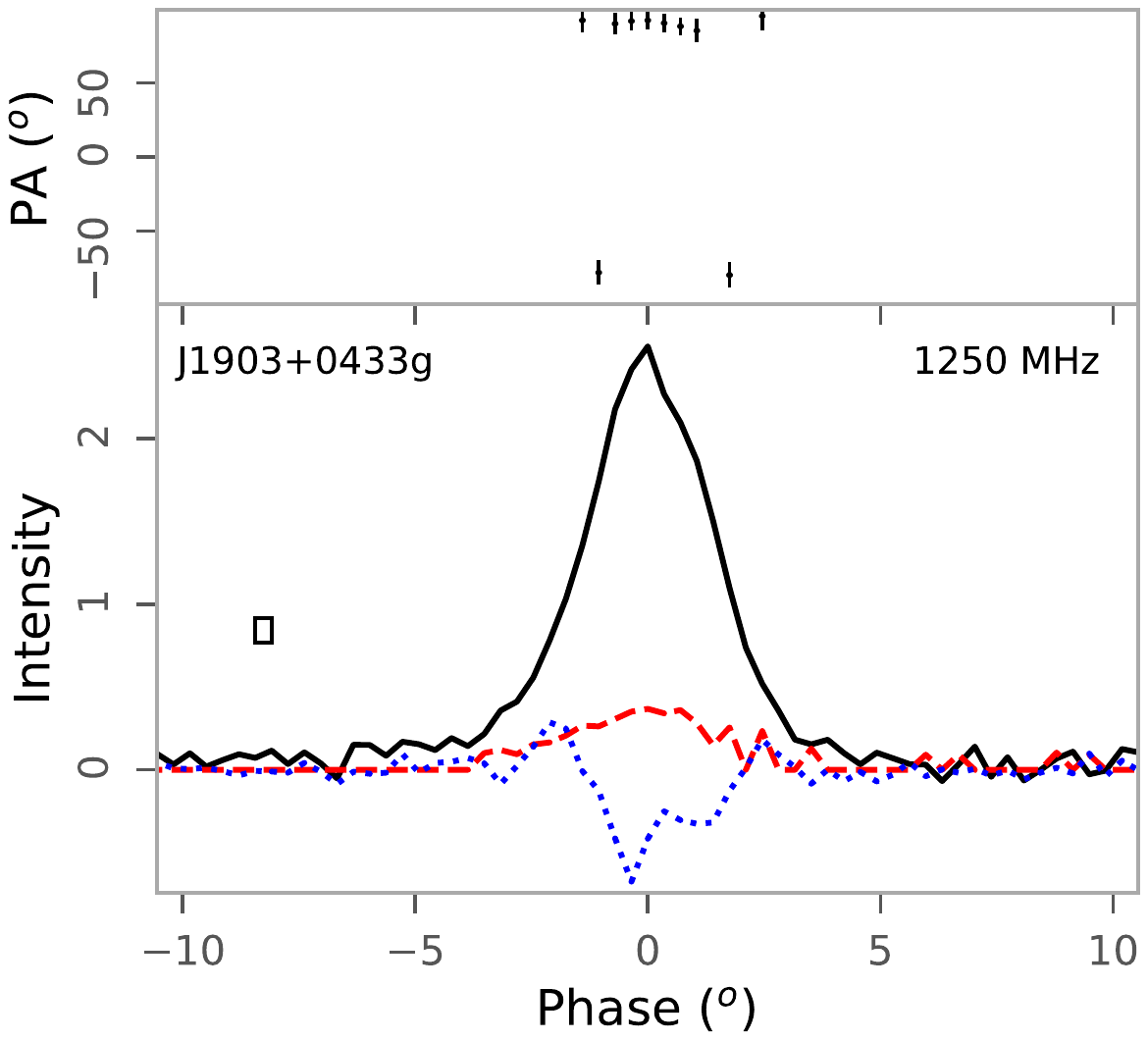}
  \caption{\baselineskip 3.8mm Integrated polarization profiles of eight
    newly discovered
    pulsars by FAST. PSRs B0355+54 and B1237+25 are taken as tests for
    the data processing pipeline. For each pulsar, the upper panel is
    polarization angles, and the bottom panel includes total intensity
    ({\it thick line}), linear ({\it dashed line}) and circular ({\it dotted line})
    polarization with the positive values for the left-hand sense. The
    phase is defined as 0$^{\circ}$--360$^{\circ}$ for a period, starting at the
    peak of the pulse.}
  \label{fig20_poln}
\end{figure}

\subsection{Follow-up Observations for Polarization Profiles of
  Newly Discovered Pulsars}

The L-band 19-beam receiver mounted on FAST has excellent stable
performance in terms of polarization \citep{lwm+20}. During the verification
observation, the polarization data were recorded, which are very useful
to get the polarization profiles when a candidate is verified, without
further costs of valuable FAST observation time. For this purpose, as
mentioned above, the calibration signal was on-off and the signals for
four polarization channels were recorded for 2 minutes before and after
each verification session.

The polarization data and observation parameters in each fits file are
carefully checked, and then data are calibrated and processed by using
the package PSRCHIVE \citep{hvm04}. We obtained the polarization
profiles (see Fig.~\ref{fig20_poln}) and also the Faraday RMs
(see examples in Table~\ref{table:poln}). The RMs have been
corrected for the ionosphere contribution. To verify the results, FAST
data on two pulsars, PSR B0355+54 and PSR B1237+25, were processed with
the same procedures, and results are consistent with polarization
profiles at L-band previously published in \citet{gl98},
\citet{scr+84} and \citet{jhv+05}. For newly discovered pulsars in the
GPPS survey, the polarization profiles of only a few pulsars are
presented here. More results will be presented by Wang P.F. et
al. (2021, in preparation).

The polarization profiles of PSR J1852$-$0023g (gpps0042) presented
here look like a typical unresolved cone-core pulsar \citep{ran93},
with a sense reversal of circular polarization at the profile
center. PSR J1848+0127g (gpps0035) has a highly polarized trailing
component \citep[i.e. the second class in][]{wh16} and a very steep
polarization angle sweep in the leading unpolarized component, much
like a partial cone discussed by \citet{lm88}. PSR J1926+1857g
(gpps0025) is highly polarized with a linearly declining
polarization angle and a steep trailing edge of mean pulse, which
indicates its cone nature of the emission. Three other pulsars, PSRs
J1849-0014g (gpps0027), J1852+0056g (gpps0014) and J2052+4421g
(gpps0019), obviously show long tails for scattered emission, in
which scattering reduces the linear polarization and results in a
flat polarization angle curve. Such a scattering effect on
polarization profiles was first outlined by \citet{lh03} and later
confirmed by \citet{kj08}.

Polarization profiles of two long period pulsars mentioned in Sect.~\ref{longP}
are also presented in Figure~\ref{fig20_poln}. PSR J1856+0211g (gpps0158) is highly
polarized with a strong left-hand circular polarization in the leading edge.
PSR J1903+0433g (gpps0090, with the second longest period) is mildly polarized.

\begin{table*}
  \centering

  \begin{minipage}{9cm}
  \caption{Some Parameters are Updated for 64 Known Pulsars}
  \label{para64psr}\end{minipage}

  \fns
  \tabcolsep 0.7mm
  \begin{tabular}{lllrllclcrcc}
    \hline\noalign{\smallskip}
PSR name      & Ref. & {$P$} in Ref. & DM  &  RA(2000)    &  \multicolumn{1}{c}{Dec(2000)}    & Updated &  \multicolumn{1}{c}{{$P$} ($\sigma$)} & Epoach for {$P$}  & DM ($\sigma$) &  RA(2000) &  Dec(2000)  \\
&      & \multicolumn{1}{c}{(s)} &  & hh:mm:ss.s & \multicolumn{1}{c}{dd:mm:ss}& items & \multicolumn{1}{c}{(s) }   &   (MJD)   &  & hh:mm:ss.s &\multicolumn{1}{c}{dd:mm} \\
(1) & (2) & \multicolumn{1}{c}{(3)} & (4) &\multicolumn{1}{c}{ (5)} & \multicolumn{1}{c}{ (6)} & (7) &\multicolumn{1}{c}{ (8)} & (9) &\multicolumn{1}{c}{ (10)} & (11) & (12)  \\
 \hline\noalign{\smallskip}
J0611+1436  &    [1]      &  0.270329   &  45.7    &  06:11:18.65   &  +14:36:52     &  DM     &  0.27032969(38)   &  58847.665398  &  43.7(2)    &  06:11:17  &  +14:37    \\
J0658+0022  &    [2]      &  0.563295   &  122.0   &  06:58:15.21   &  +00:22:35.3   &  DM     &  0.5632972(16)    &  58839.756478  &  115.6(5)   &  06:58:14  &  +00:23    \\
J1832+0029  &    [3]      &  0.533917   &  28.3    &  18:32:50.7    &  +00:29:27     &  DM     &  0.5339170(16)    &  58765.445867  &  32.7(4)    &  18:32:48  &  +00:30    \\
J1843+01    &    [8]      &  1.26702    &  248.0   &  18:43         &  +01           &  Posi   &  1.266990(12)     &  59198.252593  &  251.9(16)  &  18:43:28  &  +01:19    \\
J1847+01    &    [4]      &  0.00346    &  20.0    &  18:47         &  +01           &  Posi   &  0.0034630711(1)  &  59093.545385  &  20.100(5)  &  18:47:05  &  +01:13    \\
J1848-0023  &    [5]      &  0.537624   &  30.6    &  18:48:37.89   &  $-$00:23:17   &  DM     &  0.53762431(52)   &  59077.578982  &  34.6(4)    &  18:48:40  &  $-$00:24  \\
J1849+01    &    [4]      &  1.832      &  213.6   &  18:49         &  +01           &  Posi   &  1.832250(18)     &  58909.025657  &  217.2(15)  &  18:49:56  &  +01:06    \\
J1849+0409  &    [3]      &  0.761194   &  56.1    &  18:49:3.47    &  +04:09:42.3   &  DM     &  0.7612033(32)    &  58720.600254  &  64.1(6)    &  18:49:03  &  +04:09    \\
J1849+04    &    [4]      &  0.42111    &  191.8   &  18:49         &  +04           &  Posi   &  0.4211266(11)    &  58720.585427  &  188.2(4)   &  18:49:40  &  +04:30    \\
J1849-0040  &    [5]      &  0.672481   &  1234.9  &  18:49:10.25   &  $-$00:40:20   &  DM     &  0.672479(8)      &  59244.140909  &  1267.6(18) &  18:49:11  &  $-$00:40  \\
J1850-0006  &    [6]      &  2.191498   &  570.0   &  18:50:47.93   &  $-$00:06:26.1 &  DM     &  2.1915070(88)    &  59082.572228  &  655.0(18)  &  18:50:47  &  $-$00:07  \\
J1851+00    &    [4]      &  0.02283    &  107.4   &  18:51         &  +00           &  Posi   &  0.0228466631(27) &  58902.066301  &  107.60(2)  &  18:51:02  &  +00:10    \\
J1851+0242  &    [4]      &  1.49714    &  534.1   &  18:51:22      &  +02:42:37     &  P      &  4.4912(2)        &  58997.792192  &  519.3(71)  &  18:51:19  &  +02:42    \\
J1852+00    &    [4]      &  1.92066    &  590.4   &  18:52         &  +00           &  Posi   &  1.920653(21)     &  58923.007492  &  590.0(15)  &  18:52:39  &  +00:00    \\
J1853+0009  &    [4]      &  0.03341    &  192.0   &  18:53         &  +00:09        &  Posi   &  0.033395062(2)   &  58939.971197  &  192.40(3)  &  18:53:39  &  +00:08    \\
J1853+0029  &    [4]      &  0.93832    &  230.0   &  18:53         &  +00:29        &  Posi,P &  1.8767568(63)    &  59112.477816  &  227.1(15)  &  18:53:18  &  +00:29    \\
J1853+03    &    [7]      &  0.58553    &  290.2   &  18:53         &  +03:00        &  Posi   &  0.5855545(32)    &  58977.860290  &  295.8(8)   &  18:53:12  &  +03:00    \\
J1853+04    &    [8]      &  1.32065    &  549.3   &  18:53         &  +04           &  Posi   &  1.320623(13)     &  59004.780117  &  548.9(13)  &  18:53:46  &  +04:27    \\
J1854+00    &    [7]      &  0.76733    &  532.9   &  18:54         &  +00:00        &  Posi   &  0.767278(1)      &  58940.968884  &  525.0(6)   &  18:54:43  &  +00:50    \\
J1854+0317  &    [9]      &  1.36645    &  404.0   &  18:54:29.06   &  +03:17:31     &  DM     &  1.366465(11)     &  58991.813118  &  390.0(11)  &  18:54:29  &  +03:17    \\
J1855+0422  &    [10]     &  1.678106   &  438.0   &  18:55:41.37   &  +04:22:47     &  DM     &  1.678105(15)     &  59187.292726  &  455.6(13)  &  18:55:41  &  +04:22    \\
J1858+02    &    [7]      &  0.19765    &  492.1   &  18:58         &  +02:00        &  Posi   &  0.19764740(24)   &  58900.051935  &  492.8(2)   &  18:58:23  &  +02:41    \\
J1853+0853  &    [3]      &  3.914658   &  214.0   &  18:53:22.07   &  +08:53:17     &  DM     &  3.914696(79)     &  59253.091997  &  236.6(31)  &  18:53:20  &  +08:54    \\
J1859+00    &    [11],[5] &  0.559634   &  420.0   &  18:59:46      &  +00:35        &  Posi   &  0.5596363(18)    &  58982.867548  &  424.0(4)   &  18:59:47  &  +00:38    \\
J1859+03    &    [4]      &  1.51171    &  555.1   &  18:59         &  +03           &  Posi   &  1.511506(28)     &  58867.217455  &  558.7(26)  &  18:59:10  &  +03:46    \\
J1901+00    &    [11],[5] &  0.777662   &  345.5   &  19:01:32      &  +00:26        &  Posi   &  0.7776620(32)    &  58977.880799  &  340.3(6)   &  19:01:32  &  +00:32    \\
J1901+0435  &    [3]      &  0.690576   &  1042.6  &  19:01:32.2    &  +04:35:23     &  DM     &  0.690605(11)     &  58565.032593  &  920.0(25)  &  19:01:31  &  +04:35    \\
J1902+02    &    [7]      &  0.41532    &  281.2   &  19:02         &  +02:00        &  Posi   &  0.4153927(11)    &  59233.161342  &  280.4(4)   &  19:02:31  &  +02:35    \\
J1903+0415  &    [7]      &  1.15139    &  473.5   &  19:03         &  +04:15        &  Posi   &  1.151411(11)     &  58902.108022  &  481.9(14)  &  19:03:30  &  +04:15    \\
J1903+09    &    [4]      &  0.16631    &  362.9   &  19:03         &  +09           &  Posi   &  0.16631690(28)   &  59164.332493  &  362.9(3)   &  19:03:39  &  +09:12    \\
J1905+10    &    [4]      &  1.72688    &  165.7   &  19:05         &  +10           &  Posi   &  1.726813(17)     &  58852.191366  &  163.7(14)  &  19:05:19  &  +10:34    \\
J1906+0725  &    [7]      &  1.53651    &  480.4   &  19:06         &  +07:25        &  Posi   &  1.536442(14)     &  58770.485688  &  476.4(12)  &  19:06:23  &  +07:25    \\
J1907+05    &    [7]      &  0.16868    &  456.7   &  19:07         &  +05:00        &  Posi   &  0.16867706(35)   &  58982.878143  &  457.1(3)   &  19:08:02  &  +05:59    \\
J1909+12    &    [4]      &  1.229338   &  292.5   &  19:09:57.7    &  +12:04:55.3   &  DM     &  1.229319(8)      &  59189.269571  &  302.5(10)  &  19:09:51  &  +12:06    \\
J1910+1027  &    [7]      &  0.53147    &  705.7   &  19:10         &  +10:27        &  Posi   &  0.5315526(26)    &  59093.531512  &  715.3(7)   &  19:10:48  &  +10:27    \\
J1910+07    &    [4]      &  0.53869    &  256.5   &  19:10         &  +07           &  Posi   &  0.5386463(54)    &  58911.070131  &  255.3(15)  &  19:10:06  &  +07:11    \\
J1911+09    &    [7]      &  0.27371    &  334.7   &  19:11         &  +09:00        &  Posi   &  0.27370588(47)   &  59067.582251  &  340.7(8)   &  19:11:47  &  +09:22    \\
J1911+10    &    [7]      &  0.19089    &  446.2   &  19:11         &  +10:00        &  Posi   &  0.19087553(57)   &  59155.408562  &  445.0(5)   &  19:11:43  &  +10:52    \\
J1913+05    &    [4]      &  0.662      &  330.1   &  19:13         &  +05           &  Posi   &  0.6619994(40)    &  59185.290377  &  335.9(9)   &  19:13:22  &  +05:24    \\
J1914+08    &    [12]     &  0.440048   &  285.0   &  19:14:18      &  +08:45        &  Posi   &  0.44003996(35)   &  59090.587469  &  290.6(4)   &  19:14:25  &  +08:39    \\
J1916+1023  &    [5]      &  1.618339   &  329.8   &  19:16:36.91   &  +10:23:03     &  DM     &  1.618330(14)     &  58909.104157  &  341.8(13)  &  19:16:37  &  +10:23    \\
J1918+1541  &    [13]     &  0.370883   &  13.0    &  19:18:7.7     &  +15:41:15.2   &  DM     &  0.37088498(71)   &  59198.231761  &  11.4(3)    &  19:18:10  &  +15:42    \\
J1920+1110  &    [10]     &  0.509886   &  182.0   &  19:20:13.31   &  +11:10:59     &  DM     &  0.5098862(16)    &  58977.910791  &  188.4(4)   &  19:20:14  &  +11:11    \\
J1924+1628  &    [7]      &  0.37509    &  542.9   &  19:24         &  +16:28        &  Posi   &  0.37508191(88)   &  58808.402942  &  541.7(3)   &  19:24:45  &  +16:29    \\
J1924+17    &    [7]      &  0.75843    &  527.4   &  19:24         &  +17:00        &  Posi   &  0.7584369(11)    &  59069.686122  &  540.9(7)   &  19:24:32  &  +17:14    \\
J1924+2040  &    [14],[15]&  0.23779    &  213.0   &  19:24:40      &  +20:40:03     &  DM     &  0.2377930(3)     &  58843.315936  &  226.6(2)   &  19:24:35  &  +20:40    \\
J1926+1613  &    [7]      &  0.3083     &  32.9    &  19:26         &  +16:13        &  DM,Posi&  0.308304(1)      &  58936.061255  &  24.5(5)    &  19:26:51  &  +16:14    \\
J1927+1852  &    [16]     &  0.482766   &  254.0   &  19:27:10.42   &  +18:52:8.5    &  DM     &  0.48276601(42)   &  58940.066550  &  264.5(4)   &  19:27:07  &  +18:51    \\
J1929+1905  &    [17]     &  0.339243   &  553.9   &  19:29:31.7    &  +19:05:43.4   &  DM     &  0.33921584(21)   &  59174.390702  &  528.4(3)   &  19:29:14  &  +19:10    \\
J1930+14    &    [7]      &  0.42571    &  209.2   &  19:30         &  +14:00        &  Posi   &  0.4257200(22)    &  58808.383858  &  214.8(7)   &  19:30:19  &  +14:09    \\
J1930+17    &    [18]     &  1.60969    &  201.0   &  19:30:44      &  +17:25        &  Posi   &  1.609723(13)     &  59157.422738  &  197.4(13)  &  19:30:31  &  +17:23    \\
J1934+19    &    [7]      &  0.23099    &  97.6    &  19:34         &  +19:00        &  Posi   &  0.23098462(95)   &  58995.887279  &  98.4(5)    &  19:34:18  &  +19:26    \\
J1936+18    &    [4]      &  0.05835    &  125.5   &  19:36         &  +18           &  Posi   &  0.05834513(18)   &  58997.893701  &  125.9(4)   &  19:36:03  &  +18:05    \\
J1936+20    &    [7]      &  1.39088    &  205.1   &  19:36         &  +20:00        &  DM,Posi&  1.3907326(35)    &  59102.591721  &  195.9(11)  &  19:36:29  &  +20:41    \\
J1936+21    &    [18]     &  0.642932   &  264.0   &  19:36:29      &  +21:12        &  Posi   &  0.6429599(22)    &  58944.052936  &  262.4(5)   &  19:36:11  &  +21:09    \\
J1936+21a   &    [4]      &  0.0316     &  73.9    &  19:36         &  +21           &  Posi   &  0.03158187(3)    &  59155.467443  &  75.0(2)    &  19:37:00  &  +21:43    \\
J1938+14    &    [19]     &  2.90251    &  74.2    &  19:38:19      &  +14:42        &  Posi   &  2.902504(51)     &  59209.202485  &  75.4(23)   &  19:38:07  &  +15:06    \\
J1939+26    &    [4]      &  0.46593$*$ &  48.1    &  19:39         &  +26           &  Posi   &  0.4669615(11)    &  59253.183278  &  47.5(4)    &  19:39:42  &  +26:09    \\
J1940+26    &    [4]      &  0.00481    &  171.6   &  19:40         &  +26           &  Posi   &  0.0048135292(14) &  59253.175813  &  171.61(4)  &  19:40:13  &  +26:01    \\
J1955+29    &    [4]      &  1.07377    &  206.0   &  19:55         &  +29           &  Posi   &  1.0738820(62)    &  58735.618100  &  212.4(9)   &  19:55:07  &  +29:31    \\
J2000+29    &    [8]      &  3.0737     &  132.5   &  20:00         &  +29           &  Posi   &  3.073809(58)     &  58901.180519  &  131.3(25)  &  20:00:14  &  +29:21    \\
J2003+29    &    [4]      &  1.00987    &  208.7   &  20:03         &  +29           &  Posi   &  1.0098800(56)    &  58943.068686  &  209.5(8)   &  20:03:02  &  +29:16    \\
J2010+31    &    [4]      &  1.55147    &  251.0   &  20:10         &  +31           &  Posi   &  1.551535(24)     &  58910.171365  &  251.8(23)  &  20:10:35  &  +31:50    \\
J2138+4911  &    [20]     &  0.696      &  168.0   &  21:38         &  +49:11        &  Posi   &  0.696171(3)      &  58941.118314  &  168.0(6)   &  21:38:15  &  +49:12    \\
\hline\noalign{\smallskip}
 \end{tabular}

\parbox{17cm}{(1) Reference in Column~(2):
[1]: \citet{lyne2017};
[2]: \citet{Burgay2013};
[3]: \citet{Lorimer2006};
[4]: {\it \url{http://www.naic.edu/~palfa/newpulsars/}};
[5]: \citet{hfs+04};
[6]: \citet{kel+09};
[7]: \citet{lbh+15};
[8]: \citet{patel2018};
[9]: \citet{Eatough2013};
[10]: \citet{mhl+02};
[11]: \citet{camilo1996}
[12]: \citet{kbj+18};
[13]: \citet{Camilo1995};
[14]: \citet{Weisberg1981};
[15]: \citet{vms83};
[16]: \citet{Lorimer2002};
[17]: {\it\url{http://astro.phys.wvu.edu/dmb/}};
[18]: \citet{Lorimer2013};
[19]: \citet{Deneva2016};
[20]: \citet{HRK2008}.  \\
  (2) Columns (3)--(6) are paraemeters in the reference, and
  Cols.~(7), (9), (10) and (11) are paraemters obtained in the GPPS survey; \\
   DM has {the} unit of cm$^{-3}$~pc. Newly measured {$P$} and DM have an uncertainty in brackets. $*$: mistyping in the reference.}
\end{table*}

%
\begin{figure*}[htp!!!]
\begin{tabular}{rrrrrr}
\includegraphics[width=38mm]{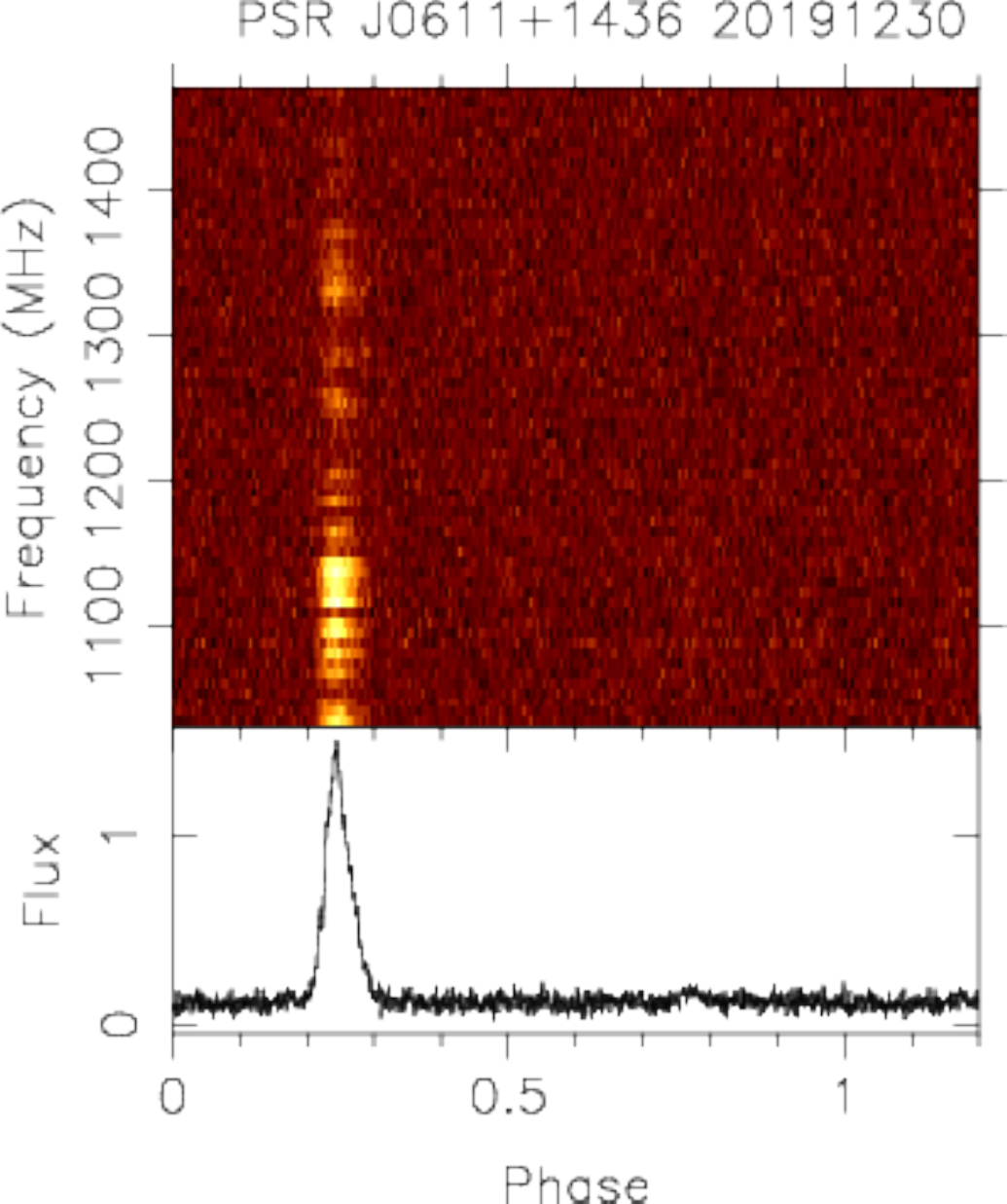}&
\includegraphics[width=38mm]{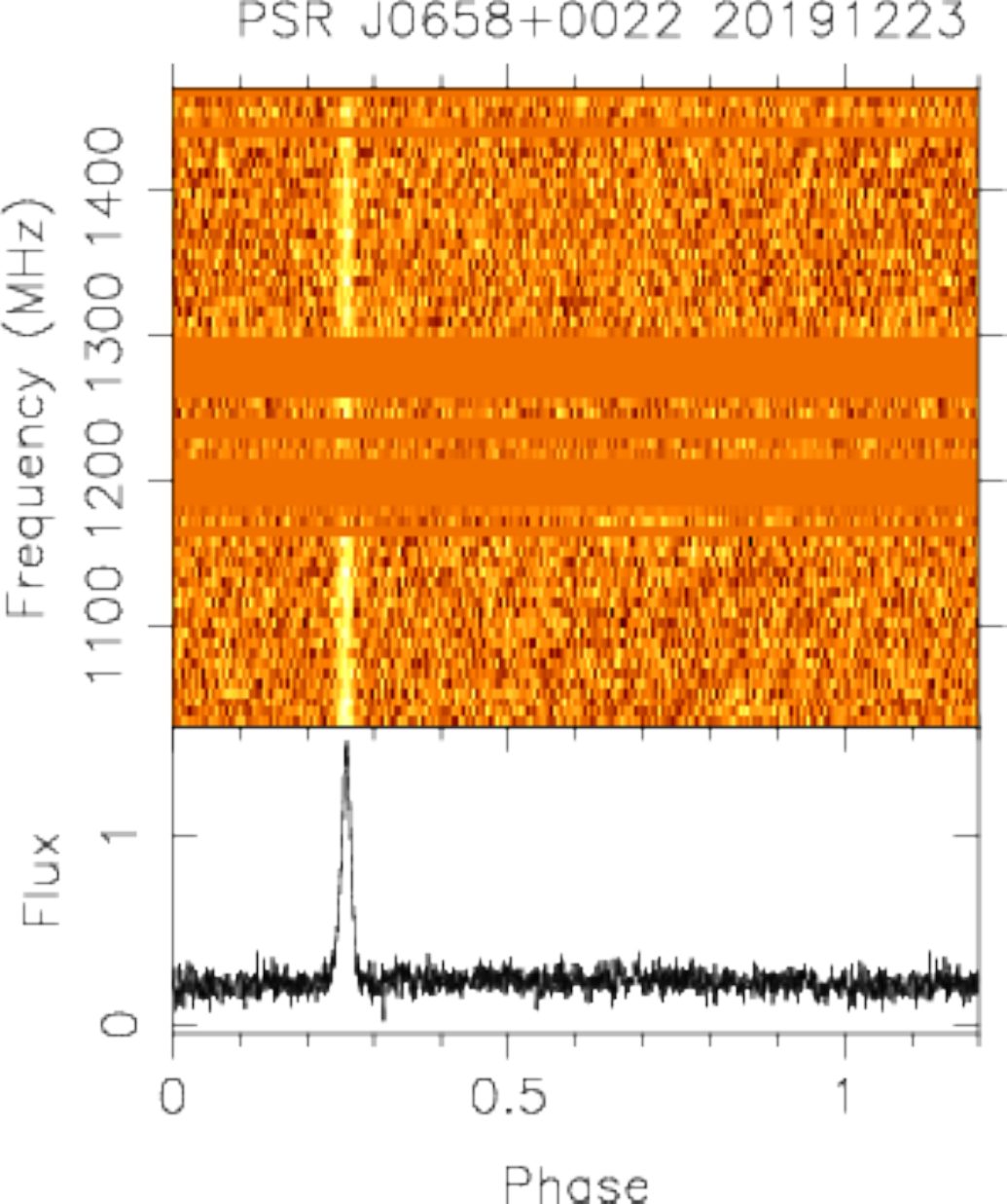}&
\includegraphics[width=38mm]{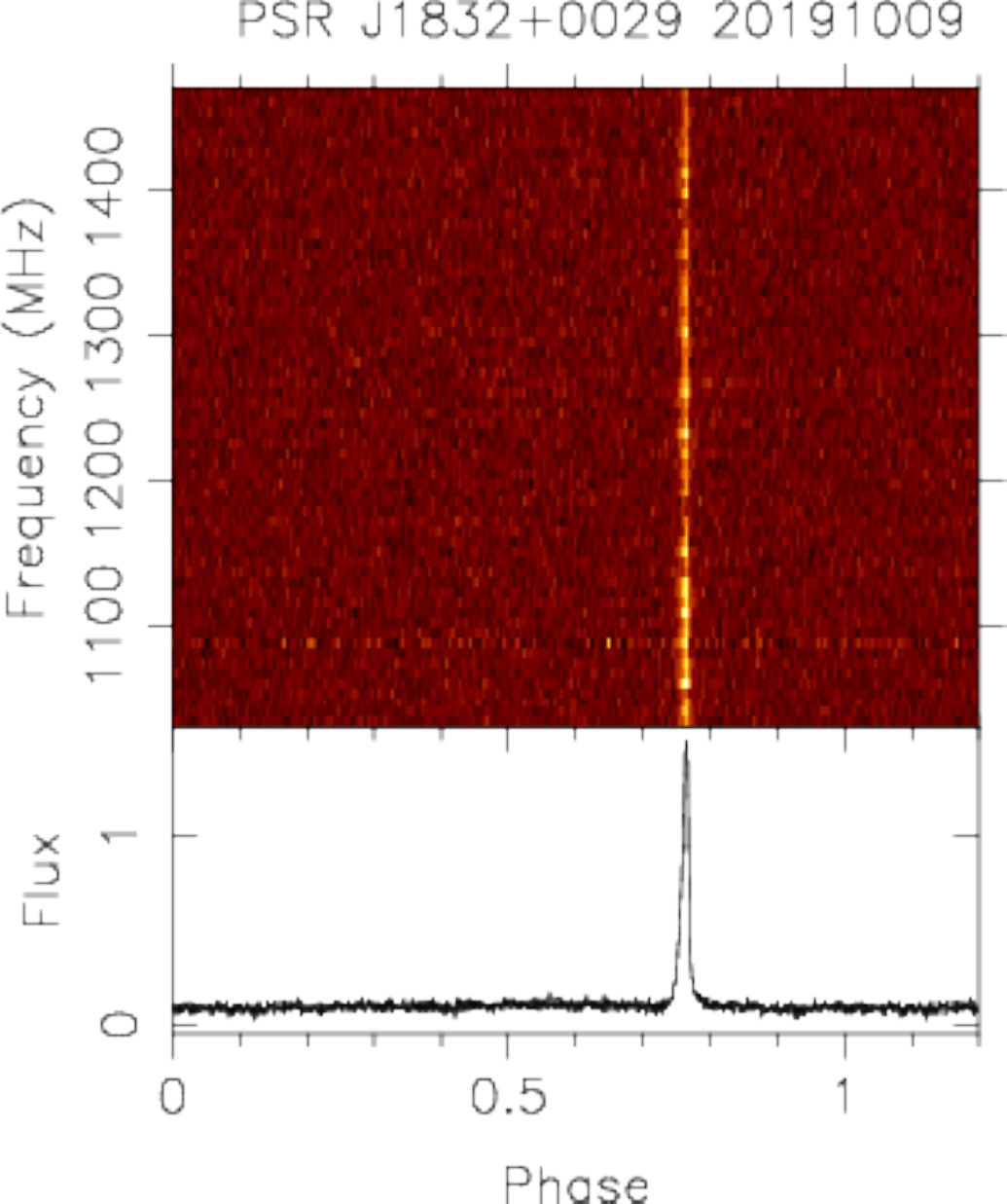}&
\includegraphics[width=38mm]{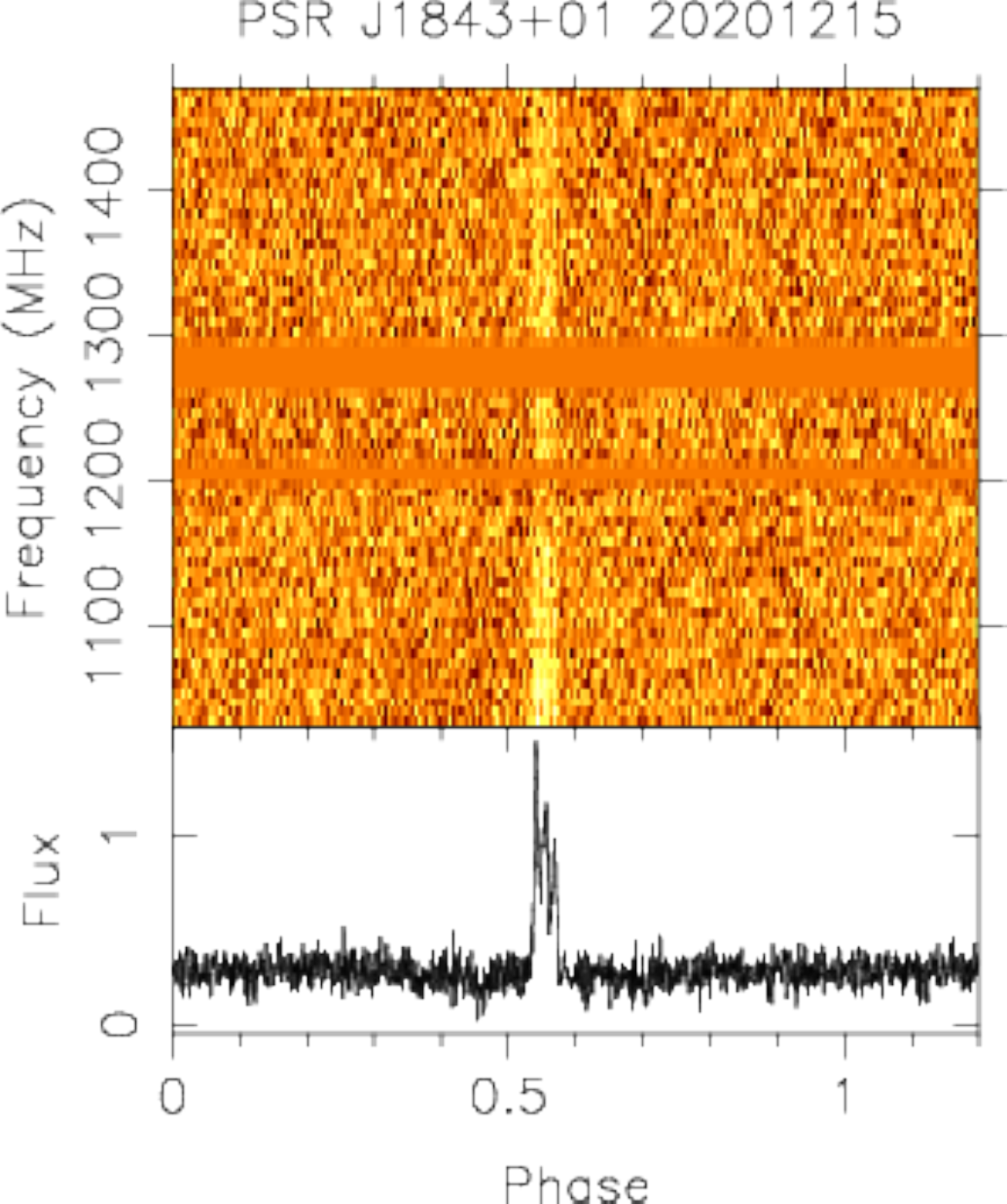}\\[2mm]
\includegraphics[width=38mm]{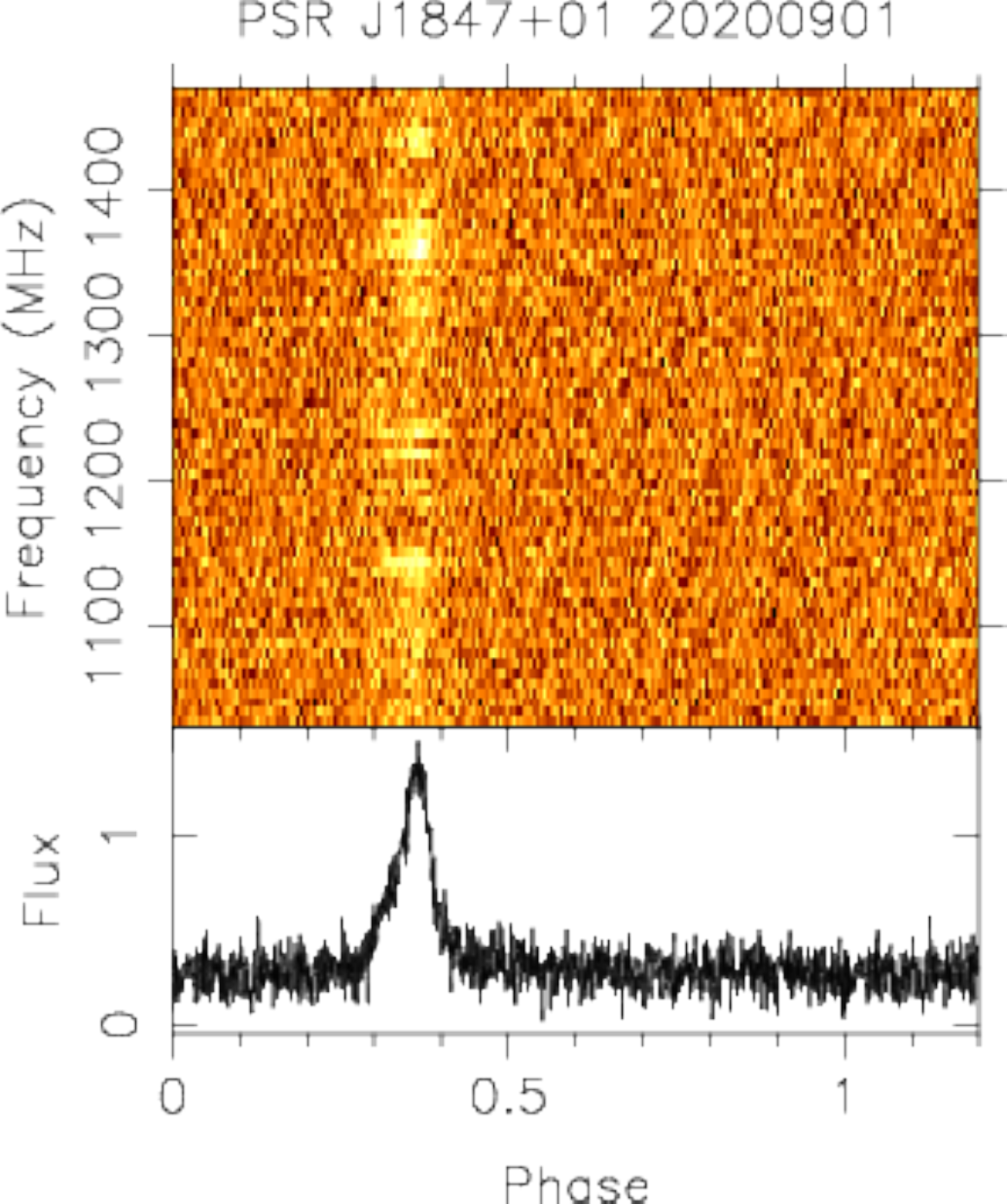}&
\includegraphics[width=38mm]{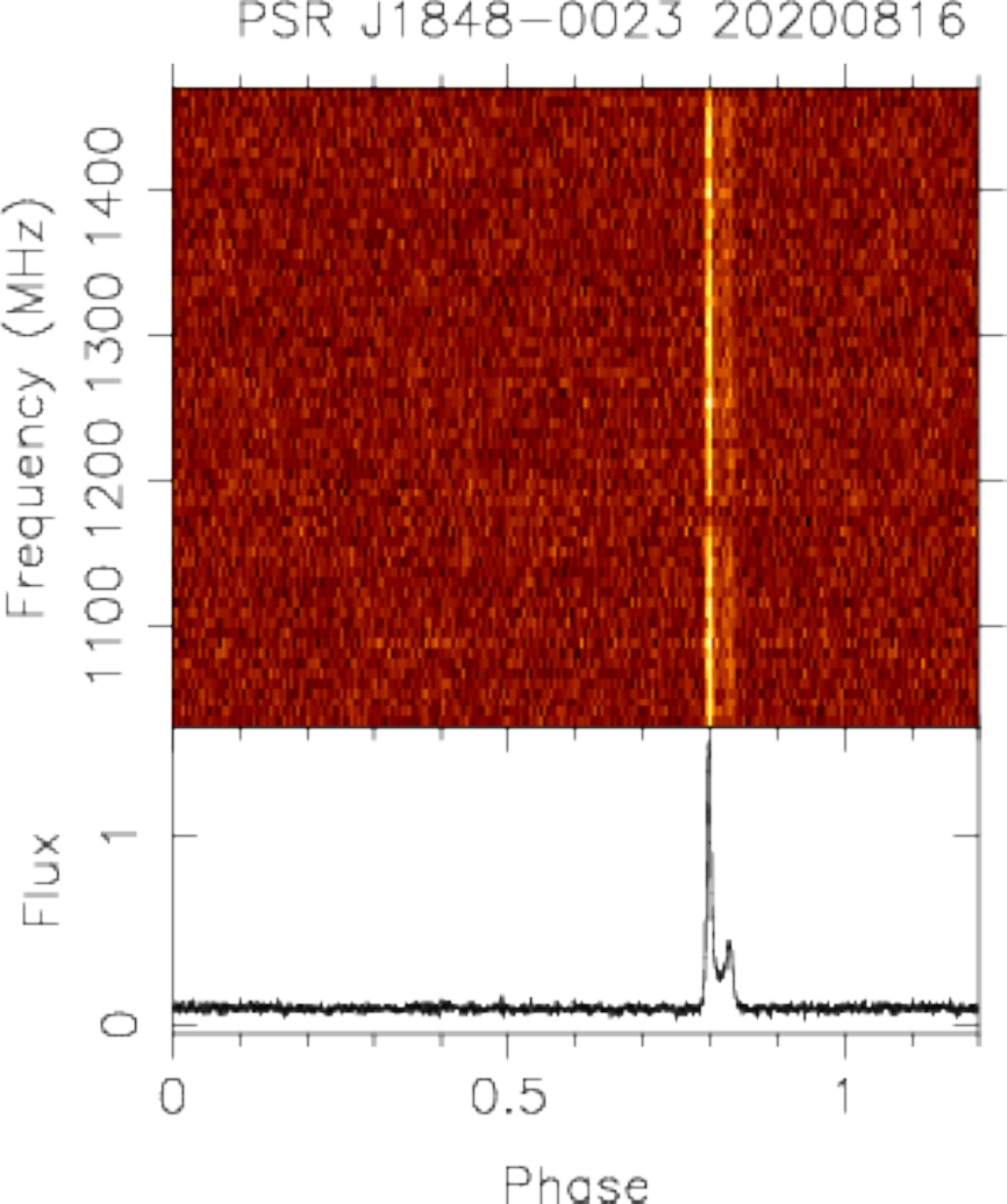}&
\includegraphics[width=38mm]{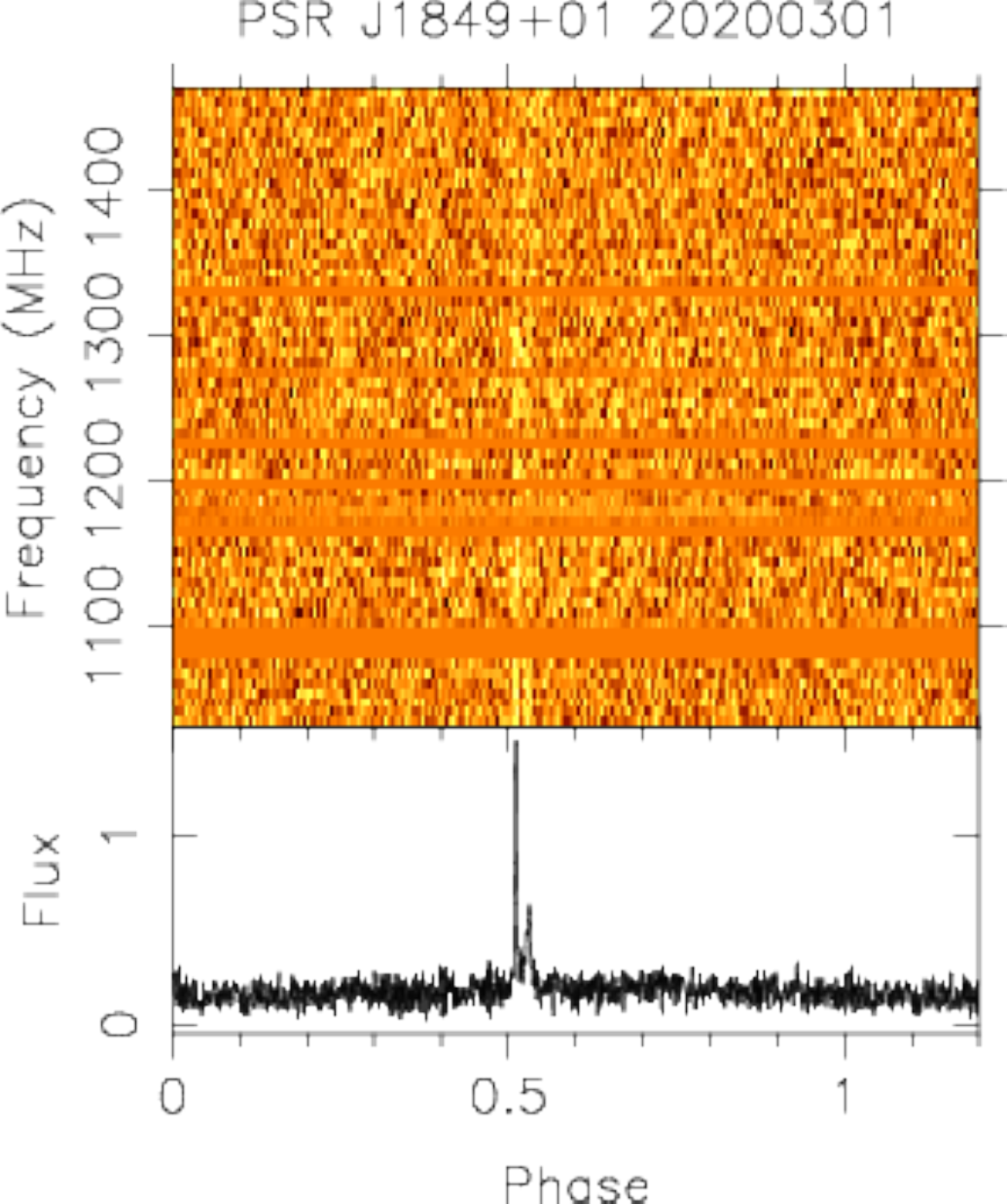}&
\includegraphics[width=38mm]{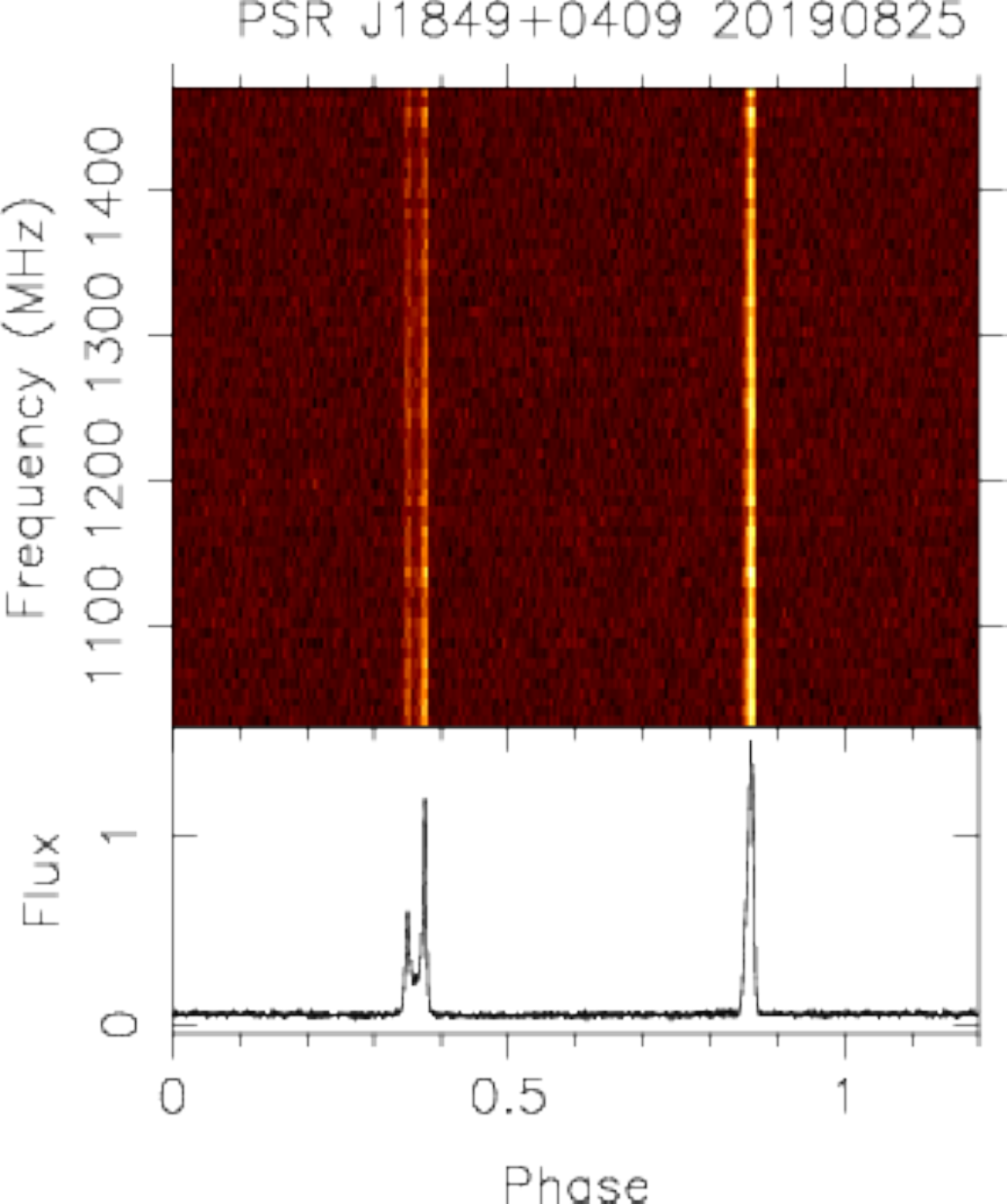}\\[2mm]
\includegraphics[width=38mm]{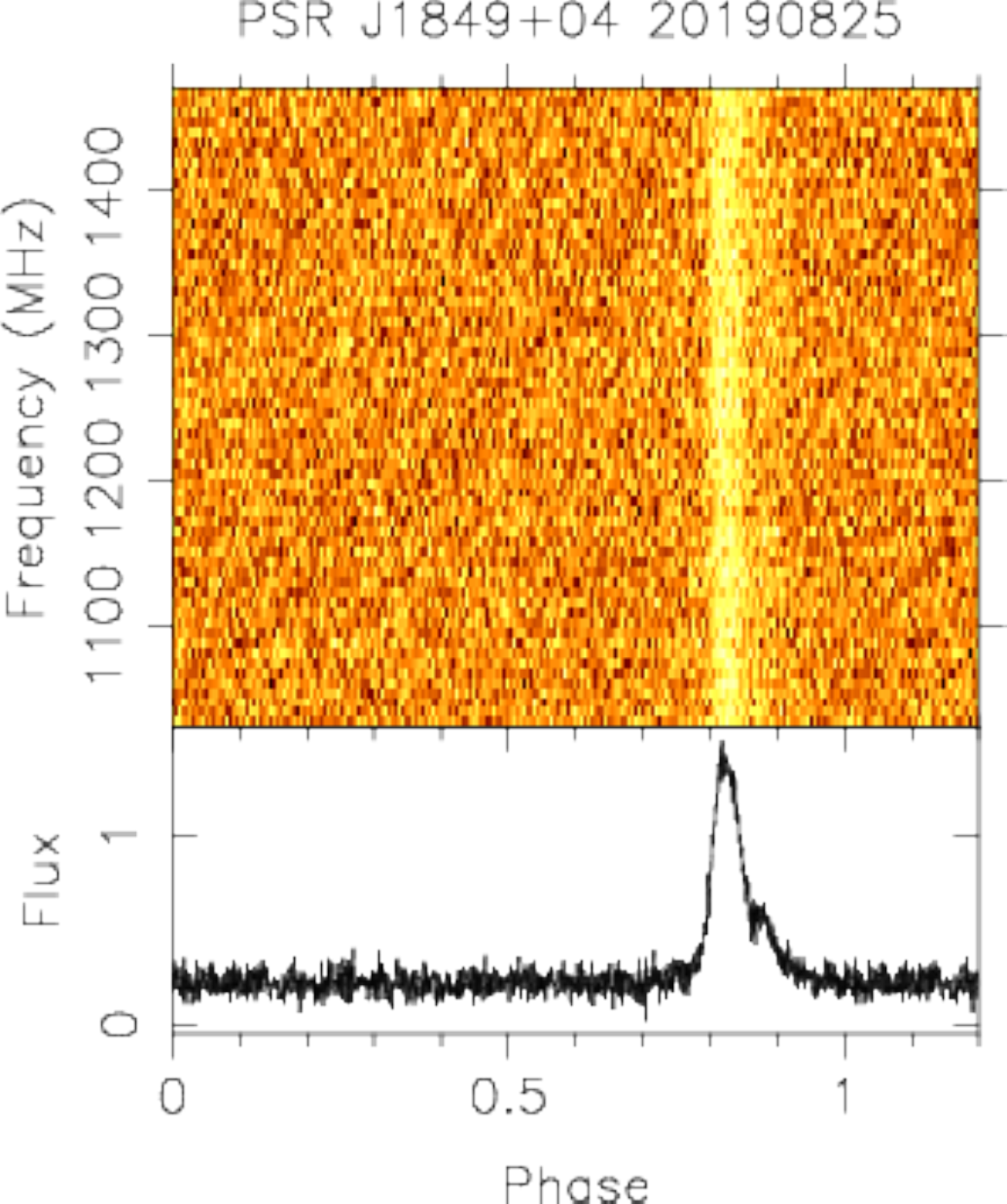}&
\includegraphics[width=38mm]{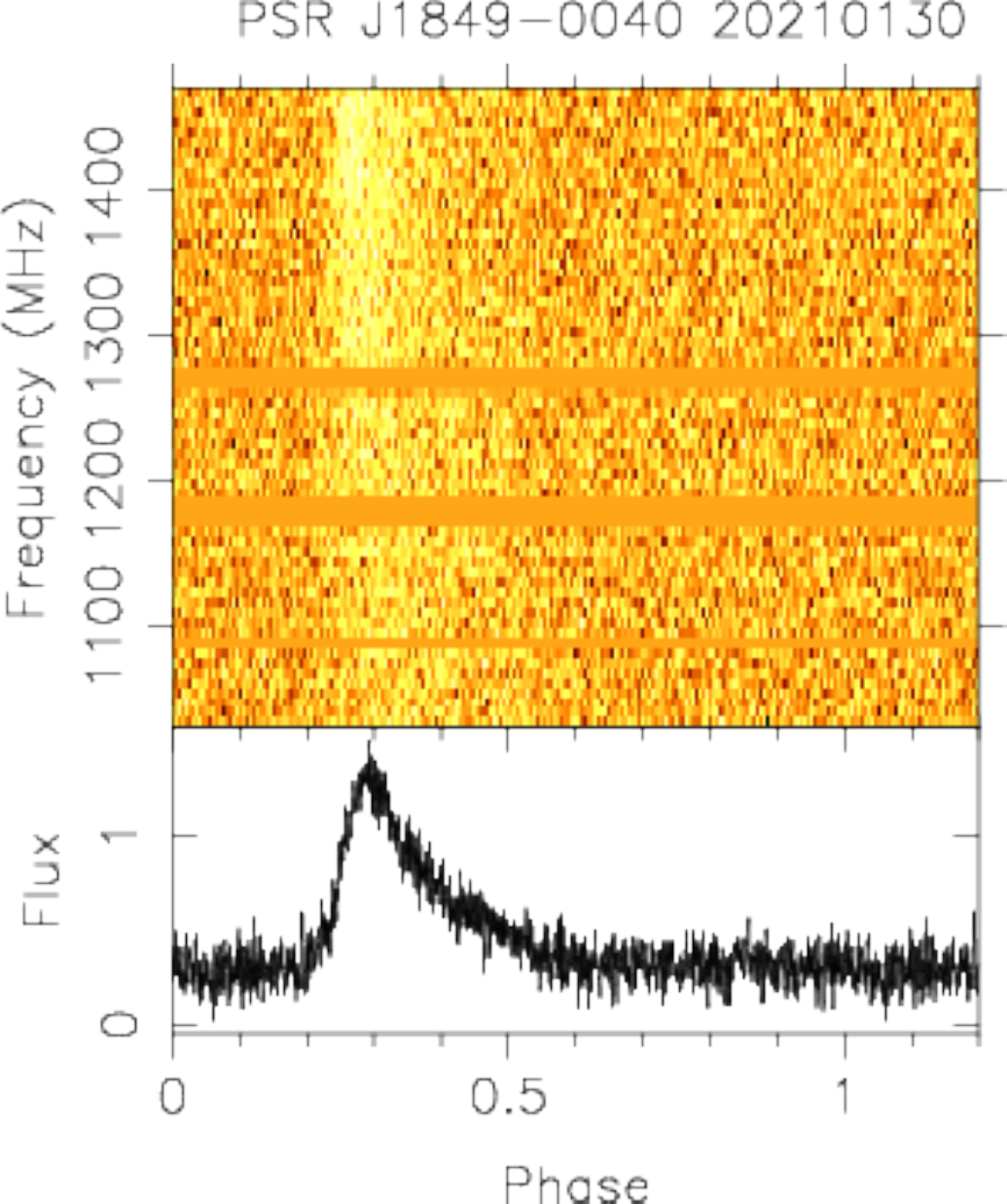}&
\includegraphics[width=38mm]{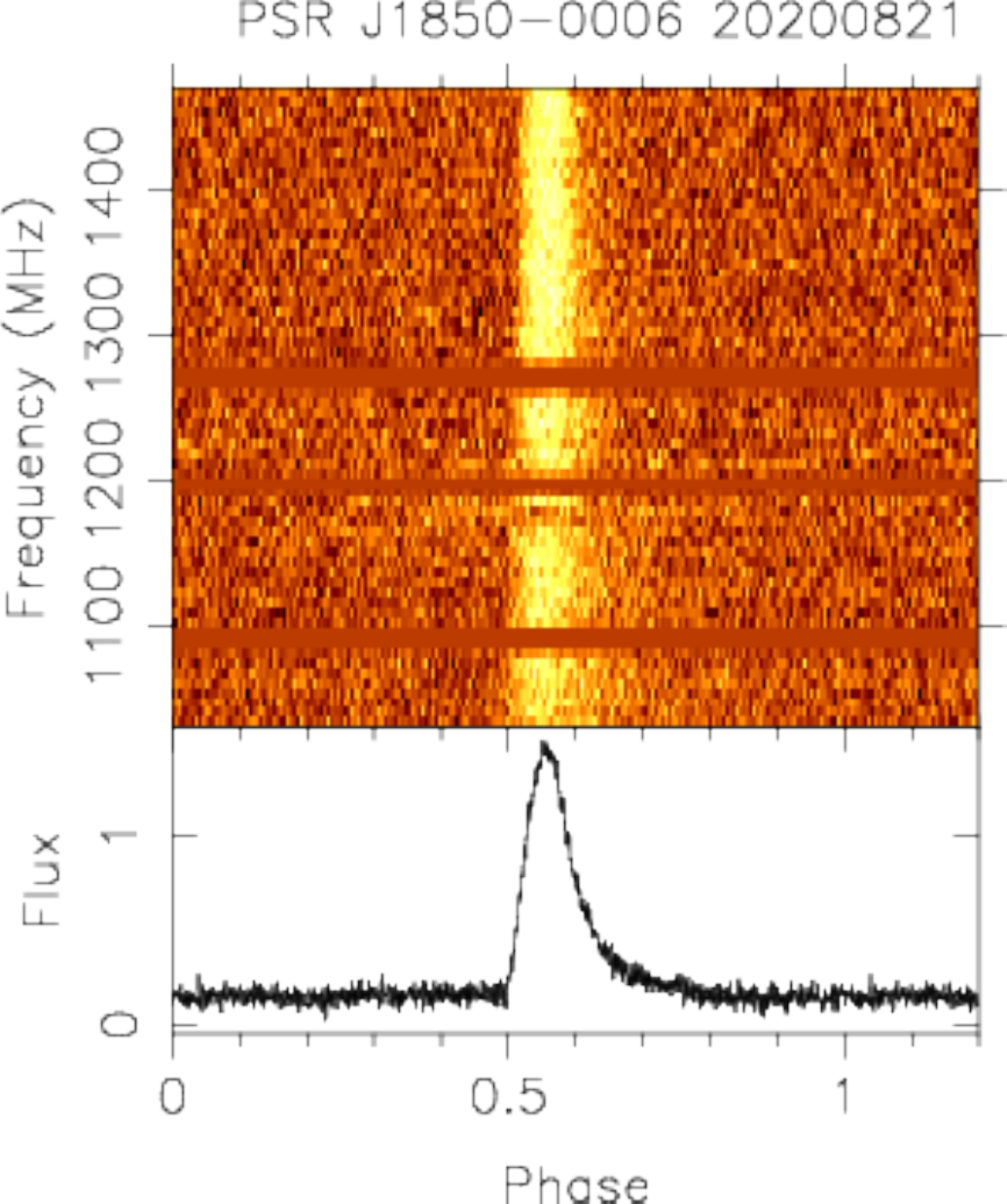}&
\includegraphics[width=38mm]{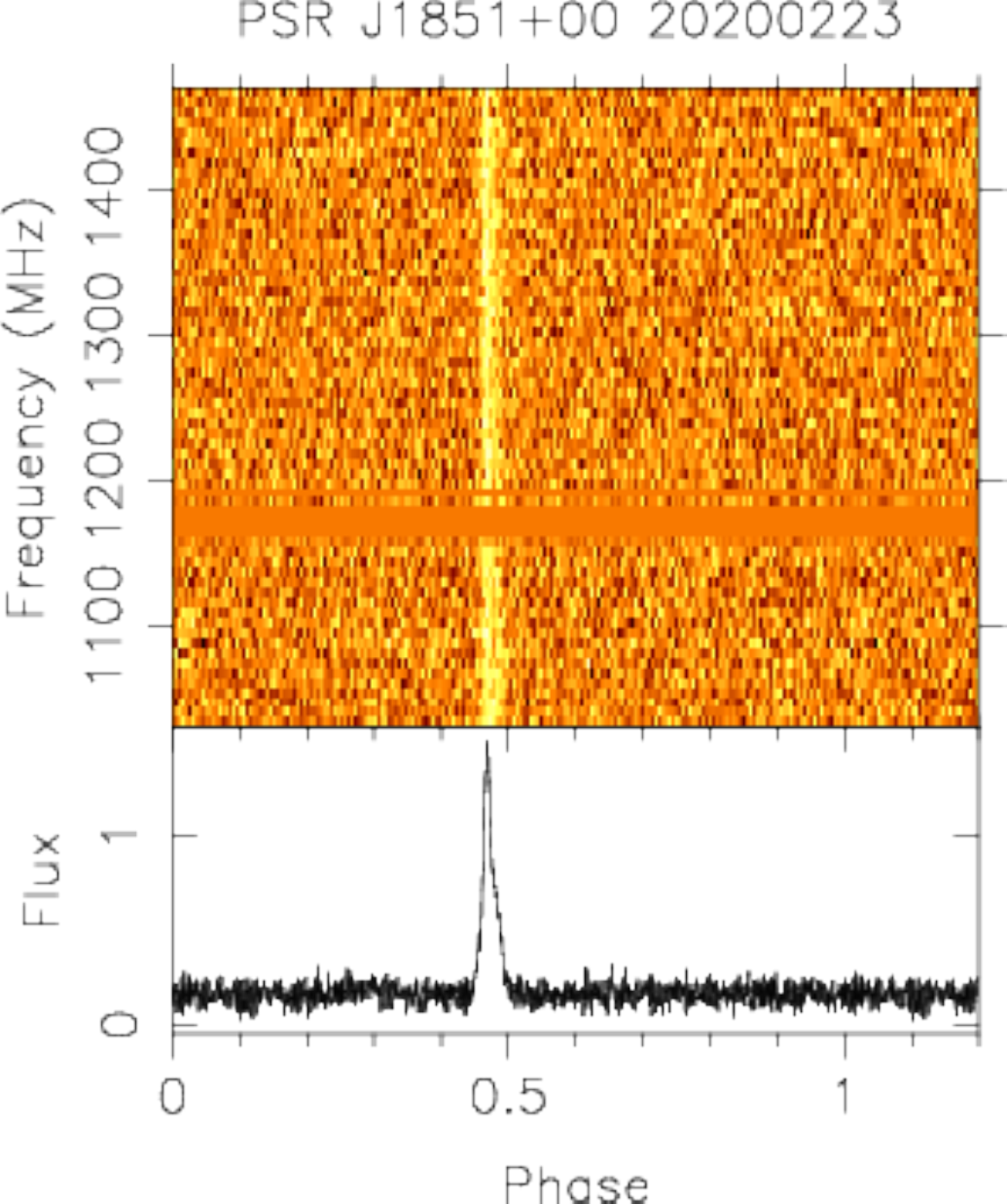}\\[2mm]
\includegraphics[width=38mm]{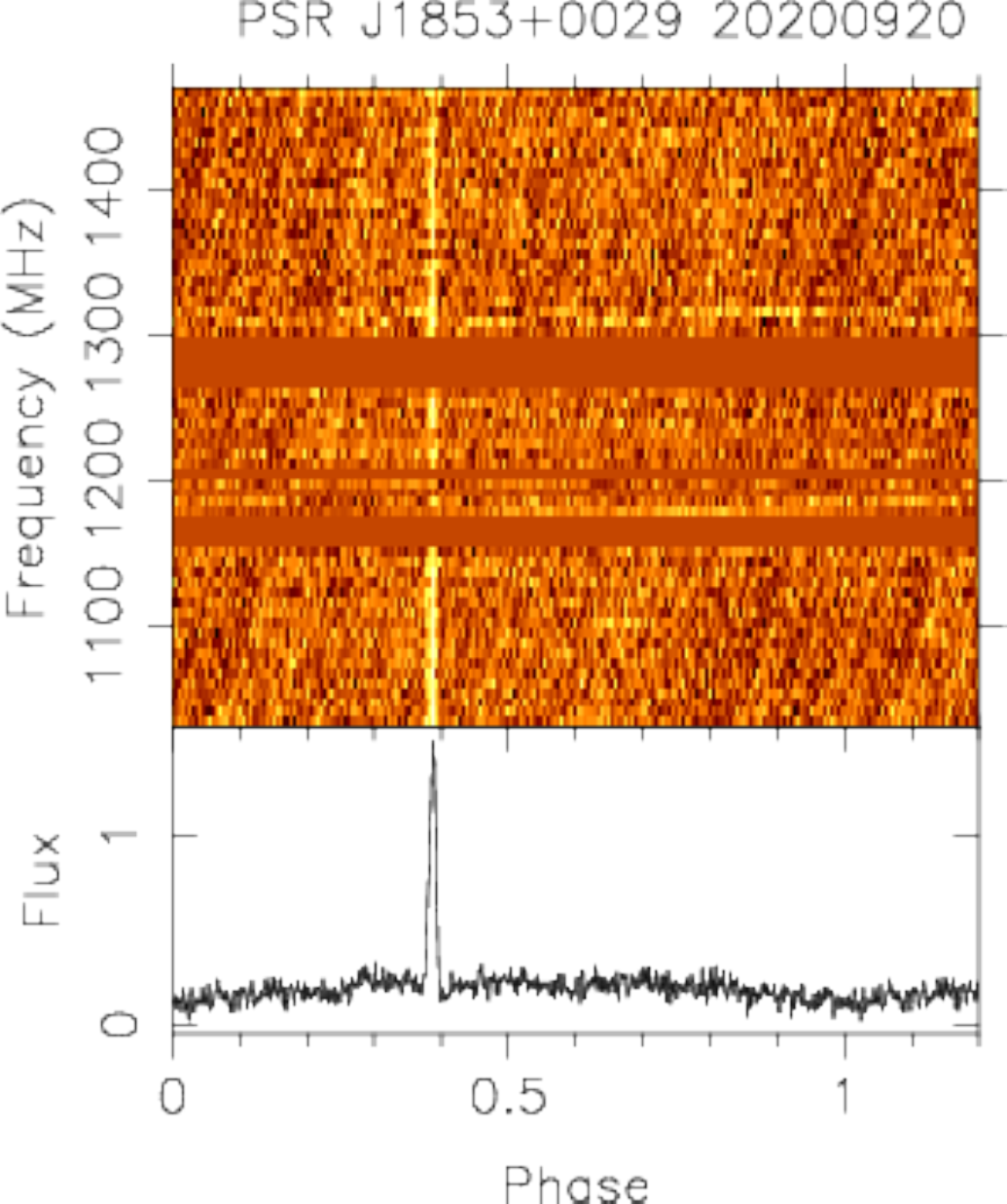}&
\includegraphics[width=38mm]{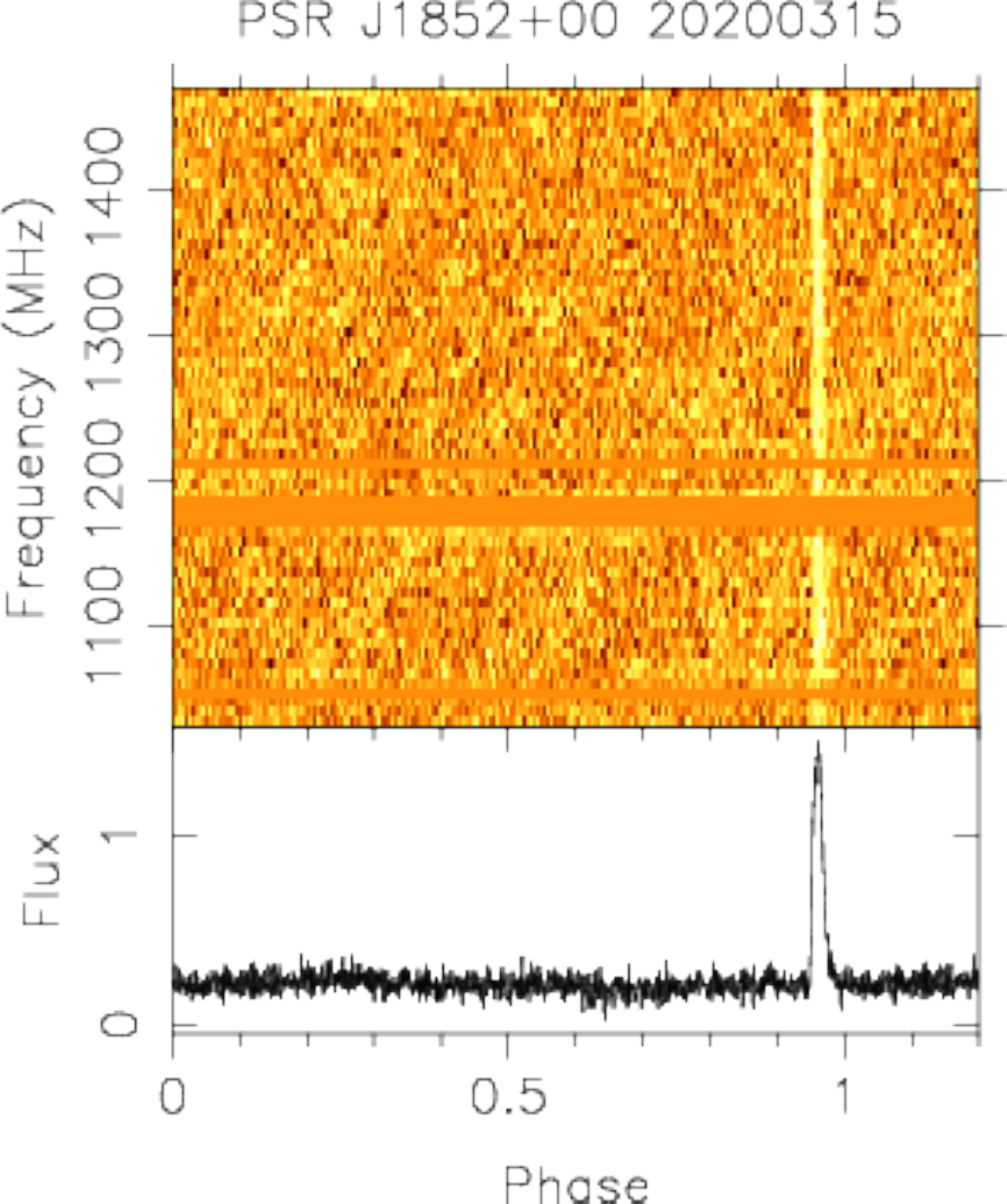}&
\includegraphics[width=38mm]{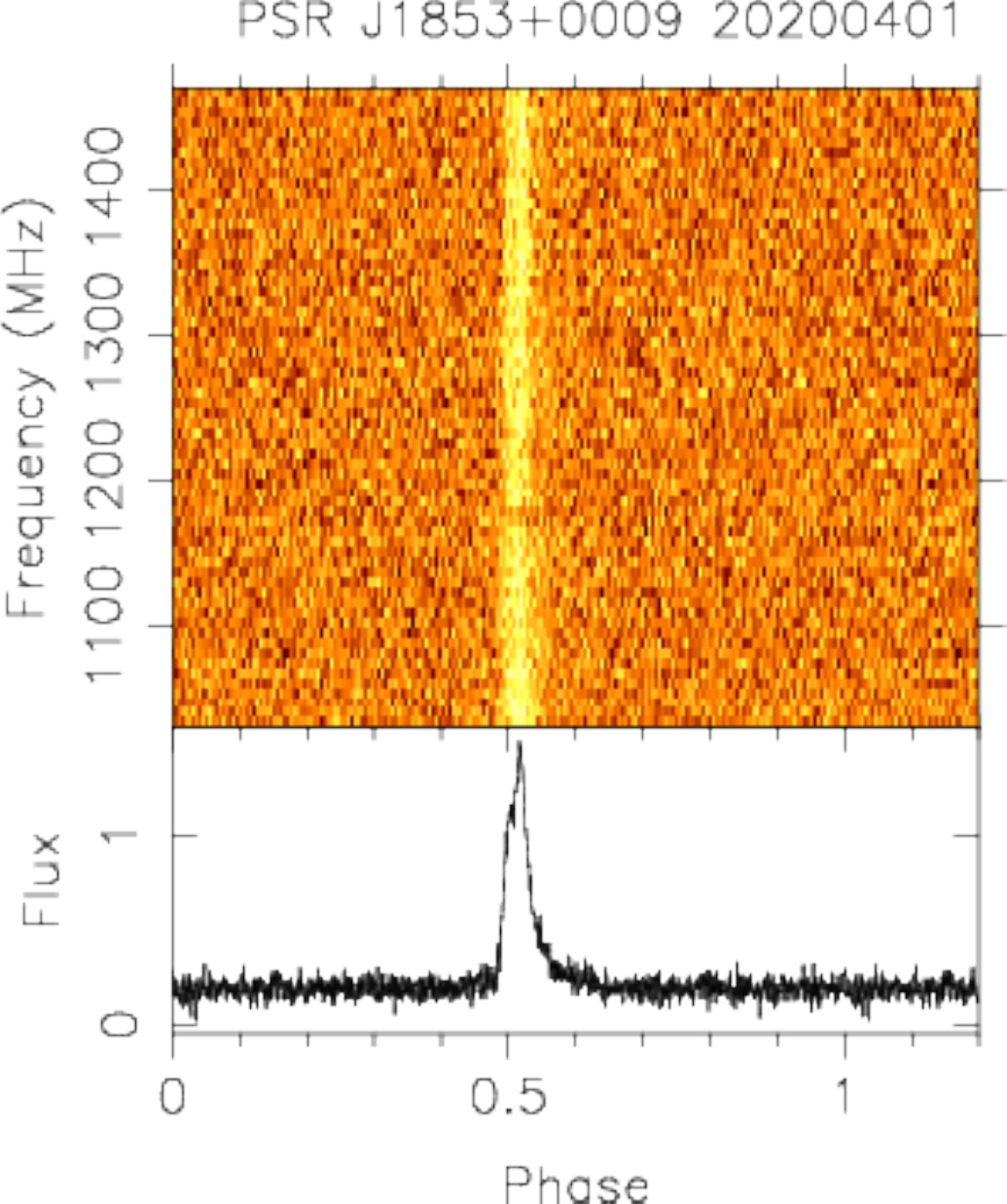}&
\includegraphics[width=38mm]{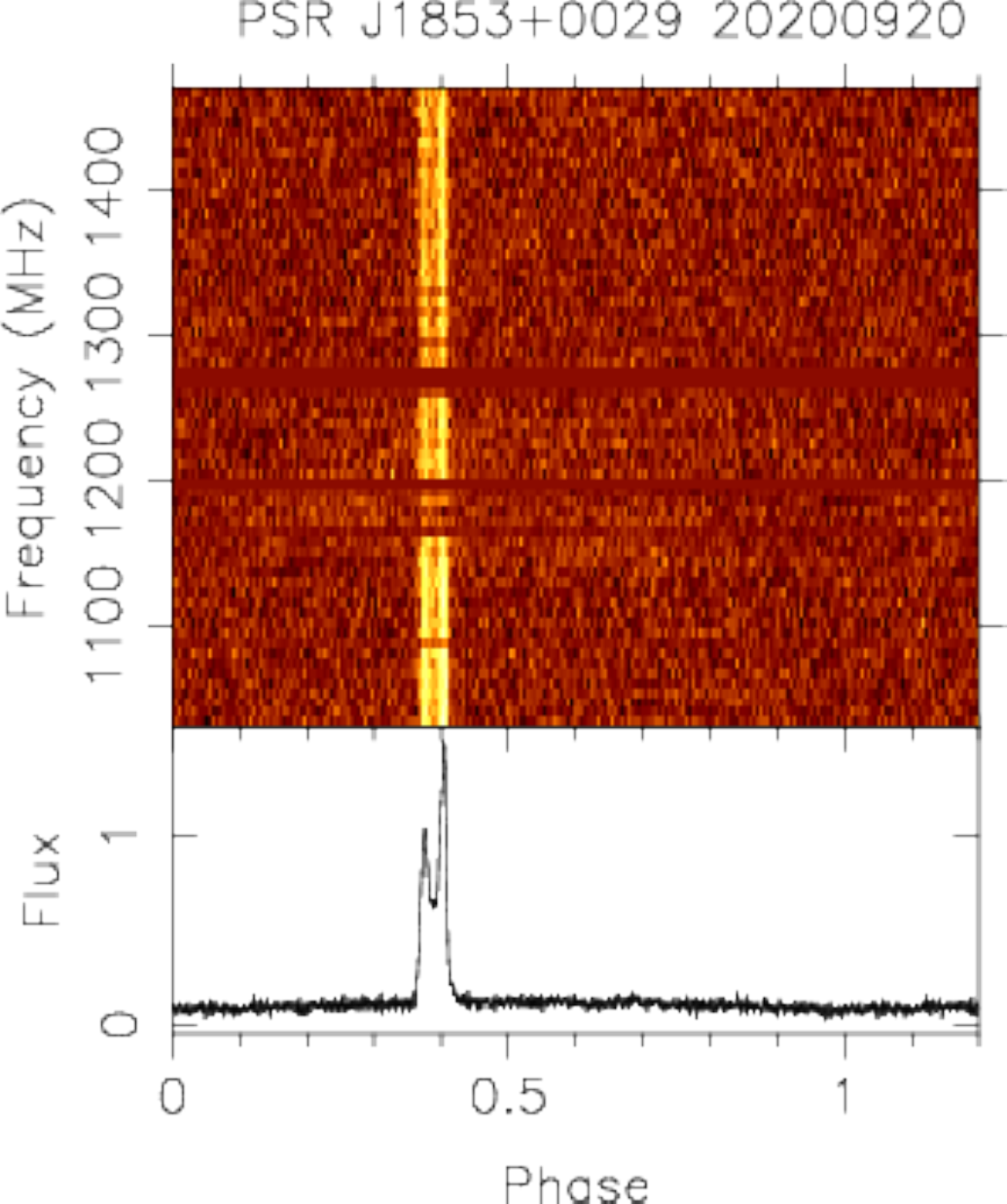}\\[2mm]
\includegraphics[width=38mm]{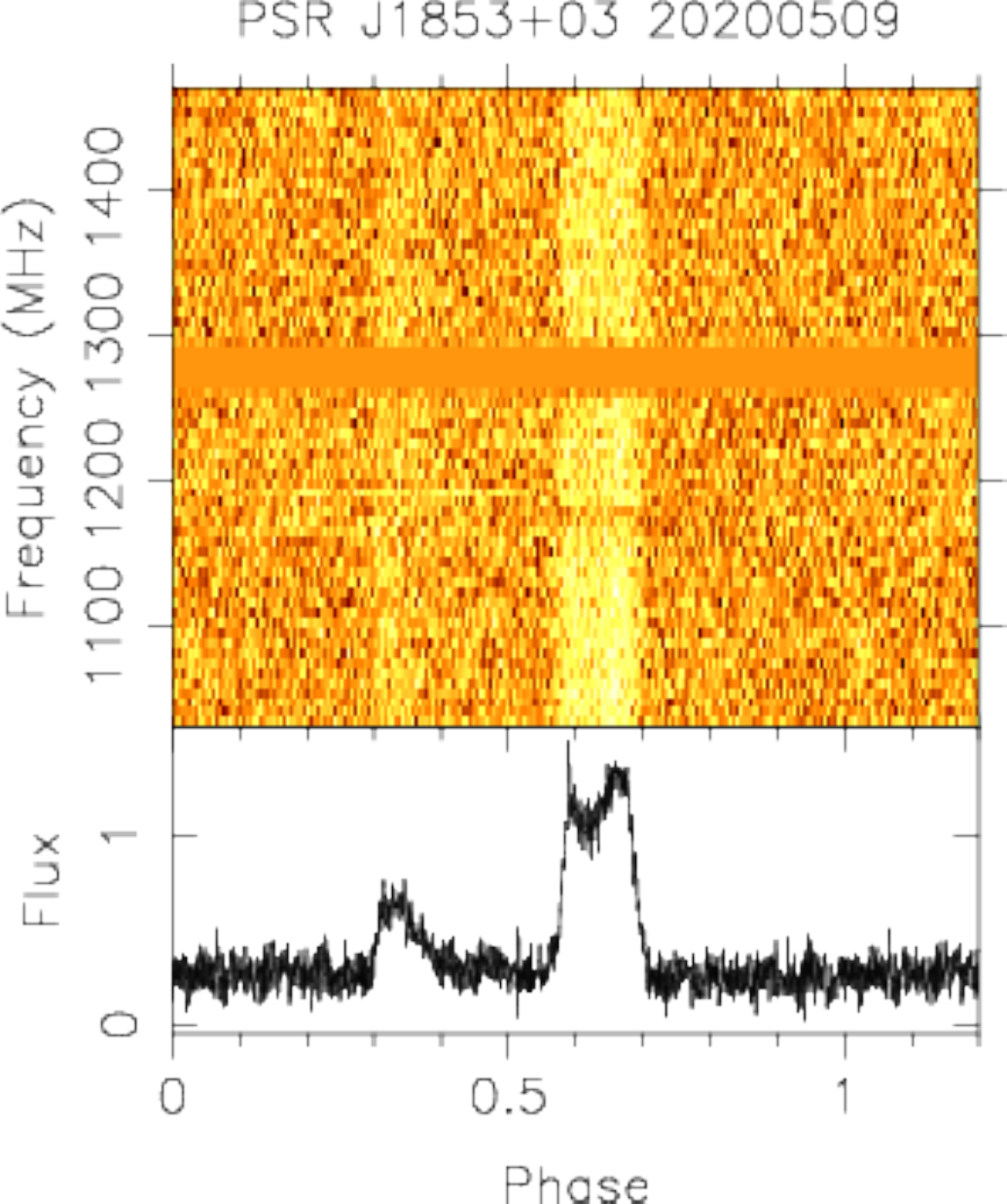}&
\includegraphics[width=38mm]{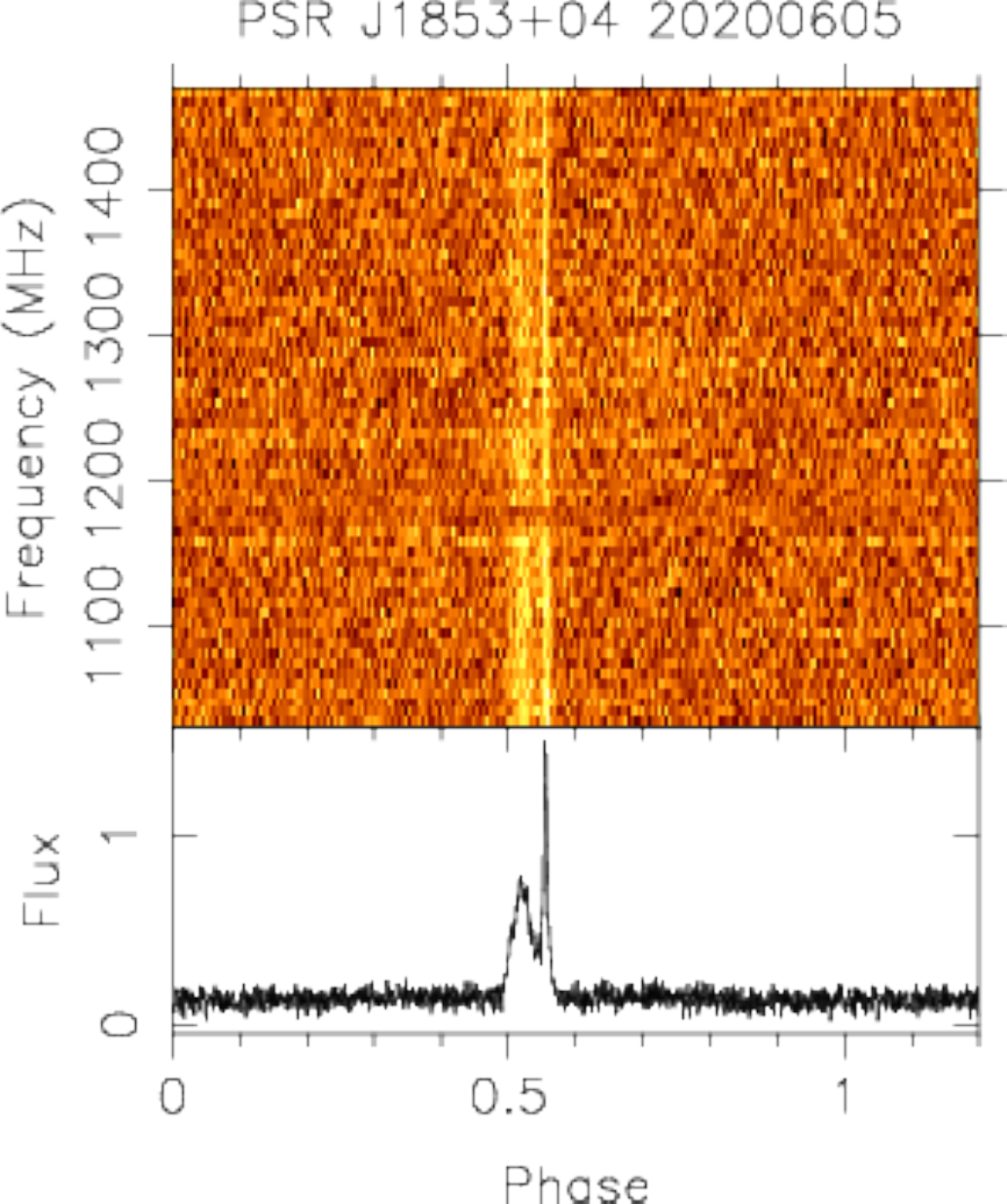}&
\includegraphics[width=38mm]{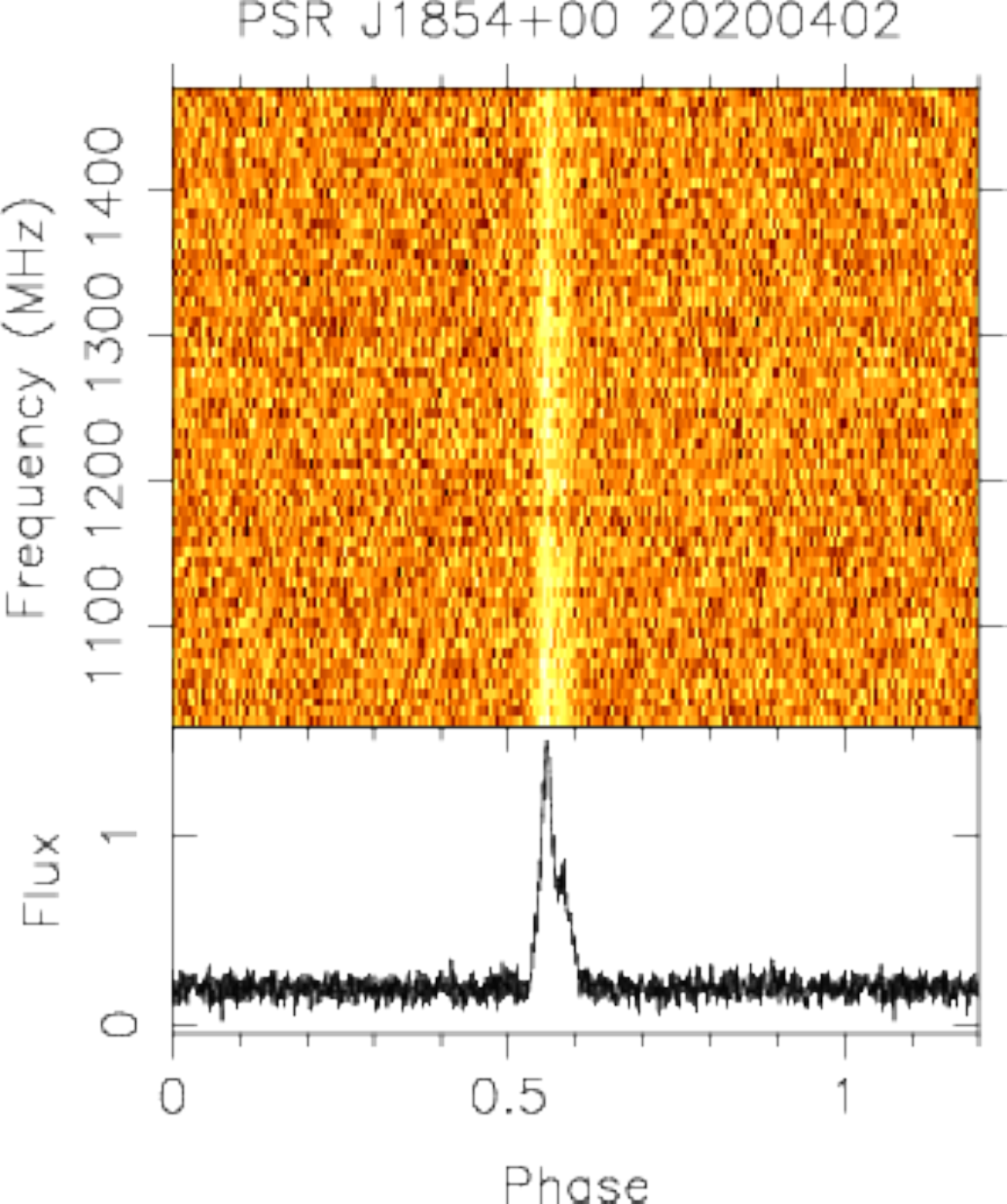}&
\includegraphics[width=38mm]{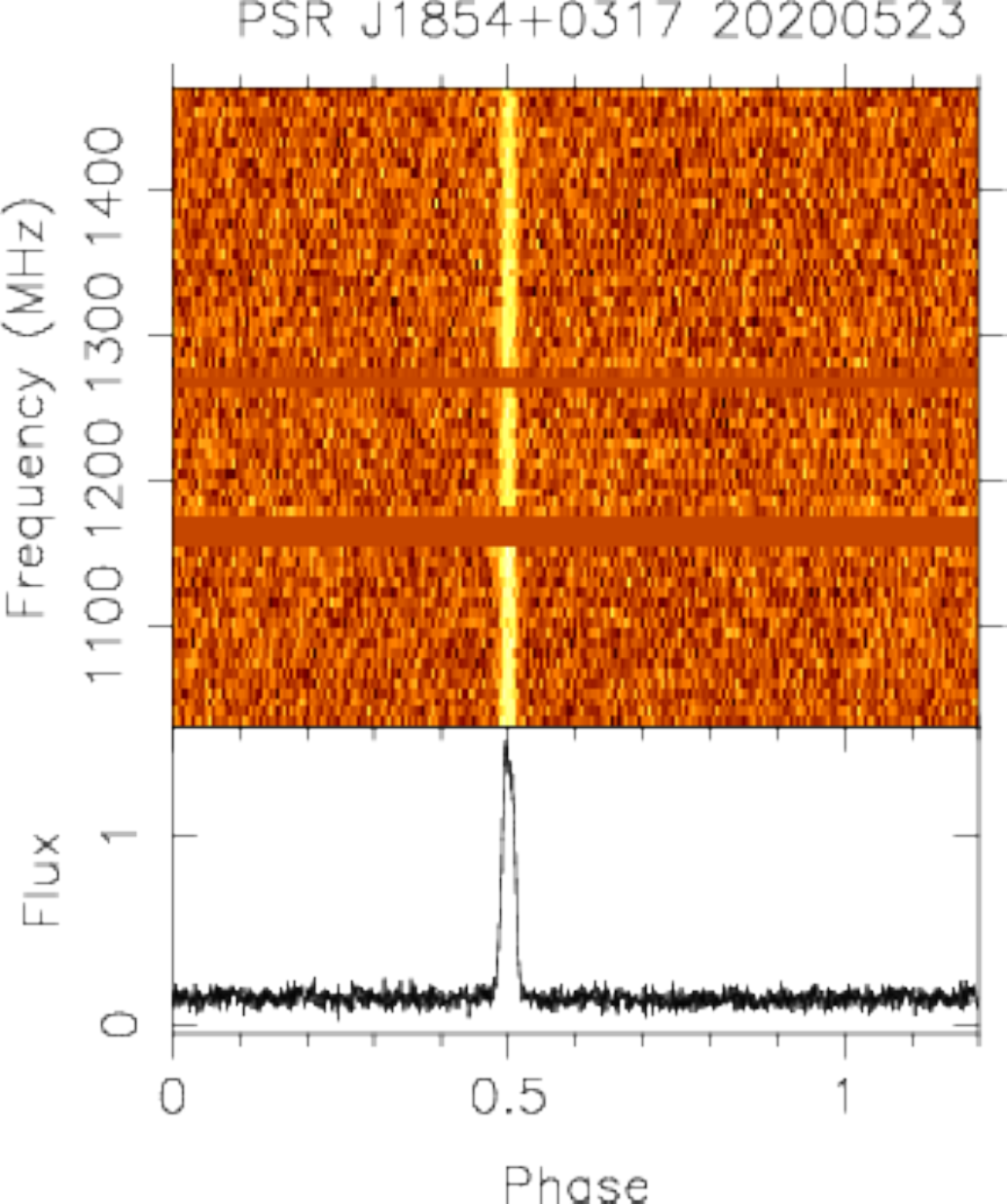}
\end{tabular}%

\caption[]{\baselineskip 3.8mm The phase-frequency plot and the integrated profile for
  64 previously known pulsars for which the GPPS survey gives updated
  parameters. Channels with RFI are cleaned or suppressed. }
\label{fig21_knownPSRs}
\addtocounter{figure}{-1}
\end{figure*}%
\begin{figure*}[htp!!!]
\centering
\begin{tabular}{rrrrrr}
\includegraphics[width=38mm]{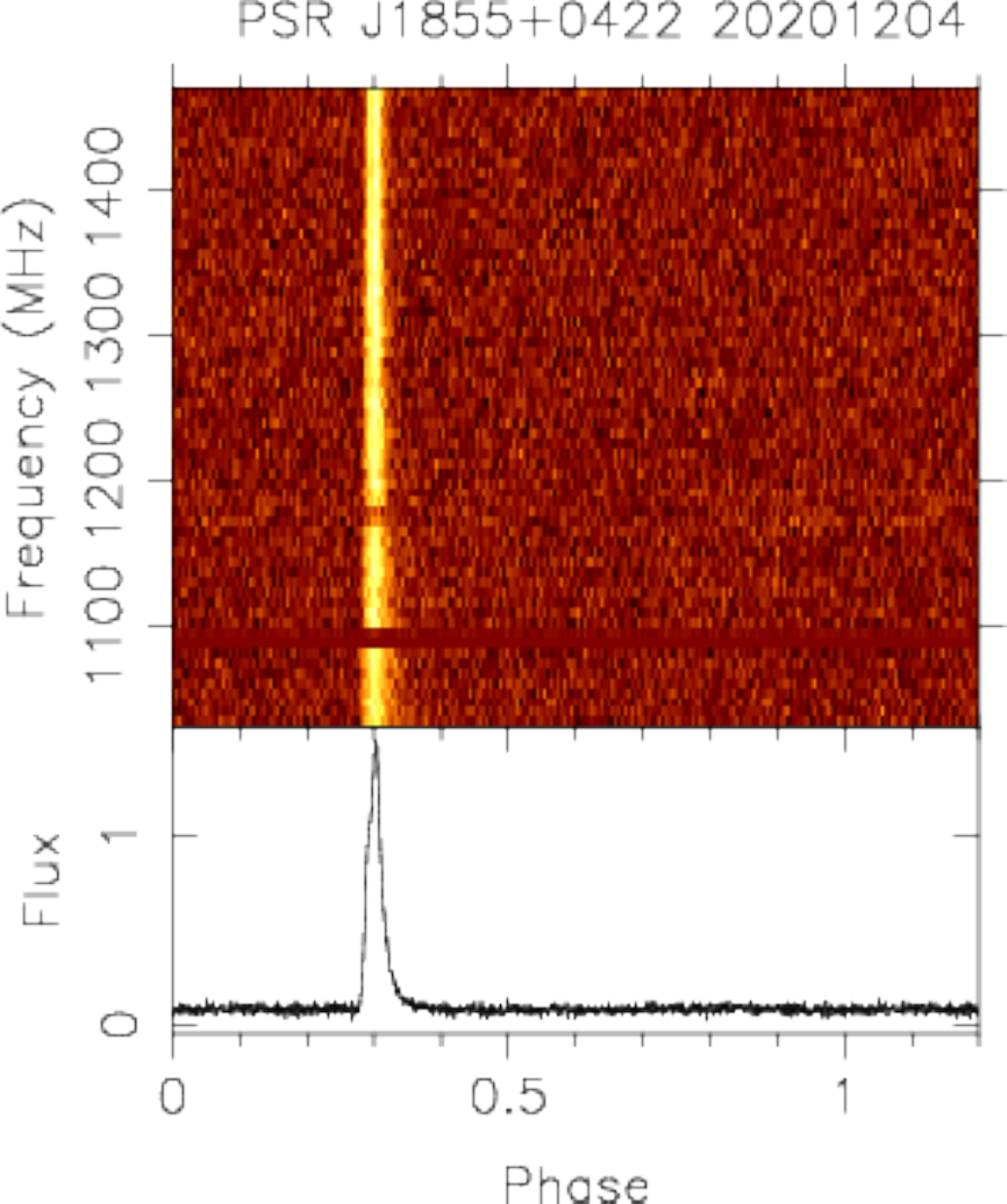}&
\includegraphics[width=38mm]{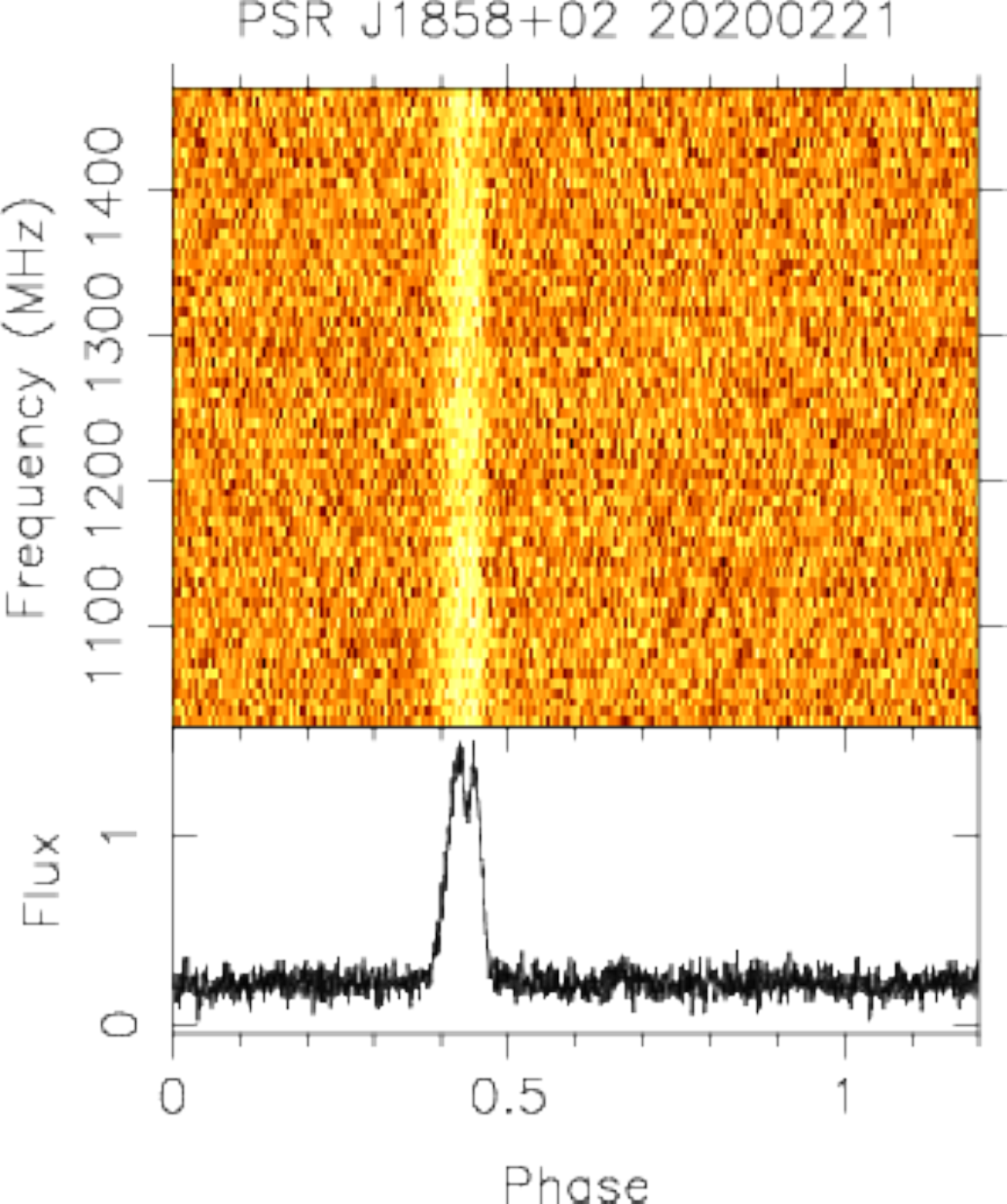}&
\includegraphics[width=38mm]{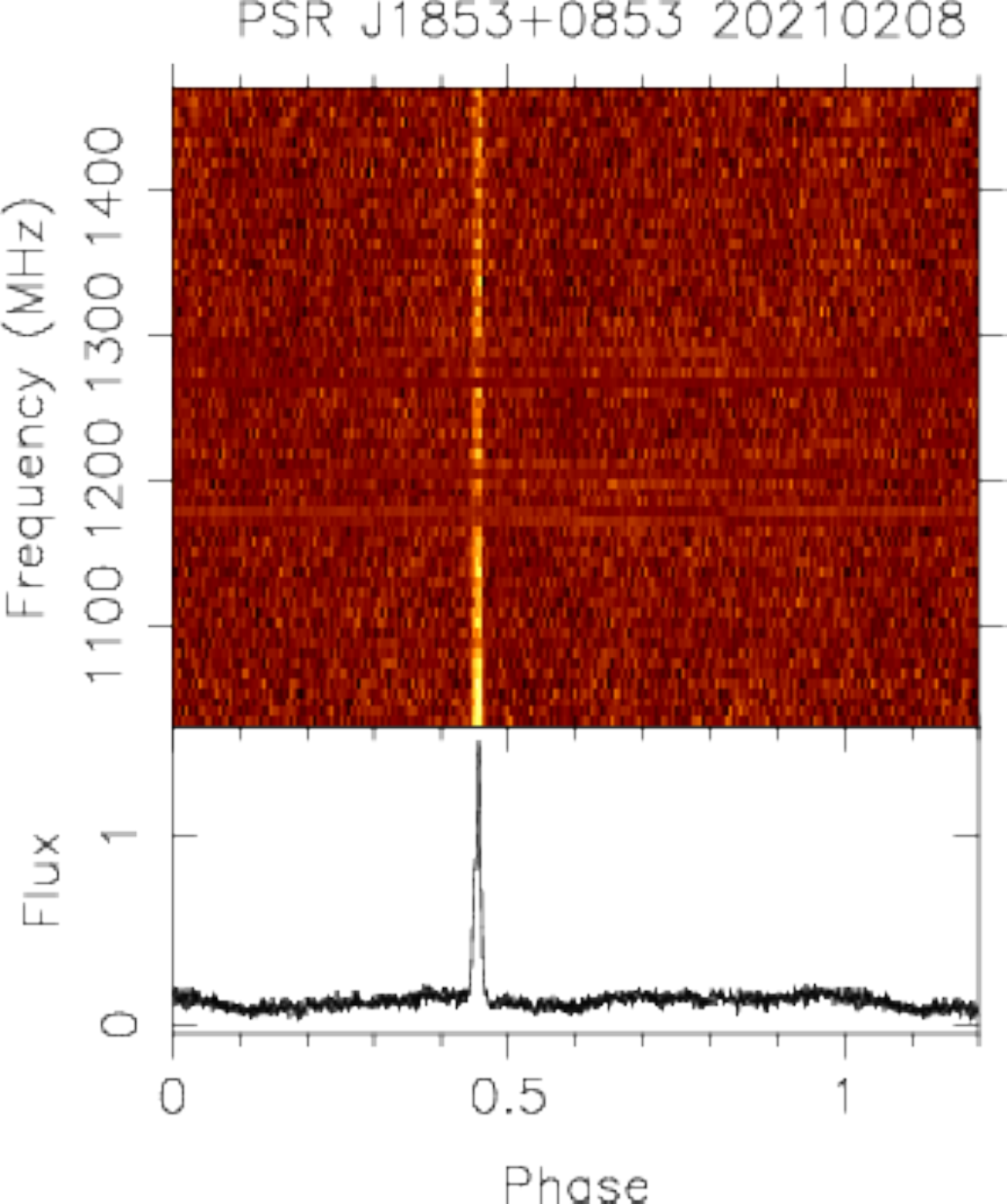}&
\includegraphics[width=38mm]{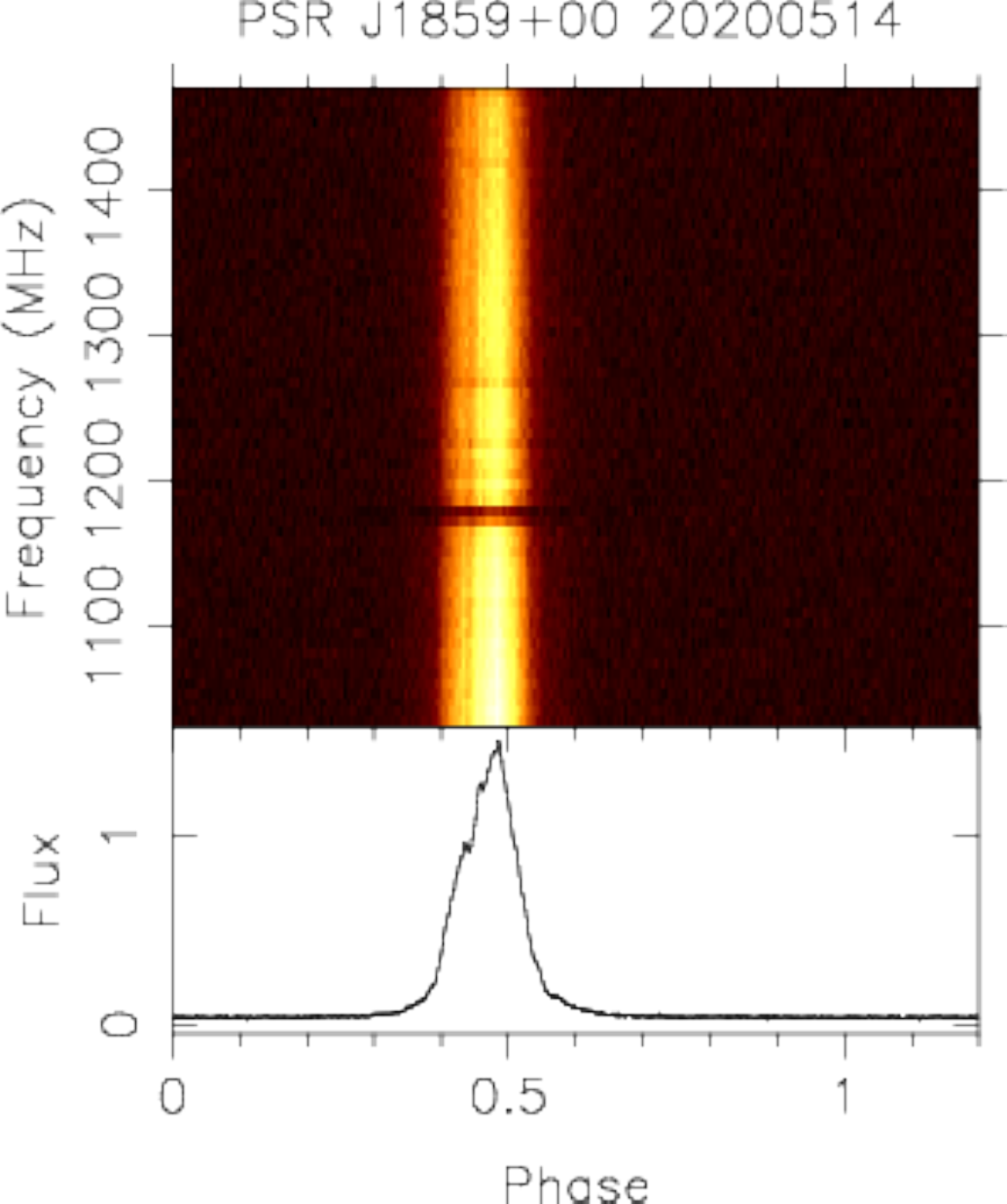}\\[2mm]
\includegraphics[width=38mm]{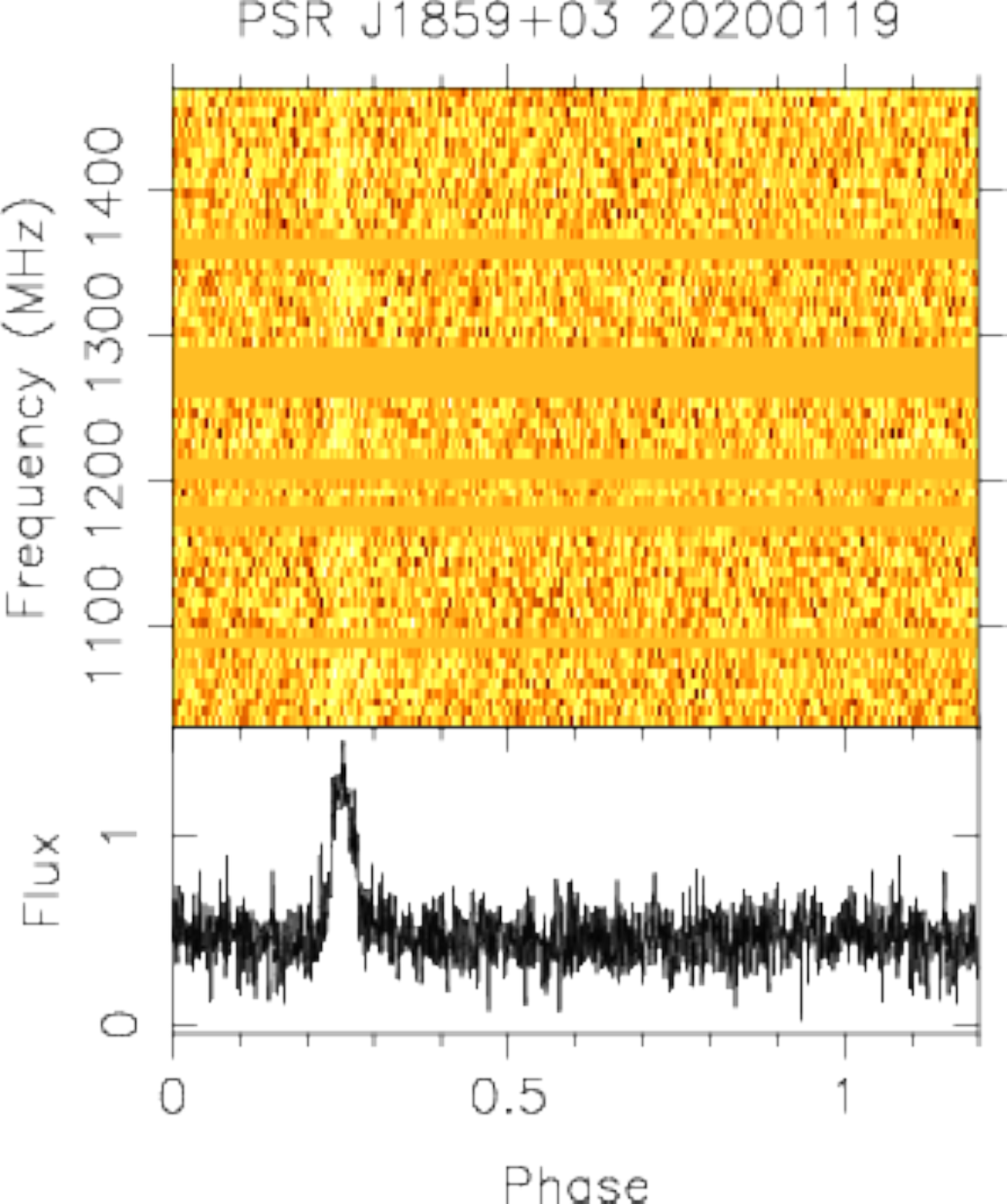}&
\includegraphics[width=38mm]{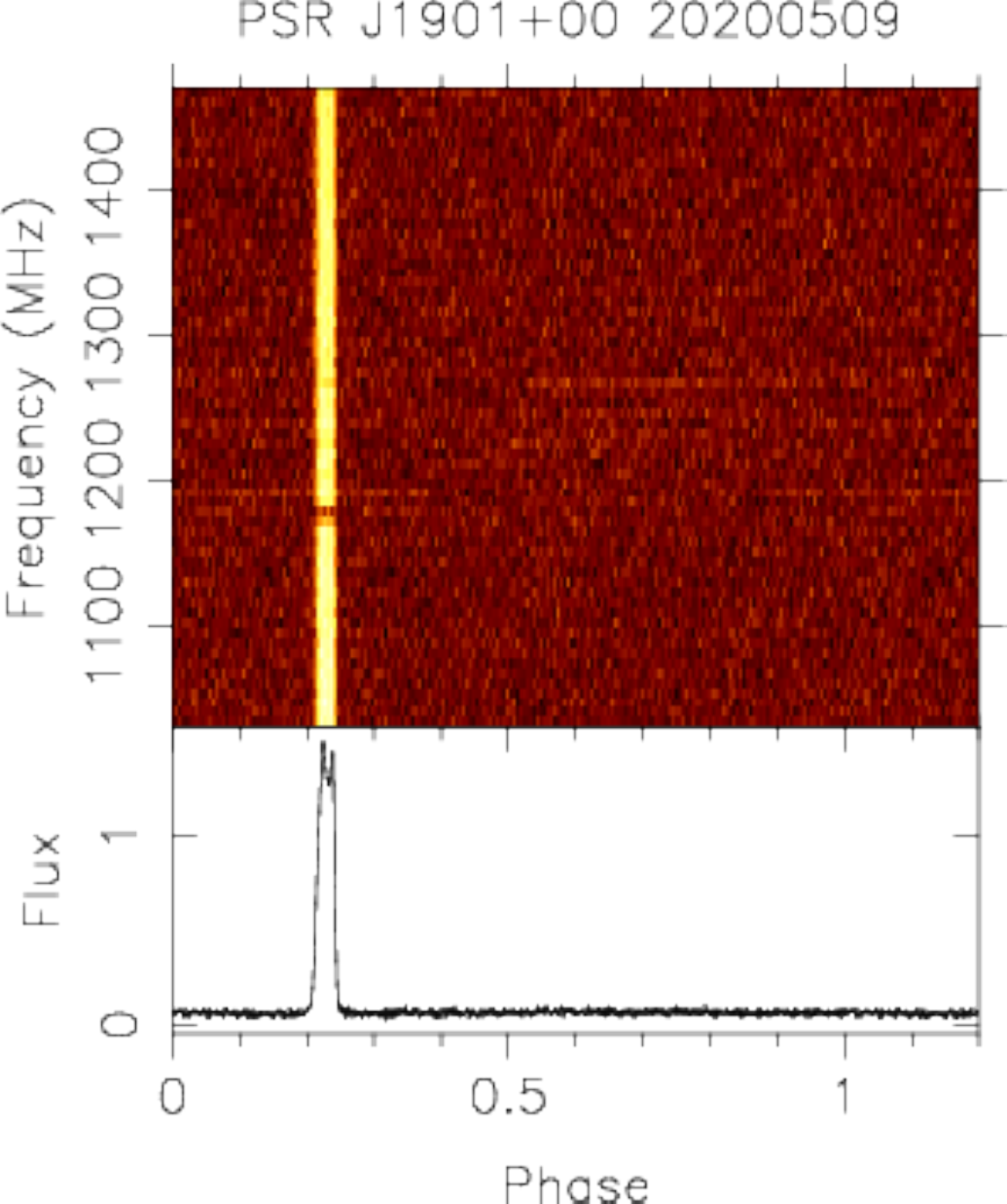}&
\includegraphics[width=38mm]{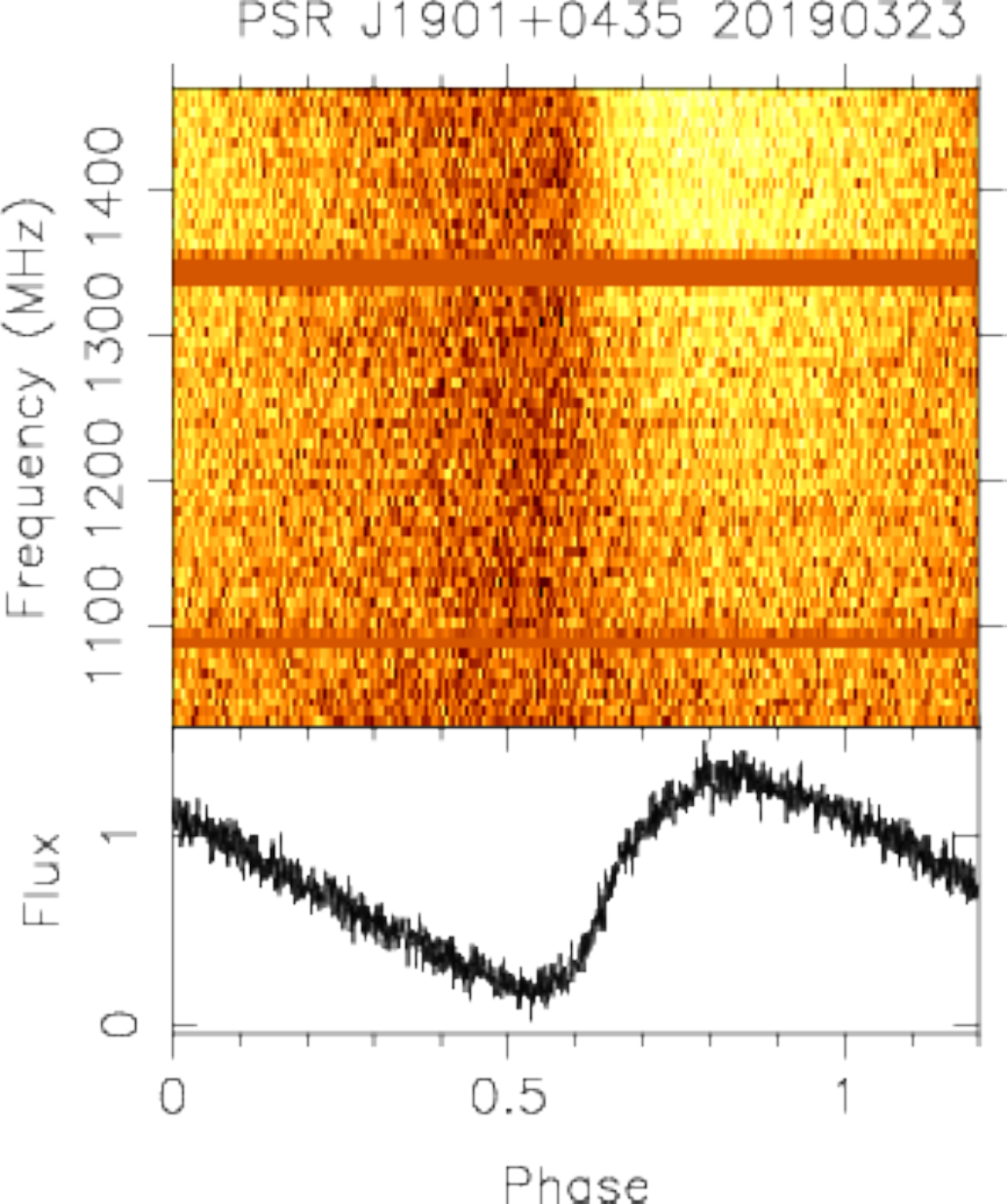}&
\includegraphics[width=38mm]{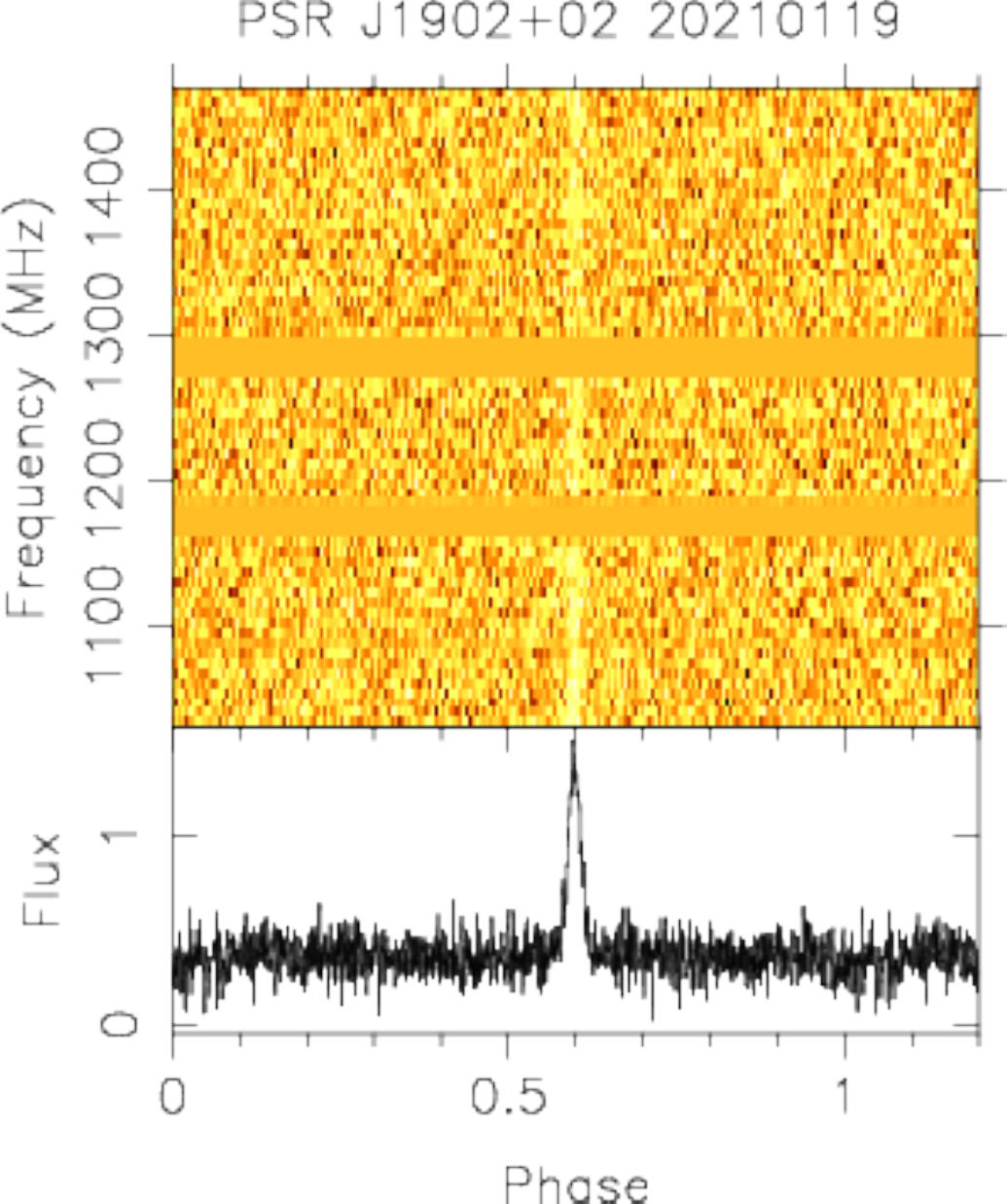}\\[2mm]
\includegraphics[width=38mm]{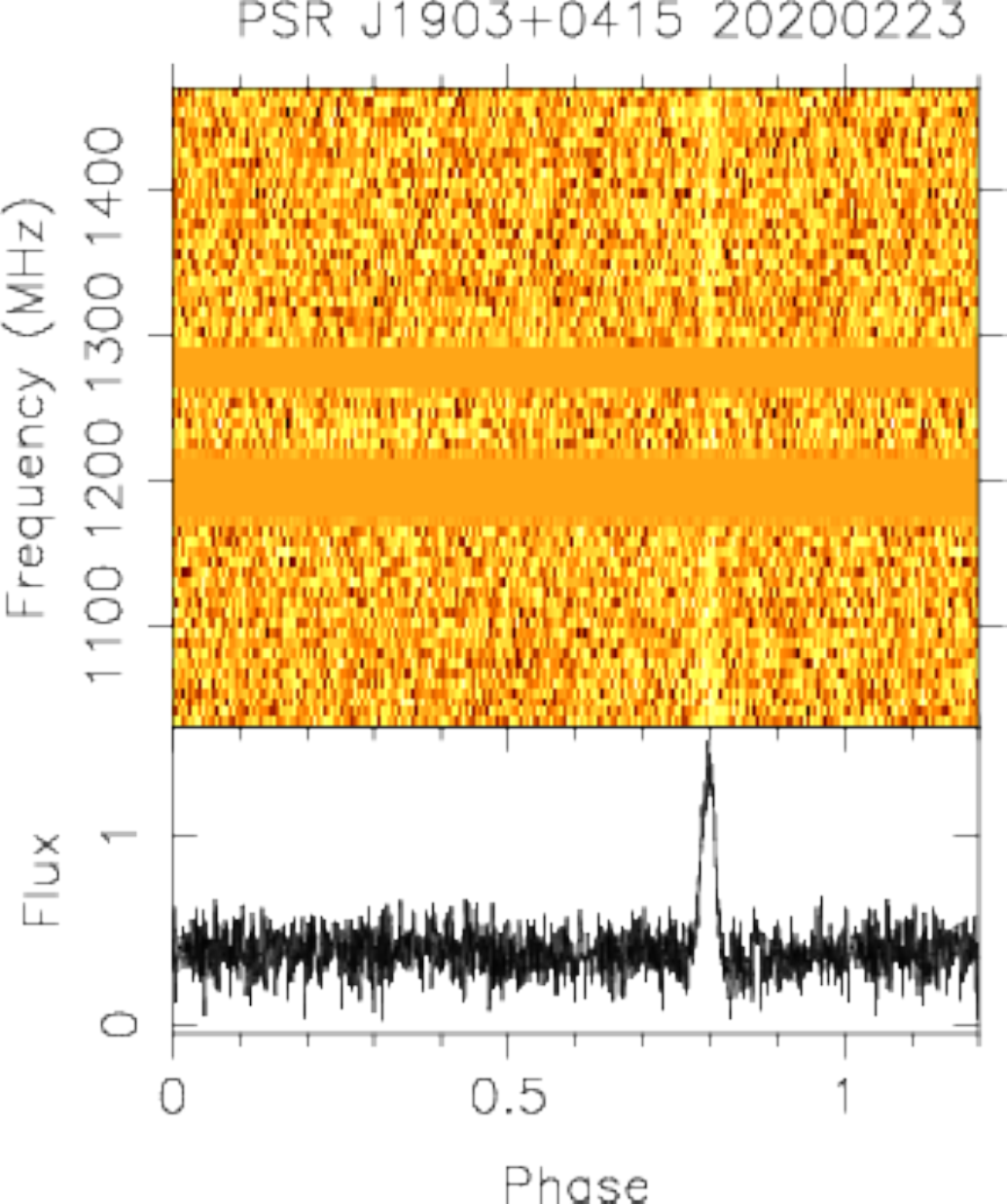}&
\includegraphics[width=38mm]{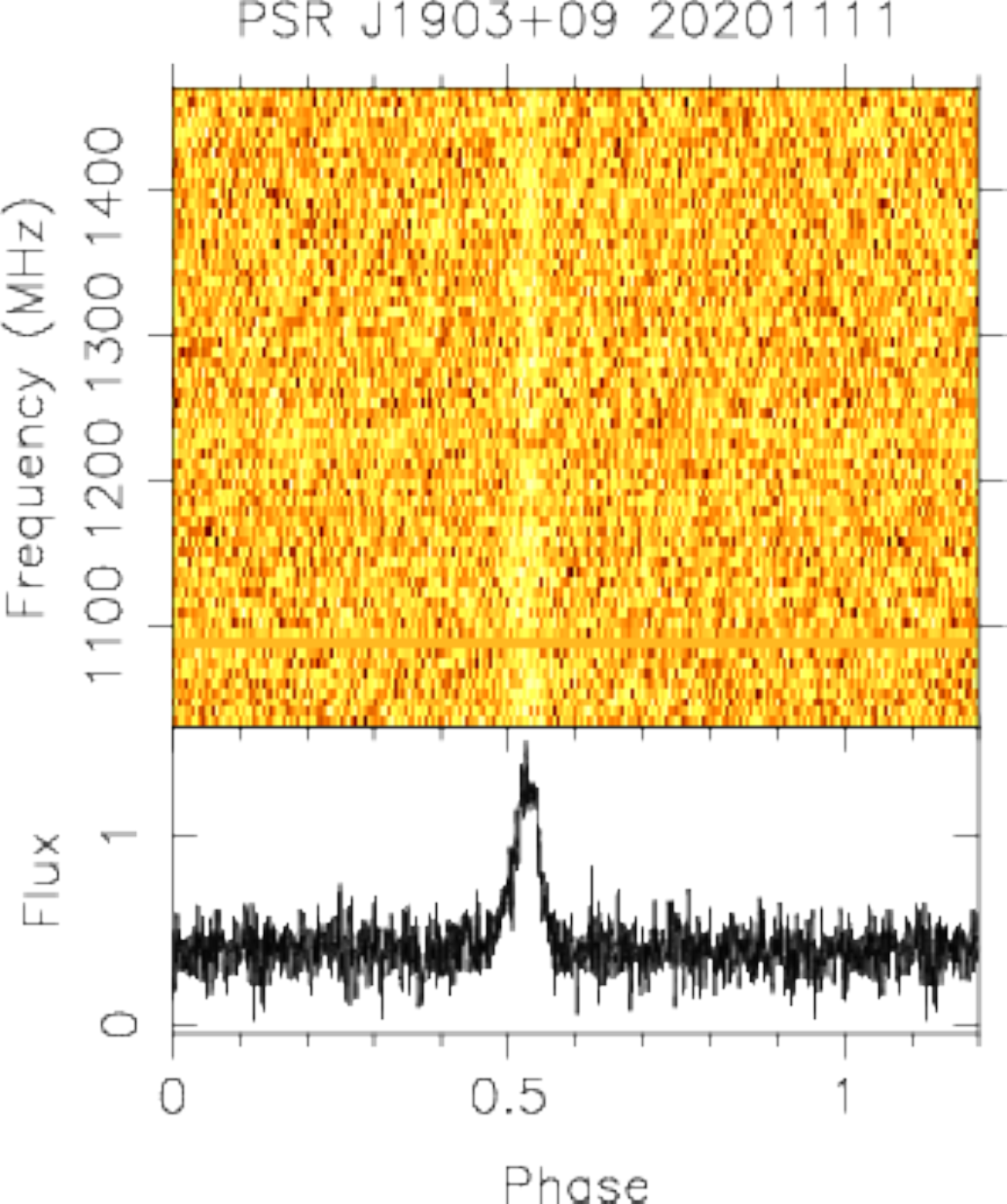}&
\includegraphics[width=38mm]{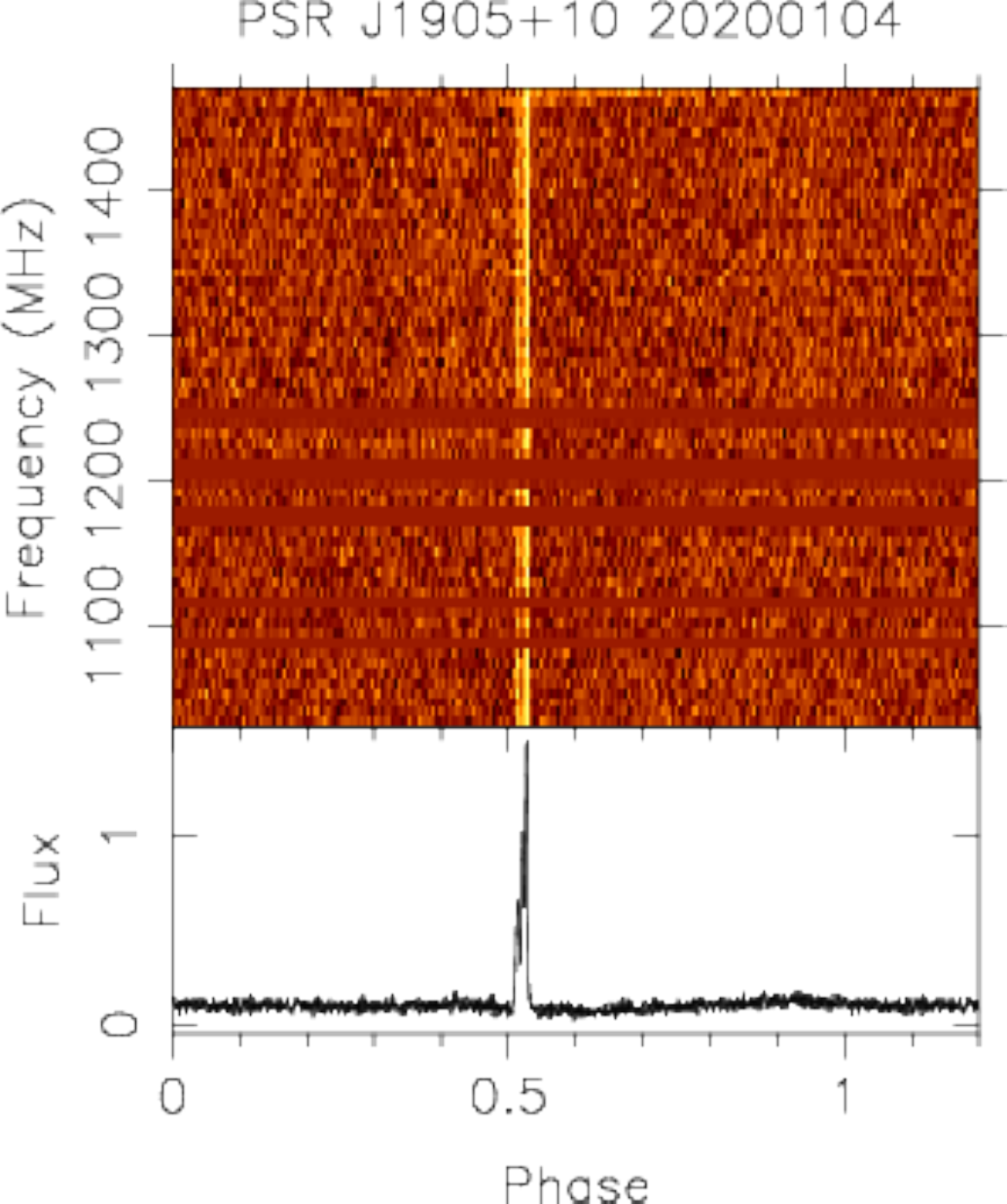}&
\includegraphics[width=38mm]{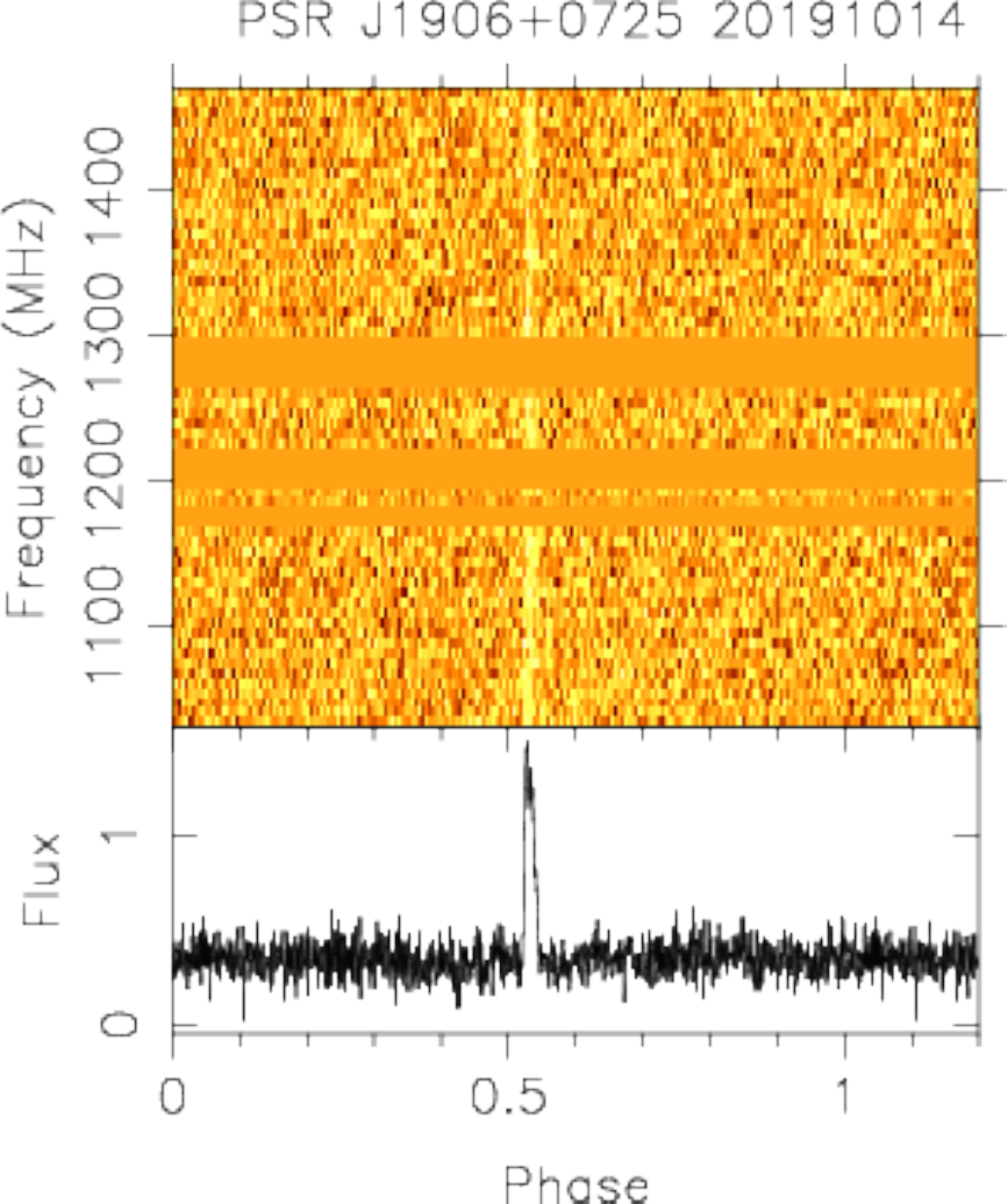}\\[2mm]
\includegraphics[width=38mm]{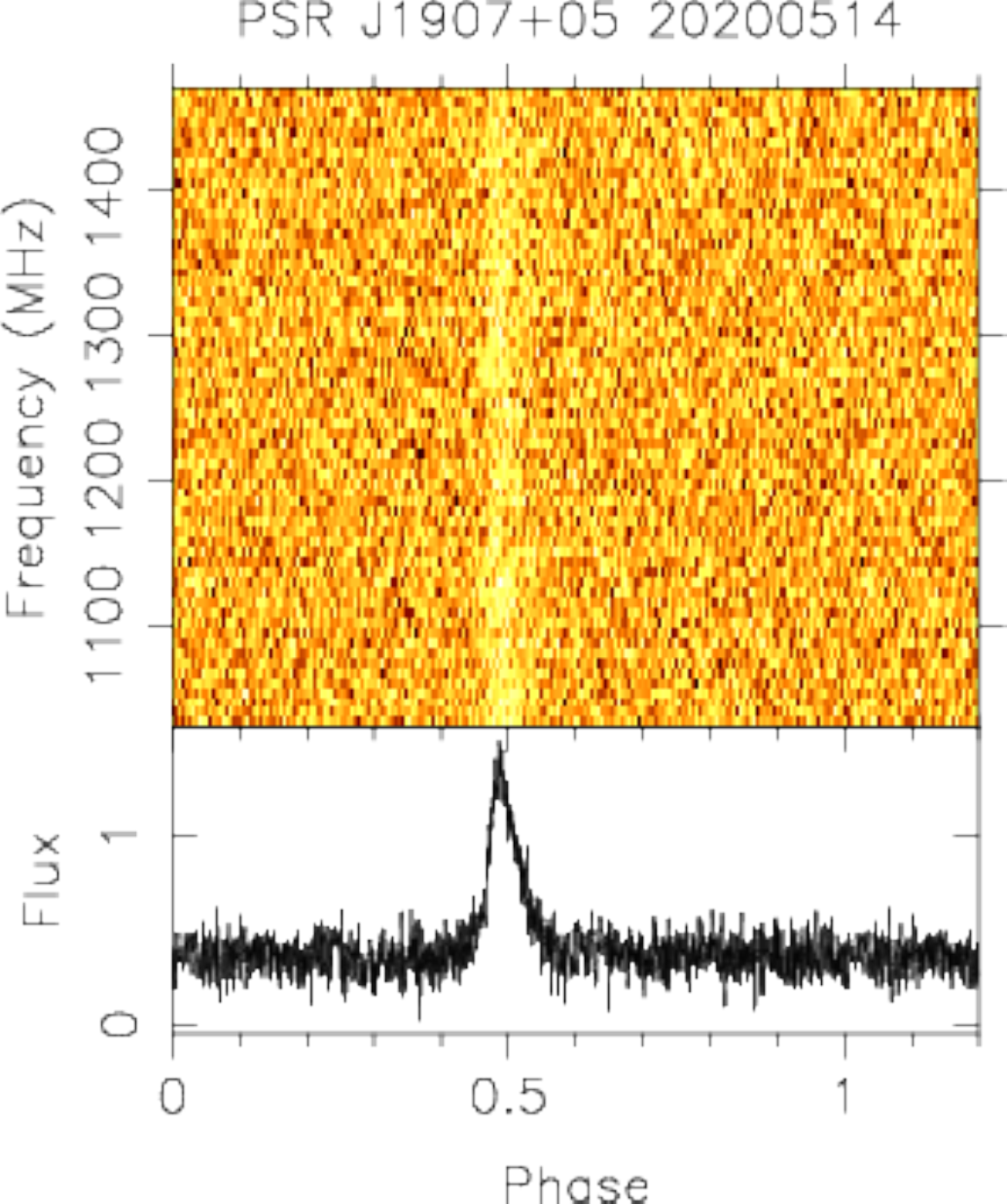}&
\includegraphics[width=38mm]{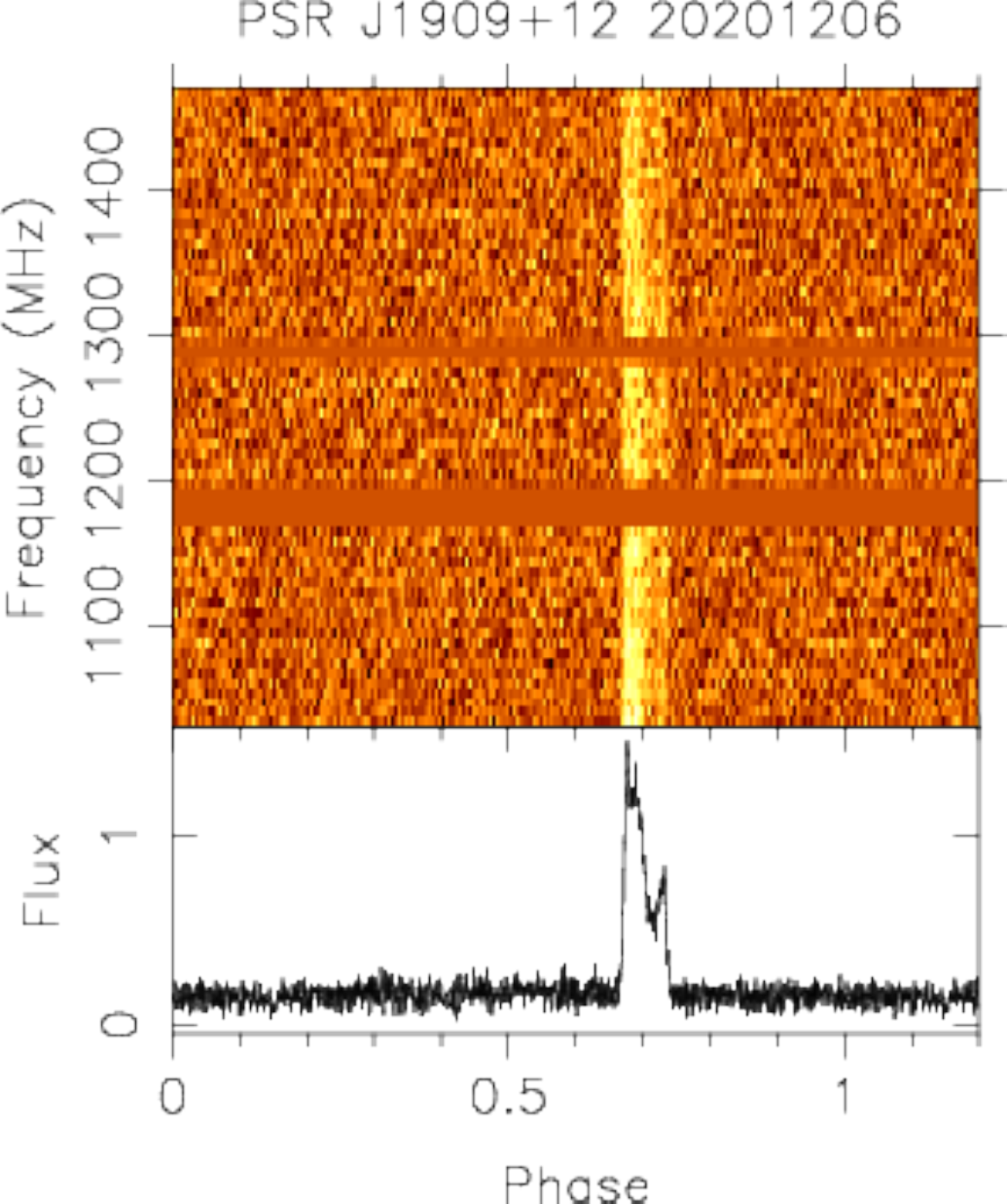}&
\includegraphics[width=38mm]{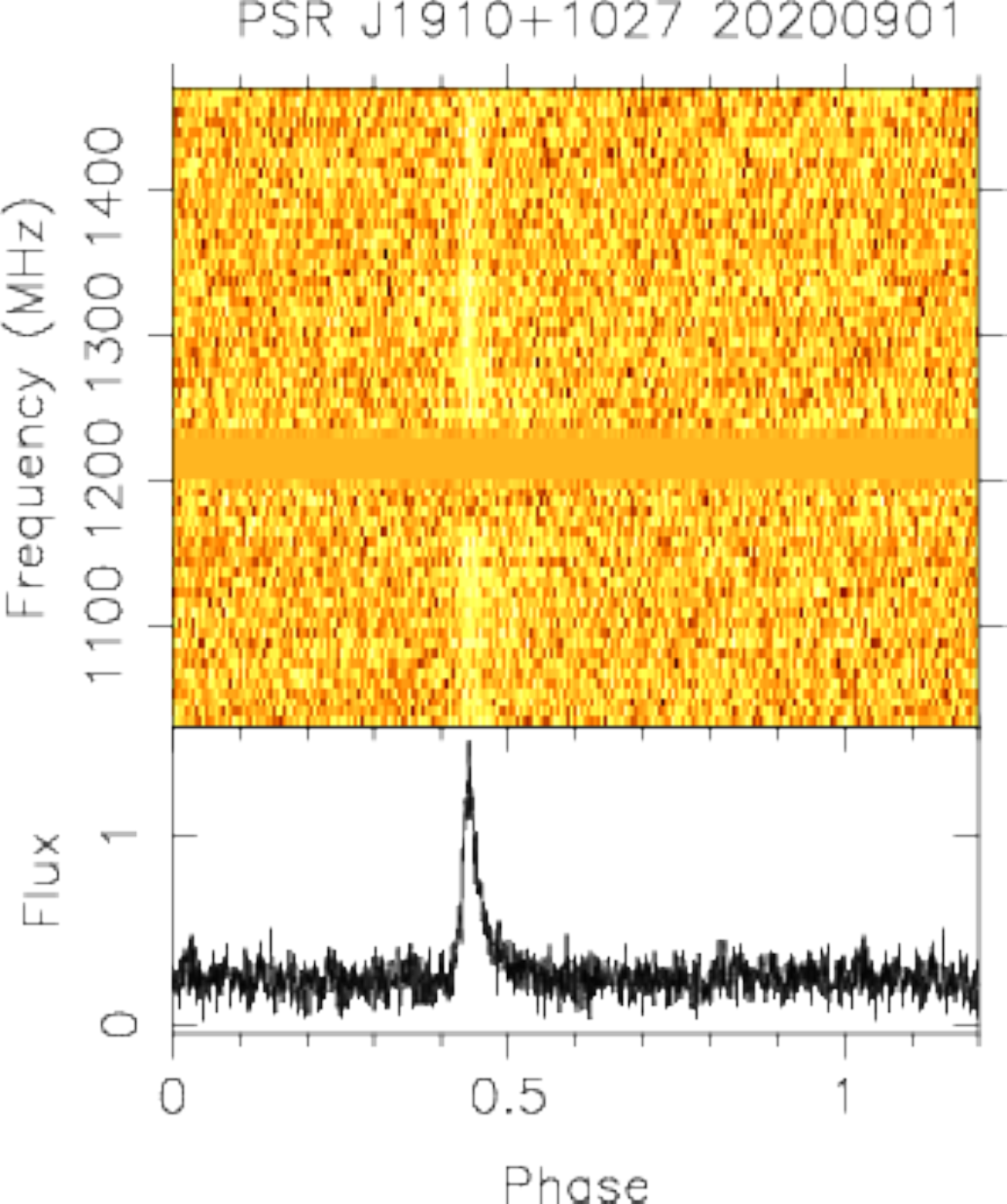}&
\includegraphics[width=38mm]{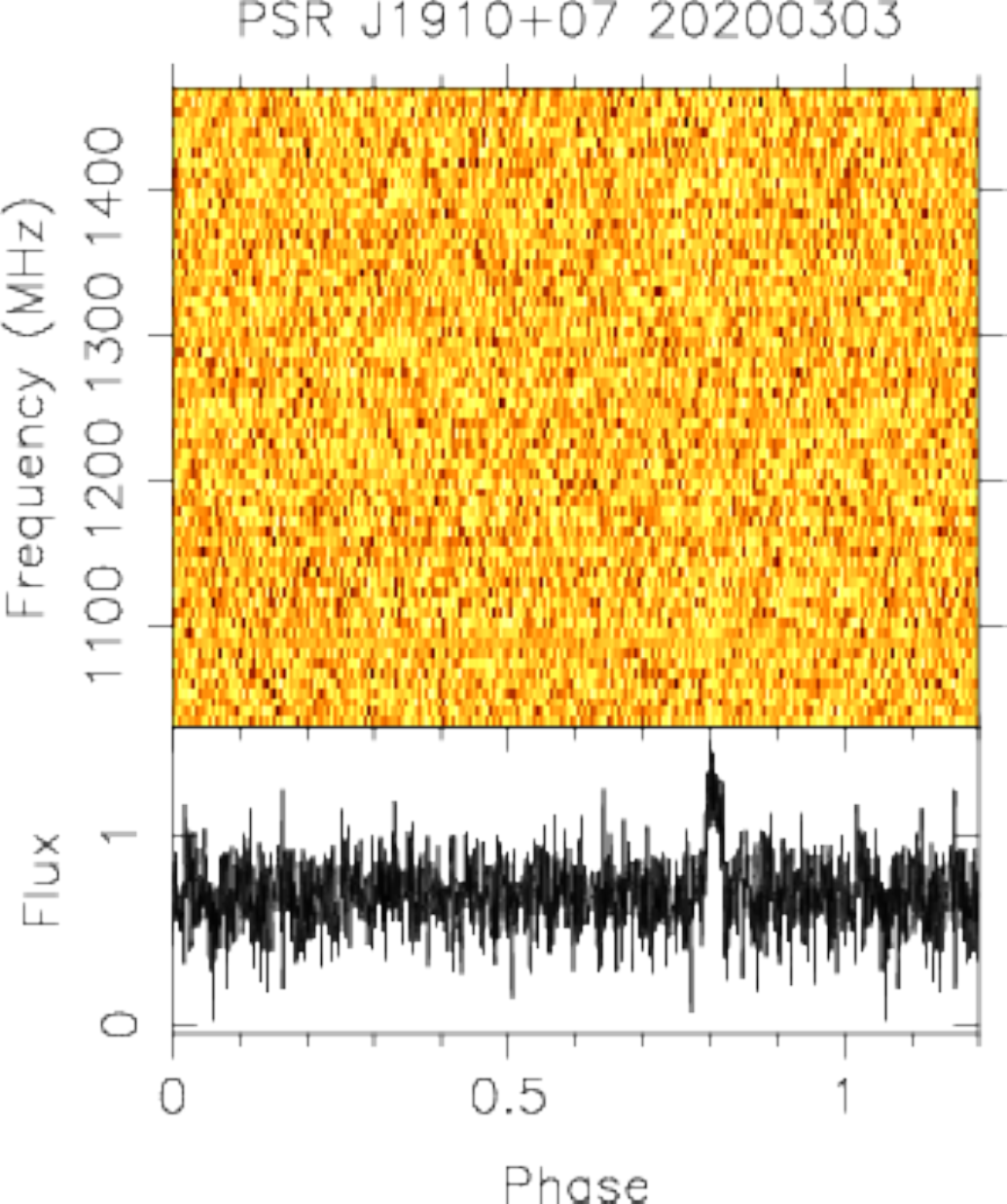}\\[2mm]
\includegraphics[width=38mm]{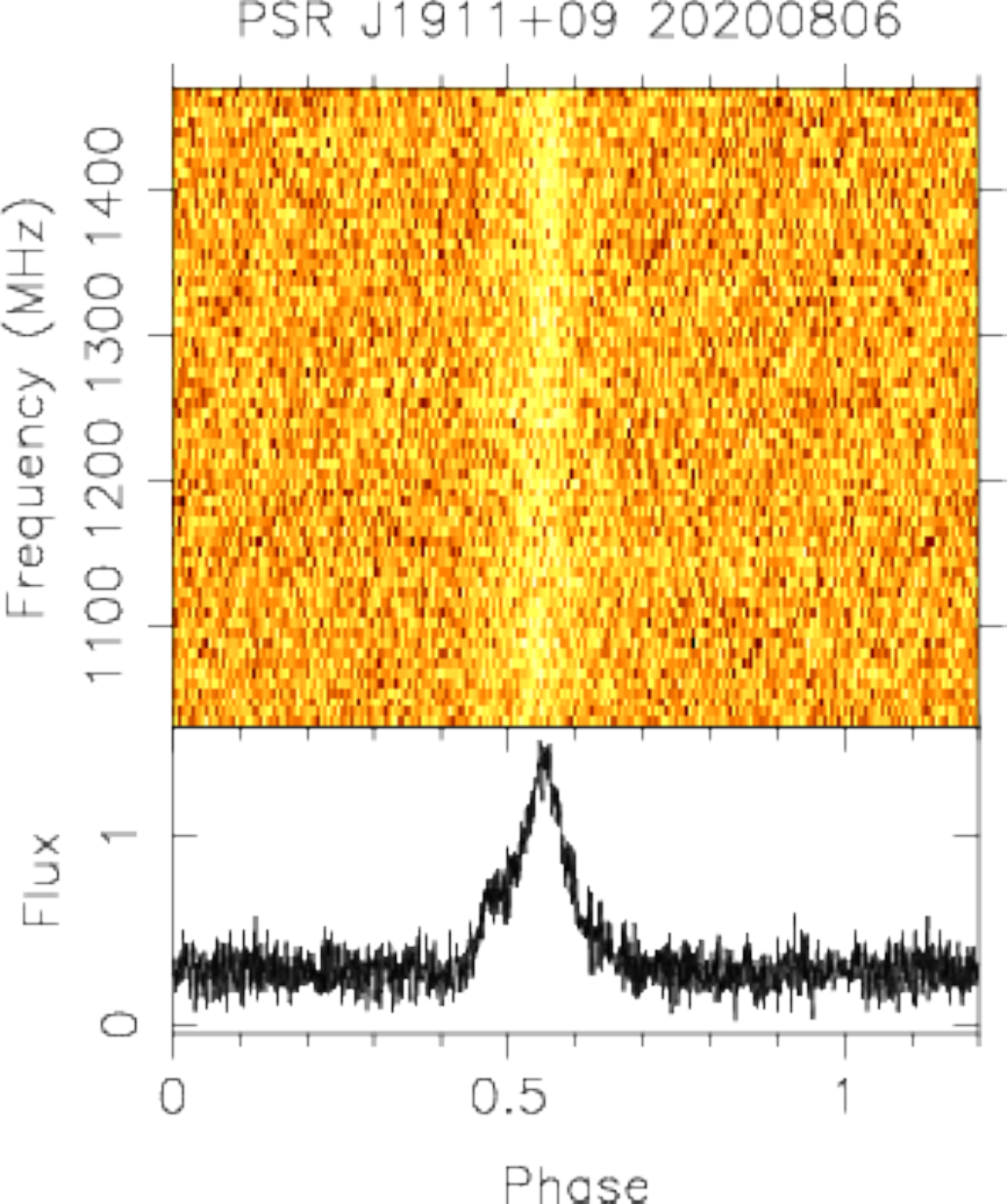}&
\includegraphics[width=38mm]{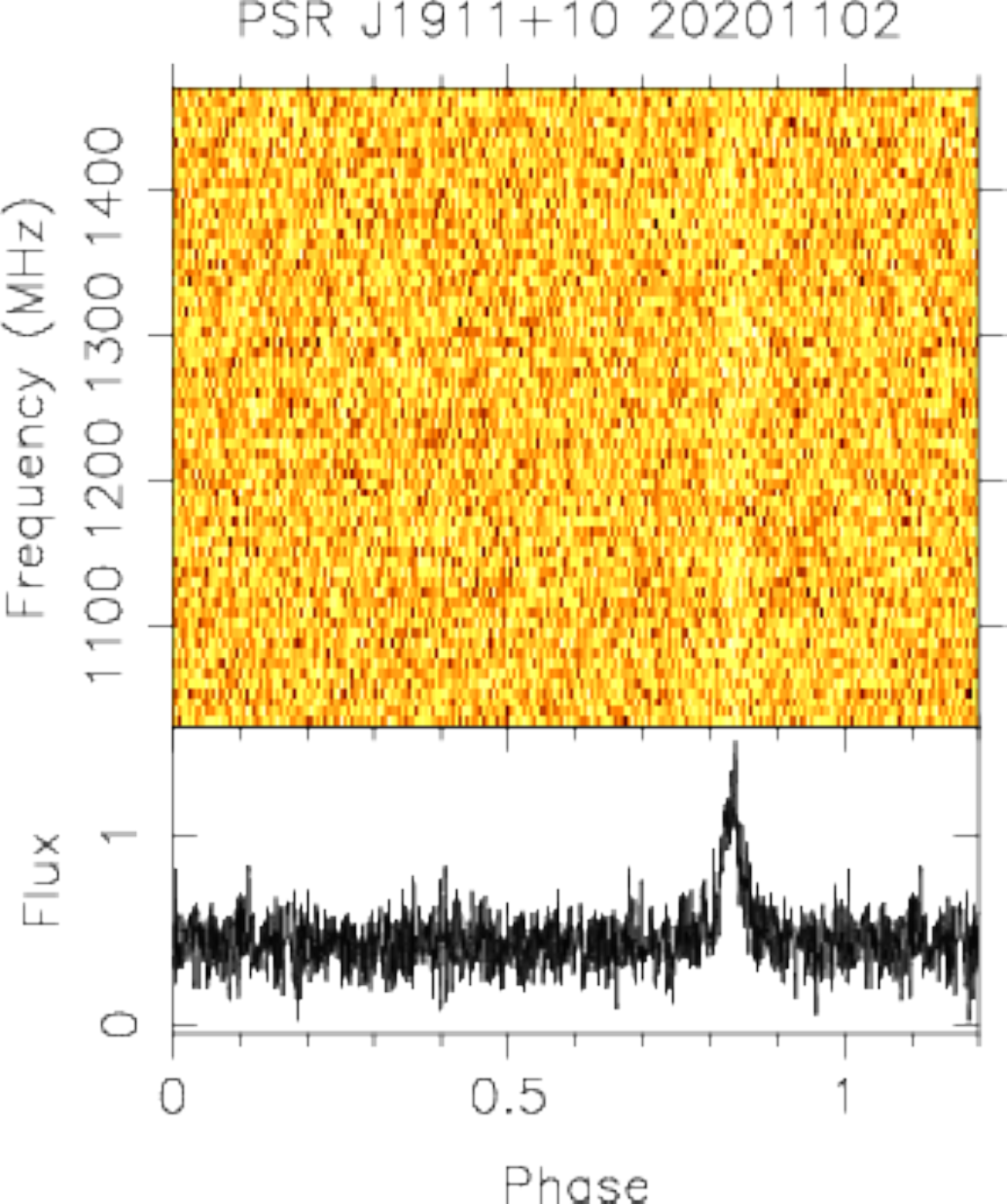}&
\includegraphics[width=38mm]{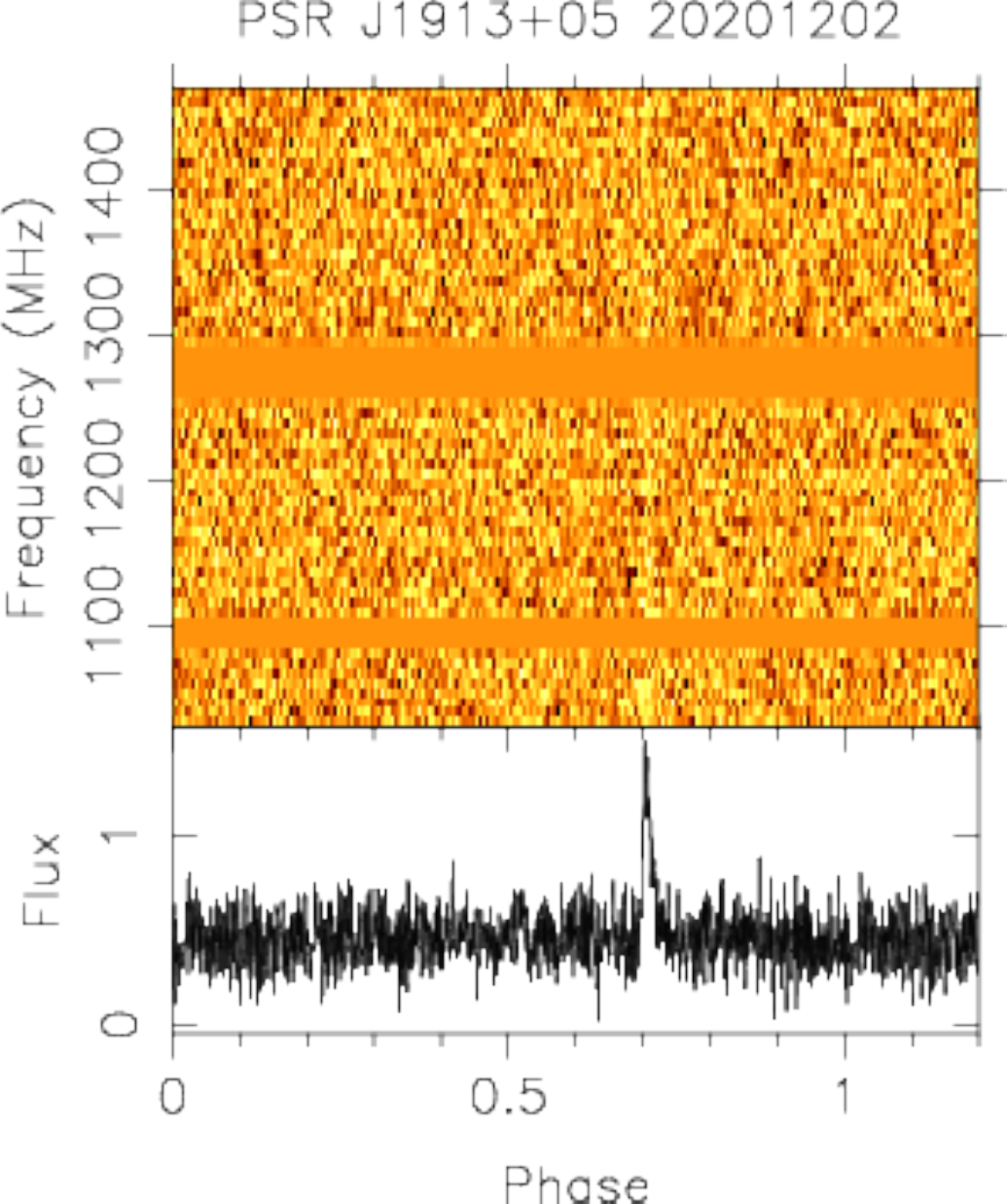}&
\includegraphics[width=38mm]{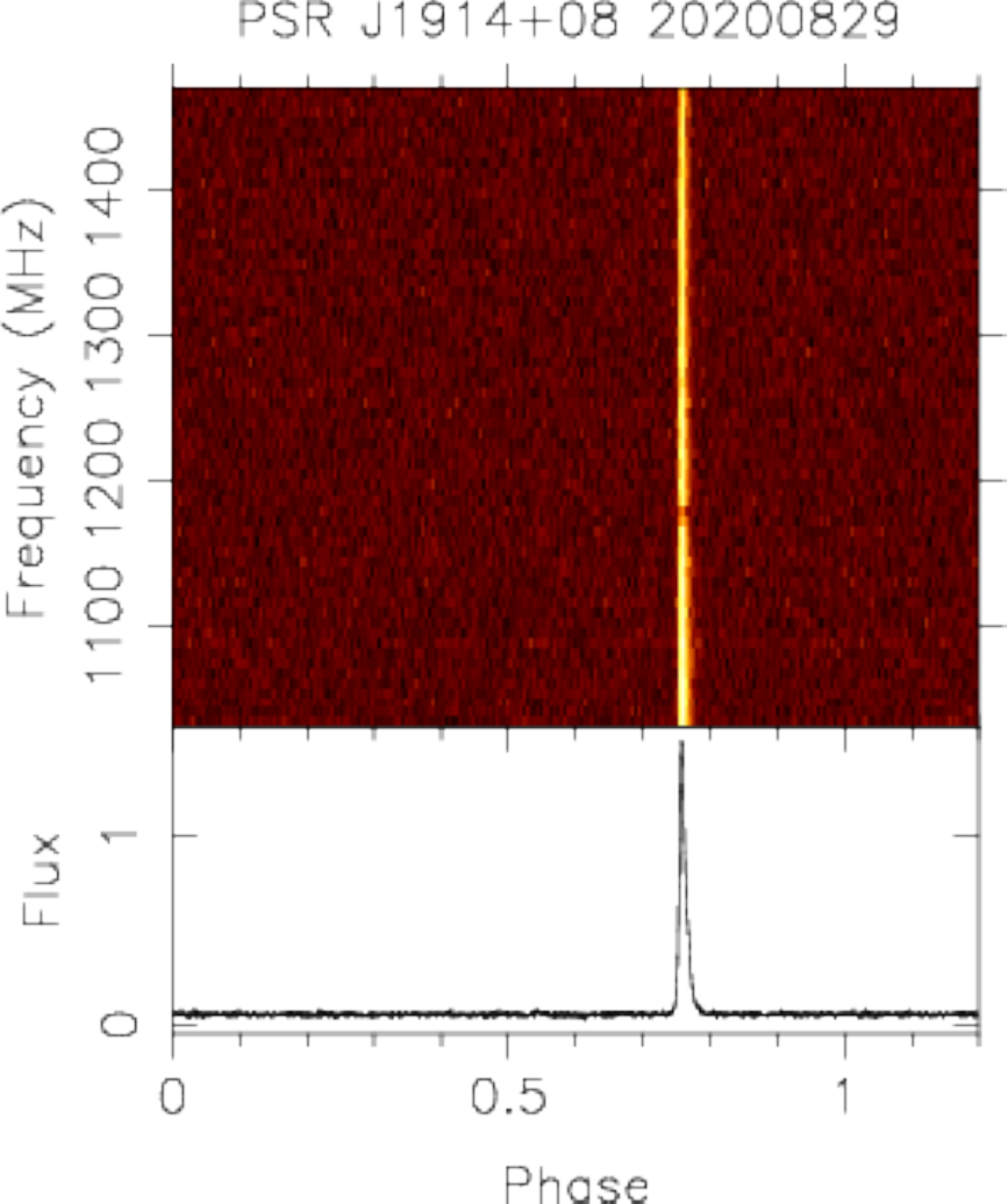}
\end{tabular}%

\begin{minipage}{3cm}
\caption[]{-- {\it Continued}.}\end{minipage}
\addtocounter{figure}{-1}
\end{figure*}%
\begin{figure*}[htp!!!]
\centering
\begin{tabular}{rrrrrr}
\includegraphics[width=38mm]{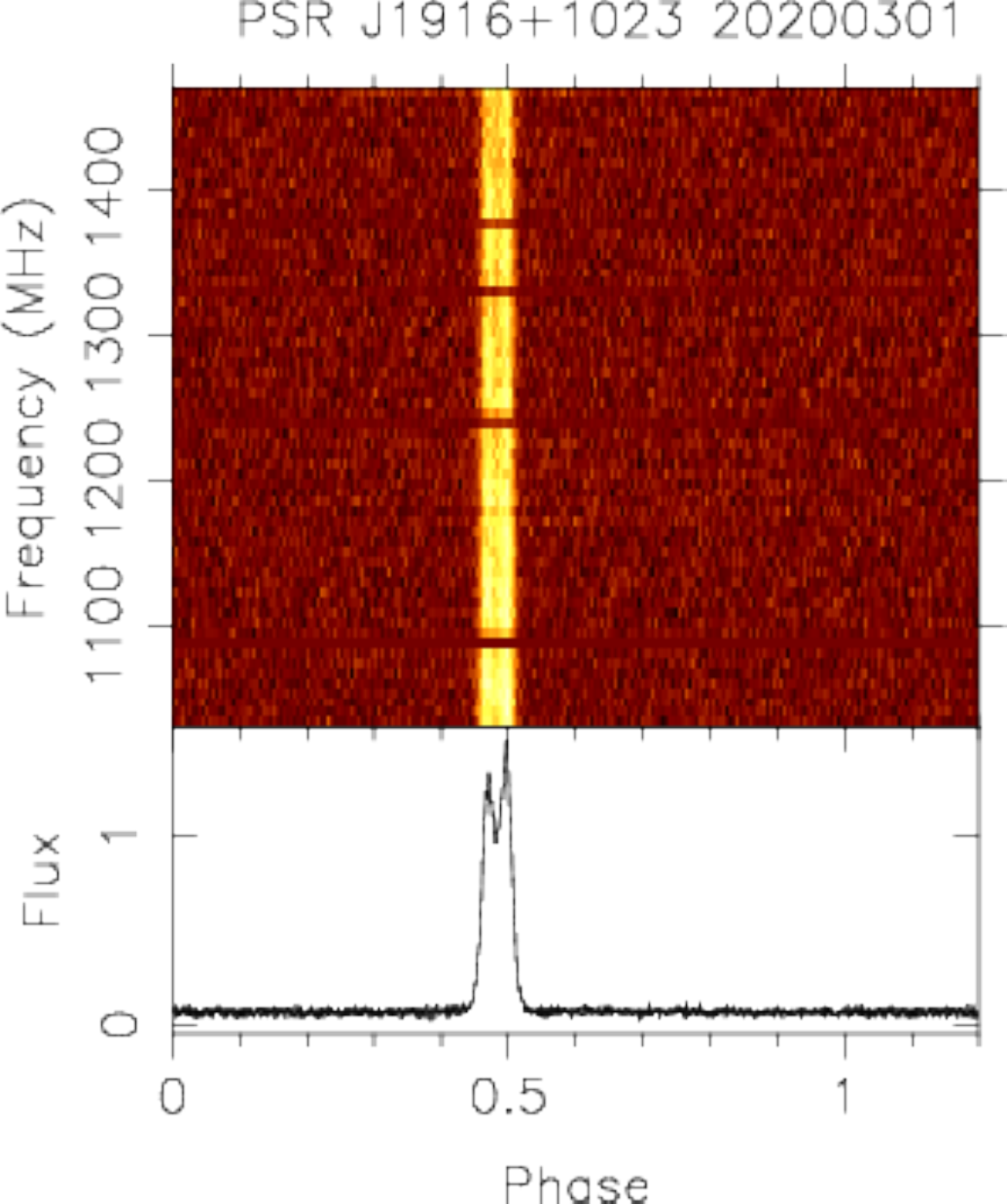}&
\includegraphics[width=38mm]{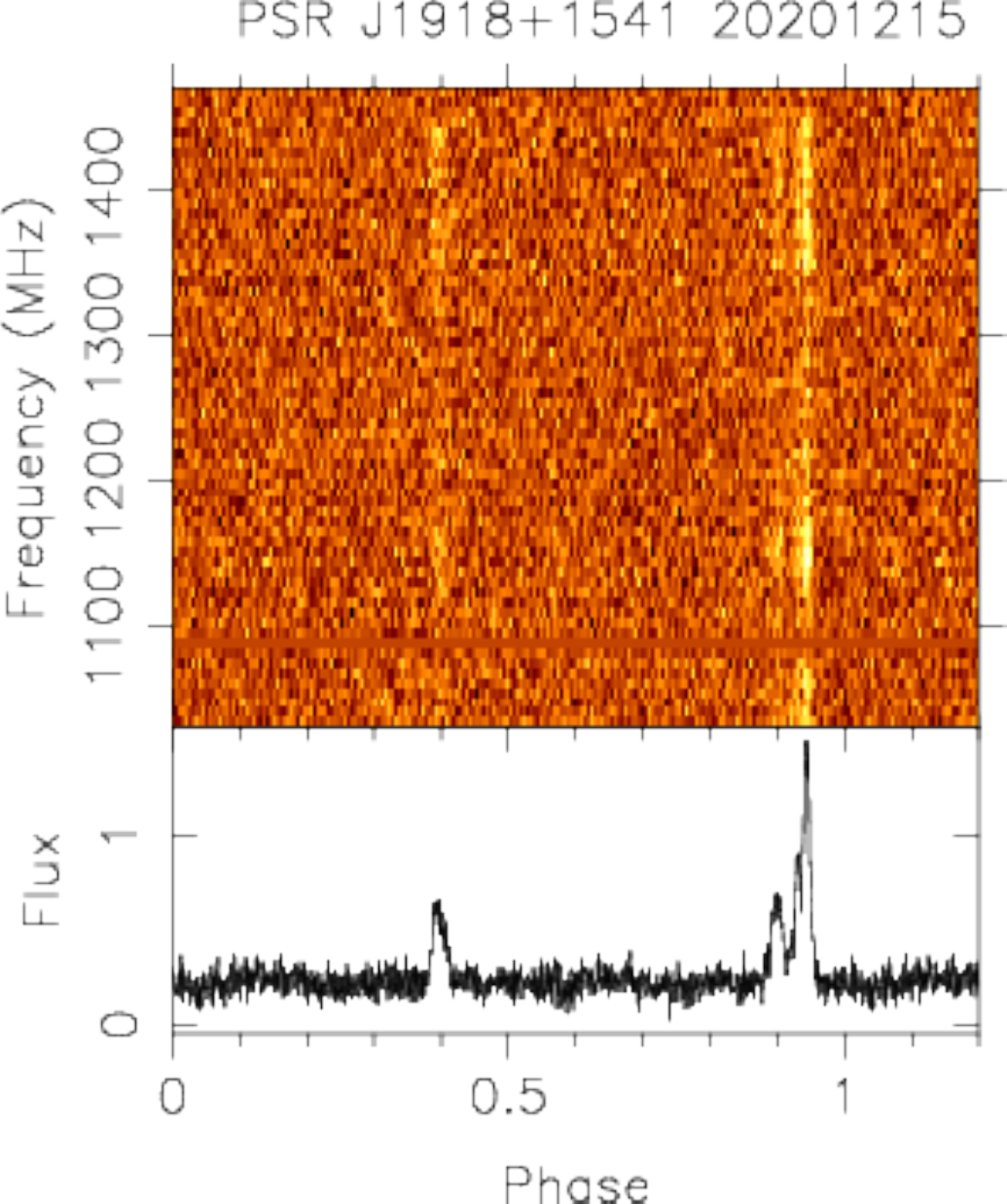}&
\includegraphics[width=38mm]{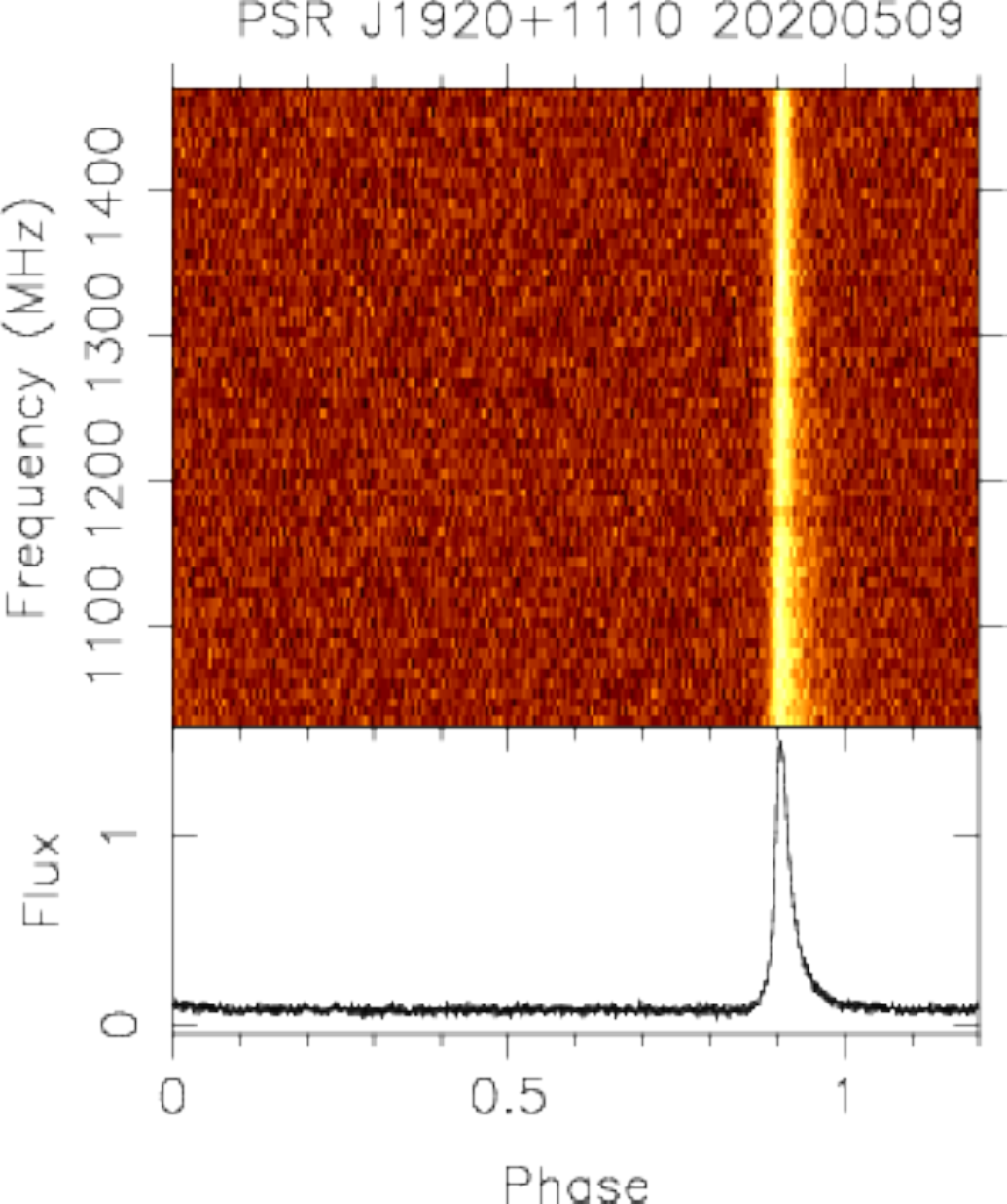}&
\includegraphics[width=38mm]{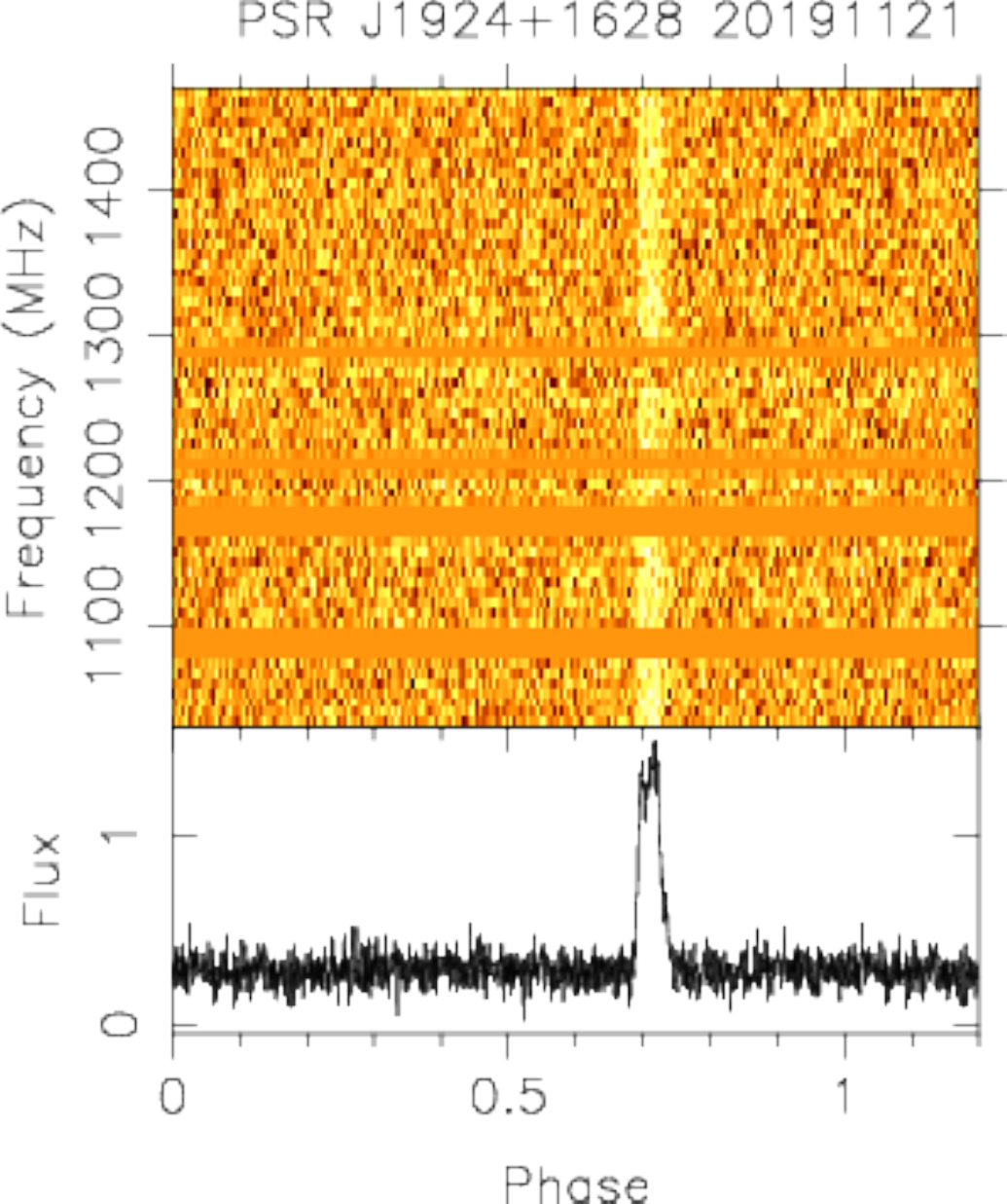}\\[2mm]
\includegraphics[width=38mm]{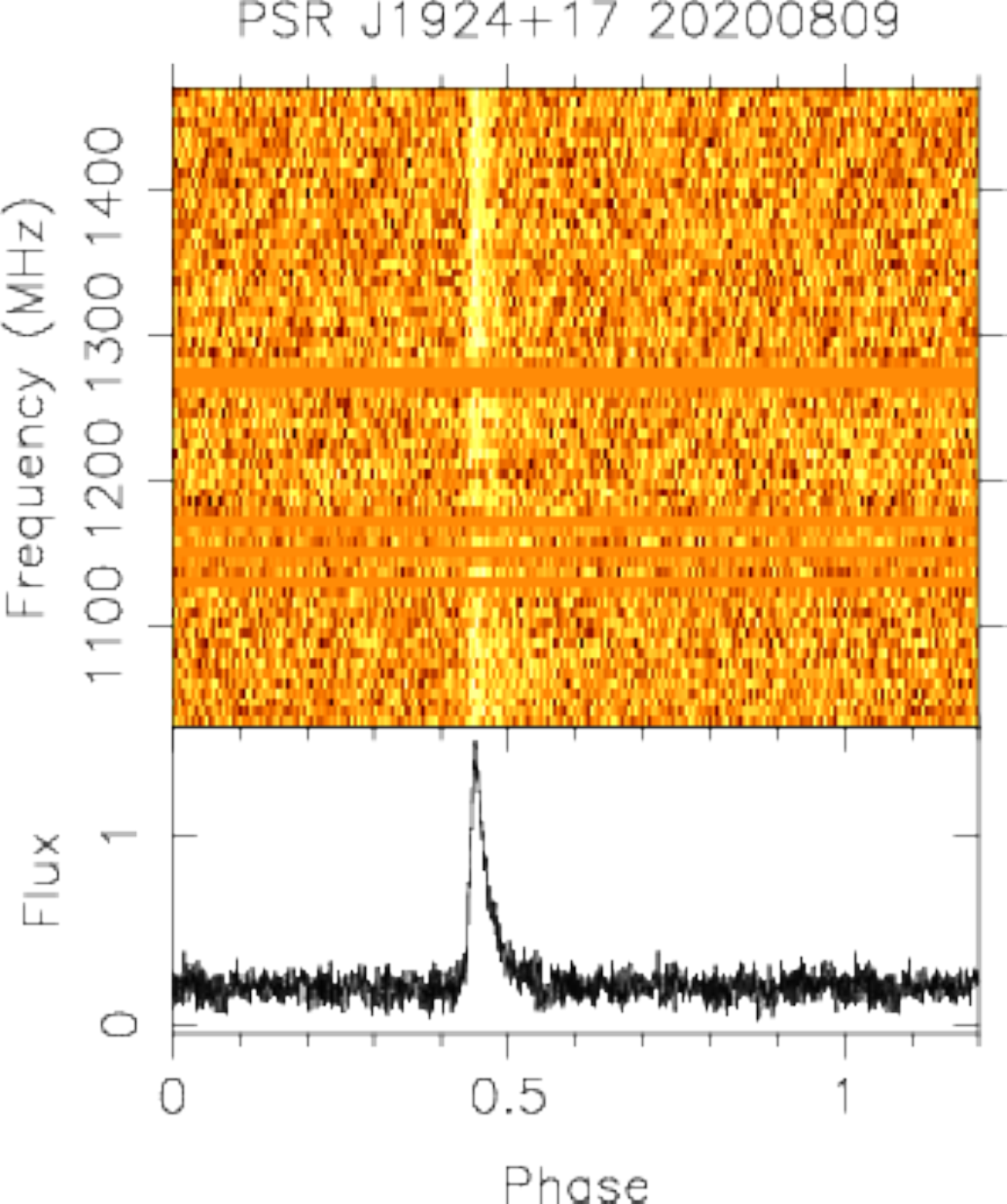}&
\includegraphics[width=38mm]{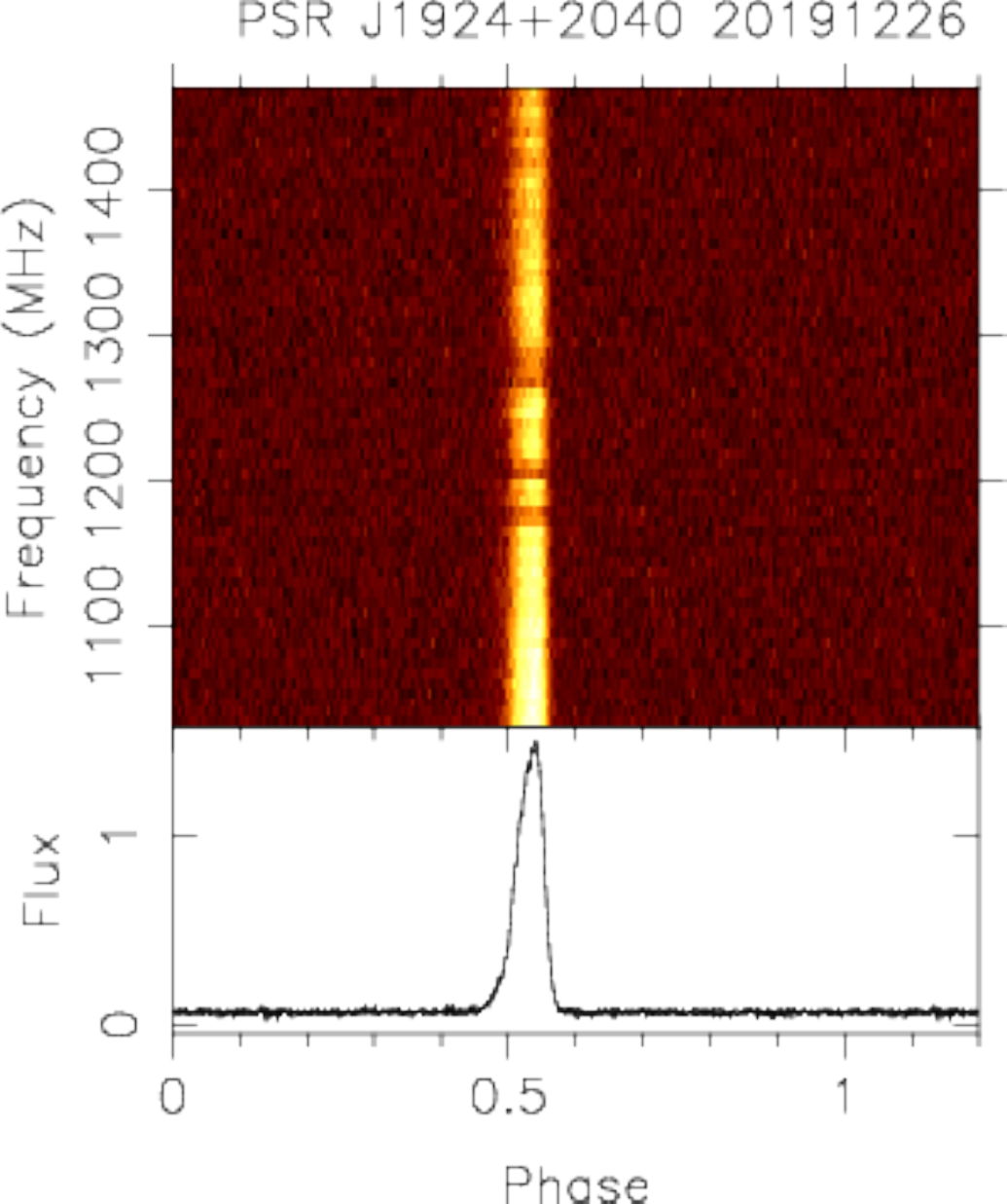}&
\includegraphics[width=38mm]{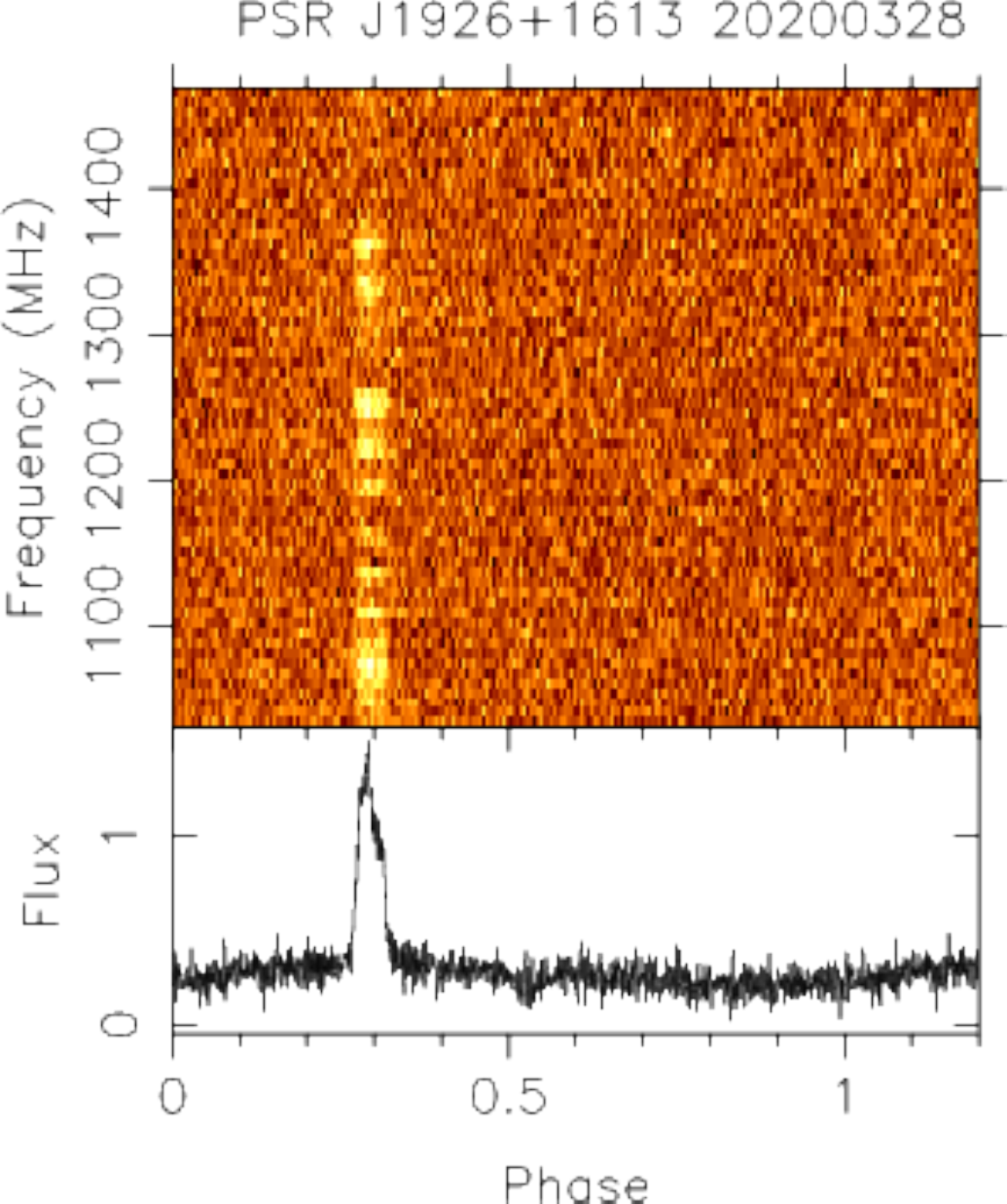}&
\includegraphics[width=38mm]{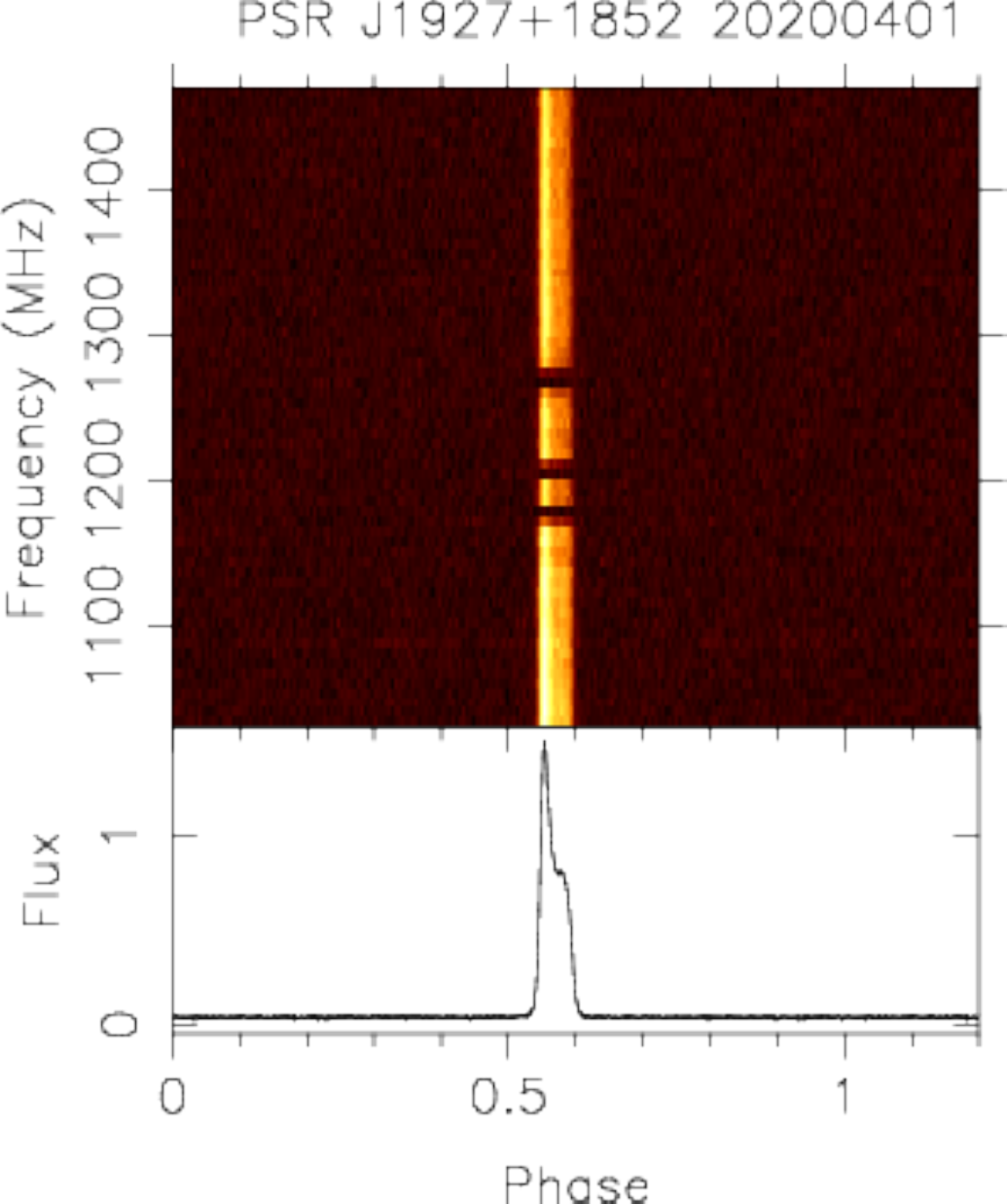}\\[2mm]
\includegraphics[width=38mm]{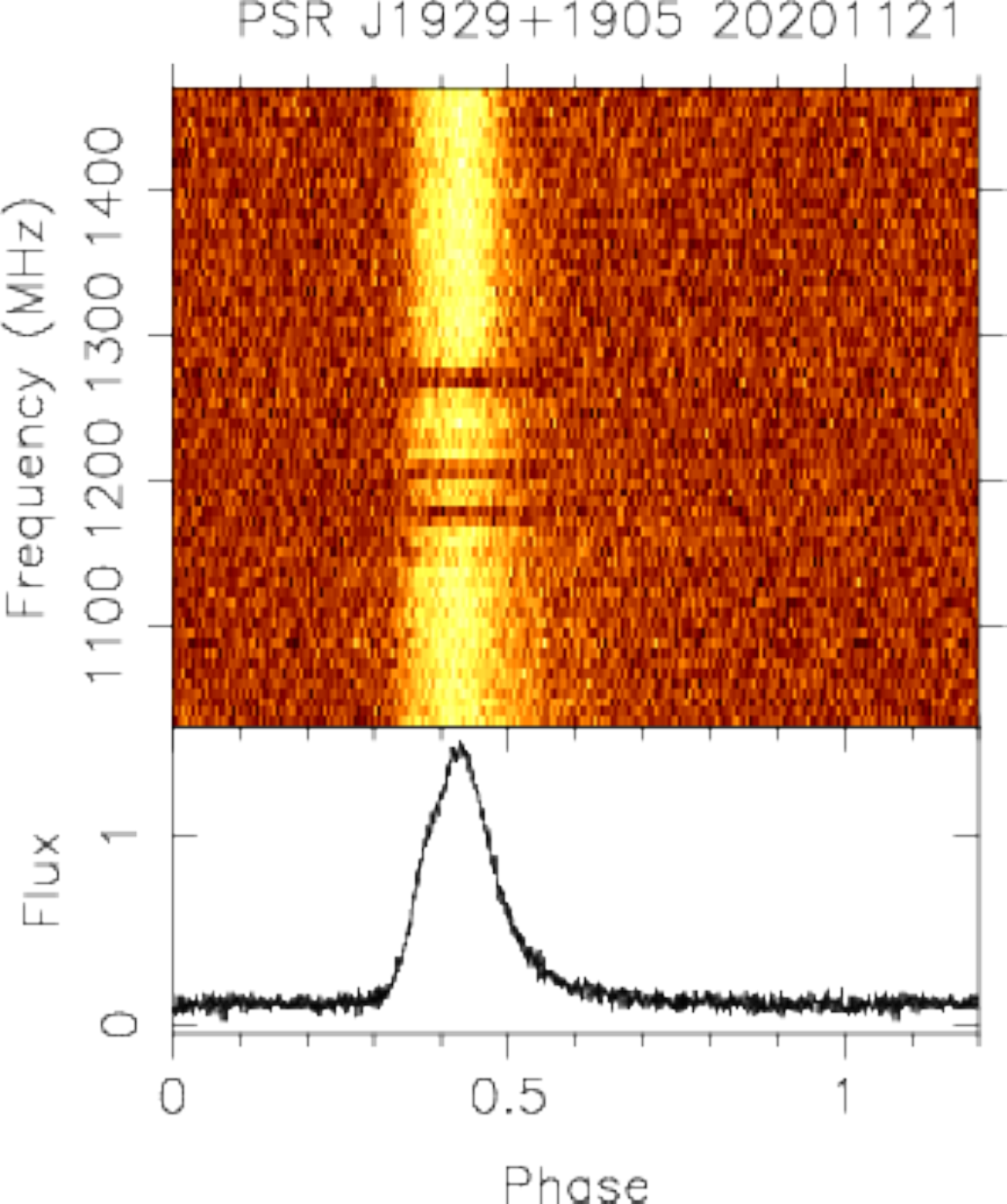}&
\includegraphics[width=38mm]{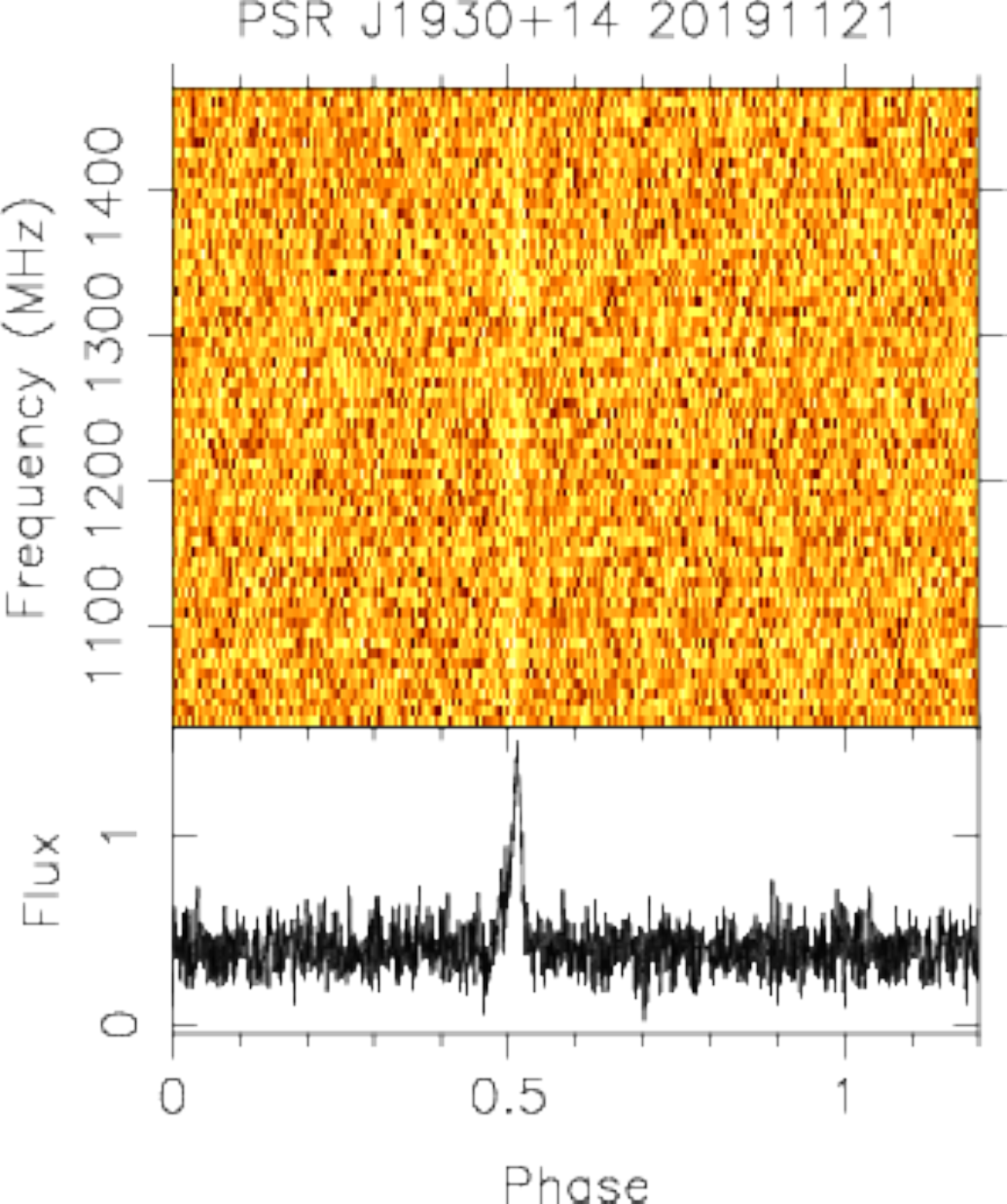}&
\includegraphics[width=38mm]{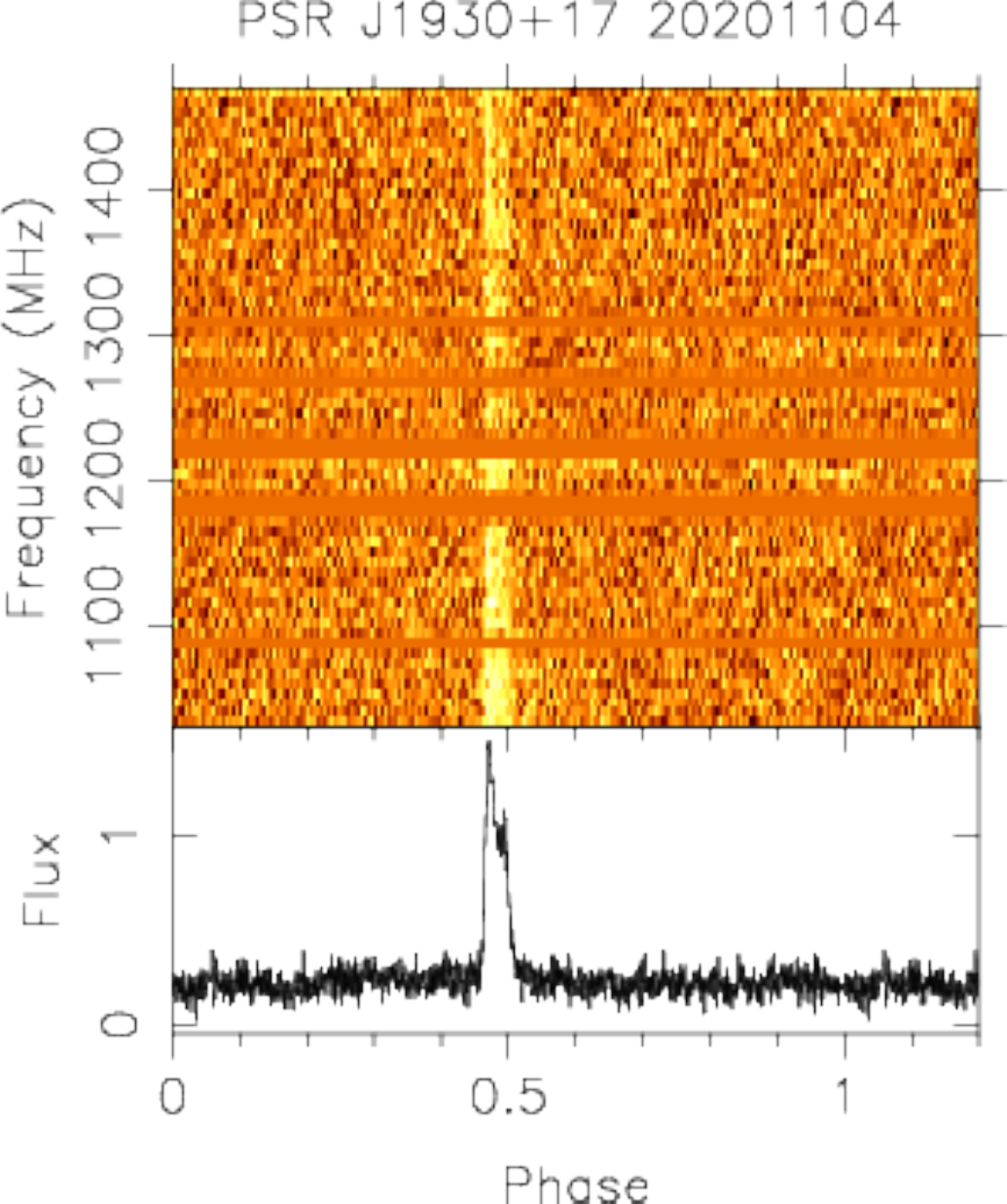}&
\includegraphics[width=38mm]{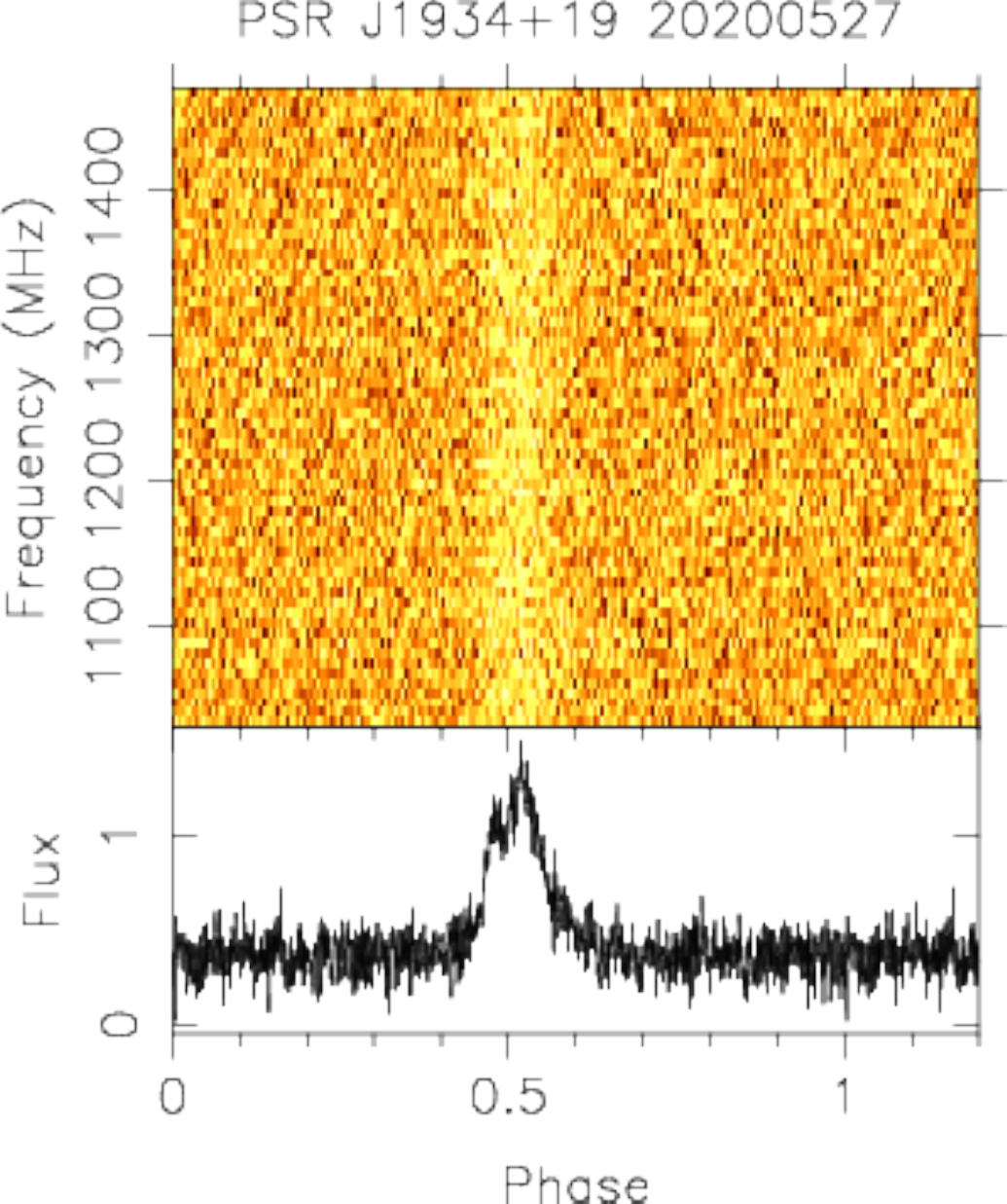}\\[2mm]
\includegraphics[width=38mm]{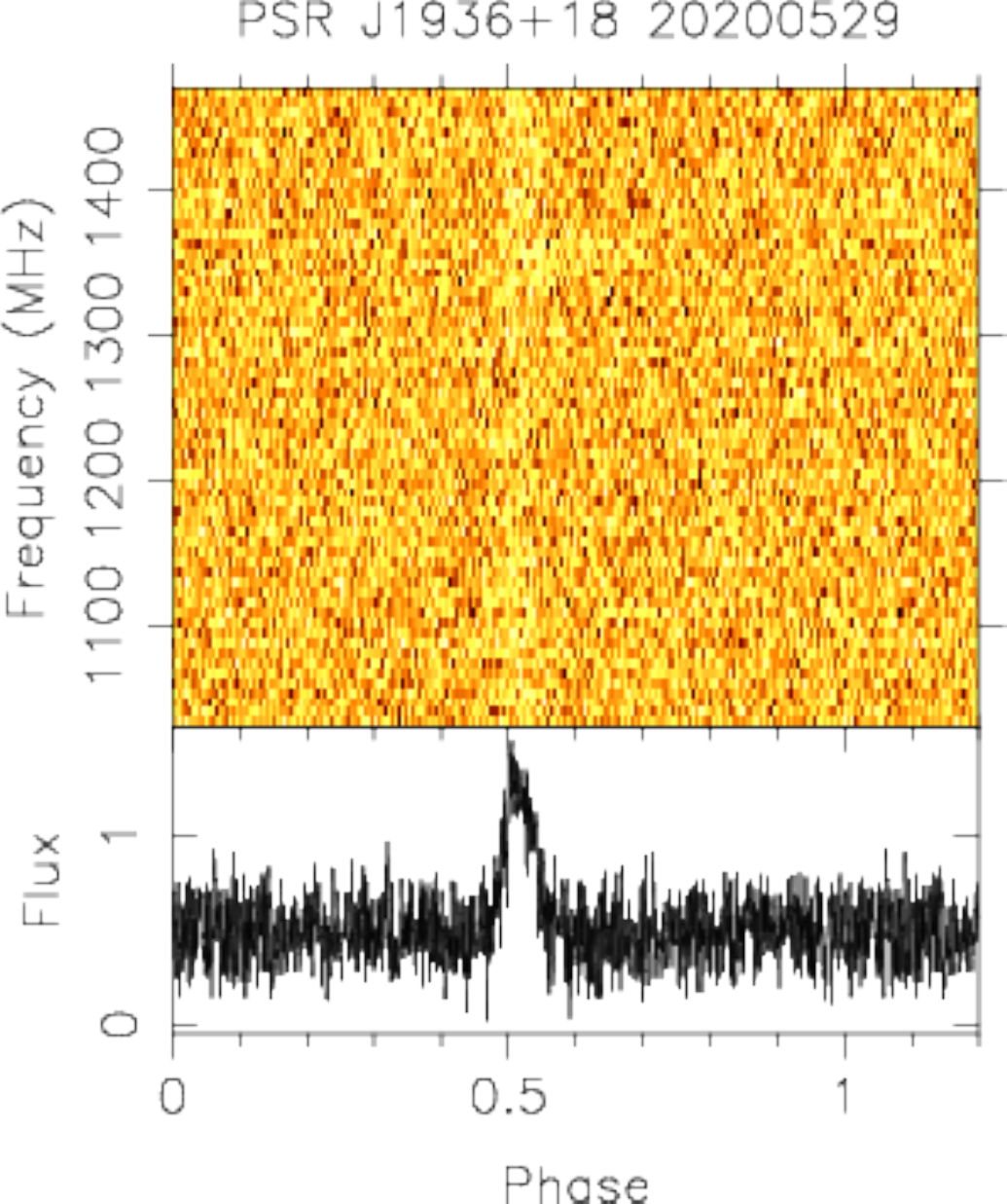}&
\includegraphics[width=38mm]{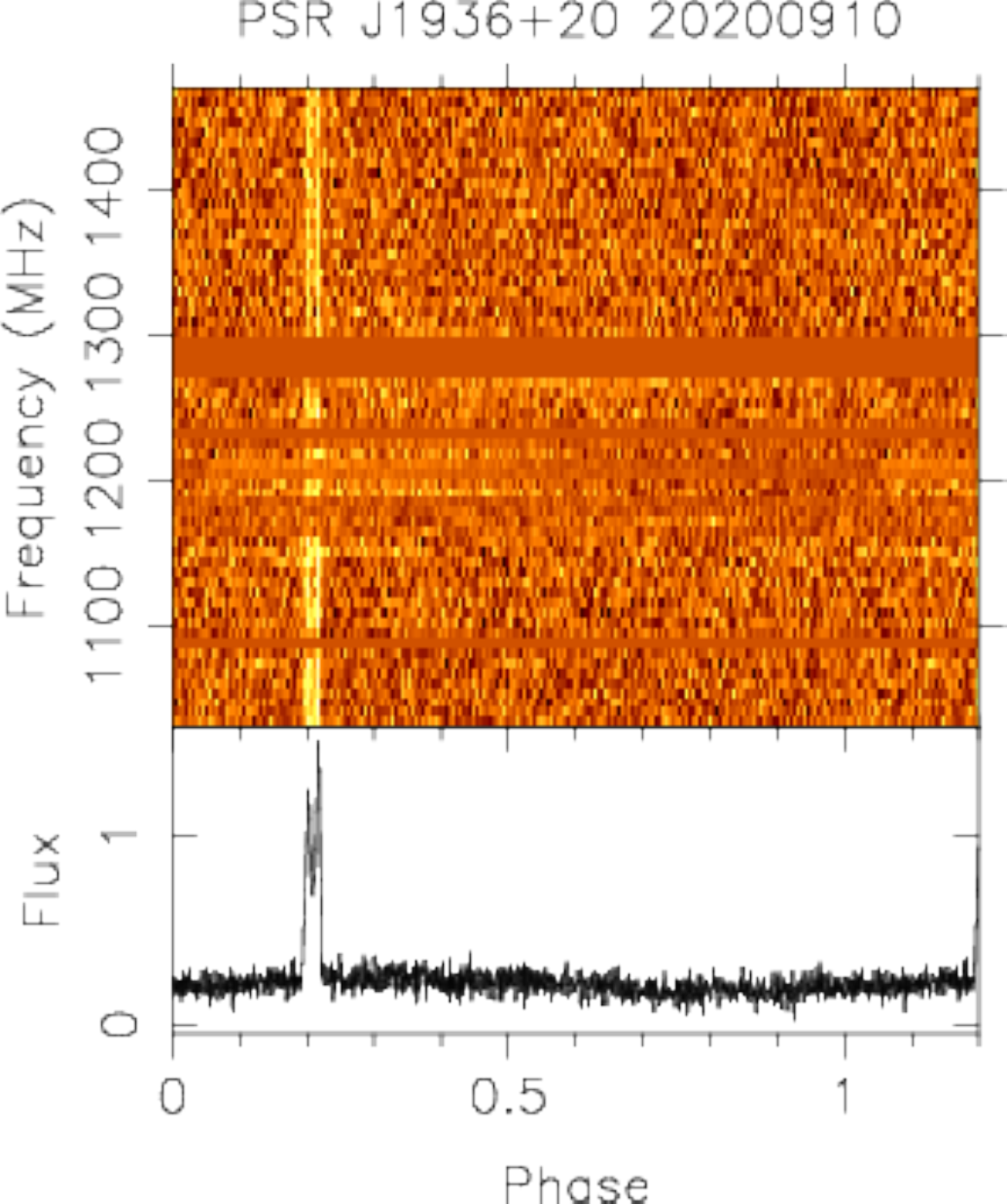}&
\includegraphics[width=38mm]{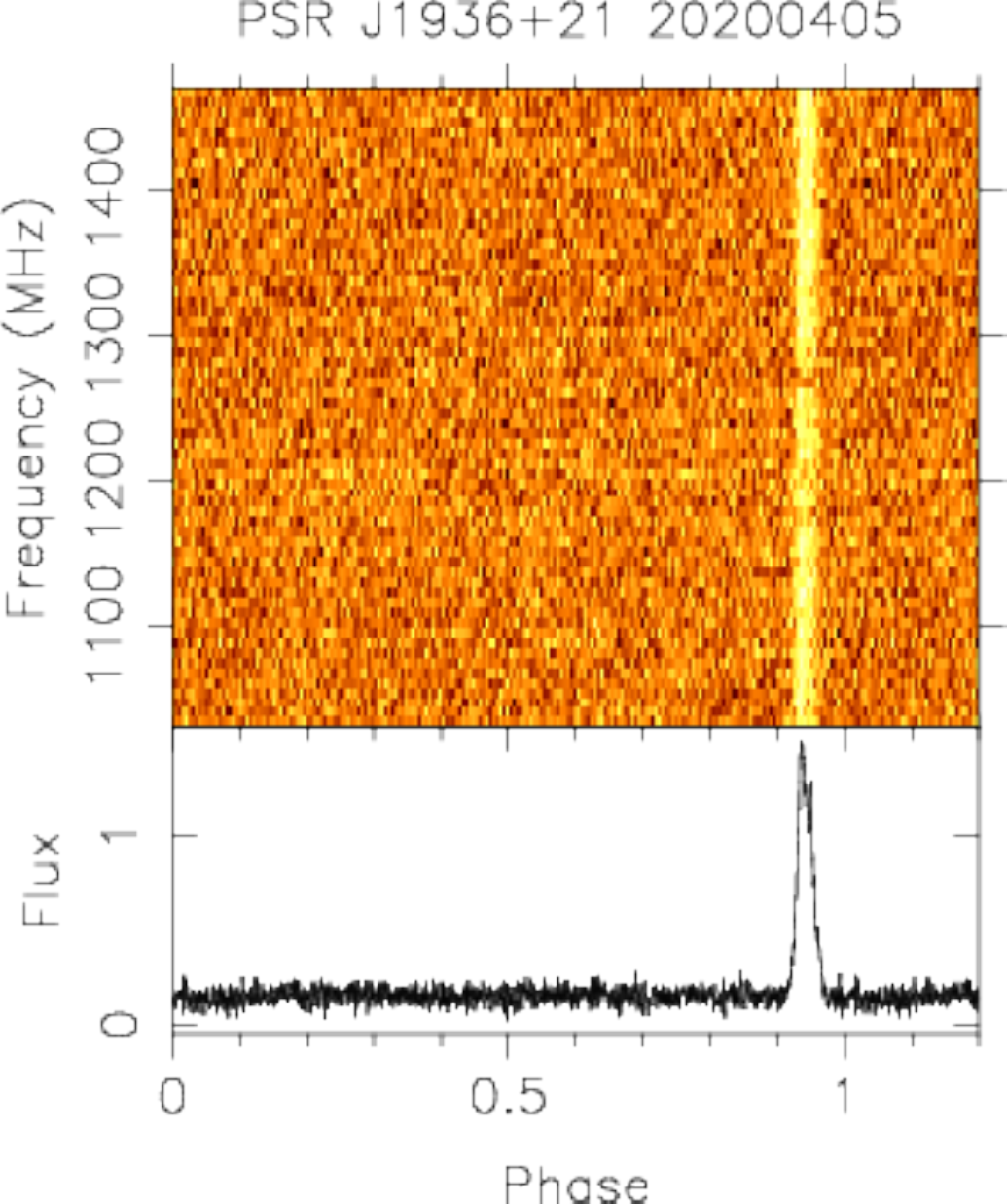}&
\includegraphics[width=38mm]{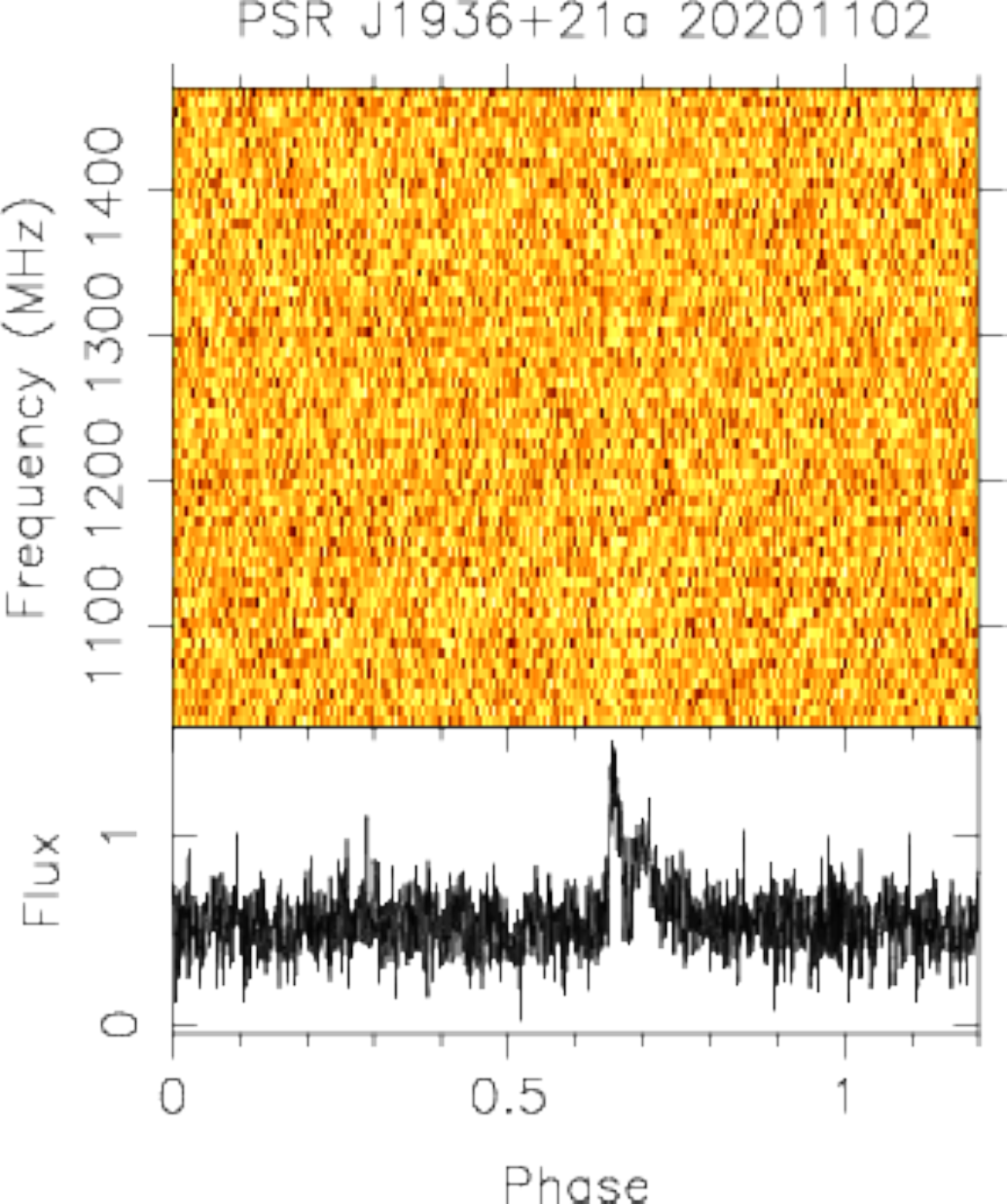}\\[2mm]
\includegraphics[width=38mm]{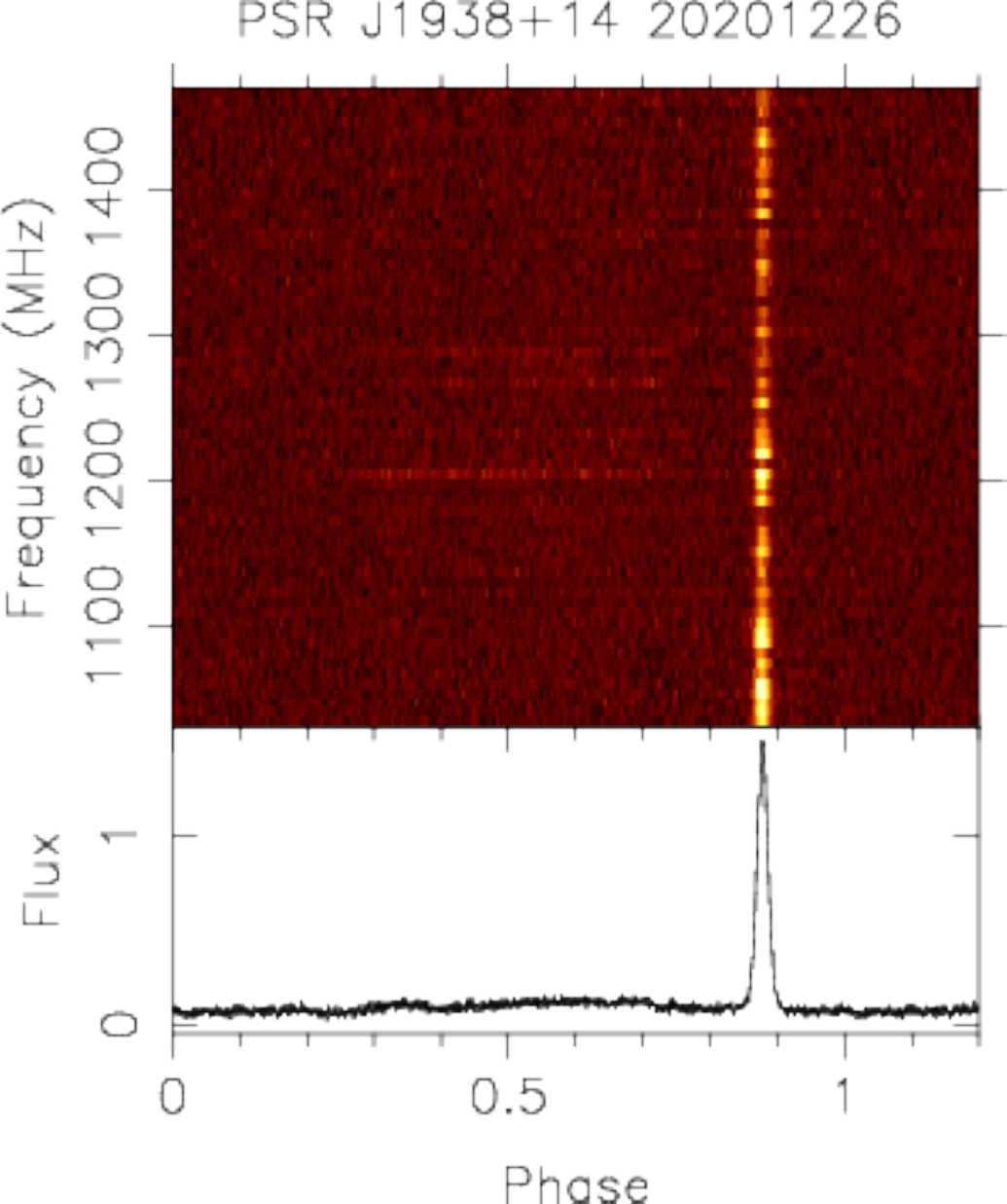}&
\includegraphics[width=38mm]{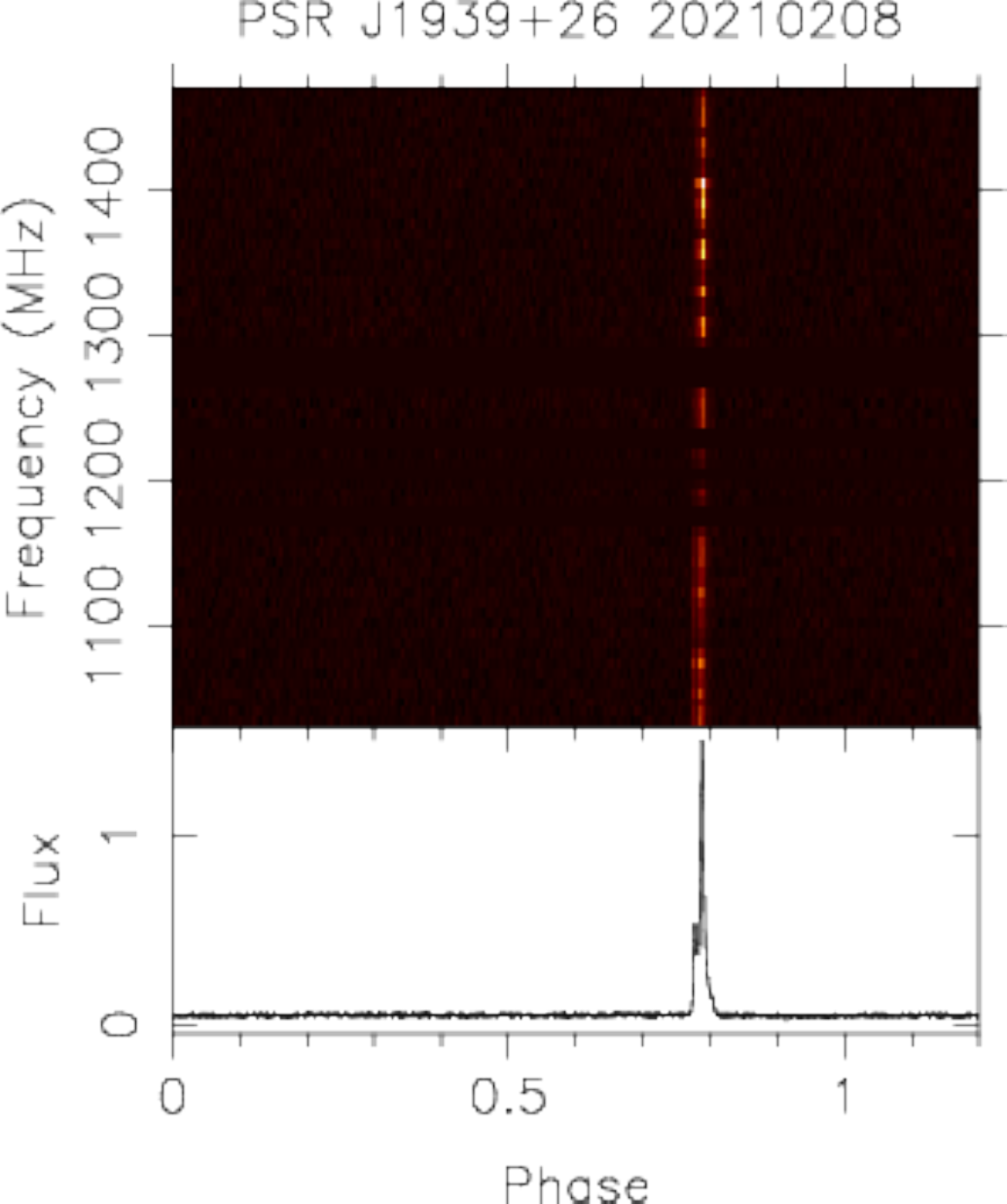}&
\includegraphics[width=38mm]{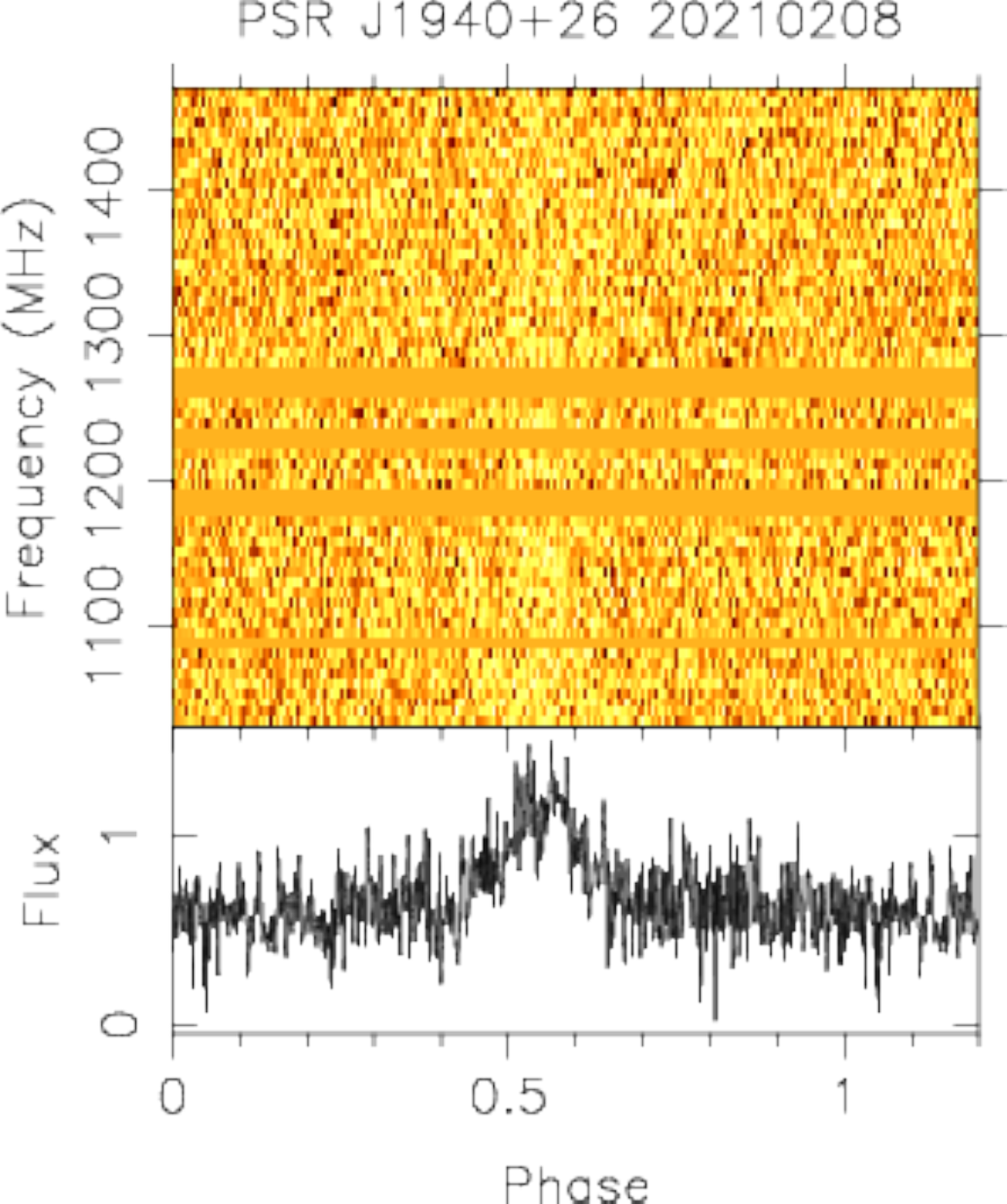}&
\includegraphics[width=38mm]{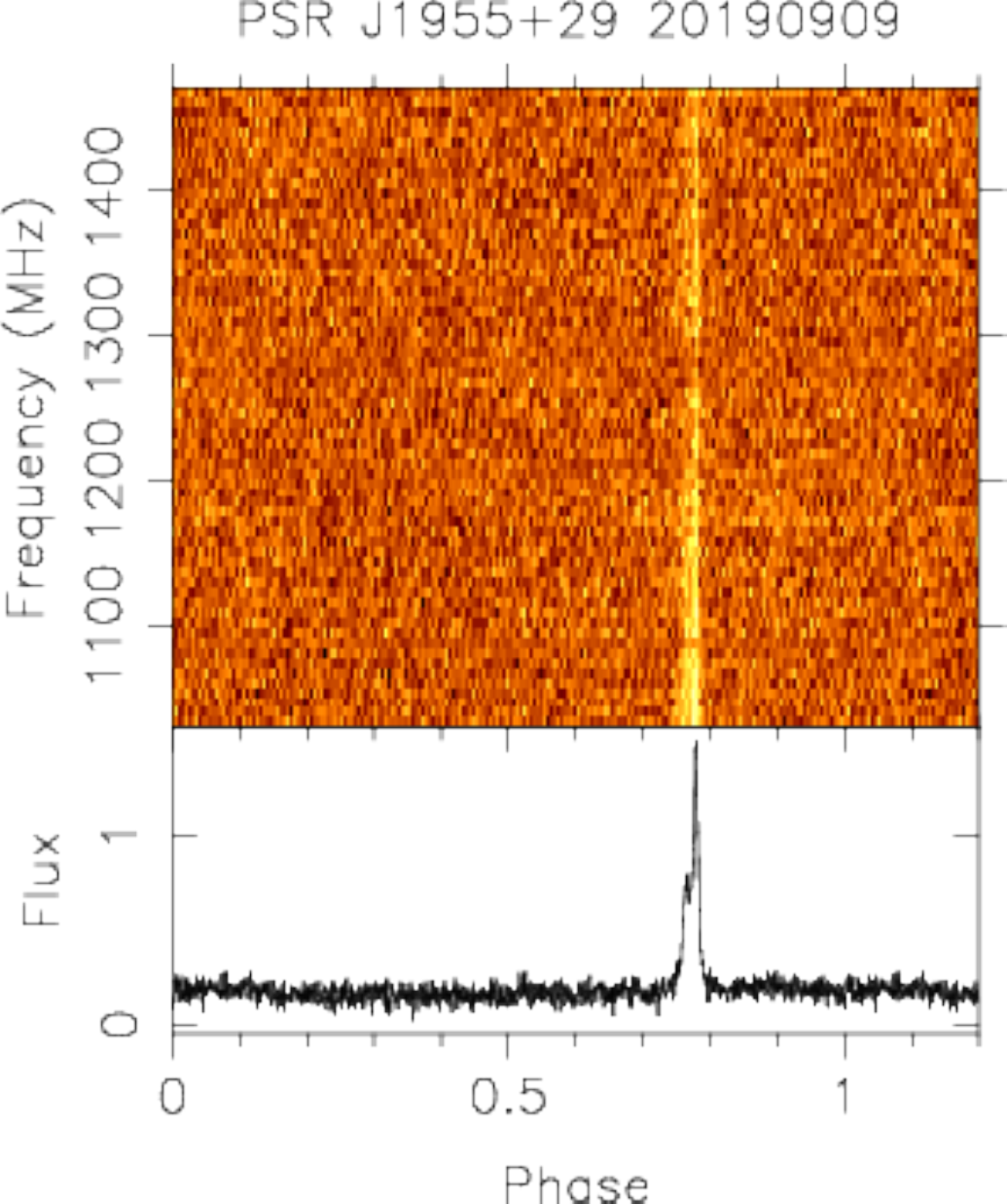}
\end{tabular}%

\begin{minipage}{3cm}
\caption[]{-- {\it Continued}.}\end{minipage}
\addtocounter{figure}{-1}
\end{figure*}%

\begin{figure}[htp!!!]
\centering
\begin{tabular}{rrrrrr}
\includegraphics[width=38mm]{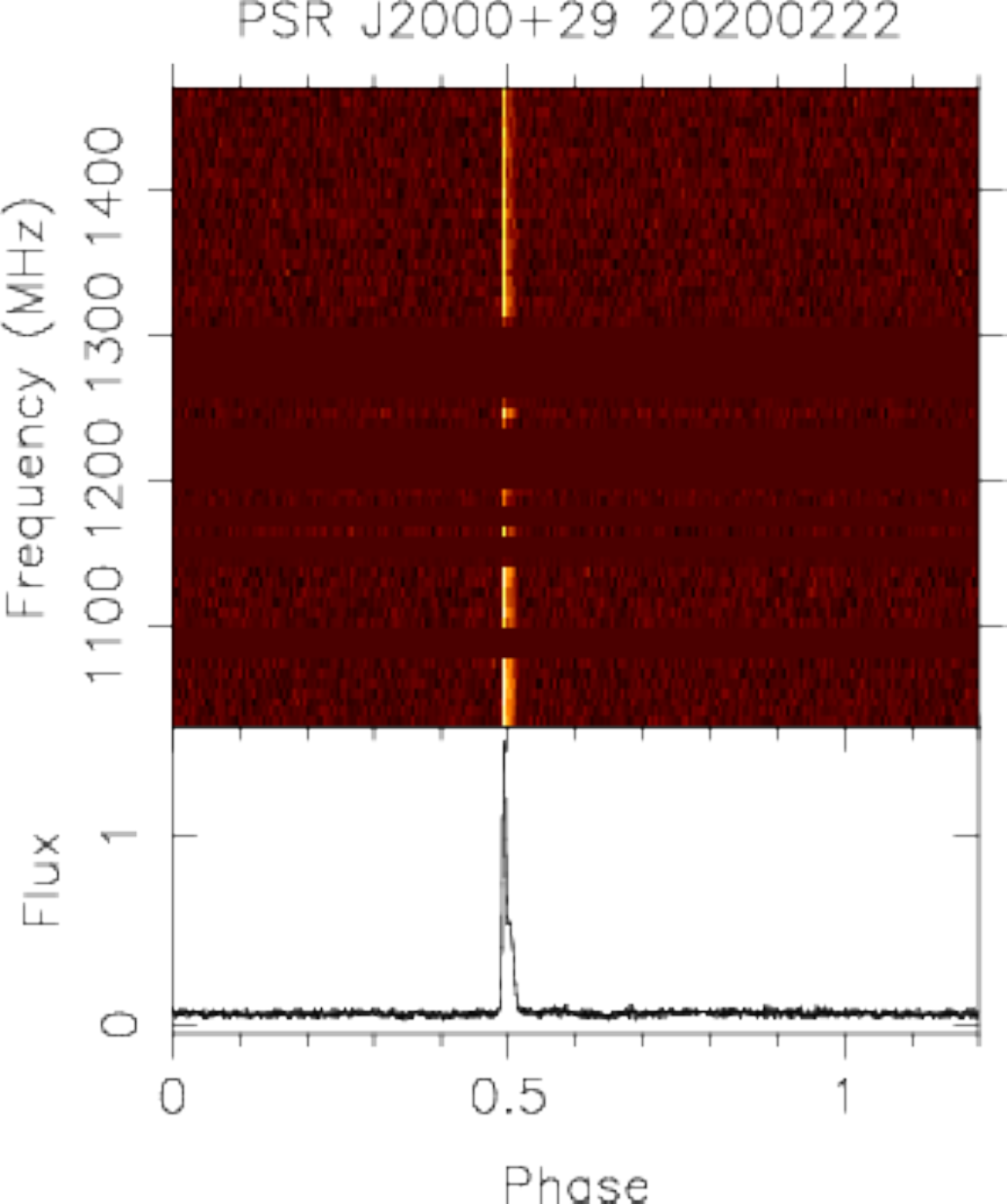}&
\includegraphics[width=38mm]{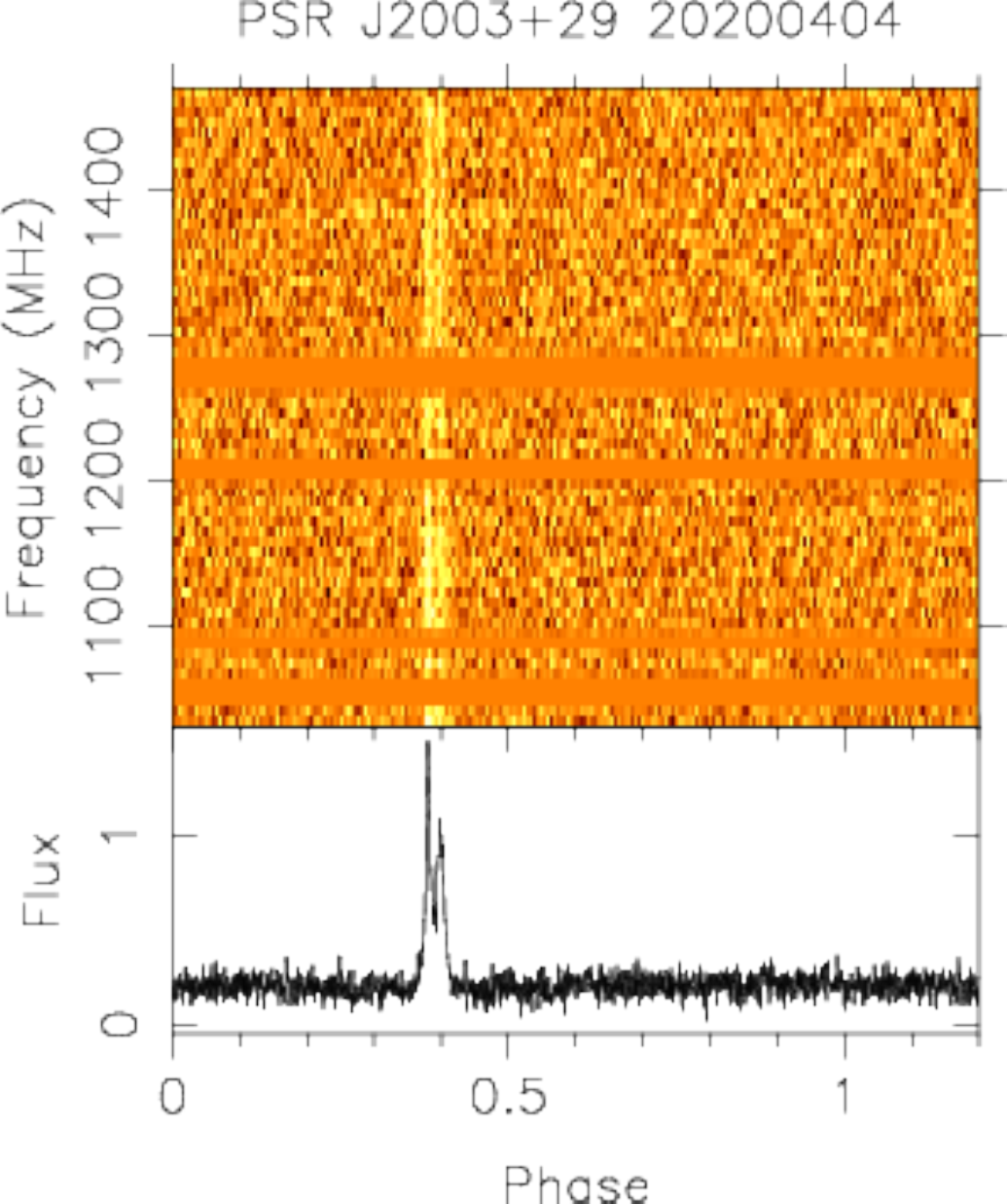}\\[2mm]
\includegraphics[width=38mm]{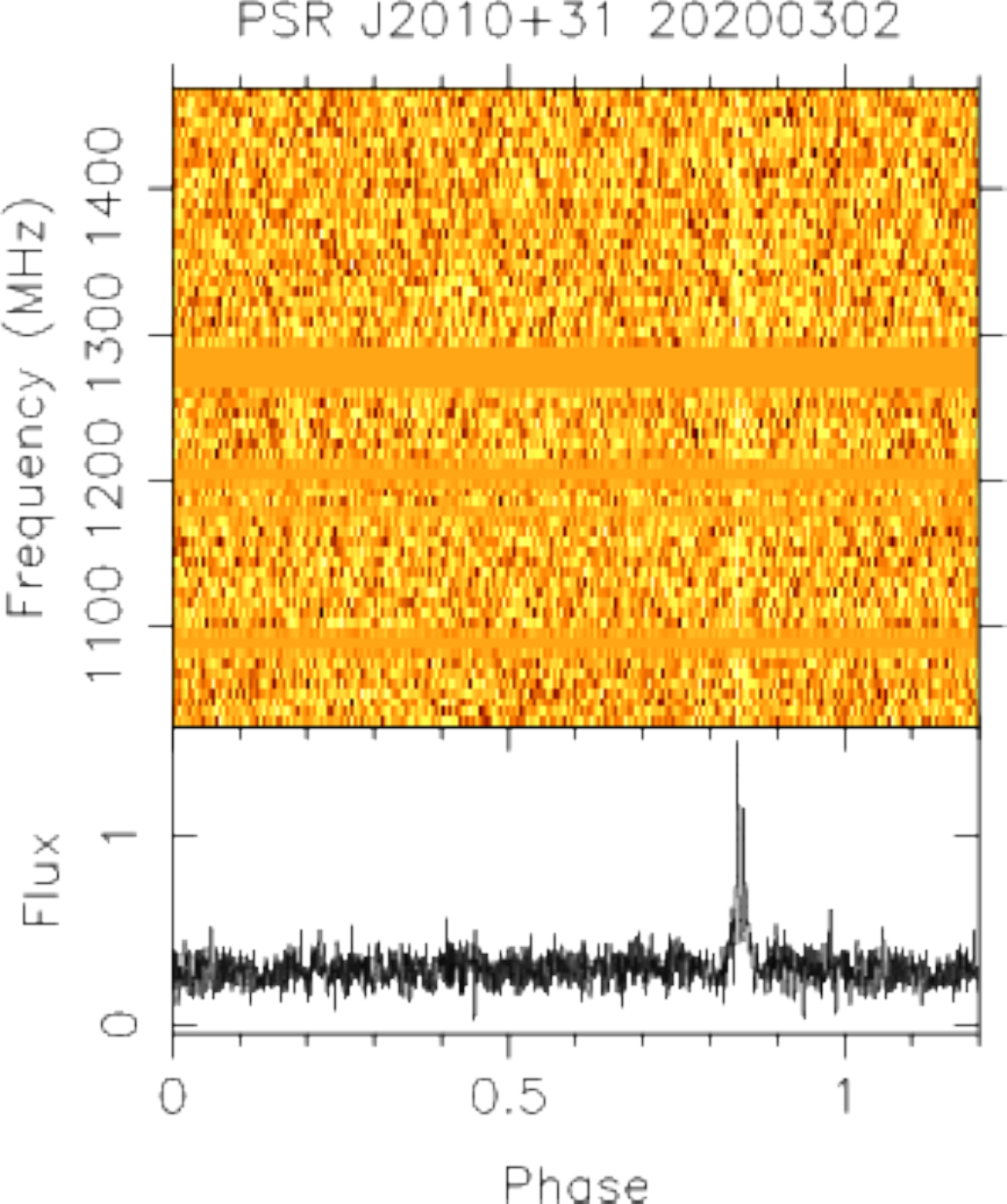}&
\includegraphics[width=38mm]{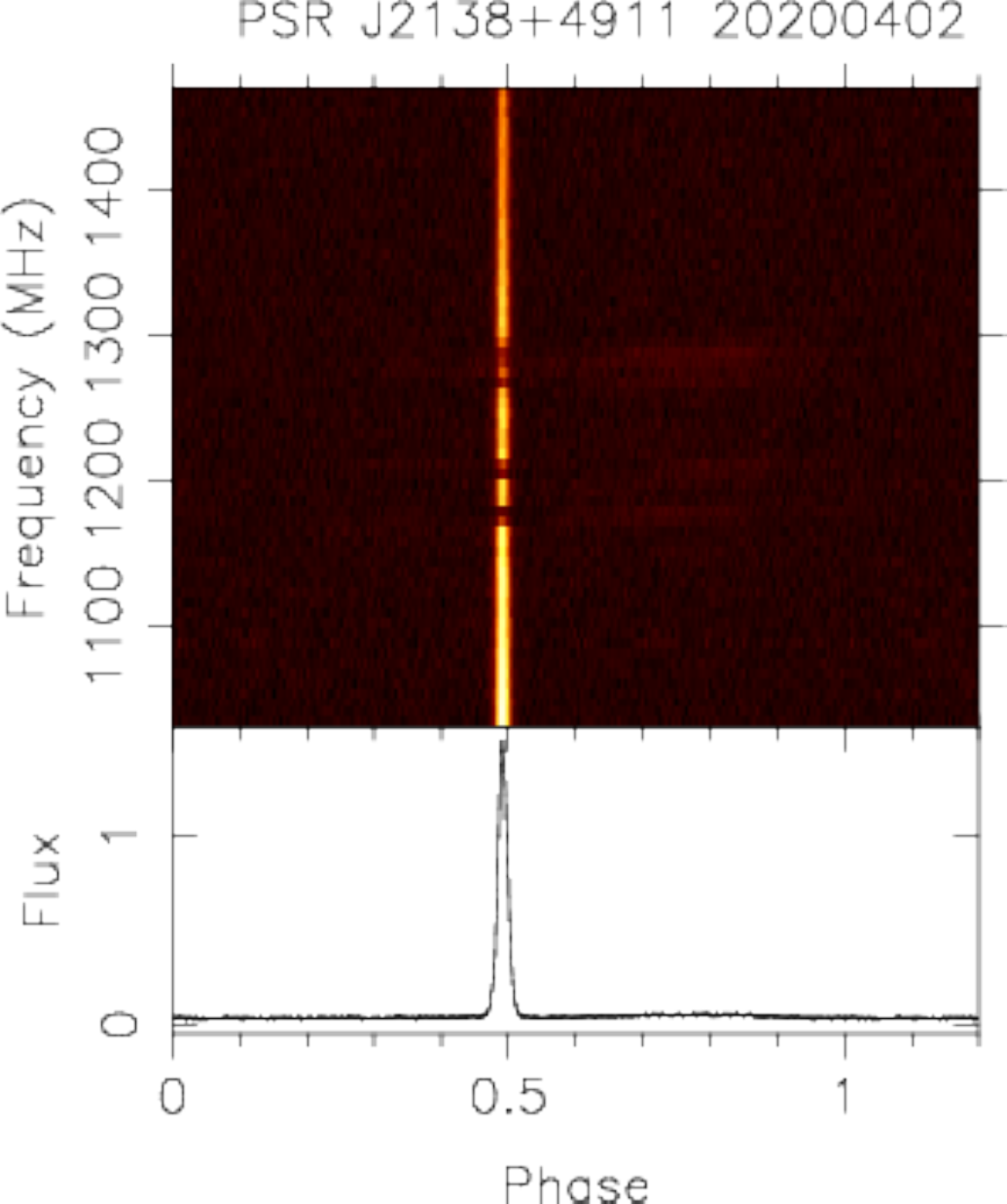}
\end{tabular}%

\begin{minipage}{3cm}
\caption[]{-- {\it Continued}.}\end{minipage}
\addtocounter{figure}{-1}
\end{figure}%

\section{Improved parameters for known pulsars}
\label{Sect5}

In addition to the new discoveries, the GPPS survey detected more than
330 previously known pulsars, and most of them have a very good S/N. A
full list of these pulsars and detailed studies will be published
elsewhere. Here we present the integrated profiles with the
phase-frequency plots (see Fig.~\ref{fig21_knownPSRs}) for only 64
pulsars whose parameters are significantly improved by the GPPS survey
(see Table~\ref{para64psr}). The RFI affected channels have been
removed, as displayed in the phase-frequency plots which in fact
reflects the RFI situation at the FAST site. For pulsars observed at
nighttime, the RFI occupies very few channels and we cannot see the
related influences. Nevertheless, RFI in daytime could occasionally
affect about 20\% of channels or more. The lower DM pulsars are more
easily affected so that the baseline of the profiles cannot be
flattened even after pulses are averaged over all frequency channels.

For many pulsars in Table~\ref{para64psr}, their positions
available in literature or on webpages have large uncertainties, even as
low as 1$^{\circ}$. The snapshot observations that rely on the FAST
L-band 19-beam receiver can detect a known strong pulsar in several
beams, so that the position can be determined with an accuracy of
better than $1'$ according to the positions of beam centers and
signal-to-noise ratios of obtained profiles. See the updated positions
for 45 pulsars in Table~\ref{para64psr}.

Sensitive observations of two pulsars (PSRs J1851+0242 and J1853+0029) acquired by
FAST confirm that their previously available periods are just
harmonics. The corrected periods are expressed in Table~\ref{para64psr}.

An accurate DM can be determined by a higher S/N of profiles. The GPPS
survey improved the DM values for 20 pulsars, as given in
Table~\ref{para64psr}.

In addition to these improved parameters, some remarkable features are
noticed as follows.

(1) Eight low-DM pulsars: PSRs
J0611+1436 (DM = 43.7\,cm$^{-3}${\,}pc),
J1832+0029 (DM = 32.7\,cm$^{-3}${\,}pc),
J1847+01 (DM = 20.1\,cm$^{-3}${\,}pc),
J1910+1256 (DM = 38.1\,cm$^{-3}${\,}pc),
J1918+1541 (DM = 13.0\,cm$^{-3}${\,}pc),
J1926+1613 (DM = 24.5\,cm$^{-3}${\,}pc),
J1938+14 (DM = 75.4\,cm$^{-3}${\,}pc) and
J1939+26 (DM = 47.5\,cm$^{-3}${\,}pc),
manifest obvious scintillation features in their
phase-frequency plots.

(2) Five high DM pulsars: PSRs
  J1849$-$0040 (DM = 1267.6\,cm$^{-3}${\,}pc),
  J1850$-$0006 (DM = 655.0\,cm$^{-3}${\,}pc),
  J1855+0422 (DM = 455.6\,cm$^{-3}${\,}pc),
  J1920+1110 (DM = 188.4\,cm$^{-3}${\,}pc) and
  J1929+1905 (DM = 528.4\,cm$^{-3}${\,}pc),
exhibit scattering features, i.e. emission seen in a
wider phase range towards the lower frequency end.

(3) PSR J1901+0435 has a very large DM ($920.0\pm2.5$\,cm$^{-3}${\,}pc) and
very wide profile, which was discovered in the Parkes multi-beam
survey \citep{lfl+06} and is located behind the Scutum spiral arm of
the Milky Way. The much stronger pulses are seen in the high frequency
end of the band, and much weaker emission in the lower end,
effectively looking like an inverted steep spectrum, which is simply
caused by the scattering effect rather than the intrinsic emission
feature. The scattering caused by ionized gas clouds in the spiral
arms redistribute much more emission at lower frequencies to other
directions, so that the sources or the pulses look weaker and pulsed
emission is much more delayed at lower frequencies, even not appearing as
pulsed emission. In such a case, caution should be taken for the DM
determination due to such scattering. Simple alignment of pulsed emission
peaks in a wide range of frequency would cause an overestimated DM.
In fact, any Galactic pulsars \citep{klm+11,brkl18} or extragalactic
radio sources with such an inverted spectrum at lower frequencies
\citep{gnhb02,gsm+14,mgd+19,mgp+19} can be so explained by the
scattering effect caused by very cloudy ionized gas in front of or in
the environment of a strong emission source. The flux densities and
polarization properties may vary if the source and the intervening clouds
are moving relatively in any transverse direction.

\section{Perspectives}          
\label{sect:fin}

In the Milky Way, over 10\,000 pulsars can potentially be found by upcoming surveys
\citep{lpr+19}, and currently only about 3000 pulsars are known. FAST,
as the most sensitive radio telescope currently in the world, can
survey a limited part of the Milky Way and should be able to discover
about 1000 pulsars, depending on available observation time
\citep{slk+09}. The GPPS survey is so-far the deepest survey for the
FAST-accessible Galactic plane, and the results presented here are the
first of many expected from this highly sensitive system. Among the
list of newly discovered pulsars, about 20\% are MSPs. The most
interesting are binaries awaiting for timing observations and are
valuable for excellent science on tests on theories of gravity. The
long-term timing of more MSPs discovered by the GPPS survey could
enlarge the chance for possible detection of ultra-low-frequency
gravitational waves by the Chinese Pulsar Timing Array.

The GPPS survey provides sensitive observations of newly discovered
and previously known pulsars, and can obtain their polarization
profiles and spectra with excellent quality, which can act as a
fundamental database for pulsar astrophysics, such as exploring the
emission process, emission region and emission mechanisms. For
example, as clearly revealed by FAST observations for pulsars with a
strong scattering effect, their inverted spectrum can be interpreted
due to missing flux densities at lower frequencies caused by the
scattering. Therefore, the results presented in this paper are only
the tip of the iceberg for FAST pulsar observations.

\normalem
\begin{acknowledgements}

We thank Prof. R.T. Gangadhara and the referees, Prof. R.N. Manchester
and Prof. Jim Cordes, for helpful comments.
This project, as one of five key projects, is being carried out by using
FAST, a Chinese national mega-science facility built and operated by
the National Astronomical Observatories, Chinese Academy of Sciences.
J.L. Han is supported by the National Natural Science Foundation of
China (NSFC, Nos. 11988101 and 11833009) and the Key Research
Program of the Chinese Academy of Sciences (Grant
No. QYZDJ-SSW-SLH021);
C. Wang is partially supported by NSFC No. U1731120;
X.Y. Gao is partially supported by NSFC No. U1831103;
P.F. Wang is partially supported by the NSFC No. 11873058 and the National SKA program of China No. 2020SKA0120200.
Jun Xu is partially supported by NSFC No. U2031115;
H.G. Wang  is partially supported by the National SKA program of China (No. 2020SKA0120100).
R. Yuen is partly supported by Xiaofeng Yang's Xinjiang Tianchi Bairen
project and CAS Pioneer Hundred Talents Program.
L.G. Hou thanks the support from the Youth Innovation Promotion
Association CAS.
%
\end{acknowledgements}

\section*{Authors contributions}
The GPPS survey is a key science project of FAST and is being carried
out via teamwork. J.~L. Han initially proposed the survey project
by using the snapshot mode, coordinated the teamwork, then realized
the pipeline for data processing, coordinated the computational resources
and processed most of the data, and finally was in charge of writing this
paper; Chen Wang designed the observation plan, and fed targets for
each observation session and monitored the progress of observations,
tested and realized the snapshot mode together with Jing-Hai Sun,
verified the data integrity, and initialized the data preparing module
with Tao Wang; P.~F. Wang realized the initial PRESTO searching
pipeline, contributed to processing the data, realized the polarization
processing pipeline, obtained the relevant results for this paper and
also contributed to verifying pulsar candidates; Tao Wang and D.~J.
Zhou realized the data preparing module, and also contributed to the
candidates-checking module; Jun Xu realized the Sigproc pipeline;
D.~J. Zhou developed the single pulse module and later was in charge
of processing data for picking up single pulses, contributed to manual
verification of newly discovered pulsars and calculated pulsar flux;
Yi Yan was in charge of checking the detected known pulsars, and
obtained proper positions for previously known pulsars and also newly
discovered pulsars from the signal-to-noise ratios from several nearby
beams, and preparing the relevant results for this paper; Xue Chen was
in charge of checking the RFI of data; Wei-Cong Jing and Wei-Qi Su were
in charge of checking the results for newly discovered pulsars:
Wei-Cong Jing checked the period, Wei-Qi Su folded the pulsar profiles,
and they prepared the relevant results in this paper and also
the webpage; Li-Gang Hou was in charge of checking the beam pattern, and
X.~Y. Gao was in charge of making a log of the observations, and
prepared the plots for coincidence of SNRs and pulsars; Jun Xu also
contributed to analyzing the RMs of pulsars for this paper. K.J. Lee,
N. Wang, P. Jiang, R.~X. Xu and J. Yan jointly planned the project and
coordinated the commissioning, and P. Jiang and Chun Sun coordinated
FAST observations; other people jointly made the formal proposal for
the project and/or ensured the proper operations of observation
systems.

\section*{Data availability}
All pulsar profile data presented in this paper are available on the
webpage of the GPPS survey {\it\url{http://zmtt.bao.ac.cn/GPPS/}}.


\end{document}